\documentclass[12pt, a4paper,twoside,openright]{report}
\pdfoutput=1
\usepackage[utf8]{inputenc}

\usepackage[ngerman, english]{babel}
\usepackage[T1]{fontenc}
\usepackage[titletoc]{appendix}
\usepackage{titletoc}
\usepackage{cite}
\usepackage{graphicx}
\usepackage{subcaption} 
\usepackage{xcolor}
\usepackage{amsmath,amssymb,amsthm}
\usepackage{bm}
\usepackage{float}
\usepackage[format=plain,labelfont={bf,it},textfont=it]{caption}
\usepackage{multirow}
\usepackage{slashed} 
\usepackage{mathtools} 
\usepackage{tabularx}
\usepackage{bbold}
\usepackage{enumitem}
\usepackage{tikz}
\usetikzlibrary{positioning,arrows,calc}
\usetikzlibrary{arrows.meta}
\usetikzlibrary{decorations.pathmorphing}
\usetikzlibrary{decorations.markings}
\usetikzlibrary{shapes.geometric}
\usetikzlibrary{patterns}
\tikzset{
  snake it/.style={
    decorate, 
    decoration=snake,
    segment length=3
  }
}
\usepackage{xparse}
\usepackage{pdfpages} 
\usepackage{titlesec} 
\usepackage{fancyhdr} 
\usepackage{emptypage} 
\usepackage{chngcntr} 
\usepackage{etoolbox} 

\usepackage{dsfont} 
\usepackage[colorlinks = true,linktocpage=true,linkcolor=DarkBlueGrey,citecolor=DarkBlueGrey]{hyperref}
\definecolor{DarkBlueGrey}{RGB}{76,94,107}
\definecolor{MediumBlueGrey}{RGB}{110,135,153}
\definecolor{LightBlueGrey}{RGB}{134,163,184}
\definecolor{WCOrange}{RGB}{242,146,29}
\definecolor{SCRed}{RGB}{179,48,48}
\definecolor{VertexColor}{RGB}{242,146,29}
\definecolor{GluonColor}{RGB}{255,172,172}
\definecolor{SEColor}{RGB}{134,163,184}

\definecolor{BGBox}{RGB}{255,254,230}
\definecolor{PlaneColor}{RGB}{230,230,230}
\definecolor{BlobColor}{RGB}{190,180,230}

\newcommand\frontmatter{%
    \cleardoublepage
  \pagenumbering{roman}}

\newcommand\mainmatter{%
    \cleardoublepage
  \pagenumbering{arabic}}

\newcommand{\newparallel}{{\mkern3mu\vphantom{\perp}\vrule depth 0pt\mkern2mu\vrule depth 0pt\mkern3mu}}

\titlecontents{section}
  [3.2em]
  {\vspace{.20\baselineskip} \normalfont\bfseries}
  {\contentslabel{1.5em}\enspace}
  {}
  {\titlerule*[8pt]{.}\contentspage}
\titlecontents*{subsubsection}
  [7em]
  {\small\itshape}
  {\thecontentslabel}
  {}
  {\hspace{-.1cm}\contentspage}
  [\hspace{.75cm} 
   ]
  []
\makeatletter
\def\@chapter[#1]#2{\ifnum \c@secnumdepth >\m@ne
                       \if@mainmatter
                         \refstepcounter{chapter}%
                         \typeout{\@chapapp\space\thechapter.}%
                         \addcontentsline{toc}{chapter}%
                                   {\protect\numberline{\thechapter}\uppercase{#1}}%
                       \else
                         \addcontentsline{toc}{chapter}{\uppercase{#1}}%
                       \fi
                    \else
                      \addcontentsline{toc}{chapter}{\uppercase{#1}}%
                    \fi
                    \chaptermark{#1}%
                    \addtocontents{lof}{\protect\addvspace{10\p@}}%
                    \addtocontents{lot}{\protect\addvspace{10\p@}}%
                    \if@twocolumn
                      \@topnewpage[\@makechapterhead{#2}]%
                    \else
                      \@makechapterhead{#2}%
                      \@afterheading
                    \fi}
                    
\titleformat{\chapter}[display]
  {\bfseries\Large}
  {\filright\MakeUppercase{\chaptertitlename} \Huge\thechapter}
  {1ex}
  {\titlerule\vspace{1ex}\filleft}
  [\vspace{1ex}\titlerule
  ]
  
\renewcommand{\chaptermark}[1]{\markboth{#1}{}}
\renewcommand{\sectionmark}[1]{\markright{#1}}
\counterwithout{equation}{section} 
\counterwithin{equation}{chapter}  

\newcommand{\pd}{\partial}
\newcommand{\spd}{\slashed{\partial}}
\newcommand{\sx}{\slashed{x}}
\DeclareMathOperator{\sgn}{sgn}
\DeclareMathOperator{\tr}{tr}
\DeclareMathOperator{\Disc}{Disc}
\DeclareMathOperator{\dDisc}{dDisc}
\DeclareMathOperator{\Res}{Res}

\newcommand{\Li}{{\normalfont\text{Li}}}

\newcommand{\vvev}[1]{\langle\!\langle\, #1 \, \rangle\!\rangle}
\newcommand{\vev}[1]{\langle\, #1 \, \rangle}

\newcommand{\chib}{\bar{\chi}}

\newcommand{\wb}{\bar{w}}

\def\veps{\varepsilon}

\newcommand{\Wl}{\mathcal{W}_\ell}
\newcommand{\Wc}{\mathcal{W}_c}
\newcommand{\WC}{\mathcal{W}_C}

\newcommand{\Op}{\mathcal{O}}
\newcommand{\Oh}{\hat{\mathcal{O}}}
\newcommand{\Dh}{\hat{\Delta}}
\newcommand{\uh}{\hat{u}}
\newcommand{\xd}{\dot{x}}

\newcommand{\tlambda}{\tilde{\lambda}}
\newcommand{\tg}{\tilde{g}}

\newcommand\Fb{{\bar{F}}}
\newcommand{\nh}{\hat{n}}
\newcommand{\AdSfive}{AdS$_5 \times S^5$}

\newcommand{\kh}{\hat{k}}

\newcommand{\bt}{\tilde{b}}

\newcommand{\sigmab}{\bar{\sigma}}

\newcommand{\fh}{\hat{f}}
\newcommand{\gammah}{\hat{\gamma}}
\newcommand{\Fh}{\hat{F}}
\newcommand{\ellb}{\bar{\ell}}

\newcommand{\Gh}{\hat{\mathcal{G}}}

\newcommand{\OR}{\sigma}

\def\Am{{\mathcal{A}}}
\def\Bm{{\mathcal{B}}}
\def\Cm{{\mathcal{C}}}
\def\Dm{{\mathcal{D}}}

\def\Fm{{\mathcal{F}}}
\def\Gm{{\mathcal{G}}}

\def\Km{{\mathcal{K}}}
\def\Lm{{\mathcal{L}}}
\def\Mm{{\mathcal{M}}}
\def\Nm{{\mathcal{N}}}
\def\Om{{\mathcal{O}}}
\def\Pm{{\mathcal{P}}}

\def\Dds{{\mathds{D}}}

\def\Fds{{\mathds{F}}}

\def\Ids{{\mathds{I}}}

\def\Zds{{\mathds{Z}}}

\usepackage{bbm}

\def\eps{\epsilon}
\def\veps{\varepsilon}
\newcommand{\Disp}{\mathbb{D}}
\newcommand\psib{{\bar{\psi}}}
\newcommand\cb{{\bar{c}}}
\newcommand\pexp{{\Pm \exp}\ }
\newcommand\dx{{\dot{x}}}

\newcommand\Nf{{N_f}}

\newcommand\zb{{\bar{z}}}

\newif\ifstartcompletesineup
\newif\ifendcompletesineup
\pgfkeys{
    /pgf/decoration/.cd,
    start up/.is if=startcompletesineup,
    start up=true,
    start up/.default=true,
    start down/.style={/pgf/decoration/start up=false},
    end up/.is if=endcompletesineup,
    end up=true,
    end up/.default=true,
    end down/.style={/pgf/decoration/end up=false}
}
\pgfdeclaredecoration{complete sines}{initial}
{
    \state{initial}[
        width=+0pt,
        next state=upsine,
        persistent precomputation={
            \ifstartcompletesineup
                \pgfkeys{/pgf/decoration automaton/next state=upsine}
                \ifendcompletesineup
                    \pgfmathsetmacro\matchinglength{
                        0.5*\pgfdecoratedinputsegmentlength / (ceil(0.5* \pgfdecoratedinputsegmentlength / \pgfdecorationsegmentlength) )
                    }
                \else
                    \pgfmathsetmacro\matchinglength{
                        0.5 * \pgfdecoratedinputsegmentlength / (ceil(0.5 * \pgfdecoratedinputsegmentlength / \pgfdecorationsegmentlength ) - 0.499)
                    }
                \fi
            \else
                \pgfkeys{/pgf/decoration automaton/next state=downsine}
                \ifendcompletesineup
                    \pgfmathsetmacro\matchinglength{
                        0.5* \pgfdecoratedinputsegmentlength / (ceil(0.5 * \pgfdecoratedinputsegmentlength / \pgfdecorationsegmentlength ) - 0.4999)
                    }
                \else
                    \pgfmathsetmacro\matchinglength{
                        0.5 * \pgfdecoratedinputsegmentlength / (ceil(0.5 * \pgfdecoratedinputsegmentlength / \pgfdecorationsegmentlength ) )
                    }
                \fi
            \fi
            \setlength{\pgfdecorationsegmentlength}{\matchinglength pt}
        }] {}
    \state{downsine}[width=\pgfdecorationsegmentlength,next state=upsine]{
        \pgfpathsine{\pgfpoint{0.5\pgfdecorationsegmentlength}{0.5\pgfdecorationsegmentamplitude}}
        \pgfpathcosine{\pgfpoint{0.5\pgfdecorationsegmentlength}{-0.5\pgfdecorationsegmentamplitude}}
    }
    \state{upsine}[width=\pgfdecorationsegmentlength,next state=downsine]{
        \pgfpathsine{\pgfpoint{0.5\pgfdecorationsegmentlength}{-0.5\pgfdecorationsegmentamplitude}}
        \pgfpathcosine{\pgfpoint{0.5\pgfdecorationsegmentlength}{0.5\pgfdecorationsegmentamplitude}}
}
    \state{final}{}
}

\tikzset{
corner/.style={line width=1pt,dashed,draw=black,dash pattern=on 6pt off 4pt},
scalar/.style={line width=1pt,draw=black},
gluon/.style={line width=1pt,decorate, draw=GluonColor,
    decoration={complete sines,aspect=0,amplitude=1.25mm,segment length=1.5mm,start up,end up}},
gluontwo/.style={line width=1pt,decorate, draw=GluonColor,
    decoration={complete sines,aspect=0,amplitude=.7mm,segment length=1mm,start up,end up}},
ghost/.style={line width=1pt,loosely dotted,draw=black},
wilson/.style={line width=2pt,draw=black},
 }
\NewDocumentCommand\semiloop{O{black}mmmO{}O{above}}
{%
\draw[#1] let \p1 = ($(#3)-(#2)$) in (#3) arc (#4:({#4+180}):({0.5*veclen(\x1,\y1)})node[midway, #6] {#5};)
}

\newcommand\x{.75}

\newcommand\vertexsize{.07}
\newcommand\scaleCC{1.3}

\newcommand{\TreeLevel}[1]{\draw[scalar] (#1,.2) circle (.2);
\draw[fill=white,white] (#1-.22,.22) rectangle (#1+.22,-.02);
\draw[scalar] (#1-.2,.22) -- (#1-.2,0);
\draw[scalar] (#1+.2,.22) -- (#1+.2,0);
\node[] at (#1,.2) {\footnotesize $t$};}
\newcommand{\SinglePropagator}[1]{\draw[fill=white,scalar] (#1,.2) circle (.5);
\draw[fill=white,white] (#1-.52,.22) rectangle (#1+.52,-.02);
\draw[scalar] (#1-.5,.22) -- (#1-.5,0);
\draw[scalar] (#1+.5,.22) -- (#1+.5,0);}
\newcommand{\SinglePropagatorSmall}[1]{\draw[fill=white,scalar] (#1,.2) circle (.4);
\draw[fill=white,white] (#1-.42,.22) rectangle (#1+.42,-.02);
\draw[scalar] (#1-.4,.22) -- (#1-.4,0);
\draw[scalar] (#1+.4,.22) -- (#1+.4,0);}
\newcommand{\SinglePropagatorBig}[1]{\draw[fill=white,scalar] (#1,-.3) circle (1.05);
}

\newcommand{\ExampleLineDefect}{\begin{tikzpicture}[baseline={([yshift=0ex]current bounding box.center)},scale=.95]
\draw[scalar] (0,0) -- (5,0);
\draw[scalar] (0,0) -- (0,2.5);
\draw[scalar] (0,0) -- (1,1.5);
\draw[-Latex] (4.5,0)--(5.1,0);
\draw[-Latex] (0,2)--(0,2.6);
\draw[-Latex] (.44,.675)--(1.04,1.575);
\node[] at (5.3,0) {\footnotesize $x$};
\node[] at (1.15,1.75) {\footnotesize $y$};
\node[] at (0,2.8) {\footnotesize $\tau$};
\draw[scalar,LightBlueGrey] (0+.7,0+.4) -- (5,0+.4);
\draw[scalar,LightBlueGrey] (0+.7+.25,0+.375+.4) -- (5+.25,0+.375+.4);
\draw[scalar,LightBlueGrey] (0+.7+.5,0+.75+.4) -- (5+.5,0+.75+.4);
\draw[scalar,LightBlueGrey] (0+.7+.75,0+.75+.375+.4) -- (5+.75,0+.75+.375+.4);
\draw[scalar,LightBlueGrey] (0+.7+1,1.5+.4) -- (5+1,1.5+.4);
\draw[scalar,LightBlueGrey] (0+.7,0+.4) -- (1+.7,1.5+.4);
\draw[scalar,LightBlueGrey] (0+.7+1.075,0+.4) -- (1+.7+1.075,1.5+.4);
\draw[scalar,LightBlueGrey] (0+.7+2.15,0+.4) -- (1+.7+2.15,1.5+.4);
\draw[scalar,LightBlueGrey] (0+.7+2.15+1.075,0+.4) -- (1+.7+2.15+1.075,1.5+.4);
\draw[scalar,LightBlueGrey] (0+.7+4.3,0+.4) -- (1+.7+4.3,1.5+.4);
\draw[wilson,SCRed,fill=SCRed,opacity=.75] (0+.7+.5+8+2.15,0+.75+.4) -- (0+.7+.5+8+2.15,0+.75+.4-1);
\draw[scalar] (0+8,0) -- (5+8,0);
\draw[scalar] (0+8,0) -- (0+8,2.5);
\draw[scalar] (0+8,0) -- (1+8,1.5);
\draw[-Latex] (4.5+8,0)--(5.1+8,0);
\draw[-Latex] (0+8,2)--(0+8,2.6);
\draw[-Latex] (.44+8,.675)--(1.04+8,1.575);
\node[] at (5.3+8,0) {\footnotesize $x$};
\node[] at (1.15+8,1.75) {\footnotesize $y$};
\node[] at (0+8,2.8) {\footnotesize $\tau$};
\draw[scalar,LightBlueGrey] (0+.7+8,0+.4) -- (5+8,0+.4);
\draw[scalar,LightBlueGrey] (0+.7+.25+8,0+.375+.4) -- (5+8+.25,0+.375+.4);
\draw[scalar,LightBlueGrey] (0+.7+.5+8,0+.75+.4) -- (5+8+.5,0+.75+.4);
\draw[scalar,LightBlueGrey] (0+.7+.75+8,0+.75+.375+.4) -- (5+8+.75,0+.75+.375+.4);
\draw[scalar,LightBlueGrey] (0+.7+1+8,1.5+.4) -- (5+8+1,1.5+.4);
\draw[scalar,LightBlueGrey] (0+.7+8,0+.4) -- (1+8+.7,1.5+.4);
\draw[scalar,LightBlueGrey] (0+.7+8+1.075,0+.4) -- (1+8+.7+1.075,1.5+.4);
\draw[scalar,LightBlueGrey] (0+.7+8+2.15,0+.4) -- (1+8+.7+2.15,1.5+.4);
\draw[scalar,LightBlueGrey] (0+.7+8+2.15+1.075,0+.4) -- (1+8+.7+2.15+1.075,1.5+.4);
\draw[scalar,LightBlueGrey] (0+.7+8+4.3,0+.4) -- (1+8+.7+4.3,1.5+.4);
\draw[SCRed,fill=SCRed] (0+.7+.5+8+2.15,0+.75+.4) circle (.075);
\draw[wilson,SCRed,fill=SCRed] (0+.7+.5+8+2.15,0+.75+.4) -- (0+.7+.5+8+2.15,0+.75+.4+1+.25);
\end{tikzpicture}}

\newcommand{\TwoPointFunctionSetup}{\begin{tikzpicture}[baseline={([yshift=-10ex]current bounding box.center)},scale=1.3]
\draw[scalar,fill=PlaneColor,color=PlaneColor] (0,2) -- (2,1) -- (4,2) -- (2,3) -- cycle ;
\draw[wilson] (0,0) -- (0,4);
\draw[scalar] (0,2) -- (3,.5);
\draw[-Latex] (2.5,.75)--(3.1,.45);
\draw[scalar] (0,2) -- (3,3.5);
\draw[-Latex] (2.5,3.25)--(3.1,3.55);
\draw[ghost] (1.35,2.7) -- (2.25,2.25);
\draw[ghost] (.85,1.55) -- (2.25,2.25);
\draw[color=DarkBlueGrey,fill=DarkBlueGrey,line width=1pt] (2,1) circle (.1);
\draw[color=DarkBlueGrey,fill=DarkBlueGrey,line width=1pt] (2.25,2.25) circle (.1);
\node[] at (3.35,.4) {\footnotesize $x$};
\node[] at (3.35,3.6) {\footnotesize $y$};
\node[color=DarkBlueGrey] at (1.2,.75) {\footnotesize $\mathcal{O}(x_1)$};
\node[color=DarkBlueGrey] at (3,2) {\footnotesize $\mathcal{O}(x_2)$};
\end{tikzpicture}}

\newcommand{\BulkCrossing}{\begin{tikzpicture}[baseline={([yshift=0ex]current bounding box.center)},scale=1.25]
\draw[wilson,SCRed,fill=SCRed] (.5,1) -- (1,1);
\draw[scalar] (0,.5) -- (.5,1);
\draw[scalar] (0,1.5) -- (.5,1);
\draw[scalar] (1.5,.5) -- (1,1);
\draw[scalar] (1.5,1.5) -- (1,1);
\draw[wilson,SCRed,fill=SCRed] (.5+2.5,1.25) -- (.5+2.5,.75);
\draw[scalar] (0+2.5,1.5+.25) -- (.5+2.5,1+.25);
\draw[scalar] (1+2.5,1.5+.25) -- (.5+2.5,1+.25);
\draw[scalar] (.5+2.5,.5+.25) -- (0+2.5,0+.25);
\draw[scalar] (.5+2.5,.5+.25) -- (1+2.5,0+.25);
\node[] at (2,.95) {\Large $=$};
\end{tikzpicture}}

\newcommand{\DefectCrossing}{\begin{tikzpicture}[baseline={([yshift=0ex]current bounding box.center)},scale=1.25]
\draw[wilson,SCRed,fill=SCRed] (.5,1) -- (1,1);
\draw[wilson] (1,.5) -- (1,1.5);
\draw[scalar] (0,.5) -- (.5,1);
\draw[scalar] (0,1.5) -- (.5,1);
\draw[scalar,white] (2,1.25) rectangle (3,.5);
\draw[wilson] (.75+2,.5) -- (.75+2,1.5););
\draw[wilson,SCRed,fill=SCRed] (.75+2,.75) -- (.75+2,1.25);
\draw[scalar] (0+2,.5) -- (.75+2,.75);
\draw[scalar] (0+2,1.5) -- (.75+2,1.25);
\node[] at (1.5,.95) {\Large $=$};
\end{tikzpicture}}

\newcommand{\AharonovBohm}{\begin{tikzpicture}[baseline={([yshift=0ex]current bounding box.center)},scale=.8]
\draw[wilson,LightBlueGrey] (2,-2.2) -- (2,1.45);
\draw[-{Latex[length=9,width=6]},LightBlueGrey] (2,1.325+.4)--(2,1.375+.4);
\node[MediumBlueGrey] at (2.75,1) {\large $\vec{B}$};
\draw[scalar] (0,0) -- (1,-.25);
\draw[-Latex] (.5,-.125) -- (.6,-.125-.025);
\draw[scalar] (3,-.75) -- (4,-1);
\draw[-Latex] (3.5,-.75-.125) -- (3.6,-.75-.125-.025);
\draw[scalar] (1,-.25) to[out=20,in=120] (3,-.75);
\draw[-Latex] (1.6,-.78) -- (1.7,-.79-.025);
\draw[-Latex] (2.3,-.18) -- (2.4,-.18-.025);
\node[cylinder, draw=DarkBlueGrey,cylinder uses custom fill, cylinder body fill=MediumBlueGrey,cylinder end fill=LightBlueGrey,minimum width=15, minimum height=75,shape border rotate=90,opacity=.4] (c) at (2,-.5) {};
\draw[scalar] (1,-.25) to[out=-60,in=-160] (3,-.75);
\node[] at (3.25+.05+.1,-.75-.125+.4) {\footnotesize $\Delta\varphi$};
\end{tikzpicture}}

\newcommand{\Branes}{\begin{tikzpicture}[baseline={([yshift=0ex]current bounding box.center)},scale=.7]
\draw[scalar,white] (0,2) rectangle (5,0);
\draw[scalar,DarkBlueGrey, fill=LightBlueGrey,opacity=.35] (0,0) -- (2,1) -- (2,-1) -- (0,-2) -- (0,0);
\draw[scalar,DarkBlueGrey, fill=LightBlueGrey,opacity=.35] (0+.5,0) -- (2+.5,1) -- (2+.5,-1) -- (0+.5,-2) -- (0+.5,0);
\draw[scalar,DarkBlueGrey, fill=LightBlueGrey,opacity=.35] (0+1,0) -- (2+1,1) -- (2+1,-1) -- (0+1,-2) -- (0+1,0);
\draw[scalar,DarkBlueGrey, fill=LightBlueGrey,opacity=.35] (0+1.5,0) -- (2+1.5,1) -- (2+1.5,-1) -- (0+1.5,-2) -- (0+1.5,0);
\draw[SCRed,fill=SCRed] (1+1.5,-.5) circle (.075);
\draw[scalar,DarkBlueGrey, fill=LightBlueGrey,opacity=.35] (0+5,0) -- (2+5,1) -- (2+5,-1) -- (0+5,-2) -- (0+5,0);
\draw[SCRed,fill=SCRed] (1+5,-.5) circle (.075);
\draw[wilson,SCRed,opacity=.5] (2.5,-.5) to[out=10,in=140] (2.5+1.75,-.75) to[out=-50,in=180] (6,-.5);
\node[MediumBlueGrey] at (1.75,-2.5) {\large $N$};
\node[MediumBlueGrey] at (6,-2.5) {\large $N+1$};
\end{tikzpicture}}

\newcommand{\Worldsheet}{\begin{tikzpicture}[baseline={([yshift=0ex]current bounding box.center)}]
\draw[scalar,DarkBlueGrey, fill=LightBlueGrey,opacity=.35] (0,0) -- (1,1) -- (1,-1) -- (0,-2) -- (0,0);
\draw[wilson,SCRed] (.5,-.5) ellipse (.175 and .5);
\draw[wilson,SCRed,opacity=.5] (.5,0) to[out=0,in=90] (1.5,-.5) to[out=-90,in=0] (.5,-1);
\node[SCRed] at (.25,-1.25) {\footnotesize $\Cm$};
\end{tikzpicture}}

\newcommand{\MultipointCorrelators}{\begin{tikzpicture}[baseline={([yshift=-2ex]current bounding box.center)}]
\draw[wilson] (-5,0) -- (3,0);
\draw[fill=white] (-4,0) circle (.1);
\draw[fill=white] (-2.5,0) circle (.1);
\draw[fill=white] (0.5,0) circle (.1);
\draw[fill=white] (2,0) circle (.1);
\node[] at (-1,.4) {$\dots$};
\node[] at (-4,.4) {$\tau_1$};
\node[] at (-2.5,.4) {$\tau_2$};
\node[] at (0.5,.4) {$\tau_{n-1}$};
\node[] at (2,.4) {$\tau_{n}$};
\node[] at (-4,-.6) {$\Oh_{\Dh_1}$};
\node[] at (-2.5,-.6)  {$\Oh_{\Dh_2}$};
\node[] at (0.5,-.6)  {$\Oh_{\Dh_{n-1}}$};
\node[] at (2,-.6)  {$\Oh_{\Dh_n}$};
\node[] at (3.5,0) {$\Wl$};
\end{tikzpicture}}

\newcommand{\SingleTraceHalfBPS}{\begin{tikzpicture}[baseline={([yshift=-.85ex]current bounding box.center)}]
\draw[fill=white] (0,0) circle (.18);
\draw[fill=white] (-.13,-.1) circle (.075);
\draw[fill=white] (.13,-.1) circle (.075);
\end{tikzpicture}}

\newcommand{\ScalarPropagator}{\begin{tikzpicture}[baseline={([yshift=-.85ex]current bounding box.center)}]
\draw[scalar] (0,.5) -- (1,.5);
\draw[fill=white] (0,.5) circle (.075);
\draw[fill=white] (1,.5) circle (.075);
\node[] at (0,.1) {\footnotesize \makebox[\widthof{$\smash{\mu,a}$}][c]{$\smash{I,a}$}};
\node[] at (0,.9) {\footnotesize $1$};
\node[] at (1,.1) {\footnotesize \makebox[\widthof{$\smash{\mu,a}$}][c]{$\smash{J,b}$}};
\node[] at (1,.9) {\footnotesize $2$};
\end{tikzpicture}}
\newcommand{\ScalarPropagatorStrong}{\begin{tikzpicture}[baseline={([yshift=-.4ex]current bounding box.center)}]
\draw[scalar] (0,.5) -- (1,.5);
\draw[fill=white] (0,.5) circle (.075);
\draw[fill=white] (1,.5) circle (.075);
\node[] at (0,-.025) {\footnotesize \makebox[\widthof{$\smash{\mu}$}][c]{$\smash{I}$}};
\node[] at (0,.9) {\footnotesize $1$};
\node[] at (1,-.025) {\footnotesize \makebox[\widthof{$\smash{\mu}$}][c]{$\smash{J}$}};
\node[] at (1,.9) {\footnotesize $2$};
\end{tikzpicture}}
\newcommand{\ScalarPropagatorY}{\begin{tikzpicture}[baseline={([yshift=-.5ex]current bounding box.center)}]
\draw[scalar] (0,.5) -- (1,.5);
\draw[fill=white] (0,.5) circle (.075);
\draw[fill=white] (1,.5) circle (.075);
\node[] at (0,0) {\footnotesize \makebox[\widthof{$\smash{I}$}][c]{$\smash{I}$}};
\node[] at (0,.9) {\footnotesize $1$};
\node[] at (1,0) {\footnotesize \makebox[\widthof{$\smash{I}$}][c]{$\smash{J}$}};
\node[] at (1,.9) {\footnotesize $2$};
\end{tikzpicture}}
\newcommand{\GluonPropagator}{\begin{tikzpicture}[baseline={([yshift=-.85ex]current bounding box.center)}]
\draw[gluon] (0,.5) -- (1,.5);
\draw[fill=white] (0,.5) circle (.075);
\draw[fill=white] (1,.5) circle (.075);
\node[] at (0,.1) {\footnotesize \makebox[\widthof{$\mu,a$}][c]{$\smash{\mu,a}$}};
\node[] at (0,.9) {\footnotesize $1$};
\node[] at (1,.1) {\footnotesize \makebox[\widthof{$\nu,b$}][c]{$\smash{\nu,b}$}};
\node[] at (1,.9) {\footnotesize $2$};
\end{tikzpicture}}
\newcommand{\FermionPropagator}{\begin{tikzpicture}[baseline={([yshift=-.85ex]current bounding box.center)}]
\draw[scalar] (0,.5) -- (1,.5);
\draw[fill=white] (0,.5) circle (.075);
\draw[fill=white] (1,.5) circle (.075);
\draw[-Latex] (.62,.5)--(.63,.5);
\node[] at (0,.1) {\footnotesize \makebox[\widthof{$\smash{\mu,a}$}][c]{$\smash{a}$}};
\node[] at (0,.9) {\footnotesize $1$};
\node[] at (1,.1) {\footnotesize \makebox[\widthof{$\smash{\nu,b}$}][c]{$\smash{b}$}};
\node[] at (1,.9) {\footnotesize $2$};
\end{tikzpicture}}
\newcommand{\FermionPropagatorY}{\begin{tikzpicture}[baseline={([yshift=-.75ex]current bounding box.center)}]
\draw[scalar] (0,.5) -- (1,.5);
\draw[fill=white] (0,.5) circle (.075);
\draw[fill=white] (1,.5) circle (.075);
\draw[-Latex] (.62,.5)--(.63,.5);
\node[] at (0,.1) {\footnotesize \makebox[\widthof{$\smash{a}$}][c]{$\smash{a}$}};
\node[] at (0,.9) {\footnotesize $1$};
\node[] at (1,.1) {\footnotesize \makebox[\widthof{$\smash{a}$}][c]{$\smash{b}$}};
\node[] at (1,.9) {\footnotesize $2$};
\end{tikzpicture}}
\newcommand{\GhostPropagator}{\begin{tikzpicture}[baseline={([yshift=-.85ex]current bounding box.center)}]
\draw[ghost] (0,.5) -- (1,.5);
\draw[fill=white] (0,.5) circle (.075);
\draw[fill=white] (1,.5) circle (.075);
\node[] at (0,.1) {\footnotesize \makebox[\widthof{$\smash{\mu,a}$}][c]{$\smash{a}$}};
\node[] at (0,.9) {\footnotesize $1$};
\node[] at (1,.1) {\footnotesize \makebox[\widthof{$\smash{\nu,b}$}][c]{$\smash{b}$}};
\node[] at (1,.9) {\footnotesize $2$};
\end{tikzpicture}}

\newcommand{\VertexScalarScalarGluon}{\begin{tikzpicture}[baseline={([yshift=-.35ex]current bounding box.center)}]
\draw[gluon] (0,.5) -- (.75,.5);
\draw[scalar] (.75,.5) -- (1.25,1);
\draw[scalar] (.75,.5) -- (1.25,0);
\draw[fill=VertexColor,VertexColor] (.75,.5) circle (\vertexsize);
\draw[fill=white] (0,.5) circle (.075);
\draw[fill=white] (1.25,1) circle (.075);
\draw[fill=white] (1.25,0) circle (.075);
\end{tikzpicture}}
\newcommand{\VertexFermionFermionScalar}{\begin{tikzpicture}[baseline={([yshift=-.45ex]current bounding box.center)}]
\draw[scalar] (0,.5) -- (.75,.5);
\draw[scalar] (.75,.5) -- (1.25,1);
\draw[-Latex] (1.05,.8)--(.95,.7);
\draw[scalar] (.75,.5) -- (1.25,0);
\draw[-Latex] (1,.25)--(1.1,.15);
\draw[fill=VertexColor,VertexColor] (.75,.5) circle (\vertexsize);
\draw[fill=white] (0,.5) circle (.075);
\draw[fill=white] (1.25,1) circle (.075);
\draw[fill=white] (1.25,0) circle (.075);
\end{tikzpicture}}
\newcommand{\VertexFermionFermionScalarY}{\begin{tikzpicture}[baseline={([yshift=-.35ex]current bounding box.center)}]
\draw[scalar] (0,.5) -- (.75,.5);
\draw[scalar] (.75,.5) -- (1.25,1);
\draw[-Latex] (1.05,.8)--(.95,.7);
\draw[scalar] (.75,.5) -- (1.25,0);
\draw[-Latex] (1,.25)--(1.1,.15);
\draw[fill=VertexColor,VertexColor] (.75,.5) circle (\vertexsize);
\draw[fill=white] (0,.5) circle (.075);
\draw[fill=white] (1.25,1) circle (.075);
\draw[fill=white] (1.25,0) circle (.075);
\node[anchor=west] at (1.3,1) {\footnotesize $a$};
\node[] at (1.25,1.3) {\footnotesize $1$};
\node[anchor=west] at (1.3,0) {\footnotesize $b$};
\node[] at (1.25,-.3) {\footnotesize $2$};
\node[] at (0,.1) {\footnotesize $I$};
\node[] at (0,.9) {\footnotesize $3$};
\end{tikzpicture}}
\newcommand{\VertexFermionFermionGluon}{\begin{tikzpicture}[baseline={([yshift=-.35ex]current bounding box.center)}]
\draw[gluon] (0,.5) -- (.75,.5);
\draw[scalar] (.75,.5) -- (1.25,1);
\draw[-Latex] (1.05,.8)--(.95,.7);
\draw[scalar] (.75,.5) -- (1.25,0);
\draw[-Latex] (1,.25)--(1.1,.15);
\draw[fill=VertexColor,VertexColor] (.75,.5) circle (\vertexsize);
\draw[fill=white] (0,.5) circle (.075);
\draw[fill=white] (1.25,1) circle (.075);
\draw[fill=white] (1.25,0) circle (.075);
\end{tikzpicture}}
\newcommand{\VertexGhostGhostGluon}{\begin{tikzpicture}[baseline={([yshift=-.35ex]current bounding box.center)}]
\draw[gluon] (0,.5) -- (.75,.5);
\draw[ghost] (.75,.5) -- (1.25,1);
\draw[ghost] (.75,.5) -- (1.25,0);
\draw[fill=VertexColor,VertexColor] (.75,.5) circle (\vertexsize);
\draw[fill=white] (0,.5) circle (.075);
\draw[fill=white] (1.25,1) circle (.075);
\draw[fill=white] (1.25,0) circle (.075);
\end{tikzpicture}}
\newcommand{\VertexGluonGluonGluon}{\begin{tikzpicture}[baseline={([yshift=-.35ex]current bounding box.center)}]
\draw[gluon] (0,.5) -- (.75,.5);
\draw[gluon] (.75,.5) -- (1.25,1);
\draw[gluon] (.75,.5) -- (1.25,0);
\draw[fill=VertexColor,VertexColor] (.75,.5) circle (\vertexsize);
\draw[fill=white] (0,.5) circle (.075);
\draw[fill=white] (1.25,1) circle (.075);
\draw[fill=white] (1.25,0) circle (.075);
\end{tikzpicture}}
\newcommand{\VertexFourScalars}{\begin{tikzpicture}[baseline={([yshift=-.35ex]current bounding box.center)}]
\draw[scalar] (0,0) -- (1.25,1);
\draw[scalar] (0,1) -- (1.25,0);
\draw[fill=VertexColor,VertexColor] (.625,.5) circle (\vertexsize);
\draw[fill=white] (0,0) circle (.075);
\draw[fill=white] (0,1) circle (.075);
\draw[fill=white] (1.25,1) circle (.075);
\draw[fill=white] (1.25,0) circle (.075);
\end{tikzpicture}}
\newcommand{\VertexFourScalarsY}{\begin{tikzpicture}[baseline={([yshift=-.35ex]current bounding box.center)}]
\draw[scalar] (0,0) -- (1.25,1);
\draw[scalar] (0,1) -- (1.25,0);
\draw[fill=VertexColor,VertexColor] (.625,.5) circle (\vertexsize);
\draw[fill=white] (0,0) circle (.075);
\draw[fill=white] (0,1) circle (.075);
\draw[fill=white] (1.25,1) circle (.075);
\draw[fill=white] (1.25,0) circle (.075);
\node[anchor=west] at (1.3,1) {\footnotesize $I$};
\node[] at (1.25,1.3) {\footnotesize $1$};
\node[anchor=west] at (1.3,0) {\footnotesize $J$};
\node[] at (1.25,-.3) {\footnotesize $2$};
\node[anchor=east] at (-0.05,1) {\footnotesize $K$};
\node[] at (0,1.3) {\footnotesize $3$};
\node[anchor=east] at (-0.05,0) {\footnotesize $L$};
\node[] at (0,-.3) {\footnotesize $4$};
\end{tikzpicture}}

\newcommand{\VertexScalarScalarGluonGluon}{\begin{tikzpicture}[baseline={([yshift=-.35ex]current bounding box.center)}]
\draw[scalar] (.625,.5) -- (1.25,1);
\draw[scalar] (.625,.5) -- (1.25,0);
\draw[gluon] (0,0) -- (.625,.5);
\draw[gluon] (0,1) -- (.625,.5);
\draw[fill=VertexColor,VertexColor] (.625,.5) circle (\vertexsize);
\draw[fill=white] (0,0) circle (.075);
\draw[fill=white] (0,1) circle (.075);
\draw[fill=white] (1.25,1) circle (.075);
\draw[fill=white] (1.25,0) circle (.075);
\end{tikzpicture}}
\newcommand{\VertexGluonGluonGluonGluon}{\begin{tikzpicture}[baseline={([yshift=-.35ex]current bounding box.center)}]
\draw[gluon] (.625,.5) -- (1.25,1);
\draw[gluon] (.625,.5) -- (1.25,0);
\draw[gluon] (0,0) -- (.625,.5);
\draw[gluon] (0,1) -- (.625,.5);
\draw[fill=VertexColor,VertexColor] (.625,.5) circle (\vertexsize);
\draw[fill=white] (0,0) circle (.075);
\draw[fill=white] (0,1) circle (.075);
\draw[fill=white] (1.25,1) circle (.075);
\draw[fill=white] (1.25,0) circle (.075);
\end{tikzpicture}}

\newcommand{\SelfEnergyNoText}{\begin{tikzpicture}[baseline={([yshift=-.35ex]current bounding box.center)}]
\draw[scalar] (0,.5) -- (1,.5);
\draw[fill=white] (0,.5) circle (.075);
\draw[fill=white] (1,.5) circle (.075);
\draw[fill=SEColor,SEColor] (.5,.5) circle (.15);
\end{tikzpicture}}
\newcommand{\SelfEnergyDiagramOne}{\begin{tikzpicture}[baseline={([yshift=-.4ex]current bounding box.center)}]
\draw[scalar] (0,.5) -- (2,.5);
\draw[gluon] (.5,.5) arc (180:-20:.5);
\draw[fill=white] (0,.5) circle (.075);
\draw[fill=white] (2,.5) circle (.075);
\draw[fill=VertexColor,VertexColor] (.5,.5) circle (\vertexsize);
\draw[fill=VertexColor,VertexColor] (1.5,.5) circle (\vertexsize);
\end{tikzpicture}}
\newcommand{\SelfEnergyDiagramTwo}{\begin{tikzpicture}[baseline={([yshift=-.5ex]current bounding box.center)}]
\draw[scalar] (0,.5) -- (.5,.5);
\draw[scalar] (1.5,.5) -- (2,.5);
\draw[scalar,postaction={decorate}] (.5,.5) arc (180:0:.5);
\draw[-Latex] (1.1,1)--(1.11,1);
\draw[scalar,postaction={decorate}] (.5,.5) arc (-180:0:.5);
\draw[-Latex] (.9,0)--(.89,0);
\draw[fill=white] (0,.5) circle (.075);
\draw[fill=white] (2,.5) circle (.075);
\draw[fill=VertexColor,VertexColor] (.5,.5) circle (\vertexsize);
\draw[fill=VertexColor,VertexColor] (1.5,.5) circle (\vertexsize);
\end{tikzpicture}}
\newcommand{\SelfEnergyDiagramThree}{\begin{tikzpicture}[baseline={([yshift=-.75ex]current bounding box.center)}]
\draw[scalar] (0,.5) -- (2,.5);
\draw[scalar] (1,1) circle (.5);
\draw[fill=white] (0,.5) circle (.075);
\draw[fill=white] (2,.5) circle (.075);
\draw[fill=VertexColor,VertexColor] (1,.5) circle (\vertexsize);
\end{tikzpicture}}
\newcommand{\SelfEnergyDiagramFour}{\begin{tikzpicture}[baseline={([yshift=-1.05ex]current bounding box.center)}]
\draw[scalar] (0,.5) -- (2,.5);
\draw[gluon] (1,1) circle (.5);
\draw[fill=white] (0,.5) circle (.075);
\draw[fill=white] (2,.5) circle (.075);
\draw[fill=VertexColor,VertexColor] (1,.5) circle (\vertexsize);
\end{tikzpicture}}

\newcommand{\DefectVertexZeroPoint}{\begin{tikzpicture}[baseline={([yshift=.5ex]current bounding box.center)}]
\draw[wilson] (0,0) -- (1.5,0);
\end{tikzpicture}}
\newcommand{\DefectVertexOnePointScalarInf}{\begin{tikzpicture}[baseline={([yshift=-1ex]current bounding box.center)}]
\draw[wilson] (0,0) -- (1.5,0);
\draw[scalar] (.75,0) -- (.75,1);
\draw[fill=white] (.75,1) circle (.075);
\draw[fill=VertexColor,VertexColor] (.75,0) circle (\vertexsize);
\node[anchor=east] at (.7,1) {\footnotesize $I$};
\node[] at (1.05,1) {\footnotesize $1$};
\node[] at (-.45,0) {\footnotesize $- \infty$};
\node[] at (1.8,0) {\footnotesize $\infty$};
\end{tikzpicture}}
\newcommand{\DefectVertexOnePointScalarLeft}{\begin{tikzpicture}[baseline={([yshift=-1ex]current bounding box.center)}]
\draw[wilson] (0,0) -- (1.5,0);
\draw[scalar] (.5,0) -- (.5,1);
\draw[fill=white] (.5,1) circle (.075);
\draw[fill=white] (1,0) circle (.075);
\draw[fill=VertexColor,VertexColor] (.5,0) circle (\vertexsize);
\node[anchor=east] at (.45,1) {\footnotesize $I$};
\node[] at (.8,1) {\footnotesize $1$};
\node[] at (-.45,0) {\footnotesize $- \infty$};
\node[] at (1,-.3) {\footnotesize $2$};
\node[] at (1.8,0) {\phantom{\footnotesize $\infty$}};
\end{tikzpicture}}
\newcommand{\DefectVertexOnePointScalarCenter}{\begin{tikzpicture}[baseline={([yshift=-1ex]current bounding box.center)}]
\draw[wilson] (0,0) -- (1.5,0);
\draw[scalar] (.75,0) -- (.75,1);
\draw[fill=white] (.75,1) circle (.075);
\draw[fill=white] (.3,0) circle (.075);
\draw[fill=white] (1.2,0) circle (.075);
\draw[fill=VertexColor,VertexColor] (.75,0) circle (\vertexsize);
\node[anchor=east] at (.7,1) {\footnotesize $I$};
\node[] at (1.05,1) {\footnotesize $1$};
\node[] at (.3,-.3) {\footnotesize $2$};
\node[] at (1.2,-.3) {\footnotesize $3$};
\node[] at (-.45,0) {\phantom{\footnotesize $- \infty$}};
\node[] at (1.8,0) {\phantom{\footnotesize $\infty$}};
\end{tikzpicture}}
\newcommand{\DefectVertexOnePointScalarRight}{\begin{tikzpicture}[baseline={([yshift=-1ex]current bounding box.center)}]
\draw[wilson] (0,0) -- (1.5,0);
\draw[scalar] (1,0) -- (1,1);
\draw[fill=white] (1,1) circle (.075);
\draw[fill=white] (.5,0) circle (.075);
\draw[fill=VertexColor,VertexColor] (1,0) circle (\vertexsize);
\node[anchor=east] at (.95,1) {\footnotesize $I$};
\node[] at (1.3,1) {\footnotesize $1$};
\node[] at (1.8,0) {\footnotesize $\infty$};
\node[] at (.5,-.3) {\footnotesize $2$};
\node[] at (-.45,0) {\phantom{\footnotesize $- \infty$}};
\end{tikzpicture}}

\newcommand{\HolographicTwoPointWeakLO}{\begin{tikzpicture}[baseline={([yshift=-1.5ex]current bounding box.center)}]
\draw[wilson] (0,0) -- (2,0);
\draw[scalar] (.55,0) -- (.55,1);
\draw[scalar] (1.45,0) -- (1.45,1);
\draw[scalar] (.55,1) -- (1.45,1);
\draw[fill=white] (.55,1) circle (.18);
\draw[fill=white] (1.45,1) circle (.18);
\draw[fill=white] (.55,.82) circle (.075);
\draw[fill=white] (.73,1) circle (.075);
\draw[fill=white] (1.45,.82) circle (.075);
\draw[fill=white] (1.27,1) circle (.075);
\draw[fill=VertexColor,VertexColor] (.55,0) circle (\vertexsize);
\draw[fill=VertexColor,VertexColor] (1.45,0) circle (\vertexsize);
\end{tikzpicture}}
\newcommand{\HolographicTwoPointWeakNLOFtwo}{\begin{tikzpicture}[baseline={([yshift=-1.5ex]current bounding box.center)}]
\draw[wilson] (0,0) -- (2,0);
\draw[scalar] (.42,0) -- (.42,1);
\draw[scalar] (.68,0) -- (.68,1);
\draw[scalar] (1.32,0) -- (1.32,1);
\draw[scalar] (1.58,0) -- (1.58,1);
\draw[fill=white] (.55,1) circle (.18);
\draw[fill=white] (1.45,1) circle (.18);
\draw[fill=white] (.42,.88) circle (.075);
\draw[fill=white] (.68,.88) circle (.075);
\draw[fill=white] (1.32,.88) circle (.075);
\draw[fill=white] (1.58,.88) circle (.075);
\draw[fill=VertexColor,VertexColor] (.42,0) circle (\vertexsize);
\draw[fill=VertexColor,VertexColor] (.68,0) circle (\vertexsize);
\draw[fill=VertexColor,VertexColor] (1.32,0) circle (\vertexsize);
\draw[fill=VertexColor,VertexColor] (1.58,0) circle (\vertexsize);
\end{tikzpicture}}

\newcommand{\WittenDiagramWilsonLoop}{\begin{tikzpicture}[baseline={([yshift=-.55ex]current bounding box.center)}]
 \pgfmathsetmacro{\x}{.8*sqrt(2)/2}
 \pgfmathsetmacro{\y}{.8*sqrt(2)/2}
 \pgfmathsetmacro{\z}{.8*.5}
 \draw [wilson, GluonColor] (-\x,\y) to[out=-60,in=60] (-\x,-\y);
 \draw [thick] (0,0) circle [radius=.8];
\end{tikzpicture}}
\newcommand{\WittenDiagramOnePoint}{\begin{tikzpicture}[baseline={([yshift=-.55ex]current bounding box.center)}]
 \pgfmathsetmacro{\x}{.8*sqrt(2)/2}
 \pgfmathsetmacro{\y}{.8*sqrt(2)/2}
 \pgfmathsetmacro{\z}{.8*.5}
 \draw [wilson, GluonColor] (-\x,\y) to[out=-60,in=60] (-\x,-\y);
 \draw [thick] (0,0) circle [radius=.8];
 \draw [scalar] (-\z, 0) -- (.8, 0);
 \draw[fill=SCRed,SCRed] (-.4,0) circle (\vertexsize);
  \draw[fill=white] (.8, 0) circle (.075);
\end{tikzpicture}}
\newcommand{\WittenDiagramTwoPointLO}{\begin{tikzpicture}[baseline={([yshift=-.55ex]current bounding box.center)}]
 \pgfmathsetmacro{\x}{.8*sqrt(2)/2}
 \pgfmathsetmacro{\y}{.8*sqrt(2)/2}
 \pgfmathsetmacro{\z}{.8*.55}
 \draw [wilson, GluonColor] (-\x,\y) to[out=-60,in=60] (-\x,-\y);
 \draw [thick] (0,0) circle [radius=.8];
 \draw [scalar] (-\z, -.3) -- (.65, -.45);
 \draw [scalar] (-\z, .3) -- (.65, .45);
 \draw[fill=SCRed,SCRed] (-.8*.55,-.3) circle (\vertexsize);
 \draw[fill=SCRed,SCRed] (-.8*.55,.3) circle (\vertexsize);
 \draw[fill=white] (.65, -.45) circle (.075);
 \draw[fill=white] (.65, .45) circle (.075);
\end{tikzpicture}}
\newcommand{\WittenDiagramTwoPointNLOOne}{\begin{tikzpicture}[baseline={([yshift=-.55ex]current bounding box.center)}]
 \pgfmathsetmacro{\x}{.8*sqrt(2)/2}
 \pgfmathsetmacro{\y}{.8*sqrt(2)/2}
 \pgfmathsetmacro{\z}{.8*.5}
 \draw [wilson, GluonColor] (-\x,\y) to[out=-60,in=60] (-\x,-\y);
 \draw [thick] (0,0) circle [radius=.8];
 \draw [scalar] (-\z, 0) -- (.2, 0);
 \draw [scalar] (.2, 0) -- (.65, -.45);
 \draw [scalar] (.2, 0) -- (.65, .45);
 \draw[fill=SCRed,SCRed] (-.4,0) circle (\vertexsize);
 \draw[fill=SCRed,SCRed] (.2,0) circle (\vertexsize);
  \draw[fill=white] (.65, -.45) circle (.075);
 \draw[fill=white] (.65, .45) circle (.075);
\end{tikzpicture}}
\newcommand{\WittenDiagramTwoPointNLOTwo}{\begin{tikzpicture}[baseline={([yshift=-.55ex]current bounding box.center)}]
  \pgfmathsetmacro{\x}{.8*sqrt(2)/2}
 \pgfmathsetmacro{\y}{.8*sqrt(2)/2}
 \pgfmathsetmacro{\z}{.8*.55}
 \draw [wilson, GluonColor] (-\x,\y) to[out=-60,in=60] (-\x,-\y);
 \draw [white,fill=white] (-\z, -.3) rectangle (-\z+.2,.3);
 \draw[wilson, SCRed,fill=SCRed] (-.8*.55,-.3) -- (-.8*.55,.3);
 \draw [thick] (0,0) circle [radius=.8];
 \draw [scalar] (-\z, -.3) -- (.65, -.45);
 \draw [scalar] (-\z, .3) -- (.65, .45);
 \draw[fill=SCRed,SCRed] (-.8*.55,-.3) circle (\vertexsize);
 \draw[fill=SCRed,SCRed] (-.8*.55,.3) circle (\vertexsize);
  \draw[fill=white] (.65, -.45) circle (.075);
 \draw[fill=white] (.65, .45) circle (.075);
\end{tikzpicture}}
\newcommand{\WittenDiagramTwoPointNLOThree}{\begin{tikzpicture}[baseline={([yshift=-.55ex]current bounding box.center)}]
 \pgfmathsetmacro{\x}{.8*sqrt(2)/2}
 \pgfmathsetmacro{\y}{.8*sqrt(2)/2}
 \pgfmathsetmacro{\z}{.8*.5}
 \draw [wilson, GluonColor] (-\x,\y) to[out=-60,in=60] (-\x,-\y);
 \draw [thick] (0,0) circle [radius=.8];
 \draw [scalar] (-\z, 0) -- (.65, -.45);
 \draw [scalar] (-\z, 0) -- (.65, .45);
 \draw[fill=SCRed,SCRed] (-.4,0) circle (\vertexsize);
  \draw[fill=white] (.65, -.45) circle (.075);
 \draw[fill=white] (.65, .45) circle (.075);
\end{tikzpicture}}
\newcommand{\WittenDiagramThreePoint}{\begin{tikzpicture}[baseline={([yshift=-.55ex]current bounding box.center)}]
 \pgfmathsetmacro{\x}{.8*1/2}
 \pgfmathsetmacro{\y}{.8*sqrt(3)/2}
 \pgfmathsetmacro{\z}{.8*0.5}
 \pgfmathsetmacro{\r}{-.8*0.3}
 \pgfmathsetmacro{\q}{.8*1.3}
 \draw [thick] (0,0) circle [radius=.8];
 \draw [scalar]            (+\x,+\y) -- (0, 0);
 \draw [scalar]            (+\x,-\y) -- (0, 0);
 \draw [scalar]            (  0,  0) -- (-.8, 0);
 \draw[fill=SCRed,SCRed] (0,0) circle (\vertexsize);
  \draw[fill=white] (+\x,+\y) circle (.075);
 \draw[fill=white] (+\x,-\y) circle (.075);
 \draw[fill=white] (-.8, 0) circle (.075);
\end{tikzpicture}}

\newcommand{\IntegrationContours}{\begin{tikzpicture}[baseline, scale=1.7]
  \draw[-stealth] (-1, 0) -- (3.5, 0);
  \draw[-stealth] (0, -1) -- (0, 3);
  \draw[snake it] (0, 0) -- (1, 0);
  \draw[snake it] (2, 0) -- (3.3, 0); 
  \draw (3.05, 3) -- (3.05, 2.55) -- (3.45, 2.55);
  \node at (1, -0.35) {$r$};
  \node at (2, -0.4) {$\frac{1}{r}$};
  \node at (3.3, 2.8) {$w'$};
  \node at (1.5, 1.5) {$w$};
  \node[circle, draw=black, fill=black, inner sep=0.75pt] at (1.5, 1.3) {};
  \draw[
    decoration={markings, mark=at position 0 with {\arrow{stealth}}},
    postaction={decorate}
  ] (1.5,1.4) circle (0.5);
  \draw[ 
    decoration={markings, mark=at position 0.5 with {\arrow{stealth}}},
    postaction={decorate}
  ]  (0, 0.15) -- (1, 0.15);
  \draw (1, 0.15) to[out=0,in=0] (1, -0.15);
  \draw[ 
    decoration={markings, mark=at position 0.5 with {\arrow{stealth}}},
    postaction={decorate}
  ]  (1, -0.15) -- (0, -0.15);
  \draw (0, -0.15) to[out=180, in=180] (0, 0.15);
  \draw[ 
    decoration={markings, mark=at position 0.5 with {\arrow{stealth}}},
    postaction={decorate}
  ]  (3.3, -0.15) -- (2, -0.15);
  \draw (2, -0.15) to[out=180, in=180] (2, 0.15);
  \draw[ 
    decoration={markings, mark=at position 0.5 with {\arrow{stealth}}},
    postaction={decorate}
  ]  (2, 0.15) -- (3.3, 0.15);
  \node at (1.0, 2.00) {$C_0$};
  \node at (0.5, 0.35) {$C_1$};
  \node at (2.5, 0.35) {$C_2$};
 \end{tikzpicture}}
 
 \newcommand{\BulkDefectDefectLO}{\begin{tikzpicture}[baseline={([yshift=-1ex]current bounding box.center)}]
\draw[wilson] (0,0) -- (1.1,0);
\draw[scalar] (.42,0) -- (.42,1);
\draw[scalar] (.68,0) -- (.68,1);
\draw[fill=white] (.55,1) circle (.18);
\draw[fill=white] (.42,.88) circle (.075);
\draw[fill=white] (.68,.88) circle (.075);
\draw[fill=white] (.42,0) circle (.075);
\draw[fill=white] (.68,0) circle (.075);
\end{tikzpicture}}

\newcommand{\BulkDefectDefectNLOSEOne}{\begin{tikzpicture}[baseline={([yshift=-1ex]current bounding box.center)}]
\draw[wilson] (0,0) -- (1.1,0);
\draw[scalar] (.2,0) -- (.42,.88);
\draw[scalar] (.9,0) -- (.68,.88);
\draw[fill=white] (.55,1) circle (.18);
\draw[fill=white] (.42,.88) circle (.075);
\draw[fill=white] (.68,.88) circle (.075);
\draw[fill=white] (.2,0) circle (.075);
\draw[fill=white] (.9,0) circle (.075);
\draw[fill=SEColor,SEColor] (.315,.44) circle (.15);
\end{tikzpicture}}
\newcommand{\BulkDefectDefectNLOSETwo}{\begin{tikzpicture}[baseline={([yshift=-1ex]current bounding box.center)}]
\draw[wilson] (0,0) -- (1.1,0);
\draw[scalar] (.2,0) -- (.42,.88);
\draw[scalar] (.9,0) -- (.68,.88);
\draw[fill=white] (.55,1) circle (.18);
\draw[fill=white] (.42,.88) circle (.075);
\draw[fill=white] (.68,.88) circle (.075);
\draw[fill=white] (.2,0) circle (.075);
\draw[fill=white] (.9,0) circle (.075);
\draw[fill=SEColor,SEColor] (.785,.44) circle (.15);
\end{tikzpicture}}
\newcommand{\BulkDefectDefectNLOX}{\begin{tikzpicture}[baseline={([yshift=-1ex]current bounding box.center)}]
\draw[wilson] (0,0) -- (1.1,0);
\draw[scalar] (.35,0) -- (.68,.88);
\draw[scalar] (.75,0) -- (.42,.88);
\draw[fill=white] (.55,1) circle (.18);
\draw[fill=white] (.42,.88) circle (.075);
\draw[fill=white] (.68,.88) circle (.075);
\draw[fill=white] (.35,0) circle (.075);
\draw[fill=white] (.75,0) circle (.075);
\draw[fill=VertexColor,VertexColor] (.55,.53) circle (\vertexsize);
\end{tikzpicture}}
\newcommand{\BulkDefectDefectNLOH}{\begin{tikzpicture}[baseline={([yshift=-1ex]current bounding box.center)}]
\draw[wilson] (0,0) -- (1.1,0);
\draw[scalar] (.2,0) -- (.42,.88);
\draw[scalar] (.9,0) -- (.68,.88);
\draw[gluon] (.31,.44) -- (.79,.44);
\draw[fill=white] (.55,1) circle (.18);
\draw[fill=white] (.42,.88) circle (.075);
\draw[fill=white] (.68,.88) circle (.075);
\draw[fill=white] (.2,0) circle (.075);
\draw[fill=white] (.9,0) circle (.075);
\draw[fill=VertexColor,VertexColor] (.31,.44) circle (\vertexsize);
\draw[fill=VertexColor,VertexColor] (.79,.44) circle (\vertexsize);
\end{tikzpicture}}
\newcommand{\BulkDefectDefectNLOYOne}{\begin{tikzpicture}[baseline={([yshift=-1ex]current bounding box.center)}]
\draw[wilson] (0,0) -- (1.1,0);
\draw[scalar] (.35,0) -- (.42,.88);
\draw[scalar] (.75,0) -- (.68,.88);
\draw[gluon] (.385,.44) -- (0,.44);
\draw[fill=white] (.55,1) circle (.18);
\draw[fill=white] (.42,.88) circle (.075);
\draw[fill=white] (.68,.88) circle (.075);
\draw[fill=white] (.35,0) circle (.075);
\draw[fill=white] (.75,0) circle (.075);
\draw[fill=VertexColor,VertexColor] (.385,.44) circle (\vertexsize);
\draw[fill=VertexColor,VertexColor] (.15,0) circle (\vertexsize);
\draw[fill=VertexColor,VertexColor] (.95,0) circle (\vertexsize);
\draw[fill=VertexColor,VertexColor] (.55,0) circle (\vertexsize);
\end{tikzpicture}}
\newcommand{\BulkDefectDefectNLOYTwo}{\begin{tikzpicture}[baseline={([yshift=-1ex]current bounding box.center)}]
\draw[wilson] (0,0) -- (1.1,0);
\draw[scalar] (.35,0) -- (.42,.88);
\draw[scalar] (.75,0) -- (.68,.88);
\draw[gluon] (.715,.44) -- (1.1,.44);
\draw[fill=white] (.55,1) circle (.18);
\draw[fill=white] (.42,.88) circle (.075);
\draw[fill=white] (.68,.88) circle (.075);
\draw[fill=white] (.35,0) circle (.075);
\draw[fill=white] (.75,0) circle (.075);
\draw[fill=VertexColor,VertexColor] (.715,.44) circle (\vertexsize);
\draw[fill=VertexColor,VertexColor] (.15,0) circle (\vertexsize);
\draw[fill=VertexColor,VertexColor] (.95,0) circle (\vertexsize);
\draw[fill=VertexColor,VertexColor] (.55,0) circle (\vertexsize);
\end{tikzpicture}}
\newcommand{\BulkDefectDefectNLOFoneOne}{\begin{tikzpicture}[baseline={([yshift=-1ex]current bounding box.center)}]
\draw[wilson] (-.5,0) -- (1.6,0);
\draw[scalar] (-.12,0) -- (.42,.88);
\draw[scalar] (.95,.44) -- (.68,.88);
\draw[fill=white] (.55,1) circle (.18);
\draw[fill=white] (.42,.88) circle (.075);
\draw[fill=white] (.68,.88) circle (.075);
\draw[fill=white] (.35,0) circle (.075);
\draw[fill=white] (.75,0) circle (.075);
\draw[fill=VertexColor,VertexColor] (-.12,0) circle (\vertexsize);
\draw[fill=VertexColor,VertexColor] (-.35,0) circle (\vertexsize);
\draw[fill=VertexColor,VertexColor] (.125,0) circle (\vertexsize);
\draw[fill=VertexColor,VertexColor] (1.2,0) circle (\vertexsize);
\path[clip] (-.4,0) rectangle (1.9,1.2);
\draw[scalar] (.55,0) circle (.2);
\draw[fill=white] (.55,1) circle (.18);
\draw[fill=white] (.42,.88) circle (.075);
\draw[fill=white] (.68,.88) circle (.075);
\draw[fill=white] (.35,0) circle (.075);
\draw[fill=white] (.75,0) circle (.075);
\draw[fill=VertexColor,VertexColor] (-.12,0) circle (\vertexsize);
\draw[fill=VertexColor,VertexColor] (-.35,0) circle (\vertexsize);
\draw[fill=VertexColor,VertexColor] (.125,0) circle (\vertexsize);
\draw[fill=VertexColor,VertexColor] (1.2,0) circle (\vertexsize);
\end{tikzpicture}}
\newcommand{\BulkDefectDefectNLOFoneTwo}{\begin{tikzpicture}[baseline={([yshift=-.1ex]current bounding box.center)}]
\draw[wilson] (-.25,0) -- (1.35,0);
\draw[scalar] (.42,0) -- (.42,.88);
\draw[scalar] (.68,.44) -- (.68,.88);
\draw[fill=white] (.55,1) circle (.18);
\draw[fill=white] (.42,.88) circle (.075);
\draw[fill=white] (.68,.88) circle (.075);
\draw[fill=white] (0,0) circle (.075);
\draw[fill=white] (1.1,0) circle (.075);
\draw[fill=VertexColor,VertexColor] (.42,0) circle (\vertexsize);
\draw[fill=VertexColor,VertexColor] (.2,0) circle (\vertexsize);
\draw[fill=VertexColor,VertexColor] (.72,0) circle (\vertexsize);
\path[clip] (-.4,0) rectangle (1.9,-.41);
\draw[scalar] (.55,.2) circle (.6);
\draw[wilson] (-.1,0) -- (1.2,0);
\draw[scalar] (.42,0) -- (.42,.88);
\draw[scalar] (.68,.44) -- (.68,.88);
\draw[fill=white] (.55,1) circle (.18);
\draw[fill=white] (.42,.88) circle (.075);
\draw[fill=white] (.68,.88) circle (.075);
\draw[fill=white] (0,0) circle (.075);
\draw[fill=white] (1.1,0) circle (.075);
\draw[fill=VertexColor,VertexColor] (.42,0) circle (\vertexsize);
\draw[fill=VertexColor,VertexColor] (.2,0) circle (\vertexsize);
\draw[fill=VertexColor,VertexColor] (.72,0) circle (\vertexsize);
\end{tikzpicture}}
\newcommand{\BulkDefectDefectNLOFoneThree}{\begin{tikzpicture}[baseline={([yshift=-1ex]current bounding box.center)}]
\draw[wilson] (-.5,0) -- (1.6,0);
\draw[scalar] (.15,.44) -- (.42,.88);
\draw[scalar] (1.22,0) -- (.68,.88);
\draw[fill=white] (.55,1) circle (.18);
\draw[fill=white] (.42,.88) circle (.075);
\draw[fill=white] (.68,.88) circle (.075);
\draw[fill=white] (.35,0) circle (.075);
\draw[fill=white] (.75,0) circle (.075);
\draw[fill=VertexColor,VertexColor] (1.22,0) circle (\vertexsize);
\draw[fill=VertexColor,VertexColor] (1.45,0) circle (\vertexsize);
\draw[fill=VertexColor,VertexColor] (1.1-.125,0) circle (\vertexsize);
\draw[fill=VertexColor,VertexColor] (-.1,0) circle (\vertexsize);
\path[clip] (-.4,0) rectangle (1.9,1.2);
\draw[scalar] (.55,0) circle (.2);
\draw[fill=white] (.55,1) circle (.18);
\draw[fill=white] (.42,.88) circle (.075);
\draw[fill=white] (.68,.88) circle (.075);
\draw[fill=white] (.35,0) circle (.075);
\draw[fill=white] (.75,0) circle (.075);
\draw[fill=VertexColor,VertexColor] (1.22,0) circle (\vertexsize);
\draw[fill=VertexColor,VertexColor] (1.45,0) circle (\vertexsize);
\draw[fill=VertexColor,VertexColor] (1.1-.125,0) circle (\vertexsize);
\draw[fill=VertexColor,VertexColor] (-.1,0) circle (\vertexsize);
\end{tikzpicture}}

\newcommand{\DefectSSSSLOOne}{\begin{tikzpicture}[baseline={([yshift=-2ex]current bounding box.center)}]
\draw[wilson] (0,0) -- (2,0);
\draw[fill=white] (.25,0) circle (.075);
\draw[fill=white] (.75,0) circle (.075);
\draw[fill=white] (1.25,0) circle (.075);
\draw[fill=white] (1.75,0) circle (.075);
\path[clip] (0,0) rectangle (2,1);
\draw[scalar] (.5,.085) circle (.28);
\draw[scalar] (1.5,.085) circle (.28);
\draw[wilson] (0,0) -- (2,0);
\draw[fill=white] (.25,0) circle (.075);
\draw[fill=white] (.75,0) circle (.075);
\draw[fill=white] (1.25,0) circle (.075);
\draw[fill=white] (1.75,0) circle (.075);
\end{tikzpicture}}
\newcommand{\DefectSSSSLOTwo}{\begin{tikzpicture}[baseline={([yshift=-2ex]current bounding box.center)}]
\draw[wilson] (0,0) -- (2,0);
\draw[fill=white] (.25,0) circle (.075);
\draw[fill=white] (.75,0) circle (.075);
\draw[fill=white] (1.25,0) circle (.075);
\draw[fill=white] (1.75,0) circle (.075);
\path[clip] (0,0) rectangle (2,1);
\draw[scalar] (1,-.2) circle (.8);
\draw[scalar] (1,-.025) circle (.27);
\draw[wilson] (0,0) -- (2,0);
\draw[fill=white] (.25,0) circle (.075);
\draw[fill=white] (.75,0) circle (.075);
\draw[fill=white] (1.25,0) circle (.075);
\draw[fill=white] (1.75,0) circle (.075);
\end{tikzpicture}}
\newcommand{\DefectSSSSOneLoopSEOne}{\begin{tikzpicture}[baseline={([yshift=-2ex]current bounding box.center)}]
\draw[wilson] (0,0) -- (2,0);
\draw[fill=white] (.25,0) circle (.075);
\draw[fill=white] (.75,0) circle (.075);
\draw[fill=white] (1.25,0) circle (.075);
\draw[fill=white] (1.75,0) circle (.075);
\draw[fill=SEColor,SEColor] (.5,.28+.085) circle (.15);
\path[clip] (0,0) rectangle (2,1);
\draw[scalar] (.5,.085) circle (.28);
\draw[scalar] (1.5,.085) circle (.28);
\draw[wilson] (0,0) -- (2,0);
\draw[fill=white] (.25,0) circle (.075);
\draw[fill=white] (.75,0) circle (.075);
\draw[fill=white] (1.25,0) circle (.075);
\draw[fill=white] (1.75,0) circle (.075);
\draw[fill=SEColor,SEColor] (.5,.28+.085) circle (.15);
\path[clip] (0,.365) rectangle (2,1);
\draw[fill=SEColor,SEColor] (.5,.28+.085) circle (.15);
\end{tikzpicture}}
\newcommand{\DefectSSSSOneLoopSETwo}{\begin{tikzpicture}[baseline={([yshift=-2ex]current bounding box.center)}]
\draw[wilson] (0,0) -- (2,0);
\draw[fill=white] (.25,0) circle (.075);
\draw[fill=white] (.75,0) circle (.075);
\draw[fill=white] (1.25,0) circle (.075);
\draw[fill=white] (1.75,0) circle (.075);
\draw[fill=SEColor,SEColor] (1.5,.28+.085) circle (.15);
\path[clip] (0,0) rectangle (2,1);
\draw[scalar] (.5,.085) circle (.28);
\draw[scalar] (1.5,.085) circle (.28);
\draw[wilson] (0,0) -- (2,0);
\draw[fill=white] (.25,0) circle (.075);
\draw[fill=white] (.75,0) circle (.075);
\draw[fill=white] (1.25,0) circle (.075);
\draw[fill=white] (1.75,0) circle (.075);
\draw[fill=SEColor,SEColor] (1.5,.28+.085) circle (.15);
\path[clip] (0,.365) rectangle (2,1);
\draw[fill=SEColor,SEColor] (1.5,.28+.085) circle (.15);
\end{tikzpicture}}
\newcommand{\DefectSSSSOneLoopSEThree}{\begin{tikzpicture}[baseline={([yshift=-2ex]current bounding box.center)}]
\draw[wilson] (0,0) -- (2,0);
\draw[fill=white] (.25,0) circle (.075);
\draw[fill=white] (.75,0) circle (.075);
\draw[fill=white] (1.25,0) circle (.075);
\draw[fill=white] (1.75,0) circle (.075);
\draw[fill=SEColor,SEColor] (1,.6) circle (.15);
\path[clip] (0,0) rectangle (2,1);
\draw[scalar] (1,-.2) circle (.8);
\draw[scalar] (1,-.025) circle (.27);
\draw[wilson] (0,0) -- (2,0);
\draw[fill=white] (.25,0) circle (.075);
\draw[fill=white] (.75,0) circle (.075);
\draw[fill=white] (1.25,0) circle (.075);
\draw[fill=white] (1.75,0) circle (.075);
\draw[fill=SEColor,SEColor] (1,.6) circle (.15);
\end{tikzpicture}}
\newcommand{\DefectSSSSOneLoopSEFour}{\begin{tikzpicture}[baseline={([yshift=-2ex]current bounding box.center)}]
\draw[wilson] (0,0) -- (2,0);
\draw[fill=white] (.25,0) circle (.075);
\draw[fill=white] (.75,0) circle (.075);
\draw[fill=white] (1.25,0) circle (.075);
\draw[fill=white] (1.75,0) circle (.075);
\draw[fill=SEColor,SEColor] (1,.27-.025) circle (.15);
\path[clip] (0,0) rectangle (2,1);
\draw[scalar] (1,-.2) circle (.8);
\draw[scalar] (1,-.025) circle (.27);
\draw[wilson] (0,0) -- (2,0);
\draw[fill=white] (.25,0) circle (.075);
\draw[fill=white] (.75,0) circle (.075);
\draw[fill=white] (1.25,0) circle (.075);
\draw[fill=white] (1.75,0) circle (.075);
\draw[fill=SEColor,SEColor] (1,.27-.025) circle (.15);
\end{tikzpicture}}
\newcommand{\DefectSSSSOneLoopX}{\begin{tikzpicture}[baseline={([yshift=-2ex]current bounding box.center)}]
\draw[wilson] (0,0) -- (2,0);
\draw[fill=white] (.25,0) circle (.075);
\draw[fill=white] (.75,0) circle (.075);
\draw[fill=white] (1.25,0) circle (.075);
\draw[fill=white] (1.75,0) circle (.075);
\draw[fill=VertexColor,VertexColor] (1,.43) circle (\vertexsize);
\path[clip] (0,0) rectangle (2,1);
\draw[scalar] (.75,0) circle (.5);
\draw[scalar] (1.25,0) circle (.5);
\draw[wilson] (0,0) -- (2,0);
\draw[fill=white] (.25,0) circle (.075);
\draw[fill=white] (.75,0) circle (.075);
\draw[fill=white] (1.25,0) circle (.075);
\draw[fill=white] (1.75,0) circle (.075);
\draw[fill=VertexColor,VertexColor] (1,.43) circle (\vertexsize);
\end{tikzpicture}}
\newcommand{\DefectSSSSOneLoopHOne}{\begin{tikzpicture}[baseline={([yshift=-2ex]current bounding box.center)}]
\draw[wilson] (0,0) -- (2,0);
\draw[fill=white] (.25,0) circle (.075);
\draw[fill=white] (.75,0) circle (.075);
\draw[fill=white] (1.25,0) circle (.075);
\draw[fill=white] (1.75,0) circle (.075);
\draw[fill=VertexColor,VertexColor] (.5,.365) circle (\vertexsize);
\draw[fill=VertexColor,VertexColor] (1.5,.365) circle (\vertexsize);
\path[clip] (0,0) rectangle (2,1);
\draw[scalar] (.5,.085) circle (.28);
\draw[scalar] (1.5,.085) circle (.28);
\draw[wilson] (0,0) -- (2,0);
\draw[fill=white] (.25,0) circle (.075);
\draw[fill=white] (.75,0) circle (.075);
\draw[fill=white] (1.25,0) circle (.075);
\draw[fill=white] (1.75,0) circle (.075);
\draw[fill=VertexColor,VertexColor] (.5,.365) circle (\vertexsize);
\draw[fill=VertexColor,VertexColor] (1.5,.365) circle (\vertexsize);
\path[clip] (0,.365) rectangle (2,1);
\draw[gluon] (1,0) circle (.62);
\draw[fill=VertexColor,VertexColor] (.5,.365) circle (\vertexsize);
\draw[fill=VertexColor,VertexColor] (1.5,.365) circle (\vertexsize);
\end{tikzpicture}}
\newcommand{\DefectSSSSOneLoopHTwo}{\begin{tikzpicture}[baseline={([yshift=-2ex]current bounding box.center)}]
\draw[wilson] (0,0) -- (2,0);
\draw[fill=white] (.25,0) circle (.075);
\draw[fill=white] (.75,0) circle (.075);
\draw[fill=white] (1.25,0) circle (.075);
\draw[fill=white] (1.75,0) circle (.075);
\draw[fill=VertexColor,VertexColor] (1,.6) circle (\vertexsize);
\draw[fill=VertexColor,VertexColor] (1,.245) circle (\vertexsize);
\path[clip] (0,0) rectangle (2,1);
\draw[scalar] (1,-.2) circle (.8);
\draw[scalar] (1,-.025) circle (.27);
\draw[wilson] (0,0) -- (2,0);
\draw[fill=white] (.25,0) circle (.075);
\draw[fill=white] (.75,0) circle (.075);
\draw[fill=white] (1.25,0) circle (.075);
\draw[fill=white] (1.75,0) circle (.075);
\draw[fill=VertexColor,VertexColor] (1,.6) circle (\vertexsize);
\draw[fill=VertexColor,VertexColor] (1,.245) circle (\vertexsize);
\path[clip] (.8,.245) rectangle (1.2,.6);
\draw[gluon] (1,.6) -- (1, .245);
\draw[fill=VertexColor,VertexColor] (1,.6) circle (\vertexsize);
\draw[fill=VertexColor,VertexColor] (1,.245) circle (\vertexsize);
\end{tikzpicture}}

\newcommand{\DefectSSSSOneLoopYOne}{\begin{tikzpicture}[baseline={([yshift=-2ex]current bounding box.center)}]
\draw[wilson] (-.2,0) -- (2.2,0);
\draw[fill=white] (.25,0) circle (.075);
\draw[fill=white] (.75,0) circle (.075);
\draw[fill=white] (1.25,0) circle (.075);
\draw[fill=white] (1.75,0) circle (.075);
\draw[fill=VertexColor,VertexColor] (.5,.365) circle (\vertexsize);
\draw[fill=VertexColor,VertexColor] (0,0) circle (\vertexsize);
\draw[fill=VertexColor,VertexColor] (.5,0) circle (\vertexsize);
\draw[fill=VertexColor,VertexColor] (1,0) circle (\vertexsize);
\draw[fill=VertexColor,VertexColor] (2,0) circle (\vertexsize);
\path[clip] (0,0) rectangle (2,1);
\draw[scalar] (.5,.085) circle (.28);
\draw[scalar] (1.5,.085) circle (.28);
\draw[wilson] (0,0) -- (2,0);
\draw[fill=white] (.25,0) circle (.075);
\draw[fill=white] (.75,0) circle (.075);
\draw[fill=white] (1.25,0) circle (.075);
\draw[fill=white] (1.75,0) circle (.075);
\draw[fill=VertexColor,VertexColor] (.5,.365) circle (\vertexsize);
\draw[fill=VertexColor,VertexColor] (0,0) circle (\vertexsize);
\draw[fill=VertexColor,VertexColor] (.5,0) circle (\vertexsize);
\draw[fill=VertexColor,VertexColor] (1,0) circle (\vertexsize);
\draw[fill=VertexColor,VertexColor] (2,0) circle (\vertexsize);
\path[clip] (0,.365) rectangle (2,1);
\draw[gluon] (.5,.365) -- (.5,.7);
\draw[fill=VertexColor,VertexColor] (.5,.365) circle (\vertexsize);
\draw[fill=VertexColor,VertexColor] (0,0) circle (\vertexsize);
\draw[fill=VertexColor,VertexColor] (.5,0) circle (\vertexsize);
\draw[fill=VertexColor,VertexColor] (1,0) circle (\vertexsize);
\draw[fill=VertexColor,VertexColor] (2,0) circle (\vertexsize);
\end{tikzpicture}}
\newcommand{\DefectSSSSOneLoopYTwo}{\begin{tikzpicture}[baseline={([yshift=-2ex]current bounding box.center)}]
\draw[wilson] (-.2,0) -- (2.2,0);
\draw[fill=white] (.25,0) circle (.075);
\draw[fill=white] (.75,0) circle (.075);
\draw[fill=white] (1.25,0) circle (.075);
\draw[fill=white] (1.75,0) circle (.075);
\draw[fill=VertexColor,VertexColor] (1.5,.365) circle (\vertexsize);
\draw[fill=VertexColor,VertexColor] (0,0) circle (\vertexsize);
\draw[fill=VertexColor,VertexColor] (1.5,0) circle (\vertexsize);
\draw[fill=VertexColor,VertexColor] (1,0) circle (\vertexsize);
\draw[fill=VertexColor,VertexColor] (2,0) circle (\vertexsize);
\path[clip] (0,0) rectangle (2,1);
\draw[scalar] (.5,.085) circle (.28);
\draw[scalar] (1.5,.085) circle (.28);
\draw[wilson] (0,0) -- (2,0);
\draw[fill=white] (.25,0) circle (.075);
\draw[fill=white] (.75,0) circle (.075);
\draw[fill=white] (1.25,0) circle (.075);
\draw[fill=white] (1.75,0) circle (.075);
\draw[fill=VertexColor,VertexColor] (1.5,.365) circle (\vertexsize);
\draw[fill=VertexColor,VertexColor] (0,0) circle (\vertexsize);
\draw[fill=VertexColor,VertexColor] (1.5,0) circle (\vertexsize);
\draw[fill=VertexColor,VertexColor] (1,0) circle (\vertexsize);
\draw[fill=VertexColor,VertexColor] (2,0) circle (\vertexsize);
\path[clip] (0,.365) rectangle (2,1);
\draw[gluon] (1.5,.365) -- (1.5,.7);
\draw[fill=VertexColor,VertexColor] (1.5,.365) circle (\vertexsize);
\draw[fill=VertexColor,VertexColor] (0,0) circle (\vertexsize);
\draw[fill=VertexColor,VertexColor] (1.5,0) circle (\vertexsize);
\draw[fill=VertexColor,VertexColor] (1,0) circle (\vertexsize);
\draw[fill=VertexColor,VertexColor] (2,0) circle (\vertexsize);
\end{tikzpicture}}
\newcommand{\DefectSSSSOneLoopYThree}{\begin{tikzpicture}[baseline={([yshift=-2ex]current bounding box.center)}]
\draw[wilson] (-.2,0) -- (2.2,0);
\draw[fill=white] (.25,0) circle (.075);
\draw[fill=white] (.75,0) circle (.075);
\draw[fill=white] (1.25,0) circle (.075);
\draw[fill=white] (1.75,0) circle (.075);
\draw[fill=VertexColor,VertexColor] (1,.245) circle (\vertexsize);
\draw[fill=VertexColor,VertexColor] (.5,0) circle (\vertexsize);
\draw[fill=VertexColor,VertexColor] (1,0) circle (\vertexsize);
\draw[fill=VertexColor,VertexColor] (1.5,0) circle (\vertexsize);
\path[clip] (0,0) rectangle (2,1);
\draw[scalar] (1,-.2) circle (.8);
\draw[scalar] (1,-.025) circle (.27);
\draw[wilson] (0,0) -- (2,0);
\draw[fill=white] (.25,0) circle (.075);
\draw[fill=white] (.75,0) circle (.075);
\draw[fill=white] (1.25,0) circle (.075);
\draw[fill=white] (1.75,0) circle (.075);
\draw[fill=VertexColor,VertexColor] (1,.245) circle (\vertexsize);
\draw[fill=VertexColor,VertexColor] (.5,0) circle (\vertexsize);
\draw[fill=VertexColor,VertexColor] (1,0) circle (\vertexsize);
\draw[fill=VertexColor,VertexColor] (1.5,0) circle (\vertexsize);
\path[clip] (.8,.245) rectangle (1.2,.6);
\draw[gluon] (1,.5) -- (1, .245);
\draw[fill=VertexColor,VertexColor] (1,.245) circle (\vertexsize);
\draw[fill=VertexColor,VertexColor] (.5,0) circle (\vertexsize);
\draw[fill=VertexColor,VertexColor] (1,0) circle (\vertexsize);
\draw[fill=VertexColor,VertexColor] (1.5,0) circle (\vertexsize);
\end{tikzpicture}}
\newcommand{\DefectSSSSOneLoopYFour}{\begin{tikzpicture}[baseline={([yshift=-2ex]current bounding box.center)}]
\draw[wilson] (-.2,0) -- (2.2,0);
\draw[fill=white] (.25,0) circle (.075);
\draw[fill=white] (.75,0) circle (.075);
\draw[fill=white] (1.25,0) circle (.075);
\draw[fill=white] (1.75,0) circle (.075);
\draw[fill=VertexColor,VertexColor] (1,.6) circle (\vertexsize);
\draw[fill=VertexColor,VertexColor] (0,0) circle (\vertexsize);
\draw[fill=VertexColor,VertexColor] (.5,0) circle (\vertexsize);
\draw[fill=VertexColor,VertexColor] (1.5,0) circle (\vertexsize);
\draw[fill=VertexColor,VertexColor] (2,0) circle (\vertexsize);
\draw[gluon] (1,.6) -- (1, .85);
\path[clip] (0,0) rectangle (2,1);
\draw[scalar] (1,-.2) circle (.8);
\draw[scalar] (1,-.025) circle (.27);
\draw[wilson] (0,0) -- (2,0);
\draw[fill=white] (.25,0) circle (.075);
\draw[fill=white] (.75,0) circle (.075);
\draw[fill=white] (1.25,0) circle (.075);
\draw[fill=white] (1.75,0) circle (.075);
\draw[fill=VertexColor,VertexColor] (1,.6) circle (\vertexsize);
\draw[fill=VertexColor,VertexColor] (0,0) circle (\vertexsize);
\draw[fill=VertexColor,VertexColor] (.5,0) circle (\vertexsize);
\draw[fill=VertexColor,VertexColor] (1.5,0) circle (\vertexsize);
\draw[fill=VertexColor,VertexColor] (2,0) circle (\vertexsize);
\draw[gluon] (1,.6) -- (1, .85);
\path[clip] (.8,.245) rectangle (1.2,.6);
\draw[gluon] (1,.6) -- (1, .85);
\draw[fill=VertexColor,VertexColor] (1,.6) circle (\vertexsize);
\draw[fill=VertexColor,VertexColor] (0,0) circle (\vertexsize);
\draw[fill=VertexColor,VertexColor] (.5,0) circle (\vertexsize);
\draw[fill=VertexColor,VertexColor] (1.5,0) circle (\vertexsize);
\draw[fill=VertexColor,VertexColor] (2,0) circle (\vertexsize);
\end{tikzpicture}}

\newcommand{\RSymmetryChannelsIllustration}{\begin{tikzpicture}[baseline={([yshift=-2ex]current bounding box.center)}]
\draw[wilson] (-.5-.66,0) -- (2.5,0);
\draw[fill=white] (-.66,0) circle (.2);
\draw[fill=white] (0,0) circle (.2);
\draw[fill=white] (.66,0) circle (.2);
\draw[fill=white] (1.33,0) circle (.2);
\draw[fill=white] (2,0) circle (.2);
\node[] at (-.66,0) {\tiny $1$};
\node[] at (0,0) {\tiny $1$};
\node[] at (.66,0) {\tiny $1$};
\node[] at (1.33,0) {\tiny $1$};
\node[] at (2,0) {\tiny $2$};
\node[] at (3,-.025) {\large $=$};
\node[] at (3.6,0) {\large $3$};
\draw[wilson] (4,0) -- (4.5+.66+.66+.5,0);
\draw[fill=white] (4.5,0) circle (.2);
\draw[fill=white] (4.5+.66,0) circle (.2);
\draw[fill=white] (4.5+.66+.66,0) circle (.2);
\node[] at (4.5,0) {\tiny $1$};
\node[] at (4.5+.66,0) {\tiny $1$};
\node[] at (4.5+.66+.66,0) {\tiny $2$};
\node[] at (4.5+.66+.66+.5+.4,0) {\large $+$};
\draw[wilson] (6.72+.4,0) -- (6.72+.4+.5+.66+.66+.66+.5,0);
\draw[fill=white] (6.72+.4+.5,0) circle (.2);
\draw[fill=white] (6.72+.4+.5+.66,0) circle (.2);
\draw[fill=white] (6.72+.4+.5+.66+.66,0) circle (.2);
\draw[fill=white] (6.72+.4+.5+.66+.66+.66,0) circle (.2);
\node[] at (6.72+.4+.5,0) {\tiny $1$};
\node[] at (6.72+.4+.5+.66,0) {\tiny $1$};
\node[] at (6.72+.4+.5+.66+.66,0) {\tiny $1$};
\node[] at (6.72+.4+.5+.66+.66+.66,0) {\tiny $1$};
\node[] at (3,-1) {\large $=$};
\node[] at (3.6,-1) {\large $3$};
\draw[wilson] (4,-1) -- (4.5+.66+.5,-1);
\draw[fill=white] (4.5,-1) circle (.2);
\draw[fill=white] (4.5+.66,-1) circle (.2);
\node[] at (4.5,-1) {\tiny $1$};
\node[] at (4.5+.66,-1) {\tiny $1$};
\node[] at (4.5+.66+.5+.4,-1) {\large $+$};
\node[] at (4.5+.66+.5+.4+.5,-1) {\large $3$};
\draw[wilson] (4.5+.66+.5+.4+.4+.4,-1) -- (4.5+.66+.5+.4+.4+.4+.5+.66+.5,-1);
\draw[fill=white] (4.5+.66+.5+.4+.4+.4+.5,-1) circle (.2);
\draw[fill=white] (4.5+.66+.5+.4+.4+.4+.5+.66,-1) circle (.2);
\node[] at (4.5+.66+.5+.4+.4+.4+.5,-1) {\tiny $1$};
\node[] at (4.5+.66+.5+.4+.4+.4+.5+.66,-1) {\tiny $1$};
\node[] at (3,-2) {\large $=$};
\node[] at (3.6,-2) {\large $6$};
\end{tikzpicture}}

\newcommand{\CombChannel}{\scalebox{\scaleCC}{\begin{tikzpicture}[baseline={([yshift=-0ex]current bounding box.center)}]
\draw[scalar] (-4.5,0) -- (0.5,0);
\draw[scalar] (-3.75,0) -- (-3.75,0.7);
\draw[scalar] (-2.75,0) -- (-2.75,0.7);
\draw[scalar] (-1.25,0) -- (-1.25,0.7);
\draw[scalar] (-0.25,0) -- (-0.25,0.7);
\node[] at (-3.2,.2) {\tiny $\Dh_1$};
\node[] at (-.75,.2) {\tiny $\Dh_{n-3}$};
\node[] at (-2,.5) {$\ldots$};
\draw[MediumBlueGrey,fill=MediumBlueGrey] (-3.75,0) circle (.05);
\draw[MediumBlueGrey,fill=MediumBlueGrey] (-2.75,0) circle (.05);
\draw[MediumBlueGrey,fill=MediumBlueGrey] (-1.25,0) circle (.05);
\draw[MediumBlueGrey,fill=MediumBlueGrey] (-0.25,0) circle (.05);
\end{tikzpicture}}}
\newcommand{\SnowflakeChannel}{\scalebox{\scaleCC}{\begin{tikzpicture}[baseline={([yshift=-0ex]current bounding box.center)}]
\draw[scalar] (0,0) -- (0,.7);
\draw[scalar] (0,.7) -- (.7,1.15);
\draw[scalar] (0,.7) -- (-.7,1.15);
\draw[scalar] (0,0) -- (-.7,-.45);
\draw[scalar] (-.7,-.45) -- (-1.4,0);
\draw[scalar] (-.7,-.45) -- (-.7,-1.15);
\draw[scalar] (0,0) -- (.7,-.45);
\draw[scalar] (.7,-.45) -- (1.4,0);
\draw[scalar] (.7,-.45) -- (.7,-1.15);
\node[] at (-.5,0) {\tiny $\Dh_1$};
\node[] at (.2,-.4) {\tiny $\Dh_2$};
\node[] at (.25,.35) {\tiny $\Dh_3$};
\draw[MediumBlueGrey,fill=MediumBlueGrey] (0,0) circle (.05);
\draw[MediumBlueGrey,fill=MediumBlueGrey] (0,.7) circle (.05);
\draw[MediumBlueGrey,fill=MediumBlueGrey] (-.7,-.45) circle (.05);
\draw[MediumBlueGrey,fill=MediumBlueGrey] (.7,-.45) circle (.05);
\end{tikzpicture}}}

\newcommand{\RecursionLOEven}{\begin{tikzpicture}[baseline={([yshift=-.75ex]current bounding box.center)}]
\draw[wilson] (-.3,0) -- (2.05,0);
\draw[fill=white] (.1,0) circle (.075);
\draw[fill=white] (1.1,0) circle (.075);
\node[] at (1.2,-.3) {\tiny $2j+2$};
\path[clip] (-.3,0) rectangle (2.05,1);
\SinglePropagator{.6}
\TreeLevel{.6};
\TreeLevel{1.6};
\draw[wilson] (-.3,0) -- (2.05,0);
\draw[fill=white] (.1,0) circle (.075);
\draw[fill=white] (1.1,0) circle (.075);
\end{tikzpicture}}
\newcommand{\TreeLevelInsertion}{\begin{tikzpicture}[baseline={([yshift=-.8ex]current bounding box.center)}]
\TreeLevel{0};
\draw[wilson] (-.3,0) -- (.3,0);
\end{tikzpicture}}
\newcommand{\RecursionLOOddOne}{\begin{tikzpicture}[baseline={([yshift=-.75ex]current bounding box.center)}]
\draw[wilson] (-.85,0) -- (2.05,0);
\draw[fill=white] (1.1,0) circle (.075);
\draw[fill=VertexColor,VertexColor] (.1,0) circle (\vertexsize);
\node[] at (-.2,-.3) {\tiny $2j$};
\node[] at (1.2,-.3) {\tiny $i$};
\path[clip] (-.85,0) rectangle (2.05,1);
\SinglePropagator{.6}
\TreeLevel{.-.4};
\TreeLevel{.6};
\TreeLevel{1.6};
\draw[wilson] (-.85,0) -- (2.05,0);
\draw[fill=white] (1.1,0) circle (.075);
\draw[fill=VertexColor,VertexColor] (.1,0) circle (\vertexsize);
\end{tikzpicture}}
\newcommand{\RecursionLOOddTwo}{\begin{tikzpicture}[baseline={([yshift=-.75ex]current bounding box.center)}]
\draw[wilson] (-.85,0) -- (2.05,0);
\draw[fill=VertexColor,VertexColor] (1.1,0) circle (\vertexsize);
\draw[fill=white] (.1,0) circle (.075);
\node[] at (.8,-.3) {\tiny $2j$};
\node[] at (.1,-.3) {\tiny $i$};
\path[clip] (-.85,0) rectangle (2.05,1);
\SinglePropagator{.6}
\TreeLevel{.-.4};
\TreeLevel{.6};
\TreeLevel{1.6};
\draw[wilson] (-.85,0) -- (2.05,0);
\draw[fill=VertexColor,VertexColor] (1.1,0) circle (\vertexsize);
\draw[fill=white] (.1,0) circle (.075);
\end{tikzpicture}}
\newcommand{\RecursionLOOddThree}{\begin{tikzpicture}[baseline={([yshift=-.75ex]current bounding box.center)}]
\draw[wilson] (-.85,0) -- (2.05,0);
\draw[fill=white] (.1,0) circle (.075);
\draw[fill=white] (1.1,0) circle (.075);
\node[] at (.1,-.3) {\tiny $i$};
\node[] at (1.1,-.3) {\tiny $j$};
\path[clip] (-.85,0) rectangle (2.05,1);
\SinglePropagator{.6}
\TreeLevel{-.4};
\TreeLevel{.6};
\TreeLevel{1.6};
\draw[wilson] (-.85,0) -- (2.05,0);
\draw[fill=white] (.1,0) circle (.075);
\draw[fill=white] (1.1,0) circle (.075);
\end{tikzpicture}}

\newcommand{\RecursionNLOBulk}{\begin{tikzpicture}[baseline={([yshift=-.5ex]current bounding box.center)}]
\TreeLevel{.4};
\TreeLevel{1.2};
\TreeLevel{2};
\TreeLevel{2.8};
\TreeLevel{3.6};
\draw[wilson] (0,0) -- (4,0);
\draw[fill=white] (.8,0) circle (.075);
\draw[fill=white] (1.6,0) circle (.075);
\draw[fill=white] (2.4,0) circle (.075);
\draw[fill=white] (3.2,0) circle (.075);
\node[] at (.8,-.3) {\tiny $i$};
\node[] at (1.6,-.3) {\tiny $j$};
\node[] at (2.4,-.3) {\tiny $k$};
\node[] at (3.2,-.3) {\tiny $l$};
\path[clip] (0,0) rectangle (4,1);
\draw[scalar] (1.6,0) circle (.8);
\draw[scalar] (2.4,0) circle (.8);
\draw[fill=white] (2,.7) circle (.2);
\node[] at (2,.7) {\tiny $4$};
\draw[wilson] (0,0) -- (4,0);
\draw[fill=white] (.8,0) circle (.075);
\draw[fill=white] (1.6,0) circle (.075);
\draw[fill=white] (2.4,0) circle (.075);
\draw[fill=white] (3.2,0) circle (.075);
\end{tikzpicture}}
\newcommand{\RecursionNLOBulkBridge}{\begin{tikzpicture}[baseline={([yshift=-1ex]current bounding box.center)}]
\TreeLevel{.4};
\TreeLevel{3.6};
\draw[wilson] (0,0) -- (4,0);
\draw[fill=white] (.9,0) circle (.075);
\draw[fill=white] (3.1,0) circle (.075);
\node[] at (.9,-.3) {\tiny $i$};
\node[] at (3.1,-.3) {\tiny $j$};
\path[clip] (0,0) rectangle (4,1.2);
\draw[scalar] (2,0) circle (1.1);
\draw[scalar] (2,0) circle (.8);
\node[] at (2,.25) {\tiny $A_{\text{NLO,bulk}}$};
\draw[wilson] (0,0) -- (4,0);
\draw[fill=white] (.9,0) circle (.075);
\draw[fill=white] (3.1,0) circle (.075);
\end{tikzpicture}}
\newcommand{\RecursionNLOBulkFour}{\begin{tikzpicture}[baseline={([yshift=-1ex]current bounding box.center)}]
\draw[wilson] (0,0) -- (2,0);
\draw[fill=white] (.25,0) circle (.075);
\draw[fill=white] (.75,0) circle (.075);
\draw[fill=white] (1.25,0) circle (.075);
\draw[fill=white] (1.75,0) circle (.075);
\draw[fill=white] (1,.43) circle (.2);
\node[] at (1,.43) {\tiny $4$};
\node[] at (.25,-.3) {\tiny $i$};
\node[] at (.75,-.3) {\tiny $j$};
\node[] at (1.25,-.3) {\tiny $k$};
\node[] at (1.75,-.3) {\tiny $l$};
\path[clip] (0,0) rectangle (2,1);
\draw[scalar] (.75,0) circle (.5);
\draw[scalar] (1.25,0) circle (.5);
\draw[wilson] (0,0) -- (2,0);
\draw[fill=white] (.25,0) circle (.075);
\draw[fill=white] (.75,0) circle (.075);
\draw[fill=white] (1.25,0) circle (.075);
\draw[fill=white] (1.75,0) circle (.075);
\draw[fill=white] (1,.43) circle (.2);
\node[] at (1,.43) {\tiny $4$};
\end{tikzpicture}}

\newcommand{\RecursionNLOBdryOne}{\begin{tikzpicture}[baseline={([yshift=-.5ex]current bounding box.center)}]
\draw[wilson] (-1.85,0) -- (2.05,0);
\draw[fill=white] (.1,0) circle (.075);
\draw[fill=white] (1.1,0) circle (.075);
\node[] at (.1,-.3) {\tiny $i$};
\node[] at (1.1,-.3) {\tiny $j$};
\draw[fill=VertexColor,VertexColor] (.6,.7) circle (\vertexsize);
\draw[fill=VertexColor,VertexColor] (-.9,0) circle (\vertexsize);
\path[clip] (-1.85,0) rectangle (2.05,1);
\draw[gluon] (0,0) circle (.875);
\SinglePropagator{.6}
\TreeLevel{-1.4};
\TreeLevel{-.4};
\TreeLevel{.6};
\TreeLevel{1.6};
\draw[wilson] (-1.85,0) -- (2.05,0);
\draw[fill=white] (.1,0) circle (.075);
\draw[fill=white] (1.1,0) circle (.075);
\draw[fill=VertexColor,VertexColor] (.6,.7) circle (\vertexsize);
\draw[fill=VertexColor,VertexColor] (-.9,0) circle (\vertexsize);
\end{tikzpicture}}
\newcommand{\RecursionNLOBdryTwo}{\begin{tikzpicture}[baseline={([yshift=-.5ex]current bounding box.center)}]
\draw[wilson] (-1.85,0) -- (2.05,0);
\draw[fill=white] (-.9,0) circle (.075);
\draw[fill=white] (1.1,0) circle (.075);
\node[] at (-.9,-.3) {\tiny $i$};
\node[] at (1.1,-.3) {\tiny $j$};
\draw[fill=VertexColor,VertexColor] (.1,.75) circle (\vertexsize);
\draw[fill=VertexColor,VertexColor] (.1,0) circle (\vertexsize);
\path[clip] (-1.85,0) rectangle (2.05,1);
\SinglePropagatorBig{.1};
\draw[gluon] (.1,0) -- (.1,.75);
\TreeLevel{-1.4};
\TreeLevel{-.4};
\TreeLevel{.6};
\TreeLevel{1.6};
\draw[wilson] (-1.85,0) -- (2.05,0);
\draw[fill=white] (-.9,0) circle (.075);
\draw[fill=white] (1.1,0) circle (.075);
\draw[fill=VertexColor,VertexColor] (.1,.75) circle (\vertexsize);
\draw[fill=VertexColor,VertexColor] (.1,0) circle (\vertexsize);
\end{tikzpicture}}
\newcommand{\RecursionNLOBdryThree}{\begin{tikzpicture}[baseline={([yshift=-.5ex]current bounding box.center)}]
\draw[wilson] (-1.85,0) -- (2.05,0);
\draw[fill=white] (-.9,0) circle (.075);
\draw[fill=white] (.1,0) circle (.075);
\node[] at (-.9,-.3) {\tiny $i$};
\node[] at (.1,-.3) {\tiny $j$};
\draw[fill=VertexColor,VertexColor] (-.4,.7) circle (\vertexsize);
\draw[fill=VertexColor,VertexColor] (1.1,0) circle (\vertexsize);
\path[clip] (-1.85,0) rectangle (2.05,1);
\draw[gluon] (.175,0) circle (.875);
\SinglePropagator{-.4}
\TreeLevel{-1.4};
\TreeLevel{-.4};
\TreeLevel{.6};
\TreeLevel{1.6};
\draw[wilson] (-1.85,0) -- (2.05,0);
\draw[fill=white] (-.9,0) circle (.075);
\draw[fill=white] (.1,0) circle (.075);
\draw[fill=VertexColor,VertexColor] (-.4,.7) circle (\vertexsize);
\draw[fill=VertexColor,VertexColor] (1.1,0) circle (\vertexsize);
\end{tikzpicture}}
\newcommand{\RecursionNLOBdryBridge}{\begin{tikzpicture}[baseline={([yshift=-1ex]current bounding box.center)}]
\TreeLevel{.4};
\TreeLevel{3.6};
\draw[wilson] (0,0) -- (4,0);
\draw[fill=white] (.9,0) circle (.075);
\draw[fill=white] (3.1,0) circle (.075);
\node[] at (.9,-.3) {\tiny $i$};
\node[] at (3.1,-.3) {\tiny $j$};
\path[clip] (0,0) rectangle (4,1.2);
\draw[scalar] (2,0) circle (1.1);
\draw[scalar] (2,0) circle (.8);
\node[] at (2,.25) {\tiny $A_{\text{NLO,bdry}}$};
\draw[wilson] (0,0) -- (4,0);
\draw[fill=white] (.9,0) circle (.075);
\draw[fill=white] (3.1,0) circle (.075);
\end{tikzpicture}}
\newcommand{\RecursionNLOBdryExtra}{\begin{tikzpicture}[baseline={([yshift=-2.5ex]current bounding box.center)}]
\draw[wilson] (0,0) -- (4,0);
\draw[fill=white] (.8,0) circle (.075);
\draw[fill=white] (2.4,0) circle (.075);
\draw[fill=VertexColor,VertexColor] (1.6,0) circle (\vertexsize);
\draw[fill=VertexColor,VertexColor] (3.2,0) circle (\vertexsize);
\path[clip] (0,0) rectangle (4,1);
\SinglePropagatorSmall{1.2};
\SinglePropagatorSmall{2.8};
\TreeLevel{.4};
\TreeLevel{1.2};
\TreeLevel{2};
\TreeLevel{2.8};
\TreeLevel{3.6};
\draw[wilson] (0,0) -- (4,0);
\draw[fill=white] (.8,0) circle (.075);
\draw[fill=white] (2.4,0) circle (.075);
\draw[fill=VertexColor,VertexColor] (1.6,0) circle (\vertexsize);
\draw[fill=VertexColor,VertexColor] (3.2,0) circle (\vertexsize);
\end{tikzpicture}}

\newcommand{\DefectSSNLOSeven}{\begin{tikzpicture}[baseline={([yshift=-2ex]current bounding box.center)}]
\draw[wilson] (0,0) -- (2,0);
\draw[fill=white] (.25,0) circle (.075);
\draw[fill=white] (.75,0) circle (.075);
\draw[fill=VertexColor,VertexColor] (1.25,0) circle (\vertexsize);
\draw[fill=VertexColor,VertexColor] (1.75,0) circle (\vertexsize);
\path[clip] (0,0) rectangle (2,1);
\draw[scalar] (1,-.2) circle (.8);
\draw[scalar] (1,-.025) circle (.27);
\draw[wilson] (0,0) -- (2,0);
\draw[fill=white] (.25,0) circle (.075);
\draw[fill=white] (.75,0) circle (.075);
\draw[fill=VertexColor,VertexColor] (1.25,0) circle (\vertexsize);
\draw[fill=VertexColor,VertexColor] (1.75,0) circle (\vertexsize);
\end{tikzpicture}}

\newcommand{\DefectSSSSTwoLoopsSelfEnergyOne}{\begin{tikzpicture}[baseline={([yshift=-2ex]current bounding box.center)}]
\draw[wilson] (0,0) -- (2,0);
\draw[fill=white] (.25,0) circle (.075);
\draw[fill=white] (.75,0) circle (.075);
\draw[fill=white] (1.25,0) circle (.075);
\draw[fill=white] (1.75,0) circle (.075);
\draw[fill=VertexColor,VertexColor] (1,.43) circle (\vertexsize);
\path[clip] (0,0) rectangle (2,1);
\draw[scalar] (.75,0) circle (.5);
\draw[scalar] (1.25,0) circle (.5);
\draw[wilson] (0,0) -- (2,0);
\draw[fill=white] (.25,0) circle (.075);
\draw[fill=white] (.75,0) circle (.075);
\draw[fill=white] (1.25,0) circle (.075);
\draw[fill=white] (1.75,0) circle (.075);
\draw[fill=VertexColor,VertexColor] (1,.43) circle (\vertexsize);
\draw[fill=SEColor,SEColor] (.5,.4) circle (.1);
\end{tikzpicture}}
\newcommand{\DefectSSSSTwoLoopsSelfEnergyTwo}{\begin{tikzpicture}[baseline={([yshift=-2ex]current bounding box.center)}]
\draw[wilson] (0,0) -- (2,0);
\draw[fill=white] (.25,0) circle (.075);
\draw[fill=white] (.75,0) circle (.075);
\draw[fill=white] (1.25,0) circle (.075);
\draw[fill=white] (1.75,0) circle (.075);
\draw[fill=VertexColor,VertexColor] (1,.43) circle (\vertexsize);
\path[clip] (0,0) rectangle (2,1);
\draw[scalar] (.75,0) circle (.5);
\draw[scalar] (1.25,0) circle (.5);
\draw[wilson] (0,0) -- (2,0);
\draw[fill=white] (.25,0) circle (.075);
\draw[fill=white] (.75,0) circle (.075);
\draw[fill=white] (1.25,0) circle (.075);
\draw[fill=white] (1.75,0) circle (.075);
\draw[fill=VertexColor,VertexColor] (1,.43) circle (\vertexsize);
\draw[fill=SEColor,SEColor] (.82,.24) circle (.1);
\end{tikzpicture}}
\newcommand{\DefectSSSSTwoLoopsSelfEnergyThree}{\begin{tikzpicture}[baseline={([yshift=-2ex]current bounding box.center)}]
\draw[wilson] (0,0) -- (2,0);
\draw[fill=white] (.25,0) circle (.075);
\draw[fill=white] (.75,0) circle (.075);
\draw[fill=white] (1.25,0) circle (.075);
\draw[fill=white] (1.75,0) circle (.075);
\draw[fill=VertexColor,VertexColor] (1,.43) circle (\vertexsize);
\path[clip] (0,0) rectangle (2,1);
\draw[scalar] (.75,0) circle (.5);
\draw[scalar] (1.25,0) circle (.5);
\draw[wilson] (0,0) -- (2,0);
\draw[fill=white] (.25,0) circle (.075);
\draw[fill=white] (.75,0) circle (.075);
\draw[fill=white] (1.25,0) circle (.075);
\draw[fill=white] (1.75,0) circle (.075);
\draw[fill=VertexColor,VertexColor] (1,.43) circle (\vertexsize);
\draw[fill=SEColor,SEColor] (1.18,.24) circle (.1);
\end{tikzpicture}}
\newcommand{\DefectSSSSTwoLoopsSelfEnergyFour}{\begin{tikzpicture}[baseline={([yshift=-2ex]current bounding box.center)}]
\draw[wilson] (0,0) -- (2,0);
\draw[fill=white] (.25,0) circle (.075);
\draw[fill=white] (.75,0) circle (.075);
\draw[fill=white] (1.25,0) circle (.075);
\draw[fill=white] (1.75,0) circle (.075);
\draw[fill=VertexColor,VertexColor] (1,.43) circle (\vertexsize);
\path[clip] (0,0) rectangle (2,1);
\draw[scalar] (.75,0) circle (.5);
\draw[scalar] (1.25,0) circle (.5);
\draw[wilson] (0,0) -- (2,0);
\draw[fill=white] (.25,0) circle (.075);
\draw[fill=white] (.75,0) circle (.075);
\draw[fill=white] (1.25,0) circle (.075);
\draw[fill=white] (1.75,0) circle (.075);
\draw[fill=VertexColor,VertexColor] (1,.43) circle (\vertexsize);
\draw[fill=SEColor,SEColor] (1.5,.4) circle (.1);
\end{tikzpicture}}

\newcommand{\DefectSSSSTwoLoopsXXOne}{\begin{tikzpicture}[baseline={([yshift=-2ex]current bounding box.center)}]
\draw[scalar] (.5,.365) -- (1.5,.365);
\draw[wilson] (0,0) -- (2,0);
\draw[fill=white] (.25,0) circle (.075);
\draw[fill=white] (.75,0) circle (.075);
\draw[fill=white] (1.25,0) circle (.075);
\draw[fill=white] (1.75,0) circle (.075);
\draw[fill=VertexColor,VertexColor] (.5,.365) circle (\vertexsize);
\draw[fill=VertexColor,VertexColor] (1.5,.365) circle (\vertexsize);
\path[clip] (0,0) rectangle (2,1);
\draw[scalar] (.5,.085) circle (.28);
\draw[scalar] (1.5,.085) circle (.28);
\draw[wilson] (0,0) -- (2,0);
\draw[fill=white] (.25,0) circle (.075);
\draw[fill=white] (.75,0) circle (.075);
\draw[fill=white] (1.25,0) circle (.075);
\draw[fill=white] (1.75,0) circle (.075);
\draw[fill=VertexColor,VertexColor] (.5,.365) circle (\vertexsize);
\draw[fill=VertexColor,VertexColor] (1.5,.365) circle (\vertexsize);
\path[clip] (0,.365) rectangle (2,1);
\draw[scalar] (1,0) circle (.62);
\draw[fill=VertexColor,VertexColor] (.5,.365) circle (\vertexsize);
\draw[fill=VertexColor,VertexColor] (1.5,.365) circle (\vertexsize);
\end{tikzpicture}}
\newcommand{\DefectSSSSTwoLoopsXXTwo}{\begin{tikzpicture}[baseline={([yshift=-2ex]current bounding box.center)}]
\draw[wilson] (0,0) -- (2,0);
\draw[fill=white] (.25,0) circle (.075);
\draw[fill=white] (.75,0) circle (.075);
\draw[fill=white] (1.25,0) circle (.075);
\draw[fill=white] (1.75,0) circle (.075);
\draw[fill=VertexColor,VertexColor] (1,.6) circle (\vertexsize);
\draw[fill=VertexColor,VertexColor] (1,.245) circle (\vertexsize);
\path[clip] (0,0) rectangle (2,1);
\draw[scalar] (1,-.2) circle (.8);
\draw[scalar] (1,-.025) circle (.27);
\draw[wilson] (0,0) -- (2,0);
\draw[fill=white] (.25,0) circle (.075);
\draw[fill=white] (.75,0) circle (.075);
\draw[fill=white] (1.25,0) circle (.075);
\draw[fill=white] (1.75,0) circle (.075);
\draw[fill=VertexColor,VertexColor] (1,.6) circle (\vertexsize);
\draw[fill=VertexColor,VertexColor] (1,.245) circle (\vertexsize);
\path[clip] (.8,.245) rectangle (1.2,.6);
\draw[scalar] (.8,.4225) circle (.27);
\draw[scalar] (1.2,.4225) circle (.27);
\draw[fill=VertexColor,VertexColor] (1,.6) circle (\vertexsize);
\draw[fill=VertexColor,VertexColor] (1,.245) circle (\vertexsize);
\end{tikzpicture}}

\newcommand{\DefectSSSSTwoLoopsXHOne}{\begin{tikzpicture}[baseline={([yshift=-2ex]current bounding box.center)}]
\draw[gluontwo] (.4,.36) -- (.81,.225);
\draw[wilson] (0,0) -- (2,0);
\draw[fill=white] (.25,0) circle (.075);
\draw[fill=white] (.75,0) circle (.075);
\draw[fill=white] (1.25,0) circle (.075);
\draw[fill=white] (1.75,0) circle (.075);
\draw[fill=VertexColor,VertexColor] (1,.43) circle (\vertexsize);
\draw[fill=VertexColor,VertexColor] (.4,.36) circle (\vertexsize);
\draw[fill=VertexColor,VertexColor] (.81,.225) circle (\vertexsize);
\path[clip] (0,0) rectangle (2,1);
\draw[scalar] (.75,0) circle (.5);
\draw[scalar] (1.25,0) circle (.5);
\draw[wilson] (0,0) -- (2,0);
\draw[fill=white] (.25,0) circle (.075);
\draw[fill=white] (.75,0) circle (.075);
\draw[fill=white] (1.25,0) circle (.075);
\draw[fill=white] (1.75,0) circle (.075);
\draw[fill=VertexColor,VertexColor] (1,.43) circle (\vertexsize);
\draw[fill=VertexColor,VertexColor] (.4,.36) circle (\vertexsize);
\draw[fill=VertexColor,VertexColor] (.81,.225) circle (\vertexsize);
\end{tikzpicture}}
\newcommand{\DefectSSSSTwoLoopsXHTwo}{\begin{tikzpicture}[baseline={([yshift=-2ex]current bounding box.center)}]
\draw[gluontwo] (.81,.225) -- (1.19,.225);
\draw[wilson] (0,0) -- (2,0);
\draw[fill=white] (.25,0) circle (.075);
\draw[fill=white] (.75,0) circle (.075);
\draw[fill=white] (1.25,0) circle (.075);
\draw[fill=white] (1.75,0) circle (.075);
\draw[fill=VertexColor,VertexColor] (1,.43) circle (\vertexsize);
\draw[fill=VertexColor,VertexColor] (.81,.225) circle (\vertexsize);
\draw[fill=VertexColor,VertexColor] (1.19,.225) circle (\vertexsize);
\path[clip] (0,0) rectangle (2,1);
\draw[scalar] (.75,0) circle (.5);
\draw[scalar] (1.25,0) circle (.5);
\draw[wilson] (0,0) -- (2,0);
\draw[fill=white] (.25,0) circle (.075);
\draw[fill=white] (.75,0) circle (.075);
\draw[fill=white] (1.25,0) circle (.075);
\draw[fill=white] (1.75,0) circle (.075);
\draw[fill=VertexColor,VertexColor] (1,.43) circle (\vertexsize);
\draw[fill=VertexColor,VertexColor] (.81,.225) circle (\vertexsize);
\draw[fill=VertexColor,VertexColor] (1.19,.225) circle (\vertexsize);
\end{tikzpicture}}
\newcommand{\DefectSSSSTwoLoopsXHThree}{\begin{tikzpicture}[baseline={([yshift=-2ex]current bounding box.center)}]
\draw[gluontwo] (1.6,.36) -- (1.19,.225);
\draw[wilson] (0,0) -- (2,0);
\draw[fill=white] (.25,0) circle (.075);
\draw[fill=white] (.75,0) circle (.075);
\draw[fill=white] (1.25,0) circle (.075);
\draw[fill=white] (1.75,0) circle (.075);
\draw[fill=VertexColor,VertexColor] (1,.43) circle (\vertexsize);
\draw[fill=VertexColor,VertexColor] (1.6,.36) circle (\vertexsize);
\draw[fill=VertexColor,VertexColor] (1.19,.225) circle (\vertexsize);
\path[clip] (0,0) rectangle (2,1);
\draw[scalar] (.75,0) circle (.5);
\draw[scalar] (1.25,0) circle (.5);
\draw[wilson] (0,0) -- (2,0);
\draw[fill=white] (.25,0) circle (.075);
\draw[fill=white] (.75,0) circle (.075);
\draw[fill=white] (1.25,0) circle (.075);
\draw[fill=white] (1.75,0) circle (.075);
\draw[fill=VertexColor,VertexColor] (1,.43) circle (\vertexsize);
\draw[fill=VertexColor,VertexColor] (1.6,.36) circle (\vertexsize);
\draw[fill=VertexColor,VertexColor] (1.19,.225) circle (\vertexsize);
\end{tikzpicture}}
\newcommand{\DefectSSSSTwoLoopsXHFour}{\begin{tikzpicture}[baseline={([yshift=-2ex]current bounding box.center)}]
\draw[wilson] (0,0) -- (2,0);
\draw[fill=white] (.25,0) circle (.075);
\draw[fill=white] (.75,0) circle (.075);
\draw[fill=white] (1.25,0) circle (.075);
\draw[fill=white] (1.75,0) circle (.075);
\draw[fill=VertexColor,VertexColor] (1,.43) circle (\vertexsize);
\draw[fill=VertexColor,VertexColor] (.4,.36) circle (\vertexsize);
\draw[fill=VertexColor,VertexColor] (1.6,.36) circle (\vertexsize);
\path[clip] (0,0) rectangle (2,1);
\draw[scalar] (.75,0) circle (.5);
\draw[scalar] (1.25,0) circle (.5);
\draw[wilson] (0,0) -- (2,0);
\draw[fill=white] (.25,0) circle (.075);
\draw[fill=white] (.75,0) circle (.075);
\draw[fill=white] (1.25,0) circle (.075);
\draw[fill=white] (1.75,0) circle (.075);
\draw[fill=VertexColor,VertexColor] (.4,.36) circle (\vertexsize);
\draw[fill=VertexColor,VertexColor] (1.6,.36) circle (\vertexsize);
\path[clip] (0,.36) rectangle (2,1);
\draw[gluontwo] (1,.15) circle (.65);
\draw[fill=VertexColor,VertexColor] (1,.43) circle (\vertexsize);
\draw[fill=VertexColor,VertexColor] (.4,.36) circle (\vertexsize);
\draw[fill=VertexColor,VertexColor] (1.6,.36) circle (\vertexsize);
\end{tikzpicture}}

\newcommand{\DefectSSSSTwoLoopsSpider}{\begin{tikzpicture}[baseline={([yshift=-2ex]current bounding box.center)}]
\draw[wilson] (0,0) -- (2,0);
\draw[fill=white] (.25,0) circle (.075);
\draw[fill=white] (.75,0) circle (.075);
\draw[fill=white] (1.25,0) circle (.075);
\draw[fill=white] (1.75,0) circle (.075);
\draw[fill=VertexColor,VertexColor] (1,.43) circle (\vertexsize);
\path[clip] (0,0) rectangle (2,1);
\draw[scalar] (.75,.1) circle (.52);
\draw[scalar] (1.25,.1) circle (.52);
\draw[wilson] (0,0) -- (2,0);
\draw[fill=white] (.25,0) circle (.075);
\draw[fill=white] (.75,0) circle (.075);
\draw[fill=white] (1.25,0) circle (.075);
\draw[fill=white] (1.75,0) circle (.075);
\draw[scalar, fill=white] (1,.45) circle (.275);
\draw[-Latex] (1,.725)--(1.1,.725);
\draw[-Latex] (1,.18)--(.9,.18);
\draw[-Latex] (.725,.43)--(.725,.53);
\draw[-Latex] (1.275,.44)--(1.275,.34);
\draw[fill=VertexColor,VertexColor] (.775,.615) circle (\vertexsize);
\draw[fill=VertexColor,VertexColor] (.785,.25) circle (\vertexsize);
\draw[fill=VertexColor,VertexColor] (1.225,.615) circle (\vertexsize);
\draw[fill=VertexColor,VertexColor] (1.215,.25) circle (\vertexsize);
\end{tikzpicture}}

\newcommand{\DefectSSSSTwoLoopsXYOne}{\begin{tikzpicture}[baseline={([yshift=-2ex]current bounding box.center)}]
\draw[wilson] (-.2,0) -- (2.2,0);
\draw[gluon] (.4,.36) -- (.1,.66);
\draw[fill=white] (.25,0) circle (.075);
\draw[fill=white] (.75,0) circle (.075);
\draw[fill=white] (1.25,0) circle (.075);
\draw[fill=white] (1.75,0) circle (.075);
\draw[fill=VertexColor,VertexColor] (1,.43) circle (\vertexsize);
\draw[fill=VertexColor,VertexColor] (.4,.36) circle (\vertexsize);
\draw[fill=VertexColor,VertexColor] (0,0) circle (\vertexsize);
\draw[fill=VertexColor,VertexColor] (.5,0) circle (\vertexsize);
\draw[fill=VertexColor,VertexColor] (2,0) circle (\vertexsize);
\path[clip] (0,0) rectangle (2,1);
\draw[scalar] (.75,0) circle (.5);
\draw[scalar] (1.25,0) circle (.5);
\draw[wilson] (0,0) -- (2,0);
\draw[fill=white] (.25,0) circle (.075);
\draw[fill=white] (.75,0) circle (.075);
\draw[fill=white] (1.25,0) circle (.075);
\draw[fill=white] (1.75,0) circle (.075);
\draw[fill=VertexColor,VertexColor] (1,.43) circle (\vertexsize);
\draw[fill=VertexColor,VertexColor] (.4,.36) circle (\vertexsize);
\draw[fill=VertexColor,VertexColor] (0,0) circle (\vertexsize);
\draw[fill=VertexColor,VertexColor] (.5,0) circle (\vertexsize);
\draw[fill=VertexColor,VertexColor] (2,0) circle (\vertexsize);
\end{tikzpicture}}
\newcommand{\DefectSSSSTwoLoopsXYTwo}{\begin{tikzpicture}[baseline={([yshift=-2ex]current bounding box.center)}]
\draw[wilson] (0,0) -- (2,0);
\draw[gluon] (.81,.225) -- (.46,.225);
\draw[fill=white] (.25,0) circle (.075);
\draw[fill=white] (.75,0) circle (.075);
\draw[fill=white] (1.25,0) circle (.075);
\draw[fill=white] (1.75,0) circle (.075);
\draw[fill=VertexColor,VertexColor] (1,.43) circle (\vertexsize);
\draw[fill=VertexColor,VertexColor] (.81,.225) circle (\vertexsize);
\draw[fill=VertexColor,VertexColor] (.5,0) circle (\vertexsize);
\draw[fill=VertexColor,VertexColor] (1,0) circle (\vertexsize);
\path[clip] (0,0) rectangle (2,1);
\draw[scalar] (.75,0) circle (.5);
\draw[scalar] (1.25,0) circle (.5);
\draw[wilson] (0,0) -- (2,0);
\draw[fill=white] (.25,0) circle (.075);
\draw[fill=white] (.75,0) circle (.075);
\draw[fill=white] (1.25,0) circle (.075);
\draw[fill=white] (1.75,0) circle (.075);
\draw[fill=VertexColor,VertexColor] (1,.43) circle (\vertexsize);
\draw[fill=VertexColor,VertexColor] (.81,.225) circle (\vertexsize);
\draw[fill=VertexColor,VertexColor] (.5,0) circle (\vertexsize);
\draw[fill=VertexColor,VertexColor] (1,0) circle (\vertexsize);
\end{tikzpicture}}
\newcommand{\DefectSSSSTwoLoopsXYThree}{\begin{tikzpicture}[baseline={([yshift=-2ex]current bounding box.center)}]
\draw[wilson] (0,0) -- (2,0);
\draw[gluon] (1.19,.225) -- (1.54,.225);
\draw[fill=white] (.25,0) circle (.075);
\draw[fill=white] (.75,0) circle (.075);
\draw[fill=white] (1.25,0) circle (.075);
\draw[fill=white] (1.75,0) circle (.075);
\draw[fill=VertexColor,VertexColor] (1,.43) circle (\vertexsize);
\draw[fill=VertexColor,VertexColor] (1.19,.225) circle (\vertexsize);
\draw[fill=VertexColor,VertexColor] (1.5,0) circle (\vertexsize);
\draw[fill=VertexColor,VertexColor] (1,0) circle (\vertexsize);
\path[clip] (0,0) rectangle (2,1);
\draw[scalar] (.75,0) circle (.5);
\draw[scalar] (1.25,0) circle (.5);
\draw[wilson] (0,0) -- (2,0);
\draw[fill=white] (.25,0) circle (.075);
\draw[fill=white] (.75,0) circle (.075);
\draw[fill=white] (1.25,0) circle (.075);
\draw[fill=white] (1.75,0) circle (.075);
\draw[fill=VertexColor,VertexColor] (1,.43) circle (\vertexsize);
\draw[fill=VertexColor,VertexColor] (1.19,.225) circle (\vertexsize);
\draw[fill=VertexColor,VertexColor] (1.5,0) circle (\vertexsize);
\draw[fill=VertexColor,VertexColor] (1,0) circle (\vertexsize);
\end{tikzpicture}}
\newcommand{\DefectSSSSTwoLoopsXYFour}{\begin{tikzpicture}[baseline={([yshift=-2ex]current bounding box.center)}]
\draw[wilson] (-.2,0) -- (2.2,0);
\draw[gluon] (1.6,.36) -- (1.9,.66);
\draw[fill=white] (.25,0) circle (.075);
\draw[fill=white] (.75,0) circle (.075);
\draw[fill=white] (1.25,0) circle (.075);
\draw[fill=white] (1.75,0) circle (.075);
\draw[fill=VertexColor,VertexColor] (1,.43) circle (\vertexsize);
\draw[fill=VertexColor,VertexColor] (1.6,.36) circle (\vertexsize);
\draw[fill=VertexColor,VertexColor] (0,0) circle (\vertexsize);
\draw[fill=VertexColor,VertexColor] (1.5,0) circle (\vertexsize);
\draw[fill=VertexColor,VertexColor] (2,0) circle (\vertexsize);
\path[clip] (0,0) rectangle (2,1);
\draw[scalar] (.75,0) circle (.5);
\draw[scalar] (1.25,0) circle (.5);
\draw[wilson] (0,0) -- (2,0);
\draw[fill=white] (.25,0) circle (.075);
\draw[fill=white] (.75,0) circle (.075);
\draw[fill=white] (1.25,0) circle (.075);
\draw[fill=white] (1.75,0) circle (.075);
\draw[fill=VertexColor,VertexColor] (1,.43) circle (\vertexsize);
\draw[fill=VertexColor,VertexColor] (1.6,.36) circle (\vertexsize);
\draw[fill=VertexColor,VertexColor] (0,0) circle (\vertexsize);
\draw[fill=VertexColor,VertexColor] (1.5,0) circle (\vertexsize);
\draw[fill=VertexColor,VertexColor] (2,0) circle (\vertexsize);
\end{tikzpicture}}

\newcommand{\OPESymmetric}{\begin{tikzpicture}[baseline={([yshift=0ex]current bounding box.center)}]
\draw [wilson] (0,0) -- (1,0);
\draw [wilson, SCRed] (1,0) -- (2.5,0);
\draw [scalar] (2.5,0) -- (3.5,0);
\draw [wilson] (1,0) -- (1,1);
\draw [wilson] (2.5,.5) -- (2,1);
\draw [wilson] (2.5,.5) -- (3,1);
\draw [wilson, SCRed] (2.5,.5) -- (2.5,0);
\draw[fill=MediumBlueGrey,MediumBlueGrey] (1,0) circle (\vertexsize);
\draw[fill=MediumBlueGrey,MediumBlueGrey] (2.5,0) circle (\vertexsize);
\draw[fill=MediumBlueGrey,MediumBlueGrey] (2.5,.5) circle (\vertexsize);
\node[] at (-.4,.05) {$\Oh_1$};
\node[] at (1,1.4) {$\Oh_1$};
\node[] at (2,1.4) {$\Oh_1$};
\node[] at (3,1.4) {$\Oh_1$};
\node[] at (4,.05) {$\Oh_2$};
\end{tikzpicture}}
\newcommand{\OPEAsymmetric}{\begin{tikzpicture}[baseline={([yshift=0ex]current bounding box.center)}]
\draw [wilson] (0,0) -- (1,0);
\draw [wilson, SCRed] (1,0) -- (2,0);
\draw [wilson, WCOrange] (2,0) -- (3,0);
\draw [wilson] (3,0) -- (4,0);
\draw [wilson] (1,0) -- (1,1);
\draw [scalar] (2,0) -- (2,1);
\draw [wilson] (3,0) -- (3,1);
\draw[fill=MediumBlueGrey,MediumBlueGrey] (1,0) circle (\vertexsize);
\draw[fill=MediumBlueGrey,MediumBlueGrey] (2,0) circle (\vertexsize);
\draw[fill=MediumBlueGrey,MediumBlueGrey] (3,0) circle (\vertexsize);
\node[] at (-.4,.05) {$\Oh_1$};
\node[] at (1,1.4) {$\Oh_1$};
\node[] at (2,1.4) {$\Oh_1$};
\node[] at (3,1.4) {$\Oh_1$};
\node[] at (4.5,.05) {$\Oh_2$};
\end{tikzpicture}}

\newcommand{\WittenDiagramFivePointLOOne}{\begin{tikzpicture}[baseline={([yshift=-.55ex]current bounding box.center)}]
\draw [thick] (0,0) circle [radius=.8];
\draw [scalar] (-.625,.5) -- (-.75,-.25);
\draw [scalar] (0,-.8) -- (.625,.5);
\draw [scalar] (.75,-.25) -- (.625,.5);
\draw[fill=white] (-.75,-.25) circle (.075);
\draw[fill=white] (-.625,.5) circle (.075);
\draw[fill=white] (.75,-.25) circle (.075);
\draw[fill=white] (.625,.5) circle (.075);
\draw[fill=white] (0,-.8) circle (.075);
\end{tikzpicture}}
\newcommand{\WittenDiagramFivePointLOTwo}{\begin{tikzpicture}[baseline={([yshift=-.55ex]current bounding box.center)}]
\draw [thick] (0,0) circle [radius=.8];
\draw [scalar] (-.625,.5) -- (0,-.8);
\draw [white,fill=white] (-.375,-.05) circle (.1);
\draw [scalar] (-.75,-.25) -- (.625,.5);
\draw [scalar] (.75,-.25) -- (.625,.5);
\draw[fill=white] (-.75,-.25) circle (.075);
\draw[fill=white] (-.625,.5) circle (.075);
\draw[fill=white] (.75,-.25) circle (.075);
\draw[fill=white] (.625,.5) circle (.075);
\draw[fill=white] (0,-.8) circle (.075);
\end{tikzpicture}}
\newcommand{\WittenDiagramFivePointLOThree}{\begin{tikzpicture}[baseline={([yshift=-.55ex]current bounding box.center)}]
\draw [thick] (0,0) circle [radius=.8];
\draw [scalar] (-.625,.5) -- (.75,-.25);
\draw [white,fill=white] (0,.15) circle (.1);
\draw [white,fill=white] (.375,-.05) circle (.1);
\draw [scalar] (-.75,-.25) -- (.625,.5);
\draw [scalar] (0,-.8) -- (.625,.5);
\draw[fill=white] (-.75,-.25) circle (.075);
\draw[fill=white] (-.625,.5) circle (.075);
\draw[fill=white] (.75,-.25) circle (.075);
\draw[fill=white] (.625,.5) circle (.075);
\draw[fill=white] (0,-.8) circle (.075);
\end{tikzpicture}}
\newcommand{\WittenDiagramFivePointLOFour}{\begin{tikzpicture}[baseline={([yshift=-.55ex]current bounding box.center)}]
\draw [thick] (0,0) circle [radius=.8];
\draw [scalar] (-.625,.5) -- (.625,.5);
\draw [scalar] (-.75,-.25) -- (0,-.8);
\draw [scalar] (.75,-.25) -- (.625,.5);
\draw[fill=white] (-.75,-.25) circle (.075);
\draw[fill=white] (-.625,.5) circle (.075);
\draw[fill=white] (.75,-.25) circle (.075);
\draw[fill=white] (.625,.5) circle (.075);
\draw[fill=white] (0,-.8) circle (.075);
\end{tikzpicture}}
\newcommand{\WittenDiagramFivePointLOFive}{\begin{tikzpicture}[baseline={([yshift=-.55ex]current bounding box.center)}]
\draw [thick] (0,0) circle [radius=.8];
\draw [scalar] (-.625,.5) -- (.625,.5);
\draw [scalar] (-.75,-.25) -- (.75,-.25);
\draw [white,fill=white] (.25,-.25) circle (.1);
\draw [scalar] (0,-.8) -- (.625,.5);
\draw[fill=white] (-.75,-.25) circle (.075);
\draw[fill=white] (-.625,.5) circle (.075);
\draw[fill=white] (.75,-.25) circle (.075);
\draw[fill=white] (.625,.5) circle (.075);
\draw[fill=white] (0,-.8) circle (.075);
\end{tikzpicture}}
\newcommand{\WittenDiagramFivePointLOSix}{\begin{tikzpicture}[baseline={([yshift=-.55ex]current bounding box.center)}]
\draw [thick] (0,0) circle [radius=.8];
\draw [scalar] (-.625,.5) -- (.625,.5);
\draw [scalar] (-.75,-.25) -- (.625,.5);
\draw [scalar] (0,-.8) -- (.75,-.25);
\draw[fill=white] (-.75,-.25) circle (.075);
\draw[fill=white] (-.625,.5) circle (.075);
\draw[fill=white] (.75,-.25) circle (.075);
\draw[fill=white] (.625,.5) circle (.075);
\draw[fill=white] (0,-.8) circle (.075);
\end{tikzpicture}}

\newcommand{\DefectFFFFOneLoopSEOne}{\begin{tikzpicture}[baseline={([yshift=-2ex]current bounding box.center)}]
\draw[wilson] (0,0) -- (2,0);
\draw[fill=white] (.25,0) circle (.075);
\draw[fill=white] (.75,0) circle (.075);
\draw[fill=white] (1.25,0) circle (.075);
\draw[fill=white] (1.75,0) circle (.075);
\draw[-Latex] (.265,.24)--(.305,.315);
\draw[-Latex] (.755,.195)--(.795,.12);
\draw[-Latex] (1.5,.365)--(1.6,.365);
\draw[fill=SEColor,SEColor] (.5,.28+.085) circle (.15);
\path[clip] (0,0) rectangle (2,1);
\draw[scalar] (.5,.085) circle (.28);
\draw[scalar] (1.5,.085) circle (.28);
\draw[wilson] (0,0) -- (2,0);
\draw[fill=white] (.25,0) circle (.075);
\draw[fill=white] (.75,0) circle (.075);
\draw[fill=white] (1.25,0) circle (.075);
\draw[fill=white] (1.75,0) circle (.075);
\draw[fill=SEColor,SEColor] (.5,.28+.085) circle (.15);
\path[clip] (0,.365) rectangle (2,1);
\draw[fill=SEColor,SEColor] (.5,.28+.085) circle (.15);
\end{tikzpicture}}
\newcommand{\DefectFFFFOneLoopSETwo}{\begin{tikzpicture}[baseline={([yshift=-2ex]current bounding box.center)}]
\draw[wilson] (0,0) -- (2,0);
\draw[fill=white] (.25,0) circle (.075);
\draw[fill=white] (.75,0) circle (.075);
\draw[fill=white] (1.25,0) circle (.075);
\draw[fill=white] (1.75,0) circle (.075);
\draw[-Latex] (1.265,.24)--(1.305,.315);
\draw[-Latex] (1.755,.195)--(1.795,.12);
\draw[-Latex] (.5,.365)--(.6,.365);
\draw[fill=SEColor,SEColor] (1.5,.28+.085) circle (.15);
\path[clip] (0,0) rectangle (2,1);
\draw[scalar] (.5,.085) circle (.28);
\draw[scalar] (1.5,.085) circle (.28);
\draw[wilson] (0,0) -- (2,0);
\draw[fill=white] (.25,0) circle (.075);
\draw[fill=white] (.75,0) circle (.075);
\draw[fill=white] (1.25,0) circle (.075);
\draw[fill=white] (1.75,0) circle (.075);
\draw[fill=SEColor,SEColor] (1.5,.28+.085) circle (.15);
\path[clip] (0,.365) rectangle (2,1);
\draw[fill=SEColor,SEColor] (1.5,.28+.085) circle (.15);
\end{tikzpicture}}
\newcommand{\DefectFFFFOneLoopHOne}{\begin{tikzpicture}[baseline={([yshift=-2ex]current bounding box.center)}]
\draw[wilson] (0,0) -- (2,0);
\draw[fill=white] (.25,0) circle (.075);
\draw[fill=white] (.75,0) circle (.075);
\draw[fill=white] (1.25,0) circle (.075);
\draw[fill=white] (1.75,0) circle (.075);
\draw[-Latex] (.265,.24)--(.305,.315);
\draw[-Latex] (.755,.195)--(.795,.12);
\draw[-Latex] (1.265,.24)--(1.305,.315);
\draw[-Latex] (1.755,.195)--(1.795,.12);
\draw[fill=VertexColor,VertexColor] (.5,.365) circle (\vertexsize);
\draw[fill=VertexColor,VertexColor] (1.5,.365) circle (\vertexsize);
\path[clip] (0,0) rectangle (2,1);
\draw[scalar] (.5,.085) circle (.28);
\draw[scalar] (1.5,.085) circle (.28);
\draw[wilson] (0,0) -- (2,0);
\draw[fill=white] (.25,0) circle (.075);
\draw[fill=white] (.75,0) circle (.075);
\draw[fill=white] (1.25,0) circle (.075);
\draw[fill=white] (1.75,0) circle (.075);
\draw[fill=VertexColor,VertexColor] (.5,.365) circle (\vertexsize);
\draw[fill=VertexColor,VertexColor] (1.5,.365) circle (\vertexsize);
\path[clip] (0,.365) rectangle (2,1);
\draw[scalar] (1,0) circle (.62);
\draw[fill=VertexColor,VertexColor] (.5,.365) circle (\vertexsize);
\draw[fill=VertexColor,VertexColor] (1.5,.365) circle (\vertexsize);
\end{tikzpicture}}
\newcommand{\DefectFFFFOneLoopHTwo}{\begin{tikzpicture}[baseline={([yshift=-2ex]current bounding box.center)}]
\draw[wilson] (0,0) -- (2,0);
\draw[fill=white] (.25,0) circle (.075);
\draw[fill=white] (.75,0) circle (.075);
\draw[fill=white] (1.25,0) circle (.075);
\draw[fill=white] (1.75,0) circle (.075);
\draw[-Latex] (.265,.24)--(.305,.315);
\draw[-Latex] (.755,.195)--(.795,.12);
\draw[-Latex] (1.265,.24)--(1.305,.315);
\draw[-Latex] (1.755,.195)--(1.795,.12);
\draw[fill=VertexColor,VertexColor] (.5,.365) circle (\vertexsize);
\draw[fill=VertexColor,VertexColor] (1.5,.365) circle (\vertexsize);
\path[clip] (0,0) rectangle (2,1);
\draw[scalar] (.5,.085) circle (.28);
\draw[scalar] (1.5,.085) circle (.28);
\draw[wilson] (0,0) -- (2,0);
\draw[fill=white] (.25,0) circle (.075);
\draw[fill=white] (.75,0) circle (.075);
\draw[fill=white] (1.25,0) circle (.075);
\draw[fill=white] (1.75,0) circle (.075);
\draw[fill=VertexColor,VertexColor] (.5,.365) circle (\vertexsize);
\draw[fill=VertexColor,VertexColor] (1.5,.365) circle (\vertexsize);
\path[clip] (0,.365) rectangle (2,1);
\draw[gluon] (1,0) circle (.62);
\draw[fill=VertexColor,VertexColor] (.5,.365) circle (\vertexsize);
\draw[fill=VertexColor,VertexColor] (1.5,.365) circle (\vertexsize);
\end{tikzpicture}}

\newcommand{\DefectFFFFOneLoopYOne}{\begin{tikzpicture}[baseline={([yshift=-2ex]current bounding box.center)}]
\draw[wilson] (-.2,0) -- (2.2,0);
\draw[fill=white] (.25,0) circle (.075);
\draw[fill=white] (.75,0) circle (.075);
\draw[fill=white] (1.25,0) circle (.075);
\draw[fill=white] (1.75,0) circle (.075);
\draw[-Latex] (.265,.24)--(.305,.315);
\draw[-Latex] (.755,.195)--(.795,.12);
\draw[-Latex] (1.5,.365)--(1.6,.365);
\draw[fill=VertexColor,VertexColor] (.5,.365) circle (\vertexsize);
\draw[fill=VertexColor,VertexColor] (0,0) circle (\vertexsize);
\draw[fill=VertexColor,VertexColor] (.5,0) circle (\vertexsize);
\draw[fill=VertexColor,VertexColor] (1,0) circle (\vertexsize);
\draw[fill=VertexColor,VertexColor] (2,0) circle (\vertexsize);
\path[clip] (0,0) rectangle (2,1);
\draw[scalar] (.5,.085) circle (.28);
\draw[scalar] (1.5,.085) circle (.28);
\draw[wilson] (0,0) -- (2,0);
\draw[fill=white] (.25,0) circle (.075);
\draw[fill=white] (.75,0) circle (.075);
\draw[fill=white] (1.25,0) circle (.075);
\draw[fill=white] (1.75,0) circle (.075);
\draw[-Latex] (.265,.24)--(.305,.315);
\draw[-Latex] (.755,.195)--(.795,.12);
\draw[-Latex] (1.5,.365)--(1.6,.365);
\draw[fill=VertexColor,VertexColor] (.5,.365) circle (\vertexsize);
\draw[fill=VertexColor,VertexColor] (0,0) circle (\vertexsize);
\draw[fill=VertexColor,VertexColor] (.5,0) circle (\vertexsize);
\draw[fill=VertexColor,VertexColor] (1,0) circle (\vertexsize);
\draw[fill=VertexColor,VertexColor] (2,0) circle (\vertexsize);
\path[clip] (0,.365) rectangle (2,1);
\draw[scalar] (.5,.365) -- (.5,.7);
\draw[fill=VertexColor,VertexColor] (.5,.365) circle (\vertexsize);
\draw[fill=VertexColor,VertexColor] (0,0) circle (\vertexsize);
\draw[fill=VertexColor,VertexColor] (.5,0) circle (\vertexsize);
\draw[fill=VertexColor,VertexColor] (1,0) circle (\vertexsize);
\draw[fill=VertexColor,VertexColor] (2,0) circle (\vertexsize);
\end{tikzpicture}}
\newcommand{\DefectFFFFOneLoopYTwo}{\begin{tikzpicture}[baseline={([yshift=-2ex]current bounding box.center)}]
\draw[wilson] (-.2,0) -- (2.2,0);
\draw[fill=white] (.25,0) circle (.075);
\draw[fill=white] (.75,0) circle (.075);
\draw[fill=white] (1.25,0) circle (.075);
\draw[fill=white] (1.75,0) circle (.075);
\draw[-Latex] (.265,.24)--(.305,.315);
\draw[-Latex] (.755,.195)--(.795,.12);
\draw[-Latex] (1.5,.365)--(1.6,.365);
\draw[fill=VertexColor,VertexColor] (.5,.365) circle (\vertexsize);
\draw[fill=VertexColor,VertexColor] (0,0) circle (\vertexsize);
\draw[fill=VertexColor,VertexColor] (.5,0) circle (\vertexsize);
\draw[fill=VertexColor,VertexColor] (1,0) circle (\vertexsize);
\draw[fill=VertexColor,VertexColor] (2,0) circle (\vertexsize);
\path[clip] (0,0) rectangle (2,1);
\draw[scalar] (.5,.085) circle (.28);
\draw[scalar] (1.5,.085) circle (.28);
\draw[wilson] (0,0) -- (2,0);
\draw[fill=white] (.25,0) circle (.075);
\draw[fill=white] (.75,0) circle (.075);
\draw[fill=white] (1.25,0) circle (.075);
\draw[fill=white] (1.75,0) circle (.075);
\draw[fill=VertexColor,VertexColor] (.5,.365) circle (\vertexsize);
\draw[fill=VertexColor,VertexColor] (0,0) circle (\vertexsize);
\draw[fill=VertexColor,VertexColor] (.5,0) circle (\vertexsize);
\draw[fill=VertexColor,VertexColor] (1,0) circle (\vertexsize);
\draw[fill=VertexColor,VertexColor] (2,0) circle (\vertexsize);
\path[clip] (0,.365) rectangle (2,1);
\draw[gluon] (.5,.365) -- (.5,.7);
\draw[fill=VertexColor,VertexColor] (.5,.365) circle (\vertexsize);
\draw[fill=VertexColor,VertexColor] (0,0) circle (\vertexsize);
\draw[fill=VertexColor,VertexColor] (.5,0) circle (\vertexsize);
\draw[fill=VertexColor,VertexColor] (1,0) circle (\vertexsize);
\draw[fill=VertexColor,VertexColor] (2,0) circle (\vertexsize);
\end{tikzpicture}}
\newcommand{\DefectFFFFOneLoopYThree}{\begin{tikzpicture}[baseline={([yshift=-2ex]current bounding box.center)}]
\draw[wilson] (-.2,0) -- (2.2,0);
\draw[fill=white] (.25,0) circle (.075);
\draw[fill=white] (.75,0) circle (.075);
\draw[fill=white] (1.25,0) circle (.075);
\draw[fill=white] (1.75,0) circle (.075);
\draw[-Latex] (1.265,.24)--(1.305,.315);
\draw[-Latex] (1.755,.195)--(1.795,.12);
\draw[-Latex] (.5,.365)--(.6,.365);
\draw[fill=VertexColor,VertexColor] (1.5,.365) circle (\vertexsize);
\draw[fill=VertexColor,VertexColor] (0,0) circle (\vertexsize);
\draw[fill=VertexColor,VertexColor] (1.5,0) circle (\vertexsize);
\draw[fill=VertexColor,VertexColor] (1,0) circle (\vertexsize);
\draw[fill=VertexColor,VertexColor] (2,0) circle (\vertexsize);
\path[clip] (0,0) rectangle (2,1);
\draw[scalar] (.5,.085) circle (.28);
\draw[scalar] (1.5,.085) circle (.28);
\draw[wilson] (0,0) -- (2,0);
\draw[fill=white] (.25,0) circle (.075);
\draw[fill=white] (.75,0) circle (.075);
\draw[fill=white] (1.25,0) circle (.075);
\draw[fill=white] (1.75,0) circle (.075);
\draw[-Latex] (1.265,.24)--(1.305,.315);
\draw[-Latex] (1.755,.195)--(1.795,.12);
\draw[-Latex] (.5,.365)--(.6,.365);
\draw[fill=VertexColor,VertexColor] (1.5,.365) circle (\vertexsize);
\draw[fill=VertexColor,VertexColor] (0,0) circle (\vertexsize);
\draw[fill=VertexColor,VertexColor] (1.5,0) circle (\vertexsize);
\draw[fill=VertexColor,VertexColor] (1,0) circle (\vertexsize);
\draw[fill=VertexColor,VertexColor] (2,0) circle (\vertexsize);
\path[clip] (0,.365) rectangle (2,1);
\draw[scalar] (1.5,.365) -- (1.5,.7);
\draw[fill=VertexColor,VertexColor] (1.5,.365) circle (\vertexsize);
\draw[fill=VertexColor,VertexColor] (0,0) circle (\vertexsize);
\draw[fill=VertexColor,VertexColor] (1.5,0) circle (\vertexsize);
\draw[fill=VertexColor,VertexColor] (1,0) circle (\vertexsize);
\draw[fill=VertexColor,VertexColor] (2,0) circle (\vertexsize);
\end{tikzpicture}}
\newcommand{\DefectFFFFOneLoopYFour}{\begin{tikzpicture}[baseline={([yshift=-2ex]current bounding box.center)}]
\draw[wilson] (-.2,0) -- (2.2,0);
\draw[fill=white] (.25,0) circle (.075);
\draw[fill=white] (.75,0) circle (.075);
\draw[fill=white] (1.25,0) circle (.075);
\draw[fill=white] (1.75,0) circle (.075);
\draw[-Latex] (1.265,.24)--(1.305,.315);
\draw[-Latex] (1.755,.195)--(1.795,.12);
\draw[-Latex] (.5,.365)--(.6,.365);
\draw[fill=VertexColor,VertexColor] (1.5,.365) circle (\vertexsize);
\draw[fill=VertexColor,VertexColor] (0,0) circle (\vertexsize);
\draw[fill=VertexColor,VertexColor] (1.5,0) circle (\vertexsize);
\draw[fill=VertexColor,VertexColor] (1,0) circle (\vertexsize);
\draw[fill=VertexColor,VertexColor] (2,0) circle (\vertexsize);
\path[clip] (0,0) rectangle (2,1);
\draw[scalar] (.5,.085) circle (.28);
\draw[scalar] (1.5,.085) circle (.28);
\draw[wilson] (0,0) -- (2,0);
\draw[fill=white] (.25,0) circle (.075);
\draw[fill=white] (.75,0) circle (.075);
\draw[fill=white] (1.25,0) circle (.075);
\draw[fill=white] (1.75,0) circle (.075);
\draw[fill=VertexColor,VertexColor] (1.5,.365) circle (\vertexsize);
\draw[fill=VertexColor,VertexColor] (0,0) circle (\vertexsize);
\draw[fill=VertexColor,VertexColor] (1.5,0) circle (\vertexsize);
\draw[fill=VertexColor,VertexColor] (1,0) circle (\vertexsize);
\draw[fill=VertexColor,VertexColor] (2,0) circle (\vertexsize);
\path[clip] (0,.365) rectangle (2,1);
\draw[gluon] (1.5,.365) -- (1.5,.7);
\draw[fill=VertexColor,VertexColor] (1.5,.365) circle (\vertexsize);
\draw[fill=VertexColor,VertexColor] (0,0) circle (\vertexsize);
\draw[fill=VertexColor,VertexColor] (1.5,0) circle (\vertexsize);
\draw[fill=VertexColor,VertexColor] (1,0) circle (\vertexsize);
\draw[fill=VertexColor,VertexColor] (2,0) circle (\vertexsize);
\end{tikzpicture}}

\newcommand{\AnalyticContinuation}{\scalebox{1.04}{\begin{tikzpicture}[baseline={([yshift=-2ex]current bounding box.center)}]
\draw[scalar] (0,0) rectangle (13,6);
\draw[LightBlueGrey,fill=LightBlueGrey] (6.2,4) -- (6.8,4) -- (6.5,4.5) -- (6.2,4);
\draw[LightBlueGrey,fill=LightBlueGrey] (6.4,2) rectangle (6.6,4);
\draw[LightBlueGrey,fill=LightBlueGrey] (6.2,2) -- (6.8,2) -- (6.5,2-.5) -- (6.2,2);
\draw[LightBlueGrey,fill=LightBlueGrey] (3.25,2.7) -- (3.25,3.3) -- (3.25-.5,3) -- (3.25,2.7);
\draw[LightBlueGrey,fill=LightBlueGrey] (3.25,2.9) rectangle (6.5+3.25,3.1);
\draw[LightBlueGrey,fill=LightBlueGrey] (6.5+3.25,2.7) -- (6.5+3.25,3.3) -- (6.5+3.25+.5,3) -- (6.5+3.25,2.7);
\draw[wilson] (3.25,1.5) circle (.75);
\draw[wilson] (6.5+3.25,1.5) circle (.75);
\draw[wilson] (3.25,3+.75) -- (3.25,3+.75+1.5);
\draw[wilson] (6.5+3.25,3+.75) -- (6.5+3.25,3+.75+1.5);
\draw[fill=white] (6.5+3.25,3+.75+.75+.45) circle (\vertexsize);
\draw[fill=white] (6.5+3.25,3+.75+.75+.15) circle (\vertexsize);
\draw[fill=white] (6.5+3.25,3+.75+.75-.15) circle (\vertexsize);
\draw[fill=white] (6.5+3.25,3+.75+.75-.45) circle (\vertexsize);
\draw[fill=white] (6.5+3.25-.525,1.5+.525) circle (\vertexsize);
\draw[fill=white] (6.5+3.25-.525,1.5-.525) circle (\vertexsize);
\draw[fill=white] (6.5+3.25+.525,1.5+.525) circle (\vertexsize);
\draw[fill=white] (6.5+3.25+.525,1.5-.525) circle (\vertexsize);
\draw[fill=white] (3.25-.25,3+.75+.75-.15) circle (\vertexsize);
\draw[fill=white] (3.25+.25,3+.75+.75+.15) circle (\vertexsize);
\draw[fill=white] (3.25-.25,1.5-.15) circle (\vertexsize);
\draw[fill=white] (3.25+.25,1.5+.15) circle (\vertexsize);
\draw[scalar,fill=white] (6.5-.5,3.55-.25) rectangle (6.5+.5,3.55+.25);
\draw[scalar,fill=white] (6.5+1.5-.65,3-.25) rectangle (6.5+1.5+.65,3+.25);
\node[] at (6.5,5.5) {LINE DEFECT};
\node[] at (6.5,4.9) {\small ${\color{WCOrange} p=1}, {\color{SCRed} q=3}$};
\node[] at (6.5,.5) {BOUNDARY};
\node[] at (6.5,1.1) {\small ${\color{SCRed} p=3}, {\color{WCOrange} q=1}$};
\node[rotate=90,anchor=north] at (.2, 3) {TWO-POINT FUNCTIONS};
\node[rotate=-90,anchor=north] at (12.8, 3) {FOUR-POINT FUNCTIONS};
\node[WCOrange] at (2.4,6-.6) {\small $r,w$};
\node[SCRed] at (2.4+6.5-4.8,6-.6) {\small $\sigma$};
\node[SCRed] at (2.4,.6) {\small $r$};
\node[WCOrange] at (2.4+6.5-4.8,.6) {\small $\sigma, \sigmab$};
\node[SCRed] at (2.4+6.5,6-.6) {\small $\chi$};
\node[WCOrange] at (2.4+6.5+6.5-4.8,6-.6) {\small $r,s$};
\node[WCOrange] at (2.4+6.5,.6) {\small $\chi, \chib$};
\node[SCRed] at (2.4+6.5+6.5-4.8,.6) {\small $r$};
\node[] at (6.5,3.55) {\tiny $p \leftrightarrow q$};
\node[] at (6.5+1.5,3) {\tiny ST $\leftrightarrow$ $R$};
\end{tikzpicture}}}

\newcommand{\DefectTwoScalarsYOne}{\begin{tikzpicture}[baseline={([yshift=-2ex]current bounding box.center)}]
\draw[wilson] (0,0) -- (2,0);
\draw[fill=white] (.3,0) circle (.075);
\draw[fill=white] (1.7,0) circle (.075);
\path[clip] (0,0) rectangle (2,1);
\draw[scalar] (1,-.25) circle (.75);
\draw[wilson] (0,0) -- (2,0);
\draw[fill=white] (.3,0) circle (.075);
\draw[fill=white] (1.7,0) circle (.075);
\end{tikzpicture}}
\newcommand{\DefectTwoScalarsYTwo}{\begin{tikzpicture}[baseline={([yshift=-2ex]current bounding box.center)}]
\draw[fill=white] (1,.45) circle (.3);
\draw[scalar] (1,.45) circle (.3);
\draw[-Latex] (1,.75)--(1.1,.75);
\draw[-Latex] (.99,.15)--(.89,.15);
\draw[wilson] (0,0) -- (2,0);
\draw[fill=white] (.3,0) circle (.075);
\draw[fill=white] (1.7,0) circle (.075);
\draw[fill=VertexColor,VertexColor] (.7,.42) circle (\vertexsize);
\draw[fill=VertexColor,VertexColor] (1.3,.42) circle (\vertexsize);
\path[clip] (0,0) rectangle (2,1);
\draw[scalar] (1,-.25) circle (.75);
\draw[fill=white] (1,.45) circle (.3);
\draw[scalar] (1,.45) circle (.3);
\draw[-Latex] (1,.75)--(1.1,.75);
\draw[-Latex] (.99,.15)--(.89,.15);
\draw[wilson] (0,0) -- (2,0);
\draw[fill=white] (.3,0) circle (.075);
\draw[fill=white] (1.7,0) circle (.075);
\draw[fill=VertexColor,VertexColor] (.7,.42) circle (\vertexsize);
\draw[fill=VertexColor,VertexColor] (1.3,.42) circle (\vertexsize);
\end{tikzpicture}}
\newcommand{\DefectTwoScalarsYThree}{\begin{tikzpicture}[baseline={([yshift=-2ex]current bounding box.center)}]
\draw[scalar] (.75,0) -- (1,.5);
\draw[scalar] (1.25,0) -- (1,.5);
\draw[wilson] (0,0) -- (2,0);
\draw[fill=white] (.3,0) circle (.075);
\draw[fill=white] (1.7,0) circle (.075);
\draw[fill=VertexColor,VertexColor] (.75,0) circle (\vertexsize);
\draw[fill=VertexColor,VertexColor] (1.25,0) circle (\vertexsize);
\draw[fill=VertexColor,VertexColor] (1,.5) circle (\vertexsize);
\path[clip] (0,0) rectangle (2,1);
\draw[scalar] (1,-.25) circle (.75);
\draw[scalar] (.75,0) -- (1,.5);
\draw[scalar] (1.25,0) -- (1,.5);
\draw[wilson] (0,0) -- (2,0);
\draw[fill=white] (.3,0) circle (.075);
\draw[fill=white] (1.7,0) circle (.075);
\draw[fill=VertexColor,VertexColor] (.75,0) circle (\vertexsize);
\draw[fill=VertexColor,VertexColor] (1.25,0) circle (\vertexsize);
\draw[fill=VertexColor,VertexColor] (1,.5) circle (\vertexsize);
\end{tikzpicture}}

\newcommand{\DefectTwoFermionsYOne}{\begin{tikzpicture}[baseline={([yshift=-2ex]current bounding box.center)}]
\draw[wilson] (0,0) -- (2,0);
\draw[fill=white] (.3,0) circle (.075);
\draw[fill=white] (1.7,0) circle (.075);
\draw[-Latex] (.99,.5)--(.89,.5);
\path[clip] (0,0) rectangle (2,1);
\draw[scalar] (1,-.25) circle (.75);
\draw[wilson] (0,0) -- (2,0);
\draw[fill=white] (.3,0) circle (.075);
\draw[fill=white] (1.7,0) circle (.075);
\end{tikzpicture}}
\newcommand{\DefectTwoFermionsYTwo}{\begin{tikzpicture}[baseline={([yshift=-2ex]current bounding box.center)}]
\draw[wilson] (0,0) -- (2,0);
\draw[fill=white] (.3,0) circle (.075);
\draw[fill=white] (1.7,0) circle (.075);
\draw[scalar] (1,0) -- (1,.5);
\draw[fill=VertexColor,VertexColor] (1,.5) circle (\vertexsize);
\draw[fill=VertexColor,VertexColor] (1,0) circle (\vertexsize);
\draw[-Latex] (.59,.38)--(.49,.33);
\draw[-Latex] (1.44,.35)--(1.34,.42);
\path[clip] (0,0) rectangle (2,1);
\draw[scalar] (1,-.25) circle (.75);
\draw[wilson] (0,0) -- (2,0);
\draw[fill=white] (.3,0) circle (.075);
\draw[fill=white] (1.7,0) circle (.075);
\draw[fill=VertexColor,VertexColor] (1,.5) circle (\vertexsize);
\draw[fill=VertexColor,VertexColor] (1,0) circle (\vertexsize);
\end{tikzpicture}}
\newcommand{\DefectTwoFermionsYThree}{\begin{tikzpicture}[baseline={([yshift=-2ex]current bounding box.center)}]
\draw[fill=white] (1,.45) circle (.3);
\draw[scalar] (1,.45) circle (.3);
\draw[-Latex] (.47,.27)--(.37,.18);
\draw[-Latex] (1.56,.25)--(1.46,.35);
\draw[-Latex] (.99,.15)--(.89,.15);
\draw[wilson] (0,0) -- (2,0);
\draw[fill=white] (.3,0) circle (.075);
\draw[fill=white] (1.7,0) circle (.075);
\draw[fill=VertexColor,VertexColor] (.7,.42) circle (\vertexsize);
\draw[fill=VertexColor,VertexColor] (1.3,.42) circle (\vertexsize);
\path[clip] (0,0) rectangle (2,1);
\draw[scalar] (1,-.25) circle (.75);
\draw[fill=white] (1,.45) circle (.3);
\draw[scalar] (1,.45) circle (.3);
\draw[-Latex] (.99,.15)--(.89,.15);
\draw[wilson] (0,0) -- (2,0);
\draw[fill=white] (.3,0) circle (.075);
\draw[fill=white] (1.7,0) circle (.075);
\draw[fill=VertexColor,VertexColor] (.7,.42) circle (\vertexsize);
\draw[fill=VertexColor,VertexColor] (1.3,.42) circle (\vertexsize);
\end{tikzpicture}}
\newcommand{\DefectTwoFermionsYFour}{\begin{tikzpicture}[baseline={([yshift=-2ex]current bounding box.center)}]
\draw[scalar] (.7,0) -- (.7,.42);
\draw[scalar] (1.3,0) -- (1.3,.42);
\draw[-Latex] (.47,.27)--(.37,.18);
\draw[-Latex] (1.56,.25)--(1.46,.35);
\draw[-Latex] (.99,.5)--(.89,.5);
\draw[wilson] (0,0) -- (2,0);
\draw[fill=white] (.3,0) circle (.075);
\draw[fill=white] (1.7,0) circle (.075);
\draw[fill=VertexColor,VertexColor] (.7,.42) circle (\vertexsize);
\draw[fill=VertexColor,VertexColor] (1.3,.42) circle (\vertexsize);
\draw[fill=VertexColor,VertexColor] (.7,0) circle (\vertexsize);
\draw[fill=VertexColor,VertexColor] (1.3,0) circle (\vertexsize);
\path[clip] (0,0) rectangle (2,1);
\draw[scalar] (1,-.25) circle (.75);
\draw[-Latex] (.99,.5)--(.89,.5);
\draw[wilson] (0,0) -- (2,0);
\draw[fill=white] (.3,0) circle (.075);
\draw[fill=white] (1.7,0) circle (.075);
\draw[fill=VertexColor,VertexColor] (.7,.42) circle (\vertexsize);
\draw[fill=VertexColor,VertexColor] (1.3,.42) circle (\vertexsize);
\draw[fill=VertexColor,VertexColor] (.7,0) circle (\vertexsize);
\draw[fill=VertexColor,VertexColor] (1.3,0) circle (\vertexsize);
\end{tikzpicture}}

\newcommand{\DefectThreeScalarsY}{\begin{tikzpicture}[baseline={([yshift=-2ex]current bounding box.center)}]
\draw[scalar] (.75,0) -- (1,.5);
\draw[scalar] (1.25,0) -- (1,.5);
\draw[wilson] (0,0) -- (2,0);
\draw[fill=white] (.3,0) circle (.075);
\draw[fill=white] (.75,0) circle (.075);
\draw[fill=white] (1.25,0) circle (.075);
\draw[fill=VertexColor,VertexColor] (1.7,0) circle (\vertexsize);
\draw[fill=VertexColor,VertexColor] (1,.5) circle (\vertexsize);
\path[clip] (0,0) rectangle (2,1);
\draw[scalar] (1,-.25) circle (.75);
\draw[scalar] (.75,0) -- (1,.5);
\draw[scalar] (1.25,0) -- (1,.5);
\draw[wilson] (0,0) -- (2,0);
\draw[fill=white] (.3,0) circle (.075);
\draw[fill=white] (.75,0) circle (.075);
\draw[fill=white] (1.25,0) circle (.075);
\draw[fill=VertexColor,VertexColor] (1.7,0) circle (\vertexsize);
\draw[fill=VertexColor,VertexColor] (1,.5) circle (\vertexsize);
\end{tikzpicture}}

\newcommand{\DefectOneScalarTwoDisplacementsYTwo}{\begin{tikzpicture}[baseline={([yshift=-2ex]current bounding box.center)}]
\draw[scalar] (1,0) -- (1,.5);
\draw[scalar] (1.7,0) -- (1,.5);
\draw[wilson] (0,0) -- (2,0);
\draw[fill=white] (.3,0) circle (.075);
\draw[fill=white] (1,0) circle (.075);
\draw[fill=white] (1.7,0) circle (.075);
\draw[fill=VertexColor,VertexColor] (1,.5) circle (\vertexsize);
\path[clip] (0,0) rectangle (2,1);
\draw[scalar] (1,-.25) circle (.75);
\draw[wilson] (0,0) -- (2,0);
\draw[fill=white] (.3,0) circle (.075);
\draw[fill=white] (1,0) circle (.075);
\draw[fill=white] (1.7,0) circle (.075);
\draw[fill=VertexColor,VertexColor] (1,.5) circle (\vertexsize);
\end{tikzpicture}}

\newcommand{\DefectTwoFermionsOneScalarY}{\begin{tikzpicture}[baseline={([yshift=-2ex]current bounding box.center)}]
\draw[wilson] (0,0) -- (2,0);
\draw[scalar] (1,0) -- (1,.5);
\draw[fill=white] (.3,0) circle (.075);
\draw[fill=white] (1,0) circle (.075);
\draw[fill=white] (1.7,0) circle (.075);
\draw[fill=VertexColor,VertexColor] (1,.5) circle (\vertexsize);
\draw[-Latex] (.59,.38)--(.49,.33);
\draw[-Latex] (1,.25)--(1,.35);
\path[clip] (0,0) rectangle (2,1);
\draw[scalar] (1,-.25) circle (.75);
\draw[wilson] (0,0) -- (2,0);
\draw[fill=white] (.3,0) circle (.075);
\draw[fill=white] (1,0) circle (.075);
\draw[fill=white] (1.7,0) circle (.075);
\draw[fill=VertexColor,VertexColor] (1,.5) circle (\vertexsize);
\end{tikzpicture}}

\newcommand{\DefectFourFermionsHOne}{\begin{tikzpicture}[baseline={([yshift=-2ex]current bounding box.center)}]
\draw[scalar] (.75,0) -- (.75,.42+.075);
\draw[scalar] (1.25,0) -- (1.25,.42+.075);
\draw[-Latex] (.47,.27+.075)--(.37,.18+.075);
\draw[-Latex] (1.65-.075,.15+.075+.075)--(1.55-.075,.25+.075+.075);
\draw[-Latex] (.75,.25)--(.75,.35);
\draw[-Latex] (1.25,.25)--(1.25,.15);
\draw[wilson] (0,0) -- (2,0);
\draw[fill=white] (.25,0) circle (.075);
\draw[fill=white] (.75,0) circle (.075);
\draw[fill=white] (1.25,0) circle (.075);
\draw[fill=white] (1.75,0) circle (.075);
\draw[fill=VertexColor,VertexColor] (.75,.42+.075) circle (\vertexsize);
\draw[fill=VertexColor,VertexColor] (1.25,.42+.075) circle (\vertexsize);
\path[clip] (0,0) rectangle (2,1);
\draw[scalar] (1,-.25) circle (.8);
\draw[wilson] (0,0) -- (2,0);
\draw[fill=white] (.25,0) circle (.075);
\draw[fill=white] (.75,0) circle (.075);
\draw[fill=white] (1.25,0) circle (.075);
\draw[fill=white] (1.75,0) circle (.075);
\draw[fill=VertexColor,VertexColor] (.75,.42+.075) circle (\vertexsize);
\draw[fill=VertexColor,VertexColor] (1.25,.42+.075) circle (\vertexsize);
\end{tikzpicture}}
\newcommand{\DefectFourFermionsHTwo}{\begin{tikzpicture}[baseline={([yshift=-2ex]current bounding box.center)}]
\draw[wilson] (0,0) -- (2,0);
\draw[fill=white] (.25,0) circle (.075);
\draw[fill=white] (.75,0) circle (.075);
\draw[fill=white] (1.25,0) circle (.075);
\draw[fill=white] (1.75,0) circle (.075);
\draw[-Latex] (.47,.27+.075+.05)--(.37,.18+.075+.05);
\draw[-Latex] (1.65-.075,.15+.075+.075+.05)--(1.55-.075,.25+.075+.075+.05);
\draw[-Latex] (.37+.44,.18-.0175)--(.47+.44,.27-.0175);
\draw[-Latex] (1.55-.4,.25-.05)--(1.65-.4,.15-.05);
\draw[fill=VertexColor,VertexColor] (1,.6) circle (\vertexsize);
\draw[fill=VertexColor,VertexColor] (1,.245) circle (\vertexsize);
\path[clip] (0,0) rectangle (2,1);
\draw[scalar] (1,-.2) circle (.8);
\draw[scalar] (1,-.025) circle (.27);
\draw[wilson] (0,0) -- (2,0);
\draw[fill=white] (.25,0) circle (.075);
\draw[fill=white] (.75,0) circle (.075);
\draw[fill=white] (1.25,0) circle (.075);
\draw[fill=white] (1.75,0) circle (.075);
\draw[fill=VertexColor,VertexColor] (1,.6) circle (\vertexsize);
\draw[fill=VertexColor,VertexColor] (1,.245) circle (\vertexsize);
\path[clip] (.8,.245) rectangle (1.2,.6);
\draw[scalar] (1,.6) -- (1, .245);
\draw[fill=VertexColor,VertexColor] (1,.6) circle (\vertexsize);
\draw[fill=VertexColor,VertexColor] (1,.245) circle (\vertexsize);
\end{tikzpicture}}

\newcommand{\DefectFFSSHOne}{\begin{tikzpicture}[baseline={([yshift=-2ex]current bounding box.center)}]
\draw[scalar] (.75,0) -- (.75,.42+.075);
\draw[scalar] (1.25,0) -- (1.25,.42+.075);
\draw[-Latex] (.47,.27+.075)--(.37,.18+.075);
\draw[-Latex] (1.56,.25+.075)--(1.46,.35+.075);
\draw[-Latex] (.99,.5+.055)--(.89,.5+.055);
\draw[wilson] (0,0) -- (2,0);
\draw[fill=white] (.25,0) circle (.075);
\draw[fill=white] (.75,0) circle (.075);
\draw[fill=white] (1.25,0) circle (.075);
\draw[fill=white] (1.75,0) circle (.075);
\draw[fill=VertexColor,VertexColor] (.75,.42+.075) circle (\vertexsize);
\draw[fill=VertexColor,VertexColor] (1.25,.42+.075) circle (\vertexsize);
\path[clip] (0,0) rectangle (2,1);
\draw[scalar] (1,-.25) circle (.8);
\draw[-Latex] (.99,.5+.055)--(.89,.5+.055);
\draw[wilson] (0,0) -- (2,0);
\draw[fill=white] (.25,0) circle (.075);
\draw[fill=white] (.75,0) circle (.075);
\draw[fill=white] (1.25,0) circle (.075);
\draw[fill=white] (1.75,0) circle (.075);
\draw[fill=VertexColor,VertexColor] (.75,.42+.075) circle (\vertexsize);
\draw[fill=VertexColor,VertexColor] (1.25,.42+.075) circle (\vertexsize);
\end{tikzpicture}}
\newcommand{\DefectFFSSHTwo}{\begin{tikzpicture}[baseline={([yshift=-2ex]current bounding box.center)}]
\draw[scalar] (.75,0) -- (1.25,.42+.075);
\draw[white,fill=white] (1,.21+.0325) circle (.075);
\draw[scalar] (1.25,0) -- (.75,.42+.075);
\draw[-Latex] (.47,.27+.075)--(.37,.18+.075);
\draw[-Latex] (1.56,.25+.075)--(1.46,.35+.075);
\draw[-Latex] (.99,.5+.055)--(.89,.5+.055);
\draw[wilson] (0,0) -- (2,0);
\draw[fill=white] (.25,0) circle (.075);
\draw[fill=white] (.75,0) circle (.075);
\draw[fill=white] (1.25,0) circle (.075);
\draw[fill=white] (1.75,0) circle (.075);
\draw[fill=VertexColor,VertexColor] (.75,.42+.075) circle (\vertexsize);
\draw[fill=VertexColor,VertexColor] (1.25,.42+.075) circle (\vertexsize);
\path[clip] (0,0) rectangle (2,1);
\draw[scalar] (1,-.25) circle (.8);
\draw[-Latex] (.99,.5+.055)--(.89,.5+.055);
\draw[wilson] (0,0) -- (2,0);
\draw[fill=white] (.25,0) circle (.075);
\draw[fill=white] (.75,0) circle (.075);
\draw[fill=white] (1.25,0) circle (.075);
\draw[fill=white] (1.75,0) circle (.075);
\draw[fill=VertexColor,VertexColor] (.75,.42+.075) circle (\vertexsize);
\draw[fill=VertexColor,VertexColor] (1.25,.42+.075) circle (\vertexsize);
\end{tikzpicture}}

\newcommand{\BulkSSOnePointOne}{\begin{tikzpicture}[baseline={([yshift=-2ex]current bounding box.center)}]
\draw[wilson] (.25,0) -- (1.75,0);
\draw[scalar] (.55,0) -- (1,1);
\draw[scalar] (1.45,0) -- (1,1);
\draw[fill=white] (1,1) circle (.075);
\draw[fill=VertexColor,VertexColor] (.55,0) circle (\vertexsize);
\draw[fill=VertexColor,VertexColor] (1.45,0) circle (\vertexsize);
\end{tikzpicture}}
\newcommand{\BulkSSOnePointTwo}{\begin{tikzpicture}[baseline={([yshift=-2ex]current bounding box.center)}]
\draw[wilson] (.25,0) -- (1.75,0);
\draw[scalar] (.55,0) -- (1,1);
\draw[scalar] (1.45,0) -- (1,1);
\draw[fill=white,scalar] (.775,.5) circle (.2);
\draw[-Latex] (.45+.025+.1275,.5+.05+.025)--(.45-.025+.1275,.5-.05+.025);
\draw[-Latex] (1.095-.025-.1225,.5-.05-.025)--(1.095+.025-.1225,.5+.05-.025);
\draw[fill=white] (1,1) circle (.075);
\draw[fill=VertexColor,VertexColor] (.55,0) circle (\vertexsize);
\draw[fill=VertexColor,VertexColor] (1.45,0) circle (\vertexsize);
\draw[fill=VertexColor,VertexColor] (.775-.08,.3+.025) circle (\vertexsize);
\draw[fill=VertexColor,VertexColor] (.775+.08,.7-.025) circle (\vertexsize);
\end{tikzpicture}}
\newcommand{\BulkSSOnePointThree}{\begin{tikzpicture}[baseline={([yshift=-2ex]current bounding box.center)}]
\draw[wilson] (.25,0) -- (1.75,0);
\draw[scalar] (.55,0) -- (1,1);
\draw[scalar] (1.45,0) -- (1,1);
\draw[fill=white,scalar] (1.225,.5) circle (.2);
\draw[-Latex] (.45+.45-.025+.1275+.04,.5+.05-.125)--(.45+.45+.025+.1275+.04,.5-.05-.125);
\draw[-Latex] (.45+1.095+.025-.1225-.04,.5-.05-.025+.15)--(.45+1.095-.025-.1225-.04,.5+.05-.025+.15);
\draw[fill=white] (1,1) circle (.075);
\draw[fill=VertexColor,VertexColor] (.55,0) circle (\vertexsize);
\draw[fill=VertexColor,VertexColor] (1.45,0) circle (\vertexsize);
\draw[fill=VertexColor,VertexColor] (1.225+.08,.3+.025) circle (\vertexsize);
\draw[fill=VertexColor,VertexColor] (1.225-.08,.7-.025) circle (\vertexsize);
\end{tikzpicture}}
\newcommand{\BulkSSOnePointFour}{\begin{tikzpicture}[baseline={([yshift=-2ex]current bounding box.center)}]
\draw[wilson] (.25,0) -- (1.75,0);
\draw[scalar] (.55,0) -- (1,1);
\draw[scalar] (1.45,0) -- (1,1);
\draw[scalar] (.775,0) -- (.775,.5);
\draw[scalar] (1,0) -- (.775,.5);
\draw[fill=white] (1,1) circle (.075);
\draw[fill=VertexColor,VertexColor] (.55,0) circle (\vertexsize);
\draw[fill=VertexColor,VertexColor] (.775,0) circle (\vertexsize);
\draw[fill=VertexColor,VertexColor] (1,0) circle (\vertexsize);
\draw[fill=VertexColor,VertexColor] (.775,.5) circle (\vertexsize);
\draw[fill=VertexColor,VertexColor] (1.45,0) circle (\vertexsize);
\end{tikzpicture}}
\newcommand{\BulkSSOnePointFive}{\begin{tikzpicture}[baseline={([yshift=-2ex]current bounding box.center)}]
\draw[wilson] (.25,0) -- (1.75,0);
\draw[scalar] (.55,0) -- (1,1);
\draw[scalar] (1.45,0) -- (1,1);
\draw[scalar] (1,0) -- (1.225,.5);
\draw[scalar] (1.225,0) -- (1.225,.5);
\draw[fill=white] (1,1) circle (.075);
\draw[fill=VertexColor,VertexColor] (.55,0) circle (\vertexsize);
\draw[fill=VertexColor,VertexColor] (1,0) circle (\vertexsize);
\draw[fill=VertexColor,VertexColor] (1.225,0) circle (\vertexsize);
\draw[fill=VertexColor,VertexColor] (1.45,0) circle (\vertexsize);
\draw[fill=VertexColor,VertexColor] (1.225,.5) circle (\vertexsize);
\end{tikzpicture}}
\newcommand{\BulkSSOnePointSix}{\begin{tikzpicture}[baseline={([yshift=-2ex]current bounding box.center)}]
\draw[wilson] (.25,0) -- (1.75,0);
\draw[scalar] (1,.75) circle (.25);
\draw[scalar] (.55,0) -- (1,.5);
\draw[scalar] (1.45,0) -- (1,.5);
\draw[fill=white] (1,1) circle (.075);
\draw[fill=VertexColor,VertexColor] (.55,0) circle (\vertexsize);
\draw[fill=VertexColor,VertexColor] (1.45,0) circle (\vertexsize);
\draw[fill=VertexColor,VertexColor] (1,.5) circle (\vertexsize);
\end{tikzpicture}}

\newcommand{\BulkFFOnePoint}{\begin{tikzpicture}[baseline={([yshift=-2ex]current bounding box.center)}]
\draw[wilson] (.5,0) -- (1.5,0);
\draw[scalar] (1,.75) circle (.25);
\draw[-Latex] (.75,.71+.05) -- (.75,.71-.05);
\draw[-Latex] (1.245,.71-.05+.09) -- (1.245,.71+.05+.09);
\draw[scalar] (1,0) -- (1,.5);
\draw[fill=white] (1,1) circle (.075);
\draw[fill=VertexColor,VertexColor] (1,0) circle (\vertexsize);
\draw[fill=VertexColor,VertexColor] (1,.5) circle (\vertexsize);
\end{tikzpicture}}

\newcommand{\BulkDefectTwoPointOne}{\begin{tikzpicture}[baseline={([yshift=-2ex]current bounding box.center)}]
\draw[wilson] (.25,0) -- (1.75,0);
\draw[scalar] (1,0) -- (1,1);
\draw[fill=white] (1,1) circle (.075);
\draw[fill=white] (1,0) circle (.075);
\end{tikzpicture}}
\newcommand{\BulkDefectTwoPointTwo}{\begin{tikzpicture}[baseline={([yshift=-2ex]current bounding box.center)}]
\draw[wilson] (.25,0) -- (1.75,0);
\draw[scalar] (1,0) -- (1,1);
\draw[scalar] (.5,0) -- (1,.5);
\draw[scalar] (1.5,0) -- (1,.5);
\draw[fill=white] (1,1) circle (.075);
\draw[fill=white] (1,0) circle (.075);
\draw[fill=VertexColor,VertexColor] (.5,0) circle (\vertexsize);
\draw[fill=VertexColor,VertexColor] (1.5,0) circle (\vertexsize);
\draw[fill=VertexColor,VertexColor] (1,.5) circle (\vertexsize);
\end{tikzpicture}}
\newcommand{\BulkDefectTwoPointThree}{\begin{tikzpicture}[baseline={([yshift=-2ex]current bounding box.center)}]
\draw[wilson] (.25,0) -- (1.75,0);
\draw[scalar] (1,0) -- (1,1);
\draw[fill=white] (1,1) circle (.075);
\draw[fill=white] (1,0) circle (.075);
\draw[fill=white,scalar] (1,.75-.25) circle (.25);
\draw[-Latex] (.75,.71+.05-.25) -- (.75,.71-.05-.25);
\draw[-Latex] (1.245,.71-.05+.09-.25) -- (1.245,.71+.05+.09-.25);
\draw[fill=VertexColor,VertexColor] (1,.25) circle (\vertexsize);
\draw[fill=VertexColor,VertexColor] (1,.75) circle (\vertexsize);
\end{tikzpicture}}

\newcommand{\TwoPointNLO}{\begin{tikzpicture}[baseline={([yshift=-1ex]current bounding box.center)}]
\draw[wilson] (0,0) -- (2,0);
\draw[scalar] (1.5,0) -- (.5,1);
\draw[scalar] (.5,0) -- (1.5,1);
\draw[fill=white] (.5,1) circle (.075);
\draw[fill=white] (1.5,1) circle (.075);
\draw[fill=VertexColor,VertexColor] (1,.5) circle (\vertexsize);
\draw[fill=VertexColor,VertexColor] (.5,0) circle (\vertexsize);
\draw[fill=VertexColor,VertexColor] (1.5,0) circle (\vertexsize);
\end{tikzpicture}}

\newcommand{\TwoPointFFLO}{\begin{tikzpicture}[baseline={([yshift=-1ex]current bounding box.center)}]
\draw[wilson] (.5,0) -- (1.5,0);
\draw[scalar] (1,.5) -- (.5,1);
\draw[-Latex] (.75+.05-.025,.75-.05+.025) -- (.75-.05-.025,.75+.05+.025);
\draw[-Latex] (1.25+.05-.025,.75+.05-.025) -- (1.25-.05-.025,.75-.05-.025);
\draw[scalar] (1,.5) -- (1.5,1);
\draw[scalar] (1,.5) -- (1,0);
\draw[fill=white] (.5,1) circle (.075);
\draw[fill=white] (1.5,1) circle (.075);
\draw[fill=VertexColor,VertexColor] (1,.5) circle (\vertexsize);
\draw[fill=VertexColor,VertexColor] (1,0) circle (\vertexsize);
\end{tikzpicture}}

\newcommand{\TwoPointFFNLO}{\begin{tikzpicture}[baseline={([yshift=-1ex]current bounding box.center)}]
\draw[wilson] (.25,0) -- (1.75,0);
\draw[scalar] (.5,0) -- (.5,1.5);
\draw[scalar] (1.5,0) -- (1.5,1.5);
\draw[scalar] (.5,.75) -- (1.5,.75);
\draw[-Latex] (.5,.75-.05+.05+.375) -- (.5,.75+.05+.05+.375);
\draw[-Latex] (1.5,.75+.05-.1+.375) -- (1.5,.75-.05-.1+.375);
\draw[-Latex] (1+.05-.075,.5+.25) -- (1-.05-.075,.5+.25);
\draw[fill=white] (.5,1.5) circle (.075);
\draw[fill=white] (1.5,1.5) circle (.075);
\draw[fill=VertexColor,VertexColor] (.5,0) circle (\vertexsize);
\draw[fill=VertexColor,VertexColor] (1.5,0) circle (\vertexsize);
\draw[fill=VertexColor,VertexColor] (.5,.75) circle (\vertexsize);
\draw[fill=VertexColor,VertexColor] (1.5,.75) circle (\vertexsize);
\end{tikzpicture}}

\newcommand{\TwoPointNoClosedFormOne}{\begin{tikzpicture}[baseline={([yshift=-1ex]current bounding box.center)}]
\draw[wilson] (.25,0) -- (1.75,0);
\draw[scalar] (.5,0) -- (.5,1.5);
\draw[scalar] (1.5,0) -- (1.5,1.5);
\draw[gluon] (.5,.75) -- (1.5,.75);
\draw[fill=white] (.5,1.5) circle (.075);
\draw[fill=white] (1.5,1.5) circle (.075);
\draw[fill=VertexColor,VertexColor] (.5,0) circle (\vertexsize);
\draw[fill=VertexColor,VertexColor] (1.5,0) circle (\vertexsize);
\draw[fill=VertexColor,VertexColor] (.5,.75) circle (\vertexsize);
\draw[fill=VertexColor,VertexColor] (1.5,.75) circle (\vertexsize);
\end{tikzpicture}}
\newcommand{\TwoPointNoClosedFormTwo}{\begin{tikzpicture}[baseline={([yshift=-1ex]current bounding box.center)}]
\draw[wilson] (.25,0) -- (1.75,0);
\draw[scalar] (.5,.75) -- (.5,1.5);
\draw[scalar] (1.5,.75) -- (1.5,1.5);
\draw[gluon] (.5,0) -- (.5,.75);
\draw[gluon] (1.5,0) -- (1.5,.75);
\draw[scalar] (.5,.75) -- (1.5,.75);
\draw[fill=white] (.5,1.5) circle (.075);
\draw[fill=white] (1.5,1.5) circle (.075);
\draw[fill=VertexColor,VertexColor] (.5,0) circle (\vertexsize);
\draw[fill=VertexColor,VertexColor] (1.5,0) circle (\vertexsize);
\draw[fill=VertexColor,VertexColor] (.5,.75) circle (\vertexsize);
\draw[fill=VertexColor,VertexColor] (1.5,.75) circle (\vertexsize);
\end{tikzpicture}}

\newcommand{\FourPointSSFFNLOOne}{\begin{tikzpicture}[baseline={([yshift=-.5ex]current bounding box.center)}]
\draw[scalar] (.5,0) -- (.5,1.5);
\draw[scalar] (1.5,0) -- (1.5,1.5);
\draw[scalar] (.5,.75) -- (1.5,.75);
\draw[-Latex] (.5,.75-.05+.05+.375) -- (.5,.75+.05+.05+.375);
\draw[-Latex] (1.5,.75+.05-.1+.375) -- (1.5,.75-.05-.1+.375);
\draw[-Latex] (1+.05-.075,.5+.25) -- (1-.05-.075,.5+.25);
\draw[fill=white] (.5,1.5) circle (.075);
\draw[fill=white] (1.5,1.5) circle (.075);
\draw[fill=white] (.5,0) circle (.075);
\draw[fill=white] (1.5,0) circle (.075);
\draw[fill=VertexColor,VertexColor] (.5,.75) circle (\vertexsize);
\draw[fill=VertexColor,VertexColor] (1.5,.75) circle (\vertexsize);
\end{tikzpicture}}
\newcommand{\FourPointSSFFNLOTwo}{\begin{tikzpicture}[baseline={([yshift=-.5ex]current bounding box.center)}]
\draw[scalar] (.5,.75) -- (.5,1.5);
\draw[scalar] (1.5,.75) -- (1.5,1.5);
\draw[scalar] (.5,.75) -- (1.5,.75);
\draw[scalar] (.5,.75) -- (1.5,0);
\draw[white,fill=white] (1,.375) circle (.075);
\draw[scalar] (.5,0) -- (1.5,.75);
\draw[-Latex] (.5,.75-.05+.05+.375) -- (.5,.75+.05+.05+.375);
\draw[-Latex] (1.5,.75+.05-.1+.375) -- (1.5,.75-.05-.1+.375);
\draw[-Latex] (1+.05-.075,.5+.25) -- (1-.05-.075,.5+.25);
\draw[fill=white] (.5,1.5) circle (.075);
\draw[fill=white] (1.5,1.5) circle (.075);
\draw[fill=white] (.5,0) circle (.075);
\draw[fill=white] (1.5,0) circle (.075);
\draw[fill=VertexColor,VertexColor] (.5,.75) circle (\vertexsize);
\draw[fill=VertexColor,VertexColor] (1.5,.75) circle (\vertexsize);
\end{tikzpicture}}

\newcommand{\XIntegral}{\begin{tikzpicture}[baseline={([yshift=-.35ex]current bounding box.center)}]
\draw[scalar] (0,0) -- (1.25,1);
\draw[scalar] (0,1) -- (1.25,0);
\draw[fill=VertexColor,VertexColor] (.625,.5) circle (\vertexsize);
\draw[fill=white] (0,0) circle (.075);
\draw[fill=white] (0,1) circle (.075);
\draw[fill=white] (1.25,1) circle (.075);
\draw[fill=white] (1.25,0) circle (.075);
\end{tikzpicture}}
\newcommand{\YIntegral}{\begin{tikzpicture}[baseline={([yshift=-.45ex]current bounding box.center)}]
\draw[scalar] (0,.5) -- (.75,.5);
\draw[scalar] (.75,.5) -- (1.25,1);
\draw[scalar] (.75,.5) -- (1.25,0);
\draw[fill=VertexColor,VertexColor] (.75,.5) circle (\vertexsize);
\draw[fill=white] (0,.5) circle (.075);
\draw[fill=white] (1.25,1) circle (.075);
\draw[fill=white] (1.25,0) circle (.075);
\end{tikzpicture}}
\newcommand{\HIntegral}{\begin{tikzpicture}[baseline={([yshift=-1ex]current bounding box.center)}]
\draw[scalar] (0,0) -- (1.25,0);
\draw[scalar] (0,1) -- (1.25,1);
\draw[scalar] (.625,0) -- (.625,1);
\draw[fill=VertexColor,VertexColor] (.625,0) circle (\vertexsize);
\draw[fill=VertexColor,VertexColor] (.625,1) circle (\vertexsize);
\draw[fill=white] (0,0) circle (.075);
\draw[fill=white] (0,1) circle (.075);
\draw[fill=white] (1.25,1) circle (.075);
\draw[fill=white] (1.25,0) circle (.075);
\end{tikzpicture}}
\newcommand{\IntegralYOneTwoTwo}{\scalebox{\x}{\begin{tikzpicture}[baseline={([yshift=-.55ex]current bounding box.center)}]
\draw[scalar] (0,.5) -- (1,.5);
\draw[scalar] (1.5,.5) circle (.5);
\draw[fill=white] (0,.5) circle (.075);
\draw[fill=white] (2,.5) circle (.075);
\draw[fill=VertexColor,VertexColor] (1,.5) circle (\vertexsize);
\end{tikzpicture}}}
\newcommand{\IntegralXOneTwoThreeThree}{\scalebox{\x}{\begin{tikzpicture}[baseline={([yshift=-.55ex]current bounding box.center)}]
\draw[scalar] (.5,1) -- (1,.5);
\draw[scalar] (.5,0) -- (1,.5);
\draw[scalar] (1.5,.5) circle (.5);
\draw[fill=white] (.5,1) circle (.075);
\draw[fill=white] (.5,0) circle (.075);
\draw[fill=white] (2,.5) circle (.075);
\draw[fill=VertexColor,VertexColor] (1,.5) circle (\vertexsize);
\end{tikzpicture}}}
\newcommand{\IntegralXOneOneTwoTwo}{\scalebox{\x}{\begin{tikzpicture}[baseline={([yshift=-.55ex]current bounding box.center)}]
\draw[scalar] (.5,.5) circle (.5);
\draw[scalar] (1.5,.5) circle (.5);
\draw[fill=white] (0,.5) circle (.075);
\draw[fill=white] (2,.5) circle (.075);
\draw[fill=VertexColor,VertexColor] (1,.5) circle (\vertexsize);
\end{tikzpicture}}}
\newcommand{\IntegralFOneThreeTwoThree}{\scalebox{\x}{\begin{tikzpicture}[baseline={([yshift=-.55ex]current bounding box.center)}]
\draw[scalar] (0,0) -- (.625,0);
\draw[scalar] (0,1) -- (.625,1);
\draw[scalar] (.625,0) arc (270:450:.5);
\draw[scalar] (.625,0) -- (.625,1);
\draw[fill=VertexColor,VertexColor] (.625,0) circle (\vertexsize);
\draw[fill=VertexColor,VertexColor] (.625,1) circle (\vertexsize);
\draw[fill=white] (0,0) circle (.075);
\draw[fill=white] (0,1) circle (.075);
\draw[fill=white] (1.125,.5) circle (.075);
\end{tikzpicture}}}

\newcommand{\KiteIntegral}{\begin{tikzpicture}[baseline={([yshift=-.35ex]current bounding box.center)}]
\draw[scalar] (0,0) -- (2,0);
\draw[scalar] (.5,0) -- (1,1);
\draw[scalar] (.5,0) -- (1,-1);
\draw[scalar] (1.5,0) -- (1,1);
\draw[scalar] (1.5,0) -- (1,-1);
\draw[ghost] (1,-1) -- (1,1);
\draw[fill=VertexColor,VertexColor] (.5,0) circle (\vertexsize);
\draw[fill=VertexColor,VertexColor] (1.5,0) circle (\vertexsize);
\draw[fill=white] (0,0) circle (.075);
\draw[fill=white] (2,0) circle (.075);
\draw[fill=white] (1,1) circle (.075);
\draw[fill=white] (1,-1) circle (.075);
\node[] at (-.3,0) {{\footnotesize $2$}};
\node[] at (2.3,0) {{\footnotesize $4$}};
\node[] at (1,1.3) {{\footnotesize $1$}};
\node[] at (1,-1.3) {{\footnotesize $3$}};
\end{tikzpicture}}

\newcommand{\DefectVertexOnePointScalarCenterBis}{\scalebox{.85}{\begin{tikzpicture}[baseline={([yshift=-1ex]current bounding box.center)}]
\draw[wilson] (0,0) -- (1.5,0);
\draw[scalar] (.75,0) -- (.75,1);
\draw[fill=white] (.75,1) circle (.075);
\draw[fill=white] (.3,0) circle (.075);
\draw[fill=white] (1.2,0) circle (.075);
\draw[fill=VertexColor,VertexColor] (.75,0) circle (\vertexsize);
\node[] at (1.05,1) {\footnotesize $1$};
\node[] at (.3,-.3) {\footnotesize $2$};
\node[] at (1.2,-.3) {\footnotesize $3$};
\end{tikzpicture}}}
\newcommand{\TDivergent}{\scalebox{.85}{\begin{tikzpicture}[baseline={([yshift=-1ex]current bounding box.center)}]
\draw[scalar,fill=white] (1,.3) circle (.3);
\draw[white, fill=white] (.7,.3) rectangle (1.3,0);
\draw[scalar] (.7,.3) -- (.7,0);
\draw[scalar] (1.3,.3) -- (1.3,0);
\draw[gluon] (1,.6) -- (1,1);
\draw[wilson] (0,0) -- (2,0);
\draw[fill=white] (.2,0) circle (.075);
\draw[fill=white] (.7,0) circle (.075);
\draw[fill=white] (1.3,0) circle (.075);
\draw[fill=white] (1.8,0) circle (.075);
\draw[fill=VertexColor,VertexColor] (.45,0) circle (\vertexsize);
\draw[fill=VertexColor,VertexColor] (1,0) circle (\vertexsize);
\draw[fill=VertexColor,VertexColor] (1.55,0) circle (\vertexsize);
\draw[fill=VertexColor,VertexColor] (1,.6) circle (\vertexsize);
\end{tikzpicture}}}
\newcommand{\UIntegral}{\scalebox{.85}{\begin{tikzpicture}[baseline={([yshift=-1ex]current bounding box.center)}]
\draw[scalar] (1,0) -- (1,.75);
\draw[wilson] (0,0) -- (2,0);
\draw[fill=white] (.35,0) circle (.075);
\draw[fill=white] (1,0) circle (.075);
\draw[fill=white] (1.65,0) circle (.075);
\draw[fill=VertexColor,VertexColor] (.35+.325,0) circle (\vertexsize);
\draw[fill=VertexColor,VertexColor] (1.65-.325,0) circle (\vertexsize);
\end{tikzpicture}}}
\graphicspath{{./figures/}}

\begin{document}

\includepdf[pages=-]{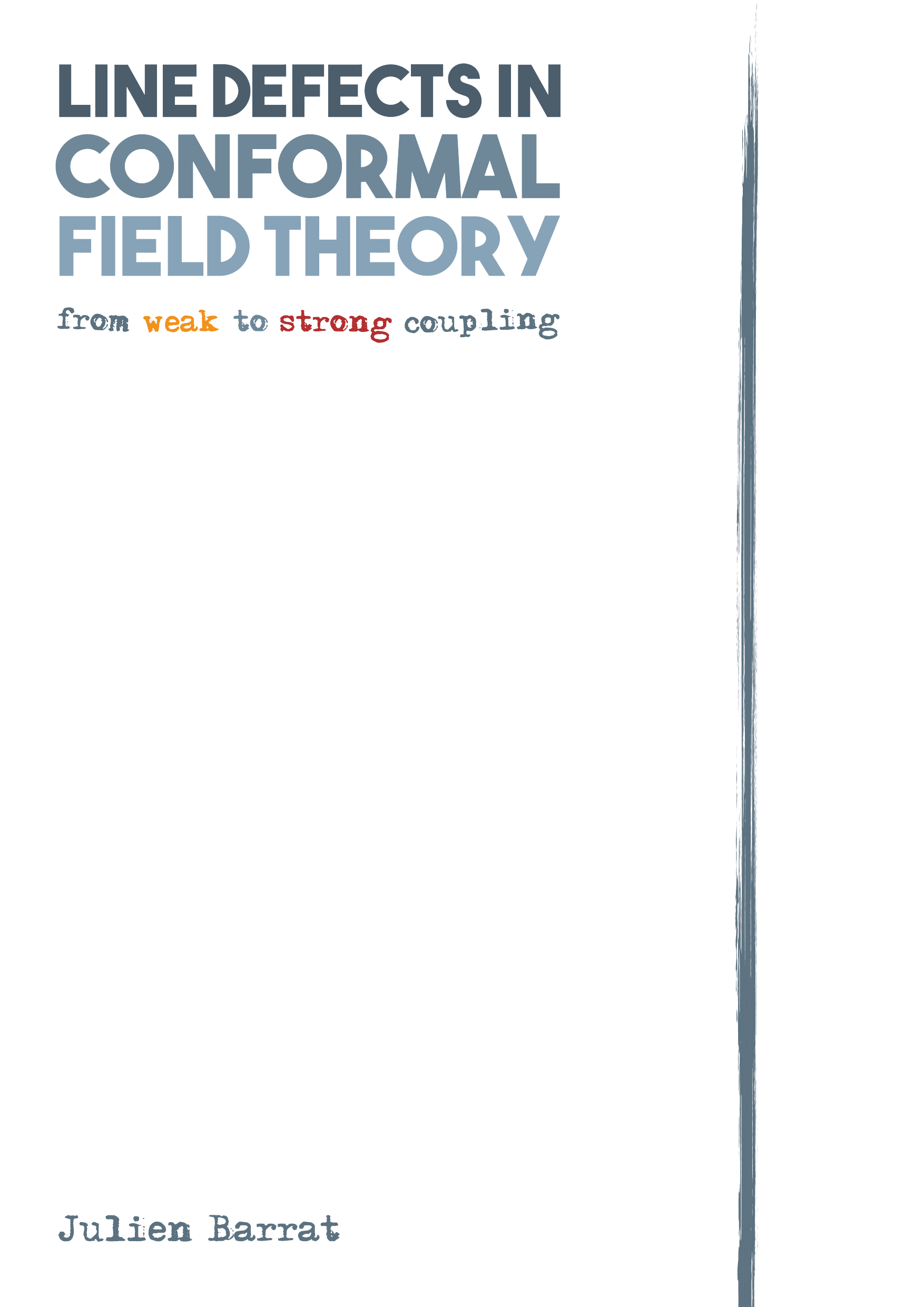}

\frontmatter

\includepdf[pages=-]{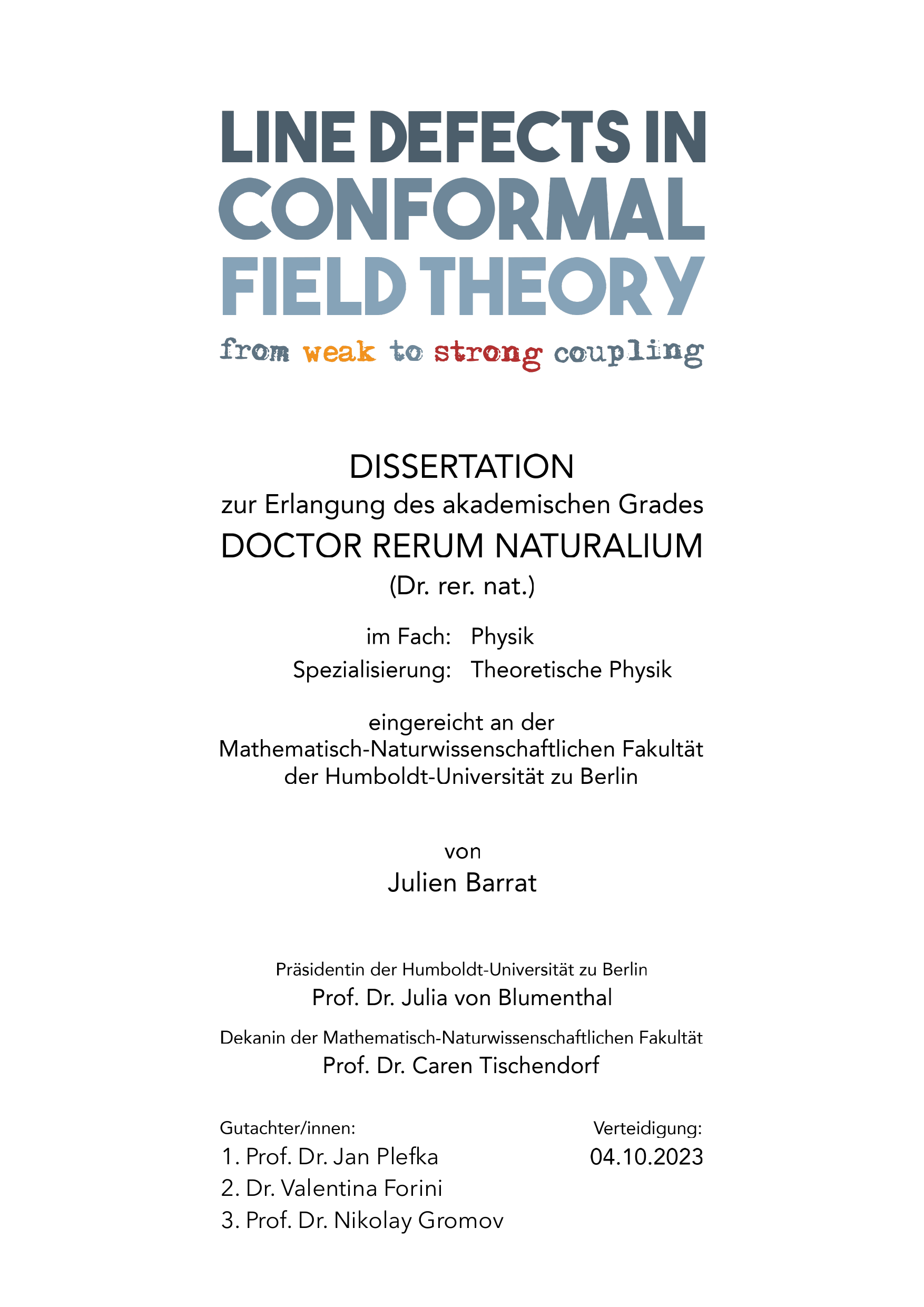}
	
\chapter*{Abstract}

Conformal field theory finds applications across diverse fields, from statistical systems at criticality to quantum gravity through the AdS/CFT correspondence.
These theories are subject to strong constraints, enabling a systematic non-perturbative analysis.
Conformal defects provide a controlled means of breaking the symmetry, introducing new physical phenomena while preserving crucial benefits of the underlying conformal symmetry.
This thesis investigates conformal line defects in both the weak- and strong-coupling regimes.
Two distinct classes of models are studied.
First, we focus on the supersymmetric Wilson line in $\Nm=4$ Super Yang--Mills, which serves as an ideal testing ground for the development of innovative techniques such as the analytic conformal bootstrap.
The second class consists of magnetic lines in Yukawa models, which have fascinating applications in $3d$ condensed-matter systems.
These systems have the potential to emulate phenomena observed in the Standard Model in a low-energy setting.

\vspace{-.6em}

\begin{center}
$\star$
\end{center}

\vspace{-.6em}

{\emergencystretch 3em
Die konforme Feldtheorie findet in verschiedenen Bereichen Anwendungen, von statistischen Systemen in der Nähe kritischer Punkte bis hin zur Quantengravitation durch die AdS/CFT-Korrespondenz.
Diese Theorien unterliegen starken Einschränkungen, die eine systematische nicht-perturbative Analyse ermöglichen.
Konforme Defekte bieten eine kontrollierte Möglichkeit, die Symmetrie zu brechen und neue physikalische Phänomene einzuführen, während wichtige Vorteile der zugrunde liegenden konformen Symmetrie erhalten bleiben.
Diese Dissertation untersucht konforme Liniendefekte sowohl im schwachen als auch im starken Kopplungsregimes.
Es werden zwei verschiedene Klassen von Modellen untersucht.
Wir konzentrieren uns zuerst auf die supersymmetrische Wilson-Linie in $\Nm=4$ Super Yang--Mills, die als ideales Testfeld für die Entwicklung innovativer Techniken wie dem analytischen konformen Bootstrap dient.
Die zweite Klasse besteht aus magnetische Linien in Yukawa-Modellen, die faszinierende Anwendungen in $3d$ kondensierten Materiesystemen haben.
Diese Systeme haben das Potenzial, Phänomene des Standardmodells in einem Niedrigenergieszenario nachzubilden.
}

\chapter*{Acknowledgements}

First and foremost, I want to express my deepest gratitude to my two supervisors, \textsc{Jan Plefka} and \textsc{Valentina Forini}, for their continuous support throughout these three beautiful years.
I am especially thankful to Jan, whose availability, kindness, and trust have been invaluable.
His guidance and expertise have consistently provided me with the necessary support and knowledge to progress in my research.

I am greatly indebted to \textsc{Pedro Liendo} for introducing me to the world of conformal defects and analytic bootstrap.
Without his guidance and mentorship, none of the works presented here would have been possible.
Our countless discussions about physics, artsy movies, and vintage video games have been both enjoyable and enlightening.

I would like to thank the additional members of my PhD committee, \textsc{Benjamin Lindner}, \textsc{Nikolay Gromov}, and \textsc{Agostino Patella}, for showing dedication in reading my thesis as well as for making insightful comments.

\bigskip

I wish to extend my sincere gratitude to the remarkable individuals with whom I have had the privilege of collaborating:
\textsc{Daniele \mbox{Artico}}, \textsc{Gabriel \mbox{Bliard}}, \textsc{Ilija \mbox{Burić}}, \textsc{Pietro \mbox{Ferrero}}, \textsc{Aleix Gimenez-Grau}, \textsc{Carlo \mbox{Meneghelli}}, \textsc{Giulia \mbox{Peveri}}, \textsc{Volker \mbox{Schomerus}}, and \textsc{Philine van \mbox{Vliet}}. 
I have learned something special from each of you, and I am eagerly looking forward to continuing these fruitful collaborations in the future.
I wish you all the best in your future endeavors, both professionally and personally.

\bigskip

Being in the research training group \textsc{Rethinking Quantum Field Theory} has been a truly enjoyable experience, and I would like to thank everyone in this program for that.
The friendships that have formed there go beyond physics, and I am especially grateful for the camaraderie and support from \textsc{Ilaria Costa}, \textsc{Tomás Codina}, \textsc{Felipe Diaz-Jaramillo}, \textsc{Moritz Kade}, \textsc{Allison Pinto}, and \textsc{Jasper Roosmale Nepveu}.
I deeply enjoyed the environment at HU, and I have also greatly benefited from the interactions with \textsc{Roberto Bonezzi}, \textsc{Luke Corcoran}, \textsc{Michele Galli}, \textsc{Olaf Hohm}, \textsc{Emanuel Malek}, \textsc{Agostino Patella}, and all the other members of the string theory and mathematical physics groups.
Our limited lunch options have brought us all closer!

\bigskip

I wish to express my gratitude to the team of \textsc{Non-Standard Models}, our precious outreach project: \textsc{Daniele Artico}, \textsc{Ilaria Costa}, \textsc{Michele Galli}, \textsc{Claudio Iuliano}, \textsc{Giulia Peveri}, and \textsc{Allison Pinto}.
Working with all of you on this project has brought me a great sense of fulfillment, and I will always be proud of what we achieved.
I would like to also address special thanks to Jan and Agostino for always supporting us, even when our ideas seemed unrealistic.
I want to express my appreciation to \textsc{Alessandro Cotellucci}, \textsc{Tim Meier}, \textsc{Davide Scazzuso}, and others mentioned earlier for their occasional contributions to this project.

\bigskip

{\emergencystretch 6em
I would also like to thank the individuals across the world who have welcomed me to their institutions:
\textsc{T. ~Bargheer} (DESY), \textsc{D.~Chicherin} (LAPTh), \textsc{N.~Drukker} (King's College), \textsc{P.~Ferrero} (Stony Brook), \textsc{F.~Galvagno} (ETH), \textsc{S.~Giombi} (Princeton), \textsc{V. Rosenhaus} (CUNY), \textsc{A.~Tseytlin} (Imperial College), and \textsc{K.~Zarembo} (Nordita).
}

I have also enjoyed insightful discussions with \textsc{A.~Cavagliá} (King's College), \textsc{N.~Gromov} (King's College), \textsc{J.~Henn} (MPI), \textsc{J.~Julius} (King's College), \textsc{E.~Pomoni} (DESY), and \textsc{M.~Preti} (King's College).

\bigskip

I want to thank my friends scattered all over the world and my family for their unwavering support.
There are too many names to mention here, but each of you holds a special place in my heart.
I should not forget to express my gratitude to \textsc{Leonie}, my cat, for waking me up at 5 am when I was already overworked.

\bigskip

Finally, I wish to express my deepest gratitude and heartfelt appreciation to my sunshine, \textsc{Anne Reitz}, for bringing immense joy and giving meaning to everything I do.
I cherish all of our moments together and the infinite happiness that you bring into my world.

\chapter*{Publications}

This thesis is based on the following publications.
\begin{table}[h]
\begin{tabularx}{\textwidth}{l X}
\cite{Barrat:2021yvp} & J. Barrat, A. Gimenez-Grau, and P. Liendo, \textit{Bootstrapping holographic defect correlators in $ \mathcal{N} $ = 4 super Yang-Mills}, \href{https://link.springer.com/article/10.1007/JHEP04(2022)093}{JHEP \textbf{04} (2022) 093}; \\[1.5em]
\cite{Barrat:2021tpn} & J. Barrat, P. Liendo, G. Peveri, and J. Plefka, \textit{Multipoint correlators on the supersymmetric Wilson line defect CFT}, \href{https://link.springer.com/article/10.1007/JHEP08(2022)067}{JHEP \textbf{08} (2022) 067}; \\[1.5em]
\cite{Barrat:2022psm} & J. Barrat, A. Gimenez-Grau, and P. Liendo, \textit{A dispersion relation for defect CFT}, \href{https://link.springer.com/article/10.1007/JHEP02(2023)255}{JHEP \textbf{02} (2023) 255}; \\[1.5em]
\cite{Barrat:2022eim} & J. Barrat, P. Liendo, and G. Peveri, \textit{Multipoint correlators on the supersymmetric Wilson line defect CFT II: Unprotected operators}, \href{https://arxiv.org/abs/2210.14916}{2210.14916}; \\[1.5em]
\cite{Barrat:2023ivo} & J. Barrat, P. Liendo, and P. van Vliet, \textit{Line defect correlators in fermionic CFTs}, \href{https://arxiv.org/abs/2304.13588}{2304.13588}.
\end{tabularx}
\end{table}

\noindent Some of the content to appear in the following works is also presented:
\begin{table}[h]
\begin{tabularx}{\textwidth}{l X}
\cite{Barrat:2023ta1} & D. Artico, J. Barrat, and G. Peveri, \textit{Four-point functions at two-loop on the Wilson line defect CFT}; \\[1.5em]
\cite{Barrat:2023ta2} & J. Barrat, G. Bliard, P. Ferrero, C. Meneghelli, and G. Peveri, \textit{Bootstrapping multipoint correlators on the Wilson line defect CFT}; \\[1.5em]
\cite{Barrat:2023ta3} & J. Barrat, I. Burić, P. Liendo, V. Schomerus, and P. van Vliet, \textit{Scalar-fermion correlators in Yukawa CFT across dimensions}.
\end{tabularx}
\end{table}

\noindent Finally, the unpublished notes \cite{Barrat:2021un} are presented in Section \ref{sec:TheMicrobootstrap}.
Parts of the author's master's thesis are also included:
\begin{table}[h]
\begin{tabularx}{\textwidth}{l X}
\cite{Barrat:2020vch} & J. Barrat, P. Liendo, and J. Plefka, \textit{Two-point correlator of chiral primary operators with a Wilson line defect in $ \mathcal{N} $ = 4 SYM}, \href{https://link.springer.com/article/10.1007/JHEP05(2021)195}{JHEP \textbf{05} (2021) 195}.
\end{tabularx}
\end{table}

\noindent Each paper has received equal contributions from all the authors.

\chapter*{Declaration of authorship}

I hereby certify that the thesis that I am submitting is entirely my own original work, except where otherwise stated.
I am aware of the University’s regulations concerning plagiarism, including those regulations concerning disciplinary actions that may result from plagiarism.
Any use of the works of any other author, in any form, is properly acknowledged at their point of use.

\vspace{2\baselineskip}

\begin{flushright}
{\scshape Julien Barrat}
\end{flushright}

\tableofcontents
\pagenumbering{gobble}

\pagestyle{fancy}

\mainmatter

\chapter{Introduction}
\label{chapter:Introduction}

Conformal field theories (CFTs) are of central importance in modern theoretical physics, finding applications across various fields ranging from high-energy to condensed-matter physics.
The investigation of conformal defects has recently emerged as a focal point, yielding a wealth of new insights and discoveries.
This thesis is devoted to the study of line defects within the framework of conformal field theories.
In this chapter, we present an introduction to the themes that motivate this work.

\section{An invitation to conformal field theory}
\label{sec:AnInvitationToConformalFieldTheory}

Conformal field theories are quantum field theories (QFTs) possessing \textit{scale invariance}, meaning that physical quantities remain unchanged under the rescaling of space and time.
More generally, they provide a framework to describe the behavior of fields and their interactions in a way that is invariant under \textit{conformal transformations}, an extension of the Lorentz transformations used to describe relativistic physics \cite{Einstein:1905ve}.
In the following, we motivate the study of conformal field theory through three applications: understanding phase transitions in statistical systems, characterizing the space of quantum field theories, and exploring the AdS/CFT correspondence as a realization of the holographic principle.

\subsection{Critical phenomena}
\label{subsec:CriticalPhenomena}

Quantum field theories in Euclidean spacetime provide a framework to describe and understand the collective behavior and statistical properties of large ensembles of particles in classical statistical physics.\footnote{See for instance \cite{Polchinski:1987dy} and the textbooks \cite{das1997finite,abrikosov2012methods,landau2013statistical}.}
Critical phenomena, including phase transitions, often exhibit scale-invariant physics and can be analyzed using conformal field theories \cite{ma2018modern}.

\subsubsection{Phase transitions}
\label{subsubsec:PhaseTransitions}

Phase transitions are fundamental phenomena found in various physical systems when a sudden change occurs in their \textit{macroscopic} properties, such as the emergence of long-range order or the appearance of new collective behavior.
Consider water in its liquid and gas states as an example.
Despite having the same elementary components, these two phases exhibit distinct behaviors.
The liquid phase is typically simulated using molecular dynamics, while the gas phase is described by the ideal-gas model, where the water molecules are considered as non-interacting point particles.

Conformal field theory plays a crucial role in understanding phase transitions and critical phenomena.
Second-order phase transitions exhibit scale invariance, meaning that the physics of the system remains the same at all length scales.
Physical quantities, such as correlation functions and thermodynamic properties, exhibit scaling behavior with respect to specific variables, such as temperature or distance from the critical point.
These scaling relations allow for the determination of \textit{critical exponents} that characterize the critical behavior of the system.
In most cases, the scale invariance is intimately connected to the presence of conformal symmetry, making the critical point correspond to a Euclidean conformal field theory.

\subsubsection{Universality}
\label{subsubsec:Universality}

\begin{table}[t]
\centering
\caption{Measurements and theoretical estimations of the critical exponents $\eta$ and $\nu$ in fluid and magnet phase transitions (on the experimental side) and the $3d$ Ising model (on the theoretical side), illustrating universality.
This table reproduces excerpts of Table 1 in \cite{Henriksson:2022rnm}, with the results based on \cite{damay1998universal,sullivan2000small,belanger1987neutron,Kos:2016ysd,Hasenbusch:2021tei,Shalaby:2020xvv}.}
\begin{tabular}{lccccc}
\hline
& \multicolumn{2}{c}{Experiment} & \multicolumn{3}{c}{Theory} \\ \hline
& Liquid-gas & Ferro-para & Bootstrap & Monte Carlo & $\veps$-exp. \\ \hline
$\eta$ & 0.042(6) & & 0.0362978(20) & 0.036284(40) & 0.03653(65) \\
$\nu$ & 0.62(3) & 0.64(1) & 0.629971(4) & 0.62998(5) & 0.62977(22) \\ \hline
\end{tabular}
\label{table:Universality}
\end{table}

\begin{figure}[t]
\centering
  \includegraphics[width=1\linewidth]{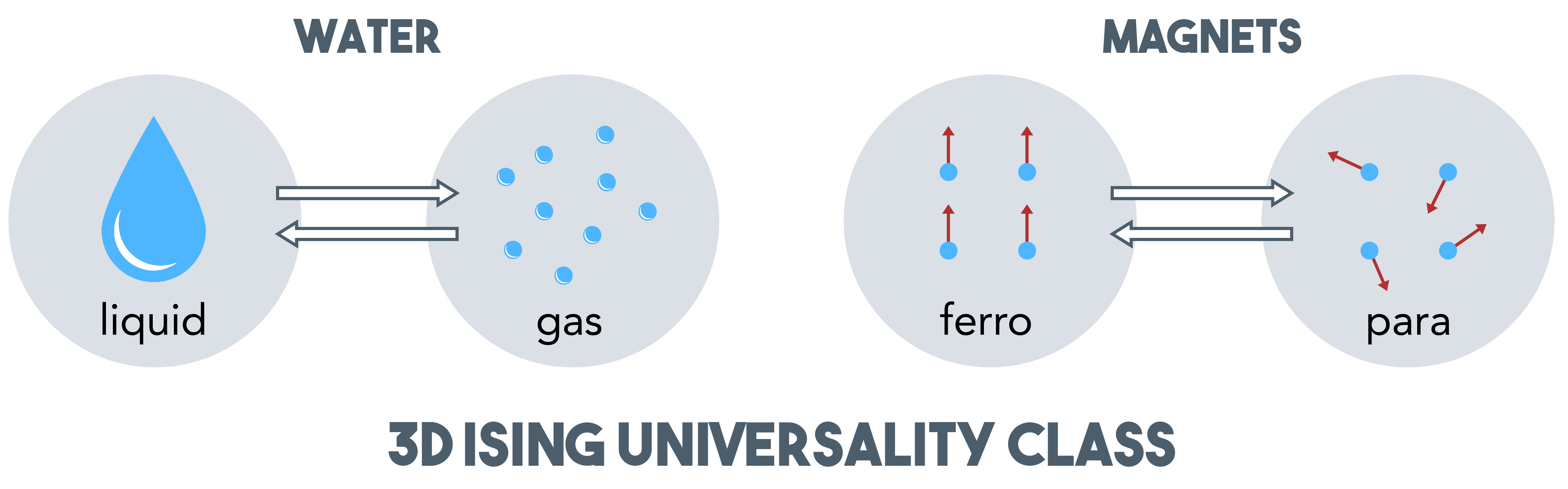}
\caption{The $3d$ Ising universality class is known to describe the phase transitions of various systems, such as the change from liquid to gas of water (left) and the ferromagnetic-paramagnetic transition of uniaxial magnets (right).
}
\label{fig:3dIsingUniversalityClass}
\end{figure}

At critical points, different physical systems can exhibit the \textit{same} behavior, implying that a single conformal field theory can describe the observables of very different systems \cite{Kadanoff:1971pc,kadanoff1990scaling}.
Despite the apparent fundamental differences between water and magnets,
it turns out that the liquid-gas phase transition in water shares the same critical exponents as the ferromagnetic-paramagnetic phase transition observed in uniaxial magnets  (see Table \ref{table:Universality} and \cite{cardy1996scaling}).
This intriguing connection categorizes water and magnets into the same \textit{$3d$ Ising universality class}, which is also known to describe phase transitions in binary mixtures in polymers \cite{binder2005polymer,muller1997homogeneous,pipich2005ginzburg} and alloys \cite{binder1974kinetic,zunger2002obtaining}.
This is illustrated in Figure \ref{fig:3dIsingUniversalityClass}.

The concept of universality extends well beyond the $3d$ Ising universality class and encompasses various categories of critical behavior.
Universality implies that microscopic details become irrelevant at critical points, and only a set of robust properties is necessary to describe the physics.
The study and understanding of conformal field theory thus provide essential tools for the description and classification of critical phenomena.

\subsection{Signposts in the space of quantum field theories}
\label{subsec:SignpostsInTheSpaceOfQuantumFieldTheories}

The space of quantum field theories is vast and diverse, and finding effective methods for classifying them is crucial.
Due to the correspondence between Euclidean quantum field theories and statistical systems, we can reformulate the principles of criticality described above in the language of QFT.
There exists a remarkable common feature among all members of this family.
In the context of the \textit{renormalization group} (RG), conformal field theories serve as distinctive \textit{signposts} that facilitate the classification of quantum field theories \textit{flowing} into CFTs.

\subsubsection{The RG flow}
\label{subsubsec:TheRGFlow}

\begin{figure}
\centering
\begin{subfigure}{.55\textwidth}
  \raggedright
  \includegraphics[width=.8\linewidth]{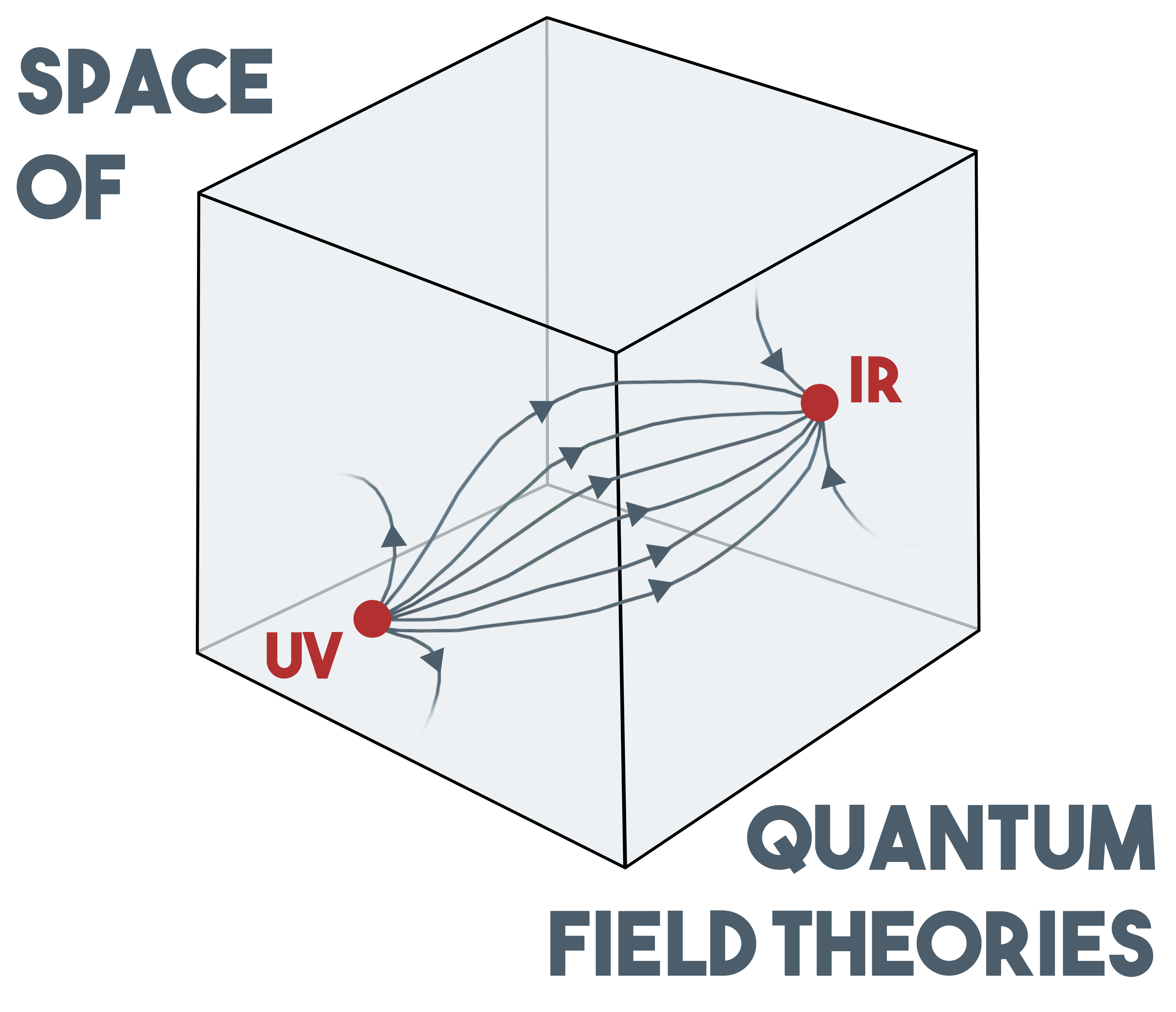}
\end{subfigure}%
\begin{subfigure}{.45\textwidth}
  \raggedleft
  \includegraphics[width=.8\linewidth]{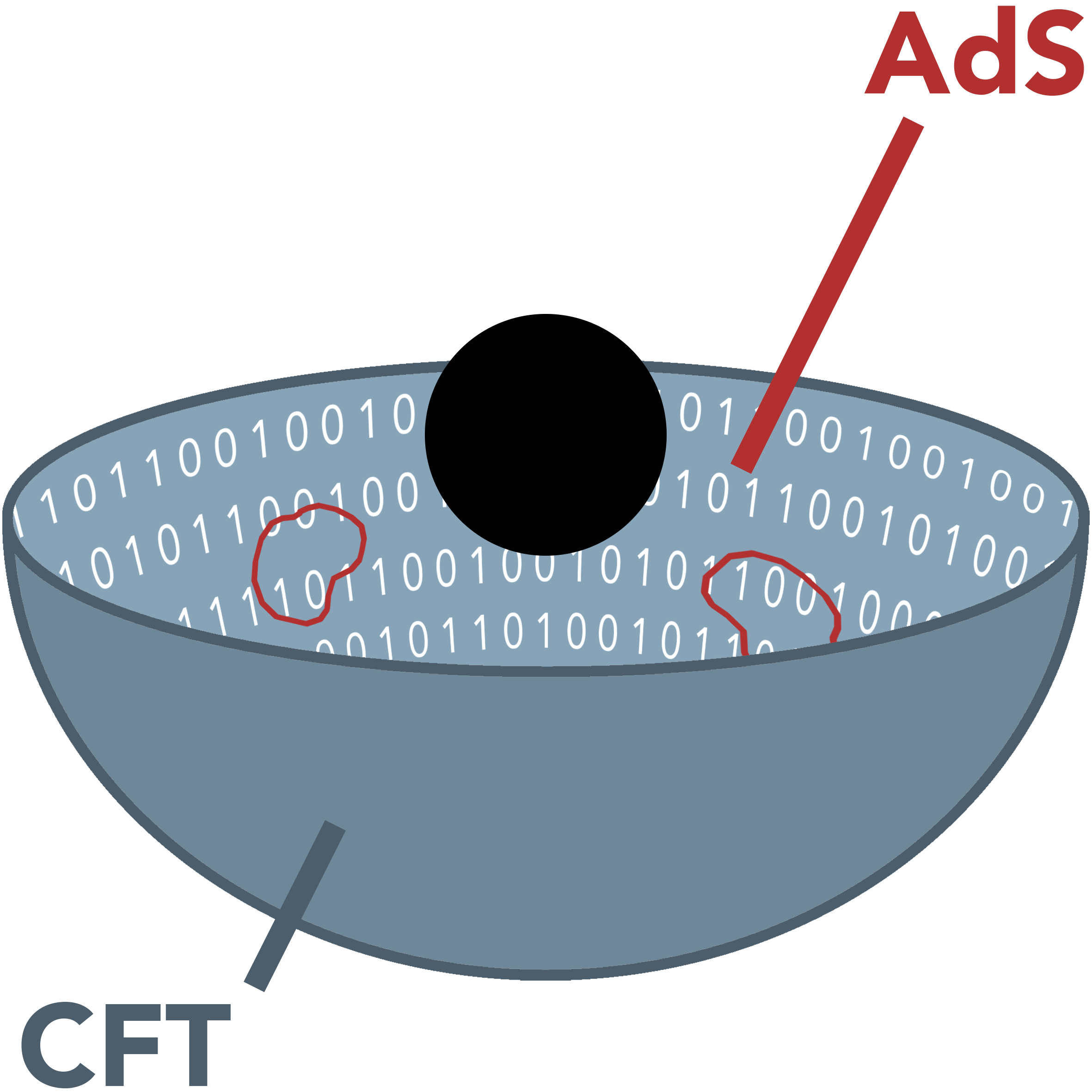}
\end{subfigure}
\caption{The image on the left represents the flow of models in the space of quantum field theories.
Conformal field theories serve as fixed points in the RG flow.
The holographic principle is depicted on the right.
The information contained in the $d$-dimensional gravitational theory in AdS space is encoded on its boundary, where a conformal field theory resides.
}
\label{fig:RGFlow_AdSCFT}
\end{figure}

Quantum field theories are characterized by a finite set of parameters, such as masses and coupling constants, which can be modified to study the behavior of the theory.
This gives rise to the concept of \textit{RG flow}, illustrated in Figure \ref{fig:RGFlow_AdSCFT}, which describes how theories evolve as these parameters are varied \cite{stuckelberg1953,Wilson:1971bg,Wilson:1971dh,Wilson:1973jj,pelissetto2002critical}.
The trajectories traced out in the space of theories converge towards \textit{fixed points}, representing special models which, remarkably, exhibit scale invariance.
These fixed points correspond to conformal field theories and serve as infrared (IR) attractors towards which specific quantum field theories flow from their ultraviolet (UV) descriptions.
The RG flow provides a powerful tool for understanding the underlying symmetries of quantum field theories across different energy scales.

\subsubsection{Universality again}
\label{subsubsec:UniversalityAgain}

From this perspective, the space of quantum field theories can be viewed as a landscape of RG flows, with the IR fixed points serving as the destinations where the flows converge.
This statement is fully equivalent to the discussion of the previous section about the $3d$ Ising universality class.
We encounter again the principle of universality since distinct quantum field theories can share common critical behavior, therefore belonging to the same class.
In the IR phase, the microscopic details of individual models become irrelevant, and what dictates the physics are the universal properties and symmetries encoded by the fixed point.
This remarkable structure connects seemingly disparate models in their behavior at long distances, giving conformal symmetry a special place in the realm of quantum field theories.

\subsection{From CFT to gravity}
\label{subsec:FromCFTToGravity}

The Standard Model of particle physics stands as a remarkable achievement in science, yielding unprecedented agreement between experimental observations and theoretical predictions \cite{fan2023measurement}.
It is widely recognized however that the Standard Model is not complete.
It necessitates the ad hoc fixing of $19$ free parameters, does not provide a unified description of the electroweak and strong forces, and fails to account for dark matter and dark energy, which constitute about $95{\%}$ of the Universe (see for instance the reviews \cite{troitsky2012unsolved,nagashima2014beyond}).
Furthermore, the theory lacks a quantum version of gravity.
We now explore how conformal field theory plays a fundamental role in the modern approach to resolving this open problem.

\subsubsection{The holographic principle}
\label{subsubsec:TheHolographicPrinciple}

The origins of this story lie in the theoretical study of \textit{black holes}, where an intriguing connection between gravity, spacetime, and information was discovered in the early 1970s \cite{Bekenstein:1972tm,Bekenstein:1973ur,Hawking:1975vcx}.
It was observed that the entropy of a black hole, which measures its internal disorder, is proportional to its \textit{surface area} rather than its volume.
In other words, the information contained within a black hole is encoded on its \textit{boundary}, much like a hologram captures the essence of a three-dimensional object on a two-dimensional surface.
More generally, the \textit{holographic principle} postulates that a higher-dimensional theory \textit{with} gravity can be fully described by a lower-dimensional theory \textit{without} gravity \cite{Susskind:1994vu}.

\subsubsection{The AdS/CFT conjecture}
\label{subsubsec:TheAdSCFTConjecture}

The first concrete realization of the holographic principle emerged with the development of the \textit{AdS/CFT correspondence}.
It was conjectured in \cite{Maldacena:1997re} that $\Nm=4$ Super Yang--Mills (SYM) theory in \textit{four} dimensions is dual to a (super)string theory in a \textit{five}-dimensional anti-de Sitter (AdS) space.
In other words, the gravitational physics of the bulk theory is fully encoded by the quantum field theory residing on the \textit{boundary} of the AdS space.
$\Nm=4$ SYM stands out among quantum field theories due to its remarkable features, including exact quantum conformal symmetry, hidden integrability in the large $N$ limit, and profound dualities such as AdS/CFT and $S$-duality \cite{Montonen:1977sn,Seiberg:1994rs,Seiberg:1994pq,DHoker:2002nbb,Ammon:2015wua,Beisert:2010kp}.
Since then, numerous other instances of AdS/CFT dualities have been discovered \cite{Aharony:2008ug,aharony2000large,Gubser:1998bc,Klebanov:2002ja,Witten:2007kt,Witten:1998qj}, providing an exciting application of conformal symmetry for the study of quantum gravity and black holes.

The AdS/CFT correspondence can be used to address another important problem apart from the nature of quantum gravity \cite{Sonnenschein:1999re,Girardello:1999hj,Babington:2003vm,Sakai:2004cn,Sakai:2005yt,Polyakov:2008zx,Klebanov:2009zz}.
This duality establishes a connection between the gravitational theory at \textit{weak} coupling to the quantum field theory at \textit{strong} coupling.
The strong coupling regime is notoriously difficult to analyze, as it cannot be accessed through small perturbations of a free theory.
This regime is important for instance in the study of quantum chromodynamics (QCD) to understand features such as confinement \cite{Wilson:1974sk,Cahill:1978cva,Muta:2010xua,Greensite:2011zz}.
The AdS/CFT correspondence unlocks access to this regime, by perturbing the gravitational theory starting from a free model.
This has led to numerous advancements in the fields of particle physics \cite{Barbon:2004dq,Evans:2004ia,Erdmenger:2007bn,Burrington:2007qd,Erdmenger:2007cm,Papantonopoulos:2011zz,Callebaut:2011ab} and condensed matter \cite{Evans:2001ab,Hartnoll:2008kx,Kachru:2008yh,Nakamura:2009tf,Liu:2009dm,Cubrovic:2009ye,Hartnoll:2010gu,Charmousis:2010zz,Faulkner:2010zz,anderson2013strange,Fitzpatrick:2014xwa}.

\section{Conformal defects}
\label{sec:ConformalDefects}

Extended operators, or \textit{defects}, are important objects in quantum field theory for probing new dynamics that would otherwise be inaccessible using local operators alone.
Defects serve as an ideal bridge between low- and high-energy physics, facilitating the connection between experimental observations and theoretical considerations.
They find wide-range applications in both condensed-matter systems and high-energy physics, offering opportunities for shedding light on phenomena such as quark confinement in the strong force \cite{Wilson:1974sk,Giles:1981ej,Aharony:2013hda} and the AdS/CFT correspondence in supersymmetric theories \cite{Maldacena:1998im,Ooguri:1999ta,Drukker:1999zq,Erickson:2000af,Drukker:2000rr,Plefka:2001bu,Arutyunov:2001hs,Klebanov:2002ja,Alday:2007hr,Alday:2007he,Alday:2008yw,Beisert:2010gn,Alday:2010zy,Drukker:2011za,Giombi:2012ep,Buchbinder:2012vr,Giombi:2017cqn,Giombi:2018qox,Giombi:2018hsx,Beccaria:2020ykg,Giombi:2020amn,Beccaria:2021ksw,Giombi:2021zfb,Giombi:2022anm,Giombi:2023vzu}.

\subsection{Breaking the symmetry}
\label{subsec:BreakingTheSymmetry}

The presence of defects breaks the symmetry of the original system.
There exists however in conformal field theory a special class of defects, called \textit{conformal}, which retain crucial aspects of the underlying symmetry.
This class of defects is interesting, as the concepts presented above for defect-free conformal field theory find analogous counterparts in \textit{defect} conformal field theory (dCFT).

\subsubsection{That's what defects do}
\label{subsubsec:ThatsWhatDefectsDo}

In a (Euclidean) $d$-dimensional conformal field theory, the presence of a conformal defect introduces a controlled breaking of the conformal group $SO(d+1,1)$ \cite{Billo:2016cpy,Bianchi:2015liz,Cardy:1984bb,Jensen:2015swa,Frohlich:2006ch,DeWolfe:2001pq,Aharony:2003qf,deLeeuw:2015hxa,Billo:2013jda}.
These defects are called conformal because they preserve a subset of the conformal symmetry:
\begin{equation}
\underbrace{SO(d+1, 1)}_{\substack{\text{conformal group} \\ \text{in $d$ dimensions}}} \longrightarrow \underbrace{SO(p+1,1)}_{\substack{\text{conformal group} \\ \text{in $p$ dimensions}}} \times \underbrace{SO(q)}_{\substack{\text{orthogonal} \\\text{rotations}}}\,,
\label{eq:Intro_SymmetryBreaking}
\end{equation}
where $p$ denotes the \textit{dimension} of the defect and $q:=d-p$ its \textit{codimension}.

Defects break translation invariance in the transverse directions, leading to non-vanishing one-point functions of bulk operators.
This feature proves valuable for instance in numerical simulations of critical theories \cite{assaad2013pinning,toldin2017critical}.
While the study of spontaneous symmetry breaking in statistical systems typically relies on the behavior of two-point functions, the quadratic suppression of these functions can pose challenges.
The introduction of a defect offers a solution to increase the accuracy of the simulations.
Despite the symmetry breaking induced by the defect, the parameters of the bulk theory remain unchanged far from the defect, enabling an efficient investigation of the critical behavior through one-point functions.

\subsubsection{The bootstrap program}
\label{subsubsec:TheBootstrapProgram}

The \textit{defect conformal bootstrap} program investigates the implications of the residual symmetry in quantum systems.
In bulk theories, the bootstrap concept was initially proposed in $S$-matrix theory to explain hadronic structures \cite{Chew:1962mpd,Chew:1962eu}.
It was eventually shadowed by the advent of quantum chromodynamics.
However, it has experienced since then a revival in the context of conformal field theory \cite{Rattazzi:2008pe}, achieving remarkable success in the analysis of quantum field theories beyond the weak-coupling limit \cite{Belavin:1984vu,Polyakov:1988am,Kos:2013tga,Beem:2014zpa,Kos:2014bka,Kos:2015mba,Lemos:2015awa,Iliesiu:2015qra,Kos:2016ysd,Lemos:2016xke,Poland:2016chs,Iliesiu:2017nrv,Poland:2018epd,Atanasov:2018kqw,Chester:2019wfx,Anand:2020gnn,Li:2020bnb,Liendo:2021egi,Lemos:2021azv,Albayrak:2021xtd,Atanasov:2022bpi,Erramilli:2022kgp,Poland:2023vpn}.
The bootstrap program has notably provided the most precise determination of critical exponents for the $3d$ Ising model to date (see \cite{El-Showk:2012cjh,El-Showk:2014dwa,Kos:2016ysd} and Table \ref{table:Universality}).

In recent years, significant strides have been made in the \textit{analytic} conformal bootstrap, which relies on the behavior of operators at large spin and has culminated into the development of the Lorentzian inversion formula \cite{Caron-Huot:2017vep,Simmons-Duffin:2017nub}.
This formula, applicable to \textit{any} conformal field theory in \textit{any} dimension, is a non-perturbative result that facilitates the derivation of strong-coupling correlation functions under suitable conditions (see \cite{Caron-Huot:2018kta,Albayrak:2019gnz,Chester:2020dja,Caron-Huot:2020adz,Zhou:2020ptb,Paulos:2020zxx,Bissi:2021spj,Alday:2021vfb,Bissi:2022mrs,Bertucci:2022ptt,Caron-Huot:2022sdy,Goncalves:2023oyx} and references therein).

The application of bootstrap ideas to defects has been actively pursued \cite{McAvity:1995zd,Liendo:2012hy,Gliozzi:2015qsa,Rastelli:2017ecj,Lemos:2017vnx,Bissi:2018mcq,Gimenez-Grau:2019hez,Kaviraj:2018tfd,Mazac:2018biw,Liendo:2019jpu,Bianchi:2020hsz,Gimenez-Grau:2020jvf,Gimenez-Grau:2021wiv,Bianchi:2021piu,Dey:2020jlc,Kusuki:2021gpt,GimenezGrau:2022izf,Kusuki:2022ozk,Gimenez-Grau:2022czc,Cavaglia:2022yvv,Gimenez-Grau:2022ebb,Bianchi:2022sbz,Meneghelli:2022gps,Gimenez-Grau:2023fcy,SoderbergRousu:2023ucv,Krishnan:2023cff,Niarchos:2023lot,SoderbergRousu:2023nvd}.
Since the presence of a defect does not significantly alter local physics, the fundamental properties that enable the effectiveness of the bootstrap at strong coupling in the parent theory remain intact.
This program offers a promising avenue for the study of correlation functions in the presence of defects across various regimes.

\subsection{Line defects}
\label{subsec:LineDefects}

In this thesis, our primary focus is on \textit{line} defects, characterized by $p=1$ in \eqref{eq:Intro_SymmetryBreaking}.
This type of defect can be found in various condensed-matter systems, including thin films \cite{hu1997defect,kou2011tunable,yang2017effects,yun2018unique,kim2022defect,tian2022valleytronics} and layered materials \cite{wen2007analysis,enyashin2013line,schweizer2019manipulation,mazza2022defect}.
Another example of line defects is a magnetic field localized along an infinite line \cite{allais2014magnetic,ParisenToldin:2016szc,wu2020topological,Cuomo:2021kfm,Rodriguez-Gomez:2022gif,Rodriguez-Gomez:2022gbz,Gimenez-Grau:2022czc,Gimenez-Grau:2022ebb,Bianchi:2022sbz,Giombi:2022vnz}.
In the context of gauge theories, the supersymmetric Wilson line has played a central role in the study of the AdS/CFT correspondence (see references above).

\subsubsection{As a point in space}
\label{subsubsec:AsAPointInSpace}

Let us gain some intuitive understanding of line defects.
In quantum field theory, line defects that extend along the time direction can be visualized as physical phenomena that manifest as \textit{points} within the spatial domain.
These points serve as sources of disturbance for the surrounding system and exert an influence on the nearby operators and observables.

To provide an analogy for this concept, we can imagine observing a serene lake from an aerial perspective.
Suppose now that we drop a small pebble into the water, creating a localized disturbance.
This disturbance affects the propagation of nearby ripples, altering the overall pattern of motion in the lake.
From our vantage point above, we perceive this event as a point dynamically influencing its environment.
A line defect in a quantum system operates similarly, influencing the dynamics of its parent theory.
The pebble can be considered as a \textit{probe}, allowing measurements of specific properties of the lake through its response to the disturbance.
Similarly, line defects can serve as probes to investigate the properties of the bulk theory.

\subsubsection{As a brane}
\label{subsubsec:AsABrane}

The AdS/CFT correspondence provides a completely different perspective on the line defects described above.
One notable example of a line defect is the supersymmetric Wilson line operator in the $\Nm=4$ SYM theory in four dimensions.
This particular defect captures the interaction between a heavy probe particle and the gauge fields of the theory.
In the holographic dual description, the Wilson line corresponds to a fundamental string or a $D3$-brane that stretches into the extra dimension of the AdS space.

Certain correlation functions involving the Wilson line can be computed \textit{exactly}, which has enabled explicit verification of the AdS/CFT conjecture.
Conversely, the AdS interpretation of the Wilson line offers valuable handles to constrain the strong-coupling behavior of correlation functions involving local operators associated with the defect.
This setup therefore not only facilitates the examination of the AdS/CFT correspondence but also presents an appealing opportunity for the defect bootstrap program.

\section{Structure of the thesis}
\label{sec:StructureOfTheThesis}

The topics explored in this thesis are intimately connected to the themes discussed above.
With these foundations in mind, we now present the structure of the thesis.

\subsection{Outline}
\label{subsec:Outline}

The chapters are organized as follows.
\textbf{Chapter \ref{chapter:Foundations}: Foundations} provides a compact review of the necessary concepts for the main chapters of the thesis.
After a general introduction to conformal defects, the specific examples of the Wilson-line defect CFT in $\Nm=4$ SYM and of the magnetic line in Yukawa CFTs are discussed.

In \textbf{Chapter \ref{chapter:BootstrappingHolographicDefectCorrelators}: Bootstrapping holographic defect correlators}, we present the results of \cite{Barrat:2021yvp,Barrat:2022psm} as well as the unpublished notes \cite{Barrat:2021un}.
In $\Nm=4$ SYM, correlation functions of operators in the presence of a Wilson-line defect are computed at strong coupling using analytic bootstrap methods and input from the AdS/CFT correspondence.

\textbf{Chapter \ref{chapter:MultipointCorrelatorsInTheWilsonLineDefectCFT}: Multipoint correlators in the Wilson-line defect CFT} is dedicated to the study of excitations along the defect, and summarizes the findings of \cite{Barrat:2021tpn,Barrat:2022eim,Barrat:2023ta1,Barrat:2023ta2}.
Correlation functions of multiple operators are considered, both in the weak- and strong-coupling regimes.

The case of a magnetic line in Yukawa CFTs is considered in \textbf{Chapter \ref{chapter:LineDefectCorrelatorsInFermionicCFT}: Line defect correlators in fermionic CFT}, which is based on \cite{Barrat:2023ivo} and presents preliminary results to appear in \cite{Barrat:2023ta3}.
Perturbative techniques are used to calculate the correlation functions of $3d$ fermionic systems, which are potentially relevant for condensed-matter systems and the study of certain aspects of the Standard Model.

Chapters \ref{chapter:BootstrappingHolographicDefectCorrelators}-\ref{chapter:LineDefectCorrelatorsInFermionicCFT} begin with an invitation to the topic, and are concluded with an outlook sketching potential follow-ups.

\textbf{Chapter \ref{chapter:Conclusions}: Conclusions} provides a concise summary of the results and discusses future research directions related to the topics encountered throughout the thesis.

\subsection{Conventions}
\label{subsec:Conventions}

We consider spacetime to be \textit{Euclidean}, i.e., the metric is given by
\begin{equation}
g_{\mu\nu} (x) := \delta_{\mu\nu}\,,
\end{equation}
with $\mu\,, \nu = 0\,, 1\,, \ldots\,, d-1$.
The direction $x^0 := \tau$ is referred to as the \textit{Euclidean time}.
All the line defects studied here are chosen to extend in the $\tau$-direction.
Quantities are given in natural units, i.e., $c = \hbar = G = 1$.

Note that the conventions used here might differ from \cite{Barrat:2021yvp,Barrat:2021tpn,Barrat:2022psm,Barrat:2022eim,Barrat:2023ivo,Barrat:2023ta1,Barrat:2023ta2,Barrat:2023ta3,Barrat:2021un,Barrat:2020vch}.
Changes have been made to ensure consistency throughout the thesis.

\chapter{Foundations}
\label{chapter:Foundations}

The presence of conformal symmetry imposes significant constraints on the observables of a conformal field theory, making them more tractable compared to general quantum field theories while retaining key features.
Conformal defects provide a way to break conformal symmetry in a controlled manner.
They introduce non-trivial boundary conditions on the fields that give rise to new interesting physical phenomena.
Despite this symmetry breaking, correlation functions associated with defects remain highly constrained, occasionally allowing exact results to be derived.
Defects play a crucial role in holography, where a plethora of results have been obtained in the past two decades \cite{Erickson:2000af,Drukker:2000rr,Semenoff:2001xp,Semenoff:2002kk,Drukker:2006xg,Drukker:2007yx,Drukker:2010jp,Drukker:2011za,Drukker:2012de,Cagnazzo:2017sny,Cooke:2017qgm,Cooke:2018obg,Drukker:2020dcz,Drukker:2020swu,Drukker:2020atp,Kristjansen:2020mhn,Linardopoulos:2021rfq,Medina-Rincon:2018wjs,Drukker:2022txy,Kristjansen:2023ysz}, but also in condensed-matter physics, where they characterize critical phenomena in systems with impurities \cite{blavats2003critical,kropman2010interaction,kimura2014current,bernard2015non,liu2021magnetic,wang2022defect}.

This chapter serves as an introduction to line defects in conformal field theories.
We review how the correlation functions of a bulk theory are impacted by the presence of a defect, and how the residual symmetry can be used to derive results in both the weak- and strong-coupling regimes.
Our focus lies on two distinct classes of models: \textit{$\Nm=4$ Super Yang--Mills} with a supersymmetric Wilson-line defect, and \textit{Yukawa CFTs} in the presence of a magnetic line.
The first system serves as an ideal laboratory for the development of analytic conformal bootstrap techniques, due to its abundance of symmetries.
On top of conformal symmetry, $\Nm=4$ SYM possesses supersymmetry and is believed to be integrable, while its holographic dual has been identified and studied extensively.
On the other hand, Yukawa CFTs are interesting because they are closer to real-world physical phenomena, such as phase transitions in graphene sheets in three dimensions \cite{Herbut:2009qb,roy2011multicritical,Janssen:2014gea,Classen:2015ssa,ebert2016phase,Gracey:2018cff,Ray:2021moi}, which exhibit chiral symmetry breaking akin to the Standard Model.
Here we take these models across dimensions to study strongly-coupled systems as a perturbation from free theory.

\bigskip

This chapter draws on various resources, which are acknowledged at the beginning of each section and in the text.
We aim to present the material in an accessible and engaging way, carefully selecting topics and methods that are particularly relevant to the focus of this thesis and complement the existing literature.

\section{A brief introduction to conformal defects}
\label{sec:ABriefIntroductionToConformalDefects}

We begin our journey in the world of conformal defects with an overview of how the behavior of conformal field theories is impacted by their presence.
In this section, we provide a review of the key concepts of conformal symmetry in $d$ dimensions, with a special focus on the correlation functions of local operators.
We then analyze how the constraints of conformal symmetry are modified in the presence of a defect.
Although the concepts discussed here apply to more general defects, we specialize some of our formulas to the case of line defects.
By the end of this section, we hope to have established a solid foundation for the topics presented in this thesis.

\subsection{Conformal field theory}
\label{subsec:ConformalFieldTheory}

Before considering defects, we revisit the general principles of conformal field theory and familiarize ourselves with the structure of correlation functions.
The content of this section is mostly based on \cite{DiFrancesco:1997nk,Ammon:2015wua}.\footnote{See also the lectures \cite{Ginsparg:1988ui,Rychkov:2016iqz}.}

\subsubsection{Conformal symmetry\ }
\label{subsubsec:ConformalSymmetry}

\begin{figure}
\centering
\includegraphics[width=.45\linewidth]{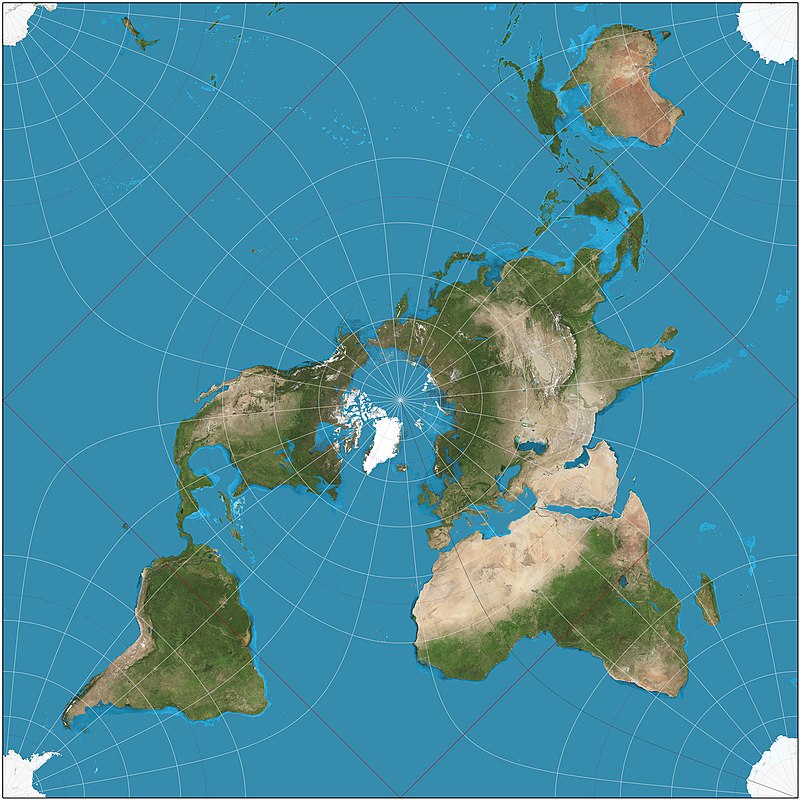}
\caption{An example of a conformal transformation:
the Peirce quincuncial projection \cite{peirce1879quincuncial} of the surface of the Earth shows the world in a square while preserving the original angles.
}
\label{fig:ExampleConformalTransformations}
\end{figure}

Conformal symmetry is a rich mathematical concept that can be intuitively understood as the group of transformations that \textit{preserve angles}.
The example of a conformal transformation is presented in Figure \ref{fig:ExampleConformalTransformations}.
Conformal symmetry extends the fundamental \textit{Poincaré} group, which underlies our understanding of spacetime symmetries.

For a transformation to be angle-preserving, the transformed metric must be a \textit{scaling} of the original metric by a conformal factor:
\begin{equation}
g'_{\mu\nu} (x') = \kappa(x) g_{\mu\nu} (x).
\label{eq:ScalingOfTheMetric}
\end{equation}
The Poincaré group can be restored by setting $\kappa(x) = 1$.
The \textit{conformal group} in $d$ dimensions consists of $(d+1)(d+2)/2$ generators, which can be classified into four categories:
\begin{equation}
\begin{alignedat}{2}
&P_\mu := - i \pd_\mu && \qquad \text{($d$ translations)}\,,\\
&(J_{\rho\sigma})_{\mu\nu} := - i (\delta_{\rho\nu} \delta_{\mu\sigma} - \delta_{\sigma\nu} \delta_{\mu\rho}) && \qquad \text{($d(d-1)/2$ LTs)}\,,\\
&D := - i x_\mu \pd_\mu && \qquad \text{($1$ scalings)}\,,\\
&K_\mu := i (x^2 \pd_\mu - 2 x_\mu x_\nu \pd_\nu) && \qquad \text{($d$ SCTs)}\,,
\end{alignedat}
\label{eq:Generators}
\end{equation}
where LTs refers to \textit{Lorentz transformations} and SCTs to \textit{special conformal transformations}.
SCTs can be seen as the compositions of an \textit{inversion}, a \textit{translation}, and another \textit{inversion}, while $D$ rescales distances.
The question of whether scaling invariance automatically implies invariance under SCTs has been the subject of much debate.
While there have been several counterexamples found in which scaling symmetry was preserved alone, these examples are limited and have so far been constrained to the case $d=2$ (see \cite{Riva:2005gd,Dorigoni:2009ra,Nakayama:2013is,Nakayama:2016cyh} and references therein). 

The \textit{conformal algebra} is formed by promoting the generators of the conformal group to operators satisfying specific commutation relations:
\begin{align}
\begin{split}
[D, K_\mu] &= - i K_\mu\,, \\
[D, P_\mu] &= i P_\mu\,, \\
[K_\mu, P_\nu] &= 2i (\delta_{\mu\nu} D - J_{\mu\nu})\,, \\
[K_\mu, J_{\nu\rho}] &= i (\delta_{\mu\nu} K_\rho - \delta_{\mu\rho} K_\nu)\,,
\end{split}
\label{eq:Commutators}
\end{align}
which have to be complemented by the commutators of the Poincaré algebra (see for instance \cite{Peskin:1995ev,Weinberg:1995mt}).
All other possible commutators vanish, leading to a simple and elegant algebraic structure.\footnote{In fact, the conformal algebra in $d$ dimensions can be expressed as a Poincaré algebra in $d+2$ dimensions, known as the \textit{embedding formalism} (see for instance \cite{osborn2019lectures}).
This approach offers an interesting perspective on conformal field theory, but will not be treated in this thesis.}

\subsubsection{Representations\ }
\label{subsubsec:Representations}

In this section, we direct our attention towards the \textit{representations} of the conformal algebra.
These are also representations of the Lorentz algebra, and we refer the reader to \cite{Peskin:1995ev} for a review of the \textit{scalar}, \textit{spinor}, and \textit{vector} representations in this context.
We denote the corresponding (local) fields by $\phi (x)$ (scalar), $\psi(x)$, $\psib(x)$ (spinors) and $A_\mu(x)$ (vector).
In the following, we primarily focus on scalar fields, while the case of spinors is treated in Chapter \ref{chapter:LineDefectCorrelatorsInFermionicCFT} to some extent.

In a CFT, scalar fields undergo new transformation laws in addition to the Poincaré transformations. These are given by
\begin{align}
\begin{split}
D \phi(x) &= i (x_\mu \pd_\mu + \Delta) \phi(x)\,, \\
K_\mu \phi(x) &= i (x^2 \pd_\mu - 2 x_\mu x_\nu \pd_\nu - 2 x_\mu \Delta) \phi(x)\,.
\end{split}
\label{eq:NewTransformationLaws}
\end{align}
$\Delta$ is known as the \textit{scaling dimension} of the field $\phi(x)$, and it is defined by the action of a scaling transformation on the spacetime coordinates:
\begin{equation}
\phi(\lambda x) = \lambda^{-\Delta} \phi(x)\,,
\label{eq:DefinitionOfScalingDimension}
\end{equation}
with $\lambda$ a real number.

Vector and spinor representations also carry scaling dimensions.
These quantities are important for the classification of representations in a conformal field theory.
At the classical level, they are typically integer-valued.
However, at the quantum level, they can be subject to corrections.
Small deviations from classical behavior are commonly referred to as \textit{anomalous dimensions}, which can arise in both the weak- and strong-coupling regimes, where they can be expanded as
\begin{equation}
\Delta = \Delta^{(0)} + \sum_{\ell=0}^{\infty} g^\ell\, \gamma^{(\ell)}\,,
\label{eq:Delta_CouplingExpansion}
\end{equation}
with $g$ a coupling constant.
{\emergencystretch 3em
Anomalous dimensions play a central role throughout this thesis.
}

We specialize our analysis to the case of \textit{unitary} CFTs.
For these theories, there exists a fundamental lower bound for the scaling dimension $\Delta$, known as the \textit{unitarity bound}. 
The generators $P_\mu$ \textit{increase} $\Delta$ while the $K_\mu$ \textit{decrease} it, and as a consequence the fields that saturate the lower bound are the ones satisfying
\begin{equation}
[K_\mu, \phi(0)] = 0\,.
\label{eq:ConformalPrimaries}
\end{equation}
Such fields are called \textit{conformal primaries}.
A \textit{conformal multiplet} consists of a conformal primary and all its \textit{descendants}, which are the fields with higher dimension $\Delta$ that can be constructed by applying $P_\mu$ an arbitrary number of times on the primary.

\subsubsection{Correlation functions\ }
\label{subsubsec:CorrelationFunctions1}

Conformal symmetry imposes strong constraints on \textit{correlation functions} of the local operators presented in the previous section.
The symmetry uniquely determines the kinematics of two- and three-point functions, while higher-order correlators are constrained to depend on a limited number of cross-ratios.
One-point functions of any operator vanish since that would break the translation invariance of the vacuum.\footnote{As mentioned in Section \ref{subsec:BreakingTheSymmetry}, this will not be true any longer once we add defects to the theory.}
In the following, we review the correlation functions of scalar fields.

Let us begin with the two-point functions.
It is straightforward to show that the correlators of two operators $\Op_{\Delta_1}$ and $\Op_{\Delta_2}$ take the form
\begin{equation}
\vev{\Op_{\Delta_1} (x_1) \Op_{\Delta_2} (x_2)} = n_{\Delta} \frac{\delta^{\Delta_1 \Delta_2}}{x_{12}^{2\Delta}}\,,
\label{eq:ConformalTwoPointFunctions}
\end{equation}
with $x_{ij} := x_i - x_j$, and where we have defined $\Delta := \Delta_1 = \Delta_2$, since all other two-point functions are orthogonal and thus vanish.
The constant $n_{\Delta}$ is known as the \textit{normalization constant}.
In this thesis, we choose the operators to be \textit{unit-normalized}, i.e., they satisfy
\begin{equation}
\vev{\Op_{\Delta_1} (x_1) \Op_{\Delta_2} (x_2)} = \frac{\delta^{\Delta_1 \Delta_2}}{x_{12}^{2\Delta}}\,.
\label{eq:UnitNormalization}
\end{equation}
This can always be achieved by rescaling the operators:
\begin{equation}
\Op_{\Delta} \longrightarrow \frac{1}{\sqrt{n_\Delta}} \Op_{\Delta}\,.
\label{eq:RescalingToUnitNormalization}
\end{equation}
Although $n_\Delta$ is a convention-dependent quantity (and, as such, unphysical), determining it is still crucial for writing the operators explicitly.
In a perturbative setting, the two-point functions are also essential for obtaining the anomalous dimensions of a given operator.

Conformal symmetry also places constraints on three-point functions, which take the form
\begin{equation}
\vev{\Op_{\Delta_1} (x_1) \Op_{\Delta_2} (x_2) \Op_{\Delta_3} (x_3)} = \frac{\lambda_{\Delta_1 \Delta_2 \Delta_3}}{x_{12}^{2\Delta_{123}} x_{23}^{2\Delta_{231}} x_{31}^{2\Delta_{312}}}\,,
\label{eq:ConformalThreePointFunctions}
\end{equation}
where $\Delta_{ijk} := \Delta_i + \Delta_j - \Delta_k$. 
The three-point functions \eqref{eq:ConformalThreePointFunctions} are therefore completely determined by the scaling dimensions $\Delta_i$ and the constants $\lambda_{\Delta_1 \Delta_2 \Delta_3}$, which we refer to as \textit{OPE coefficients} for reasons that will soon become clear.

The kinematic dependence of higher-point functions is no longer fully determined by conformal symmetry. In $d>1$ dimensions, the four-point functions of scalars depend on \textit{two} cross-ratios $u$ and $v$:
\begin{equation}
\vev{\Op_{\Delta_1} (x_1) \Op_{\Delta_2} (x_2) \Op_{\Delta_3} (x_3) \Op_{\Delta_4} (x_4)} = \Km\, F(u,v)\,,
\label{eq:ConformalFourPointFunctions}
\end{equation}
where $\Km$ represents a trivial \textit{conformal prefactor}, which will be defined precisely later.
The cross-ratios are defined as
\begin{equation}
u := \frac{x_{12}^2 x_{34}^2}{x_{13}^2 x_{24}^2}\,, \qquad v := \frac{x_{14}^2 x_{23}^2}{x_{13}^2 x_{24}^2}\,.
\label{eq:SpacetimeCrossRatiosuv}
\end{equation}
This choice of variables for expressing the \textit{reduced correlator} $F(u,v)$ is not unique and depends on the application.
Another set of cross-ratios is often used:
\begin{equation}
\chi \chib := u\,, \qquad (1-\chi)(1-\chib) := v\,.
\label{eq:SpacetimeCrossRatiosChi}
\end{equation}

In line defect CFTs, the case $d=1$ is particularly important.
In this case, the cross-ratios are not independent of each other.
Instead, the reduced correlators depend on a \textit{single} cross-ratio $\chi = \chib$, leading to the mappings
\begin{equation}
u \overset{1d}{\longrightarrow} \chi^2\,, \qquad v \overset{1d}{\longrightarrow} (1-\chi)^2\,.
\label{eq:SpacetimeCrossRatio1d}
\end{equation}
This fact is important for the study of the correlation functions of operators inserted along a line defect.

The observations made above about the four-point functions can be extended straightforwardly to the higher-point functions. Generally, a correlator consisting of $n$ operators depends on $n (n-3)/2$ cross-ratios\footnote{This number holds for \textit{small} $n$. For $n \geq d+1$, it would be $n d - \frac{1}{2}(d+1)(d+2)$, where the first term corresponds to the naive number of degrees of freedom, while the second term is the number of generators.} and can be written as
\begin{equation}
\vev{\Op_{\Delta_1} (x_1) \ldots \Op_{\Delta_n} (x_n)} = \Km\,  F( \lbrace u\,, v\,, w \rbrace )\,,
\label{eq:HigherPointFunctions}
\end{equation}
where $\lbrace u\,, v\,, w \rbrace$ refers to the following set of cross-ratios:
\begin{equation}
\begin{alignedat}{2}
u_i &:= \frac{x^2_{1(i+1)} x^2_{(n-1)n}}{x^2_{1(n-1)} x^2_{(i+1)n}} &&\qquad \left(n-3 \text{ cross-ratios}\right)\,,\\
v_i &:= \frac{x^2_{1 n} x^2_{(i+1)(n-1)}}{x^2_{1(n-1)} x^2_{(i+1)n}} &&\qquad \left(n-3 \text{ cross-ratios}\right)\,, \\
w_{ij} &:= \frac{x^2_{1(n-1)} x^2_{(i+1)n} x^2_{(j+1)n}}{x^2_{1n} x^2_{(n-1)n} x^2_{(i+1)(j+1)}} && \qquad \left(\frac{(n-3)(n-4)}{2} \text{ cross-ratios}\right)\,.
\end{alignedat}
\label{eq:SpacetimeCrossRatios}
\end{equation}

Conformal field theory possesses a remarkable feature known as the \textit{operator product expansion} (OPE), which is a powerful tool for understanding the behavior of local operators and the correlation functions introduced above.
The OPE allows to expand a product of two local operators in terms of all the possible local operators in the theory, and it takes the form\footnote{See for instance \cite{DiFrancesco:1997nk} for a detailed derivation of the OPE.}
\begin{equation}
\Op_{\Delta_1} (x_1) \Op_{\Delta_2} (x_2) \sim \sum_{\Op_\Delta \text{ prim.}} \lambda_{\Delta_1 \Delta_2 \Delta} C_{\Delta_1 \Delta_2 \Delta} (x_{12}\,, \pd_2)\, \Op_\Delta (x_2)\,,
\label{eq:OPEv2}
\end{equation}
where the sum runs over all the primaries of the theory.
$C_{\Delta_1 \Delta_2 \Delta}$ is a differential operator encoding the construction of descendants, while $\lambda_{\Delta_1 \Delta_2 \Delta}$ is the three-point function coefficient introduced in \eqref{eq:ConformalThreePointFunctions}.

While the idea of an OPE is not unique to CFT, conformal symmetry guarantees that the expansion converges with a non-zero radius \cite{Pappadopulo:2012jk,Rychkov:2015lca}, making it an especially valuable tool.
This expansion was first introduced in \cite{Wilson:1972ee} several decades ago, and since then it has become a fundamental concept in the study of CFTs.

To understand the power of the operator product expansion, let us look at the four-point functions defined in \eqref{eq:ConformalFourPointFunctions}.
By considering the pairs of operators $\Op_{\Delta_1} \Op_{\Delta_2}$ and $\Op_{\Delta_3} \Op_{\Delta_4}$, it can be expressed as a sum over conformal primaries, the \textit{exchanged} operators $\Op_\Delta$.
This gives rise to \textit{conformal blocks} $f_{\Op}^{12,34}(u,v)$, which are fully characterized by the symmetry group.
The reduced correlator $F(u,v)$ can then be written as
\begin{equation}
F(u,v) = \sum_{\Op} \lambda_{\Delta_1 \Delta_2 \Delta} \lambda_{\Delta_3 \Delta_4 \Delta}\, f^{12,34}_{\Op} (u,v)\,.
\label{eq:ExpansionInConformalBlocks}
\end{equation}
The conformal blocks offer an expansion of correlation functions in terms of basis functions, which are determined by a corresponding \textit{Casimir equation} fixed by the symmetry group.
The explicit form of the blocks in four dimensions can be found for instance with the \textsc{Mathematica} notebook attached to \cite{Cuomo:2017wme}.
Furthermore, the OPE coefficients and the scaling dimensions provide the necessary information to fully determine the remaining parts of the correlator.
Together, this information is commonly referred to as \textit{CFT data}, and it provides a powerful tool for analyzing and characterizing CFTs.

Before moving our attention to conformal defects, let us notice an interesting consequence of \eqref{eq:OPEv2}.
There is another OPE channel that can be considered to express the correlator \eqref{eq:ConformalFourPointFunctions}, which is to pair the operators as $\Op_{\Delta_1} \Op_{\Delta_4}$ and $\Op_{\Delta_2} \Op_{\Delta_3}$.
This yields an identical correlator, but expressed as a sum over different coefficients, giving a consistency relation known as \textit{crossing symmetry}:
\begin{equation}
\sum_{\Op} \lambda_{\Delta_1 \Delta_2 \Delta} \lambda_{\Delta_3 \Delta_4 \Delta}\, f_{\Op}^{12,34} (u,v)
=
\sum_{\Op'} \lambda_{\Delta_1 \Delta_4 \Delta'} \lambda_{\Delta_2 \Delta_3 \Delta'}\, f_{\Op'}^{14,23} (u,v)\,.
\label{eq:Bulk_CrossingSymmetry}
\end{equation}
This relation is represented graphically in Figure \ref{fig:DefectTwoPoint_Setup}.
It plays a crucial role in the numerical conformal bootstrap, and a defect counterpart will be presented shortly.

\subsection{Conformal defects}
\label{subsec:ConformalDefects}

Conformal symmetry imposes strong constraints on the observables of a conformal field theory.
One way to introduce new physical phenomena while preserving the benefits of conformal symmetry is to incorporate \textit{conformal defects} into the theory.
In this section, we explore the concept of conformal defect and examine the implications for the underlying bulk CFT.

\subsubsection{Symmetry breaking\ }
\label{subsubsec:SymmetryBreaking}

\begin{figure}
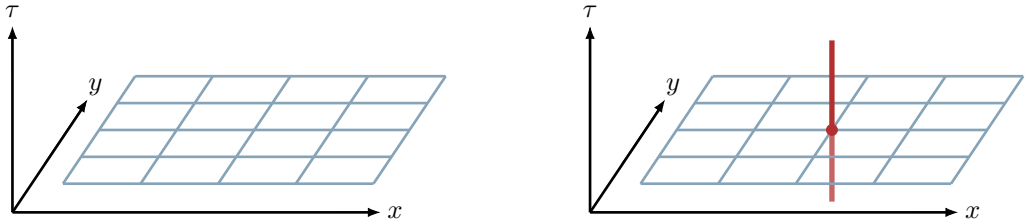

\centering
\ExampleLineDefect
\caption{An example of a line defect extending in the time direction in a $(2+1)$-dimensional spacetime.
The grid in both figures represents the two-dimensional space.
The left figure shows the bulk theory without any defects, while the right figure depicts the introduction of a line defect.
As explained in Section \ref{subsec:LineDefects}, line defects can be understood as point-like impurities residing in the two-dimensional space.}
\label{fig:ExampleLineDefect}
\end{figure}

Conformal defects are non-local extensions of a bulk theory that deliberately break conformal symmetry in a controlled manner.
These defects preserve a specific subgroup of the original symmetry of the parent theory.
For instance, in a $d \geq 2$ CFT with a line defect extending in the time direction, translations are broken in the orthogonal directions while remaining preserved in the time direction (see Figure \ref{fig:ExampleLineDefect}).

As we have already seen in \eqref{eq:Intro_SymmetryBreaking}, a conformal defect can be viewed as a deformation that breaks conformal symmetry in the following way:
\begin{equation}
SO(d+1,1) \longrightarrow SO(p+1,1) \times SO(q)\,,
\label{eq:SymmetryBreaking}
\end{equation}
where $p$ is the \textit{dimension} of the defect and $q$ its \textit{codimension}, such that $p+q=d$.
This symmetry breaking leads to constrained correlation functions in the \textit{defect CFT} compared to a homogeneous CFT.
Although conformal symmetry is preserved in the $p$-dimensional subspace $SO(p+1,1)$, it is broken in the full $d$-dimensional space, with the second subgroup $SO(q)$ representing $q$-dimensional rotations around the defect.

Not all quantum field theories can be described by a Lagrangian, and this is also true for defect CFTs.
However, in fortunate instances, it is possible to express the bulk theory and the defect in terms of an action, given by\footnote{This expression should be understood as schematic.
In the case of non-Abelian gauge theories, it has to be adapted to accommodate the gauge group.}
\begin{equation}
S_{\text{defect}} := S_{\text{bulk}} + h \int d^p x\, \Op(x^\newparallel)\,,
\label{eq:DefectAction}
\end{equation}
where $h$ is the \textit{defect coupling constant} and $\Op(x^\newparallel)$ depends solely on the $p$-dimensional coordinates $x^{\newparallel} := (x^0, \ldots, x^p)$.
To ensure consistency, $\Op(x)$ must have mass dimension $[\Op] = p$ if $h$ is dimensionless.
The directions orthogonal to the defect are denoted by $x^\perp$.
This thesis primarily focuses on defects with dimension $p=1$,\footnote{The main exception is the dispersion relation in Section \ref{sec:ADispersionRelationForDefectCFT}, which is valid for all defects of codimension $q \geq 2$.}
which have codimension $q=d-1$.
Following the conventions given in \ref{subsec:Conventions}, we choose these line defects to extend along the (Euclidean) time direction $x^0=\tau$.

We conclude this section with a brief discussion on \textit{boundary} defects, which have codimension $q=1$ and preserve the largest subgroup of conformal symmetry.
They can be seen as a separation in the $d$-dimensional space.
Although we do not consider boundaries specifically here, they are a fascinating setup with numerous physical applications, particularly in condensed-matter physics, where they arise naturally in the study of finite-temperature systems \cite{Dowker:1978md,kennedy1982finite,Witten:1998zw,Kapusta:2006pm,Iliesiu:2018fao,Alday:2020eua,Marchetto:2023fcw}.
Boundary defects have also been used to describe impurities in systems at criticality, interfaces between different phases, and have applications in the AdS/CFT correspondence.
Boundary defects present several conceptual and technical challenges, such as the emergence of edge modes \cite{Wen:1992vi,Carrozza:2021gju} and the need to properly define the boundary conditions.
Constraints on supersymmetric boundary defects are briefly discussed in the outlook \ref{subsec:MoreWardIdentities} of Chapter \ref{chapter:MultipointCorrelatorsInTheWilsonLineDefectCFT}.

\subsubsection{Defect correlators\ }
\label{subsubsec:DefectCorrelators}

In the presence of a defect, it is necessary to revisit some of the results that we established when considering the correlation functions of a CFT.
We previously determined that a quantum field theory with conformal symmetry was completely fixed by the CFT data $\lbrace \Delta_i, \lambda_{\Delta_i \Delta_j \Delta_k} \rbrace$.
The defect breaks this symmetry and introduces new physics, and as a result a new set of numbers, known as \textit{defect CFT data}, is needed for characterizing the defect theory, in complete analogy with our previous analysis.

As in Section \ref{subsec:ConformalFieldTheory}, we focus on scalar operators.
The correlators of $n$ local bulk operators $\Op_{\Delta_k} (x_k)$ in the presence of a defect $\Dm$ are defined as
\begin{equation}
\vvev{ \Op_{\Delta_1}(x_1) \ldots \Op_{\Delta_n}(x_n) } := \vev{ \Op_{\Delta_1}(x_1) \ldots \Op_{\Delta_n}(x_n) \Dm }\,.
\label{eq:DefectCorrelators}
\end{equation}
The first observation is that one-point functions are \textit{non-vanishing} in a defect CFT, meaning that operators acquire expectation values due to their interactions with the defect.
These one-point functions are given by
\begin{equation}
\vvev{ \Op_\Delta(x) } = \frac{a_\Delta}{|x^\perp|^\Delta}\,,
\label{eq:DefectOnePointFunctions}
\end{equation}
where the coefficients $a_\Delta$ are functions of the coupling independent of the kinematic variables, similarly to $\lambda_{\Delta_1 \Delta_2 \Delta_3}$.
Since the normalization of $\Op_\Delta$ is determined by the bulk two-point functions, $a_\Delta$ contains \textit{dynamic} information about the defect CFT that cannot be absorbed into the operator.\footnote{One may question whether the parameter $a_\Delta$ can be absorbed into the defect itself.
It is standard to normalize the defect such that the vacuum of the theory is $\vev{\Dm} = 1$.
This makes $a_\Delta$ a physical observable in its own right.}

Note that \eqref{eq:DefectOnePointFunctions} also contains information about the bulk theory.
One-point functions can be used to extract the scaling dimensions of the bulk fields.
In fact, in certain experiments and simulations, it is sometimes easier to measure defect one-point functions than bulk two-point functions.

\begin{figure}
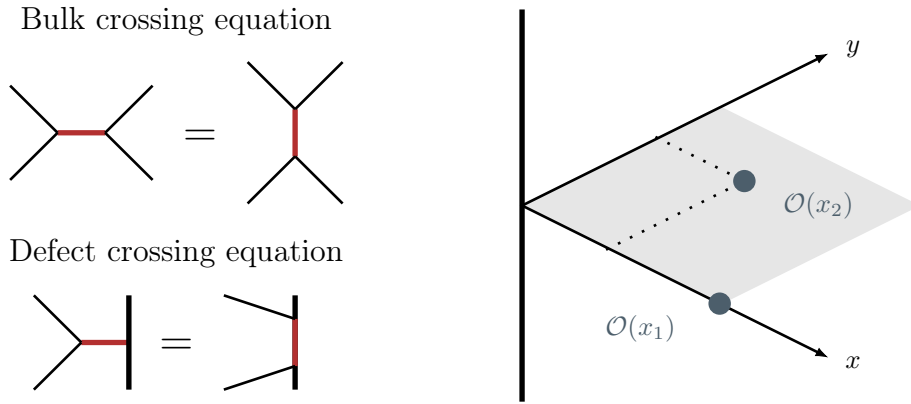

\centering
\begin{subfigure}{.425\textwidth}
\centering
Bulk crossing equation \\[.75em] \BulkCrossing \\[1em]
Defect crossing equation \\[.75em] \DefectCrossing
\end{subfigure}%
\begin{subfigure}{.5\textwidth}
\raggedleft
\TwoPointFunctionSetup
\end{subfigure}
\caption{
The bulk and defect crossing equations, given respectively in \eqref{eq:Bulk_CrossingSymmetry} and \eqref{eq:DefectTwoPoint_BlockExpansionBulk2}-\eqref{eq:DefectTwoPoint_BlockExpansionDefect}, are represented in the left figure, where the red lines correspond to the exchanged operators.
The right figure shows the two-point function in the presence of a line defect (the bold line) in the conformal frame $x_1 = (0,1,0,0)$ and $x_2 = (0,x,y,0)$.
}
\label{fig:DefectTwoPoint_Setup}
\end{figure}

In a defect CFT, two-point functions are the first correlation functions to exhibit a non-trivial dependence on kinematic variables.
For a defect of codimension $q \geq 2$, the correlators of two scalar operators take the form
\begin{equation}
\vvev{ \Op_{\Delta_1} (x_1) \Op_{\Delta_2} (x_2) } = \frac{F(r,w)}{ |x_1^\perp|^{\Delta_1} |x_2^\perp|^{\Delta_2} }\,,
\label{eq:DefectTwoPointFunctions}
\end{equation}
where the variables $r$ and $w$ are defect cross-ratios defined via
\begin{equation}
r + \frac{1}{r} = \frac{| x_{12}^\newparallel|^2 + |x_1^\perp|^2 + |x_2^\perp|^2 }{ |x_1^\perp|^2 |x_2^\perp|^2 }\,, \quad w + \frac{1}{w} = \frac{2 x_1^\perp \cdot x_2^\perp}{ |x_1^\perp|^2 |x_2^\perp|^2 }\,.
\label{eq:DefectCrossRatios_rw}
\end{equation}
Another useful set of cross-ratios $z$, $\zb$ can be defined through
\begin{equation}
z + \zb := 2  (x_1^\perp \cdot x_2^\perp)\,, \quad (1-z)(1-\zb) := x_{12}^2\,.
\label{eq:DefectCrossRatios_zzb}
\end{equation}
Geometrically, $z$ and $\zb$ are coordinates in a plane orthogonal to the defect, while $r$ corresponds to a radius and $w$ to an angle.
Without loss of generality, we can place the operator in a conformal frame by setting $x_1 = (0,1,0,0)$ and $x_2 = (0, x, y, 0)$ (see Figure \ref{fig:DefectTwoPoint_Setup}).
We can then identify
\begin{equation}
z = r w = x + i y\,, \quad \zb = \frac{r}{w} = x - i y\,.
\label{eq:DefectCrossRatios_zzb2}
\end{equation}
Note that in the Lorentzian signature, one would find instead that $z$ and $\zb$ are real and independent.

In a sense, two-point functions in the presence of a defect can be viewed as the counterpart of four-point functions \textit{without} a defect.
Both involve two cross-ratios and an expansion in conformal blocks of \eqref{eq:DefectTwoPointFunctions} can be performed using the bulk OPE given in \eqref{eq:OPEv2}:
\begin{equation}
\vvev{ \Op_1 (x_1) \Op_2 (x_2) } = \sum_{\Op_{\Delta, \ell} \text{ prim.}} \lambda_{\Delta_1 \Delta_2 \Delta} C_{\Delta_1 \Delta_2 \Delta} (x_{12}, \pd_2) \vvev{\Op_{\Delta, \ell} (x_2)}\,.
\label{eq:DefectTwoPoint_BlockExpansionBulk1}
\end{equation}
Similarly to the analysis in \eqref{eq:ExpansionInConformalBlocks}, this can be written as
\begin{equation}
F (r,w) = \left( \frac{rw}{(r-w)(rw-1)} \right)^{\frac{\Delta_1 + \Delta_2}{2}} \sum_{\Delta, \ell} \lambda_{\Delta_1 \Delta_2 \Delta} a_{\Delta}\, f_{\Delta, \ell} (r,w)\,.
\label{eq:DefectTwoPoint_BlockExpansionBulk2}
\end{equation}
Here, the functions $f_{\Delta,\ell} (r, w)$ are conformal blocks fixed by the symmetry and the quantum numbers $\Delta$ and $\ell$ (spin) of the exchanged operators $\Op_{\Delta,\ell}$.
Their precise form can be found in \eqref{eq:ConformalBlocks_TwoPoint_Bulk}.
Note that the three- and one-point function coefficients depend on $\ell$ as well.
However, we suppress this dependence here to avoid cluttering.
We reinstate it whenever confusion may arise.

Another similarity between \eqref{eq:DefectTwoPointFunctions} and \eqref{eq:ConformalFourPointFunctions} is that bulk four-point functions depend on a single cross-ratio in $d=1$.
This is also true for defect two-point functions with codimension $q=1$, i.e., for boundaries.
In a more general way, there exist interesting group-theoretical relations between bulk and defect theories, which are briefly discussed in Section \ref{subsec:MoreWardIdentities} for a supersymmetric case.

The higher-point functions of a defect CFT can be derived from lower-point functions through iterative application of the OPE \eqref{eq:OPEv2}.
As a result, all these correlation functions can be written in terms of the CFT data, which now includes the scaling dimensions of the local operators $\Delta_i$, their one-point function coefficients $a_{\Delta_i}$, and the bulk OPE coefficients $\lambda_{\Delta_i \Delta_j \Delta_k}$.

\subsubsection[Defect \\ representations\ ]{Defect representations}
\label{subsubsec:DefectRepresentations}

There exists another OPE that one can take to expand the defect two-point functions, which involves bringing one bulk operator close to the defect and expanding around that limit. This is called the \textit{defect OPE}, leading to a second block expansion of \eqref{eq:DefectTwoPointFunctions}
\begin{equation}
F(r,w) = \sum_{\Dh, s} b_{\Delta_1 \Dh} b_{\Delta_1 \Dh} \fh_{\Dh, s} (r,w)\,.
\label{eq:DefectTwoPoint_BlockExpansionDefect}
\end{equation}
The defect blocks $\fh_{\Dh, s} (r,w)$ (with $s$ the $SO(q)$ quantum number) can be found in \eqref{eq:ConformalBlocks_TwoPoint_Defect}.
Similarly to \eqref{eq:DefectTwoPoint_BlockExpansionBulk2}, we suppress the dependence on $s$ in $b_{\Delta \Dh}$.
Note that, throughout this thesis, hats always refer to the defect channel.

In \eqref{eq:DefectTwoPoint_BlockExpansionDefect}, the exchanged operators, denoted by $\Oh_{\Dh,s}$, are new \textit{defect} operators that only exist in the defect CFT.
They correspond to local excitations on the defect, and as such they reside in the $p$-dimensional CFT preserved by the defect.
It is important to notice that the scaling dimensions $\Dh$ of these operators are \textit{independent} of the bulk operators.
The OPE coefficients $b_{\Delta \Dh}$ correspond to bulk-defect two-point functions, defined for the case of scalars ($s=0$) as
\begin{equation}
\vvev{\Op_{\smash{\Delta_1}} (x^\perp_1) \Oh_{\smash{\Dh_2}} (x^\newparallel_2)} = \frac{b_{\Delta_1 \Dh_2}}{(|x_1^\perp|^2 + |x_{12}^\newparallel|^2)^{\Dh_2} |x_1^\perp|^{\Delta_1 - \Dh_2}}\,.
\label{eq:BulkDefectTwoPoint}
\end{equation}

As a consequence, the CFT data listed at the end of the previous section is \textit{not sufficient} for solving the defect CFT as a whole.
The correlation functions that involve defect operators $\Oh_{\Dh,s}$ form a $p$-dimensional CFT, and consequently they follow the constraints derived in \eqref{eq:ConformalTwoPointFunctions}-\eqref{eq:HigherPointFunctions}.
For a line defect ($p=1$), these correlators are \textit{one-dimensional} and depend on $n-3$ cross ratios, as explained in \eqref{eq:SpacetimeCrossRatio1d}.
All the additional bulk-defect correlators can be obtained from the knowledge of $b_{\Delta \Dh}$ and the $1d$ CFT data.

We can now provide a complete list of the CFT data necessary for fully solving a defect CFT:
\begin{equation}
\lbrace \Delta_i, \Dh_i, a_{\Delta_i}, b_{\Delta_i \Dh_j}, \lambda_{\Delta_i \Delta_j \Delta_k}, \lambda_{\Dh_i \Dh_j \Dh_k}\rbrace\,,
\label{eq:DefectCFTData}
\end{equation}
where the coefficients $\lambda_{\Dh_i \Dh_j \Dh_k}$ refer to the three-point functions of defect operators.
This thesis focuses on the study of correlation functions in specific defect CFTs, both in the weak- and strong-coupling regimes, from which an infinite amount of CFT data can be extracted by expanding them in blocks.

To conclude this section, note that the expansion \eqref{eq:DefectTwoPoint_BlockExpansionDefect} provides another parallel to the bulk four-point functions \eqref{eq:ConformalFourPointFunctions}: a \textit{defect crossing equation}.
This relation is obtained by equating \eqref{eq:DefectTwoPoint_BlockExpansionBulk2} with \eqref{eq:DefectTwoPoint_BlockExpansionDefect}, and is represented graphically in Figure \ref{fig:DefectTwoPoint_Setup}.
Notice however that there is a crucial difference with \eqref{eq:Bulk_CrossingSymmetry}: while the two sides of the equation in a bulk CFT involve the same type of OPE coefficients $\lambda_{\Delta_i \Delta_j \Delta_k}$, in the defect CFT case we have $a_{\Delta} \lambda_{\Delta_1 \Delta_2 \Delta}$ on one side and $b_{\Delta_1 \Dh} b_{\Delta_2 \Dh}$ on the other.
Therefore, there is \textit{no real crossing}, and the name is an abuse of language that comes from the study of CFTs in the bulk.
Nevertheless, the defect crossing equation is a powerful identity that strongly constrains the defect CFT data.

\subsection{From weak to strong coupling}
\label{subsec:FromWeakToStrongCoupling}

The residual conformal symmetry in defect CFTs provides strong handles on the associated correlation functions, making them potentially tractable even in regimes that would otherwise be challenging to explore using traditional techniques.
In this thesis, we study these correlators perturbatively in both the weak- and strong-coupling regimes.

This section offers an overview of the methodologies used in each context.
At weak coupling, Feynman diagrams are computed using the machinery of perturbation theory.
In the strong-coupling regime, the analytic conformal bootstrap is our preferred approach.

\subsubsection{Perturbation theory\ }
\label{subsubsec:PerturbationTheory}

The calculation of correlation functions in weakly-coupled quantum field theories is a topic that has been extensively studied and refined in the past seventy years.
There exist many introductions of great quality to this topic (see for instance \cite{Peskin:1995ev}), and here we only discuss the important notions related to the content of this thesis.

One efficient way of studying weakly-coupled systems is to consider the interacting theory as a small deviation from a free theory, which can be solved exactly.
Such computations are represented by \textit{Feynman diagrams}, which are assembled using a set of basic ingredients known as \textit{Feynman rules}.
Here we focus on $\Nm=4$ SYM and on Yukuwa CFTs, for which Feynman rules are given respectively in Sections \ref{subsec:N4SuperYangMills} and \ref{subsec:YukawaCFTsIn4EpsDimensions}.
This concept extends straightforwardly to defect CFTs, in which the coupling to the defect can be represented by Feynman rules as well.
For the models mentioned above, they are given respectively in Sections \ref{subsec:TheWilsonLineDefectCFT} and \ref{subsec:TheMagneticLine}.

Feynman diagrams give rise to \textit{Feynman integrals}, for which the difficulty increases as the order of the coupling (and hence the precision of the calculation) increases.
Such integrals have been studied in great depth, in particular for deriving precise estimations of the processes observed in high-energy colliders.
Appendix \ref{sec:FeynmanIntegrals} presents the results for various one- and two-loop integrals.
In the case of defect CFT, we encounter new integrals due to the coupling with the defect, about which much less is known.
There, the symmetry is often of great help for obtaining a full solution (see for instance \cite{Barrat:2020vch}).

Feynman integrals typically contain divergences that need to be accounted for.
As a consequence, operators with anomalous dimensions of the form given in \eqref{eq:Delta_CouplingExpansion} require a \textit{renormalization} procedure.
Expanding the two-point function at small coupling $g \sim 0$, we have
\begin{equation}
\vev{\Op_{\Delta} (x_1) \Op_{\Delta} (x_2)} = \frac{1}{x_{12}^{2\Delta^{(0)}}}
\biggl\lbrace
1 + g\, \gamma_{\Delta}^{(1)}\, \log \frac{\veps^2}{x_{12}^2} + \ldots
\biggr\rbrace\,,
\label{eq:DivergentTwoPoint_PointSplitting}
\end{equation}
with $\veps \sim 0$ encoding the logarithmic divergence.
Higher orders of $g$ produce new divergences of the form $\log^{\ell} \veps^2$.
Note that we are using in this case the \textit{point-splitting} regularization scheme, which consists of taking the limit of two operators approaching each other to regulate the divergences.
This method is explained in more detail in Appendix \ref{subsec:RegularizationSchemes}.
To cancel the divergence, we promote $\Op_\Delta$ to its renormalized version:
\begin{equation}
\Op_\Delta^R (x) := \Op_\Delta (x) \left\lbrace
1 - g\, \gamma_{\Delta}^{(1)}\, \log \frac{\veps^2}{\mu^2} + \ldots
\right\rbrace\,,
\label{eq:RenormalizedOperator}
\end{equation}
where $\mu$ corresponds to some choice of scale.
The function between the brackets is referred to as the \textit{renormalization constant} $Z^{-1}_\Delta$.
This results in a finite two-point function:
\begin{equation}
\vev{\Op^R_{\Delta} (x_1) \Op^R_{\Delta} (x_2)} = \frac{1}{x_{12}^{2\Delta^{(0)}}}
\biggl\lbrace
1 + g\, \gamma_{\Delta}^{(1)}\, \log \frac{\mu^2}{x_{12}^2} + \ldots
\biggr\rbrace\,.
\label{eq:RenormalizedTwoPoint}
\end{equation}
Note that, despite having introduced a scale, this correlation function is still conformal upon renormalization of the dilatation operator.
In the rest of this work, we drop the superscript as we always refer to renormalized operators.

In Chapter \ref{chapter:LineDefectCorrelatorsInFermionicCFT}, we study CFTs across dimensions by using \textit{dimensional regularization} (see Appendix \ref{subsec:RegularizationSchemes}).
The renormalization procedure is the same as for point-splitting, with the difference that the logarithmic divergence is encoded in inverse powers of $\veps$:
\begin{equation}
\vev{\Op_{\Delta} (x_1) \Op_{\Delta} (x_2)} = \frac{1}{x_{12}^{2\Delta^{(0)}}}
\biggl\lbrace
1 - \frac{g}{\veps} \gamma_{\Delta}^{(1)}\, \log \frac{x_{12}^2}{\mu^2} + \ldots
\biggr\rbrace\,,
\label{eq:DivergentTwoPoint_DimReg}
\end{equation}
where we have absorbed a scale in $g$ to keep it dimensionless.

\subsubsection{The conformal bootstrap\ }
\label{subsubsec:TheConformalBootstrap}

Solving a conformal field theory exactly remains a challenging task due to the \textit{infinite} number of unknowns involved in \eqref{eq:ExpansionInConformalBlocks} (for the bulk case) or \eqref{eq:DefectTwoPoint_BlockExpansionBulk2} and \eqref{eq:DefectTwoPoint_BlockExpansionDefect} (for the defect case).
In recent years, a lot of progress has been achieved in the computation of correlation functions in the strong-coupling regime with the \textit{analytic bootstrap}.
Perhaps the biggest accomplishment was the derivation of the \textit{Lorentzian inversion formula} in \cite{Caron-Huot:2017vep}, which can be used to reconstruct the CFT data from the \textit{double discontinuity} of a four-point correlator in the following way:
\begin{equation}
C_{\Delta, \ell} \sim \int dz\, d\zb\, \mu_{\Delta, \ell}(z, \zb)\, \dDisc F(z, \zb)\,,
\label{eq:InversionFormulaBulk}
\end{equation}
where the left-hand side corresponds to the product of OPE coefficients $\lambda_{\Delta_1 \Delta_2 \Delta} \lambda_{\Delta_3 \Delta_4 \Delta}$ for an exchanged operator of scaling dimension $\Delta$ and spin $\ell$, while on the right-hand side $\mu_{\Delta, \ell}$ is a known kernel related to conformal blocks.
The double discontinuity can be regarded as the \textit{absorptive} part of the correlator.
This formula is useful because the absorptive part of the correlator is often simpler than the full correlator.
In particular, the spectrum of exchanged operators is \textit{sparse} at strong coupling, and using insights from holography in combination with \eqref{eq:InversionFormulaBulk} can lead to new strong-coupling results.

We can go one step further and use \eqref{eq:InversionFormulaBulk} as an input in \eqref{eq:ExpansionInConformalBlocks}.
This leads to the following \textit{dispersion relation} for conformal four-point functions:
\begin{equation}
F(z, \zb) = \int dz'\, d\zb'\, K(z, \zb, z', \zb')\, \dDisc F(z', \zb')\,,
\label{eq:DispersionRelationBulk}
\end{equation}
where the integration kernel $K(z, \zb, z', \zb')$ was derived in \cite{Carmi:2019cub}.\footnote{See \cite{Bissi:2019kkx} for an alternative formulation with a single discontinuity.}
This is analogous to the dispersion relation for the four-particle scattering amplitude $\Mm (s,t)$, which involves the \textit{discontinuity} of $\Mm (s,t)$ and takes the form
\begin{equation}
\Mm (s,t) \sim \int \frac{dt'}{t-t'} \Disc \Mm (s,t') + (t \leftrightarrow u)\,,
\label{eq:DispersionRelationScatteringAmps}
\end{equation}
where $s$ and $t$ are the Mandelstam variables.
The relation \eqref{eq:DispersionRelationBulk} is interesting because it gives a direct way to derive the correlator from its absorptive part, without having to go through the CFT data.

Similar developments have followed for defect CFTs.
A \textit{defect Lorentzian inversion formula} was derived in \cite{Lemos:2017vnx,Liendo:2019jpu}.
The principle of this relation is to extract the bulk-defect coefficients presented in \eqref{eq:DefectTwoPoint_BlockExpansionDefect} from the two-point functions \eqref{eq:DefectTwoPointFunctions} in the presence of a defect.
This inversion formula takes the form
\begin{equation}
B(\Dh, s) \sim \int dr'\, dw'\, \hat{\mu}(r', w')\, \Disc F(r',w')\,,
\label{eq:InversionFormulaDefect}
\end{equation}
where $B(\Dh, s)$ represents the product $b_{\Delta_1 \Dh} b_{\Delta_2 \Dh}$ for an exchanged operator of scaling dimension $\Dh$ and spin $s$.
Note that a \textit{single} discontinuity suffices in this context.

It is therefore natural to expect a \textit{dispersion relation for defect CFT}\footnote{The interpretation of such a formula as a dispersion relation should be taken lightly, as the discontinuity of defect correlators does not have well-defined positivity properties, and thus it cannot be interpreted as the absorptive part of a physical process.} with a form similar to \eqref{eq:DispersionRelationBulk} and \eqref{eq:DispersionRelationScatteringAmps}.
Such a formula is derived in Chapter \ref{sec:ADispersionRelationForDefectCFT}, where it is then used to obtain novel strong-coupling results in $\Nm=4$ SYM with a Wilson-line defect.

\bigskip

Analytic bootstrap methods have also been developed for the defect correlators discussed in Section \ref{subsec:ConformalDefects}.
In Chapter \ref{chapter:MultipointCorrelatorsInTheWilsonLineDefectCFT}, we use the techniques developed in \cite{Ferrero:2019luz,Ferrero:2021bsb} for computing a five-point function at strong coupling.

The method can be summarized as follows.
The crossing symmetry \eqref{eq:Bulk_CrossingSymmetry} naturally generalizes to higher-point functions and leads to a network of relations highly constraining for the correlator.
In holographic systems, the leading order can be readily computed using \textit{Witten diagrams} and serves as an input for the bootstrap routine.
The next-to-leading order is determined by assuming a Regge behavior for the anomalous dimensions and using the leading order as an input.
In fortunate cases, the correlator is then fixed by the crossing symmetry relations.
One can then proceed iteratively to extract higher orders, by reinserting the CFT data of the lower orders.
This procedure becomes however increasingly difficult, due to the fact that operators are \textit{mixed}.
Indeed, different operators can have the same scaling dimensions at low orders in the coupling.
This mixing must be resolved to progress order by order.
The authors of \cite{Ferrero:2021bsb} were able to reach the next-to-next-to-next-to-leading order (NNNLO) at strong coupling.

\section{The Wilson-line defect CFT}
\label{sec:TheWilsonLineDefectCFT}

Amidst the vast landscape of conformal field theories, $4d$ $\Nm=4$ Super Yang--Mills is a particularly intriguing model.
It is renowned for its abundance of symmetries, which make it one of the simplest interacting theories in four dimensions.
Its importance in the study of the AdS/CFT correspondence, integrability, and the (super)conformal bootstrap has earned it the nickname of \textit{hydrogen atom} of quantum field theories.
A canonical defect to consider in $\Nm=4$ SYM is the \textit{supersymmetric Wilson loop}.
This extended operator plays a central role in holography, and as such it has been the object of intense research.
The associated defect CFT preserves many of the properties of the bulk theory, making it a prime candidate for the study of defect physics.

In the following, we introduce the bulk theory $\Nm=4$ SYM and Wilson loops, as well as the resulting Wilson-line defect CFT.
Most of the content presented here relies on \cite{Ammon:2015wua}.

\subsection{$\Nm = 4$ Super Yang--Mills}
\label{subsec:N4SuperYangMills}

We begin with a short review of the bulk theory $\Nm=4$ SYM.
On top of the properties mentioned above, this model presents the advantage of being a rare example of a CFT with a coupling constant, and thus it can be studied both in the weak- and strong-coupling regimes.
We are interested in particular in the large $N$ limit, in which the observables of $\Nm=4$ SYM correspond to a free $4d$ quantum field theory at weak coupling and to an integrable $2d$ string worldsheet at strong coupling.\footnote{The notion of weak and strong coupling is somewhat subtle at large $N$.
The appropriate coupling constant will be discussed in \eqref{eq:tHooftCoupling}.}

A comprehensive review of supersymmetry can be found for instance in \cite{wess1992supersymmetry,Weinberg:2000cr,muller2010introduction}.
For an introduction to string theory, we refer the reader to \cite{Green:1987sp,Green:1987mn,Polchinski:1998rq,Polchinski:1998rr}.

\subsubsection{The action\ }
\label{subsubsec:TheAction}

$\Nm = 4$ SYM corresponds to the \textit{maximally extended} supersymmetric theory in four dimensions.
In other words, it is the realization of supersymmetry with the largest number of supercharges $Q^A$ ($A=1, \ldots, \Nm$) with a multiplet representation of spin $\leq 1$, i.e., without a graviton.
The field content of $\Nm = 4$ SYM is the following:
\begin{align*}
\mathbf{1} \text{ gluon } \oplus\, \mathbf{4} \text{ Weyl fermions } \oplus\, \mathbf{6} \text{ scalars}\,.
\label{eq:FieldContent}
\end{align*}
As required by supersymmetry, the bosonic and fermionic degrees of freedom are equal: eight each.
The transformations between bosons and fermions are generated by the supercharges mentioned above (see for instance \cite{Ammon:2015wua}).

The action takes the following form:
\begin{equation}
\begin{split}
S_{\Nm = 4} =\ & \frac{1}{g^2} \tr \int d^4 x\, \biggl( 
\frac{1}{2} F_{\mu\nu} F_{\mu\nu}
+ D_\mu \phi^I D_\mu \phi^I
- \frac{1}{2} [ \phi^I, \phi^J ] [ \phi^I, \phi^J ] \\
& + i\, \psib \slashed{D} \psi
+ \psib \Gamma^I [ \phi^I\,, \psi ]
+ \pd_\mu \cb D_\mu c
+ \xi ( \pd_\mu A_\mu )^2
\biggr)\,,
\end{split}
\label{eq:ActionSYM}
\end{equation}
where $\mu=0, \ldots, 3$ are the spacetime directions and $I=1, \ldots, 6$ are the $R$-symmetry indices.
This notation introduces a single $16$-component Majorana fermion composed of the $4$ Weyl fermions mentioned above.
All the fields are in the \textit{adjoint} representation of $SU(N)$ and carry a generator in their definitions:
\begin{equation}
\phi := T^a \phi^a\,, \quad \psi := T^a \psi^a\,, \quad A_\mu := T^a A_\mu^a\,,
\label{eq:FieldsAndGenerators}
\end{equation}
where $a = 1, \ldots, N$ corresponds to the \textit{color} index of the $SU(N)$ gauge group.

The action \eqref{eq:ActionSYM} exhibits conformal symmetry at the quantum level, meaning that the $\beta$-function of the coupling is zero non-perturbatively.\footnote{There exists another class of vacua for $\Nm=4$ SYM.
Here we choose the \textit{superconformal phase}, where all the vacuum expectation values of the scalars vanish.
The other possibility is the \textit{Coulomb phase} \cite{Kraus:1998hv}, where at least one scalar has a non-zero expectation value.
This breaks both conformal symmetry and the gauge group.
We will not consider the Coulomb phase in this thesis.}
In addition, $\Nm=4$ SYM is UV-complete in perturbation theory and is invariant under the $S$-duality group $SL(2, \Zds)$, leading to an interesting weak/strong-coupling duality of the electric/magnetic type and known as \textit{Montonen--Olive} duality \cite{Montonen:1977sn,Seiberg:1994rs,Seiberg:1994pq}.
Although not the main topic of this thesis, $S$-duality will be briefly discussed in Section \ref{subsec:tHooftLoops}.

\subsubsection[The AdS/CFT correspondence\ ]{The AdS/CFT correspondence}
\label{subsubsec:TheAdSCFTCorrespondence}

We are now turning our attention to one of the most fascinating discoveries of modern theoretical physics: the anti-de Sitter/conformal field theory (AdS/CFT) correspondence.
This duality was discovered by Juan Maldacena in 1997 \cite{Maldacena:1997re} and relates a quantum field theory on flat spacetime (the CFT) to a string theory in anti-de Sitter space.
This is a remarkable relation since string theory is a promising candidate framework to describe quantum gravity, while quantum field theory on flat space does not appear to contain gravity.
The AdS/CFT correspondence implies that these two theories describe the same physics.

$\Nm = 4$ SYM was the first realization of the \textit{holographic principle} to be unraveled.
The holographic principle \cite{Susskind:1994vu} states that the number of degrees of freedom of a gravitational theory in a \textit{volume} $V$ scales as the \textit{surface area} $\pd V$ of that volume.
In other words, it postulates that gravity in $d+1$ dimensions can be described by a theory living on the \textit{boundary} of that space, i.e., in $d$ dimensions.
This is a generalization of the black hole example provided in Section \ref{subsec:FromCFTToGravity}.
In the case of $\Nm = 4$ SYM, the CFT is \textit{four-dimensional} and is dual to a \textit{five-dimensional} string theory in AdS space.
$\Nm = 4$ SYM can therefore be thought of as being defined on the (conformal) boundary of the AdS space.

The weakest form of the conjecture, which is also the most studied one, is that the \textit{large $N$} limit of $\Nm=4$ SYM at \textit{strong} coupling is dual to a type IIB superstring theory in AdS$_5 \times S^5$ at \textit{weak} coupling.
In this context, it is convenient to define the \textit{'t Hooft coupling} \cite{t1993planar}
\begin{equation}
\lambda := g^2 N\,.
\label{eq:tHooftCoupling}
\end{equation}

On the AdS side, the two free parameters are the \textit{string coupling} $g_s$ and the ratio $L^2/\alpha'$, where $L$ is the \textit{radius of curvature} of the AdS space and $\alpha'$ is the square of the \textit{string length}.
The correspondence establishes the following relations between the parameters of the field and string theories:
\begin{equation}
g^2 = 2 \pi g_s\,, \quad \lambda = \frac{L^4}{2 \alpha'^2}\,.
\label{eq:ParametersStringTheory}
\end{equation}
As mentioned above, one important observation is that the AdS/CFT correspondence is an example of \textit{weak/strong}-coupling duality:\footnote{Note that, with respect to $g$, the theory is \textit{weakly coupled}, even when $\lambda \gg 1$.} when the dual gravitational theory is considered at the classical level, the quantum field theory is strongly coupled.
For this reason, holographic models are prime examples for studying field theories in the strong-coupling regime.

The \textit{supergravity} (SUGRA) limit of the string theory is studied perturbatively through \textit{Witten diagrams}.
In Chapter \ref{chapter:BootstrappingHolographicDefectCorrelators}, we use insights from the dual SUGRA theory to obtain valuable input for our bootstrap computations.
More generally, holographic theories are interesting as they provide additional handles to study correlation functions in the strong-coupling regime.

\subsubsection{Feynman rules\ }
\label{subsubsec:FeynmanRules}

The action given in \eqref{eq:ActionSYM} generates correlation functions, which can be computed at weak coupling using the laws of quantum field theory and the \textit{Feynman rules} of the action.
The (free) propagators take the form
\begin{equation}
\begin{split}
\text{Scalars:} \qquad 
& \ScalarPropagator = g^2 \delta^{IJ} \delta^{ab}\, I_{12}\,, \\
\text{Gluons:} \qquad 
& \GluonPropagator = g^2 \delta_{\mu\nu} \delta^{ab}\, I_{12}\,, \\
\text{Fermions:} \qquad 
& \FermionPropagator = i g^2 \delta^{ab} \slashed{\pd}_1 I_{12}\,, \\
\text{Ghosts:} \qquad 
& \GhostPropagator = g^2 \delta^{ab} I_{12}\,,
\end{split}
\label{eq:Propagators}
\end{equation}
where we have introduced the $4d$ scalar propagator
\begin{equation}
I_{12} = \frac{1}{4\pi^2 x_{12}^2}\,.
\label{eq:PropagatorFunction4d}
\end{equation}
Note that, in our conventions, free propagators carry the (dimensionless) coupling $g^2$. 

The dynamics of a quantum field theory are controlled by its vertices.
In $\Nm=4$ SYM, we have \textit{eight} vertices, which relate to the coupling constant by a prefactor $g^{-2}$.
The cubic vertices are
\begin{equation}
\VertexGluonGluonGluon \quad
\VertexScalarScalarGluon \quad
\VertexFermionFermionGluon \quad
\VertexFermionFermionScalar \quad
\VertexGhostGhostGluon\ ,
\label{eq:CubicVertices}
\end{equation}
while the quartic ones are
\begin{equation}
\VertexFourScalars \quad
\VertexScalarScalarGluonGluon \quad
\VertexGluonGluonGluonGluon\ .
\label{eq:QuarticVertices}
\end{equation}
For brevity, we do not give here the associated expressions explicitly.
The relevant ones can be found for instance in the appendices of \cite{Barrat:2020vch,Barrat:2021tpn,Barrat:2022eim} in the form of insertion rules.
For future purposes, we give here the insertion rule for the one-loop correction of the propagator:
\begin{equation}
\begin{split}
\SelfEnergyNoText &=
\SelfEnergyDiagramOne
+
\SelfEnergyDiagramTwo
+
\SelfEnergyDiagramThree
+
\SelfEnergyDiagramFour \\
&=
- 2 g^4 N \delta^{ab} \delta^{IJ}\, Y_{112}\,,
\end{split}
\end{equation}
with the integral $Y_{112}$ log-divergent and given in \eqref{eq:YDivergent_PointSplitting}-\eqref{eq:YDivergent_DimReg}.

\subsubsection[Half-BPS operators\ ]{Half-BPS operators}
\label{subsubsec:HalfBPSOperators}

As in Section \ref{sec:ABriefIntroductionToConformalDefects}, local operators are classified according to the quantum numbers of the symmetry group.
The action \eqref{eq:ActionSYM} of $\Nm=4$ SYM corresponds to the \textit{superconformal} algebra $PSU(2,2|4)$.
The associated quantum numbers are the scaling dimension $\Delta$, the spin $\ell$ and the $SO(6)_R$ $R$-symmetry charge $k$.

In Chapter \ref{chapter:BootstrappingHolographicDefectCorrelators}, we focus on the subset of superconformal operators known as \textit{half-BPS} primaries, which satisfy the \textit{Bogomol'nyi--Prasad--Sommerfield} (BPS) condition \cite{bogomol1976stability,Prasad:1975kr}
\begin{equation}
[ Q^A_\alpha\,, \Op_\Delta (x) ]_\pm = 0\,,
\label{eq:BPSCondition}
\end{equation}
for \textit{half} of the supercharges $Q^A$, and where the subscript $\pm$ means that the operation should either be a commutator or an anticommutator.
Scalar half-BPS operators can be fully characterized by their quantum number $\Delta$, with the $R$-symmetry Dynkin labels being $[0,\Delta,0]$ while the spin is $\ell=0$.

For simplicity, we focus on \textit{single-trace} half-BPS operators, which can be defined as
\begin{equation}
\Op_{\Delta} (u,x) := \frac{1}{\sqrt{n_{\Delta}}} \tr \left( u \cdot \phi (x) \right)^\Delta\,.
\label{eq:SingleTraceHalfBPSOperators_Bulk}
\end{equation}
To satisfy the half-BPS condition, the $SO(6)_R$ vector $u^{I=1, \ldots, 6}$ must satisfy
\begin{equation}
u^2 = 0\,.
\label{eq:u_Bulk}
\end{equation}
These operators have protected scaling dimensions, meaning that the corrections $\gamma^{(\ell)}$ in \eqref{eq:Delta_CouplingExpansion} vanish \cite{Dobrev:1985qv,Arutyunov:2001mh}.
Similarly, the normalization constants are fixed at large $N$ by their leading-order values
\begin{equation}
n_{\Delta} = \frac{\Delta \lambda^\Delta}{2^{3\Delta} \pi^{2\Delta}}\,,
\label{eq:NormalizationConstantHalfBPS_Bulk}
\end{equation}
with $\lambda$ the 't Hooft coupling defined in \eqref{eq:tHooftCoupling}.
These operators are said to saturate the unitarity bound (see Section \ref{subsec:ConformalFieldTheory}) and belong to a \textit{short} multiplet.
This means that the representation has null states (i.e., states with zero norm), which can safely be discarded from the multiplet.

It is also possible to construct \textit{multi-trace} operators, which are indistinguishable from single-trace operators at the group-theoretical level. These operators can be defined as follows:
\begin{equation}
\Op_{\vec{\Delta}} (u,x) := \frac{1}{\sqrt{n_{\smash{\vec{\Delta}}}}} \tr \left( u \cdot \phi (x) \right)^{\Delta_1} \ldots \tr \left( u \cdot \phi (x) \right)^{\Delta_n}\,,
\label{eq:MultiTraceOperators_Bulk}
\end{equation}
with $\vec{\Delta} := (\Delta_{1}\,, \ldots\,, \Delta_{n})$, such that $\Delta_1 + \ldots + \Delta_n = \Delta$.
The scaling dimensions of these operators are equal to the ones of single-trace operators, but the normalization constants differ.

To conclude, note that BPS operators form a special subset of the local operators in $\mathcal{N}=4$ SYM and that the theory contains infinitely many other operators with \textit{unprotected} scaling dimensions and belonging to \textit{long} multiplets.
One important example is the \textit{Konishi} operator
\begin{equation}
\Km (x) := \tr \phi^2 (x)\,,
\label{eq:KonishiOperator}
\end{equation}
which is degenerate with $\Op_2 (u,x)$ at the classical level.
While the anomalous dimensions of the Konishi and other unprotected operators have been extensively studied, we will not be considering them directly in this thesis.
Nonetheless, these operators play an important role in the OPE of bulk operators.

\subsection{The Wilson-line defect CFT}
\label{subsec:TheWilsonLineDefectCFT}

We now come back to the study of $\Nm=4$ SYM to introduce the supersymmetric Wilson line.
As explained in Section \ref{subsec:ConformalDefects}, the presence of a (conformal) line defect breaks the full conformal group and generates a defect CFT.
In the case of $\Nm = 4$ SYM with a Wilson line, the resulting theory preserves many of the attractive properties of its parent theory.
It is supersymmetric, believed to be integrable, and the holographic dual is identified.
This offers many handles for the study of the defect as well as for the development of the bootstrap techniques presented in Section \ref{subsec:FromWeakToStrongCoupling}.

In the following, we review well-known results for Wilson loops and introduce the correlation functions of the Wilson-line defect CFT.
This section is based on many modern articles, which are acknowledged in each subsection.

\subsubsection{Supersymmetric Wilson loops\ }
\label{subsubsec:SupersymmetricWilsonLoops}

\begin{figure}
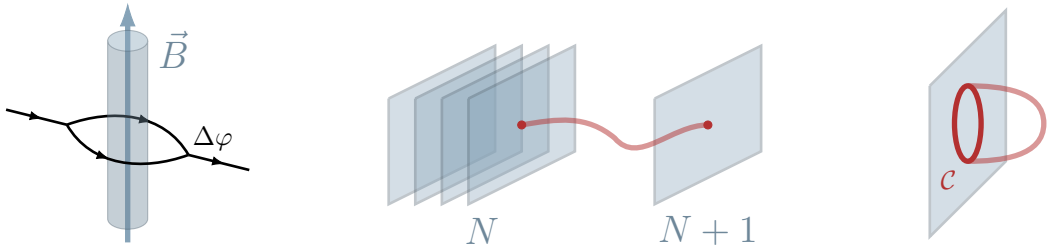

\begin{subfigure}[l]{.35\textwidth}
\flushleft
\AharonovBohm
\end{subfigure}%
\begin{subfigure}[c]{.40\textwidth}
\centering
\Branes
\end{subfigure}%
\begin{subfigure}[r]{.25\textwidth}
\flushright
\Worldsheet
\end{subfigure}
\caption{
The leftmost figure depicts the Aharonov--Bohm effect, where a charged particle experiences a phase shift depending on the path it takes close to a magnetic field $\vec{B}$.
In the center, a stack of $N$ $D3$-branes is depicted, to which an additional brane has been added at a large distance from the slack.
The open string corresponds to a massive $W$-boson, for which trajectories around a closed path correspond to the expectation value of the supersymmetric Wilson loop.
The rightmost figure represents the AdS dual of the supersymmetric Wilson loop: a string with a worldsheet ending on the path of the Wilson loop at the boundary.}
\label{fig:Aharonov}
\end{figure}

Wilson loops are observables in quantum field theory that measure the \textit{phase} accumulated by a particle carrying a quantum number associated with a gauge field, such as an electric charge (Abelian case) or a color charge (non-Abelian), when it is transported around a closed path in spacetime.
In other words, the parallel transport between two points $x$ and $y$ depends on the \textit{path} taken from $x$ to $y$.
It is related to the \textit{Aharonov--Bohm} effect in quantum mechanics \cite{Aharonov:1959fk}, where a charged particle experiences a phase shift in its wave function when it is transported around a closed path that encloses the magnetic flux (see Figure \ref{fig:Aharonov}). 

For pure Yang--Mills theory, the Wilson loop operator is defined by
\begin{equation}
\WC := \frac{1}{N} \tr \pexp i \oint_C dx_\mu\, A_\mu(x)\,,
\label{eq:WilsonLoop}
\end{equation}
where the trace is taken over the fundamental representation of the gauge group $SU(N)$.
Note that the gauge fields carry a generator as in \eqref{eq:FieldsAndGenerators}, which do not commute in general.
This leads to an ambiguity in how to order the fields in the definition of the exponential.
This is solved by defining the \textit{path-ordered} operator such that for any operators $\phi (x(s_1))$ and $\psi (x(s_2))$
\begin{equation}
\Pm ( \phi (x(s_1)) \psi (x(s_2)) ) = \begin{cases} \phi (x(s_1)) \psi (x(s_2)) & \text{for } s_1 > s_2\,, \\ \psi (x(s_2)) \phi (x(s_1)) & \text{for } s_2 > s_1\,. \end{cases}
\label{eq:PathOrdering}
\end{equation}

For certain paths $\Cm$, the expectation value $\vev{\WC}$ of the Wilson loop encodes interesting information about the gauge theory.
In QCD, Wilson loops provide an order parameter for \textit{confinement} and \textit{deconfinement}.
Moreover, a closed Wilson loop describes the propagator of a quark and an antiquark along the path.
In this context, the expectation value of the Wilson loop gives the \textit{potential energy} between the two particles \cite{Wilson:1974sk,CTEQ:1993hwr}.

We wish now to construct an object similar to the Wilson loop in $\Nm=4$ SYM.
To obtain an analogon of \eqref{eq:WilsonLoop}, we must find a way to introduce massive particles transforming in the fundamental representation to form a quark-antiquark potential.
In $\Nm = 4$ SYM, all the fields are massless and transform in the adjoint representation, and we find it useful to think of the theory as the low-energy limit of open strings ending on $D3$-branes. 
One way to introduce heavy particles is to consider an additional brane, parallel to the stack of $D3$-branes but located at a large distance $\ell$ of the stack (see Figure \ref{fig:Aharonov}).
This procedure introduces massive fundamental particles to $\Nm=4$ SYM, corresponding to the ground states of the open strings stretching between the stack of $D3$-branes and the single separated brane.
In analogy to the Higgs mechanism, we refer to these particles as \textit{$W$-bosons} of the (broken) gauge group $SU(N+1) \to SU(N) \times U(1)$.
The trajectories of these particles around a closed path $\Cm$ give rise to a phase factor, given by the expectation value of
\begin{equation}
\WC = \frac{1}{N} \tr \pexp \oint_C d\tau\, \left( i \dx_\mu A_\mu (\tau) + |\dx|\, \theta \cdot \phi (\tau) \right)\,.
\label{eq:SUSYWilsonLoop}
\end{equation}
This operator is called the supersymmetric \textit{Maldacena--Wilson loop} and was introduced in \cite{Maldacena:1998im}.
$\theta^I$ is a $SO(6)_R$ vector tuning the polarization of the scalars and satisfying $\theta^2 = 1$ for the defect to be half-BPS, which means that it satisfies the condition given in \eqref{eq:BPSCondition}.
Without loss of generality, we choose it throughout this thesis to be
\begin{equation}
\theta = (0,0,0,0,0,1)\,,
\label{eq:theta_SYM}
\end{equation}
which means that only $\phi^6$ couples to the line.

The AdS dual of the Wilson loop is very intuitive.
Following the prescription of \cite{Maldacena:1997re,Maldacena:1998im}, we replace the $N$ $D3$-branes by the spacetime \AdSfive.
The remaining brane, corresponding to the Wilson loop, is identified with the conformal boundary of AdS$_5$.
Therefore, the expectation value of the Wilson loop operator is dual to the partition function of a macroscopic string with its worldsheet ending on the path of the Wilson loop at the boundary (see Figure \ref{fig:Aharonov}).

In the fundamental representation, the expectation value of the \textit{circular} Wilson loop was computed exactly both at finite \cite{Drukker:2000rr} and at large $N$ \cite{Erickson:2000af}:
\begin{equation}
\begin{alignedat}{2}
\vev{ \Wc } &= \frac{1}{N} L_{N-1}^1 \left( - \frac{g^2}{4} \right) \exp \left( \frac{g^2}{8} \frac{N-1}{N} \right) && \qquad \text{(at finite } N)\,, \\
\vev{ \Wc } &= \frac{2}{\sqrt{\lambda}} I_1 (\sqrt{\lambda} ) && \qquad \text{(at } N \to \infty)\,,
\end{alignedat}
\label{eq:Wc_FiniteNAndLargeN}
\end{equation}
with the \textit{generalized Laguerre polynomials} \cite{sonine1880recherches}, which can be obtained via the recurrence relation
\begin{equation}
L_{k+1}^\alpha (x) = \frac{ (2k+1+\alpha-x) L_k^\alpha (x) - (k+\alpha) L_{k-1}^\alpha (x) }{ k+1 }\,,
\label{eq:LaguerrePolynomials}
\end{equation}
and the \textit{modified Bessel function of the first kind} \cite{abramowitz1988handbook}
\begin{equation}
I_\alpha (x) := \sum_{m=0}^{\infty} \frac{1}{m! \Gamma(m+\alpha+1)} \left( \frac{x}{2} \right)^{2m+\alpha}\,.
\label{eq:BesselFunction}
\end{equation}
These results were first obtained perturbatively by observing that only free contractions contribute at the first orders in perturbation theory in the seminal paper by Erickson, Semenoff, and Zarembo \cite{Erickson:2000af}.
It was then assumed that this property would extend to all orders, leaving only so-called rainbow diagrams to compute.
These could be summed over at large $N$, leading to \eqref{eq:Wc_FiniteNAndLargeN}.
This result was later confirmed rigorously using supersymmetric localization \cite{Pestun:2009nn}.

For the infinite straight line, the same results can be derived and the expectation value is just
\begin{equation}
\vev{ \Wl } = 1\,.
\label{eq:Wl_vev}
\end{equation}
The circular Wilson loop and the infinite straight Wilson line actually \textit{fully determine one another}, since one can go from one to the other through a conformal transformation.
The fact that open and closed geometries can be related via conformal transformations is not per se surprising.
Indeed, the special conformal transformations introduced in \eqref{eq:Generators} exchange a point at infinity with a point at a finite distance.
In this sense, the SCTs form a symmetry on $S^4$ (\textit{not} $\mathbb{R}^4$), and on the sphere, there is no distinction between a line and a circle.\footnote{It was shown in \cite{Drukker:2012de} that the Wilson loops on circular and straight lines are equivalent, up to a conformal anomaly equal to \eqref{eq:Wc_FiniteNAndLargeN}.}

\subsubsection[Defect Feynman \\ rules\ ]{Defect Feynman rules}
\label{subsubsec:DefectFeynmanRules}

In Chapters \ref{chapter:BootstrappingHolographicDefectCorrelators} and \ref{chapter:MultipointCorrelatorsInTheWilsonLineDefectCFT}, we consider correlation functions of local operators with a Wilson line.
We can derive a set of \textit{defect Feynman rules} useful for weak-coupling computations.

\begingroup
\allowdisplaybreaks

When the supersymmetric Wilson line is disconnected from the rest, the contribution is just
\begin{equation}
\DefectVertexZeroPoint\ = \vev{\Wl} = 1\,.
\label{eq:DefectVertex_ZeroPoint}
\end{equation}
If we expand the defect up to the first order, we obtain insertion rules for scalars and gluons.
To account for correlators of bulk as well as defect operators, we need to consider several situations.
For scalars, we have
\begin{align}
\begin{split}
& \DefectVertexOnePointScalarInf\ = 0\,, \\
& \DefectVertexOnePointScalarLeft\ = \frac{1}{N} \theta^I \tr\, ( T^{a_1} T^{a_2} \ldots ) \int_{-\infty}^{\tau_2} d\tau_3\ I_{13}\,, \\
& \DefectVertexOnePointScalarCenter\ = \frac{1}{N} \theta^I \tr\, ( T^{a_2} T^{a_1} T^{a_3} \ldots ) \int_{\tau_2}^{\tau_3} d\tau_4\ I_{14}\,, \\
& \DefectVertexOnePointScalarRight\ = \frac{1}{N} \theta^I \tr\, ( T^{a_2} T^{a_1} \ldots ) \int_{\tau_2}^{\infty} d\tau_3\ I_{13}\,.
\end{split}
\label{eq:DefectVertices_OnePoint}
\end{align}

Similar rules can be derived for gluons by doing the replacement $\theta^I \to i \dot{x}_\mu$.
In the above, the diagram with limits of integration $-\infty$ and $+\infty$ vanishes due to the tracelessness of the generators of $SU(N)$.
Insertion rules with a larger number of insertions can be found in \cite{Barrat:2020vch}, but they will not be needed for this thesis.

\endgroup

\subsubsection[Correlation functions\ ]{Correlation functions}
\label{subsubsec:CorrelationFunctions2}

We have seen in Section \ref{subsec:ConformalDefects} that correlation functions in the presence of a defect are fixed by conformal symmetry to take a specific form.
In a supersymmetric theory, we must account for the $R$-symmetry dependence and promote them to supersymmetric versions.

One-point functions of half-BPS operators can be written as
\begin{equation}
\vvev{ \Op_\Delta (u,x) } = a_\Delta \frac{(u \cdot \theta)^\Delta}{|x^\perp|^\Delta}\,,
\label{eq:OnePointFunction_HalfBPS}
\end{equation}
where $u$ and $\theta$ are the $R$-symmetry variables associated respectively to the local operator (see \eqref{eq:SingleTraceHalfBPSOperators_Bulk}) and to the Wilson line (see \eqref{eq:SUSYWilsonLoop}).

For the two-point functions, we now have
\begin{equation}
\vvev{ \Op_{\Delta_1} (u_1, x_1) \Op_{\Delta_2} (u_2, x_2) } = \frac{ (u_1 \cdot \theta)^{\Delta_1} (u_2 \cdot \theta)^{\Delta_2} }{ |x_1^\perp|^{\Delta_1} |x_2^\perp|^{\Delta_2} } \Fm (z, \zb; \sigma)\,,
\label{eq:TwoPointFunctions_HalfBPS}
\end{equation}
with $z$ and $\zb$ defined as in \eqref{eq:DefectCrossRatios_zzb}.
We can set $\Delta_1 \leq \Delta_2$ without loss of generality.
Again, the conformal prefactor has been promoted to a superconformal one.
Note also that the reduced correlator $\Fm$ depends on an additional \textit{$R$-symmetry cross-ratio} $\sigma$, defined to be
\begin{equation}
\sigma := \frac{u_1 \cdot u_2}{(u_1 \cdot \theta)(u_2 \cdot \theta)}\,.
\label{eq:TwoPoint_RSymmetryCrossRatio}
\end{equation}
In terms of $\sigma$, the correlator $\Fm (z, \zb; \sigma)$ is a polynomial of degree $\Delta_1$, with each power corresponding to a different \textit{$R$-symmetry channel}.
This can be expressed as
\begin{equation}
\Fm (z, \zb; \sigma) = \sum_{j=0}^{\Delta_1} \Omega^j\, F_{\Delta_1 - j} (z, \zb)\,,
\label{eq:TwoPoint_RSymmetryChannels}
\end{equation}
where we have defined for future convenience
\begin{equation}
\Omega := \frac{z \zb\, \sigma}{(1-z)(1-\zb)}\,.
\label{eq:Omega}
\end{equation}
This reflects the different ways in which $u_1$, $u_2$ and $\theta$ can be contracted.

Similarly to \eqref{eq:DefectTwoPointFunctions}, $\Fm (z, \zb; \sigma)$ can be expanded in \textit{superconformal blocks} (or \textit{superblocks}) $\Gm (z, \zb; \sigma)$ \cite{Liendo:2016ymz}.
In the bulk channel, the expansion takes the form
\begin{equation}
\Fm (z, \zb; \sigma) = \Omega^{\frac{\Delta_1 + \Delta_2}{2}} \sum_{\Delta, \ell, k} \lambda_{\Delta_1 \Delta_2 \Delta} a_\Delta\, \Gm_{\Delta,\ell, k} (z, \zb; \sigma)\,.
\label{eq:TwoPoint_SuperconformalBlocksBulk}
\end{equation}
The difference with \eqref{eq:DefectTwoPoint_BlockExpansionBulk2} is that the superblocks account for the $R$-symmetry quantum number $k$ as well, and as a consequence, the exchanged operators are the superprimaries of $PSU(2,2|4)$ with quantum numbers $\Delta$, $\ell$ and $k$.
As in \eqref{eq:ExpansionInConformalBlocks}, the superblocks depend on the external dimensions $\Delta_1$ and $\Delta_2$, which can be seen explicitly in \eqref{eq:ConformalBlocks_TwoPoint_BulkHalfBPS}.

The defect channel \eqref{eq:DefectTwoPoint_BlockExpansionDefect} becomes
\begin{equation}
\Fm (z, \zb; \sigma) = \sum_{\Dh, s, \kh} b_{\Delta_1 \Dh} b_{\Delta_2 \Dh}\, \Gm_{\Dh,s, \kh} (z, \zb; \sigma)\,,
\label{eq:TwoPoint_SuperconformalBlocksDefect}
\end{equation}
where the exchanged operators are superprimaries of $OSP(4^*|4)$ with the quantum numbers $\Dh$, $s$, and $\kh$, which will be introduced shortly.
Note that in \eqref{eq:TwoPoint_SuperconformalBlocksBulk} and \eqref{eq:TwoPoint_SuperconformalBlocksDefect}, the OPE coefficients depend on all the quantum numbers, not just $\Delta$ and $\Dh$.
We do not indicate this dependence explicitly to streamline the notation, and it is specified only in cases where confusion might arise.
These superblocks will not be needed in this work, however, we point out that they are known in closed form \cite{Liendo:2016ymz}.

Let us now discuss the \textit{defect operators} of our defect CFT.
Following \eqref{eq:SymmetryBreaking}, the Wilson line breaks the $4d$ conformal symmetry to the product $SO(2,1) \times SO(3)$.
The first piece corresponds to a one-dimensional CFT, to which we assign the defect scaling dimension $\Dh$.
The second part refers to rotations orthogonal to the defect, for which the quantum number is $s$ (spin).
The $R$-symmetry is also broken by the presence of the defect: the $SO(6)_R$ group becomes $SO(5)_R$, with $\kh$ the quantum number associated with the defect algebra.

Altogether, the symmetry $PSU(2,2|4)$ of $\Nm = 4$ SYM breaks into the superalgebra $OSP(4^*|4)$ and, as for the bulk, these quantum numbers characterize the representations (the \textit{defect} operators) of the superconformal algebra.
They live in the one-dimensional CFT induced by the Wilson line, and in this context, they form a realization of $\Nm = 8$ superconformal quantum mechanics \cite{Giombi:2017cqn,Bellucci:2003hn}.
In the following, we introduce the half-BPS operators that are the pendant of \eqref{eq:SingleTraceHalfBPSOperators_Bulk} along the defect.

We focus on \textit{single-trace}\footnote{The meaning of single-trace differs from \eqref{eq:SingleTraceHalfBPSOperators_Bulk}.
Here it refers to the fact that the correlation functions defined in \eqref{eq:1dCorrelators} contain a single-trace \textit{for all} the operators, as opposed to the bulk theory in which each single-trace operator carries its own trace.} operators for simplicity.
They are defined as
\begin{equation}
\Oh_{\Dh} (\uh, \tau) := \frac{1}{\sqrt{\nh_{\Dh}}} \Wl [ (\uh \cdot \phi(\tau) )^{\Dh} ]\,,
\label{eq:SingleTraceHalfBPS_Defect}
\end{equation}
where the vector $\uh^{i=1, \ldots, 5}$ is a $SO(5)_R$ null-vector that parallels the $u$ defined in \eqref{eq:u_Bulk} for bulk operators.
It satisfies
\begin{equation}
\uh^2 = 0 \quad \text{and} \quad \uh \cdot \theta = 0\,.
\label{eq:uh}
\end{equation}
This ensures that the representation \eqref{eq:SingleTraceHalfBPS_Defect} is symmetric traceless and does not contract with the field $\phi^6$ present in the definition \eqref{eq:SUSYWilsonLoop} of the Wilson line.
The operation $\Wl \bigl[ \ldots \bigr]$ means that the scalar fields are to be inserted inside the trace of the Wilson line. 
This is explicitly defined in \eqref{eq:1dCorrelators}.

As usual by now, the normalization constant $\nh_{\Dh}$ is chosen such that the two-point function of operators $\Oh$ is unit-normalized.
For the case $\Dh = 1$, it was determined to be
\begin{equation}
\nh_{\Dh = 1} = 2 \mathds{B} (\lambda) = \frac{\sqrt{\lambda}}{\vphantom{I_1(\sqrt{\lambda})}2 \pi^2} \frac{I_2 (\sqrt{\lambda})}{I_1 (\sqrt{\lambda})}\,,
\label{eq:NormalizationConstant_Defect}
\end{equation}
where $\mathds{B} (\lambda)$ is called the \textit{Bremsstrahlung} function.
The name refers to the fact that the defect scalar field represents the emission of \textit{soft particles} from the line, similar to how charged particles emit radiation when they undergo acceleration \cite{Sommerfeld:1931qaf,Maldacena:1998im,Erickson:1999qv,Pineda:2007kz,Forini:2010ek,Drukker:2011za,Correa:2012nk,Henn:2012qz,Gromov:2013qga,Prausa:2013qva,Gromov:2015dfa,Gromov:2016rrp}.
This expression can be expanded both at weak and strong coupling, yielding
\begin{equation}
\begin{alignedat}{2}
\nh_{1} &= \frac{\lambda}{8\pi^2} - \frac{\lambda^2}{192 \pi^2} + \Op(\lambda^3) && \quad \text{ for } \lambda \sim 0\,, \\
\nh_{1} &= \frac{\sqrt{\lambda}}{2 \pi^2} - \frac{3}{4\pi^2} + \Op \left( \frac{1}{\sqrt{\lambda}} \right) && \quad \text{ for } \lambda \gg 1\,.
\end{alignedat}
\label{eq:NormalizationConstant_Defect_WeakStrong}
\end{equation}

We now come back to the correlators involving defect operators.
The supersymmetric version of the bulk-defect two-point functions presented in \eqref{eq:BulkDefectTwoPoint} is
\begin{equation}
\vvev{\Op_{\smash{\Delta_1}} (u, x_1) \Oh_{\smash{\Dh_2}} (\uh, \tau_2)} = b_{\Delta_1 \Dh_2} \frac{(u_1 \cdot \uh_2)^{\Dh_2} (u_1 \cdot \theta)^{\Delta_1 - \Dh_2}}{x_{12}^{2\Dh_2} |x_1^\perp|_{\vphantom{12}}^{\Delta_1 - \Dh_2}}\,.
\label{eq:BulkDefectTwoPointHalfBPS}
\end{equation}

\begin{figure}
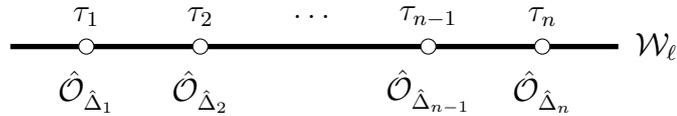

\centering
\MultipointCorrelators
\caption{Representation of the correlation functions \eqref{eq:1dCorrelators} in the $1d$ defect CFT defined by inserting operators along the supersymmetric Wilson line, following \eqref{eq:SUSYWilsonLoop}.}
\label{fig:MultipointCorrelators}
\end{figure}

Correlators of the operators defined in \eqref{eq:SingleTraceHalfBPS_Defect} can be constructed by \textit{inserting} the fields in $\Wl[ \ldots ]$ inside the trace of the Wilson loop \cite{Alday:2007hr,Giombi:2017cqn,Cooke:2017qgm}, i.e.,
\begin{equation}
\vev{\Oh_1 \ldots \Oh_n}_{1d} := \frac{1}{N} \vev{ \tr \Pm \bigl[ \Oh_1 \ldots \Oh_n \exp \int d\tau\, \left( i \xd_\mu A_\mu + |\xd| \phi^6 \right) \bigr]}_{4d}\,,
\label{eq:1dCorrelators}
\end{equation}
where we suppressed the dependence on $\tau_1\,, \ldots\,, \tau_n$ (for the local insertions) and on $\tau$ (for the Wilson line itself) for compactness.
Without loss of generality, we consider the $\tau$'s to be ordered, i.e., $\tau_1 < \tau_2 < \ldots < \tau_n$. 
This type of correlator is illustrated in Figure \ref{fig:MultipointCorrelators}.
As indicated by the subscripts, the expectation value on the left-hand side refers to the correlators of the $1d$ CFT, while the one on the right-hand side corresponds to correlators in the $4d$ $\Nm=4$ SYM theory. Throughout this thesis, correlators with single brackets are always meant to be understood as $1d$ correlators, and so from now on we drop the subscripts.

Correlation functions take the form given in Section \ref{subsec:ConformalFieldTheory} for the one-dimensional case.
In particular, two- and three-point functions are fixed by the conformal symmetry, while $n$-point functions depend on $n-3$ cross-ratios.
\textit{Multipoint} correlators in this $1d$ CFT are the subject of Chapter \ref{chapter:MultipointCorrelatorsInTheWilsonLineDefectCFT}.

To conclude this section, we introduce the useful notion of \textit{pinching}.
For a given correlation function, one can bring two operators or more together to produce single-trace operators with a higher length.
For instance, for elementary scalar fields, we can pinch as follows:
\begin{equation}
\vev{\ldots\, \underbrace{\phi^{I_{k-1}} (\tau_{k-1}) \phi^{I_k} (\tau_k)}_{\text{two operators of length $1$}} \ldots} \overset{\tau_k \to \tau_{k-1}}{\longrightarrow} \vev{\ldots\, \underbrace{\phi^{I_{k-1}} (\tau_{k-1}) \phi^{I_k} (\tau_{k-1})}_{\text{one operator of length $2$}} \ldots}\,.
\label{eq:Pinching}
\end{equation}
This pinching technique corresponds to the strict OPE limit and allows to construct any \textit{composite} single-trace scalar operator made of fundamental scalar fields from correlation functions involving operators of length $1$.\footnote{Note that this is \textit{not} the case in the bulk theory, where the pinching of two single-trace operators produces a \textit{double-trace} operator since each operator carries its own trace.}

\section[Line defects in fermionic CFTs across dimensions]{Line defects in fermionic CFTs across \\ dimensions}
\label{sec:LineDefectsInFermionicCFTsAcrossDimensions}

We now turn our attention to a more general class of models: fermionic conformal field theories in $2 < d < 4$ dimensions.
These models describe a variety of physical phenomena at the \textit{Wilson--Fisher--Yukawa} (WFY) critical point \cite{Fitzpatrick:2013rfa,Fitzpatrick:2013mja,Gracey:2018ame,Giombi:2021cnr,Zhou:2022fdn,Croney:2023gwy,Herbut:2023xgz}.
At $d=3$, the associated quantum field theories do not depend on a coupling parameter that one could expand around with the use of perturbation theory.
The $\veps$-expansion was introduced in \cite{Wilson:1971dc} to enable the computation of physical quantities at $d=3$, by starting at $d=4-\veps$ and expanding around $\veps \sim 0$ using the usual elements of perturbation theory.
It remains today one of the most powerful tools at our disposal for studying strongly-coupled quantum field theories (see for instance Table \ref{table:Universality}).

In this section, we introduce Yukawa CFTs in the context of the $\veps$-expansion.
We present the Feynman rules for these theories, as well as the values of the coupling constants at the WFY fixed point.
Magnetic lines are natural defects to consider in this context, and the critical value of the defect coupling is provided.
The content presented here draws on various resources acknowledged throughout the text.

\subsection{Yukawa CFTs in $4-\veps$ dimensions}
\label{subsec:YukawaCFTsIn4EpsDimensions}

We are studying a general class of Yukawa models with $\Nf$ fermions and $O(N)$ symmetry.
These theories are interesting for many reasons: they exhibit chiral symmetry breaking \cite{Gross:1974jv,Nambu:1961tp,Nambu:1961fr}, possess emergent supersymmetry at the fixed point \cite{Fei:2016sgs,Mihaila:2017ble,Ihrig:2018hho,Zhou:2021dfc}, and apply to certain condensed-matter systems such as graphene sheets \cite{roy2011multicritical,ebert2016phase,Gracey:2018cff,Ray:2021moi}.
Moreover, they can be seen as the UV completion of purely fermionic theories with four-fermion interactions through a reformulation using an auxiliary scalar field (see Section \ref{sec:AnInvitation3}).

\subsubsection{Yukawa CFTs\ }
\label{subsubsec:YukawaCFTs}

Massless Yukawa models are described by the following action in $d$-dimensional Euclidean space, with $2<d<4$:
\begin{equation}
S_\text{Y} = \int d^d x\,\left( \frac{1}{2} \partial_\mu \phi^I \partial_\mu \phi^I
+ i\, \bar{\psi}^a \slashed{\partial} \psi^a
+ g\, \bar{\psi}^a (\Sigma^I)^{ab} \phi^I \psi^b
+ \frac{\lambda}{4!} (\phi^I \phi^I)^2 \right)\,.
\label{eq:ActionYukawa}
\end{equation}
The spacetime dimensions are $\mu = 0, \ldots, d-1$, with $x^0 = \tau$ the Euclidean time direction as usual.
We choose here the fermions to be $4$-component Dirac spinors with \textit{flavor} index $a = 1, \ldots, \Nf$ and with $\psib := \psi^\dagger \gamma^0$.
The spinor conventions are gathered in Appendix \ref{app:Spinors}.
There are $N$ real scalar fields, which are denoted with the $O(N)$ index $I=1, \ldots, N$.
It is important to note that, although we keep $N$ arbitrary here, we do not consider models with $N > 3$ as the fixed point becomes unstable \cite{Varnashev:1999ze}. 

In \eqref{eq:ActionYukawa}, the matrix $\Sigma^I$ defines how the field $\phi^I$ interacts with fermions.
A choice of $\Sigma^I$ for a given $N$ corresponds to a choice of model.
This is discussed in detail in Section \ref{sec:AnInvitation3}.
For all values of $N$, we choose the matrix $\Sigma^1$ to be
\begin{equation}
\Sigma^1 = \mathds{1}\,,
\label{eq:Sigma1}
\end{equation}
while we keep the matrices $\Sigma^{i = 2, \ldots, N}$ arbitrary, i.e., the fields $\phi^{i}$ can behave either as \textit{scalars} or as \textit{pseudoscalars}:
\begin{equation}
\Sigma^{i} \gamma_\mu =
\begin{cases}
+ \gamma_\mu \Sigma^{i} \quad & \text{ if } \phi^i \text{ is a scalar,} \\
- \gamma_\mu \Sigma^{i} \quad & \text{ if } \phi^i \text{ is a pseudoscalar.}
\end{cases}
\label{eq:Sigma_ScalarsAndPseudoscalars}
\end{equation}
Our choice of $\phi^1$ behaves as a scalar, since \eqref{eq:Sigma1} trivially commutes with all $\gamma$ matrices. 
In Section \ref{subsec:TheMagneticLine}, we select this field to be the one coupling to the line defect.

Note that there is some ambiguity on how to take $\Sigma$ and $\gamma$ across dimensions because of the open indices.
Instead of using their concrete representations, we will assume that the following identities hold for $2<d<4$:
\begin{align}
\begin{split}
\tr\, (\Sigma^I)^{ac} \gamma_\mu (\Sigma^J)^{cb} \gamma_\nu &= 4\, \delta^{ab} \delta^{IJ} \delta_{\mu\nu}\,,  \\
\gamma_\mu (\Sigma^I)^{ac} \gamma_\nu (\Sigma^I)^{cb} \gamma_\rho &= N \delta^{ab} (\delta_{\mu \nu} \gamma_\rho + \delta_{\nu \rho} \gamma_\mu - \delta_{\mu \rho} \gamma_\nu + i\, \eps_{\sigma \mu \nu \rho} \gamma_\sigma \gamma^5)\,,
\end{split}
\label{eq:Sigma_Identities}
\end{align}
where $\gamma^5$ anticommutes with all $\gamma$ matrices.
Its explicit definition can be found in \eqref{eq:Spinors_Gamma5}.

\subsubsection{Feynman rules\ }
\label{subsubsec:Feynman rules}

We gather here the conventions and the Feynman rules associated with \eqref{eq:ActionYukawa}.
The free propagators in $d$ dimensions are given by
\begin{equation}
\begin{split}
\text{Scalars:} \qquad 
& \ScalarPropagatorY\ =
\delta^{IJ}\, I_{12}\,,\\
\text{Fermions:} \qquad 
&\FermionPropagatorY\ =
i \delta^{ab}\, \slashed{\partial}_1 I_{12}\,,
\end{split}
\label{eq:Propagators_Yukawa}
\end{equation}
where here we keep the propagator function in $d=4-\veps$ dimensions:
\begin{equation}
I_{12} := \frac{\Gamma \left( 1-\frac{\veps}{2} \right)}{4 \pi^{2-\veps/2}_{\phantom{12}} x_{12}^{2 \left( 1- \veps/2 \right)}}\,,
\label{eq:PropagatorFunction}
\end{equation}
with $x_{12} := x_1 - x_2$.
At $d=4$, \eqref{eq:PropagatorFunction} reduces to \eqref{eq:PropagatorFunction4d}.
The scalar propagator satisfies Green's equation
\begin{equation}
\partial_1^2 I_{12} = - \delta^{(d)} (x_{12})\,,
\label{eq:GreensEq}
\end{equation}
where $\delta^{(d)} (x)$ refers to the $d$-dimensional Dirac delta function.

The interaction terms yield the following insertion rules:
\begin{align}
\begin{split}
\VertexFourScalarsY\ &=
- \frac{\lambda}{2}
 (\delta^{IJ}\delta^{KL} + \delta^{IK}\delta^{JL} + \delta^{IL}\delta^{JK})
 X_{1234} \, , \\
\VertexFermionFermionScalarY\ &=
  g\, \slashed{\partial}_1 (\Sigma^I)^{ab} \slashed{\partial}_3 Y_{123}\, ,
\end{split}
\label{eq:Vertices_Yukawa}
\end{align}
with the $X$- and $Y$-integrals given in Appendix \ref{subsec:FiniteIntegrals}.
Note that one has to add a factor $1/n!$ when $n$ vertices of the same kind are being inserted, while symmetry factors have been taken into account.

\subsection{The Wilson--Fisher--Yukawa fixed point}
\label{subsec:TheWilsonFisherYukawaFixedPoint}

The action \eqref{eq:ActionYukawa} accepts \textit{three} fixed points in three dimensions: the \textit{trivial} one $g = \lambda = 0$, the \textit{Wilson--Fisher} (WF) fixed point $g=0, \lambda=\lambda_\star$ corresponding to the (conformal) $O(N)$ models, and the \textit{Wilson--Fisher--Yukawa} (WFY) fixed point.
Here we present the $\beta$-functions of the couplings and the value they take at the WFY fixed point.
The lowest-lying scalar and fermionic operators are also discussed.

\subsubsection{The critical point\ }
\label{subsubsec:TheCriticalPoint}

The $\beta-$functions of the couplings in \eqref{eq:ActionYukawa} are known up to several loop orders for each model \cite{Karkkainen:1993ef,Rosenstein:1993zf,Herbut:2009vu,Zerf:2017zqi}.\footnote{We restrict ourselves from now on to the models that preserve the $O(N)$ symmetry.}
For general Yukawa and four-scalar couplings, they can be found up to two loops for instance in the appendix of \cite{Fei:2016sgs}.
Although we keep $N$ arbitrary, the expressions given here should be considered with care and not extended beyond $N=3$.

The $\beta-$functions are given by
\begin{align}
\begin{split}
\beta_{\lambda} &= - \veps \lambda
+ \frac{1}{(4 \pi )^2}
\left(
8 g^2 \lambda \Nf -48 g^4 \Nf + \frac{N+8}{3} \lambda^2
\right)
+ \ldots\,, \\
\beta_{g} &= - \veps \frac{g}{2}
+ \frac{\kappa_1 g^3}{(4 \pi )^2}
+ \ldots\,,
\end{split}
\label{eq:BetaFunctions_Bulk}
\end{align}
with $\kappa_1$ defined in \eqref{eq:Kappas}.
The WFY fixed point can be reached for the following values of the couplings at one loop in $\veps:=4-d$:
\begin{align}
 \frac{\lambda_{\star}}{(4 \pi )^2} =  \frac{3 \kappa_{2} }{2 \kappa_{1}(N+8)} \veps + \Om(\veps^2) \,, \qquad \frac{g^{2}_\star}{(4 \pi)^2} =  \frac{\veps }{2 \kappa_{1}} + \Om(\veps^2)\,,
\label{eq:WFY_FixedPoint}
\end{align}
where we see that $g \sim \Op(\sqrt{\veps})$, while $\lambda \sim \Op(\veps)$. 
Furthermore, we have defined
\begin{align}
\begin{split}
\kappa_1 &:= 2 \Nf - N + 4\,, \\
\kappa_2 &:= 2(4-N)-\kappa_1 + \sqrt{12 \left(N^2+16\right)+\kappa_1 (\kappa_1 +12 (N+4))}\,.
\end{split}
\label{eq:Kappas}
\end{align}
Note that all the dependence on $\Nf$ is contained in $\kappa_1$.
At $\veps = 0$, the WFY fixed point corresponds to the \textit{free} theory, which flows at $\veps = 1$ to an \textit{interacting} model.

\subsubsection{Bulk operators\ }
\label{subsubsec:BulkOperators}

We now turn our attention to the lowest-lying bulk operators associated with the action \eqref{eq:ActionYukawa} for arbitrary values of $N$ and $\Nf$.
In Section \ref{sec:CorrelatorsOfBulkOperatorsWithADefect}, we calculate correlation functions involving the fundamental scalars and fermions in the presence of a line defect.
As we expand these correlators using conformal blocks, operators with higher scaling dimensions naturally emerge.
To ensure the consistency of our analysis, we provide here a selection of relevant operators that are useful for conducting sanity checks.

The lowest-lying scalar operators are the fundamental scalars, which we refer to as
\begin{equation}
\Op^I (x) := \frac{1}{\sqrt{n_I}} \phi^I (x)\,,
\label{eq:BulkScalar}
\end{equation}
with $I = 1, \ldots, N$ the $O(N)$ index.
The normalization constant $n_I$ is chosen as usual such that the two-point function is unit-normalized.
The scaling dimension $\Delta_I$ of this operator was computed in \cite{Zhou:2022fdn} at the WFY fixed point \eqref{eq:WFY_FixedPoint} and found to be
\begin{equation}
\Delta_I = 1 + \frac{N-4}{2\kappa_1} \veps + \Op(\veps^2)\,.
\label{eq:BulkScalar_Delta}
\end{equation}
The explicit expression for the normalization constant is given by
\begin{equation}
n_I = \frac{1}{4 \pi^2} \left( 1
- \frac{2 \kappa_1 + (N-4)(1 + \aleph)}{2\kappa_1} \veps + \Om(\veps^2) \right)\,.
\label{eq:BulkScalar_NormalizationConstant}
\end{equation}
In the above, $\kappa_1$ is defined in \eqref{eq:Kappas} while $\aleph$ is a constant arising from dimensional regularization and given in \eqref{eq:Aleph}.

The next set of operators that we consider are the fermions.
We denote fermionic operators as $\Op_{\ellb, \ell}$, with $\ell$ corresponding to the number of fermions $\psi^a$ and $\ellb$ to the number of antifermions $\psib^a$.
The variables $s$ and $\bar{s}$ are \textit{polarization spinors}, introduced to avoid the cluttering of indices and defined via \eqref{eq:Spinors_PolarizationSpinors}.
Using this notation, we denote the elementary fermions as
\begin{equation}
\Op_{1,0}^a (\bar{s}, x) := \frac{1}{\sqrt{n_a}} \psib^a (\bar{s}, x)\,, \quad \Op_{0,1}^a (s, x) := \frac{1}{\sqrt{n_a}} \psi^a (s, x)\,.
\label{eq:BulkFermions}
\end{equation}
The scaling dimension of these operators has also been computed in \cite{Zhou:2022fdn} at the WFY fixed point, yielding
\begin{equation}
\Delta_a = \frac{3}{2} - \frac{2\kappa_1 -N}{4\kappa_1} \veps + \Om(\veps^2)\,.
\label{eq:BulkFermions_Delta}
\end{equation}

As mentioned earlier, the block expansions of the operators defined in \eqref{eq:BulkScalar} and \eqref{eq:BulkFermions} include operators with higher leading-order scaling dimensions in $\veps$.
Particularly, for $\Delta \sim 2$, we have the following operators:
\begin{equation}
\Op^2 := \frac{1}{\sqrt{n_{\Op^2}}} \phi^2\,, \qquad T^{IJ} := \frac{1}{\sqrt{n_T}} \left( \phi^I \phi^J - \frac{\delta^{IJ}}{N} \phi^2 \right)\,,
\label{eq:BulkSquaredScalar}
\end{equation}
where, as usual, $n_{\Op^2}$ and $n_T$ are the respective normalization constants.

The number of operators grows rapidly as we consider higher combinations of fundamental fields.
For future purposes, we conclude this section by introducing the \textit{fermion bilinear}, defined as
\begin{equation}
\Op_{1,1}^{aa} (x) := \frac{1}{\sqrt{n_{\Op_{1,1}^{aa}}}} \psib^a (x) \psi^a (x)\,,
\label{eq:BulkBilinear}
\end{equation}
with its corresponding scaling dimension given by
\begin{equation}
\Delta_{ \Op_{1,1}^{aa} } = 3 + \Om(\veps)\,.
\label{eq:BulkBilinear_Delta}
\end{equation}

The couplings and operators get renormalized.
We define the \textit{bare} couplings and (elementary) fields as
\begin{equation}
\begin{alignedat}{2}
\lambda_0 &:= \mu^{\veps} \lambda Z_\lambda\,,
\qquad
&& \Op^I_0 := Z_{I} \Op^I\,, \\
g_0 &:= \mu^{\veps/2} g Z_{g}\,,
\qquad
&& (\Op_{\bar{\ell}, \ell}^a)_0 := Z_{a} \Op_{\bar{\ell}, \ell}^a\,,
\end{alignedat}
\label{eq:RenormalizedCouplingsAndFields}
\end{equation}
where we introduced rescaled couplings $g \to \mu^{\veps/2} g, \lambda \to \mu^{\veps}\lambda$ to ensure that the couplings in the renormalized Lagrangian are dimensionless, and with $\ell + \bar{\ell} = 1$. 
The expressions for the renormalization factors $Z$ up to $\Om(\veps^2)$ can be found in Appendix B of \cite{Barrat:2023ivo}.

The renormalization factors allow us to obtain the anomalous dimensions $\gamma_{I}$ and $\gamma_{a}$ for the scalar and fermionic fields, which are given here to first order in the couplings:
\begin{align}
\begin{split}
\gamma_{I} &= \frac{d \log Z_{I}}{d \log \mu} = \frac{2 g^2 \Nf}{(4 \pi )^2} + \ldots\,,\\
\gamma_{a} &= \frac{d \log Z_{a}}{d \log \mu} = \frac{g^2 N}{2(4 \pi )^2} + \ldots\,.
\end{split}
  \label{eq:AnomalousDimensions_Bulk}
\end{align}
This leads to the following values for the conformal dimensions evaluated at the WFY fixed point defined in \eqref{eq:WFY_FixedPoint}:
\begin{align}
\begin{split}
\Delta_{I} &= 1 - \frac{\veps}{2} + \gamma_{I} = 1- \veps \frac{4-N}{2 \kappa_1} + \Om(\veps^2)\,, \\
\Delta_{a} &= \frac{3}{2} - \frac{\veps}{2} + \gamma_{a} = \frac{3}{2}-\frac{\veps}{4} \left(2 - \frac{N}{\kappa_1} \right) + \Om(\veps^2)\,.
\end{split}
\label{eq:BulkDeltas}
\end{align}

\subsection{The magnetic line}
\label{subsec:TheMagneticLine}

Magnetic lines are natural defects to consider in the $O(N)$ model, which corresponds to \eqref{eq:ActionYukawa} setting $g=0$.
It was shown in \cite{Giombi:2022vnz} that a defect CFT continues to exist for this type of defect at non-zero values of the Yukawa coupling.
In the following, we present the necessary toolkit for describing such a model.
We start with the defect action and Feynman rules, then present the critical point for the defect coupling.
We then discuss the defect operators, as well as the renormalization of the defect coupling.

\subsubsection{Defect Feynman rules\ }
\label{subsubsec:DefectFeynmanRules2}

Adding a magnetic line defect to the action results in a breaking of the conformal group $SO(d+1,1) \to SO(2,1) \times SO(d-1)$, following the concepts of Section \ref{subsec:ConformalDefects}.
The action reads
\begin{equation}
S_{\text{defect}} := S_{\text{Y}} + h \int_{-\infty}^\infty d\tau\, \theta \cdot \phi (\tau)\,,
\label{eq:Sdefect}
\end{equation}
where the scalar fields $\phi^I$ must have a scaling dimension below $1$ for the deformation to be relevant.
Here $h$ is the coupling of the defect, which extends in the Euclidean time direction $\tau$, while $\theta^I$ is an $O(N)$ vector that defines the polarization of the defect. We choose here $\phi^1$ to be the scalar coupling to the defect, i.e.,
\begin{equation}
\theta = (1, 0, \ldots, 0)\,.
\label{eq:theta_Yukawa}
\end{equation}
Note that one could also choose the line defect to couple to a pseudoscalar, and this would not introduce additional technical difficulties.

The defect introduces a new vertex
\begin{align}
\DefectVertexOnePointScalarInf\ := - h\, \delta^{I1} \int_{-\infty}^\infty d \tau_2\, I_{12} \, ,
\label{eq:DefectVertex_OnePointScalar}
\end{align}
with $x_2 := (\tau_2,0,0,0)$ the point on the line that has to be integrated over.
As for the bulk Feynman rules, one should add a factor $1/n!$ when $n$ vertices are inserted, while symmetry factors have to be accounted for if more couplings to the line are being considered.\footnote{These rules differ from \eqref{eq:DefectVertices_OnePoint} due to the absence of gauge symmetry.
In other words, there is no path ordering on the defect.}

\subsubsection{The critical point\ }
\label{subsubsec:TheCriticalPoint2}

In the following, we show that there exists a non-trivial critical value for the coupling $h$ such that the defect CFT exists and can be taken across dimensions by varying $\veps$, just as in the bulk.
This has been done for the $O(N)$ model in \cite{Allais:2014fqa,Cuomo:2021kfm}, while correlation functions were considered in \cite{Gimenez-Grau:2022czc}.
With fermions, this was considered in \cite{Giombi:2022vnz} and \cite{Pannell:2023pwz} for the cases $N=1,2,3$. 

We can compute the $\beta$-function $\beta_h$ by calculating the divergences of the one-point function of the (renormalized) scalar $\Op^I$.
We obtain \cite{Giombi:2022vnz,Pannell:2023pwz}
\begin{equation}
\beta_h = -\frac{\veps h}{2} + \frac{1}{(4\pi)^2} \frac{\lambda h^3}{6} + \ldots\,.
\label{eq:Beta_h}
\end{equation}
Using the values for $\lambda$ and $g$ at the WFY fixed point given in \eqref{eq:WFY_FixedPoint}, we find the corresponding defect fixed point
\begin{equation}
h^{2}_\star = -\frac{2 (N-4) (N+8)}{\kappa_2} + \Om(\veps)\,,
\label{eq:h_FixedPoint}
\end{equation}
where the $\Om(\veps)$ term is given in Appendix B of \cite{Barrat:2023ivo}.
It is interesting to note that there exists a non-trivial critical point even at $\veps = 0$.
This can be understood by the fact that free theories can also accommodate conformal defects.

If we include the finite part of the one-point function, we can extract the one-point coefficient $a_I$ :
\begin{equation}
\vvev{\Op^I(x)} = \frac{a_I}{|x^{\bot}|^{\Delta_I}}\:, \quad a_I^2 = -\frac{(N-4) (N+8)}{2 \kappa_2} + \Om(\veps)\:.
\label{eq:OnePoint_PhiBulk}
\end{equation}
The $\Om(\veps)$ term is lengthy and given in the \textsc{Mathematica} notebook attached to \cite{Barrat:2023ivo}.

\subsubsection[\\ Defect operators\ ]{Defect operators}
\label{subsubsec:DefectOperators}

The fundamental fields discussed in \ref{subsec:TheWilsonFisherYukawaFixedPoint} give rise to a rich set of defect operators on the magnetic line.
In the defect CFT, the $O(N)$ symmetry breaks down to $O(N-1)$, resulting in the emergence of \textit{two} fundamental scalars:
\begin{equation}
\Op^{I=1, \ldots, N} \longrightarrow \lbrace \Oh^1\,, \Oh^i \rbrace\,.
\label{eq:BulkToDefect_SymmetryBreaking}
\end{equation}
Here the indices split as $I = \lbrace 1, i \rbrace$, with $i = 2\,, \ldots\,, N$ introduced in Section \ref{subsec:TheMagneticLine}.
Specifically, the fundamental defect scalars are defined as
\begin{equation}
\Oh^1 (\tau) := \frac{1}{\sqrt{\nh_1}} \phi^1 (\tau)\,, \quad \Oh^i (\tau) := \frac{1}{\sqrt{\nh_i}} \phi^i (\tau)\,.
\label{eq:DefectScalar}
\end{equation}

Recall that, according to \eqref{eq:theta_Yukawa}, $\phi^1$ is the operator that couples to the line defect, and it plays a role similar to $\phi^6$ in the case of the Wilson line.
The conformal dimension of $\Oh^1$ can be obtained from the $\beta$-function of the defect coupling:
\begin{equation}
\Dh_1 = \left.  1 + \frac{\partial \beta_h}{\partial h} \right|_{h = h_{\star}} = 1 - \frac{N-4}{\kappa_1} \veps + \Om(\veps^2)\,.
\label{eq:DefectScalar1_Delta}
\end{equation}

On the other hand, $\Oh^i$ plays the same role as $\Oh_{\Dh = 1}$ (see \eqref{eq:SingleTraceHalfBPS_Defect}) in the case of the $\Nm=4$ SYM theory.
And similarly, the scaling dimension of $\Oh^i$ is protected:
\begin{equation}
\Dh_i = 1\,.
\label{eq:DefectTilt_Delta}
\end{equation}
This operator is commonly referred to as the \textit{tilt} operator.
Note that for $N=1$, there will be no tilt operator, but only $\Oh^1 = \Oh$ on the defect.

The next scalar operator of interest is the \textit{displacement} operator, denoted by $\Disp_\mu$.
It is related to the bulk stress-energy tensor through the Ward identity
\begin{equation}
\partial_{\mu} T_{\mu \nu} = \delta^{d-1}(x^{\bot}) \Disp_\nu\,,
\label{eq:WardIdentity}
\end{equation}
where the index $\nu$ on the right-hand side represents the \textit{transverse} components, i.e., $\nu =1,\ldots, d$.
The displacement has a transverse spin $s = 1$ and a protected conformal dimension
\begin{equation}
\Dh_{\Disp} = 2\,.
\label{eq:Displacement_Delta}
\end{equation}
It can be constructed by taking a transverse derivative of the field $\Oh^1$:
\begin{equation}
\Disp_\mu := \frac{1}{\sqrt{\nh_{\Disp}}} \partial^{\bot}_\mu \Op^1\:.
\label{eq:Displacement}
\end{equation}
While there exist additional operators $\partial^{\bot}_\mu \Op^i$ corresponding to taking the transverse derivative of the tilt operator, we will not be considering them here.
Their correlators can however be computed similarly to those involving the displacement operator. 

At $\veps \sim 0$, we encounter non-scalar operators that lie between the fundamental scalars and the displacement.
These are the elementary defect Dirac fermions $\Oh_{1,0}$ and $\Oh_{0,1}$, defined similarly to \eqref{eq:BulkFermions} as
\begin{equation}
\Oh_{1,0}^a (\bar{s}, \tau) := \frac{1}{\sqrt{\nh_a}} \hat{\psib}^a (\bar{s},\tau)\,, \quad \Oh_{0,1}^a (s, \tau) := \frac{1}{\sqrt{\nh_a}} \hat{\psi}^a (s,\tau)\,.
\label{eq:DefectFermions}
\end{equation}
The conformal dimensions of these operators are given by
\begin{equation}
\Dh_a = \frac{3-\veps}{2} + \gammah_a\:,
\label{eq:DefectFermions_Delta}
\end{equation}
where the anomalous dimension $\gammah_a$ can be extracted from the two-point function.
This will be discussed in Section \ref{subsec:KinematicallyFixedCorrelators2}.

We renormalize the defect coupling in a similar way to the bulk couplings.
We define the bare coupling $h_0$ in terms of the renormalized coupling $h$ as
\begin{equation}
h_0 = \mu^{\veps/2} \, h \, Z_h\,.
\label{eq:Renormalizedh}
\end{equation}
We require that \eqref{eq:Beta_h} vanishes, or equivalently that
\begin{equation}
\vvev{\Op^I (x)} = \text{finite}\,.
\label{eq:OnePointFinite}
\end{equation}
$Z_h$ is given in Appendix B of \cite{Barrat:2023ivo} up to $\Op(\veps^{-2})$.

Note that the one-point function of a single fermion $\Op_{1,0}^a$ is zero.  
In Section \ref{subsec:KinematicallyFixedCorrelators3}, we show that the one-point function of the fermion bilinear is also finite at $\Om (\veps)$, and therefore the perturbative defect CFT generated by the magnetic line forms a consistent model.

\chapter[Bootstrapping holographic \\ defect correlators]{Bootstrapping holographic defect correlators}
\chaptermark{Bootstrapping holographic defect correlators}
\label{chapter:BootstrappingHolographicDefectCorrelators}

One of our general goals is to understand how the dynamics of a bulk theory are affected by the presence of a defect.
When the symmetry is broken, correlation functions of local operators do not obey the rules of conformal field theory anymore.
New degrees of freedom arise: the operators acquire a vacuum expectation value, while two-point functions are not fixed anymore kinematically.
Such systems have received a lot of attention in recent years \cite{Billo:2016cpy,Liendo:2016ymz,Gadde:2016fbj,Chiodaroli:2016jod,deLeeuw:2017dkd,Lemos:2017vnx,Gimenez-Grau:2020jvf,Gimenez-Grau:2022czc,Gimenez-Grau:2022ebb,Bianchi:2022sbz}, as they are relevant in different fields of theoretical physics ranging from holography to condensed matter.

In this chapter, we study two-point functions of scalar operators in the presence of a line defect, using modern analytic bootstrap methods.
In this context, the Lorentzian inversion formula presented in \eqref{eq:InversionFormulaBulk} has led to spectacular results in recent years.
If one is interested in the correlator itself, it is desirable to bypass the inversion formula and replace it with a \textit{dispersion relation}.
The result is the position-space formula \eqref{eq:DispersionRelationBulk} reminiscent of the dispersion relation \eqref{eq:DispersionRelationScatteringAmps} familiar from scattering amplitudes.
Here we wish to develop a similar formula for defect CFT correlators.
We show in particular that two-point functions of scalar operators can be reconstructed from a \textit{single} discontinuity.
This is in contrast to the bulk case, where the analogous relation contains a \textit{double} discontinuity.

We apply the dispersion relation to the special case where the bulk theory is four-dimensional $\Nm=4$ Super Yang--Mills, with the defect being the supersymmetric Wilson line introduced in Section \ref{sec:TheWilsonLineDefectCFT}.
The reasons for this choice are simple: $\Nm = 4$ SYM is rich in symmetries and offers powerful handles to conduct an analytical study.
Moreover, the intricate web of dualities surrounding it has provided insights into various regimes that may be inaccessible in other models.
There exists in particular a plethora of exact results for half-BPS operators, while Wilson lines play a special role in the context of the AdS/CFT correspondence.
It is therefore natural to study the dynamics of half-BPS operators and Wilson lines together in the $\Nm = 4$ SYM theory.

The content of this chapter is mostly based on the publications \cite{Barrat:2021yvp,Barrat:2022psm}.
Section \ref{sec:TheMicrobootstrap} presents the results of the unpublished notes \cite{Barrat:2021un}.

\section{An invitation}
\label{sec:AnInvitation}

The defect CFT generated by the supersymmetric Wilson line in $\Nm=4$ SYM has emerged as a subject of considerable attention in recent years.
To situate the computations performed here within the ongoing research, we begin our study by providing a concise overview of the relevant literature.
We collect useful results exact in the coupling constant, and present our setup of interest: the two-point functions of half-BPS single-trace operators.
As explained in Section \ref{subsec:ConformalDefects}, this is the first correlator to have non-trivial kinematics when considering bulk operators only in the presence of the defect.
This correlator has been considered perturbatively in the weak- and strong-coupling regimes, and here we summarize the known results to prepare the ground for the rest of this chapter.

\subsection{Exact results}
\label{subsec:ExactResults}

Numerous results have been obtained in $\Nm = 4$ SYM, thanks to conformal symmetry, supersymmetry, and integrability.
In particular, the half-BPS operators defined in \eqref{eq:SingleTraceHalfBPSOperators_Bulk} have been the objects of intense study since the early days of the AdS/CFT correspondence.
Their scaling dimensions are protected and they do not need to be renormalized.
This makes it possible to obtain \textit{exact} results, even for finite parameters $g$ and $N$.

From the conformal block point of view, we have seen in \eqref{eq:DefectTwoPoint_BlockExpansionBulk2} that the OPE coefficients $a_\Delta$ and $\lambda_{\Delta_1 \Delta_2 \Delta_3}$ play a prominent role in the defect two-point functions.
In the following, we list the exact results for these coefficients in the coupling $\lambda$, while taking the limit $N \to \infty$.
Moreover, we present the strong-coupling limits of bulk-defect two-point functions of half-BPS operators.
The fact that such expressions are reachable makes $\Nm = 4$ SYM a natural playground for bootstrap techniques, due to the abundance of symmetry and the amount of information already present in the literature.

\subsubsection{Three-point functions}
\label{subsubsec:ThreePointFunctions}

Three-point functions of half-BPS operators are the first correlators to consider.
Like the two-point functions without defects, they do not receive quantum corrections.
At large $N$, their values were determined in \cite{Lee:1998bxa}:
\begin{equation}
\lambda_{\Delta_1 \Delta_2 \Delta_3} = \frac{\sqrt{\Delta_1 \Delta_2 \Delta_3}}{N}\,,
\label{eq:HalfBPSThreePointFunctions}
\end{equation}
following the definitions of Section \ref{subsec:ConformalFieldTheory} adapted for supersymmetric theories.
To the best of our knowledge, there does not exist a closed-form expression for three-point functions at finite $N$.
However, since they are protected, it is easy to compute them case by case using the trace algebra of $SU(N)$ as well as elementary Wick combinatorics.

In Section \ref{sec:TwoPointFunctionsAtStrongCoupling}, we compare the outputs of our computations with three-point functions of single- and double-trace operators.
For this purpose, we note the special case
\begin{equation}
\lambda_{\Delta_1 \Delta_2 ( \Delta_1, \Delta_2 )} = 1\,,
\label{eq:ThreePointDoubleTrace}
\end{equation}
where the third operator in the subscript should be understood as the double-trace operator $\Op_{\Delta_1, \Delta_2}$ defined in \eqref{eq:MultiTraceOperators_Bulk}.
Note that this formula is only valid at large $N$.

\subsubsection{One-point functions}
\label{subsubsec:OnePointFunctions}

As we have seen in Section \ref{subsec:ConformalDefects}, operators acquire an expectation value in the presence of a defect.
For the case of a supersymmetric Wilson-line defect, one-point function coefficients have been computed with localization techniques \cite{Semenoff:2001xp,Giombi:2009ds,Giombi:2012ep}, and at large $N$ they read
\begin{equation}
a_{\Delta} = \frac{\sqrt{\lambda \Delta}}{\vphantom{I_1 (\sqrt{\lambda})}2^{\Delta/2 + 1} N} \frac{I_{\Delta} (\sqrt{\lambda})}{I_1 (\sqrt{\lambda})}\,,
\label{eq:HalfBPSOnePointFunctions}
\end{equation}
with $I_{\alpha} (z)$ the Bessel function defined in \eqref{eq:BesselFunction}.
We follow the conventions defined in \eqref{eq:OnePointFunction_HalfBPS}.
\eqref{eq:HalfBPSOnePointFunctions} can be expanded in the weak- and strong-coupling regimes, yielding
\begin{equation}
\begin{split}
a_\Delta & \overset{\lambda \sim 0}{=} \frac{ \sqrt{\Delta \lambda^\Delta} }{2^{3\Delta/2} N \Gamma(\Delta+1)} \left( 1 - \lambda \frac{\Delta-1}{8(\Delta+1)} + \Op(\lambda^2)  \right)\,, 
\\
a_\Delta & \overset{\lambda \gg 1}{=} \frac{\sqrt{\lambda \Delta}}{2^{\Delta/2+1} N} \left( 1 - \frac{\Delta^2 - 1}{2\sqrt{\lambda}} + \Op(1/\lambda) \right)\,.
\end{split}
\label{eq:HalfBPSOnePoint_WeakStrong}
\end{equation}
Note that analytical expressions were also derived at finite $N$ \cite{Okuyama:2006jc}, where the Bessel functions are replaced by Laguerre polynomials, in a way similar to \eqref{eq:Wc_FiniteNAndLargeN}.

The expression \eqref{eq:HalfBPSOnePointFunctions} generalizes straightforwardly to multi-trace operators.
In particular, we need for later use the case where the operator is a double-trace:
\begin{equation}
a_{(\Delta_1, \Delta_2)} = \frac{\sqrt{ \Delta_1 \Delta_2 } \lambda}{2^{\frac{\Delta_1 + \Delta_2 + 4}{2}} N^2} \frac{ I_{\Delta_1 + \Delta_2 - 1} ( \sqrt{\lambda} ) }{\vphantom{2^{\frac{\Delta_1 + \Delta_2 + 4}{2}}} I_1 (\sqrt{\lambda}) }\,.
\label{eq:OnePointDoubleTrace}
\end{equation}

\subsubsection[\\ Bulk-defect two-point functions]{Bulk-defect two-point functions}
\label{eq:BulkDefectTwoPointFunctions}

We conclude this section by giving the strong-coupling limit of bulk-defect two-point functions of half-BPS operators.
These correlators are defined in \eqref{eq:BulkDefectTwoPointHalfBPS}, and the OPE coefficients at strong coupling and in the planar limit read \cite{Giombi:2018hsx}
\begin{equation}
\begin{split}
b_{\Delta_1 \Dh_2} =\ & \frac{\lambda^{\frac{2-\Dh_2}{4}}}{2^{\frac{\Delta_1 + 2 - 3 \Dh_2}{2}} N} \frac{\sqrt{\Delta_1}}{\sqrt{\Dh_2!}} \frac{\Gamma \left( \frac{\Delta_1 + \Dh_2 + 1}{2} \right)}{\Gamma \left( \frac{\Delta_1 - \Dh_2 + 1}{2} \right)} \\
& \times\left( 1
+ \frac{4 (1- \Delta_1^2) + \Dh_2 (5 + \Dh_2)}{8 \sqrt{\lambda}}
+ \Om(\lambda^{-1})
\right)\,.
\end{split}
\label{eq:BulkDefectTwoPoint_Strong}
\end{equation}
In the course of our computation, we encounter the OPE coefficients corresponding to bulk-defect two-point functions with double-trace defect operators, defined as
\begin{equation}
\Oh_{0,\Dh} (\uh, \tau) := \Wl [\ ]\, \tr ( \uh \cdot \phi(x) )^{\Dh}\,,
\label{eq:DoubleTrace_Defect}
\end{equation}
with $\Wl[\ ]$ meaning that no field is inserted on the Wilson line.
To the best of our knowledge, the coefficients $b_{\Delta_1 (0,\Dh_2)}$ have not appeared in the literature yet.
An interesting outcome of our analysis is a prediction for these values.
We should stress however that our correlators contain information about infinitely many unprotected operators as well.

\subsection{Two-point functions of half-BPS operators}
\label{subsec:TwoPointFunctionsOfHalfBPSOperators}

We are interested here in the \textit{connected} two-point function, defined as
\begin{equation}
\begin{split}
\vvev{\Op_{\Delta_1} (u_1, x_1) \Op_{\Delta_2} (u_2, x_2)} :=\ & \vev{\Op_{\Delta_1}  (u_1, x_1) \Op_{\Delta_2} (u_2, x_2) \Wl} \\
&- \vev{\Op_{\Delta_1} (u_1, x_1) \Op_{\Delta_2} (u_2, x_2) } \vev{\Wl}\,.
\end{split}
\label{eq:ConnectedTwoPoint}
\end{equation}
The explicit definition of the Wilson line is given in \eqref{eq:SUSYWilsonLoop}, while the half-BPS operators are given in \eqref{eq:SingleTraceHalfBPSOperators_Bulk}.

The structure of the correlator can be found in \eqref{eq:TwoPointFunctions_HalfBPS}.
Without loss of generality, we consider as before $\Delta_1 \leq \Delta_2$ to streamline the notation.
The reduced correlator $\Fm (z, \zb; \sigma)$ depends on \textit{two} spacetime cross-ratios and \textit{one} $R$-symmetry variable $\sigma$.
Determining this function at strong coupling is the goal of this chapter.

Note that the second term on the right-hand side of \eqref{eq:ConnectedTwoPoint} (the \textit{disconnected} part) is only non-zero for the case $\Delta_1 = \Delta_2$, as implied by the form of conformal two-point functions given in \eqref{eq:ConformalTwoPointFunctions}.
We should stress that the reduced correlator $\Fm (z, \zb; \sigma)$ depends on the external dimensions $\Delta_1, \Delta_2$, although not explicitly indicated.

\subsubsection{Superconformal Ward identities}
\label{subsubsec:SuperconformalWardIdentities}

We have seen in \eqref{eq:TwoPoint_RSymmetryChannels} that the reduced correlator $\Fm (z, \zb; \sigma)$ is a polynomial of degree $\Delta_1$ (for $\Delta_1 \leq \Delta_2$) in $\sigma$, with each power corresponding to a different $R$-symmetry channel.
For example, in the simplest case $\Delta_1 = \Delta_2 = 2$, we have
\begin{equation}
\Fm (z, \zb; \sigma) = \Omega^2\, F_0 (z,\zb) + \Omega\, F_1 (z,\zb) + F_2(z,\zb)\,,
\label{eq:RSymmetryChannels_22}
\end{equation}
where the definition of $\Omega \sim \sigma$ can be found in \eqref{eq:Omega}.
As a consequence, the full correlator is known if the individual $R$-symmetry channels $F_j (z, \zb)$ are known.
We often use the case $\Delta_1 = \Delta_2 = 2$ as an example to illustrate our methods in a simple setting.

The $R$-symmetry channels are related to each other by symmetry.
This can be described concretely with the help of the \textit{superconformal Ward identities} \cite{Liendo:2016ymz}
\begin{equation}
\begin{split}
\left. \left( \pd_z + \frac{1}{2} \pd_\alpha \right) \Fm(z, \zb; \sigma) \right|_{z=\alpha} &= 0\,, \\
\left. \left( \pd_{\zb} + \frac{1}{2} \pd_\alpha \right) \Fm(z, \zb; \sigma) \right|_{\zb=\alpha} &= 0\,,
\end{split}
\label{eq:WardIdentities}
\end{equation}
where we have defined for convenience the help $R$-symmetry variable $\alpha$, which relates to $\sigma$ in the following way:
\begin{equation}
\sigma := - \frac{(1-\alpha)^2}{2\alpha} \,.
\label{eq:AlphaToSigma}
\end{equation}
The identities \eqref{eq:WardIdentities} impose powerful constraints since the differential operators are of first order in the variables.
They will be particularly helpful in Section \ref{sec:TwoPointFunctionsAtStrongCoupling} to fix the remaining open coefficients of our correlators.

\subsubsection{Crossing equation}
\label{subsubsec:CrossingEquation}

As introduced in Section \ref{subsec:TheWilsonLineDefectCFT}, the two-point functions of half-BPS operators can be expanded in \textit{superblocks}.
Superblocks are useful, as they encode the contributions of superdescendants in the block expansion.
The expansions in the \textit{bulk} and \textit{defect} channels \eqref{eq:TwoPoint_SuperconformalBlocksBulk} and \eqref{eq:TwoPoint_SuperconformalBlocksDefect} lead to a \textit{supersymmetric crossing equation}, which takes the form
\begin{equation}
\Omega^{\frac{\Delta_1 + \Delta_2}{2}} \sum_{\vphantom{\Oh}\Op} \lambda_{\Delta_1 \Delta_2 \Op} a_\Op\, \Gm_{\Op} (z, \zb, \sigma) = \sum_{\Oh} b_{\Delta_1 \Oh} b_{\Delta_2 \Oh}\, \Gh_{\Oh} (z, \zb, \sigma)\,.
\label{eq:CrossingEquation}
\end{equation}
This equation relates the bulk OPE coefficients $\lambda_{\Delta_1 \Delta_2 \Op}$ to the defect CFT data composed of one-point functions $a_\Op$ and bulk-defect correlators $b_{\Delta \Oh}$.
Here we adjusted the notation of \eqref{eq:TwoPoint_SuperconformalBlocksBulk} and \eqref{eq:TwoPoint_SuperconformalBlocksDefect} such that the sum is over the operators $\Op$ and $\Oh$, instead of the quantum numbers that describe these operators, namely $\Delta$, $\ell$ and $k$ on the left-hand side and $\Dh$, $s$ and $\kh$ on the right-hand side.

\subsubsection[\\ The topological subsector]{The topological subsector}
\label{subsubsec:TheTopologicalSubsector}

The superconformal Ward identities \eqref{eq:WardIdentities} admit an interesting special limit of the kinematical variables.
If one sets $z = \zb = \alpha$ in \eqref{eq:RSymmetryChannels_22}, we find that the sum of the $R$-symmetry channels is equal to
\begin{equation}
\Fm (x, x, x) = \sum_{j=0}^{\Delta_1} \Omega_{\text{topo}}^j\, F_{\Delta_1 - j} (x, x) = c_-\, \Theta(-x) + c_+\, \Theta(x)\,,
\label{eq:TopologicalSector}
\end{equation}
where we have defined the topological variable
\begin{equation}
\Omega_\text{topo} := \left. \Omega \right|_{z = \zb = \alpha = x} = \frac{1}{2} \sgn (x)\,.
\label{eq:OmegaTopo}
\end{equation}
This is called the \textit{topological sector} since the correlator is constant on both sides of the point $x=0$.
The $c_\pm$ are functions of the coupling constant $\lambda$ only, and they have been determined to all orders using localization techniques (see (A.2) in \cite{Giombi:2018hsx}).

The study of the topological sector in the block expansion leads to an interesting truncation of the bulk and defect OPEs.
The resulting crossing relations are called \textit{microboostrap} equations.
We take a closer look at the topological sector in Section \ref{sec:TheMicrobootstrap}.

\subsection{A first glimpse at weak coupling}
\label{subsec:AFirstGlimpseAtWeakCoupling}

To familiarize ourselves with the setup, we start by considering the two-point functions of half-BPS operators in the weak-coupling regime.
This was considered in \cite{Barrat:2020vch} for the simplest case $\Delta_1 = \Delta_2 = 2$, and we review the results at the lower orders for gaining some physical intuition.

\subsubsection{Leading order}
\label{subsubsec:LeadingOrder}

The first connected contribution was found to be
\begin{equation}
\HolographicTwoPointWeakLO = \frac{\lambda}{4 N^2} \Omega\,.
\label{eq:ConnectedTwoPointWeakLO}
\end{equation}
The half-BPS operators are represented by \SingleTraceHalfBPS\, with the smaller circles indicating the number of scalars inside the trace (here two per operator).

In terms of the $R$-symmetry channels defined in \eqref{eq:RSymmetryChannels_22}, this is equivalent to
\begin{equation}
F_0 (z, \zb) = F_2 (z, \zb) = 0\,, \quad F_1 (z, \zb) = \frac{\lambda}{4 N^2}\,.
\label{eq:RSymmetryChannelsLO}
\end{equation}
In the topological limit, we observe as anticipated in \eqref{eq:TopologicalSector} that all the dependence on the kinematic variables drops apart from the sign function:
\begin{equation}
\Fm (x, x, x) = - \frac{\lambda}{4 N^2} \sgn (x)\,.
\label{eq:ConnectedTwoPointWeakLOTopological}
\end{equation}

\subsubsection{Next-to-leading order}
\label{subsubsec:NextToLeadingOrder}

At next-to-leading order (NLO), only the channel without bulk vertices could be determined in a closed form:
\begin{align}
F_2 (z, \zb) &= \HolographicTwoPointWeakNLOFtwo \notag \\
& = \frac{\lambda}{32 \pi^2 N^2}
\biggl\lbrace
\frac{\pi^2}{4}
+ 2 \log 2 \left( H_{-1} (r) + H_{1} (r) \right) \notag \\
& \phantom{=\ }
+ \left( H_{-1,0} (r) + H_{1,0} (r) \right)
-2 \left( H_{-1,-1} (r) + H_{1,-1} (r) \right)
\biggr\rbrace\,,
\label{eq:ConnectedTwoPointWeakNLOF2}
\end{align}
where the functions $H_{\vec{a}} (x)$ are harmonic polylogarithm (HPL) functions, defined in \eqref{eq:HPL_Definition} and for which important properties are listed in Appendix \ref{subsec:HarmonicPolylogarithms}.
It was possible to compute this diagram, because in the conformal frame \eqref{eq:DefectCrossRatios_zzb} it depends only on the distance $r = z \zb$ between the defect and one of the operators.
It is interesting to note that this expression contains only transcendental functions of weight $w=2$, without rational functions in front.

It has not been possible to determine the other two $R$-symmetry channels $F_0$ and $F_1$ in a closed form.
Instead, the OPE coefficients of the block expansions were fixed by using the constraints imposed by the superconformal Ward identities \eqref{eq:WardIdentities}.
The remaining degrees of freedom were fixed by high-precision numerical integration.
In this sense, a full solution of the correlator was found, although it would be desirable to have an analytical expression in terms of transcendental functions like \eqref{eq:ConnectedTwoPointWeakNLOF2}.
In fact, little is known in general about such correlators at weak coupling.
In the $O(N)$ model, it was shown recently that the two-point function of elementary scalars is equal to the derivative of a bulk block at NLO \cite{Gimenez-Grau:2022ebb,Bianchi:2022sbz}.
These blocks are not known in closed form, and it might be the case that the two problems are intimately connected.\footnote{This point is discussed in more detail in Section \ref{subsec:TwoPointFunctionsOfScalars}.}

To summarize, the two-point functions in the presence of the line defect at weak coupling turned out to be a significantly more complicated object to study than the four-point function of bulk operators.
We now consider the strong-coupling end of the correlator, with the hope that the situation is more favorable in this regime.

\subsection{The supergravity perspective}
\label{subsec:TheSupergravityPerspective}

The AdS/CFT correspondence is introduced in Section \ref{subsec:N4SuperYangMills}.
There we have seen that the strong-coupling limit of $\Nm = 4$ SYM admits a description in terms of classical IIB supergravity.
Although this formulation will not be used to carry out our calculations, it is useful for understanding the perturbative structure of our correlator in the strong-coupling regime.

The dual of a supersymmetric Wilson loop is a string worldsheet extending inside AdS$_5$, whose boundary corresponds to the path of the loop \cite{Maldacena:1998im,Rey:1998ik}.
It can be represented graphically as the \textit{Witten diagram}
\begin{equation}
\vev{ \Wl } =\ \WittenDiagramWilsonLoop\ = 1\,.
\label{eq:WittenDiagramWilsonLine}
\end{equation}
Here the red curved line corresponds to the string worldsheet, while the black circle represents the boundary of the AdS spacetime.
In Section \ref{subsec:TheWilsonLineDefectCFT}, we discussed the fact that $\vev{ \Wl }$ does not receive corrections and thus \eqref{eq:WittenDiagramWilsonLine} remains exact, even at higher-loop orders.

\subsubsection{Leading order}
\label{subsubsec:LeadingOrder2}

We are interested in the interplay between the supersymmetric Wilson line and local operators.
In the holographic description, the single-trace half-BPS operators $\Op_\Delta$ are dual to certain Kaluza-Klein modes arising from the compactification of the IIB action on $S^5$.
The two-point function consists of two of these modes, sourced at the boundary of AdS$_5$ and propagating through the bulk before being absorbed by the string worldsheet.
Since we subtract in \eqref{eq:ConnectedTwoPoint} the disconnected contributions, the first diagram at strong coupling corresponding to this process is
\begin{align}
\WittenDiagramTwoPointLO\ &= \vvev{\Op_{\Delta_1} (u_1, x_1)} \vvev{\Op_{\Delta_2} (u_2, x_2) } \notag \\
& = \frac{\lambda}{\vphantom{2^{\frac{\Delta_1 + \Delta_2+4}{2}}}N^2} \frac{\sqrt{\Delta_1 \Delta_2}}{2^{\frac{\Delta_1 + \Delta_2+4}{2}}} \frac{(u_1 \cdot \theta)^{\Delta_1} (u_2 \cdot \theta)^{\Delta_2}}{|x_1^{\perp}|^{\Delta_2} |x_2^{\perp}|^{\Delta_1}}\,,
\label{eq:WittenDiagramTwoPointLO}
\end{align}
where the lines propagating in the bulk space correspond to $\Delta_j$ propagators.
This is a \textit{factorized} diagram of one-point functions since the operators do not interact in the bulk.
The one-point functions of $\Op_\Delta$ in the presence of the Wilson line are
\begin{equation}
\WittenDiagramOnePoint\ = \vvev{\Op_\Delta (u,x)} = \frac{\sqrt{\lambda \Delta}}{2^{\frac{\Delta+2}{2}} N} \frac{(u \cdot \theta)^\Delta}{\vphantom{2^{\frac{\Delta+2}{2}}} |x^\perp|^\Delta}\,.
\label{eq:WittenDiagramOnePointLO}
\end{equation}
The diagrams \eqref{eq:WittenDiagramTwoPointLO} and \eqref{eq:WittenDiagramOnePointLO} have been computed in \cite{Giombi:2009ek,Giombi:2012ep,Buchbinder:2012vr,Billo:2018oog}, where they were shown to match the strong-coupling limit obtained through localization.

\subsubsection{Next-to-leading order}
\label{subsubsec:NextToLeadingOrder2}

The first non-trivial correction to the two-point function contains an interaction vertex in the bulk.
In the exchanged line, one sums over all Kaluza-Klein modes that can couple to two $\Op_\Delta$'s.
Schematically, we have the \textit{bulk-exchange} diagram
\begin{equation}
\WittenDiagramTwoPointNLOOne\ .
\label{eq:WittenDiagramTwoPointNLOOne}
\end{equation}
There are two additional diagrams
\begin{equation}
\WittenDiagramTwoPointNLOTwo\ , \qquad \WittenDiagramTwoPointNLOThree\ ,
\label{eq:WittenDiagramTwoPointNLOTwoThree}
\end{equation}
which are respectively \textit{defect-exchange} and \textit{contact} diagrams.

A quantitative description of these diagrams can be found in \cite{Gimenez-Grau:2023fcy}.
Instead of an explicit calculation using the effective action in AdS$_5 \times S^5$, here we bootstrap the result using the dispersion relation derived in Section \ref{sec:ADispersionRelationForDefectCFT}.
This formula reconstructs a correlator from its singular part, which is mathematically captured by a discontinuity:
\begin{equation}
\WittenDiagramTwoPointNLOOne\ \sim \int \Disc\ \WittenDiagramTwoPointNLOOne\ .
\label{eq:WittenDiagramDisc}
\end{equation}
A crucial property of holographic CFTs is that the discontinuity is dramatically simpler than the correlator.
In particular, we show in the course of the computation that the discontinuity receives corrections only from a finite number of single-trace operators exchanged in the bulk.
Each of these contributions is schematically a product of a (tree-level) one-point function and a three-point function:
\begin{equation}
\WittenDiagramTwoPointNLOOne\ \sim \sum_{\substack{\text{single} \\ \text{traces}}} \int \WittenDiagramOnePoint\ \times\ \WittenDiagramThreePoint\ .
\label{eq:WittenDiagramTwoPointNLO2}
\end{equation}
The one- and three-point functions of half-BPS operators are known from localization, as we have seen in Section \ref{subsec:ExactResults}, and so all that is left to do is to compute a certain integral and sum over finitely many single-trace contributions.
Throughout this chapter, we translate this pictorial representation into a concrete bootstrap algorithm that fully fixes the two-point correlators with minimal external input.

The reader may wonder why we include in the left-hand side of \eqref{eq:WittenDiagramDisc}-\eqref{eq:WittenDiagramTwoPointNLO2} only the diagram presented in \eqref{eq:WittenDiagramTwoPointNLOOne}.
The reason is that the dispersion relation can only reconstruct part of the correlator.
The formula presented in Section \ref{subsec:MainResultAndDerivation} suffers from low-spin ambiguities, which in practice means here that both the defect-exchange and the contact diagrams are missed.
Nevertheless, the correlator can be \textit{entirely} reconstructed by using the superconformal Ward identities \eqref{eq:WardIdentities} and consistency conditions.

\section{A dispersion relation for defect CFT}
\label{sec:ADispersionRelationForDefectCFT}

We begin our analysis by considering a dispersion relation specific to defect CFT as our starting point.
Our objective is to derive a formula that resembles \eqref{eq:DispersionRelationBulk} but is tailored to the context of defect CFT.
Importantly, the result applies to defects of any codimension $q>2$.\footnote{It is worth mentioning that a dispersion relation for boundaries was introduced in \cite{Bianchi:2022ppi}.}

We employ similar principles to the bulk case to derive a dispersion relation for two-point functions of scalar operators in the presence of a defect.
For arbitrary scalar operators, the dispersion relation can be expressed as follows:
\begin{equation}
F(r,w) = \int dw' \int dr'\, K(r,w,r',w')\, \Disc\, F(r',w')\,,
\label{eq:DispersionRelationAnsatz}
\end{equation}
where the function $F(r,w)$, the reduced correlator, is related to the full correlator via \eqref{eq:TwoPoint_RSymmetryChannels}.
Note that the cross-ratios $r$ and $w$ are more convenient to use here than $z$ and $\zb$.
The \textit{single} discontinuity on the right-hand side should be understood as
\begin{equation}
\Disc F(r,w) := F(r, w + i0) - F(r, w - i0)\,.
\label{eq:Disc}
\end{equation}
This formula is generally non-perturbative, although subtleties related to its convergence are discussed in Section \ref{subsec:ConvergenceAndSubtractions}.
The kernel in \eqref{eq:DispersionRelationAnsatz} is yet to be determined and is the main focus of this section.

Just like in the bulk case, the dispersion relation is powerful because, in numerous applications, the discontinuity is computationally simpler than the full correlator.
The discontinuity can be computed using the bulk OPE provided in \eqref{eq:OPEv2}, and for each term in the expansion, we obtain
\begin{equation*}
\Disc \left[ \left( \frac{rw}{(r-w)(rw-1)} \right)^{\frac{\Delta_1 + \Delta_2}{2}} f_{\Delta,\ell} (r,w) \right] \sim \sin \left( \frac{\pi}{2} \left( \Delta - (\Delta_1 + \Delta_2 + \ell) \right) \right)\,.
\end{equation*}
Here $\Delta_1$ and $\Delta_2$ represent the scaling dimensions of the external operators, while $\Delta$ and $\ell$ correspond respectively to the scaling dimension and spin of the exchanged operator.
In the case of bulk operators near the multi-twist dimension, given by $\Delta = \Delta_1 + \Delta_2 + \ell + 2n + \gamma$, we observe that the contribution to the discontinuity typically scales with a small anomalous dimension, i.e., schematically $\Disc \sim \gamma$. 
For CFTs with a small coupling $g$, the anomalous dimensions usually follow the coupling, meaning $\gamma \sim g$.
Consequently, many contributions in the discontinuity are suppressed by powers of the coupling.
Even in the case of non-perturbative CFTs, the lightcone bootstrap \cite{Fitzpatrick:2012yx,Komargodski:2012ek} implies that the anomalous dimension $\gamma$ is small for large spins, leading to a suppression of contributions from operators with large spins by powers of $\ell^{-1}$.

\subsection{Main result and derivation}
\label{subsec:MainResultAndDerivation}

This section is dedicated to the derivation of the dispersion relation for defect CFT.
The analytic structure of the correlator $F(r,w)$ allows the use of Cauchy's integral formula for obtaining the kernel of \eqref{eq:DispersionRelationAnsatz}.
In the course of the derivation, an assumption about the behavior of the correlator is made, and consequences are discussed in Section \ref{subsec:ConvergenceAndSubtractions}.

Note that the dispersion relation can also be derived starting from the Lorentzian inversion formula.
In this case, we insert \eqref{eq:InversionFormulaDefect} in \eqref{eq:TwoPoint_SuperconformalBlocksDefect} and perform a resummation of the blocks to obtain the desired kernel.
While the first option is undoubtedly the simplest way, this approach demonstrates that our dispersion relation captures the same information as the Lorentzian inversion formula, highlighting the consistency and equivalence between the two methods.
For brevity, we do not provide this derivation here but it can be found in \cite{Barrat:2022psm}.

\subsubsection{Analytic structure}
\label{subsubsec:AnalyticStructure}

A straightforward derivation of the dispersion relation can be obtained by examining the analytic structure of the correlator $F(r,w)$ in the complex $w$ plane.
We perform an analytic continuation of the CFT to Lorentzian signature and consider configurations where $0 < r < 1$.
Upon closer examination, we observe that the limits $w \to r,\frac{1}{r}$ correspond to the local operators approaching null separation, resulting in branch-point singularities of the correlator.
Likewise, branch points arise at $w \to 0,\infty$, corresponding to double-lightcone singularities.
Furthermore, the correlator must be single-valued within the unit circle $|w|$, as these configurations exist in the Euclidean signature.
Consequently, for fixed $r$, two branch cuts are formed: $w \in (0,r)$ and $w \in (1/r, +\infty)$, as illustrated in Figure \ref{fig:DeformationOfIntegrationContours}.

\subsubsection{Main result}
\label{subsubsec:MainResult}

\begin{figure}[t!]
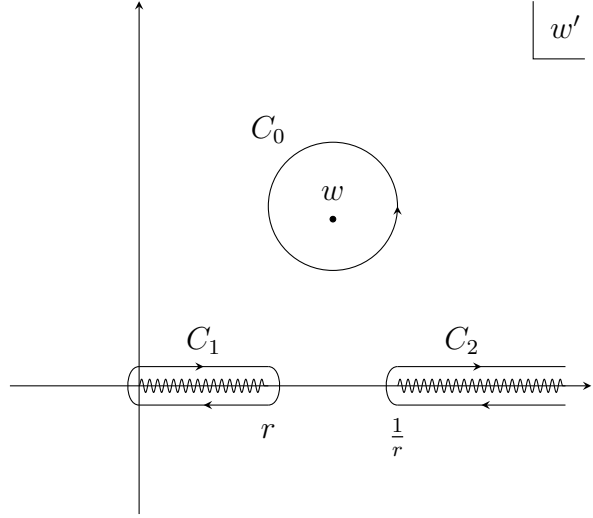

\centering
\IntegrationContours
\caption{Summary of the integration contours in the complex $w'$ plane to apply Cauchy's integral formula.
By dropping the arcs at $\infty$ and using symmetry, we can reformulate the dispersion relation as an integral over the contour $C_1$ only.
}
\label{fig:DeformationOfIntegrationContours}
\end{figure}

Keeping this information in mind, we use Cauchy's integral formula and deform the contour to encircle the two branch cuts.
This procedure is illustrated in Figure \ref{fig:DeformationOfIntegrationContours}.
At this stage, we assume that the contributions from the arcs at infinity can be neglected.
This assumption is reexamined in Section \ref{subsec:ConvergenceAndSubtractions}.

It is then straightforward to obtain the kernel of \eqref{eq:DispersionRelationAnsatz}.
Using the condition $F(r,w') = F(r,1/w')$, which is valid for defects of codimension higher than two ($q > 2$) as well as for parity-preserving codimension-two defects.
After some elementary manipulations detailed in \cite{Barrat:2022psm}, we obtain an integral over a single contour:
\begin{align}
 F(r,w)
 = \oint_{C_1} \frac{dw'}{2 \pi i}
 \frac{w (1 - w') (1 + w')}{ w' (1 - w w')}
 \frac{F(r,w')}{w'-w}\, .
 \label{eq:Cauchy3}
\end{align}
The final step involves rewriting this integral as a real integral of the discontinuity, $\Disc F(r,w')$, over the interval $w' \in (0,r)$.
This leads to the following dispersion relation:
\begin{equation}
F(r,w) = \int_0^r \frac{dw'}{2\pi i} \frac{w(1-w')(1+w')}{w'(w'-w)(1-w w')} \Disc F(r,w')\,.
\label{eq:DispersionRelation}
\end{equation}
Note that the ansatz given in \eqref{eq:DispersionRelationAnsatz} turned out to be too conservative: the final dispersion relation contains only an integral over $w'$ and not $r'$.

It is worth noting that in certain applications, the contour integral given by \eqref{eq:Cauchy3} is more advantageous than its real counterpart \eqref{eq:DispersionRelation}.
This is particularly true when the correlator $F(r,w')$ has a pole at $w' = r$ instead of a branch cut.
In such cases, the integral in \eqref{eq:Cauchy3} can be evaluated using the residue theorem:
\begin{align}
 F(r,w)
 = - \underset{w'=r}{\Res} \left[ 
 \frac{w (1 - w') (1 + w')}{ w' (1 - w w')}
 \frac{F(r,w')}{w'-w} \right] \, .
 \label{eq:PoleResidue}
\end{align}
This last formula is especially useful for the computation of the two-point functions in Section \ref{sec:TwoPointFunctionsAtStrongCoupling}.

\subsection{Convergence and subtractions}
\label{subsec:ConvergenceAndSubtractions}

We now comment on the practical use of the dispersion relation given in \eqref{eq:DispersionRelation}.
In particular, we discuss how the assumptions made in the previous section affect the dispersion relation and present a method for dealing with the related problems.

\subsubsection{Convergence}
\label{subsubsec:Convergence}

One crucial assumption for the derivation of \eqref{eq:DispersionRelation} is that the correlator $F(r,w)$ decays fast enough in the complex $w$ plane.
In realistic theories, this assumption may not hold.
In this section, we discuss the consequences related to this assumption, and how to ensure the applicability of the dispersion relation in more general scenarios.

The arcs at infinity can only be neglected if the correlator decays fast enough in this limit.
However, in general, our correlator exhibits the behavior
\begin{align}
 |F(r,w)| \le K |w|^{s_\star} \quad
 \text{as} \quad |w| \to \infty \, ,
 \label{eq:BehaviorOfTheCorrelator}
\end{align}
and the dispersion relation \eqref{eq:DispersionRelation} is valid only if $s_\star < 0$.
Furthermore, there is a singularity as $w \to 0$.
For now, we assume however that the correlator is symmetric under $F(r,w) = F(r,\frac1w)$, which means that by addressing the singularity at $w \to \infty$, we automatically resolve the issues around $w \to 0$.\footnote{It should be noted that the symmetry $w \to \frac1w$ may not hold for defects of codimension $q=2$. This special case is discussed in \cite{Barrat:2022psm}.}

Let us consider the case where $0 \leq s_\star < \infty$ instead of assuming $s_\star < 0$.
It is worth noting that in all the examples we are familiar with, the growth of the correlator can be bounded by a polynomial, resulting in a finite value of $s_\star$.
However, we do not have a rigorous argument to exclude the possibility of faster growth, such as exponential, which would require different subtraction methods.

In light of this, \textit{two} approaches can be pursued to reconstruct $F(r,w)$ from its discontinuity.
The first one is discussed in more detail in \cite{Barrat:2022psm} and consists in dividing the original correlator by a rational term such that the new function decays sufficiently fast at infinity.
The main shortcoming of this approach is that there can be new contributions to the discontinuity.
For this reason, in Section \ref{sec:TwoPointFunctionsAtStrongCoupling} we use the alternative method that is described below.

\subsubsection{Subtractions}
\label{subsubsec:Subtractions}

The second approach is called the method of \textit{subtractions}, which relies on the assumption that the correlator can be decomposed as a sum of two terms:
\begin{align}
 F(r,w)
 = \tilde F(r,w)
 + F_\star(r,w) \, .
 \label{eq:SubtractedCorrelator}
\end{align}
The first term is assumed to decay at infinity, namely,
\begin{equation}
|\tilde F(r,w)| \to 0 \quad \text{as} \quad |w| \to \infty\,,
\label{eq:BehaviorSubtractedCorrelator}
\end{equation}
and its conformal block decomposition includes transverse spins $0 \le s < +\infty$.

The second term $F_\star$ is responsible for the growth at infinity
\begin{equation}
|F_\star (r,w)| \to |w|^{s_*} \quad \text{as} \quad |w| \to \infty\,,
\label{eq:BehaviorExtraTerm}
\end{equation}
and we make the crucial assumption that its block decomposition has \textit{bounded spin} $0 \le s \le s_\star$.

Summarizing, we have
\begin{align}
\begin{split}
 \tilde F(r,w) &= \sum_{\vphantom{\Dh}s=0}^{\infty} \sum_{\Dh} \tilde\mu_{\Dh,s} \fh_{\Dh,s}(r,w) \, , \\
 F_\star (r,w) &= \sum_{\vphantom{\Dh}s=0}^{s_\star} \sum_{\Dh} \mu^\star_{\Dh,s} \fh_{\Dh,s}(r,w) \, .
\end{split}
\label{eq:SubtractedCorrelatorInBlocks}
\end{align}
Note that the two sums run over \textit{different values} of $\Dh$, and although $F_\star$ contains finitely many spins, $\Dh$ might run over infinitely many operators.
Crucially, since any individual block has zero discontinuity $\Disc \fh_{\Dh,s} =0$, the only way to generate a non-zero discontinuity is by summing infinitely many spins for $s \to \infty$.
But $F_\star$ involves finitely many spins $s \le s_\star$ by assumption, so it has zero discontinuity $\Disc F_\star=0$.
Also by assumption, $\tilde F$ decays at infinity, and it can be reconstructed with the dispersion relation.

Applying the discontinuity to $F = \tilde F + F_\star$ and recalling $\Disc F_\star=0$ gives the final formula
\begin{align}
F(r,w) =\ & \int_0^r \frac{dw'}{2 \pi i} \frac{w (1 - w') (1 + w')}{ w' (w'-w) (1 - w w')} \Disc F(r,w') \notag \\
& + \sum_{\vphantom{\Dh}s=0}^{s_\star} \sum_{\Dh}  \mu^\star_{\Dh,s} \fh_{\Dh,s}(r,w) \, .
\label{eq:DispersionRelationSubtractions}
\end{align}
In practical bootstrap calculations, one often has access to the discontinuity $\Disc F$, but determining whether additional terms are needed in the dispersion relation becomes a non-trivial task.
If additional terms are required, the challenge lies in determining the precise values of $\{\Dh, \mu^\star_{\Dh,s}\}$.
$\Nm=4$ SYM at strong coupling serves as a prime laboratory for the dispersion relation, and we demonstrate in Section \ref{sec:TwoPointFunctionsAtStrongCoupling} that it is indeed possible to determine the terms that are missed by the naive application of the dispersion relation \eqref{eq:DispersionRelation}.

\section{Two-point functions at strong coupling}
\label{sec:TwoPointFunctionsAtStrongCoupling}

The dispersion relation derived above is completely general.
We now apply it to the setup discussed in \ref{sec:AnInvitation}: the two-point functions of half-BPS operators in the presence of the supersymmetric Wilson line in $\Nm=4$ SYM.
For simplicity, we start with the case where the two operators are the lowest-lying single-trace half-BPS operators of identical scaling dimension $\Delta_1 = \Delta_2 = 2$.
We then extend our method to the general case and pave the way towards determining a closed form for all $\Delta_1$ and $\Delta_2$.

\subsection{An example: $\vvev{\Op_2 \Op_2}$}
\label{subsec:AnExample}

Our goal is to bootstrap the order $\Op(\sqrt{\lambda})$, represented by the Witten diagrams depicted in \eqref{eq:WittenDiagramTwoPointNLOOne} and \eqref{eq:WittenDiagramTwoPointNLOTwoThree}, while the leading order was computed in \cite{Giombi:2012ep} and is given in \eqref{eq:WittenDiagramTwoPointLO}.
Following \eqref{eq:RSymmetryChannels_22}, the reduced correlator can be decomposed into \textit{three} $R$-symmetry channels that fix the dependence on $\sigma$.
The goal of this section is to fix the three functions $F_j (z, \zb)$ at order $\Op(\sqrt{\lambda}/N^2)$.

\subsubsection{The method}
\label{subsubsec:TheMethod}

The computation of $\vvev{\Op_2 \Op_2}$ presented here can be decomposed into three steps:
\begin{enumerate}
\item Computing the discontinuity;
\item Applying the dispersion relation;
\item Supersymmetrizing the result.
\end{enumerate}
We give here an overview of each of these stages.

To compute the discontinuity of the correlator, we first expand the correlator in the bulk channel (see \eqref{eq:TwoPoint_SuperconformalBlocksBulk}), analyzing which of the bulk blocks contribute.
We find that very few blocks are non-vanishing, and in particular that they correspond to protected operators being exchanged.
This simplifies the computation massively and makes the discontinuity of $\Fm(z, \zb;\sigma)$ a very simple object.

In the second part, we apply the dispersion relation given in \eqref{eq:DispersionRelation}.
This means performing the integration over $w'$ to obtain the subtracted $R$-symmetry channels $\tilde{F}_j(r,w)$.
This computation is straightforward and readily gives the result for the exchange diagram \eqref{eq:WittenDiagramTwoPointNLOOne}.

Finally, we account for the missing diagrams \eqref{eq:WittenDiagramTwoPointNLOTwoThree} by imposing supersymmetry.
This is done by considering the ansatz \eqref{eq:DispersionRelationSubtractions} channel-wise and requiring that $\Fm(z,\zb ; \sigma)$ satisfies the superconformal Ward identities \eqref{eq:WardIdentities}.
There we assume that $s_\star$ takes the minimal value for the correlator to be consistent.
This assumption was later confirmed by a direct calculation in \cite{Gimenez-Grau:2023fcy}.

In the following, we describe in detail each of these steps for the two-point function $\vvev{\Op_2 \Op_2}$.
In the subsequent section, we show that this method generalizes straightforwardly to arbitrary external operators, providing an efficient way of computing correlators at strong coupling.

\subsubsection{The discontinuity}
\label{subsubsec:TheDiscontinuity}

We begin with the first step of our algorithm: the computation of the discontinuity.
In this context, the main question to elucidate is how to obtain the discontinuity despite not knowing the full correlator.

As mentioned above, the solution is to expand the two-point function using the \textit{bulk} expansion given in \eqref{eq:TwoPoint_SuperconformalBlocksBulk}.
With the help of Appendix \ref{app:ConformalBlocks}, we see that the bosonic elements of the superblocks can be written as
\begin{equation}
f_{\Delta, \ell} (z, \zb) = [(1-z)(1-\zb)]^{(\Delta-\ell)/2}\, \tilde{f}_{\Delta,\ell} (z,\zb)\,,
\label{eq:BosonicElementsOfSuperblocks}
\end{equation}
where the function $\tilde{f}_{\Delta,\ell} (z,\zb)$ has an expansion around $z, \zb \sim 1$ in positive integer powers.
As a result, only the prefactor can have a non-vanishing discontinuity.
The contribution of a single bulk operator $\Op$ to the discontinuity is simply
\begin{equation*}
\left. \Omega^{2-j} \Disc F_j (z, \zb)\right|_{\Op} \sim (z\zb)^{\Delta} (1-z)^{\frac{\Delta - (4 + \ell)}{2}} \tilde{f}_{\Delta,\ell} (z,\zb) \Disc \left[ (1-\zb)^{\frac{\Delta-(4+\ell)}{2}} \right]\,.
\end{equation*}
There are \textit{two} situations in which this discontinuity does not vanish:
\begin{enumerate}
\item If $\Delta$ is non-integer. This corresponds to $\Op$ having an anomalous dimension correcting its tree-level dimension $\Delta = 4 + \ell + 2n + \gamma$.
\item If $\Delta$ is integer, but $\Delta < 4 + \ell$. This corresponds to $\Op$ being a protected single-trace operator, whose dimension is below the double-trace threshold. Note that, even though $\Disc (1-\zb)^{-n}$ naively vanishes, for $n>0$ the singularity at $\zb = 1$ gives a finite contribution to the inversion formula.
\end{enumerate}

In our setup, the discontinuity only receives contributions of the second type. This claim can be proved by studying the superconformal bulk OPE in detail.
From the selection rules given in \cite{Liendo:2016ymz}, we observe that the following operators contribute:
\begin{equation}
\Op_2 \times \Op_2 \to \mathds{1} + \Bm_{[0,2,0]} + \Bm_{[0,4,0]} + \sum_\ell \Cm_{[0,2,0],\ell} + \sum_{\Delta, \ell} \Am_{[0,0,0],\ell}^\Delta\,.
\label{eq:BulkOPE22}
\end{equation}
Here $\Bm_{[0,\Delta,0]}$ corresponds to the half-BPS operators of Section \ref{subsec:N4SuperYangMills}.
The operators labelled as $\Cm_{[0,2,0],\ell}$ are \textit{semishorts} (with $\ell \geq 0$ even), while $\Am_{[0,0,0],\ell}^\Delta$ refers to \textit{longs} (with $\Delta \geq \ell +2$ and $\ell \geq 0$ even).
The latter are \textit{not} protected, and their contribution is particularly challenging to determine.

Let us now analyze which multiplets can contribute to the discontinuity.
The operators in the $\Bm_{[0,2,0]}$ multiplet have integer dimension $\Delta < 4$, and so they have a non-vanishing discontinuity.
The operators in the $\Bm_{[0,4,0]}$ and $\Cm_{[0,2,0],\ell}$ multiplets have integer dimension $\Delta \ge 4+\ell$, which means that they cannot contribute.
Only the unprotected multiplets $\Am_{[0,0,0],\ell}^\Delta$ remain, which can have anomalous dimensions.
The scaling dimensions of these multiplets have the following schematic structure \cite{Goncalves:2014ffa}:
\begin{equation}
\Delta = 4 + 2n + \ell + \frac{1}{N^2} \left( a + \frac{b}{\lambda^{3/2}} + \ldots \right) + O(N^{-4})\,,
\label{eq:ScalingDimensionsLongs}
\end{equation}
which means that anomalous dimensions \textit{do not contribute} at the order in perturbation theory we are working.
This implies that all the operators in the bulk OPE \eqref{eq:BulkOPE22} have integer-valued scaling dimensions at this order, and so the correlator must admit an expansion in integer powers in the limit $z, \zb \to 1$.

The main consequence of these observations is that \textit{only} the superblock $\Gm_{[0,2,0]}$ (corresponding to $\Bm_{[0,2,0]}$) has non-vanishing discontinuity:
\begin{equation}
\left. \Disc \Fm (z,\zb; \OR) \right|_{O( \sqrt{\lambda}/N^2 )} = \lambda_{222} a_2\, \Disc \Omega^2 \Gm_{[0,2,0]} (z,\zb; \OR)\,.
\label{eq:DiscSuperblock}
\end{equation}
Let us pause a moment to comment on this result.
Comparing this expression to \eqref{eq:TwoPoint_SuperconformalBlocksBulk} with \eqref{eq:BulkOPE22} shows in which proportions the discontinuity is a simpler object than the full correlator.
Indeed, instead of an infinity of blocks encoding complicated unprotected operators, the discontinuity consists of a \textit{single block} corresponding to the exchange of the half-BPS operator $\Op_2$ itself.

The superblocks $\Gm_{[0,\Delta,0]}$ are known and given in \eqref{eq:ConformalBlocks_TwoPoint_BulkHalfBPS}.
For $\Delta=2$, they take the form
\begin{equation}
\Gm_{[0,2,0]} (z,\zb;\OR) = h_2(\OR) f_{2,0}(z,\zb) + \frac{1}{180} h_0 (\OR) f_{4,2} (z,\zb)\,,
\label{eq:Superblock020}
\end{equation}
where $h_k(\sigma)$ and $f_{\Delta,\ell}(z,\zb)$ can be found in Appendix \ref{app:ConformalBlocks}.
It is easy to compute the discontinuity of these blocks, and we find
\begin{align}
\Disc \Omega^2\, F_0(z,\zb) &= - i \sigma^2\, \lambda_{222} a_2\, \sin (p \pi) \left( - \frac{1-\zb}{\sqrt{\zb}} \right)^{-p} \frac{z ( 1 - z^2 + 2z \log z )}{(1-z)^4}\,, \notag \\
\Disc \Omega\, F_1 (z,\zb) &= - 2i \sigma\, \lambda_{222} a_2\, \sin (p \pi) \left( - \frac{1-\zb}{\sqrt{\zb}} \right)^{-p} \frac{z\, \log z}{(1-z)^2}\,, 
\label{eq:DiscRSymmetryChannels22} \\
\Disc F_2 (z,\zb) &= 0\,, \notag
\end{align}
where we used $p \sim 1$ as a regulator for the results not to vanish.
The integral in \eqref{eq:DispersionRelation} is divergent at $p = 1$, and at the end of the computation the result is finite.

\subsubsection[Applying the \\ dispersion relation]{Applying the dispersion relation}
\label{subsubsec:ApplyingTheDispersionRelation}

The expressions \eqref{eq:DiscRSymmetryChannels22} can be inserted into \eqref{eq:DispersionRelation}.
The integrals are elementary and yield
\begin{align}
\begin{split}
\tilde{F}_0(z,\zb) & = - \frac{\lambda_{222} a_2}{2} \frac{(1-z)(1-\zb) }{(1-z\zb)^3} ( (1 + z\zb)(1 - z\zb) + 2 z\zb \log z\zb )\,, \\
\tilde{F}_1(z,\zb) & = - \lambda_{222} a_2 \frac{\log z\zb}{\sqrt{z\zb}(1-z\zb)}\,, \\
\tilde{F}_2(z,\zb) & = 0\,.
\end{split}
\label{eq:FTilde}
\end{align}
We put a tilde on the channels $\tilde F_n$ to anticipate that these functions do not give the final result, but instead that they should be understood as subtracted channels, following the insights of Section \ref{subsec:ConvergenceAndSubtractions}.

Note the relative ease with which \eqref{eq:FTilde} has been obtained.
To reproduce this result with the Lorentzian inversion formula, one would first extract the CFT data of the defect operators twist by twist, performing an integral of the discontinuity times a kernel.
Plugging this CFT data in the defect-channel expansion, one would then resum a double power series in $z$ and $\zb$.\footnote{This is explicitly performed in \cite{Barrat:2021yvp}.}
Instead, all we had to do now is to multiply \eqref{eq:DiscRSymmetryChannels22} with the kernel in \eqref{eq:DispersionRelation} and perform the integral, for instance by picking the residue, using \eqref{eq:PoleResidue}.

\subsubsection{Supersymmetrization}
\label{subsubsec:Supersymmetrization}

The correlation function obtained in the previous section is \textit{not} supersymmetric, i.e., the three $R$-symmetry channels given in \eqref{eq:FTilde} do not respect the superconformal Ward identities given in \eqref{eq:WardIdentities}.
This happens because the inversion formula misses contributions from low-lying spins $s \le s_\star$, as anticipated in Section \ref{subsec:ConvergenceAndSubtractions}.\footnote{ Such a phenomenon has already been observed for bulk correlators.
For example, in the bootstrap of the Wilson--Fisher fixed point, there is an ambiguity captured by a single $\ell=0$ block \cite{Alday:2016jfr,Alday:2017zzv}.
In supersymmetric theories, one expects the inversion formula to converge better than in non-supersymmetric ones. See \cite{Lemos:2021azv} for a recent discussion.
}
The value of $s_\star$ is related to the behavior of the two-point function in the Regge limit $z/\zb \to 0$ \cite{Lemos:2017vnx}, and in principle, $s_\star$ can be determined by careful analysis of the corresponding Witten diagrams.
Instead, in the present work we use the heuristic intuition that $s_\star$ should take the minimal value that generates a supersymmetric correlator.
As we show below, the resulting correlators make predictions that are in perfect agreement with the expectations from the topological sector.

To obtain a supersymmetric correlator, we add defect families with operators of dimensions $\Dh = 0, 1, 2, \ldots$, and low spin $s \le s_\star$.
The OPE coefficients of these operators are unknowns that we fix by imposing the Ward identities \eqref{eq:WardIdentities}.
We found experimentally that the minimal ansatz consists of taking $s_\star=0$ for $\tilde{F}_1(z,\zb)$ and $s_\star=1$ for $\tilde{F}_2(z,\zb)$.
To be precise, we define the final correlators $F_j(z,\zb)$ as
\begin{align}
\begin{split}
F_0 (z,\zb) &= \tilde{F}_0 (z,\zb) \,, \\
F_1 (z,\zb) &= \tilde{F}_1 (z,\zb) + \sum_{n=0}^\infty \left( k_n \hat f_{n, 0}(z, \zb) + p_n \partial_{\Dh} \hat f_{n, 0}(z, \zb) \right)\, , \\
F_2 (z,\zb) &= \tilde{F}_2 (z,\zb) + \sum_{s=0,1} \sum_{n=0}^\infty \left( q_{n,s} \hat f_{n+s, s}(z, \zb) + r_{n,s} \partial_{\Dh} \hat f_{n+s, s}(z, \zb) \right)\,.
\end{split}
\label{eq:Ambiguities22}
\end{align}
As mentioned before, the free coefficients $k_n$, $p_n$, $q_{n,s}$, and $r_{n,s}$ can be fixed by requiring that the Ward identities are satisfied.
This fixes all the coefficients in terms of $q_{0,0}$ and $k_1$.
Note that $q_{0,0}$ corresponds to the ambiguity $f_{0,0}(z,\zb)=1$, i.e., the defect identity.
We know from the Witten diagrams analysis of Section \ref{subsec:TheSupergravityPerspective} that the defect identity is given by the constant contribution $a_2^2$, and thus
\begin{align}
 q_{0,0} = a_2^2 \Big|_{O( \sqrt{\lambda}/N^2 )} \, .
 \label{eq:AmbiguityDefectId}
\end{align}
On the other hand, the unknown $k_1$ can be determined by demanding a bulk expansion that is consistent with the observations made after \eqref{eq:ScalingDimensionsLongs}, i.e., there should not appear anomalous dimensions for bulk operators.
Concretely, this means that the expansion of \eqref{eq:Ambiguities22} in the limit $z, \zb \to 1$ should take the form of a power series, without spurious $\log(1- \zb)$ terms.
Since the defect expansion in \eqref{eq:Ambiguities22} is natural around $z, \zb \sim 0$, this is only possible after 
resumming the correlator.
We were able to do so, and the remaining spurious term reads
\begin{equation}
\Omega\, F_1 (z,\zb) \sim \frac{\sigma}{2} \left(\lambda_{222} a_2 - k_1 \right) \log(1- \zb) + \ldots\,.
\label{eq:Ambiguityk1}
\end{equation}
This fixes the coefficient $k_1$ to be
\begin{equation}
k_1 = \lambda_{222} a_2\,.
\label{eq:Fixedk1}
\end{equation}

\subsubsection[\\ Final result and consistency checks]{Final result and consistency checks}
\label{subsubsec:FinalResultAndConsistencyChecks}

We now present the final result for the correlator $\vvev{\Op_2 \Op_2}$, using the input of localization for the two remaining free coefficients, namely, $a_2^2$ and $\lambda_{222} a_2$.
We can then obtain OPE coefficients of other protected operators which in turn can be checked against the localization data.

Using \eqref{eq:HalfBPSThreePointFunctions} and \eqref{eq:HalfBPSOnePoint_WeakStrong} leads to an elegant correlator without any free coefficient left:
\begin{align}
\begin{split}
F_0 (z, \zb) =\ & - \frac{\sqrt{\lambda}}{2 N^2} \frac{(1-z)(1-\zb) }{(1-z\zb)^3} ( (1 + z\zb)(1 - z\zb) + 2 z\zb \log z\zb )\,,   \\
F_1 (z,\zb) =\ & \frac{\sqrt{\lambda}}{N^2} \frac{(1-z)(1-\zb)}{\sqrt{z\zb}} \left[ \log (1 + \sqrt{z \zb}) + \frac{z \zb}{(1 - z \zb)^2} \right.  \\*
& \left. + \frac{z\zb \big( 5 z\zb -2 z^2\zb^2 +z^3\zb^3 - (z+\zb)(2-z\zb+z^2\zb^2) \big) \log z\zb}{2(1-z)(1-\zb)(1-z \zb)^3} \right]\,,  \\
F_2 (z,\zb) =\ & \frac{\sqrt{\lambda}}{8 N^2} \left[ -3 - \frac{2(z+\zb)}{\sqrt{z\zb}} + \frac{(z+\zb)(1+z\zb)-4z\zb}{(1-z\zb)^2} \right.  \\*
& \left.  + \frac{2 \big( (z+\zb)(1+z\zb)-4z\zb\big) \log (1+\sqrt{z\zb})}{z\zb} \right.  \\*
& \left. +\frac{z\zb \big( (z+\zb)(3-2z\zb+z^2 \zb^2) - 6 + 6z\zb - 4z^2 \zb^2 \big) \log z\zb}{ (1-z \zb)^3} \right] \, . 
\end{split}
\label{eq:FullCorrelator22}
\end{align}
Comparing to \eqref{eq:TwoPoint_SuperconformalBlocksBulk}, this correlator predicts the OPE coefficient of the double-trace operator $\Op_{2,2}$:
\begin{equation}
\lambda_{22(2,2)} a_{(2,2)} = \frac{\lambda}{N^2} \left( \frac18 - \frac{1}{2 \sqrt{\lambda}} + \ldots \right) \,,
\label{eq:LambdaaDoubleTrace22}
\end{equation}
which matches the localization results given in \eqref{eq:ThreePointDoubleTrace} and \eqref{eq:OnePointDoubleTrace}.
We can also extract the defect CFT data for the protected operators:
\begin{align}
\begin{split}
b_{21}^2 &= \frac{\lambda}{N^2} \left( \frac{1}{\sqrt{\lambda}} + \ldots \right)\,, \\
b_{2(0,2)}^2 &= 1 + \frac{\lambda}{N^2} \left( - \frac{1}{2 \sqrt{\lambda}} + \ldots \right)\,.
\end{split}
\label{eq:BulkDefectChecks22}
\end{align}
The OPE coefficient $b_{21}^2$ can be compared to the direct computation (see \eqref{eq:BulkDefectTwoPoint_Strong}), and we find a perfect match.
The OPE coefficient $b_{2(0,2)}^2$ corresponds to a double-trace operator $\Oh_{0,2}$ defined via \eqref{eq:DoubleTrace_Defect}.
This result is a prediction from our correlator.\footnote{The observation that $\Oh_{0,2}$ should appear in this type of correlator was first discussed in Appendix A of \cite{Giombi:2018hsx}.
In principle, the operator $\Oh_2$ should also appear, but it is not relevant at the present order in $N$.}

Moreover, using the superblocks described in \cite{Liendo:2016ymz,Barrat:2020vch}, the correlator above can be used for extracting the supersymmetric CFT data for unprotected operators. 
Although this is reserved for future work, a preliminary discussion can be found in \cite{Barrat:2021yvp}.

\subsection{The general case $\vvev{\Op_{\Delta_1} \Op_{\Delta_2}}$}
\label{subsec:TheGeneralCase}

The analysis presented above can be extended straightforwardly to correlators of general operators $\Op_{\Delta_1}$ and $\Op_{\Delta_2}$.
The calculation carries through almost unchanged.
As a concrete application of the algorithm, we present a closed form for the channel $F_0$ for arbitrary external operators, while the other channels are discussed in \ref{subsec:TheWittenDiagramBootstrap}.

\subsubsection{Generalization}
\label{eq:Generalization}

One crucial aspect of the computation of $\vvev{\Op_2 \Op_2}$ is that a single bulk block contributes to the discontinuity.
As the scaling dimensions of the external operators increase, more operators contribute to the bulk OPE, and a natural question is whether these new contributions would spoil our method.
The outcome is that, although there are indeed more blocks contributing, they all correspond to single-trace operators.
All other operators have double-trace dimensions, with zero anomalous dimensions at this order.
And, as we have seen in Section \ref{subsec:AnExample}, these double-trace operators are killed by the discontinuity.

Summarizing, the discontinuity at order $O(\sqrt{\lambda}/N^2)$ is obtained as a sum over single-trace half-BPS contributions:
\begin{equation}
\Disc \Fm (z, \zb; \sigma) 
= \Disc \Omega^{\frac{\Delta_1 + \Delta_2}{2}} \,
   \sum_{j=\frac{\Delta_{21}}{2}}^{\frac{1}{2}(\Delta_1 + \Delta_2 -2)} \!\! \lambda_{\Delta_1 \Delta_2 (2j)} \, a_{2j} \,
   \Gm_{\Om_{2j}} (z, \zb;\sigma)\, .
\label{eq:DiscBlocksGeneral}
\end{equation}
Although the sum should also include the operator $\Om_{\Delta_1+\Delta_2}$, it has exact double-trace dimension and is hence also removed by the discontinuity.\footnote{Because we choose to work with the $SU(N)$ gauge group, the operator $\Om_1 \propto \tr \phi$ is identically zero, and it should also be removed from the sum whenever it appears.}

We are now in the same position as after \eqref{eq:DiscSuperblock} for the case $\Delta_1 = \Delta_2 = 2$.
All the OPE coefficients $\lambda_{\Delta_1 \Delta_2 \Delta}$ and $a_\Delta$ can be computed using localization, and the relevant superconformal blocks are known.
All that is left to do is to use the dispersion relation \eqref{eq:DispersionRelation} to reconstruct $\tilde \Fm(z,\zb; \sigma)$, the subtracted version of the full correlator, following Section \ref{subsec:ConvergenceAndSubtractions}.
Adding low-spin contributions to the subtracted correlator and demanding consistency with the superconformal Ward identities, we can fix unambiguously the full result for any scaling dimensions $\Delta_1$, $\Delta_2$.

\subsubsection[A closed form for the highest \\ $R$-symmetry channel]{A closed form for the highest $R$-symmetry channel}
\label{eq:AClosedFormForTheHighestRSymmetryChannel}

Thanks to the efficiency of the dispersion relation, we can obtain the correlation functions for high values of $\Delta_1$ and $\Delta_2$.
Explicit results for $(\Delta_1, \Delta_2) = \lbrace (2,3), (2,4), (3,3), (3,4), (4,4) \rbrace$ can be found in \cite{Barrat:2021yvp,Barrat:2022psm}.
This allows us to guess a closed-form formula for the channel $F_0$ for arbitrary external operators, which reads
\begin{align}
F_0 (z,\zb) 
=\ & \frac{\sqrt{\Delta_1 \Delta_2}}{2^{(\Delta_{21}+2)/2} \Delta_{21}} \frac{\sqrt{\lambda}}{N^2}\,
\biggl\lbrace \Delta_{21}^2 (1-\delta_{\Delta_{21},1}) \notag \\
& - (\Delta_{21}+1) \frac{(1-z)(1-\zb)}{(1-z\zb)^2} \bigg[ 
    (\Delta_{21}+2)\, _2F_1 \Big( 1, \frac{\Delta_{21}}{2}; \Delta_{21}+2; 1-z\zb \Big) \notag \\
& + \frac12 \Delta_{21} (z \zb - 3) - 2 \bigg] \biggr\rbrace\,.
  \label{eq:ConjectureF0}
\end{align}
For any integer value of $\Delta_{21}$, the hypergeometric function can be rewritten as a sum of rational functions of the cross-ratios and $\log z \zb$.
In particular, when $\Delta_{21}$ is even the expression contains logarithms, while for $\Delta_{21}$ odd the logarithms are absent.
This is expected to be a feature of the full correlator $\Fm(z,\zb;\sigma)$.

Finally, notice that the case $\Delta_{21} = 1$ is somewhat special because the operator $\Om_1 \propto \text{tr}\, \phi$ could in principle contribute to the discontinuity \eqref{eq:DiscBlocksGeneral}.
Here we have assumed the gauge group to be $SU(N)$, in which case one should set $\Om_1 = 0$.
If we instead assume it to be $U(N)$, then $\Om_1$ contributes to the discontinuity, and the result is obtained from \eqref{eq:ConjectureF0} by setting $\delta_{\Delta_{21},1} \to 0$. 
This results in an expression that has the nice property of being \emph{analytic} in complex $\Delta_1$ and $\Delta_2$.

\section{The microbootstrap}
\label{sec:TheMicrobootstrap}

We have discussed in Section \ref{subsec:TwoPointFunctionsOfHalfBPSOperators} the fact that two-point functions of half-BPS operators contain a closed subsector that we refer to as the topological sector.
So far, we have considered the topological sector from the correlator point of view, but there exists another way to think about it that leads to the concept of \textit{microbootstrap}.
In this case, we start from the crossing equations \eqref{eq:CrossingEquation} and impose the topological limit $z = \zb = \alpha$.
This results in a web of \textit{truncated} crossing equations, which can be solved in certain cases, using the bulk CFT data as input.
In the following, we explore the microbootstrap of two-point functions and derive exact results for the cases where the gauge group is $SU(2)$ or $SU(N)$ at large $N$.

This section presents the results of the unpublished notes \cite{Barrat:2021un}.

\subsection{Microbootstrapping two-point functions}
\label{subsec:MicrobootstrappingTwoPointFunctions}

The fact that crossing equations truncate in the topological limit is a general feature of $\Nm=4$ SYM.
A similar analysis can indeed be performed for four-point functions of half-BPS operators, where in this case the supersymmetric version of \eqref{eq:Bulk_CrossingSymmetry} truncates to the half-BPS sector \cite{Lemos:un}.

Here we prepare the ground for an analysis of the defect crossing equations in this special limit.
First, we review the topological sector from the point of view of the crossing equations, before analyzing the multiplicity of the operators appearing on both sides.

\subsubsection{Another look at the topological subsector}
\label{subsubsec:AnotherLookAtTheTopologicalSubsector}

Our starting point is the topological limit \eqref{eq:TopologicalSector} of the crossing equations given in \eqref{eq:CrossingEquation}.
The result is that \textit{all the unprotected operators} are projected out, both from the bulk and defect OPE.
In other words, the crossing equation truncates into a \textit{microboostrap equation}, in which only half-BPS operators remain:
\begin{equation}
\sum_{\vphantom{\substack{\Oh \\ \text{half-BPS}}}\substack{\Op \\ \text{half-BPS}}} \Omega_\text{topo}^{\frac{\Delta_1 + \Delta_2 - \Delta}{2}} \lambda_{\Delta_1 \Delta_2 \Op} a_\Op = \sum_{\substack{\Oh \\ \text{half-BPS}}} \Omega_\text{topo}^{\Dh} b_{\Delta_1 \Oh} b_{\Delta_2 \Oh}\,.
\label{eq:MicrobootstrapEquations}
\end{equation}
The sign function present in $\Omega_\text{topo}$ (defined in \eqref{eq:OmegaTopo}) is crucial for our analysis, as it decouples each crossing equation into two independent equations.
This subsector goes well beyond our two-point functions and describes a closed subsector of the operator spectrum of the theory.
The corresponding CFT data can be obtained by solving Gaussian multi-matrix models \cite{Giombi:2018hsx}.

Here we show an alternative method for deriving this CFT data from the microboostrap equations.
Our goal is to obtain the OPE coefficients on the defect side by assuming that we know the bulk OPE coefficients $\lambda_{\Delta_1 \Delta_2 \Delta_3}$ as well as the one-point functions $a_\Delta$.
The motivation behind this technique is that solving microbootstrap equations is computationally much more efficient than matrix models, and thus it gives access to new data points that can be used for instance to guess analytical forms.

From the point of view of the two-point functions, the sign function in \eqref{eq:MicrobootstrapEquations} results in \textit{two} topologies:
\begin{equation}
\vvev{\Op_{\Delta_1} (1) \Op_{\Delta_2} (|x|)} \quad \text{and} \quad
\vvev{\Op_{\Delta_1} (1) \Op_{\Delta_2} (-|x|)}\,,
\label{eq:TwoTopologies}
\end{equation}
with $-1 < x < 1$.\footnote{The conformal frame contains a symmetry $x \leftrightarrow \frac{1}{x}$ that makes configurations beyond $|x| = 1$ equivalent to the ones at $|x| < 1$.}
Here the operators are put in the conformal frame \eqref{eq:DefectCrossRatios_zzb} and aligned, i.e., the first operator is located at $(0,1,0,0)$ while the second one is on the line $(0,x,0,0)$, either on the same side of the defect as the first operators, or on the other side.
The first setup is referred to as \textit{closed chain} in the $3$-matrix formulation of \cite{Giombi:2012ep}, while the second one is named \textit{open chain}.

Note that, in this section, we keep the connected part of the correlator inside the crossing equations, which implies that the identity operator contributes.

\subsubsection{Multiplicity}
\label{subsubsec:Multiplicity}

One complication in solving the microboostrap equations \eqref{eq:MicrobootstrapEquations} is that, in general $SU(N)$, several half-BPS operators exist for a given scaling dimension $\Delta$.
Indeed, the \textit{multi-trace} operators introduced in \eqref{eq:MultiTraceOperators_Bulk} have the same quantum numbers as the single-trace operators but different OPE coefficients.
Since the gauge group is not a symmetry per se, the conformal bootstrap is insensitive to the number of traces and keeps all these operators on the same footing as long as their quantum numbers agree.
Let us quantify this multiplicity precisely.

\begin{table}[t!]
\centering
\caption{The multiplicity of the bulk and defect operators for the gauge group $SU(N)$.
The first part of the table indicates the degeneracy of {\normalfont bulk} operators for low scaling dimensions $\Delta$.
The number of operators for arbitrary $\Delta$ is given by \eqref{eq:MultiplicityBulk} in the text.
The second part is dedicated to {\normalfont defect} operators, for which \eqref{eq:MultiplicityDefect} indicates the multiplicity for fixed scaling dimensions $\Dh$.}
\begin{tabular}{|c|cccccc|}
\hline
$\Delta$ & $0$ & $1$ & $2$ & $3$ & $4$ & \ldots \\ \hline
Operators & $\mathds{1}$ & $\emptyset$ & $\Op_2$ & $\Op_3$ & $\Op_4, \Op_{2,2}$ & \ldots \\
$m(\Delta)$ & $1$ & $0$ & $1$ & $1$ & $2$ & \ldots \\ \hline \hline
$\Dh$ & $0$ & $1$ & $2$ & $3$ & $4$ & \ldots \\ \hline
Operators & $\mathds{1}$ & $\Oh_{1}$ & $\Oh_{2}, \Oh_{0|2}$ & $\Oh_{3}, \Oh_{1|2}, \Oh_{0|3}$ & $\Oh_{4}, \Oh_{2|2}, \Oh_{1|3},$ & \ldots \\
 & & & & & $\Oh_{0|4}, \Oh_{0|2,2}$ & \\
$m(\Dh)$ & $1$ & $1$ & $2$ & $3$ & $5$ & \ldots \\ \hline
\end{tabular}
\label{table:MultiplicityOfOperators}
\end{table}

On the bulk side, operators must have a minimum of two scalar fields in their traces in order not to vanish.
As a consequence, operators are unique up to $\Delta = 3$, but at $\Delta = 4$ double-trace operators start to appear (see Table \ref{table:MultiplicityOfOperators}).
The number of bulk operators at a fixed $\Delta$ is given by
\begin{equation}
m(\Delta) = p(\Delta) - p(\Delta-1)\,,
\label{eq:MultiplicityBulk}
\end{equation}
where $p(\Delta)$ is known as the \textit{partition number}.
The partition number corresponds to the number of ways the positive integer $\Delta$ can be decomposed into smaller integers $\Delta_1, \ldots, \Delta_k$ such that their sum is equal to $\Delta$.
It is interesting to note that, although it has been widely studied, no closed form is known yet for the partition number.
It can be efficiently computed with the use of the following recursion relation \cite{ewell2004recurrences}:
\begin{equation}
p(\Delta) = \sum_{k \in \mathds{Z}^+} (-1)^{k+1} p(\Delta - k(3k-1)/2)\,.
\label{eq:PartitionNumberRecursive}
\end{equation}

We can also count the number of operators on the defect side.
The multiplicity of defect operators grows manifestly quicker than the one of bulk operators, because what we call single-trace operators corresponds in fact to insertions inside the trace of the Wilson line, and one can form non-vanishing operators with a single scalar (see Table \ref{table:MultiplicityOfOperators}).
The multiplicity in this case is exactly equal to the partition number:
\begin{equation}
m(\Dh) = p(\Dh)\,.
\label{eq:MultiplicityDefect}
\end{equation}

It turns out that the multiplicity of operators grows too rapidly for us to be able to solve the microbootstrap equations at finite $N$.
In the following, we show that it can however be done for the special cases of $N=2$ and $N \to \infty$.
It would be interesting to study the microboostrap equations of higher-point functions to see whether finite $N$ results can also be obtained with this method.

\subsection{Solving the microboostrap equations}
\label{subsec:SolvingTheMicroboostrapEquations}

We now turn our attention to concrete solvable cases.
As a warm-up, we consider the case of the gauge group being $SU(2)$, in which the situation simplifies considerably.
At large $N$, some multiplicity remains but the microbootstrap equations can nevertheless be solved for most of the operators, leading to an efficient algorithm for obtaining closed forms.

\subsubsection{A warm-up: $SU(2)$}
\label{subsubsec:AWarmUpSU2}

In $SU(2)$, operators with \textit{odd} scaling dimension $\Delta$ vanish, while the multiplicity of \textit{even} operators reduces to $1$.
These two statements are a direct consequence of the relation
\begin{equation}
T^a T^b = \frac{\delta^{ab}}{4}\, \mathds{1} + \frac{i}{4} \veps^{abc} T^c\,,
\label{eq:TraceIdentitySU2}
\end{equation}
fulfilled by the generators of $SU(2)$.
By repeatedly using \eqref{eq:TraceIdentitySU2}, one can show that multi- and single-trace operators are fully equivalent up to an overall constant, which is then reabsorbed in the normalization of the two-point function.

It is convenient to choose the \textit{maximally multi-trace} operators as the representation for a given scaling dimension:
\begin{equation}
\Op_\Delta := \frac{1}{\sqrt{n_\Delta}} \left( \tr (u \cdot \phi)^2 \right)^{\Delta/2}\,.
\label{eq:DefinitionMaximallyMultiTrace}
\end{equation}
For these operators, the microbootstrap equations become
\begin{equation}
\sum_{\substack{\Delta = \Delta_{21}\\\text{step }2}}^{\Delta_1+\Delta_2} \sgn(-x)^{\frac{\Delta_1 + \Delta_2 - \Delta}{2}}\, \lambda_{\Delta_1 \Delta_2 \Delta} a_\Delta = a_{\Delta_1} a_{\Delta_2} + \sum_{\Dh = 1}^{\Delta_1} \sgn(-x)^{\Dh}\, b_{\Delta_1 \Dh} b_{\Delta_2 \Dh}\,,
\label{eq:MicroboostrapEquationsSU2}
\end{equation}
where all the degeneracy has been lifted.
The microbootstrap equations can be solved in the form of \textit{recursion relations}, where we split the cases in which the scaling dimension $\Dh$ of the defect operator is even or odd:
\begin{equation}
\begin{split}
\left. b_{\Delta \Dh}^2 \right|_{\Dh \text{ even}} &=  \frac{1}{b_{\Dh \Dh}^2}
\Biggl(
a_\Delta a_{\Dh}
- \sum_{\substack{k = \Delta - \Dh\\\text{step }4}}^{\Delta +\Dh} \lambda_{\Dh \Delta k} a_k
+ \sum_{\substack{k = 2\\\text{step }2}}^{\Dh - 2} b_{\Dh k} b_{\Delta k}
\Biggr)^2\,, \\
\left. b_{\Delta \Dh}^2 \right|_{\Dh \text{ odd}} &= \frac{1}{b_{(\Dh+1) \Dh}^2}
\Biggl(
\sum_{\substack{k = \Delta - \Dh+1\\\text{step }4}}^{\Delta +\Dh-1} \lambda_{(\Dh+1) \Delta k} a_k
- \sum_{\substack{k = 1\\\text{step }2}}^{\Dh - 2} b_{(\Dh+1) k} b_{\Delta k}
\Biggr)^2\,.
\end{split}
\label{eq:SolutionMicrobootstrapSU2}
\end{equation}
The starting values of the recursion relations are
\begin{equation}
\begin{split}
b_{\Delta \Delta}^2 &=
1 - a_\Delta^2
+ \sum_{\substack{k = 4\\\text{step }4}}^{2\Delta} \lambda_{\Delta \Delta k} a_k
- \sum_{\substack{k = 2\\\text{step }2}}^{\Delta - 2} b_{\Delta k}^2\,,
\\
b_{\Delta (\Delta-1)}^2 &=
- \sum_{\substack{k = 2\\\text{step }4}}^{2(\Delta - 1)} \lambda_{\Delta \Delta k} a_k
- \sum_{\substack{k = 1\\\text{step }2}}^{\Dh - 2} b_{\Delta k}^2\,.
\end{split}
\label{eq:StartingValuesMicrobootstrapSU2}
\end{equation}

The OPE coefficients $b_{\Delta \Dh}^2$ can be efficiently generated for arbitrary bulk and defect scaling dimensions using these recursion relations.
They can be solved if one inputs the bulk CFT data in \eqref{eq:MicroboostrapEquationsSU2}.
In $SU(2)$, the one-point functions are given by
\begin{equation}
a_\Delta = \frac{g^2 + 8(\Delta + 1)}{g^2 + 8} \frac{g^\Delta}{8^{\Delta/2} \sqrt{\Gamma(\Delta+2)}}\,.
\label{eq:OnePointFunctionsSU2}
\end{equation}
This expression can be obtained from the finite $N$ localization computations, see for instance (4.5) in \cite{Okuyama:2006jc}.
To the best of our knowledge, there exists no known closed form for the three-point functions $\lambda_{\Delta_1 \Delta_2 \Delta_3}$, even for the case of $SU(2)$.
However, as indicated previously, these OPE coefficients are protected and are easy to generate through combinatorics since the tree-level Wick contractions suffice.
Using this data, we were able to derive a closed form analytical in $\Delta$ and $\Dh$ for the OPE coefficients $b_{\Delta \Dh}^2$:
\begin{equation}
b_{\Delta \Dh}^2 =
K_{\Delta \Dh}
\left(
(1 + \Dh)(2 + \Dh) P_0 (\Delta, \Dh) - 2 ( 2 ( 1 + \Dh ) - \Delta ) P_1 (\Delta, \Dh)
\right)^2\,,
\label{eq:BulkDefectClosedFormSU2}
\end{equation}
where we have defined the following help functions:
\begin{align}
\begin{split}
& K_{\Delta \Dh} := \frac{(-1)^{\Dh} g^{2(\Delta+2)} \sin (\pi \Delta) (1+\Delta) \Gamma(1+\Dh) \Gamma^2 (\Dh - \Delta)}{2^{3\Delta + \Dh + 4} \pi \Gamma(-\Delta) (g^2 + 8) U \bigl( - \frac{1 + \Dh}{2} , \frac{1}{2} , - \frac{g^2}{16} \bigr) U \bigl( - \frac{\Dh}{2} , \frac{1}{2} ; - \frac{g^2}{16} \bigr) }\,, \\
& P_s (\Delta, \Dh) := \sum_{k=0}^{2\Dh - 1} \frac{4^k k^s}{g^{2k} (1+\Delta - 2k) \Gamma (\Dh + 3 - 2k) (1)_k}\,,
\end{split}
\label{eq:HelpFunctionsMicrobootstrap}
\end{align}
where $U(a,b;x)$ is the confluent hypergeometric function of the second kind, defined as
\begin{equation}
U(a,b;x) = x^{-a}\ _2F_0 (a,1+a-b;;-x^{-1})\,.
\label{eq:ConfluentHyper}
\end{equation}
Note that the sum in $P_s (\Delta, \Dh)$ can be performed and gives a lengthy sum of regularized hypergeometric functions.
For compactness, we kept the formulation as a sum in \eqref{eq:HelpFunctionsMicrobootstrap}, but we emphasize that the expression is fully analytical in the scaling dimensions $\Delta$ and $\Dh$.

The topological sector itself can be expressed in an elegant closed form for the open chain:
\begin{equation}
\begin{split}
\vvev{\Op_{\Delta_1} (1) \Op_{\Delta_2} (-x)} =\ & \frac{g^{\Delta_1 + \Delta_2}}{8^{\frac{\Delta_1 + \Delta_2}{2}} (g^2 + 8) \sqrt{ \Gamma(\Delta_1 + 2) \Gamma(\Delta_2 + 2) }} \\
& \times \biggl\lbrace g^2 + 8 (\Delta_1 + 1)(\Delta_2 + 1) \\
& \times\ _3F_1 \left(-\Delta_1, -\Delta_2, 1 ; 2 ; \frac{8}{g^2} \right) \biggr\rbrace\,.
\end{split}
\label{eq:TopologicalClosedFormSU2}
\end{equation}
This is to be contrasted with the closed form given in (A.2) of \cite{Giombi:2018hsx} at large $N$, which consists of an infinite sum.
We were however not able to find such an expression for the closed chain.

\subsubsection{Large $N$}
\label{subsubsec:LargeN}

At large $N$, we also observe a reduction in the degeneracy of the operators.
On the bulk side, the truncated OPE gives
\begin{equation}
\vvev{\Op_{\Delta_1} \Op_{\Delta_2}} = \lambda_{\Delta_1 \Delta_2 (\Delta_1,\Delta_2)} a_{(\Delta_1,\Delta_2)} + \sum_{\substack{i = \Delta_{21}\\\text{step }2}}^{\Delta_1+\Delta_2-2} \Omega^{\frac{\Delta_1+\Delta_2-\Delta}{2}}_\text{topo}\, \lambda_{\Delta_1 \Delta_2 \Delta} a_{\Delta}\,.
\label{eq:BulkMicroOPELargeN}
\end{equation}
Following the notation already used throughout Section \ref{sec:TwoPointFunctionsAtStrongCoupling}, the first term refers to double-trace operators, while all the operators in the sum of the second term are single-trace.
As a result, all the degeneracy on the bulk side has been lifted.
Moreover, in the $1/N$ expansion, the first term of the sum is of order $\Op(1)$ if $\Delta_1 = \Delta_2$, while all the other terms are of order $1/N^2$.
The neglected terms are of order $\Om (1/N^{4})$.

On the defect side, the OPE takes the form
\begin{equation}
\vvev{\Op_{\Delta_1} \Op_{\Delta_2}} = a_{\Delta_1} a_{\Delta_2} + \sum_{\Dh = 1}^{\Delta_1} \Omega^{\Dh}_\text{topo}\, b_{\Delta_1 \Dh} b_{\Delta_2 \Dh} + \Omega^{\Delta_1}_\text{topo}\, b_{\Delta_1 (0,\Dh_1)} b_{\Delta_2 (0,\Dh_2)}\,.
\label{eq:DefectMicroOPELargeN}
\end{equation}
Here the notation in the last term refers to defect double-trace operators, following the definition given in \eqref{eq:DoubleTrace_Defect}.
The other terms all contain single-trace defect operators.
Contrary to the $SU(2)$ case, there remains some degeneracy for the operators of scaling dimension $\Dh = \Delta_1$.

\begingroup
\allowdisplaybreaks

Nevertheless, the microbootstrap equations can be solved for all the other operators and analytically continued to the inaccessible ones.
We obtain the following recursive expressions:
\begin{align}
\left. b_{\Delta \Dh}^2 \right|_{\Dh \text{ even}} =\ &
\frac{1}{b^2_{(\Dh + 1) \Dh}}
\Biggl(
4^{\Dh/2} \left( \lambda_{(\Dh+1) \Delta (\Delta+1, \Delta)} a_{(\Delta+1,\Delta)} - a_{\Dh+1} a_\Delta \right)  \notag \\
&+ \sum_{\substack{k = \Delta - \Dh + 1\\\text{step }4}}^{\Delta + \Dh - 3} 4^{\frac{k - \Delta + \Dh - 1}{4}} \lambda_{(\Dh+1)\Delta k} a_k
- \sum_{\substack{k = 2\\\text{step }2}}^{\Dh-2} 4^{\frac{\Dh-k}{2}} b_{(\Dh+1) k} b_{\Delta k}
\Biggr)^2\,,  \notag \\
\left. b^2_{\Delta \Dh} \right|_{\Dh \text{ odd}} =\ &
\frac{1}{b_{(\Dh + 1) \Dh}}
\Biggl(
\sum_{\substack{k = \Delta - \Dh + 1\\\text{step }4}}^{\Delta + \Dh - 1} 4^{\frac{k - \Delta + \Dh - 1}{4}} \lambda_{(\Dh+1)\Delta k} a_k \notag \\
&- \sum_{\substack{k = 1\\\text{step }2}}^{\Dh-2} 4^{\frac{\Dh-k}{2}} b_{(\Dh+1) k} b_{\Delta k}
\Biggr)^2\,.
\label{eq:SolutionMicrobootstrapLargeN}
\end{align}
These equations are valid for $\Delta > \Dh + 1$.
The case $\Delta = \Dh + 1$ corresponds to the starting values and is given by
\begin{equation}
\begin{split}
\left. b_{\Delta (\Delta-1)}^2 \right|_{\Delta \text{ even}} =\ &
\sum_{\substack{k = 2\\\text{step }4}}^{2(\Delta-1)} 4^{\frac{k-2}{4}} \lambda_{\Delta \Delta k} a_k
- \sum_{\substack{k = 1\\\text{step }2}}^{\Delta-3} 4^{\frac{\Delta-k-1}{2}} b_{\Delta k}^2\,, \\
\left. b_{\Delta (\Delta-1)}^2 \right|_{\Delta \text{ odd}} =\ &
\sum_{\substack{k = 2\\\text{step }4}}^{2(\Delta-1)} 4^{\frac{k-2}{4}} \lambda_{\Delta \Delta k} a_k
- \sum_{\substack{k = 2\\\text{step }2}}^{\Delta-3} 4^{\frac{\Delta-k-1}{2}} b_{\Delta k}^2 \\
& - 2^{\Delta - 1} a_{\Delta}^2 + 2^{\Delta - 1} \lambda_{\Delta \Delta (\Delta, \Delta)} a_{(\Delta, \Delta)}\,.
\end{split}
\label{eq:StartingValuesMicrobootstrapLargeN}
\end{equation}

\endgroup

The advantage of recursive relations over matrix models is that they are computationally extremely effective.
In particular, we make use of the closed forms known for one- and three-point functions for obtaining new defect results.
This allows us to generate results for high scaling dimensions, which in turn are useful for identifying analytical patterns.
We were not able to find a closed form for the general OPE coefficient $b_{\Delta \Dh}$, but we give here a few examples of closed forms with respect to $\Delta$ for low-lying $\Dh$.
For this purpose, it is useful to define
\begin{equation}
b_{\Delta \Dh}^2 = \frac{\Delta \sqrt{\lambda}}{\vphantom{I_1 (\sqrt{\lambda})}2^\Delta N^2} \frac{\bt_{\Delta \Dh}^2}{ I_1 (\sqrt{\lambda}) }\,.
\label{eq:DefinitionHelpFunctionMicrobootstrapLargeN}
\end{equation}
Then for $\Dh = 1$, we find
\begin{equation}
\bt_{\Delta 1}^2 = \frac{\Delta^2}{\vphantom{I_2 (\sqrt{\lambda})} 2} \frac{I^2_\Delta (\sqrt{\lambda})}{I_2 (\sqrt{\lambda})}\,.
\label{eq:BulkDefectClosedFormLargeN1}
\end{equation}
For $\Dh=2$, the recursion relations give
\begin{equation}
\bt_{\Delta 2}^2 = \frac{ \left( \sqrt{\lambda} I_3 (\sqrt{\lambda}) I_\Delta (\sqrt{\lambda}) - I_1 (\sqrt{\lambda}) \left( 2 (\Delta -1 ) I_{\Delta-1} (\sqrt{\lambda}) + \sqrt{\lambda} I_{\Delta+2} (\sqrt{\lambda}) \right) \right)^2 }{ I_1 (\sqrt{\lambda}) \left( I_1 (\sqrt{\lambda}) ( 4 I_2 (\sqrt{\lambda}) + \sqrt{\lambda} I_5 (\sqrt{\lambda}) ) \right) - \sqrt{\lambda} I^2_3 (\sqrt{\lambda}) }\,.
\label{eq:BulkDefectClosedFormLargeN2}
\end{equation}
It is easy to find the closed forms term by term for increasing $\Dh$.
The full closed form as a function of $\Delta$ and $\Dh$ is likely to merge into a single function, valid both for $\Dh$ even and odd.
Note that the expansion of these expressions at $\lambda \gg 1$ agrees with \eqref{eq:BulkDefectTwoPoint_Strong}.
We keep the search for a full analytical form for future work.

Finally, we can also write an elegant closed form for the topological sector of the connected two-point function itself.
Such a closed form was already given in (A.2) of \cite{Giombi:2018hsx}.
However, our result is slightly different, since it contains a finite sum instead of an infinite one.
We define for compactness
\begin{align}
\vvev{\Op_{\Delta_1} \Op_{\Delta_2}} = \frac{\sqrt{\Delta_1 \Delta_2} \sqrt{\lambda}}{2^{\frac{\Delta_1 + \Delta_2 + 4}{2}} N^2 I_1 (\sqrt{\lambda}) } T_{\Delta_1 \Delta_2}\,.
\label{eq:TopologicalHelpFunctionLargeN}
\end{align}
Then the connected two-point function is given by
\begin{align}
T_{\Delta_1 \Delta_2} = \sqrt{\lambda}\, I_{\Delta_1 + \Delta_2 - 1} (\sqrt{\lambda}) + 2 \sum_{\substack{k = \Delta_{21}\\\text{step }2}}^{ \Delta_1 + \Delta_2 - 2 } \sgn (x)^{\frac{\Delta_1 + \Delta_2 - k}{2}} k\, I_k (\sqrt{\lambda})\,.
\label{eq:TopologicalClosedFormLargeN}
\end{align}
We consider here the gauge group to be $U(N)$ for simplicity, which means that we include an additional operator $\Op_1$ in the spectrum.
This is useful, as it keeps the expression above analytical for the case $\Delta_2 = \Delta_1 + 1$.

\section{Summary and outlook}
\label{sec:SummaryAndOutlook}

The main result of this chapter is the dispersion relation for defect CFT given in \eqref{eq:DispersionRelation}:
\begin{equation*}
\tilde F (r,w) = \int_0^r \frac{dw'}{2\pi i} \frac{w(1-w')(1+w')}{w'(w'-w)(1-w w')} \Disc F(r,w')\,,
\end{equation*}
that expresses two-point functions as an integral of a discontinuity times a kinematically fixed kernel.
The formula is valid for general defects of codimension two or higher, along with a subtraction method to cure the potential convergence problems (see \eqref{eq:SubtractedCorrelator}):
\begin{equation*}
F(r,w) = \tilde F (r,w) + F_\star (r,w)\,.
\end{equation*}
Thanks to the simplicity of the dispersion relation, strong-coupling correlators can be computed very efficiently and we presented here as an example the results for the simplest correlator $\vvev{\Op_2 \Op_2}$ in $\Nm=4$ SYM with a supersymmetric Wilson line.
Additionally, we showed that the method generalizes straightforwardly to $\vvev{\Op_{\Delta_1} \Op_{\Delta_2}}$, and we provided a closed form for the highest-weight channel $F_0$.
The implications of the topological sector on crossing equations were also investigated, and exact results were given for the cases of $SU(2)$ and large $N$.
Although these results can also be obtained from matrix models, the microbootstrap method is significantly more efficient and can be used to determine closed forms from a large sample of examples.

The formulation of analytic bootstrap constraints in terms of dispersion relations is a promising framework, that illuminated the connection between several alternative techniques in the bulk, among which Mellin space approaches \cite{Rastelli:2016nze,Rastelli:2017udc,Alday:2020dtb,Alday:2020lbp} and exact analytic functionals \cite{Mazac:2016qev,Mazac:2018mdx,Mazac:2018ycv,Caron-Huot:2020adz}.
It would be interesting to pursue an analogous line of research in the defect setup.
The methods developed here are general and indeed go well beyond the study of supersymmetric theories, Wilson loops, and even line defects.
Surface defects in $6d$ $\Nm=(2,0)$ supersymmetric theory were recently studied using these techniques \cite{Meneghelli:2022gps}.
Also correlators in non-supersymmetric setups have been computed, such as the magnetic line in the $O(N)$ model \cite{Gimenez-Grau:2022ebb,Bianchi:2022sbz}.
The dispersion relation can also be applied to defects of higher dimension, such as monodromy defects (see \cite{Gimenez-Grau:2021wiv,Barrat:2022psm}).
In the following, we discuss specific possible directions to extend this work.

\subsection{'t Hooft loops}
\label{subsec:tHooftLoops}

Most of the considerations made in Section \ref{sec:TwoPointFunctionsAtStrongCoupling} are based on the assumptions that the external operators, including the defect, are half-BPS.
Although the supersymmetric Wilson line is undoubtedly the most studied half-BPS line defect in $\Nm=4$ SYM, it is not the only one that can be defined.
Here we provide a short discussion on 't Hooft loops and sketch how a similar computation could be performed in this case.\footnote{We thank K. Zarembo for interesting conversations on this topic.}

\subsubsection{$S$-duality}
\label{subsubsec:SDuality}

't Hooft lines are half-BPS defects related to Wilson lines through $S$-duality, which can be regarded as a generalization of the electric-magnetic duality.
This relates the two line defects via the transformation
\begin{equation}
g \leftrightarrow \frac{4\pi}{g}\,.
\end{equation}
Our bootstrap techniques are only sensitive to symmetry, and thus 't Hooft lines are seemingly on the same footing as our Wilson-line defect.
Since $S$-duality is a strong/weak-coupling duality, one expects strong-coupling results for the Wilson line to describe the 't Hooft setup at weak coupling and vice versa.
In particular, the correlators at weak coupling given in Section \ref{subsec:AFirstGlimpseAtWeakCoupling} are expected to describe 't Hooft loops at strong coupling, which we might be able to compute using our bootstrap techniques.

\subsubsection{Strong coupling}
\label{subsubsec:StrongCoupling}

It is however clear from comparing the correlators in Sections \ref{subsec:AFirstGlimpseAtWeakCoupling} and \ref{subsec:AnExample} that the structure of the functions is quite different.
At weak coupling, the two-point functions involving a Wilson line have the following schematic expansion at finite $N$:
\begin{equation}
\vvev{\Op_{\Delta_1} \Op_{\Delta_2}}_{\text{Wilson}} \sim 1 + \frac{g^2}{N} + \frac{g^4 (N^2-1)}{N^2} + \ldots
\label{eq:WilsonLineExpansion}
\end{equation}
Since half-BPS operators are self-dual under $S$-transformations, the operation
\begin{equation}
g \longrightarrow \tg := \frac{4 \pi}{g}
\label{eq:STransformation}
\end{equation}
straightforwardly translates correlators with a Wilson line to correlators with a 't Hooft line.
Thus we have the following strong-coupling expansion:
\begin{equation}
\vvev{\Op_{\Delta_1} \Op_{\Delta_2}}_{\text{'t Hooft}} \sim 1 + \frac{(4\pi)^2}{g^2 N} + \frac{(4\pi)^4 (N^2-1)}{g^4 N^2} + \ldots\,.
\label{eq:tHooftLineExpansion}
\end{equation}
This expansion is not well-defined in terms of $\lambda$ in the large $N$ limit.
One way to understand this is to realize that the strong-coupling limit $\lambda \gg 1$ while simultaneously taking the limit $N \to \infty$ is in fact a \textit{weak-coupling limit} with respect to $g$.
On the other hand, the expansion \eqref{eq:tHooftLineExpansion} truly is a strong-coupling expansion with respect to $g$.

To circumvent this issue, one can define a new coupling tailored for 't Hooft loops \cite{Pufu:2023vwo}:
\begin{equation}
\tlambda := \tg^2 N\,.
\label{eq:NewtHooftCoupling}
\end{equation}
$S$-duality requires that a strong-coupling expansion of this type exists for 't Hooft lines in the large $N$ limit.
It would be interesting to apply our methods in the limit $N \to \infty, \tlambda \gg 1$, where the only analysis to redo is the one of the scaling dimensions of the long operators provided in \eqref{eq:ScalingDimensionsLongs}.

\subsection{Bulk-defect-defect correlators}
\label{subsec:BulkDefectDefectCorrelators}

One can also look for other families of correlators to study that may be simpler than the two-point functions.
A candidate is the three-point function between \textit{one bulk} and \textit{two defect} operators.
An initial study of such operators was conducted for instance in \cite{Liendo:2015cgi}, and here we sketch how one could develop a bootstrap algorithm for the case of $\Nm = 4$ SYM, starting with the Ward identities.\footnote{We are grateful to Gabriel Bliard and Philine Van Vliet for enlightening discussions about this topic.}

\subsubsection{Kinematics}
\label{subsubsec:Kinematics}

For simplicity, we choose both the bulk and defect operators to be single-trace and half-BPS following the definitions \eqref{eq:SingleTraceHalfBPSOperators_Bulk} and \eqref{eq:SingleTraceHalfBPS_Defect}.
Conformal symmetry and supersymmetry constrain the bulk-defect-defect three-point functions to take the form
\begin{equation}
\vvev{ \Oh_{\Dh_1} (\uh_1, \tau_1) \Oh_{\Dh_2} (\uh_2, \tau_2) \Op_{\Delta_3} (u_3, x_3) } = \Km\, \Fm (\chi; \zeta)\,,
\label{eq:BulkDefectDefectForm}
\end{equation}
where we have defined the conformal prefactor
\begin{equation}
\Km := \frac{ ( \uh_1 \cdot u_3 )^{\Dh_1} ( \uh_2 \cdot u_3 )^{\Dh_2} ( u_3 \cdot \theta )^{\Delta_{312}} }{x_{13}^{2\Dh_1} x_{23}^{2\Dh_2} |x_3^\perp|_{\phantom{13}}^{2\Delta_{312}} }\,,
\label{eq:BulkDefectDefectConformalPrefactor}
\end{equation}
with
\begin{equation}
x_{i3}^2 := \tau_{i3}^2 + |x_3^\perp|^2\,,
\label{eq:Definitionxi3}
\end{equation}
and $\Delta_{312} := \Delta_3 - \Dh_1 - \Dh_2$.

The spacetime and $R$-symmetry cross-ratios are given by
\begin{equation}
\chi^2 := \frac{\tau_{12}^2 |x_3^\perp|^2}{x_{13}^2 x_{23}^2}\,, \quad \zeta^2 := \frac{ (\uh_1 \cdot \uh_2) (u_3 \cdot \theta)^2 }{ (\uh_1 \cdot u_3) (\uh_1 \cdot u_3) }\,.
\label{eq:DefinitionChiForBulkDefectDefect}
\end{equation}
Note that in \eqref{eq:BulkDefectDefectConformalPrefactor}, we have set $\Dh_1 \leq \Dh_2$ without loss of generality.
Interestingly, this family of correlators depends on only \textit{one} spacetime cross-ratio.
This is a significant simplification in comparison to the two-point functions of this chapter.

The correlator consists of $R$-symmetry channels as the two-point functions.
For the case $\Delta_3 \geq \Dh_1 + \Dh_2$, there are $\Dh_1+1$ $R$-symmetry channels, defined via
\begin{equation}
\Fm (\chi ; \zeta) = \sum_{j=0}^{\Dh_1} \left( - \frac{\zeta^2}{\chi^2} \right)^j F_j (\chi)\,.
\label{eq:BulkDefectDefectRSymmetryChannels}
\end{equation}

\subsubsection{Ward identities}
\label{subsubsec:WardIdentities}

We expect these correlators to satisfy superconformal Ward identities.
As a consequence, they should also contain a topological subsector when setting the $R$-symmetry and spacetime variables equal to each other, resulting in
\begin{equation}
\sum_{j=0}^{\Dh_1} F_j (\chi) = \text{constant}\,.
\label{eq:BulkDefectDefectTopological}
\end{equation}

Another important constraint is that the pinching of the two defect operators results in the OPE coefficient $b_{\Delta (\Dh_1 + \Dh_2)}$ after accounting for the different normalization constants:
\begin{align}
\vvev{ \Oh_{\Dh} (\uh_1, \tau_1) \Op_{\vphantom{\Dh}\Delta} (u_3, x_3) } & = \frac{\nh_1}{\sqrt{ \nh_{\Dh}} } \lim\limits_{2 \to 1}\, \vvev{ \Oh_{\Dh_1} (\uh_1, \tau_1) \Oh_{\Dh_2} (\uh_2, \tau_2) \Op_{\vphantom{\Dh_1}\Delta} (u_3, x_3) } \notag \\
& = b_{\Delta \Dh}\, \frac{ (\uh_1 \cdot u_3)^{\Dh} (u_3 \cdot \theta)^{\Delta - \Dh} }{ x_{13}^{2\Dh} |x_3^\perp|^{2(\Delta - \Dh)} }\,,
\label{eq:PinchingBulkDefectDefect}
\end{align}
with $\Dh = \Dh_1 + \Dh_2$.
These OPE coefficients are the ones that we computed in Section \ref{sec:TheMicrobootstrap} at large $N$ with the microbootstrap.

Two OPEs are relevant for these correlators, the bulk-defect OPE $\Op_\Delta \times \Oh_{\Dh}$ and the defect OPE $\Oh_{\Dh_1} \times \Oh_{\Dh_2}$.
The (bosonic) conformal blocks have been determined and can be found in \cite{Liendo:2015cgi}.
To the best of our knowledge, the superconformal Ward identities are not known explicitly but should take a form reminiscent of \eqref{eq:WardIdentities}.
They played an essential role in the bootstrap calculation performed in this chapter, and it would be interesting to determine them, for instance by using superspace techniques in the gist of \cite{Liendo:2016ymz}.
Another method is to compute a pool of perturbative correlators for different scaling dimensions, and experimentally determine the first-order differential equations that annihilate them.
In Chapter \ref{chapter:MultipointCorrelatorsInTheWilsonLineDefectCFT}, we see this method at play for determining the superconformal Ward identities for multipoint defect correlators.

\subsubsection{A perturbative result}
\label{subsubsec:APerturbativeResult}

To conclude this section, we give a weak-coupling analysis of the simplest bulk-defect-defect correlator, namely, $\vvev{\Oh_1 \Oh_1 \Op_2}$.
Perturbative computations are useful for gaining some intuition about the setup.
This correlator consists only of two $R$-symmetry channels:
\begin{equation}
\Fm (\chi; r) = F_0 (\chi) - \frac{\zeta^2}{\chi^2} F_1 (\chi)\,,
\label{eq:BulkDefectDefect112RSymmetryChannels}
\end{equation}
which according to \eqref{eq:BulkDefectDefectTopological} satisfy
\begin{equation}
F_0 (\chi) + F_1 (\chi) = \text{constant}\,.
\label{eq:BulkDefectDefect112Topological}
\end{equation}
Knowing one channel therefore fixes the correlator up to a constant.
Moreover, the pinching of the two $\Oh_1$ together should result in
\begin{align}
\vvev{\Oh_2 (\uh_1, \tau_1) \Op_2 (u_3, x_3)}  & = \frac{\nh_1}{\sqrt{\nh_2}} \lim\limits_{2 \to 1} \vvev{\Oh_1 (\uh_1, \tau_1) \Oh_1 (\uh_2, \tau_2) \Op_2 (u_3, x_3)} \notag \\
&= \left( \frac{\sqrt{2}}{N} + \frac{\lambda}{24 \sqrt{2} N} + \ldots \right) \frac{(\uh_1 \cdot u_3)}{x_{13}^2}\,,
\label{eq:Pinching112}
\end{align}
where we have used the result \eqref{eq:BulkDefectClosedFormLargeN2} obtained via the large $N$ microbootstrap.

The leading order consists of a single diagram:
\begin{equation}
\BulkDefectDefectLO\ = \left( \frac{\sqrt{2}}{N} + \frac{\lambda}{12 \sqrt{2} N} + \ldots \right) \frac{ (\uh_1 \cdot u_3) (\uh_2 \cdot u_3) }{x_{13}^2 x_{23}^2 }\,.
\end{equation}
We observe an agreement with the leading term in \eqref{eq:Pinching112}.
In particular, note that a term of order $\Op(\lambda)$ is already present because of the normalization constant $\nh_1$ (see \eqref{eq:NormalizationConstant_Defect}).

At next-to-leading order, the two channels are non-vanishing and given by the Feynman diagrams
\begin{equation}
\begin{split}
F_0^{(1)} &\sim \BulkDefectDefectNLOSEOne\ +\ \BulkDefectDefectNLOSETwo\ +\ \BulkDefectDefectNLOX\ +\ \BulkDefectDefectNLOH\ +\ \BulkDefectDefectNLOYOne\ +\ \BulkDefectDefectNLOYTwo\,, \\
F_1^{(1)} &\sim \BulkDefectDefectNLOFoneOne\ +\ \BulkDefectDefectNLOFoneTwo\ +\ \BulkDefectDefectNLOFoneThree\,,
\end{split}
\end{equation}
where as usual the interrupted lines connect to all the dots placed along the Wilson line.
These diagrams are easy to calculate in the conformal limit $\tau_2 \to \infty$, and we obtain the following non-trivial results:\footnote{Details for this calculation can be found in \cite{Barrat:2023ta4}.}
\begin{equation}
\begin{split}
F_0^{(1)} (\chi) &= \frac{\lambda}{48 \sqrt{2} \pi^2 N} ( \pi^2 + 12 \text{arccos}^2 \chi )\,, \\
F_1^{(1)} (\chi) &= - \frac{\lambda}{16 \sqrt{2} \pi^2 N} ( \pi^2 + 4 \text{arccos}^2 \chi )\,.
\end{split}
\end{equation}
As a first check, one can see that the pinching limit $\eqref{eq:Pinching112}$ is satisfied.
One can readily formulate a guess for the corresponding superconformal Ward identities:
\begin{equation}
\left. \biggl(
\pd_\chi
+
\pd_\zeta
\biggr)
\Fm(\chi; \zeta) \right|_{\zeta \to \chi}
=
0\,.
\end{equation}
It would be desirable to calculate more correlators in order to check this conjecture.
It should also be possible to derive these identities using superspace techniques.


\subsection{The Witten diagram bootstrap}
\label{subsec:TheWittenDiagramBootstrap}

In Section \ref{subsec:TheGeneralCase}, we provided a closed form for the highest-weight $R$-symmetry channel $F_0 (z, \zb)$ for arbitrary external operators $\Op_{\Delta_1}$ and $\Op_{\Delta_2}$.
We conclude this chapter with a brief review of the follow-up work \cite{Gimenez-Grau:2023fcy}, where a closed form was determined for all the channels, highlighting important connections between the dispersion relation, Witten diagrams, and Mellin space.

\subsubsection{Mellin space}
\label{subsubsec:MellinSpace}

The motivation for studying the Mellin space is the fact that in $\Nm = 4$ SYM without defects, explicit closed-form expressions for half-BPS operators take a particularly simple form \cite{Rastelli:2016nze,Rastelli:2017udc}:
\begin{equation}
M(s_{ij}) \sim \frac{6 \gamma_{13}^2 +2}{s_{13} - 2} + \frac{8 \gamma_{13}^2}{s_{13} - 4} + \frac{\gamma_{13} - 1}{s_{13} - 6} - \frac{15}{4} s_{13} + \frac{55}{2}\,,
\label{eq:MellinBulk}
\end{equation}
where $s_{13}$ and $\gamma_{13}$ are the Mellin variables.
The result is a function with simple poles located at integer values with the residues being at most a polynomial of degree $2$ in the other variable.
This follows from the constraints imposed by conformal symmetry on Mellin amplitudes.
A well-defined question is how to translate the defect correlators of Section \ref{sec:TwoPointFunctionsAtStrongCoupling} to Mellin space, where we expect the structure of the functions to simplify.

In supersymmetric bulk theories, this approach happened to be particularly successful when an explicit solution of the Ward identities could be found \cite{Meneghelli:2022gps,Lemos:2021azv}.
However, in our setup, the superconformal Ward identities \eqref{eq:WardIdentities} are more difficult to solve.
This can be understood from \cite{Dolan:2004mu}, since \eqref{eq:WardIdentities} can be recast such that they correspond to $3d$ Ward identities.
The solution is expressed with a curious differential operator of non-integer power, of which we can make sense by using Jack polynomials.
The formulation of the Ward identities in Mellin space of \cite{Gimenez-Grau:2023fcy} (see in particular (143)) might be useful to that effect.

\subsubsection{Closed form}
\label{subsubsec:ClosedForm}

In \cite{Gimenez-Grau:2023fcy}, a closed form for the two-point functions is given (see (132)) based on the bootstrap methods presented here and on explicit computations of Witten diagrams.
The result is given in terms of a finite sum of \textit{Polyakov-Regge blocks}, which can be interpreted as bulk-exchange Witten diagrams with a different normalization up to contact terms.

The author of \cite{Gimenez-Grau:2023fcy} then Mellin-transformed this result, motivated by the comments above.
This leads to drastic simplifications for the bulk-exchange and contact diagrams, which always reduce to \textit{rational} functions of the Mellin variables.
However, this is not the case for the defect-exchange diagrams, as in Mellin space they do not always reduce to rational functions.
For instance, we quote here the transformation of the correlator $\vvev{\Op_2 \Op_2}$ computed in \eqref{eq:FullCorrelator22} (see (136) in \cite{Gimenez-Grau:2023fcy}):
\begin{equation}
\begin{split}
\Mm (\delta, \rho; \sigma) =\ & \frac{\sqrt{\lambda}}{8 N^2}
\biggl\lbrace
\frac{(4 + \rho (1 - \sigma)) \sigma}{\delta-1}
- \frac{2 \rho - 4 (1- \sigma)}{\delta + \rho}
+ \frac{\rho}{\delta + \rho + 2} \\
&+ (1 - \sigma)^2
+ 2^{\delta + \rho + 2} \frac{\Gamma(- \delta - \rho)}{\Gamma^2 \left( \frac{2 - \delta - \rho}{2} \right)} \left(
\frac{\delta+2}{\delta+\rho+2} - \sigma
\right)
\biggr\rbrace\,.
\end{split}
\end{equation}
Here the $\Gamma$-functions arise from the defect-exchange amplitudes.

This result can be considered as the defect pendant of the elegant formulas for the four-point functions of half-BPS operators determined in \cite{Rastelli:2016nze,Rastelli:2017udc,Alday:2020dtb}.
In this case, the simplicity of the result was found to be related to a hidden ten-dimensional conformal symmetry \cite{Caron-Huot:2018kta,Alday:2020lbp,Caron-Huot:2021usw}.
Perhaps a similar story holds here, and if so it would be interesting to understand the origins of the symmetry.

\chapter[Multipoint correlators on the \\ Wilson-line defect CFT]{Multipoint correlators in the \\ Wilson-line defect CFT}
\chaptermark{Multipoint correlators on the Wilson-line defect CFT}
\label{chapter:MultipointCorrelatorsInTheWilsonLineDefectCFT}

So far our attention has been focused on operators localized in the bulk.
But defects themselves can be the source of local excitations.
We have seen in Section \ref{subsec:ConformalDefects} that it is indeed possible to construct correlation functions of \textit{defect} operators, and that, in the case of a line defect, their collection forms a \textit{one-dimensional} conformal field theory.

These models deserve to be investigated for several reasons.
One of them is that the resulting dynamics are expected to be considerably simpler compared to higher-$d$ CFTs.
For instance, four-point functions in these models depend on a \textit{single} spacetime cross-ratio, instead of two for $d \geq 2$.
The contrast becomes even more striking when considering multipoint correlators: while five-point functions rely only on \textit{two} cross-ratios in $1d$, this number grows to \textit{five} in higher-dimensional scenarios.

Another interesting aspect of these theories is that, despite being formulated in $1d$, they retain an intrinsic four-dimensional nature.\footnote{We work here at $d=4$, but this statement remains valid for other values of $d$.}
It is far from being obvious that these defect CFTs would form simpler models, as the fields propagate in the higher-dimensional ambient space, rendering the theory inherently \textit{non-local}.
This property is shared by all $1d$ CFTs, as the tracelessness of the stress-energy tensor leads to its complete vanishing.
Therefore, from this perspective, line defects might be the most natural way of implementing one-dimensional conformal symmetry.

Due to their simplicity, line defect CFTs serve as ideal testing grounds for the development of new methods in quantum field theory.
Indeed, many modern advancements, such as analytic functionals, initially emerged within the $1d$ framework before being extended to higher-dimensional setups \cite{Mazac:2016qev,Mazac:2018mdx,Mazac:2018ycv}.
Additionally, the conformal bootstrap proved to be a powerful tool in this context, allowing for significant progress in the computation of correlation functions at strong coupling \cite{Liendo:2018ukf,Ferrero:2021bsb}.
The case of the Wilson line is particularly favorable, as the (super)symmetry group $OSP(4^*|4)$ preserves a large portion of the group $PSU(2,2|4)$ found in $\Nm=4$ SYM.
This defect CFT is also believed to inherit integrability from its parent theory, while a holographic correspondence manifests itself in the form of an AdS$_2$/CFT$_1$ duality.

The study of multipoint correlators on the Wilson-line defect CFT has so far been limited to a special topological limit, using supersymmetric localization \cite{Giombi:2018qox}.
In this chapter, our target is to expand upon this knowledge by investigating higher-point functions in both the weak- and strong-coupling regimes.
These correlators capture a wealth of information about an infinite number of lower-point functions through the OPE.
Consequently, the study of multipoint correlators emerges as a significant focal point within the conformal bootstrap program, potentially serving as a powerful tool in advancing our understanding of conformal field theories in the future.

The content of this chapter is based on the publications \cite{Barrat:2021tpn,Barrat:2022eim}, while it also contains excerpts from the ongoing works \cite{Barrat:2023ta1,Barrat:2023ta2}.
Multiple minor results and insights that have not been presented in any of these publications are also included.

\section{An invitation}
\label{sec:AnInvitation2}

We initiate our study of correlation functions in the Wilson-line defect CFT with a comprehensive overview of the existing literature.
As mentioned above, considerable attention has been dedicated to the analysis of four-point functions, yielding significant progress in recent years.
Symmetry considerations have been essential in this success, with superconformal Ward identities assuming a pivotal role.
We review this progress here, as one of the aims of this chapter is to obtain multipoint Ward identities.
The techniques used for the strong-coupling computation of Section \ref{sec:MultipointCorrelatorsAtStrongCoupling} are presented in the context of the four-point function.
Generalities about multipoint correlators are also reviewed, and a summary of the literature for the lowest-lying unprotected operators is given.

\subsection{Four-point functions of half-BPS operators}
\label{subsec:FourPointFunctionsOfHalfBPSOperators}

The four-point functions of half-BPS operators serve as the first correlation functions that exhibit non-trivial kinematics.
These correlators manifest their kinematic dependence through a single spacetime cross-ratio, denoted as $\chi$, making them particularly simple in comparison to their higher-dimensional counterparts.
The reduced correlator $\Fm(\chi; r, s)$ is defined via the expression
\begin{equation}
\vev{ \Oh_{\Dh_1} (\uh_1, \tau_1) \Oh_{\Dh_2} (\uh_2, \tau_2) \Oh_{\Dh_3} (\uh_3, \tau_3) \Oh_{\Dh_4} (\uh_4, \tau_4) } = \Km\, \Fm(\chi; r, s)\,,
\label{eq:FourPointFunctions_ReducedCorrelator}
\end{equation}
with $\Km$ corresponding to a suitable (super)conformal prefactor.
The variables $r$ and $s$ are $R$-symmetry cross-ratios, encoding the dependence on the $SO(5)_R$ vectors $\uh$ (see \eqref{eq:uh}).
The explicit definition of $\chi$ can be found in \eqref{eq:SpacetimeCrossRatio1d}. 
It is important to note that the functions $\Km$ and $\Fm (\chi; r,s)$ depend on the external dimensions $\Dh$ of the operators, but for the sake of brevity, we refrain from explicitly expressing this dependence here.

Throughout this chapter, four-point functions play the role of reference points for the techniques that we develop and employ.
In this section, we give an overview of the results known at weak and strong coupling for these correlators.

\subsubsection{The four-point Ward identity\ }
\label{subsubsec:TheFourPointWardIdentity}

In the previous chapter, the superconformal Ward identities \eqref{eq:WardIdentities} played a fundamental role in bootstrapping correlation functions at strong coupling.
In a similar way, the reduced four-point functions $\Fm (\chi ; r, s)$ satisfy an analogous identity:
\begin{equation}
\left( \frac{1}{2} \pd_{\chi} + \alpha \pd_r - (1-\alpha) \pd_s \right)\, \Fm (\chi ; r,s) \raisebox{-1ex}{$\biggr |$}_{\raisebox{.75ex}{$\begin{subarray}{l} r = \alpha \chi\\
s = (1-\alpha)(1-\chi) \end{subarray}$}}=0\,,
\label{eq:FourPointFunctions_WardIdentities}
\end{equation}
where $\alpha$ is an arbitrary (real) parameter.

This differential equation may appear as very different from \eqref{eq:WardIdentities}, but in reality, the two Ward identities are \textit{identical}.
This can be seen by introducing alternative $R$-symmetry cross-ratios:
\begin{equation}
\zeta_1 \zeta_2 := r\,, \qquad (1-\zeta_1)(1-\zeta_2) := s\,.
\label{eq:AlternativeRSymmetryCrossRatios}
\end{equation}
Equation \eqref{eq:FourPointFunctions_WardIdentities} can then be decomposed into two independent equations:
\begin{equation}
\begin{split}
\left. \left( \frac{1}{2} \partial_\chi + \partial_{\zeta_1} \right) \Fm (\chi ; r,s) \right|_{\zeta_1 = \chi} &= 0\,, \\
\left.  \left( \frac{1}{2} \partial_\chi + \partial_{\zeta_2} \right) \Fm (\chi ; r, s) \right|_{\zeta_2 = \chi} &= 0\,.
\end{split}
\label{eq:AlternativeWardIdentities}
\end{equation}
These constraints are precisely the same as in \eqref{eq:WardIdentities} if the spacetime and $R$-symmetry variables are exchanged.
This connection should not come as a complete surprise, given that both setups possess $OSP(4^*|4)$ symmetry.
The analytic continuation between the spacetime and $R$-symmetry spaces was first identified in \cite{Liendo:2016ymz}, where a broader network of correspondences was discovered.
We come back to this point at the very end of this chapter.

The Ward identities \eqref{eq:FourPointFunctions_WardIdentities} turned out to be crucial in the bootstrap analysis of four-point functions at strong coupling \cite{Liendo:2018ukf,Ferrero:2021bsb}.
In Section \ref{sec:SuperconformalWardIdentities}, we expand upon these constraints and conjecture \textit{multipoint Ward identities}.
These new powerful identities provide a pathway to compute multipoint correlators in the strong-coupling regime through analytic bootstrap techniques.

\subsubsection{Weak coupling\ }
\label{subsubsec:WeakCoupling}

The study of multipoint correlators at weak coupling plays an essential role in obtaining these identities.
More generally, perturbative computations often offer valuable insights into the system at hand, and in lucky instances, they help to identify non-perturbative effects.

In terms of Feynman diagrams, the four-point functions of arbitrary half-BPS operators can generally be expressed as
\begin{equation}
\vev{ \Oh_{\Dh_1} \Oh_{\Dh_2} \Oh_{\Dh_3}  \Oh_{\Dh_4} } = D_{\text{bulk}} + D_{\text{bdry}}\,,
\label{eq:BulkAndBdryDiagrams}
\end{equation}
where we omitted the dependence on the variables $\uh$ and $\tau$ for brevity.
Of course, the functions on the right-hand side also depend on the external scaling dimensions $\Dh$.
At any order in perturbation theory, these correlators consist of two contributions: the \textit{bulk} and the \textit{boundary} diagrams.
It is straightforward to understand these contributions.
The bulk diagrams involve the exchange of virtual particles in the bulk spacetime only, while the boundary diagrams consist on top of that of interactions with the line defect.

Let us consider the four-point function of elementary scalars $\Oh_1$ as an example.
This correlator is particularly simple and consists of only \textit{three} $R$-symmetry channels:
\begin{equation}
\begin{split}
&\vev{ \Oh_{1} \Oh_{1} \Oh_{1}  \Oh_{1} } = \frac{(u_1 \cdot u_3)(u_2 \cdot u_4)}{\tau_{13}^2 \tau_{24}^2} \Fm (\chi; r,s)\,, \\
&\Fm (\chi; r, s) = F_0 (\chi) + \frac{r}{\chi^2} F_1 (\chi) + \frac{s}{(1-\chi)^2} F_2 (\chi)\,.
\end{split}
\label{eq:FourPointFunction_RSymmetryChannels}
\end{equation}
In the weak-coupling regime, these channels can be expanded perturbatively as
\begin{equation}
F_j (\chi) = \sum_{\ell=0}^{\infty} \lambda^{\ell} F_j^{(\ell)} (\chi)\,.
\label{eq:FourPointFunction}
\end{equation}

At the leading order, there are only bulk diagrams:\footnote{Recall that we consider only planar contributions.}
\begin{equation}
D^{(0)}_{\text{bulk}} = \DefectSSSSLOOne\ +\ \DefectSSSSLOTwo\,, \qquad D^{(0)}_{\text{bdry}} = 0\,.
\label{eq:FourPointFunction_LODiagrams}
\end{equation}
After unit-normalization, this corresponds to the $R$-symmetry channels
\begin{equation}
F^{(0)}_0 (\chi) = 0\,, \qquad F^{(0)}_1 (\chi) = F^{(0)}_2 (\chi) = 1\,.
\label{eq:FourPointFunction_LOChannels}
\end{equation}

The next-to-leading order contains both bulk and boundary diagrams.
The bulk contributions are given by
\begin{align}
D^{(1)}_{\text{bulk}} =\ & \DefectSSSSOneLoopSEOne\ +\ \DefectSSSSOneLoopSETwo\ +\ \DefectSSSSOneLoopSEThree\ + \DefectSSSSOneLoopSEFour \notag \\
&+\ \DefectSSSSOneLoopX\ +\ \DefectSSSSOneLoopHOne\ +\ \DefectSSSSOneLoopHTwo\,.
\label{eq:FourPointFunction_NLODiagramsBulk}
\end{align}
On the other hand, the boundary diagrams are
\begin{align}
\begin{split}
D^{(1)}_{\text{bdry}} =\ & \DefectSSSSOneLoopYOne\ +\ \DefectSSSSOneLoopYTwo \\
& +\ \DefectSSSSOneLoopYThree\ +\ \DefectSSSSOneLoopYFour\,.
\end{split}
\label{eq:FourPointFunction_NLODiagramsBdry}
\end{align}
Here the open gluon lines can connect to all the open vertices that are placed along the Wilson line, following the defect Feynman rules defined in \eqref{eq:DefectVertices_OnePoint} and adapted for gluon insertions.

The harmonic polylogarithms (HPLs) $H_{\vec{a}} := H_{\vec{a}} (\chi)$, introduced in Appendix \ref{subsec:HarmonicPolylogarithms}, provide a natural language for expressing the correlation functions of this defect CFT.
Using this basis of functions and the integrals provided in Appendix \ref{sec:FeynmanIntegrals}, we obtain the concise result \cite{Kiryu:2018phb}
\begin{equation}
\begin{split}
F^{(1)}_0 (\chi) &= - \frac{1}{8\pi^2} \frac{\ell^{(1)} (\chi)}{\chi}\,, \\
F^{(1)}_1 (\chi) &= \frac{1}{8 \pi^2} \left( \ell^{(1)} (\chi) + \ell^{(2)} (\chi) \right)\,, \\
F^{(1)}_2 (\chi) &= \frac{1}{8\pi^2 \chi} \left( \frac{\pi^2}{3} \chi + (1-\chi) \ell^{(1)} (\chi) - \chi \ell^{(2)} (\chi) \right)\,,
\end{split}
\label{eq:FourPointFunction_NLO}
\end{equation}
where the auxiliary functions $\ell^{(k)}$ have transcendentality weight $k$ and are defined as
\begin{equation}
\begin{split}
\ell^{(1)} (\chi) &:= \chi \left( \frac{H_0}{1 - \chi} - \frac{H_1}{\chi} \right)\,, \\
\ell^{(2)} (\chi) &:= 2 H_{0,1} - H_0  H_1\,.
\end{split}
\label{eq:TranscendentalFunctions}
\end{equation}
Note that, at order $\ell$, $F_1$ and $F_2$ have transcendentality $2\ell$ while $F_0$ is only of order $2\ell - 1$.

One may wonder looking at \eqref{eq:FourPointFunction_NLO} whether there exists an algebraic relation between the channels $F_0$, $F_1$, and $F_2$, given that they are so alike.
The answer is affirmative and can be found in the Ward identity \eqref{eq:FourPointFunctions_WardIdentities}.
The differential equation can be solved explicitly, and the reduced correlator assumes the following elegant form \cite{Liendo:2016ymz,Liendo:2018ukf}:
\begin{equation}
\Fm (\chi; r , s) = \Fds + \Dds f(\chi)\,,
\label{eq:FourPointFunction_SolutionWI}
\end{equation}
where we have introduced the differential operator
\begin{equation}
\Dds := \frac{1}{\chi^2} \left( 1 + r \frac{2-\chi}{\chi} - s \right) - \frac{1}{\chi^2} \left( (\chi + r)(1-\chi) - s \chi \right) \pd_\chi\,.
\label{eq:DifferentialOperatorWI}
\end{equation}
$\Fds$ denotes a constant term that depends solely on the coupling constant $g$.\footnote{It also depends on $N$ if it is kept finite.}
In the large $N$ limit, the topological sector of the four-point function was determined to be \cite{Giombi:2017cqn}
\begin{equation}
\Fds = F_0 (\chi) + F_1 (\chi) + F_2 (\chi) = \frac{3 I_1 (\sqrt{\lambda}) I_3 (\sqrt{\lambda})}{I_2^2 (\sqrt{\lambda})}\,,
\label{eq:FourPoint_TopologicalSector}
\end{equation}
where $I_\alpha (\sqrt{\lambda})$ represents the modified Bessel function of the first kind defined in \eqref{eq:BesselFunction}, and where $\lambda$ is the 't Hooft coupling \eqref{eq:tHooftCoupling}.
This parallels the discussion of Section \ref{subsec:TwoPointFunctionsOfHalfBPSOperators}: correlation functions that satisfy the Ward identity \eqref{eq:FourPointFunctions_WardIdentities} contain a \textit{topological sector}.
This is an interesting topic in its own right, that will be discussed further in Section \ref{subsec:CorrelatorsOfHalfBPSOperators} in the context of multipoint correlators.
In the weak coupling regime, $\Fds$ can be expanded as
\begin{equation}
\Fds = 2 + \frac{\lambda}{24} - \frac{\lambda^2}{480} + \Op(\lambda^3)\,.
\label{eq:TopologicalSectorAtWeakCoupling}
\end{equation}

In the formulation \eqref{eq:FourPointFunction_SolutionWI}, the kinematic dependence of $\vev{\Oh_1 \Oh_1 \Oh_1 \Oh_1}$ is elegantly captured by a \textit{single} function $f(\chi)$.
The leading-order result \eqref{eq:FourPointFunction_LOChannels} becomes
\begin{equation}
f^{(0)} (\chi) = \frac{\chi}{1-\chi} (1 - 2\chi)\,, \qquad \Fds^{(0)} = 2\,,
\label{eq:FourPointFunction_LittlefLO}
\end{equation}
while at next-to-leading order we reformulate \eqref{eq:FourPointFunction_NLO} as
\begin{equation}
f^{(1)} (\chi) = - \frac{1}{8\pi^2} \frac{\chi}{1-\chi} \left( \frac{2 \pi^2}{3} \chi - 2 \ell^{(2)} (\chi) \right)\,, \qquad \Fds^{(1)} = \frac{1}{24}\,.
\label{eq:FourPointFunction_LittlefNLO}
\end{equation}
We emphasize that the $R$-symmetry channels \eqref{eq:FourPointFunction_LOChannels} and \eqref{eq:FourPointFunction_NLO} are \textit{fully contained} in \eqref{eq:FourPointFunction_LittlefLO} and  \eqref{eq:FourPointFunction_LittlefNLO}.

The four-point function at next-to-next-to-leading order was investigated in \cite{Cavaglia:2022qpg}, using an innovative framework known as \textit{bootstrability}, which combines the techniques of bootstrap and integrability for computing correlation functions (see also \cite{Cavaglia:2021bnz,Cavaglia:2022yvv,Niarchos:2023lot}).
The resulting expression is
\begin{align}
f^{(2)} (\chi) =\ & \frac{1}{64 \pi^4} \frac{\chi}{1-\chi}
\biggl\lbrace
\frac{2 \pi^4}{15} \chi
- 3 \zeta_3 H_1
+ \frac{\pi^2}{3} ( H_{0,1} - H_{1,0} + H_{1,1} ) \notag \\
&+2 ( H_{0,0,1,0} - H_{0,1,0,0} + H_{0,0,1,1} - H_{1,1,0,0} + H_{1,0,1,1} - H_{1,1,0,1} ) \notag \\
&- H_{0,1,0,1} - H_{0,1,1,0} + H_{1,0,0,1} + H_{1,0,1,0}
\biggr\rbrace\,,
\label{eq:FourPointFunction_LittlefNNLO}
\end{align}
with $\Fds^{(2)} = - \frac{1}{480}$.
It is worth noting that the function $f^{(\ell)}(\chi)$ exhibits \textit{homogeneous transcendentality}, and that the HPL functions appear without a rational function coefficient, apart from an overall prefactor.
In Section \ref{subsec:TheFourPointFunctionAtNNLO}, we recompute this correlator using perturbative techniques as well as symmetry constraints, with the aim of extending this result towards four-point functions of arbitrary operators.
Note that all the functions $f^{(\ell)} (\chi)$ presented above are plotted in Figure \ref{fig:PlotsLittlef}.

We conclude this review of the weak coupling by noting that a general formula for the four-point function of arbitrary single-trace half-BPS operators was determined up to the next-to-leading order in \cite{Kiryu:2018phb}, using perturbative techniques.
This result was used to propose a generalization of the hexagonalization approach, in analogy to the bulk story \cite{Fleury:2016ykk}.

\subsubsection[\\ Strong coupling\ ]{Strong coupling}
\label{subsubsec:StrongCoupling2}

\begin{figure}
\centering
\begin{subfigure}{.5\textwidth}
  \raggedright
  \includegraphics[width=1\linewidth]{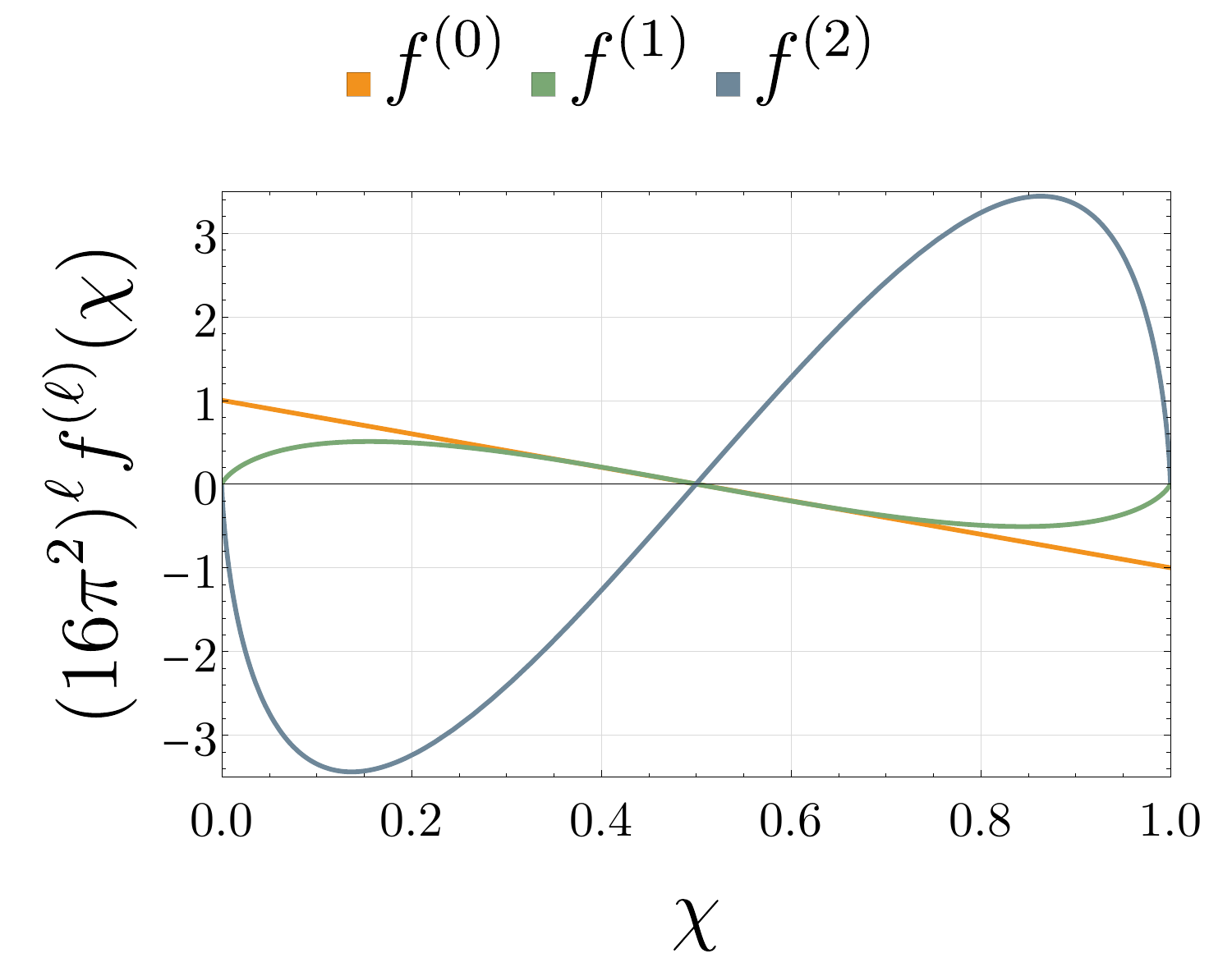}
\end{subfigure}%
\begin{subfigure}{.5\textwidth}
  \raggedleft
  \includegraphics[width=1\linewidth]{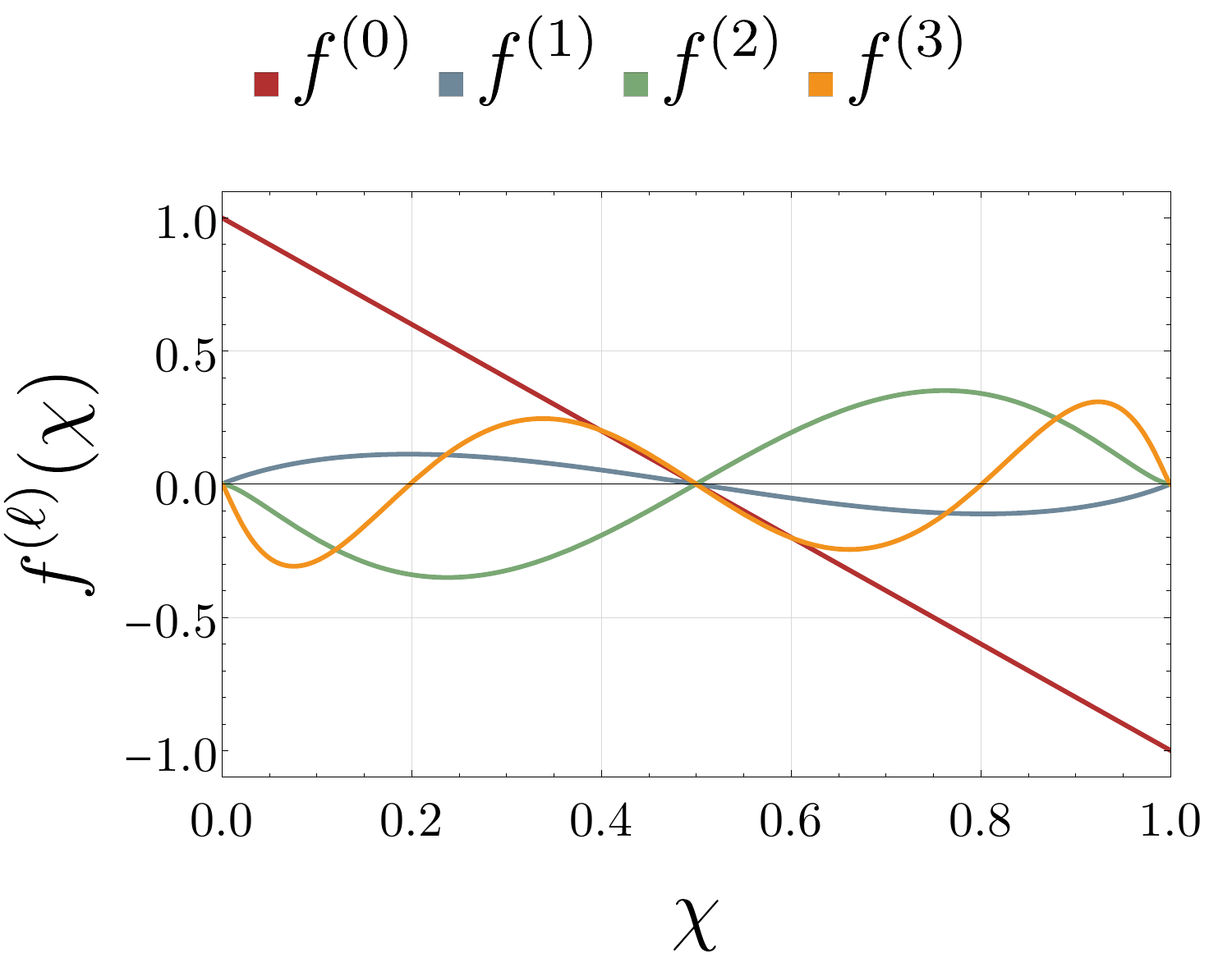}
\end{subfigure}
\caption{The two plots show the functions $f^{(\ell)}$ at weak (left) and strong (right) coupling. 
In the weak-coupling regime, the functions follow an expansion in powers of $\lambda$, while on the strong coupling side, they go as $1/\lambda^{\ell/2}$.
The analytical results can be found in the text (see \eqref{eq:FourPointFunction_LittlefLO}-\eqref{eq:FourPointFunction_LittlefNNLO} for the weak coupling and \eqref{eq:FourPointFunction_Littlef_Strong} for the strong coupling).
}
\label{fig:PlotsLittlef}
\end{figure}

The four-point function $\vev{\Oh_1 \Oh_1 \Oh_1 \Oh_1}$ has also been studied at strong coupling, using the powerful tools of the analytic bootstrap \cite{Liendo:2018ukf,Ferrero:2021bsb}.
Significant progress has been made in this regime, with perturbative results available up to four orders.
This is an exciting tour-de-force, as performing such a computation using the AdS$_2$ effective action of \cite{Giombi:2017cqn} would pose tremendous challenges.

Here the analytic bootstrap works differently than in Chapter \ref{chapter:BootstrappingHolographicDefectCorrelators}.
The basic idea, following \cite{Ferrero:2021bsb}, is to construct an ansatz based on HPLs for the solution of the Ward identity $f(\chi)$ and to impose consistency conditions coming from the crossing equation and the OPE.

\begingroup
\allowdisplaybreaks

The perturbative expansion takes the form
\begin{equation}
f(\chi) = \sum_{\ell = 0}^{\infty} \frac{f^{(\ell)} (\chi)}{\lambda^{\ell/2}}\,,
\end{equation}
and the first few orders are
\begin{equation}
\begin{split}
f^{(0)} =\ & \frac{\chi (1-2\chi)}{1-\chi}\,, \\
f^{(1)} =\ & \frac{\chi}{1-\chi} \biggl(
2\chi - 1
+ \frac{\chi^2 (2-\chi)}{1-\chi} \log \chi \\
&- \frac{(1+\chi)(1-\chi)^2}{\chi} \log (1-\chi)
\biggr)\,, \\
\ldots
\end{split}
\label{eq:FourPointFunction_Littlef_Strong}
\end{equation}
The next orders $f^{(2)} (\chi)$ and $f^{(3)} (\chi)$ are lengthy, so we do not give them explicitly here.
They can be found in \cite{Ferrero:2021bsb}.

\endgroup

Note that the expressions in \eqref{eq:FourPointFunction_Littlef_Strong} have transcendentality of maximum weight $\ell$, while at weak coupling the basis of functions was growing in steps of $2$.
This seems to be a typical behavior at strong coupling, which was also observed in Chapter \ref{chapter:BootstrappingHolographicDefectCorrelators}.

These functions are all plotted in Figure \ref{fig:PlotsLittlef}.
In Section \ref{sec:MultipointCorrelatorsAtStrongCoupling}, we aim at extending these results to \textit{five-point} functions, which depend on two cross-ratios $\chi_1$ and $\chi_2$, using multipoint Ward identities in conjunction with the analytic bootstrap.
The next section is dedicated to the review of multipoint correlators of half-BPS operators.

\subsection{Correlators of half-BPS operators}
\label{subsec:CorrelatorsOfHalfBPSOperators}

In higher-dimensional setups, the complexity of correlators escalates rapidly, as the number of cross-ratios is $n(n-3)/2$, with $n$ the number of external operators.
The situation improves significantly in $1d$, where correlators depend on $n-3$ variables only.
This makes the Wilson-line defect CFT a perfect laboratory for the study of multipoint correlators.
In general, many aspects familiar from four-point functions extend straightforwardly to multipoint correlators, where they lead to new interesting insights.

\subsubsection{General considerations\ }
\label{subsubsec:GeneralConsiderations}

The $n$-point functions of half-BPS operators $\Oh_{\Dh}$ are given by
\begin{equation}
\vev{\Oh_{\Dh_1} \ldots \Oh_{\Dh_n}} = \Km\, \Fm ( \lbrace \chi; r,s,t \rbrace )\,,
\label{eq:MultipointCorrelatorsHalfBPS}
\end{equation}
where $\lbrace \chi; r,s,t \rbrace$ represents the set of spacetime and $R$-symmetry cross-ratios that the reduced correlator depends on.
As always, the functions $\Km$ and $\Fm$ depend on the scaling dimensions of the external operators, although we omit to specify it explicitly to streamline the notation.

\begin{figure}
\centering
\RSymmetryChannelsIllustration
\caption{Example of an application of the recursion relation \eqref{eq:FormulaRSymmetryChannels} for the five-point function $\vev{\Oh_1 \Oh_1 \Oh_1 \Oh_1 \Oh_2}$, which consists of \normalfont{six} $R$-symmetry channels.}
\label{fig:RSymmetryChannelsIllustrations}
\end{figure}

Like four-point functions, multipoint correlators admit a decomposition in $R$-symmetry channels.
These channels are determined by considering all possible combinations of the vectors $\uh_1,\ldots,\uh_n$.
The number of channels depends on the scaling dimensions of the external operators and can be computed using the following recursion relation:
\begin{equation}
R_{\Dh_1, \ldots, \Dh_n} = R_{\Dh_1-1, \Dh_2-1, \ldots, \Dh_n} + \ldots + R_{\Dh_1-1,\Dh_2, \ldots, \Dh_n-1}\,,
\label{eq:FormulaRSymmetryChannels}
\end{equation}
where the function $R$ is defined by the properties
\begin{equation}
\begin{split}
& R_{\Dh_1, \ldots, \Dh_i, 0, \Dh_{i+1}, \ldots, \Dh_n} = R_{\Dh_1, \ldots, \Dh_i, \Dh_{i+1}, \ldots, \Dh_n}\,, \\
& R_{\Dh} = 0\,, \\
& R_{\Dh_1, \Dh_2} = \delta_{\Dh_1 \Dh_2}\,.
\end{split}
\label{eq:FormulaRSymmetryChannels_StartingValues}
\end{equation}
A special case arises when the $n$ external operators are elementary scalars:
\begin{equation}
R_{1,1, \ldots, 1} = (n-1)!!\,.
\label{eq:FormulaRSymmetryChannels_SpecialCase}
\end{equation}
A simple example is given in Figure \ref{fig:RSymmetryChannelsIllustrations} to gain some intuition about the recursion relation \eqref{eq:FormulaRSymmetryChannels}.

\subsubsection{Combs and snowflakes\ }
\label{subsubsec:CombsAndSnoflakes}

Much like their four-point counterparts, multipoint correlators can be expanded in terms of conformal blocks, leading to a complex network of OPE coefficients that remains largely unexplored.
In this chapter, block expansions serve two purposes: firstly, to validate the consistency of our results in the weak-coupling regime, and secondly, to perform a bootstrap calculation of a five-point function on the strong-coupling side.
We provide here an introduction to the one-dimensional multipoint block expansion, specifically focusing on the OPE channels called \textit{comb} and \textit{snowflake}.

\begin{figure}
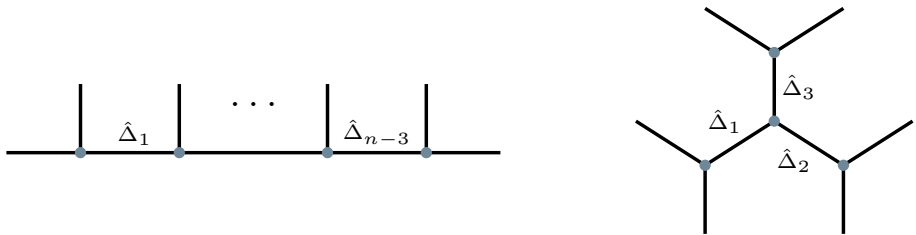

\centering
\begin{subfigure}{.55\textwidth}
  \centering
  \CombChannel
\end{subfigure}%
\begin{subfigure}{.45\textwidth}
  \centering
  \SnowflakeChannel
\end{subfigure}
\caption{Two OPE channels for multipoint correlators are depicted here.
The outer lines represent external operators, while the $\Dh_k$ refer to the exchanged (or internal) operators.
The left figure shows the \textit{comb} channel, where the OPEs are taken between an external and an internal operator, apart from the two extremities.
On the right, the \textit{snowflake} channel for a six-point function is shown, in which case the OPEs are taken pairwise between the external operators.
Both options lead to different OPE coefficients, as shown in \eqref{eq:OPECoefficientComb} and \eqref{eq:OPECoefficientSnowflake}.}
\label{fig:CombsAndSnowflakes}
\end{figure}

The comb channel consists of taking systematically the OPE between an external operator and an internal one, as depicted in Figure \ref{fig:CombsAndSnowflakes}.
While four-point blocks in $d=1$ have been known for some time \cite{Ferrara:1974ny}, it is only recently that this work was extended to higher-point functions \cite{Rosenhaus:2018zqn}.
The derivation of five-point blocks for arbitrary dimensions $d$ was achieved in \cite{Fortin:2022grf}, while the understanding of higher-point cases remains limited.

It is convenient to introduce a new set of cross-ratios to analyze the comb channel:
\begin{equation}
\eta_i := \frac{ \tau_{i(i+1)} \tau_{(i+2)(i+3)} }{ \tau_{i(i+2)} \tau_{(i+1)(i+3)} }\,.
\label{eq:CrossRatiosComb}
\end{equation}
For simplicity, we focus on correlators of $n$ elementary scalars $\phi^I$.
In this case, the block expansion takes the form
\begin{align}
F^{I_1 \ldots I_n} ( \lbrace \eta \rbrace ) = \sum_{\Dh_1, \ldots, \Dh_{n-3}} C^{I_1 \ldots I_n}_{\Dh_1 \ldots \Dh_{n-3}} f^{\phantom{I_1}}_{\Dh_1 \ldots \Dh_{n-3}} ( \lbrace \eta \rbrace )\,.
\label{eq:BlockExpansionComb}
\end{align}
This expression holds independently for each $R$-symmetry channel.
We introduced the notation
\begin{equation}
C^{I_1 \ldots I_n}_{\Dh_1 \ldots \Dh_{n-3}} := \lambda_{ \phi^{I_1} \phi^{I_2} \Dh_1 } \lambda_{\Dh_1 \phi^{I_3} \Dh_2 } \ldots \lambda_{\Dh_{n-4} \phi^{I_{n-2}} \Dh_{n-3} } \lambda_{\Dh_{n-3} \phi^{I_{n-1}} \phi^{I_n}}\,,
\label{eq:OPECoefficientComb}
\end{equation}
which corresponds to the structure of the OPE coefficients of Figure \ref{fig:CombsAndSnowflakes}.
The blocks $f_{\Dh_1 \ldots \Dh_n}$ are defined in \eqref{eq:ConformalBlocks_Multipoint_Comb}.
Note that it is easy to extend \eqref{eq:BlockExpansionComb} to operators of arbitrary scaling dimensions, using the results of \cite{Rosenhaus:2018zqn}.
The case of \textit{super}conformal blocks is considered in Section \ref{sec:MultipointCorrelatorsAtStrongCoupling}.

\bigskip

As the number of external points increases, new OPE channels emerge, offering additional insights into the CFT data.
For instance, at $n=6$, a distinct OPE limit known as the snowflake channel appears.
In this case, OPEs are performed pairwise between the external operators, allowing the OPE coefficient in the center to consist of operators that differ from the external ones (see Figure \ref{fig:CombsAndSnowflakes}).
This is in contrast to the comb channel, where at least one external operator is always involved in the three-point functions of \eqref{eq:OPECoefficientComb}.
The snowflake channel enables access to OPE coefficients beyond the reach of the comb channel, including spinning operators.
As an example, in the case of $\vev{\Oh_1 \Oh_1 \Oh_1 \Oh_1 \Oh_1 \Oh_1}$, one of the OPE coefficients is $\lambda_{\phi^6 \phi^6 \phi^6}$ in the snowflake expansion, which cannot appear in the comb channel unless at least one external operator is $\phi^6$ itself.

Since the OPE limits are different in the snowflake channel, we introduce yet another set of cross-ratios specialized for six-point functions:
\begin{equation}
z_1 := \frac{\tau_{12} \tau_{46}}{\tau_{16} \tau_{24}}\,, \quad z_2 := \frac{\tau_{26} \tau_{34}}{\tau_{23} \tau_{46}}\,, \quad z_3 := \frac{\tau_{24} \tau_{56}}{\tau_{26} \tau_{45}}\,.
\label{eq:CrossRatiosSnowflake}
\end{equation}
Once again, we focus on the scenario where the correlator consists solely of elementary scalars $\phi^I$.
The corresponding block expansion takes the form
\begin{align}
F^{I_1 \ldots I_6} ( \lbrace z \rbrace ) = \sum_{\Dh_1, \Dh_2, \Dh_3} C^{I_1 \ldots I_6}_{\Dh_1 \Dh_2 \Dh_3} f^{\phantom{I_1}}_{\Dh_1 \Dh_2 \Dh_3} ( \lbrace z \rbrace )\,,
\label{eq:BlockExpansionSnowflake}
\end{align}
where the OPE coefficients are defined as
\begin{equation}
C^{I_1 \ldots I_6}_{\Dh_1 \Dh_2 \Dh_3} := \lambda_{\phi^{I_1} \phi^{I_2} \Dh_1} \lambda_{\phi^{I_3} \phi^{I_4} \Dh_2}
\lambda_{\phi^{I_5} \phi^{I_6} \Dh_3}
\lambda_{\Dh_1 \Dh_2 \Dh_3}\,.
\label{eq:OPECoefficientSnowflake}
\end{equation}
The blocks of the snowflake channel are discussed in \eqref{eq:ConformalBlocks_Multipoint_Snowflake}.

It is important to note that, as the number of points in the correlator increases, the number of OPE channels grows accordingly.
In this study, our attention is directed toward the comb and snowflake channels exclusively.
Yet it would be interesting to extend the analysis to the structures appearing in $n >6$ correlators.
For instance, in \cite{Ferrero:2021bsb} the inclusion of four-point functions involving \textit{unprotected} operators proved necessary to disentangle the CFT data.
Perhaps a similar result can be obtained by considering higher-point functions of \textit{protected} operators combined with a systematic study of the OPE channels.

\subsubsection[\\ The topological sector\enspace]{The topological sector}
\label{subsubsec:TheTopologicalSector}

In Chapter \ref{chapter:BootstrappingHolographicDefectCorrelators}, we encountered a topological subsector in the two-point functions of half-BPS operators in the presence of the Wilson-line defect.
Similarly, the multipoint correlation functions of defect operators also possess a closed topological sector.
We already encountered in \eqref{eq:FourPoint_TopologicalSector} the topological sector $\Fds$ for the case $n=4$.
The dependence on the kinematic variables disappears when the $R$-symmetry cross-ratios are set equal to their corresponding spacetime cross-ratios.
In terms of $R$-symmetry channels, this translates into the relation
\begin{equation}
\Fds := \sum_j F_j (\chi) = \text{constant}\,,
\label{eq:MultipointTopologicalSector}
\end{equation}
which is already familiar from \eqref{eq:BulkDefectDefect112Topological} and \eqref{eq:FourPoint_TopologicalSector}.

Correlators of this kind have been extensively studied using matrix models in \cite{Giombi:2018qox}, from which exact results can be derived.
An example of interest for our purposes is the topological sector of the five-point function $\vev{\Oh_1 \Oh_1 \Oh_1 \Oh_1 \Oh_2}$, which can be expressed (at large $N$) as
\begin{equation}
\Fds = \frac{6\, \Ids_2^2}{\lambda\, \Ids_1^2} \frac{2 (\Ids_1 - 2) (\Ids_1 + 28) + \lambda ( 2 \Ids_1 - 19 ) }{ \sqrt{ 3 \lambda - (\Ids_1 - 2)(\Ids_1 + 10) } }\,.
\label{eq:FivePointTopologicalSector}
\end{equation}
Here we have defined for compactness the help function
\begin{equation}
\Ids_a := \frac{\sqrt{\lambda} I_0 (\sqrt{\lambda})}{ I_a (\sqrt{\lambda}) }\,.
\label{eq:TopologicalSector_HelpFunction}
\end{equation}
{\emergencystretch 3em
This function follows the following expansion in the weak- and strong-coupling regimes:
}
\begin{equation}
\begin{split}
\Fds &\overset{\lambda \sim 0}{\sim} 3 + \frac{7 \lambda}{48} + \Op (\lambda^2)\,, \\
\Fds &\overset{\lambda \gg 1}{\sim} 6 \sqrt{2} - \frac{33}{\sqrt{2}} \frac{1}{\sqrt{\lambda}} + \frac{189}{8 \sqrt{2}} \frac{1}{\lambda} + \Op(\lambda^{-3/2})\,.
\end{split}
\label{eq:FivePointTopologicalSector_Expansions}
\end{equation}
These results will be used in Sections \ref{subsec:Applications} and \ref{sec:MultipointCorrelatorsAtStrongCoupling} as checks for our computations of this five-point function in the weak- and strong-coupling regimes.

For the six-point function $\vev{\Oh_1 \Oh_1 \Oh_1 \Oh_1 \Oh_1 \Oh_1}$, the topological sector takes an even simpler form:
\begin{equation}
\Fds = \frac{15\, \Ids_2^3}{\lambda^{3/2}\, \Ids_1^3} ( (\lambda+24) \Ids_1 - 8(\lambda+6) )\,,
\label{eq:SixPointTopologicalSector}
\end{equation}
for which the weak- and strong-coupling expansions are
\begin{equation}
\begin{split}
\Fds &\overset{\lambda \sim 0}{\sim} 5 + \frac{\lambda}{4} + \Op (\lambda^2)\,, \\
\Fds &\overset{\lambda \gg 1}{\sim} 15 - \frac{45}{\sqrt{\lambda}} +\frac{45}{\lambda} + \Op(\lambda^{-3/2})\,.
\end{split}
\label{eq:SixPointTopologicalSector_Expansions}
\end{equation}

\subsection{Correlators of unprotected operators}
\label{subsec:CorrelatorsOfUnprotectedOperators}

Up until now, our focus has been primarily on correlation functions involving protected operators.
However, to fully understand the scalar sector, it is essential to include correlators of \textit{unprotected} operators.
In the following, we consider the lowest-lying unprotected operators.

\subsubsection{The curious case of $\phi^6$\enspace }
\label{subsubsec:TheCuriousCaseOfPhi6}

\begin{figure}
\centering
\begin{subfigure}{.5\textwidth}
  \raggedright
  \includegraphics[width=1\linewidth]{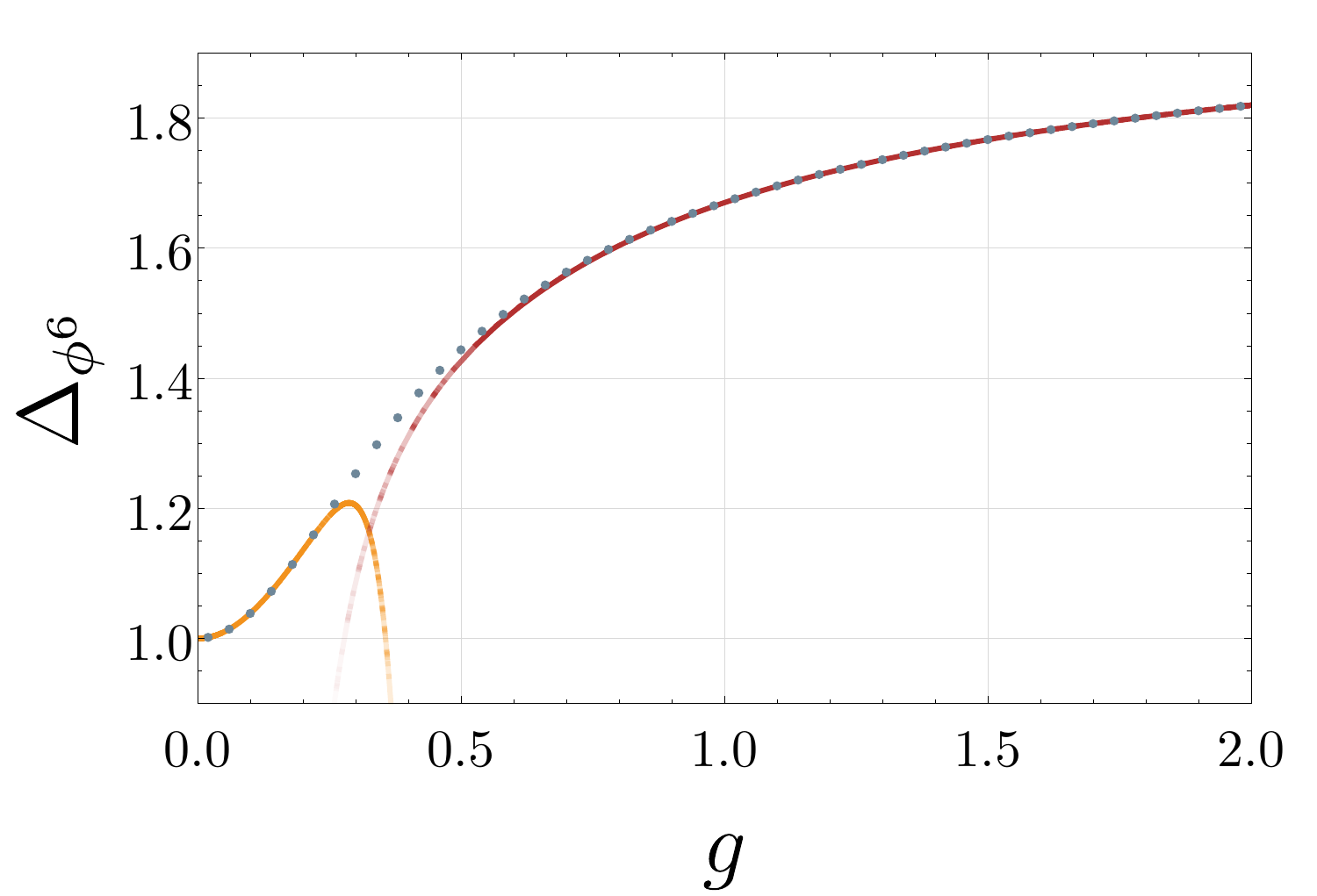}
\end{subfigure}%
\begin{subfigure}{.5\textwidth}
  \raggedleft
  \includegraphics[width=1\linewidth]{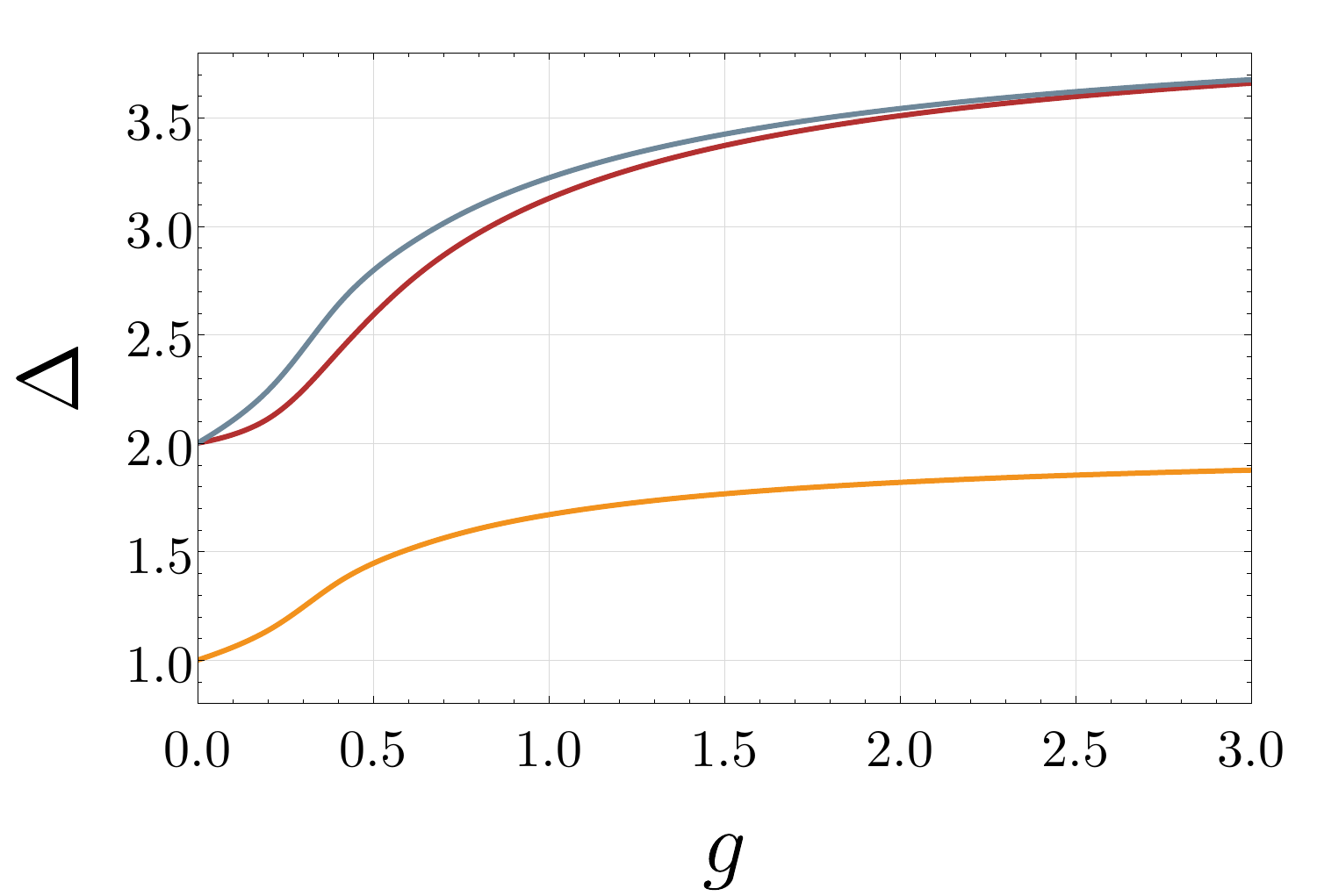}
\end{subfigure}
\caption{The two plots show the scaling dimensions of low-lying operators from weak to strong coupling.
On the left, the scaling dimension of $\phi^6$ is shown, with the blue dots displaying the numerical data obtained from integrability in \cite{Grabner:2020nis}, while the orange and red curves respectively show the analytical results \eqref{eq:Dh6_WeakCoupling} and \eqref{eq:Dh6_StrongCoupling}.
The right plot shows the integrability results of \cite{Cavaglia:2021bnz} for $\phi^6$ (the lowest curve) and of $\Oh_{\pm}$ (the two curves starting at $\Delta_0 = 2$), defined in \eqref{eq:L2Operators}.
This illustrates the fact that operators can be degenerate at the classical level.
}
\label{fig:Phi6AndLength2}
\end{figure}

The sixth scalar $\phi^6$ exhibits a distinct behavior to the other elementary scalars.\footnote{To be precise, the operator should be understood as $\frac{1}{\sqrt{\nh_{\smash{\phi^6}}}} \Wl [ \phi^6 ]$, similarly to \eqref{eq:SingleTraceHalfBPS_Defect}.}
This is a consequence of the definition of the Wilson line \eqref{eq:SUSYWilsonLoop}, which couples to $\phi^6$.
This spoils the half-BPS condition, and its scaling dimension receives corrections at the quantum level.

At weak coupling, the anomalous dimensions of $\phi^6$ were computed in \cite{Alday:2007hr,Grabner:2020nis,Cavaglia:2021bnz} using perturbation theory and integrability techniques, yielding the expression
\begin{align}
\Dh_{\smash{\phi^6}} =&\,
1
+ 4 g^2
- 16 g^4
+ \left( 128 - \frac{56 \pi^4}{45} \right) g^6 \notag \\
&+ \left( - 1280 + \frac{128}{3} \pi^2 + 126 \zeta_3 - 160 \zeta_5 - \frac{64}{3} \pi^2 \zeta_3 + \frac{272}{135} \pi^6 \right)  g^8 \notag \\
&+ \left( 14336 - \frac{2560}{3} \pi^2 - 1024 \zeta_3 - \frac{896}{45} \pi^4 - 640 \zeta_5 + \frac{512}{3} \pi^2 \zeta_3 \right. \notag \\
&\left. - \frac{64}{2835} \pi^6 - 384 \zeta_3^2 + 2688 \zeta_7 + \frac{448}{3} \pi^2 \zeta_5 + \frac{64}{3} \zeta_3 - \frac{7328}{2835} \pi^8 \right) g^{10} \notag \\
&+ \Op(g^{12})\,.
\label{eq:Dh6_WeakCoupling}
\end{align}
Note that here, the coupling $g$ is defined as
\begin{equation}
g := \frac{\sqrt{\lambda}}{4\pi}\,.
\label{eq:Definition_g}
\end{equation}

\begingroup
\allowdisplaybreaks
\endgroup

At strong coupling, $\phi^6$ flows to a \textit{two-particle} state with $\Delta_0 = 2$ in the dual AdS$_2$ theory.
This provides an interesting illustration of the AdS/CFT correspondence, where the states on the two ends of the duality are easy to interpret.
In $\Nm=4$ SYM, the operator with the lowest-lying classical dimension ($\Delta_0 = 2$) on the CFT side is the \textit{Konishi} operator, for which the anomalous dimensions are known up to $11$ loops \cite{Arutyunov:2010gb,Balog:2010xa,Gromov:2011cx,Eden:2012fe,Leurent:2012ab,Bajnok:2012bz,Marboe:2014gma}.
But contrary to $\phi^6$, the scaling dimension of the Konishi flows to $\infty$ in the strong-coupling regime \cite{Vallilo:2011fj}.

The corrections of $\Dh_{\smash{\phi^6}}$ have been computed at strong coupling using various methods, from AdS computations to bootstrap \cite{Giombi:2017cqn,Ferrero:2021bsb,Marboe:2014gma}.
The current state-of-the-art for the analytical expression is
\begin{equation}
\begin{split}
\Dh_{\smash{\phi^6}} =\ &2 - \frac{5}{\sqrt{\lambda}} + \frac{295}{24} \frac{1}{\lambda} - \frac{305}{16} \frac{1}{\lambda^{3/2}} \\
&+ \left( \frac{351845}{13824} - \frac{75}{2} \zeta_3 \right) \frac{1}{\lambda^2} + \Op(\lambda^{-5/2})\,.
\end{split}
\label{eq:Dh6_StrongCoupling}
\end{equation}
This result was derived from the expansion in conformal blocks of the four-point function $\vev{\Oh_1 \Oh_1 \Oh_1 \Oh_1}$.

The integrable structure, believed to be inherited by the Wilson-line defect CFT from its parent theory $\Nm=4$ SYM, offers a powerful tool for calculating the scaling dimension of $\phi^6$.
This made it possible to determine $\Delta_{\smash{\phi^6}}$ numerically for \textit{finite} values of the coupling constant.
This is depicted in Figure \ref{fig:Phi6AndLength2}, complementing the analytical expressions given in \eqref{eq:Dh6_WeakCoupling} and \eqref{eq:Dh6_StrongCoupling}.

One can also study the normalization constant of $\phi^6$.
Little is known about it, apart from the fact that it takes the form
\begin{equation}
\nh_{\smash{\phi^6}} = \nh_1 + \Lambda(\lambda)\,.
\label{eq:phi6_NormalizationConstant}
\end{equation}
The first term $\nh_1$ corresponds to the normalization constant of $\Oh_1$, which is known exactly and provided in \eqref{eq:NormalizationConstant_Defect}.
The second term $\Lambda (\lambda)$ encodes the contractions between $\phi^6$ and the Wilson line.
Perturbatively, it can be seen as the collection of boundary diagrams of the type
\begin{equation*}
\DefectSSNLOSeven\,.
\end{equation*}
In this type of diagram, the propagators connect the external operators to the Wilson line through the scalar term in \eqref{eq:SUSYWilsonLoop}.
The defect Feynman rules can be found in \eqref{eq:DefectVertices_OnePoint}.
Note furthermore that $\nh_{\smash{\phi^6}}$ and $\nh_1$ coincide at the leading order, i.e., $\Lambda(0) = 0$.
In Section \ref{subsec:Applications}, we compute the first non-trivial correction.

We can consider correlation functions of $\phi^6$ operators.
As discussed above \eqref{eq:CrossRatiosSnowflake}, the OPE coefficient $\lambda_{\phi^6 \phi^6 \phi^6}$ can be obtained through the snowflake channel of $\vev{\Oh_1 \Oh_1 \Oh_1 \Oh_1 \Oh_1 \Oh_1}$.
Alternatively, we can consider the expansion in blocks of the four-point function $\vev{\phi^6 \phi^6 \phi^6 \phi^6}$.
The block expansion for multipoint correlators of $\phi^6$ follows the expressions given in \eqref{eq:BlockExpansionComb} and \eqref{eq:BlockExpansionSnowflake}.
At the end of Section \ref{subsec:Applications}, we examine the CFT data associated with these correlators.

\subsubsection{More unprotected operators\enspace}
\label{subsubsec:MoreUnprotectedOperators}

\begingroup
\allowdisplaybreaks

There exists several \textit{composite} operators of length $2$ that can be built on top of the protected half-BPS operator $\Oh_2$ defined through \eqref{eq:SingleTraceHalfBPS_Defect}.
These operators can be constructed perturbatively as orthogonal eigenstates of the dilatation operator.
At the next-to-leading order, this was performed in \cite{Correa:2018fgz}, yielding the following operators:
\begin{equation}
\begin{split}
\Oh_A^{ij} &:= \frac{1}{\sqrt{\nh_{\smash{\Oh_A^{ij}}}}} \Wl \bigl[ \phi^i \phi^j - \phi^j \phi^i \bigr]\,, \\
\Oh_A^{i} &:= \frac{1}{\sqrt{\nh_{\smash{\Oh_A^{i}}}}} \Wl \bigl[ \phi^6 \phi^i - \phi^i \phi^6 \bigr]\,, \\
\Oh_S^{i} &:= \frac{1}{\sqrt{\nh_{\smash{\Oh_S^{i}}}}} \Wl \bigl[ \phi^6 \phi^i + \phi^i \phi^6 \bigr]\,, \\
\Oh_{\pm} &:= \frac{1}{\sqrt{\nh_{\smash{\Oh_{\pm}}}}} \Wl \bigl[ \phi^i \phi^i \pm \sqrt{5}\, \phi^6 \phi^6 \bigr]\,,
\end{split}
\label{eq:L2Operators}
\end{equation}
where the dependence on $\tau$ has been suppressed for the sake of brevity.
Note that $A$ denotes \textit{antisymmetric}, while $S$ represents \textit{symmetric}.
It should be noted that the last two states $\Oh_{\pm}$ are not multiplicatively renormalizable, i.e., the relative factors between the two terms may receive a correction at higher orders in the coupling constant.

The scaling dimensions of these operators have also been computed at weak coupling in \cite{Correa:2018fgz}.\footnote{In \cite{Correa:2018fgz}, a more general Wilson loop is considered, and to obtain the supersymmetric case one should set $\zeta = 1$.}
The results are
\begin{equation}
\begin{split}
\Dh_{\smash{\Oh_A^{ij}}} &= 2 + \frac{\lambda}{4\pi^2} + \Op(\lambda^2)\,, \\
\Dh_{\smash{\Oh_A^{i}}} &= 2 + \frac{3 \lambda}{8\pi^2} + \Op(\lambda^2)\,, \\
\Dh_{\smash{\Oh_S^{i}}} &= 2 + \frac{\lambda}{8\pi^2} + \Op(\lambda^2)\,, \\
\Dh_{\smash{\Oh_{\pm}}} &= 2 + \frac{5 \pm \sqrt{5}}{16\pi^2} \lambda + \Op(\lambda^2)\,. 
\end{split}
\label{eq:DhL2Operators}
\end{equation} 
The operators $\Oh_{\pm}$ have been investigated at finite $\lambda$ using an innovative mix of bootstrap and integrability techniques \cite{Cavaglia:2021bnz}.
The flow of their scaling dimensions from weak to strong coupling is depicted in Figure \ref{fig:Phi6AndLength2}.

\endgroup

Composite operators of higher lengths can be constructed using the same approach.
However, scalar operators can also be formed by combining elementary fermions, such as bilinears $\psib \psi$.
While such operators are not treated here, we discuss the potential extension of our computation to fermions in Section \ref{subsec:MultipointCorrelators of Fermions}.

\section[Correlation functions of scalars at weak coupling]{Correlation functions of scalars  at weak \\ coupling}
\sectionmark{Correlation functions of scalars at weak coupling}
\label{sec:CorrelationFunctionsOfScalarsAtWeakCoupling}

We begin our investigation of multipoint correlators by focusing on the weak coupling regime $\lambda \sim 0$.
Our main goal is to obtain correlators of scalar single-trace operators at the next-to-leading order.
This is achieved by developing an algorithmic approach based on Feynman diagrams.
The central ingredient is a set of recursion relations for the elementary scalar fields $\phi^I$, which can then be used for computing arbitrary correlators of operators made of them.
Applications are presented for correlators featuring two to six external operators.
We consider examples for the cases where the operators are elementary or composite, protected or unprotected.
Consistency checks can be performed on these results, and in particular, we check that the expansions in conformal blocks produce consistent results.
To conclude, we provide an explicit computation of the four-point function of the elementary fields $\Oh_1$ at the next-to-next-to-leading order.

This section presents the results of \cite{Barrat:2021tpn,Barrat:2022eim,Barrat:2023ta1}.

\subsection{Correlators of fundamental scalars}
\label{subsec:CorrelatorsOfFundamentalScalars}

The most elementary class of correlators that one can study consists of single scalar insertions $\phi^{I=1,\ldots, 6}$ on the Wilson line, which we define as
\begin{equation}
A^{I_1 \ldots I_n} := \vev{\phi^{I_1} (\tau_1) \ldots \phi^{I_n} (\tau_n)}\,.
\label{eq:CorrelatorOfFundamentalInsertions-Definition}
\end{equation}
Correlation functions of such operators are interesting since they can be pinched into lower-point functions of composite operators, following \eqref{eq:Pinching}.
Various quantities can be extracted from these correlators, including the four-point functions of half-BPS operators presented in \cite{Kiryu:2018phb}, as well as correlators of unprotected operators such as the ones given in \eqref{eq:L2Operators}.

The computation of these correlation functions can be efficiently performed by employing diagrammatic recursion relations specifically designed for the elementary scalars.
In our analysis, we focus on the planar limit $N \to \infty$ for simplicity.
Note however that the formulas presented in this study can be extended to the finite $N$ case if desired.

The correlators \eqref{eq:CorrelatorOfFundamentalInsertions-Definition} are deliberately kept \textit{not} unit-normalized, which here means that the operators are defined without the inclusion of the normalization constants $\nh_{\Dh}$.
This choice is made for facilitating the process of pinching operators together, as it is more convenient to perform this operation before applying unit-normalization.

At weak coupling, these correlators follow the splitting \eqref{eq:BulkAndBdryDiagrams} between bulk and boundary diagrams:
\begin{equation}
A^{I_1 \ldots I_n} = A^{I_1 \ldots I_n}_{\text{bulk}} + A^{I_1 \ldots I_n}_{\text{bdry}}\,.
\label{eq:A_BulkAndBdry}
\end{equation}
The recursion relations presented below also follow this structure.

\subsubsection{Leading order\enspace}
\label{subsubsec:LeadingOrder3}

Recursion relations are well-known for their ability to generate results efficiently, as we have already seen from the microbootstrap relations presented in Section \ref{sec:TheMicrobootstrap}.
We introduce here a recursion relation specifically designed to compute the leading-order contribution to the correlation functions defined in \eqref{eq:CorrelatorOfFundamentalInsertions-Definition}.

This formula requires to account for two distinct scenarios, determined by whether the number of insertions of $\phi^6$ is \textit{even} or \textit{odd}.
In the case of an even number of insertions, the correlation functions \eqref{eq:CorrelatorOfFundamentalInsertions-Definition} consist exclusively of bulk diagrams and can be recursively obtained at $\mathcal{O}(\lambda^{n/2})$ through
\begin{equation}
A_{\text{LO}}^{I_1 \ldots I_n} = \sum_{j=0}^{\frac{n}{2}-1}\ \RecursionLOEven\,.
\label{eq:RecursionEvenLO_Diagrams}
\end{equation}
Here, the terms $\TreeLevelInsertion$ represent the leading-order correlation functions $A_{\text{LO}}$ with the appropriate number of external scalars.
These terms correspond to lower-point insertions, making the expression recursive.
This expression was already given in (3.1) of \cite{di1997meander} for solving a related mathematical riddle.
Indeed, the problem of connecting external points without crossing belongs to the field of meanders and arch statistics.

Using \eqref{eq:RecursionEvenLO_Diagrams}, we can generate arbitrary correlation functions of scalar fields by specifying the $R$-symmetry indices, as long as the number of $\phi^6$ insertions remains even.
The recursion process begins with the following starting values, which are the vacuum expectation value of the Wilson line and the two-point functions:
\begin{equation}
A_{\text{LO}} = 1\,, \qquad A_{\text{LO}}^{I_1 I_2} = \frac{\lambda}{8\pi^2} \frac{\delta^{I_1 I_2}}{\tau_{12}^2}\,.
\label{eq:RecursionEvenLO_StartingValues}
\end{equation}

Notice that, more generally, at order $\ell$ in perturbation theory the starting values are given by the $2(\ell+1)$-point function.

We now turn our attention to the case where an odd number of $\phi^6$ insertions appears in the correlators.\footnote{It is important to note that for these correlators to be non-vanishing, the total number of protected scalars $\phi^i$ must remain even.}
This implies that at least one of them must be connected to the Wilson line, resulting in boundary diagrams exclusively.
The recursion relation for correlators with an odd number of $\phi^6$ insertions is of order $\lambda^{(n+1)/2}$ and can be expressed as
\begin{equation}
\begin{split}
A_{\text{LO}}^{I_1 \ldots I_n} =\ & \sum_{i=1}^n \sum_{j=0}^{\frac{i-1}{2}}\ \RecursionLOOddOne\ + \sum_{i=1}^n  \sum_{j=i/2}^{n/2}\ \RecursionLOOddTwo \\
&+ \sum_{i=1}^{n-1}  \sum_{\substack{j = i+2\\\text{step }2}}^{n}\ \RecursionLOOddThree\,.
\end{split}
\label{eq:RecursionOddLO_Diagrams}
\end{equation}
In the above formula, a propagator ending on the Wilson line with a colored dot indicates that it should be integrated from the previous propagator to the next, following the defect Feynman rules listed in \eqref{subsec:TheWilsonLineDefectCFT}.
These integrals, referred to as \textit{$U$-integrals}, can be evaluated using the results provided in Appendix \ref{sec:DefectIntegrals}.
It should be noted that the middle $\TreeLevelInsertion$ piece in the second line of \eqref{eq:RecursionOddLO_Diagrams} involves an odd number of $\phi^6$ fields.

To complete the recursion relation, we need an additional starting value corresponding to the vanishing of $1d$ one-point functions:
\begin{equation}
A_{\text{LO}}^I = 0\,.
\label{eq:RecursionOddLO_StartingValues}
\end{equation}

\subsubsection{Next-to-leading order\enspace}
\label{subsubsec:NextToLeadingOrder3}

At the next-to-leading order, the calculation becomes more involved as it consists of both bulk and boundary diagrams.
While we have already listed the diagrams for the four-point function in \eqref{eq:FourPointFunction_NLODiagramsBulk} and \eqref{eq:FourPointFunction_NLODiagramsBdry}, we extend here this expression to the case of multipoint correlators.

The computation of bulk diagrams at next-to-leading order follows a recursive expression that accounts for the various possibilities of connecting the scalar insertions.
This expression is given by
\begin{equation}
\begin{split}
A_{\text{NLO,bulk}}^{I_1 \ldots I_n} =\ & \sum_{i=1}^{n-3} \sum_{j=i+1}^{n-2} \sum_{k=j+1}^{n-1} \sum_{l=k+1}^{n}\ \RecursionNLOBulk \\
&+ \sum_{i=1}^{n-3} \sum_{j=i+3}^{n}\ \RecursionNLOBulkBridge\,,
\end{split}
\label{eq:RecursionNLOBulk_Diagrams}
\end{equation}
where the first line corresponds to the sum of four-point bulk diagrams given in \eqref{eq:FourPointFunction_NLODiagramsBulk}, with the appropriate index replacements.
The bulk diagrams involve the $X$-, $Y$-, and $F$-integrals, which are solved in Appendix \ref{app:Integrals}:
\begin{equation}
\begin{split}
\RecursionNLOBulkFour\ =\ & \frac{\lambda^3}{8}
\Bigl\lbrace
(2 \delta^{ik} \delta^{jl} - \delta^{ij} \delta^{kl} - \delta^{il} \delta^{jk}) X_{ijkl} \\[-.5em]
& + \delta^{ij} \delta^{kl}\, \bigl[ I_{ij} I_{kl} F_{ij,kl} - 4 ( I_{ij} Y_{kkl} + I_{kl} Y_{iij} ) \bigr] \\
& + \delta^{il} \delta^{jk}\, \bigl[ I_{il} I_{jk} F_{il,jk} - 4 ( I_{il} Y_{jjk} + I_{jk} Y_{iil} ) \bigr]
\Bigr\rbrace\,.
\end{split}
\label{eq:RecursionNLOBulk_StartingValues}
\end{equation}
These integrals are known individually, and their explicit forms can be used to compute the bulk diagrams for any correlation function $\vvev{\Oh_1 \ldots \Oh_1}$.

The formula presented in \eqref{eq:RecursionNLOBulk_Diagrams} exhibits a recursive structure in two ways.
Firstly, the tree-level insertions $\TreeLevelInsertion$ themselves follow a recursive pattern.
This means that the lower-point correlators at leading order serve as building blocks for the computation of higher-point correlators.
Additionally, the second line of the formula in \eqref{eq:RecursionNLOBulk_Diagrams} exhibits a recursive structure at the next-to-leading order level, meaning that this term requires the insertion of bulk diagrams with lower points.

A similar expression can be formulated for the boundary diagrams.
Here it is useful to first consider the diagrams involving scalars that do not couple to the line, i.e., $\phi^{i = 1, \ldots, 5}$.
In this case, the building blocks are the same as the diagrams given in \eqref{eq:FourPointFunction_NLODiagramsBdry}:
\begin{align}
A_{\text{NLO,bdry}}^{i_1 \ldots i_n} =\ & \sum_{i=1}^{n-1} \sum_{j=i+1}^{n} \biggl( \sum_{k=0}^{i-1}\ \RecursionNLOBdryOne \notag \\
&+ \sum_{k=i}^{j-1}\ \RecursionNLOBdryTwo\ + \sum_{k=j}^{n+1}\ \RecursionNLOBdryThree\ \biggr) \notag \\
&+ \sum_{i=1}^{n-3} \sum_{j=i+3}^{n}\ \RecursionNLOBdryBridge\,.
\label{eq:RecursionNLOBdry_Diagrams}
\end{align}
Using integration by parts as well as the identities listed in Appendix \ref{app:Integrals}, it is easy to show that the three first terms of \eqref{eq:RecursionNLOBdry_Diagrams} can be expressed as
\begin{equation}
\begin{split}
& \frac{\lambda^2}{4} \sum_{i=1}^{n-1} \sum_{j=i+1}^{n} \biggl\lbrace
\delta^{ij} ( T_{ij,0(n+1)} + 4 Y_{iij} )\, A_{\text{LO}}^{1\,, \ldots\,, i-1} A_{\text{LO}}^{i+1\,, \ldots\,, j-1} A_{\text{LO}}^{j+1\,, \ldots\,, n} \\
&- \delta^{ij} \frac{\lambda}{2}
\biggl( \sum_{k=1}^{i-2} \sum_{l=k+1}^{i-1}\,  \delta^{kl} I_{kl} T_{ij,kl} \Am_{klij}
+ \sum_{k=i+1}^{j-2} \sum_{l=k+1}^{j-1}\,  \delta^{kl} I_{kl} T_{ij,kl} \Am_{iklj} \\
&+ \sum_{k=j+1}^{n-1} \sum_{l=k+1}^{n}\,  \delta^{kl} I_{kl} T_{ij,kl} \Am_{ijkl}
\biggr)
\biggr\rbrace\,,
\end{split}
\label{eq:RecursionNLOBdry_StartingValues}
\end{equation}
where we have defined for compactness
\begin{equation}
\Am_{ijkl} := A_{\text{LO}}^{1\,, \ldots\,, i-1} A_{\text{LO}}^{i+1\,, \ldots\,, j-1} A_{\text{LO}}^{j+1\,, \ldots\,, k-1} A_{\text{LO}}^{k+1\,, \ldots\,, l-1} A_{\text{LO}}^{l+1\,, \ldots\,, n}\,,
\label{eq:ACompact_Definition}
\end{equation}
while the $T$-integrals of \eqref{eq:RecursionNLOBdry_StartingValues} are defined in \eqref{eq:T_Definition}.
The different orderings give rise to different integrals, and the solutions for the relevant cases are listed in \eqref{eq:T_Results}.

The cancellation of divergences between the expressions in \eqref{eq:RecursionNLOBulk_Diagrams} and \eqref{eq:RecursionNLOBdry_Diagrams} ensures that the resulting correlators are finite at the next-to-leading order.
Restricting ourselves to half-BPS operators, the correlators $A_{\text{NLO}}^{i_1 \ldots i_m}$ of arbitrary length $m$ can be used to generate $n$-point functions of \textit{composite} operators:
\begin{equation*}
\left. \vvev{ \Oh_{\Dh_1} (\uh_1, \tau_1) \ldots \Oh_{\Dh_n} (\uh_n, \tau_n) } \right|_{\sum_{i=1}^n \Dh_i = m}\,.
\end{equation*}
This can be done by pinching operators together following \eqref{eq:Pinching}.

It is worth noting that the expressions given in \eqref{eq:RecursionNLOBulk_Diagrams} and \eqref{eq:RecursionNLOBdry_Diagrams} can also be employed to generate correlators of specific unprotected operators, including the antisymmetric scalar $\Oh_A^{ij}$ defined in \eqref{eq:L2Operators}.
However, if the intention is to study correlators involving the unprotected scalar $\phi^6$, it becomes necessary to extend this framework through the inclusion of additional boundary diagrams.
Some of these terms exhibit divergences, necessitating the operators to be renormalized to obtain finite correlators that are physically meaningful.

We focus on the case where an even number of $\phi^6$ is inserted since the odd case corresponds to a higher order in perturbation theory.
The extended recursion relation is lengthy and will not be provided explicitly here.
However its structure is easy to understand: it consists of $15$ terms that encompass all the possible ways to \textit{double} the diagrams of \eqref{eq:RecursionOddLO_Diagrams}, either through the incorporation of $U$-propagators or by introducing a regular propagator enclosing an odd number of leading-order insertions $\TreeLevelInsertion$.
For instance, one such diagram is
\begin{equation}
\RecursionNLOBdryExtra\,.
\label{eq:RecursionNLOExtended_ExampleDiagram}
\end{equation}
The full formal expression can be found in (3.6) and (B.1) of \cite{Barrat:2022eim}.

It is worth mentioning that an ancillary \textsc{Mathematica} notebook is attached to \cite{Barrat:2022eim}, where all the recursion relations presented in this section have been implemented, such that arbitrary correlation functions can easily be generated.

\subsection{Applications}
\label{subsec:Applications}

The recursion relations that have been developed in the preceding sections provide a powerful tool for efficiently computing correlation functions in the Wilson-line defect CFT.
In this section, we demonstrate their practical application by computing various examples.

We begin our analysis by correlators with fixed kinematics.
These include two- and three-point functions as well as extremal correlators.
As a by-product of these computations, we determine the anomalous dimensions and normalization constants of the length $L=1,2$ operators presented in Section \ref{subsec:CorrelatorsOfUnprotectedOperators}.
Our algorithm is then employed to compute four-, five- and six-point functions involving both elementary and composite operators.
We verify the consistency of our results by comparing them to the expansions in conformal blocks.
In the case of six-point functions, we analyze two distinct OPE channels: the comb and snowflake channels of Section \ref{subsec:CorrelatorsOfHalfBPSOperators}, providing strong checks of our results.

Note that, from now on, we reinstate the unit-normalization of the operators, following \eqref{eq:SingleTraceHalfBPS_Defect}.

\subsubsection{Kinematically-fixed correlators\enspace}
\label{subsubsec:KinematicallyFixedCorrelators}

We now use the recursion relations for computing two-point functions of scalar operators.
For operators of lengths $1$ and $2$, we reproduce the scaling dimensions up to the next-to-leading order, which can be directly compared to the values provided in \eqref{eq:Dh6_WeakCoupling} and \eqref{eq:DhL2Operators}.
It is worth noting that, to the best of our knowledge, the normalization constants of the unprotected operators have not been given explicitly yet.
The approach presented in this section can be readily extended to operators of higher length consisting of the elementary fields $\phi^I$, with computational resources being the only limiting factor.

{\emergencystretch 3em
Let us begin with the half-BPS operators, which can be calculated straightforwardly by applying the recursion relations \eqref{eq:RecursionEvenLO_Diagrams}, \eqref{eq:RecursionNLOBulk_Diagrams} and \eqref{eq:RecursionNLOBdry_Diagrams}.
}
We obtain
\begin{equation}
\vev{\Oh_{\Dh} (u_1, \tau_1) \Oh_{\Dh} (u_2, \tau_2)}_\text{LO} = \frac{1}{\nh_{\Dh}} \frac{(u_1 \cdot u_2)^{\Dh}}{\tau_{12}^{2\Dh}}  \frac{\lambda^2}{64 \pi^4} \left( 1 - \frac{\lambda}{24} + \Op(\lambda^2) \right)\,.
\label{eq:TwoPointOh}
\end{equation}
The absence of logs is a sanity check that the scaling dimensions are protected and thus independent of $\lambda$.
These results match the ones of \cite{Giombi:2018qox}.

We now consider the simplest case of an unprotected operator, which is the scalar $\phi^6$ introduced in Section \ref{subsec:CorrelatorsOfUnprotectedOperators}.
Following the renormalization procedure outlined in Section \ref{subsec:FromWeakToStrongCoupling}, we find that the normalization constant takes the form anticipated in \eqref{eq:phi6_NormalizationConstant}, with
\begin{equation}
\Lambda(\lambda) = - \frac{\lambda^2}{32 \pi^4} + \Om(\lambda^3)\,,
\label{eq:Lambda_NLO}
\end{equation}
while the anomalous dimension agrees with \eqref{eq:Dh6_WeakCoupling}.

\begingroup
\allowdisplaybreaks

One can proceed similarly for the unprotected operators of \eqref{eq:L2Operators}.
The computation of the anomalous dimensions is in perfect agreement with \eqref{eq:DhL2Operators}, while the normalization constants are found to be
\begin{equation}
\begin{split}
\nh_{\Oh_A^{ij}} &= - \frac{\lambda^2}{32\pi^4} \left( 1 - \frac{\lambda}{24} + \Op(\lambda^2) \right)\,, \\
\nh_{\Oh_A^{i}} &= - \frac{\lambda^2}{32\pi^4} \left( 1 - \frac{\lambda}{24} \frac{6 + \pi^2}{\pi^2} + \Op(\lambda^2) \right)\,, \\
\nh_{\Oh_S^{i}} &= - \frac{\lambda^2}{32\pi^4} \left( 1 - \frac{\lambda}{24} + \Op(\lambda^2) \right)\,, \\
\nh_{\Oh_{\pm}} &= - \frac{5 \lambda^2}{32\pi^4} \left( 1 - \frac{\lambda}{24} \left(1 - \frac{9}{2 \pi^2} (1 \pm \sqrt{5} ) \right) + \Op(\lambda^2) \right)\,.
\end{split}
\label{eq:Length2_NormalizationConstants}
\end{equation}

\endgroup

\bigskip

Three-point functions are also fixed kinematically by conformal symmetry.
For three half-BPS operators, the OPE coefficients were computed in \cite{Giombi:2018qox,Kiryu:2018phb} and found to be
\begin{equation}
\lambda_{\Dh_1 \Dh_2 \Dh_3} = 1 + \frac{\lambda}{16} \left( 1 - \frac{2}{3} ( \delta_{\Dh_1,\Dh_2 + \Dh_3} + \delta_{\Dh_2,\Dh_1 + \Dh_3} + \delta_{\Dh_3,\Dh_1 + \Dh_2} ) \right) + \Op(\lambda^2)\,.
\label{eq:ThreePointFunctions_HalfBPS}
\end{equation}
Our formulas \eqref{eq:RecursionEvenLO_Diagrams}, \eqref{eq:RecursionNLOBulk_Diagrams} and \eqref{eq:RecursionNLOBdry_Diagrams} perfectly reproduce this result, providing further support for their validity.

Similarly, we use the recursion relations for an odd number of operators of length $1$ to compute the leading-order contributions to $\vev{\Oh_1 \Oh_1 \phi^6}$ and $\vev{\phi^6 \phi^6 \phi^6}$.
We find
\begin{equation}
\lambda_{\Oh_1 \Oh_1 \phi^6} = - \frac{\sqrt{\lambda}}{2\sqrt{2} \pi} + \Op(\lambda^{3/2})\,, \qquad \lambda_{\phi^6 \phi^6 \phi^6} = - \frac{3 \sqrt{\lambda}}{2 \sqrt{2} \pi} + \Op(\lambda^{3/2})\,.
\label{eq:ThreePointFunctions}
\end{equation}
The first result can be compared to the existing literature \cite{Cooke:2017qgm,Cavaglia:2021bnz}, while the second result appears to be new.
These three-point coefficients play a crucial role in the comb and snowflake OPE channels, where they serve as consistency checks for the higher-point functions of the subsequent sections.

\bigskip

We conclude this analysis with a certain class of kinematically-fixed correlators.
It has been demonstrated in the bulk theory that certain correlators of half-BPS operators are protected against quantum corrections \cite{Bianchi:1999ie}.
These correlators are referred to as \textit{extremal}, due to the following property of the scaling dimensions of the external operators:
\begin{equation}
\Delta_n = \sum_{i=1}^{n-1} \Delta_i\,.
\label{eq:ExtremalCorrelators_ScalingDims}
\end{equation}

Extremal correlators also exist in the Wilson-line defect CFT, in a similar fashion.
In contrast to the bulk theory, the next-to-leading order in the defect CFT is non-zero here.\footnote{We note that this result differs from (84) in \cite{Kiryu:2018phb}, which we believe does not apply to the extremal case.}
We observe however that the kinematics of these correlators are trivial.

Using the recursion relations \eqref{eq:RecursionEvenLO_Diagrams}, \eqref{eq:RecursionNLOBulk_Diagrams} and \eqref{eq:RecursionNLOBdry_Diagrams}, we find
\begin{equation}
\left. \Fm \right|_{\Dh_n = \Dh_1 + \ldots + \Dh_{n-1}} = \frac{\lambda^{\Dh_n}}{2^{3\Dh_n} \pi^{2 \Dh_n}} \left( 1 - \frac{\lambda}{24} + \Op(\lambda^2) \right)\,.
\label{eq:ExtremalCorrelators_Results}
\end{equation}
Remarkably, all the $X$-, $F$- and $T$-integrals containing transcendental functions cancel each other, leading to this simple expression.
\eqref{eq:ExtremalCorrelators_Results} can be checked against the localization results of \cite{Giombi:2018qox}, with which it agrees perfectly.

\subsubsection[Four-point \\ functions\enspace]{Four-point functions}
\label{subsubsec:FourPointFunctions}

Up until now, our focus has been on correlation functions with fixed kinematics as a consequence of conformal symmetry.
We now use our recursion relations to compute four-point functions, which exhibit non-trivial kinematics characterized by a single spacetime cross-ratio.

The four-point function of half-BPS scalars $\Oh_1$ was already given in Section \ref{subsec:FourPointFunctionsOfHalfBPSOperators}, with its leading-order and next-to-leading order expressions given respectively in \eqref{eq:FourPointFunction_LOChannels} and \eqref{eq:FourPointFunction_NLO}.
It is almost by definition that our recursion relations \eqref{eq:RecursionEvenLO_Diagrams}, \eqref{eq:RecursionNLOBulk_Diagrams} and \eqref{eq:RecursionNLOBdry_Diagrams} reproduce these results correctly.
Another important check is that they match the four-point functions of general single-trace half-BPS operators derived in \cite{Kiryu:2018phb}.
In the following, we also show that our formulas go beyond the case of protected operators, and we consider as concrete examples the correlators $\vvev{\phi^6 \phi^6 \phi^6 \phi^6}$ and $\vvev{\Oh_1 \Oh_1 \Oh^k_A \Oh^l_A}$.
The power of the recursion relations lies in the fact that they can be used to compute arbitrary correlators of scalars $\phi^I$, leading to a great variety of correlators.

For the case of four $\phi^6$ operators, the reduced correlator is given by
\begin{equation}
\vev{\phi^6 \phi^6 \phi^6 \phi^6} = \frac{F (\chi)}{\raisebox{-.6ex}{$\tau_{13}^{\smash{2\Dh_{\phi^6}}} \tau_{24}^{\smash{2\Dh_{\phi^6}}}$}}\,,
\label{eq:Phi6Phi6Phi6Phi6_ReducedCorrelator}
\end{equation}
with $F (\chi)$ the function capturing the cross-ratio dependence.
Since there are no open indices, this correlator consists of \textit{one} $R$-symmetry channel.
By applying our recursion relations \eqref{eq:RecursionEvenLO_Diagrams}, \eqref{eq:RecursionNLOBulk_Diagrams} and \eqref{eq:RecursionNLOBdry_Diagrams}, we obtain
\begin{equation}
\begin{split}
F (\chi) =\ & F_0 (\chi) + \frac{1}{\chi^2} F_1 (\chi) + \frac{1}{(1-\chi)^2} F_2(\chi)  \\
& + \frac{\lambda}{12} \frac{1}{\chi^2 (1-\chi)^2} \left( 1 - 2\chi(1-\chi) \phantom{\frac{\lambda}{\pi^2}} \right.  \\
& \left.  + \frac{3}{\pi^2} \left( 3 \chi(1-\chi) + \chi^2 \log \chi - ( 1 - \chi(2-3\chi ) \log (1-\chi) ) \right) \right) \\
& + \Op(\lambda^2)\,.
\end{split}
\label{eq:Phi6Phi6Phi6Phi6_Result}
\end{equation}
Here the functions $F_j (\chi) = F^{(0)} (\chi) + \lambda F^{(1)} (\chi) + \ldots\,$, defined via \eqref{eq:FourPointFunction_LOChannels} and \eqref{eq:FourPointFunction_NLO}, correspond to the four-point functions of the half-BPS operator $\Oh_1$.
It is worth noting that all the additional terms in the expression arise from the $U$-diagrams of type \eqref{eq:RecursionNLOExtended_ExampleDiagram}, due to the coupling of $\phi^6$ with the defect.

The second example involves a four-point function $\vvev{\Oh_1 \Oh_1 \Oh_A \Oh_A}$, which is particularly interesting because it includes an unprotected \textit{composite} operator $\Oh_A := \uh^i \Oh_A^i$.
$\Oh_A^i$ is defined in \eqref{eq:L2Operators}, and $\uh$ is introduced with the sole purpose of suppressing the $R$-symmetry index of the operator.
Since the unprotected operators are the operators located at $\tau_3$ and $\tau_4$, it is convenient to choose a conformal prefactor different to \eqref{eq:FourPointFunctions_ReducedCorrelator} and \eqref{eq:Phi6Phi6Phi6Phi6_ReducedCorrelator}:
\begin{equation}
\vev{\Oh_1 \Oh_1 \Oh_A \Oh_A} = \frac{\Fm (\chi; r,s)}{\tau_{12}^{\vphantom{2\Dh_{\smash{\Oh_A}}}2} \tau_{34}^{2\Dh_{\smash{\Oh_A}}}}\,.
\label{eq:Oh1Oh1OhAOhA_ReducedCorrelator}
\end{equation}
$\Fm (\chi; r,s)$ has the same structure as the four-point function of $\Oh_1$ operators:
\begin{equation}
\Fm (\chi) = \frac{\chi^2}{r} \left(G_0 (\chi) + \frac{r}{\chi^2} G_1 (\chi) + \frac{s}{(1-\chi)^2} G_2 (\chi) \right)\,,
\label{eq:Oh1Oh1OhAOhA_RSymmetryChannels}
\end{equation}
where the overall factor of $\chi^2/r$ accounts for the different conformal prefactor in \eqref{eq:Oh1Oh1OhAOhA_ReducedCorrelator}.

To compute this correlator, we apply the (extended) recursion relations for six scalar fields and pinch the last four operators into two $\Oh_A$ operators.
We find for the leading order
\begin{equation}
G_0^{(0)} (\chi) = 0\,, \quad G_1^{(0)} (\chi) = 1\,, \quad G_2^{(0)} (\chi) = \frac{1}{2}\,,
\label{eq:Oh1Oh1OhAOhA_LO}
\end{equation}
while at the next-to-leading order, the results can be elegantly expressed as
\begin{equation}
\begin{split}
G_0^{(1)} (\chi) &= \frac{1}{2} F_0^{(1)} (\chi)\,, \\
G_1^{(1)} (\chi) &= - \frac{\lambda}{12} F_1^{(0)} + F_1^{(1)} - \frac{\lambda}{16\pi^2} \left( \frac{8 - 9 \chi}{1 - \chi} + \log (1-\chi) \right), \\
G_2^{(1)} (\chi) &= - \frac{\lambda}{48} F_2^{(0)} (\chi) + \frac{1}{2} F_2^{(1)} (\chi) - \frac{\lambda}{16 \pi^2} ( 4 + \log (1-\chi) )\,,
\end{split}
\label{eq:Oh1Oh1OhAOhA_NLO}
\end{equation}
where the functions $F_j^{(\ell)} (\chi)$ are the $R$-symmetry channels of $\vev{\Oh_1 \Oh_1 \Oh_1 \Oh_1}$ given in \eqref{eq:FourPointFunction_LOChannels} and \eqref{eq:FourPointFunction_NLO}.

More complicated examples can be found in the ancillary notebooks of \cite{Barrat:2021tpn} and \cite{Barrat:2022eim}.
We conclude this analysis by reminding that \textit{all} four-point functions of single-trace scalar operators made of fundamental scalar fields can be obtained up to the next-to-leading order by using the recursion relations of Section \ref{subsec:CorrelatorsOfFundamentalScalars}, with computational power being the only limitation.

\subsubsection[Higher-point functions: selected \\ results\enspace]{Higher-point functions: selected results}
\label{subsubsec:HigherPointFunctionsSelectedResults}

The main benefit of the recursion relations is that they can be used for computing higher-point functions, which depend on $n-3$ spacetime cross-ratios and $n(n-3)/2$ $R$-symmetry cross-ratios for $n$ external operators.
In this section, we illustrate their power by providing two explicit examples of multipoint correlators involving half-BPS operators.

The first example that we consider is the six-point function of elementary scalars $\Oh_1$.
At the leading order, it consists of $5$ diagrams, while at the next-to-leading order, the number increases to $59$ for the planar case.
This makes it an excellent test for the efficiency and effectiveness of our method.
Recall that the recursion relations have been fully implemented in the ancillary notebook provided with the publication \cite{Barrat:2022eim}.
It suffices to give as input the external operators to generate the full result.

The correlator is defined as
\begin{equation}
\vev{\Oh_1 \Oh_1 \Oh_1 \Oh_1 \Oh_1 \Oh_1} = \Km\, \Fm ( \lbrace \chi ; r,s,t \rbrace )\,,
\label{eq:SixPoint_ReducedCorrelator}
\end{equation}
where $\lbrace \chi ; r,s,t \rbrace$ refers to the \textit{three} spacetime cross-ratios $\chi_i$ and \textit{nine} $R$-symmetry cross-ratios $r_i, s_i, t_{ij}$, following the definitions \eqref{eq:SpacetimeCrossRatios}.\footnote{$1d$ spacetime cross-ratios can be obtained from \eqref{eq:SpacetimeCrossRatios} by setting $u_i \to \chi_i$ and ignoring the rest, while for the $R$-symmetry cross-ratios we set $x_{ij}^2 \to u_i \cdot u_j$ and rename $\lbrace u, v, w \rbrace \to \lbrace r, s, t \rbrace$.}
The conformal prefactor $\Km$ is chosen to be
\begin{equation}
\Km = \frac{ (15)^2 (26) (36) (46) }{ (16) (56) }\,,
\label{eq:SixPoint_ConformalPrefactor}
\end{equation}
with
\begin{equation}
(ij) := \frac{ u_i \cdot u_j}{ \tau_{ij}^2 }\,.
\label{eq:ij_Definition}
\end{equation}
The reason for choosing this particular conformal prefactor is to have the $R$-symmetry channels follow the convenient decomposition
\begin{align}
\Fm ( \lbrace \chi ; r,s,t \rbrace ) =\ &
\frac{t_{12} }{ \chi_{12}^2} \Fb_{0} + \frac{t_{13} }{ \chi_{13}^2} \Fb_{1} + \frac{t_{23} }{ \chi_{23}^2} \Fb_{2} + \frac{ r_1 t_{23} }{ \chi_1^2 \chi_{23}^2 } \Fb_{3} + \frac{ r_2 t_{13} }{ \chi_2^2 \chi_{13}^2 } \Fb_{4} + \frac{ r_3 t_{12} }{ \chi_3^2 \chi_{12}^2 } \Fb_{5} \notag \\
& + \frac{ s_1 t_{23} }{ (1-\chi_1)^2 \chi_{23}^2 } \Fb_{6} + \frac{ s_2 t_{13} }{ (1-\chi_2)^2 \chi_{13}^2 } \Fb_{7} + \frac{ s_3 t_{12} }{ (1-\chi_3)^2 \chi_{12}^2 } \Fb_{8} \notag \\
& + \frac{ r_1 s_2 }{ \chi_1^2 (1-\chi_2)^2} \Fb_{11} + \frac{ r_1 s_3 }{ \chi_1^2 (1-\chi_3)^2} \Fb_{9}  + \frac{ r_2 s_1 }{ \chi_2^2 (1-\chi_1)^2} \Fb_{10} \notag \\
&+ \frac{ r_2 s_3 }{ \chi_2^2 (1-\chi_3)^2} \Fb_{12} + \frac{ r_3 s_1 }{ \chi_3^2 (1-\chi_1)^2} \Fb_{13} + \frac{ r_3 s_2 }{ \chi_3^2 (1-\chi_2)^2} \Fb_{14}\,,
\label{eq:SixPoint_RSymmetryChannels}
\end{align}
with $\Fb_j := \Fb_j (\chi_1, \chi_2, \chi_3)$.
Here we label the channels as $\Fb_j$ to differentiate them from the channels $F_j$ of the four-point function $\vev{\Oh_1 \Oh_1 \Oh_1 \Oh_1}$.
This is because we intend to use the results of the four-point function for compactly expressing the six-point function.

At the leading order, the five diagrams for the six-point function become
\begin{align}
\Fb_{3}^{(0)} = \Fb_{5}^{(0)} = \Fb_{6}^{(0)} = \Fb_{8}^{(0)} = \Fb_{10}^{(0)} = 1\,, \quad \Fb_j^{(0)} = 0\ \text{ otherwise}\,.
\label{eq:SixPoint_LO}
\end{align}
This corresponds to the five planar ways the external points can be Wick-contracted together.

We now compute the next-to-leading order, where eleven channels are non-vanishing.
It is worth noting that in the recursion relations \eqref{eq:RecursionNLOBulk_Diagrams} and \eqref{eq:RecursionNLOBdry_Diagrams}, higher-point diagrams are given as a four-point function multiplied by a tree-level diagram.
But these tree-level contributions are already encoded in \eqref{eq:SixPoint_ConformalPrefactor} and \eqref{eq:SixPoint_RSymmetryChannels}.
Therefore, the $R$-symmetry channels can be expressed as a sum of the $F_j (\chi)$ functions defined in \eqref{eq:FourPointFunction_NLO}, with the appropriate combinations of cross-ratios as arguments.
This provides an elegant formulation of the next-to-leading order contributions, which would otherwise cover several pages of this thesis.

\begingroup
\allowdisplaybreaks

The channels can be classified into three different groups based on the transcendentality weights.
The first group consists of channels that are made of $F_0$ and have transcendentality weight $1$:
\begin{equation}
\begin{alignedat}{2}
\Fb_{0}^{(1)} &= F_0^{(1)} (\chi_3)\,, && \qquad \Fb_{2}^{(1)} = F_0^{(1)} (\chi_1)\,, \\
\Fb_{4}^{(1)} &= F_0^{(1)} \left( \frac{\chi_1 \chi_{23}}{\chi_2 \chi_{13}} \right)\,, && \qquad\Fb_{7}^{(1)} = F_0^{(1)} \left( \frac{(1-\chi_3) \chi_{12}}{(1-\chi_2) \chi_{13}} \right)\,, \\
\Fb_{9}^{(1)} &= F_0^{(1)} \left( \frac{\chi_{32}}{1-\chi_2} \right)\,, && \qquad \Fb_{13}^{(1)} = F_0^{(1)} \left( \frac{\chi_1 }{ \chi_2 } \right)\,,
\end{alignedat}
\label{eq:SixPoint_NLOGroup1}
\end{equation}
where we defined as usual $\chi_{ij} := \chi_i - \chi_j$.

The second group consists of channels of transcendentality weight $2$, which are made of $F_1$ and $F_2$:
\begin{equation}
\begin{split}
\Fb_{3}^{(1)} &= F_1^{(1)} \left( \frac{\chi_1 \chi_{23} }{\chi_2 \chi_{13}} \right) + F_1^{(1)} \left( \frac{\chi_{32}}{1 - \chi_2} \right)\,,  \\
\Fb_{5}^{(1)} &= F_2^{(1)} \left( \frac{ \chi_1 \chi_{23} }{ \chi_2 \chi_{13} } \right) + F_1^{(1)} (\chi_3) + F_1^{(1)} \left( \frac{ \chi_{21} }{ 1-\chi_1 } \right)\,, \\
\Fb_{6}^{(1)} &= F_2^{(1)} (\chi_1) + F_2^{(1)} \left( \frac{(1-\chi_3) \chi_{12}}{(1-\chi_2) \chi_{13}} \right)\,,  \\
\Fb_{8}^{(1)} &= F_2^{(1)} \left( \frac{\chi_1}{\chi_2} \right) + F_1^{(1)} \left( \frac{ (1-\chi_3) \chi_{12} }{ (1-\chi_2) \chi_{13} } \right) + F_2^{(1)} (\chi_3)\,, \\
\Fb_{10}^{(1)} &= F_1^{(1)} \left( \frac{\chi_1}{\chi_2} \right) + F_2^{(1)} \left( \frac{ \chi_{32} }{1- \chi_2} \right)\,.
\end{split}
\label{eq:SixPoint_NLOGroup2}
\end{equation}
Note that these channels are the ones that are non-vanishing at leading order (see \eqref{eq:SixPoint_LO}).
The remaining channels are non-planar in this order and do not contribute to the correlator.

\endgroup

As explained in Section \ref{subsec:CorrelatorsOfHalfBPSOperators}, it is expected that the correlator exhibits topological behavior when all the channels are added together.
Indeed, the results above give
\begin{equation}
\Fds = \sum_{j=0}^{14} \Fb_j (\chi_1, \chi_2, \chi_3) = 5 + \frac{\lambda}{4} + \Op(\lambda^2)\,,
\label{eq:SixPoint_TopologicalSector}
\end{equation}
which agrees with \eqref{eq:SixPointTopologicalSector_Expansions}.

It is easy to derive the five-point function $\vev{\Oh_1 \Oh_1 \Oh_1 \Oh_1 \Oh_2}$ from the previous result by pinching the last two operators of the six-point function.
Specifically, we have
\begin{equation}
\vev{\Oh_1 \Oh_1 \Oh_1 \Oh_1 \Oh_2} = \frac{\nh_1}{\sqrt{\nh_2}}\, \lim\limits_{6 \to 5}\, \vev{\Oh_1 \Oh_1 \Oh_1 \Oh_1 \Oh_1 \Oh_1}\,,
\label{eq:FivePoint_Pinching}
\end{equation}
where the prefactor is here to take into account the unit-normalization.
This serves as our second example of a multipoint correlator of half-BPS operators.

We define the reduced five-point function, denoted by $\Fm (\lbrace \chi; r, s, t \rbrace)$, as
\begin{equation}
\vev{\Oh_1 \Oh_1 \Oh_1 \Oh_1 \Oh_2} = (14) (25) (35)\, \Fm ( \lbrace \chi; r, s, t \rbrace )\,.
\label{eq:FivePoint_ReducedCorrelator}
\end{equation}
Similarly to the previous case, the choice of prefactor is made to ensure a convenient decomposition in $R$-symmetry channels:
\begin{equation}
\begin{split}
\Fm ( \lbrace \chi; r, s, t \rbrace ) =\ & \Fb_0
+ \frac{r_1}{\chi_1^2} \Fb_1 + \frac{s_1}{(1-\chi_1)^2} \Fb_2  \\
&+ \frac{r_2}{\chi_2^2} \Fb_3 + \frac{s_2}{(1-\chi_2)^2} \Fb_4
+ \frac{t}{\chi_{12}^2} \Fb_5\,.
\end{split}
\label{eq:FivePoint_RSymmetryChannels}
\end{equation}
This correlator depends on \textit{two} spacetime cross-ratios and \textit{five} $R$-symmetry variables.

Pinching the last two operators of the six-point function leads to three non-vanishing diagrams at the leading order:
\begin{equation}
\Fb_1^{(0)} = \Fb_4^{(0)} = \Fb_5^{(0)} = 1\,, \qquad \Fb_0^{(0)} = \Fb_2^{(0)} = \Fb_3^{(0)} = 0\,.
\label{eq:FivePoint_LO}
\end{equation}
These terms capture the ways the external operators can be Wick contracted in the planar limit.

At the next-to-leading order, we classify the results following the grouping \eqref{eq:SixPoint_NLOGroup1}-\eqref{eq:SixPoint_NLOGroup2}.
The channels with a transcendentality weight of $1$ are
\begin{equation}
\Fb_2^{(1)} = F_0^{(1)} \left( \frac{\chi_{21}}{1-\chi_1} \right)\,, \quad
\Fb_3^{(1)} = F_0^{(1)} \left( \frac{\chi_1}{\chi_2} \right)\,.
\label{eq:FivePoint_NLOGroup1}
\end{equation}
Meanwhile, the channels with a transcendentality weight of $2$ are given by
\begin{equation}
\begin{alignedat}{2}
\Fb_1^{(1)} &= F_1^{(1)} \left( \frac{\chi_1}{\chi_2} \right) + \frac{1}{48}\,, \\
\Fb_4^{(1)} &= F_1^{(1)} \left( \frac{1-\chi_2}{1 - \chi_1} \right) + \frac{1}{48}\,, \\
\Fb_5^{(1)} &= F_1^{(1)} \left( \frac{\chi_{21}}{1 - \chi_1} \right) + F_1^{(1)} \left( \frac{\chi_{21}}{\chi_2} \right) + \frac{1}{48}\,.
\end{alignedat}
\label{eq:FivePoint_NLOGroup2}
\end{equation}
Here the additional constants are remnants of the pinching procedure.
The remaining channel $\Fb_0$ is non-planar in this order.

As usual, the topological sector of the four-point function is obtained by summing all the channels, and we find
\begin{equation}
\Fds = \sum_{j=0}^5 \Fb_j (\chi_1, \chi_2) = 3 + \frac{7 \lambda}{48} + \Op(\lambda^2)\,.
\label{eq:FivePoint_TopologicalSector}
\end{equation}
This result is consistent with the localization expression given in \eqref{eq:FivePointTopologicalSector_Expansions}.

\subsubsection{Conformal block expansion\enspace}
\label{subsubsec:ConformalBlockExpansion}

We now briefly discuss the expansion in conformal blocks of the correlators presented above, both in the comb and in the snowflake channels.
The main goal is to initiate a bootstrap analysis of multipoint correlators, although for now we only perform elementary checks of our perturbative results.
We follow the conventions of Appendix \ref{app:ConformalBlocks}.

We begin by considering the correlators of half-BPS operators $\Oh_1$ in the comb channel.
More precisely, we restrict our analysis to the highest-weight channel.
In the case of the four-point function, it is given by $F_1$ (see \eqref{eq:FourPointFunction_RSymmetryChannels}), while in the case of the six-point, it corresponds to $\Fb_3$ (see \eqref{eq:SixPoint_RSymmetryChannels}).
In the limit $\eta_i \sim 0$, the expansion in blocks \eqref{eq:BlockExpansionComb} takes the form
\begin{align}
F_{\text{hw}} (\lbrace \eta \rbrace) &= 1 + \lambda_{11\phi^6}^2\, f_{\Dh=1,0} (\lbrace \eta \rbrace) + \ldots \notag \\
&= 1 + \lambda_{11\phi^6}^2\, \eta_1 \ldots \eta_{(n-2)/2} + \ldots\,.
\label{eq:FhwExpansionBlocks}
\end{align}
The first term corresponds to the exchange of the identity operator and evaluates to $1$ due to the unit normalization of the two-point function.
The first non-trivial OPE coefficient corresponds to the correlator $\vev{\Oh_1 \Oh_1 \phi^6}$, which was explicitly computed at the relevant order in \eqref{eq:ThreePointFunctions}.
It is important to note that there is no degeneracy in this case.
However, at higher values of $\Dh$, it becomes necessary to disentangle the operators to extract their respective OPE coefficients.

Our perturbative correlators take the following form:
\begin{equation}
F_{\text{hw}} (\lbrace \eta \rbrace) = 1 + \frac{\lambda}{8 \pi^2} \eta_1 \ldots \eta_{(n-2)/2} + \ldots\,,
\label{eq:FhwExpansionCorrelator}
\end{equation}
and we observe a perfect agreement with \eqref{eq:ThreePointFunctions}.

The same analysis can be performed for $n$-point functions of unprotected operators $\phi^6$.\footnote{We restrict ourselves to even $n$ here but note that the analysis carries through unchanged for odd $n$.
This was done explicitly in \cite{Barrat:2022eim}.}
In this case, there is only one $R$-symmetry channel and the lowest OPE coefficient corresponds to $\vev{\phi^6 \phi^6 \phi^6}$, which explicit computation can be found in \eqref{eq:ThreePointFunctions}.
We find again a perfect match with the expansion of our multipoint correlators.

To conclude, we consider the snowflake channel for the six-point functions of identical elementary scalars, which can either be $\Oh_1 \sim \phi^i$ or $\phi^6$.
The expansion in blocks reads
\begin{equation}
\vev{\phi^I \phi^I \phi^I \phi^I \phi^I \phi^I} \sim 1 + \lambda_{\phi^I \phi^I \phi^6}^2\, z_1 z_2 + \ldots\,,
\label{eq:Snowflake_Expansion}
\end{equation}
where in the case of $I=1, \ldots, 5$ we limit ourselves again to the highest-weight $R$-symmetry channel.
Expanding the correlators gives
\begin{equation}
\begin{split}
\vev{\Oh_1 \Oh_1 \Oh_1 \Oh_1 \Oh_1 \Oh_1} \sim 1 + \frac{\lambda}{8 \pi^2} z_1 z_2 + \ldots\,, \\
\vev{\phi^6 \phi^6 \phi^6 \phi^6 \phi^6 \phi^6} \sim 1 + \frac{9 \lambda}{8 \pi^2} z_1 z_2 + \ldots\,,
\end{split}
\end{equation}
which match precisely the perturbative results listed in \eqref{eq:ThreePointFunctions}.

\subsection{The four-point function at NNLO}
\label{subsec:TheFourPointFunctionAtNNLO}

We conclude our analysis of the weak-coupling regime by computing the four-point function of elementary half-BPS scalars at the next-to-next-to-leading order (NNLO).
This computation serves as an alternative method for obtaining \eqref{eq:FourPointFunction_LittlefNNLO} and can be readily extended to the more general case $\vev{\Oh_1 \Oh_1 \Oh_{\Dh} \Oh_{\Dh}}$.
Ultimately, the goal is to generalize these expressions to arbitrary external dimensions.

\subsubsection{Solution of the Ward identity\enspace}
\label{subsubsec:SolutionOfTheWardIdentity}

In Section \ref{subsec:FourPointFunctionsOfHalfBPSOperators}, we discussed the superconformal Ward identity that applies to four-point functions.
This remarkable differential equation, given in \eqref{eq:FourPointFunctions_WardIdentities}, imposes stringent constraints on the correlator and allows to establish algebraic relationships among the different $R$-symmetry channels.
As a consequence, the correlator depends on a single function $f(\chi)$ and a constant function $\Fds$ (see \eqref{eq:FourPointFunction_SolutionWI}).

\begin{table}[t!]
\centering
\caption{The relevant Feynman diagrams for the computation of $F_0(\chi)$ at next-to-next-to-leading order.
The horizontal double line separates {\normalfont bulk} from {\normalfont boundary} diagrams.
In the last row, the colored dots along the Wilson line indicate where the gluon can connect.}
\begin{tabular}{lc}
\hline
Self-energy & \DefectSSSSTwoLoopsSelfEnergyOne\ \DefectSSSSTwoLoopsSelfEnergyTwo\ \DefectSSSSTwoLoopsSelfEnergyThree\ \DefectSSSSTwoLoopsSelfEnergyFour\ \\[2ex]
\hline
$XX$ & \DefectSSSSTwoLoopsXXOne\ \DefectSSSSTwoLoopsXXTwo\ \\[2ex]
\hline
$XH$ & \DefectSSSSTwoLoopsXHOne\ \DefectSSSSTwoLoopsXHTwo\ \DefectSSSSTwoLoopsXHThree\ \DefectSSSSTwoLoopsXHFour\ \\[2ex]
\hline
Spider & \DefectSSSSTwoLoopsSpider\ \\[2ex]
\hline \hline
$XY$ & \DefectSSSSTwoLoopsXYOne\ \DefectSSSSTwoLoopsXYTwo\ \DefectSSSSTwoLoopsXYThree\ \DefectSSSSTwoLoopsXYFour\ \\[2ex]
\hline
\end{tabular}
\label{table:DiagramsAtNNLO}
\end{table}

The $R$-symmetry channels can be related to these functions through the following relations:
\begin{equation}
\begin{split}
F_0 (\chi) &= \mathds{F} - \frac{f(\chi)}{\chi^2} - \frac{1-\chi}{\chi} f'(\chi)\,, \\
F_1 (\chi) &= \left( \frac{2}{\chi} - 1 \right) f(\chi) - ( 1 - \chi ) f'(\chi)\,, \\
F_2 (\chi) &= \frac{(1-\chi)^2}{\chi^2} \left( f(\chi) - \chi f'(\chi) \right)\,.
\end{split}
\label{eq:FFromLittlef}
\end{equation}
It is crucial to note that computing \textit{any one} of these channels is sufficient to determine the complete correlator.
In the following, we take advantage of this property to compute the four-point function in the planar limit, by focusing on the simplest channel from the diagrammatic point of view.

Before doing so, note that crossing symmetry imposes additional constraints on the function $f(\chi)$ and the $R$-symmetry channels.
The full correlator is invariant under the exchange of spacetime and $R$-symmetry variables $(u_2, \tau_2) \leftrightarrow (u_4, \tau_4)$.\footnote{Or, equivalently, $(u_1, \tau_1) \leftrightarrow (u_3, \tau_3)$.}
As a consequence, the function $f(\chi)$ obeys the crossing symmetry relation
\begin{equation}
\chi^2 f(1- \chi) + (1-\chi)^2 f(\chi) = 0\,,
\label{eq:CrossingSymmetryOfLittlef}
\end{equation}
from which we can read the antisymmetry of the function $f(\chi)$ observed in Figure \ref{fig:PlotsLittlef}.
In terms of the channels, this symmetry can be expressed as:
\begin{equation}
F_0 (\chi) = F_0 (1-\chi)\,, \qquad
F_1 (\chi) = F_2 (1-\chi)\,.
\label{eq:CrossingSymmetryOfF}
\end{equation}

We initiate now our analysis in terms of Feynman diagrams.
In this correlator, it is clear that $F_0$ is the simplest channel to calculate.
This was already visible at the lower orders given in \eqref{eq:FourPointFunction_LOChannels} and \eqref{eq:FourPointFunction_NLO}, and is a consequence of the large $N$ limit.

The relevant planar diagrams are listed in Table \ref{table:DiagramsAtNNLO}, separated as usual into bulk and boundary diagrams.
In total, we are facing $15$ diagrams.\footnote{There is some philosophical question as to what should be considered an individual Feynman diagram.
For instance, here we count the sum of self-energy diagrams as one.
The same holds for the $XY$-diagrams, which we count as four, although each insertion on the Wilson line could in principle be considered an individual diagram.}
If we would have computed the full correlator without using the Ward identity, we would be facing many more diagrams, including self-energy ones at next-to-next-to-leading order and complicated double couplings to the line defect.
Instead, the level of difficulty of the diagrams contributing to $F_0$ at NNLO lies somewhere between $F_1$ at NLO and $F_1$ at NNLO.

Note that we keep the Feynman diagrams not unit-normalized (i.e., we strip the prefactor $1/\sqrt{\nh_1}$ in \eqref{eq:SingleTraceHalfBPS_Defect}).
This is the natural normalization to work with when doing Wick contractions since expanding $\nh_1$ at $\lambda \sim 0$ mixes contributions from lower orders.
It is therefore more convenient to unit-normalize \textit{after} the Feynman diagrams have been computed.

\subsubsection{Bulk diagrams\enspace}
\label{subsubsec:BulkDiagrams}

The bulk diagrams can be computed using standard techniques of conformal field theory.
In particular, the integrals greatly simplify when going to the conformal frame $(\tau_1, \tau_2, \tau_3, \tau_4) \to (0, \chi, 1, \infty)$.
In the following, we present the analytical result for all the bulk diagrams of Table \ref{table:DiagramsAtNNLO} in this limit.

The self-energy diagrams form the simplest family, despite being divergent.
After doing the Wick contractions, the first diagram evaluates to
\begin{equation}
\DefectSSSSTwoLoopsSelfEnergyOne\ = - \frac{\lambda^4}{2} \int d^4 x_5\, Y_{115}\, I_{25} I_{35} I_{45}\,,
\label{eq:TwoLoopsSelfEnergyDiagramOne}
\end{equation}
where $Y_{115}$ is given in \eqref{eq:YDivergent_PointSplitting} and is log-divergent.
In the conformal frame mentioned above, the integral simplifies to
\begin{equation}
\DefectSSSSTwoLoopsSelfEnergyOne\ = - \frac{\lambda^4}{2}\, I_{24} H_{11,23} + \Op(1/\tau_4^3)\,,
\label{eq:TwoLoopsSelfEnergyDiagramOneb}
\end{equation}
with the $H$-integral defined in \eqref{eq:H1234}.
$H_{11,23}$ can in principle be computed analytically, but we will not need it here as these terms will ultimately drop from the computation.

The other self-energy diagrams can be computed analogously, and by summing them up we obtain
\begin{equation}
F_0^{\text{SE}} = - \frac{\lambda^4}{32 \pi^4 I_{13}} \left( H_{11,23} + H_{22,13} + H_{33,12} + \frac{K_{44}}{I_{24}} \right)\,.
\label{eq:F0SelfEnergy}
\end{equation}
The integral $K_{44}$ is log-divergent both in $\veps$ and in $\tau_4$ when going in the conformal frame.
For the correlator to be conformal, note that all $\log \tau_4^2$ terms should drop out from the final expression and hence we expect the $K_{44}$ terms to cancel at the level of the correlator.
This provides a strong check of our computations, on top of the fact that the $\log \veps^2$ terms must also cancel each other for the correlator to be finite.

The $XX$-diagrams are also easy to compute.
The first one reads
\begin{equation}
\DefectSSSSTwoLoopsXXOne\ = - \frac{\lambda^4}{4} \int d^4 x_5\, X_{1255}\,  I_{35} I_{45}\,,
\label{eq:TwoLoopsXXDiagramOne}
\end{equation}
with the integral $X_{1255}$ given in \eqref{eq:X1233}.

This integral can be computed in the conformal frame, and the sum of the two diagrams of Table \ref{table:DiagramsAtNNLO} give
\begin{equation}
\begin{split}
F_0^{XX} =\ & - \frac{\lambda^4}{128 \pi^4 I_{13}} \biggl(
H_{11,23} + H_{22,13} + H_{33,12} + \frac{K_{44}}{I_{24}}
- \frac{Y_{123}}{16 \pi^2} \log \tau_4^2
\biggr)\,.
\end{split}
\label{eq:F0XX}
\end{equation}
We note again that all these terms should eventually drop from the final expression since they are either divergent or non-conformal.

The $XH$-diagrams are more involved, but they can also be calculated using similar techniques.
We find
\begin{equation}
\begin{split}
F_0^{XH} =\ & \frac{\lambda^4}{128 \pi^4 I_{13}}
\biggl\lbrace
H_{11,23} + H_{22,13} + H_{33,12} + \frac{K_{44}}{I_{24}} \\
& - H_{12,13} - H_{13,23} + 2 H_{12,23} \\
&- \frac{Y_{123}}{32 \pi^2} \left( \log \frac{\chi^2 (1-\chi)^2}{\tau_4^4} - 8 \right)
\biggr\rbrace\,.
\end{split}
\label{eq:F0XH}
\end{equation}

We compute now the fermion loop, which we call the \textit{spider} diagram.
The Wick contractions yield the following challenging sixteen-dimensional integral:
\begin{equation}
\begin{split}
\DefectSSSSTwoLoopsSpider\ =\ & \frac{\lambda^4}{4} \int d^4 x_5 \int d^4 x_6 \int d^4 x_7 \int d^4 x_8\, I_{15} I_{26} I_{37} I_{48} \\
& \times \tr \spd_5 I_{56} \spd_6 I_{67} \spd_7 I_{78} \spd_8 I_{58}\,,
\end{split}
\label{eq:TwoLoopsSpiderDiagram}
\end{equation}
where the trace is acting on the $\gamma$ matrices.
We can use the fermionic star-triangle identity \eqref{eq:StarTriangle} twice to lift two integrals:
\begin{equation}
\begin{split}
\DefectSSSSTwoLoopsSpider\ =\ & - 4 \pi^4 \lambda^4 \int d^4 x_5\, I_{15} I_{25} I_{45} \int d^4 x_7\,  I_{27} I_{37} I_{47} I_{57}^2  \\
& \times \tr \sx_{25} \sx_{27} \sx_{47} \sx_{45}\,.
\end{split}
\label{eq:TwoLoopsSpiderDiagramb}
\end{equation}
The trace is easy to perform with\footnote{Note that the $\gamma$ matrices are \textit{sixteen-dimensional}.}
\begin{equation}
\tr \sx_1 \sx_2 \sx_3 \sx_4 = 16\, [ (x_1 \cdot x_2)(x_3 \cdot x_4) + (x_1 \cdot x_4)(x_2 \cdot x_3) - (x_1 \cdot x_3)(x_2 \cdot x_4) ]\,,
\label{eq:TwoLoopsTraceOfGammasb}
\end{equation}
and, after using some algebraic manipulations, the diagram reduces to
\begin{equation}
\begin{split}
\DefectSSSSTwoLoopsSpider\ =\ & - 2 \lambda^4 \frac{1}{I_{24}} \int d^4 x_5\, X_{2345}\, I_{15} I_{25} I_{45} \\
& +2 \lambda^4 \int d^4 x_5\,  \left( X_{2355}\, I_{15} I_{45} + X_{1455}\, I_{25} I_{35} \right)\,.
\end{split}
\label{eq:TwoLoopsSpiderDiagramc}
\end{equation}
The first line corresponds to the \textit{kite} integral, given in \eqref{eq:KiteIntegral}.
The second line is identical to the $XX$-diagrams and can be found in \eqref{eq:F0XX} up to an overall prefactor.

All in all, the spider diagram reads
\begin{equation}
\begin{split}
F_0^{\text{Spider}} =\ & 
\frac{\lambda^4}{64 \pi^4 I_{13}}
\biggl\lbrace
H_{11,23} + H_{22,13} + H_{33,12} + \frac{K_{44}}{I_{24}} \\
&- 2 H_{12,23}
- \frac{Y_{123}}{16 \pi^2} \log \tau_4^2
\biggr\rbrace\,.
\end{split}
\label{eq:F0Spider}
\end{equation}
This concludes our computation of the bulk diagrams, for which we were able to obtain an analytical expression in the conformal limit.

\subsubsection[\\ Boundary diagrams\enspace]{Boundary diagrams}
\label{subsubsec:BoundaryDiagrams}

\begingroup
\allowdisplaybreaks

We move now our attention to the boundary integrals.
The first diagram in the last line of Table \ref{table:DiagramsAtNNLO} reads
\begin{equation}
\begin{split}
\DefectSSSSTwoLoopsXYOne\ =\ & \frac{\lambda^4}{8} \left( \int_{-\infty}^{\tau_2} + \int_{\tau_4}^{\infty} \right) d\tau_6\ \veps (\tau_1\, \tau_3\, \tau_6) \\
&\times \int d^4 x_5\, \pd_{15} Y_{156}\, I_{25} I_{35} I_{45}\,.
\end{split}
\label{eq:TwoLoopsXYDiagramOne}
\end{equation}
This expression is straightforward to obtain from the Wick contractions.
The function $\veps (\tau_i\, \tau_j\, \tau_k)$ is defined in \eqref{eq:eps_Definition}.
This integral can be shown to give
\begin{align}
\DefectSSSSTwoLoopsXYOne\ =\ & \frac{\lambda^4}{16 \pi^2} \int d^4 x_5\, \frac{1}{|x_5^\perp|} I_{15}\, \pd_5 X_{2345} \notag \\
& \times \left\lbrace \tan^{-1} \Bigl( \frac{\tau_{45}}{|x_5^\perp|} \Bigr) + \tan^{-1} \Bigl( \frac{\tau_{25}}{|x_5^\perp|} \Bigr) - 2 \tan^{-1} \Bigl( \frac{\tau_{15}}{|x_5^\perp|} \Bigr) \right\rbrace \notag \\
& - \frac{\lambda^4}{8} \int d^4 x_5\, I_{25} I_{35} I_{45} \left( Y_{125} + Y_{145} \right) \notag \\
& + \frac{\lambda^4}{4} \int d^4 x_5\, Y_{115}\, I_{25} I_{35} I_{45}\,.
\label{eq:TwoLoopsXYDiagramOneb}
\end{align}
In the limit $\tau_4 \to \infty$, this turns into
\begin{equation}
\begin{split}
\DefectSSSSTwoLoopsXYOne\ =\ & \frac{\lambda^4}{16 \pi^2} I_{24} \int d^4 x_5\, \frac{1}{|x_5^\perp|} I_{15} \pd_5 Y_{235} \\
& \times \left\lbrace \tan^{-1} \Bigl( \frac{\tau_{25}}{|x_5^\perp|} \Bigr) - 2 \tan^{-1} \Bigl( \frac{\tau_{15}}{|x_5^\perp|} \Bigr) - \frac{\pi}{2} \right\rbrace \\
& - \frac{\lambda^4}{8} I_{24} \left( H_{12,23} - 2 H_{11,23} \right) + \Op(1/\tau_4^3)\,.
\end{split}
\label{eq:TwoLoopsXYDiagramOnec}
\end{equation}
This is as far as we can go analytically, and similar expressions can be obtained for the three other diagrams of Table \ref{table:DiagramsAtNNLO}.
Gathering the analytical expressions together, we obtain
\begin{equation}
\begin{split}
F_0^{XY, \text{a}} =\ & 
\frac{\lambda^4}{64 \pi^4 I_{13}}
\biggl\lbrace
H_{11,23} + H_{22,13} + H_{33,12} + \frac{K_{44}}{I_{24}} \\
&- \frac{1}{2} ( H_{12,13} + H_{13,23} + 2 H_{12,23} ) \\
&+ \frac{Y_{123}}{64 \pi^2} (  \log \chi^2 (1-\chi)^2 \tau_4^8 - 8)
\biggr\rbrace\,.
\end{split}
\label{eq:F0XYAnalytical}
\end{equation}

\endgroup

We now solve the remaining integrals numerically.
Before doing so, we gather the analytical terms and find that they simplify dramatically to give
\begin{equation}
\begin{split}
F_0^{\text{a}} &:= F_0^{\text{SE}} + F_0^{XX} + F_0^{XH} + F_0^{\text{Spider}} + F_0^{XY, \text{a}} \\
&\phantom{:}=
- \frac{\lambda^4}{128 \pi^2 I_{13}}
\biggl\lbrace
H_{12,13} + H_{13,23} + 2 H_{12,23}  \\
&\phantom{:=}- \frac{Y_{123}}{32 \pi^2} \left( \log \chi^2 (1-\chi)^2 - 8 \right)
\biggr\rbrace\,.
\end{split}
\label{eq:F0Analytical}
\end{equation}
We note in particular that all the divergent terms cancel each other.
The absence of $\log \tau_4^2$ is an encouraging sign that the final correlator is conformal as expected.

\subsubsection{Numerical integration\enspace}
\label{subsubsec:NumericalIntegration}

We perform a numerical integration of the sum of the remaining terms.
An efficient way to obtain the final result is to write an ansatz of transcendental functions based on the form of the other terms.
Such a method was used successfully for instance to determine \eqref{eq:FourPointFunction_LittlefNNLO} in \cite{Cavaglia:2022qpg}.
It is also similar to the ansatz made in \cite{Barrat:2020vch} to obtain (4.13) and (4.14).

A suitable ansatz is
\begin{equation}
F_0^{XY, \text{n}} = \frac{\lambda^4}{\chi(1-\chi)} \left( \sum_{\vec{a}} \alpha_{\vec{a}} H_{\vec{a}} + \chi \sum_{\vec{a}} \beta_{\vec{a}} H_{\vec{a}} \right)\,,
\label{eq:F0XYAnsatz}
\end{equation}
where we allow the sum to run up to transcendentality $3$.
The different OPE limits constrain $\vec{a}$ to contain only combinations of $0$ and $1$.

\begin{figure}[t!]
\centering
\includegraphics[width=.6\linewidth]{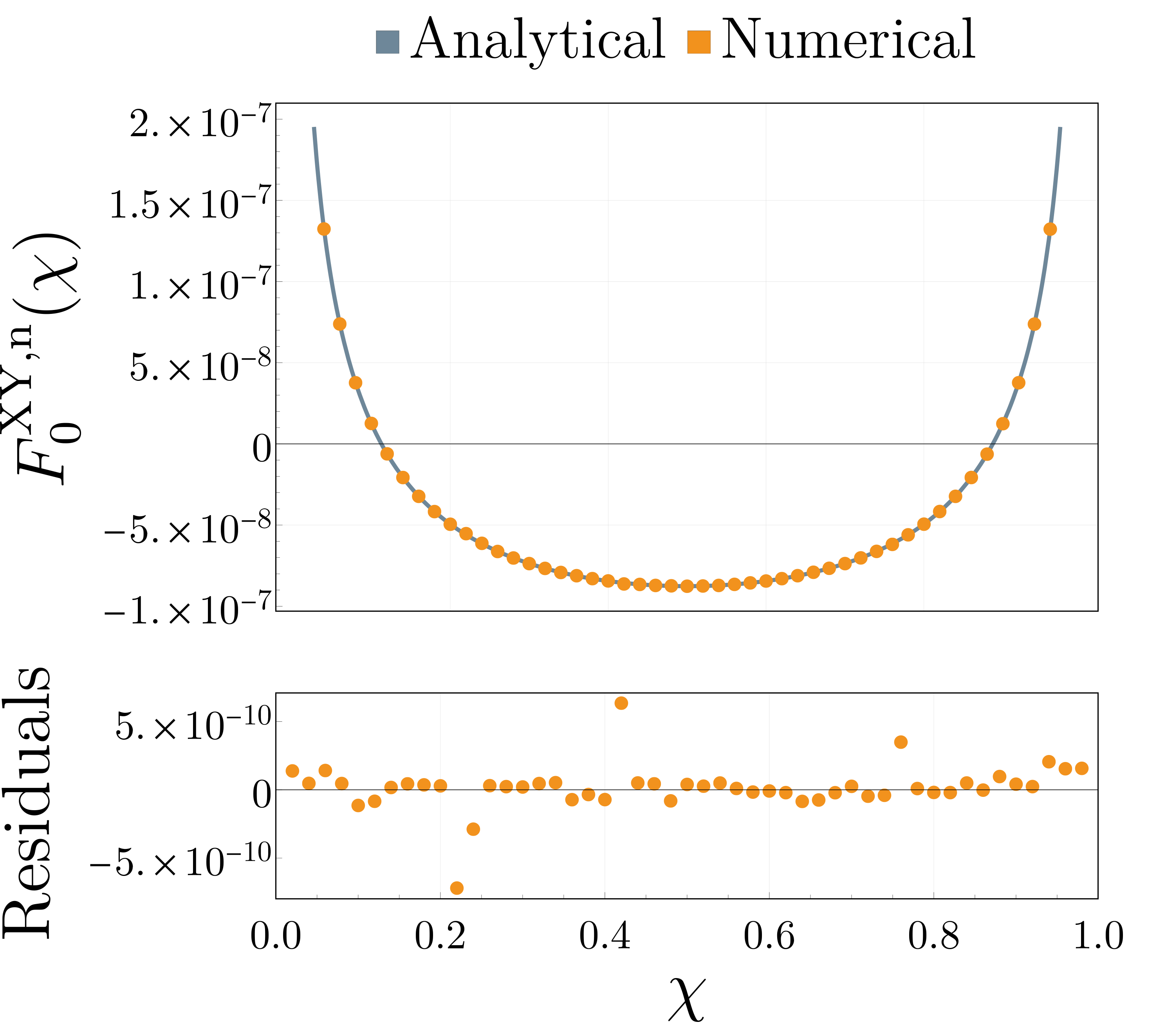}
\caption{The orange dots in this plot show the function $F^{XY,\text{n}}_0 (\chi)$, defined as the sum of terms in the $XY$-diagrams that could not be solved analytically.
The blue line corresponds to the analytical expression \eqref{eq:F0XYNumericalToAnalytical}, obtained from comparing the ansatz \eqref{eq:F0XYAnsatz} to the numerical points.
We observe a perfect agreement between the numerical and analytical results.
}
\label{fig:PlotNumerical}
\end{figure}

It turns out that the coefficients $\alpha$ and $\beta$ are simple numbers.
We find the following solution:
\begin{align}
F_0^{XY, \text{n}} =\ &
\frac{\lambda^4}{24576 \pi^8} \frac{1}{\chi(1-\chi)}
\bigl(
2 (6 - \pi^2) H_1
+ 3 ( H_{0,1} + H_{1,0} - 2 H_{1,1} ) \notag \\
&+ 3 ( 3 H_{0,0,1} - H_{0,1,0} - 2 H_{1,0,0} )
+ \chi
\bigl(
2 ( \pi^2 - 6 ) H_0 +2 ( \pi^2 - 3 ) H_1 ) \notag \\
&+ 6 ( H_{0,0} - H_{1,1} )
+ 3 ( 3 (H_{0,0,1} + H_{1,1,0}) - (H_{0,1,0} + H_{1,0,1}) \notag \\
& - 2 (H_{0,1,1} + H_{1,0,0}) )
- 9 \zeta_3
\bigr)
\bigr)\,.
\label{eq:F0XYNumericalToAnalytical}
\end{align}
The comparison between numerical data and this analytical expression is displayed in Figure \ref{fig:PlotNumerical}.
We observe a very convincing agreement between the function \eqref{eq:F0XYNumericalToAnalytical} and the numerical integration of the $XY$-diagrams.

\subsubsection[\\ Final result\enspace]{Final result}
\label{subsubsec:FinalResult}

We now put the pieces together to write down the full correlator at NNLO.
Rewriting \eqref{eq:F0Analytical} in terms of HPLs and adding \eqref{eq:F0XYNumericalToAnalytical}, we find the simple and elegant result
\begin{align}
F_0^{(2)} =\ & \frac{1}{192 \pi^4 \chi (1-\chi)}
\biggl\lbrace
\pi^2 H_1 - 3 ( H_{1,0,1} + H_{1,1,0} - 2 ( H_{0,1,0} + H_{0,1,1} - H_{1,0,0} ) ) \notag \\
& - 3 \chi
\biggl(
\frac{\pi^2}{3} H_0 - (H_{0,0,1} + H _{1,1,0}) + H_{0,1,0} + H_{1,0,1} + 3 \zeta_3
\biggr)
\biggr\rbrace\,.
\label{eq:F0FinalResult}
\end{align}
Note that here we have already unit-normalized the correlator.
Solving \eqref{eq:FFromLittlef} with \eqref{eq:F0FinalResult} yields the result \eqref{eq:FourPointFunction_LittlefNNLO}, providing a check for the bootstrability techniques developed in \cite{Cavaglia:2022qpg}.

This concludes our exploration of multipoint correlators in the weak coupling regime.
In Section \ref{sec:SummaryAndOutlook2}, we sketch possible future directions related to the techniques used throughout this section.

\section{Superconformal Ward identities}
\label{sec:SuperconformalWardIdentities}

The superconformal Ward identity \eqref{eq:FourPointFunctions_WardIdentities} plays a central role in the study of four-point functions, both in the weak- and strong-coupling regimes.
A natural question arises: do multipoint correlators of half-BPS operators also satisfy similar powerful identities?
If so, what are their specific forms?
In this section, we put forth a conjecture that such identities exist, by proposing a set of differential equations satisfied by our pool of perturbative results.
The resulting \textit{multipoint Ward identities} take a similar form to \eqref{eq:FourPointFunctions_WardIdentities} and pave the way to a strong-coupling analysis of higher-point functions.
As a practical example, we show how these Ward identities can be solved in the case of the five-point function $\vev{\Oh_1 \Oh_1 \Oh_1 \Oh_1 \Oh_2}$.

The results presented here are based on \cite{Barrat:2021tpn,Barrat:2023ta2}.

\subsection{Multipoint Ward identities}
\label{subsec:MultipointWardIdentities}

Using the recursion relations \eqref{eq:RecursionEvenLO_Diagrams}, \eqref{eq:RecursionNLOBulk_Diagrams} and \eqref{eq:RecursionNLOBdry_Diagrams}, we have computed correlators featuring up to eight external (half-BPS) operators.
Concrete examples of such correlators can be found in the supplementary \textsc{Mathematica} notebook of \cite{Barrat:2022eim}.
They can be used here for postulating differential equations satisfied by multipoint correlators.

\subsubsection{The conjecture and checks\enspace}
\label{eq:TheConjectureAndChecks}

We found experimentally that \textit{all} our correlators are annihilated by the following family of differential operators:
\begin{equation}
\sum_{k=1}^{n-3} \left( \frac{1}{2} \partial_{\chi_k} + \alpha_k \partial_{r_k} - (1-\alpha_k) \partial_{s_k} \right) \, \Fm ( \lbrace \chi; r,s,t \rbrace) \raisebox{-2ex}{$\Biggr 
|$}_{\raisebox{1.5ex}{$\begin{subarray}{l} r_i \to \alpha_i \chi_i \\
 s_i \to (1-\alpha_i)(1-\chi_i)
 \\ t_{ij} \to \alpha_{ij} \chi_{ij}\end{subarray}$}} = 0\,,
\label{eq:MultipointWardIdentities}
\end{equation}
where $\alpha_k$ are arbitrary real numbers as in \eqref{eq:FourPointFunctions_WardIdentities}.
These operators represent a natural extension of the differential operator \eqref{eq:DifferentialOperatorWI} that captures the constraints of superconformal symmetry for four-point functions.

Even though we obtained these equations from weak-coupling correlation functions, we expect these identities to also be satisfied in the strong-coupling regime.
As discussed in Section \ref{subsec:N4SuperYangMills}, this expansion is captured by a well-understood AdS dual \cite{Giombi:2017cqn}, which has been used to calculate planar correlators in the $\lambda \gg 1$ limit.
The leading order correlators are given by simple Wick contractions of the fluctuations of the dual fundamental string, which means that they correspond to the generalized free-field expressions, see for instance (4.5) in \cite{Giombi:2017cqn}.
It is easy to check that all $n$-point correlators also satisfy \eqref{eq:MultipointWardIdentities} in the extreme strong-coupling regime.
Moreover, it was independently checked through Witten-diagram computations in \cite{Bliard:2022xsm} that the next-to-leading order correlators also satisfy our conjectured multipoint Ward identities.

We have therefore \textit{four} non-trivial data points: the first two orders, both in the weak- and strong-coupling regimes.
It is then reasonable to assume that the constraint \eqref{eq:MultipointWardIdentities} is \textit{non-perturbative} and valid at all loop orders.

\subsubsection[Comments and a \\ word of caution\enspace]{Comments and a word of caution}
\label{eq:CommentsAndAWordOfCaution}

Since the superconformal Ward identities \eqref{eq:MultipointWardIdentities} are based on symmetry, they are expected to have powerful consequences, some of which we discuss here.
In our study of planar correlators, their main application is to identify the degrees of freedom of the correlator, as it was done in \eqref{eq:FourPointFunction_SolutionWI} for the four-point function.
But they are also handy for deriving superconformal blocks.
In \cite{Liendo:2016ymz}, it was shown that one can bypass solving superconformal Casimir equations by applying the Ward identity on an ansatz for the superblocks.
In Section \ref{sec:MultipointCorrelatorsAtStrongCoupling}, we apply this method for deriving the superblocks of the five-point function $\vev{\Oh_1 \Oh_1 \Oh_1 \Oh_1 \Oh_2}$.

Moreover, note that these identities go beyond the planar limit.
Superconformal constraints are insensitive to gauge symmetry, which means that they should also hold for \textit{non-planar} corrections.
This is indeed the case for the first correction in the $N$ expansion at the leading order, and for certain cases, we have checked that it also holds at the next-to-leading order.

We conclude this section with a word of caution about \eqref{eq:MultipointWardIdentities}.
We should point out that our conjecture cannot represent the \textit{full} set of superconformal constraints on the correlators.
The reason is that our analysis of protected operators only focuses on the highest-weight component while ignoring possible fermionic descendants.
Working in a suitable superspace, it is known that for four-point functions, the full superconformal correlator can be reconstructed from the highest weight.
It is thus safe to set the fermions to zero in this case.
But starting with five-point and up, one expects nilpotent invariants.\footnote{We thank P. Heslop for discussions on this topic.}
In general, for generic $n$-point functions, the Ward identities should be a collection of partial differential equations relating the components associated with each fermionic structure.
In this light, the fact that we obtained a differential operator that only acts on the highest weight and still annihilates the correlator is unexpected.
Moreover, the number of constraints seems to grow unbounded with the number of external operators.
This means that for a certain $n$, the number of constraints would exceed the number of symmetries.\footnote{We are grateful to N. Gromov for insightful comments.}
At this point, the fate of the Ward identities for a high number of external points is unclear, and it would be desirable to conduct a proper superspace analysis (in the gist of \cite{Liendo:2016ymz}) to prove that our experimental observation indeed corresponds to one of the constraints imposed by superconformal invariance.

\subsection{Solving the Ward identities}
\label{subsec:SolvingTheWardIdentities}

We now discuss how the solution \eqref{eq:FourPointFunction_SolutionWI} of the four-point Ward identity extends to higher-point functions.
We do not provide a general analysis, but instead focus on highlighting the similarities between the four-point and higher-point cases.
As a practical example, we provide the solution of the Ward identities \eqref{eq:MultipointWardIdentities} for the five-point function $\vev{\Oh_1 \Oh_1 \Oh_1 \Oh_1 \Oh_2}$.
This solution is then used for expressing the weak-coupling result compactly.
The results of this section will be crucial for the strong-coupling analysis of Section \ref{sec:MultipointCorrelatorsAtStrongCoupling}.

\subsubsection{An example: $\vev{\Oh_1 \Oh_1 \Oh_1 \Oh_1 \Oh_2}$\enspace}
\label{subsubsec:AnExample11112}

The five-point function $\vev{\Oh_1 \Oh_1 \Oh_1 \Oh_1 \Oh_2}$ is a correlator that we already considered at weak coupling, and we keep the conventions defined in \eqref{eq:FivePoint_ReducedCorrelator}-\eqref{eq:FivePoint_RSymmetryChannels}.
The only difference is that here we use new $R$-symmetry cross-ratios $\zeta$ and $\eta$ instead of $r$ and $s$, which are the five-point pendants of \eqref{eq:AlternativeRSymmetryCrossRatios}:
\begin{equation}
\begin{split}
r_1 = \zeta_1 \zeta_2\,, \quad s_1 = (1-\zeta_1) (1-\zeta_2)\,, \\
r_2 = \eta_1 \eta_2\,, \quad s_2 = (1-\eta_1) (1-\eta_2)\,.
\end{split}
\label{eq:FivePoint_CrossRatios2}
\end{equation}
The $R$-symmetry channel decomposition \eqref{eq:FivePoint_RSymmetryChannels} becomes
\begin{equation}
\begin{split}
\Fm ( \lbrace \chi; r, s, t \rbrace ) =\ & \Fb_0
+ \frac{ \zeta_1 \zeta_2 }{\chi_1^2} \Fb_1 + \frac{ (1-\zeta_1) (1-\zeta_2) }{(1-\chi_1)^2} \Fb_2 \\
&+ \frac{ \eta_1 \eta_2 }{\chi_2^2} \Fb_3 + \frac{ (1-\eta_1) (1-\eta_2) }{(1-\chi_2)^2} \Fb_4
+ \frac{t}{\chi_{12}^2} \Fb_5\,.
\end{split}
\label{eq:FivePoint_RSymmetryChannels2}
\end{equation}
This formulation is convenient, as the topological limit is reached for
\begin{align}
\zeta_i = \chi_1\,, \quad \eta_i = \chi_2\,, \quad t = \chi_{12}^2\,.
\label{eq:FivePoint_TopologicalLimit2}
\end{align}

The Ward identities can be solved by considering an ansatz based on the $R$-symmetry structure.
It produces \textit{three} constraints, which can then be used for writing the general solution
\begin{equation}
\Fm ( \lbrace \chi; r,s,t \rbrace ) = \Fds + \sum_{i=1}^3 \Dds_i f_i (\chi_1, \chi_2)\,,
\label{eq:FivePoint_SolutionWI}
\end{equation}
where we have defined the differential operators
\begin{equation}
\begin{split}
\Dds_1 &:= (v_1 + v_2) + v_1 v_2\, (\pd_{\chi_1} + \pd_{\chi_2})\,, \\
\Dds_2 &:= (w_1 + w_2) + w_1 w_2\, (\pd_{\chi_1} + \pd_{\chi_2})\,, \\
\Dds_3 &:= z + \chi_{12} ( v_1 v_2 - w_1 w_2 )\, (\pd_{\chi_1} + \pd_{\chi_2})\,,
\end{split}
\label{eq:FivePoint_DifferentialOperators}
\end{equation}
with the shorthand variables
\begin{equation}
v_i := \chi_1 - \zeta_i\,, \quad w_i := \chi_2 - \eta_i\,, \quad z := \chi_{12}^2 - t\,.
\label{eq:FivePoint_WIVariables}
\end{equation}
In terms of the $R$-symmetry channels given in \eqref{eq:FivePoint_RSymmetryChannels} and \eqref{eq:FivePoint_RSymmetryChannels2}, the $f$-functions read
\begin{equation}
\begin{split}
f_1 (\chi_1, \chi_2) &= - \frac{1}{\chi_1} F_1 (\chi_1, \chi_2) + \frac{1}{1-\chi_1} F_2 (\chi_1, \chi_2)\,, \\
f_2 (\chi_1, \chi_2) &= - \frac{1}{\chi_2} F_3 (\chi_1, \chi_2) + \frac{1}{1-\chi_2} F_4 (\chi_1, \chi_2)\,, \\
f_3 (\chi_1, \chi_2) &= - \frac{1}{\chi_{12}^2} F_5 (\chi_1, \chi_2)\,.
\end{split}
\label{eq:FivePoint_LittlefFromF}
\end{equation}
Note that $F_0$ does not appear in this decomposition.
This is because the topological sector \eqref{eq:FivePoint_TopologicalSector} fully eliminates one channel.

It may be possible to perform a more general analysis of the Ward identities and their solutions.
We reserve this for future work.

\subsubsection[Revisiting the \\ weak coupling\enspace]{Revisiting the weak coupling}
\label{subsubsec:RevisitingTheWeakCoupling}

The solution \eqref{eq:FivePoint_SolutionWI} to the Ward identity reduces the number of functions from \textit{six} $R$-symmetry channels to just \textit{three} functions $f_i$ and one constant $\Fds$, which can be computed from localization.

Our weak-coupling results \eqref{eq:FivePoint_LO} and \eqref{eq:FivePoint_NLOGroup1}-\eqref{eq:FivePoint_NLOGroup2} can be expressed in terms of these functions.
The leading order gives
\begin{equation}
\begin{alignedat}{2}
f_1^{(0)} (\chi_1, \chi_2) &= - \frac{1}{\chi_1}\,, \qquad && f_2^{(0)} (\chi_1, \chi_2) = \frac{1}{1-\chi_2}\,, \\
f_3^{(0)} (\chi_1, \chi_2) &= - \frac{1}{\chi_{12}^2}\,,\qquad && \Fds^{(0)} = 3\,.
\end{alignedat}
\label{eq:FivePoint_LittlefLO}
\end{equation}
Notice the following relations:
\begin{equation}
\begin{split}
f_1^{(0)} (\chi_1, \chi_2) & = - f_2^{(0)} (1-\chi_2, 1-\chi_1)\,, \\
f_3^{(0)} (\chi_1, \chi_2) & = f_3^{(0)} (1-\chi_2, 1-\chi_1)\,.
\end{split}
\label{eq:FivePoint_Crossing}
\end{equation}
These equations hold non-perturbatively and follow from the crossing symmetry of the correlator, given by
\begin{equation}
\Fm ( \chi_1, \chi_2 ; r_1, r_2, s_1, s_2, t ) = \Fm ( 1 - \chi_2, 1 - \chi_1 ; s_2, s_1, r_2, r_1, t )\,.
\label{eq:FivePoint_CrossingSymmetry}
\end{equation}
At the next-to-leading order, the solution of the Ward identity reads
\begin{equation}
\begin{split}
f_1^{(1)} (\chi_1, \chi_2) =\ &- \frac{1}{8\pi^2} \biggl( \ell^{(2)} \left( \frac{\chi_1}{\chi_2} \right) + \ell^{(1)} \left( \frac{\chi_1}{\chi_2} \right) \\
&+ \frac{\chi_1}{\chi_{21}} \ell^{(1)} \left( \frac{\chi_{21}}{1-\chi_1} \right) + \frac{\pi^2}{6} \biggr)\,, \\
f_2^{(1)} (\chi_1, \chi_2) =\ & - f_2^{(1)} (1-\chi_2, 1-\chi_1)\,, \\
f_3^{(1)} (\chi_1, \chi_2) =\ & - \frac{1}{8 \pi^2 \chi_{12}^2} \biggl( \ell^{(2)} \left( \frac{\chi_{21}}{1-\chi_1} \right) + \ell^{(1)} \left( \frac{\chi_{21}}{1-\chi_1} \right) \\
&+ (\chi_1 \leftrightarrow 1-\chi_2) + \frac{\pi^2}{6}
\biggr)\,, \\
\Fds^{(1)} =\ & \frac{7}{48}\,.
\end{split}
\label{eq:FivePoint_LittlefNLO}
\end{equation}
The analysis done in this section will now be used for bootstrapping this correlator in the strong-coupling regime.

\section{Multipoint correlators at strong coupling}
\label{sec:MultipointCorrelatorsAtStrongCoupling}

We have gone a long way in this chapter.
We started with multipoint correlation functions at weak coupling, from which we conjectured non-perturbative Ward identities.
This culminates now into a strong-coupling computation of the five-point function $\vev{\Oh_1 \Oh_1 \Oh_1 \Oh_1 \Oh_2}$, presenting preliminary results of \cite{Barrat:2023ta2}.

\subsection{Superconformal blocks}
\label{subsec:SuperconformalBlocks}

\begin{figure}
\centering
\begin{subfigure}{.5\textwidth}
  \centering
  \OPESymmetric
\end{subfigure}%
\begin{subfigure}{.5\textwidth}
  \centering
  \OPEAsymmetric
\end{subfigure}
\caption{The two possible ways of performing OPEs in the correlator $\vev{\Oh_1 \Oh_1 \Oh_1 \Oh_1 \Oh_2}$.
The left figure shows the symmetric channel \eqref{eq:FivePoint_BlockExpansionSymmetric}, while the asymmetric one given in \eqref{eq:FivePoint_BlockExpansionAsymmetric} is displayed on the right.
The black bold lines refer to the way the OPEs are being performed.
Red lines refer to the product of the OPE \eqref{eq:OPE_B1B1}, while the orange line corresponds to \eqref{eq:OPE_B1B2}.
}
\label{fig:OPESymmetricAsymmetric}
\end{figure}

Superconformal blocks play a crucial role in bootstrap algorithms.
In the case of the four-point functions \cite{Ferrero:2021bsb}, the superblocks were obtained with the help of the Ward identities \eqref{eq:FourPointFunctions_WardIdentities}, bypassing the use of Casimir equations \cite{Liendo:2016ymz}.
We perform a similar analysis here and show that the five-point blocks can also be entirely fixed with the Ward identities.

We consider here two ways of performing the OPEs, which are represented in Figure \ref{fig:OPESymmetricAsymmetric}.
One can either consider two pairs of $\Oh_1 \times \Oh_1$ and leave $\Oh_2$ untouched, or first perform $\Oh_1 \times \Oh_2$.
We call these channels respectively \textit{symmetric} and \textit{asymmetric}.\footnote{It might seem from Figure \ref{fig:OPESymmetricAsymmetric} that these names should be inverted.
They refer to the fact that in the symmetric channel, the two OPEs are the same ($\Oh_1 \times \Oh_1$), while in the asymmetric one, they are distinct.}
They lead to two different block expansions, which we discuss now.

\subsubsection{The symmetric channel\enspace}
\label{subsubsec:TheSymmetricChannel}

The OPE between two half-BPS operators $\Oh_1$ results in the decomposition
\begin{equation}
\Dm_1 \times \Dm_1 \longrightarrow \mathds{1} + \Dm_2 + \sum_{\Dh>1} \Lm_{0,[0,0]}^{\Dh}\,.
\label{eq:OPE_B1B1}
\end{equation}
Here we have used group-theoretical labels to designate the operators.
$\Dm_{\Dh}$ refers to half-BPS operators (not necessarily single-trace), while the operators contained in $\Lm_{0,[0,0]}^{\Dh}$ are unprotected longs with spin $s=0$ and $R$-symmetry charge $[a,b] = [0,0]$ (see \cite{Liendo:2018ukf} for more details about the notation).

This leads to the expansion in superblocks
\begin{equation}
\begin{split}
\Fm ( \lbrace \chi; r,s,t \rbrace ) =\ & \lambda_{112}^{\phantom{2}} ( \Gm_{\mathds{1}, \Dm_2} + \Gm_{\Dm_2, \mathds{1}} )
+ \lambda_{112}^2 \lambda_{222}^{\phantom{2}} \Gm_{\Dm_2, \Dm_2} \\[.8em]
&+ \sum_{\Dh} \lambda_{\vphantom{\Dh}112} \lambda_{11\Dh} \lambda_{22\Dh} ( \Gm_{\Dm_2, \Lm_{\Dh}} + \Gm_{\Lm_{\Dh}, \Dm_2} ) \\
&+ \sum_{\Dh_1, \Dh_2} \lambda_{11\Dh_1} \lambda_{11\Dh_2} \lambda_{2 \Dh_1 \Dh_2} \Gm_{\Lm_{\Dh_1},\Lm_{\Dh_2}}\,,
\end{split}
\label{eq:FivePoint_BlockExpansionSymmetric}
\end{equation}
where we omitted the dependence on the cross-ratios on the right-hand side to streamline the notation, and with $\Lm_{\Dh} := \Lm_{0,[0,0]}^{\Dh}$.

The superblocks themselves take the form \eqref{eq:ConformalBlocks_FivePointForm} and are discussed in Appendix \ref{subsec:CombChannel}.
Importantly, their form is fully fixed by \eqref{eq:MultipointWardIdentities}, up to an overall normalization constant.
This illustrates the power of the superconformal Ward identities.

Note that the expansion in superblocks largely truncates in the topological sector.
In particular, all the contributions from the longs drop out and we are left with
\begin{equation}
\Fds = 2\lambda_{112}^{\phantom{2}}  \lambda_{112}^2 \lambda_{222}^{\phantom{2}}\,.
\label{eq:FivePoint_TopologicalSectorSuperblocks1}
\end{equation}

\subsubsection{The asymmetric channel\enspace}
\label{subsubsec:TheAsymmetricChannel}

We now turn our attention to the asymmetric channel.
In this case, we need in addition to \eqref{eq:OPE_B1B1} the following OPE:
\begin{equation}
\Dm_1 \times \Dm_2 \longrightarrow \Dm_1 + \Dm_3 + \sum_{\Dh} \Lm_{0,[0,1]}^{\Dh}\,.
\label{eq:OPE_B1B2}
\end{equation}
This results in an expansion in superblocks of the following form:
\begin{equation}
\begin{split}
\Fm ( \lbrace \chi; r,s,t \rbrace ) =\ & \lambda_{112}^{\phantom{2}} \Gm_{\mathds{1}, \Dm_1}
+ \lambda_{112}^3 \Gm_{\Dm_2, \Dm_1}
+ \lambda_{112}^2 \lambda_{123}^{\phantom{2}} \Gm_{\Dm_2, \Dm_1} \\[.8em]
&+ \sum_{\Dh_1} \left(
\lambda_{11\Dh_1}^2 \Gm_{\Lm_{\Dh_1}, \Dm_1}
+ \lambda_{11\Dh_1} \lambda_{13\Dh_1} \lambda_{\vphantom{\Dh}123} \Gm_{\Lm_{\Dh_1}, \Dm_3}
\right) \\
&+ \sum_{\Dh_2} \lambda_{\vphantom{\Dh}112}^{\phantom{2}} \lambda_{12\Dh_2}^2 \Gm_{\Dm_2, \Lm_{\Dh_2}} \\
&+ \sum_{\Dh_1, \Dh_2} \lambda_{11\Dh_1} \lambda_{12\Dh_2} \lambda_{1\Dh_1\Dh_2} \Gm_{\Lm_{\Dh_1} \Lm_{\Dh_2}}\,.
\end{split}
\label{eq:FivePoint_BlockExpansionAsymmetric}
\end{equation}
Here the subscripts in $\Dh_1$ and $\Dh_2$ are more than a label; they are a shorthand notation to refer to the long operators $\Lm_{0,[0,0]}^{\Dh_1}$ and $\Lm_{0,[0,1]}^{\Dh_2}$, which have different $R$-charges.

The blocks can be obtained using the same method as for the symmetric channel.
Meanwhile, the topological sector in this channel reads
\begin{equation}
\Fds = \lambda_{112}^{\phantom{2}} + \lambda_{112}^2 + \frac{3}{2} \lambda_{112}^2 \lambda_{123}^{\phantom{2}}\,.
\label{eq:FivePoint_TopologicalSectorSuperblocks2}
\end{equation}

\subsection{Bootstrapping the correlator}
\label{subsec:BootstrappingTheCorrelator}

We are now ready to initiate the bootstrap calculation of the correlator $\vev{\Oh_1 \Oh_1 \Oh_1 \Oh_1 \Oh_2}$ at strong coupling.
In this section, we give the results for the leading and next-to-leading orders, while the higher-loop orders will be presented in \cite{Barrat:2023ta2}.

\subsubsection{Generalized free-field theory\enspace}
\label{subsubsec:GeneralizedFreeFieldTheory}

We need as an input for the higher orders the CFT data of the leading order, which corresponds to \textit{generalized free-field} (GFF) theory.
It is easy to compute the corresponding Witten diagrams, which are
\begin{equation}
\begin{split}
&\WittenDiagramFivePointLOOne \quad \WittenDiagramFivePointLOTwo \quad \WittenDiagramFivePointLOThree \\
&\WittenDiagramFivePointLOFour \quad \WittenDiagramFivePointLOFive \quad \WittenDiagramFivePointLOSix\,,
\end{split}
\end{equation}
where the AdS propagators correspond to \cite{Giombi:2017cqn}
\begin{equation}
\ScalarPropagatorStrong = \frac{\sqrt{\lambda}}{2 \pi^2} \frac{\delta^{IJ}}{\tau_{12}^2}\,.
\end{equation}
This leads to the $R$-symmetry channel decomposition (see \eqref{eq:FivePoint_RSymmetryChannels})
\begin{equation}
F_j^{(0)} = \sqrt{2}\,,
\end{equation}
which in terms of the solution of the Ward identity \eqref{eq:FivePoint_SolutionWI} translates into
\begin{equation}
\begin{alignedat}{2}
& f_1^{(0)} = - \sqrt{2} \frac{1-2\chi_1}{\chi_1 (1-\chi_1)}\,, \qquad &&
f_2^{(0)} = - \sqrt{2} \frac{1-2\chi_2}{\chi_2 (1-\chi_2)}\,, \\
& f_3^{(0)} = - \frac{\sqrt{2}}{\chi_{12}^2}\,, && \mathds{F} = 6 \sqrt{2}\,.
\end{alignedat}
\end{equation}

This result can be expanded in the superconformal blocks introduced in Section \ref{subsec:SuperconformalBlocks}.
For the symmetric channel, we find the following OPE coefficients:
\begin{equation}
\begin{split}
&\left. \lambda_{112}^{\phantom{2}} \right. = \sqrt{2}\,, \qquad\qquad \lambda_{112}^2 \lambda_{222}^{\phantom{2}} = 4 \sqrt{2}\,, \\[.8em]
&\left. \lambda_{112\vphantom{\Dh}} \lambda_{11 \Dh} \lambda_{22 \Dh} \right|_{\Dh \text{ even}} = \frac{\sqrt{\pi} (\Dh - 1) \Gamma(\Dh + 3)}{ 2^{\frac{4 \Dh + 1}{2}} \Gamma \left(\Dh + \frac{3}{2} \right)}\,, \\
&\left.  \lambda_{11\Dh_1} \lambda_{11\Dh_2} \lambda_{2 \Dh_1 \Dh_2} \right|_{\Dh_{1,2} \text{ even}} = \frac{\pi (\Dh_1)_3 (\Dh_2)_3 \Gamma(\Dh_1 + \Dh_2)}{ 2^{\frac{4 (\Dh_1 + \Dh_2) + 7}{2}} \Gamma\left( \Dh_1 + \frac{3}{2} \right) \Gamma\left( \Dh_2 + \frac{3}{2} \right)}\,.
\end{split}
\end{equation}
Note that the OPE coefficients with odd $\Dh$ vanish.

From the asymmetric channel, we obtain
\begin{equation}
\begin{split}
&\left. \lambda_{112}^{\phantom{2}} \right. = \sqrt{2}\,, \qquad \lambda_{112}^3 = 2\sqrt{2}\,, \qquad \lambda_{112}^2 \lambda_{123}^{\phantom{2}} = 4\sqrt{2}\,, \\
&\left. \lambda_{11\Dh_1}^2 \right|_{\Dh_1 \text{ even}} = \frac{\sqrt{\pi} (\Dh_1 - 1) \Gamma(\Dh_1 + 3)}{ 2^{\frac{4 \Dh_1 + 3}{2}} \Gamma \left(\Dh_1 + \frac{3}{2} \right)}\,, \\
&\left. \lambda_{112\vphantom{\Dh}}^{\phantom{2}}\lambda_{12\Dh_2}^2 \right|_{\Dh_2 \text{ even}} = \frac{ \sqrt{\pi} (\Dh_2 - 2)^2 ( \Dh_2 )_6 \Gamma (\Dh_2) }{2^{\frac{4 \Dh_2 + 7}{2}} 3 (\Dh_2 + 1) (\Dh_2 + 4) \Gamma \left( \Dh_2 + \frac{3}{2} \right) }\,, \\
&\left. \lambda_{11\Dh_1}\lambda_{12\Dh_2} \lambda_{1 \Dh_1 \Dh_2} \right|_{\substack{ \Dh_1 \text{ even} \\ \Dh_2 > \Dh_1}} = 2\sqrt{2} \Dh_{21} (\Dh_1 - 1)_{3} \\
&\qquad\qquad\qquad\qquad\qquad \times\frac{ \Gamma (\Dh_1 + 3) \Gamma (\Dh_2 - 1) \Gamma (\Dh_1 + \Dh_2 + 4) }{ \Gamma ( 2 \Dh_1 +3 ) \Gamma ( 2 \Dh_2 +3 ) }\,,
\end{split}
\end{equation}
where all the other OPE coefficients are zero.

\subsubsection{First correction\enspace}
\label{subsubsec:FirstCorrection}

We now bootstrap the next order following the method sketched in Section \ref{subsec:FromWeakToStrongCoupling}.
Our starting point is the crossing symmetry given in \eqref{eq:FivePoint_CrossingSymmetry} that relates the different functions $f_j (\chi_1, \chi_2)$.
Additional \textit{braiding} constraints can be obtained by considering different orderings of the operators in the correlator.
In this case, the functions are equal up to a phase.

We begin by writing an ansatz for $f_1$ and $f_3$\footnote{Recall that $f_2$ is fixed by \eqref{eq:FivePoint_CrossingSymmetry}.} based on the OPE limits.
They take the form
\begin{equation}
\begin{split}
f_1 (\chi_1, \chi_2) =\ & p_1 (\chi_1, \chi_2)
+ r_1 (\chi_1, \chi_2) \log \chi_1
+ r_2 (\chi_1, \chi_2) \log \chi_2 \\
&+ r_3 (\chi_1, \chi_2) \log (1-\chi_1)
+ r_4 (\chi_1, \chi_2) \log (1-\chi_2) \\
&+ r_5 (\chi_1, \chi_2) \log \chi_{21}\,,
\end{split}
\label{eq:FivePoint_Ansatzf1}
\end{equation}
and
\begin{equation}
\begin{split}
f_3 (\chi_1, \chi_2) =\ & p_3 (\chi_1, \chi_2)
+ s_1 (\chi_1, \chi_2) \log \chi_1
+ s_2 (\chi_1, \chi_2) \log \chi_2 \\
&+ s_3 (\chi_1, \chi_2) \log (1-\chi_1)
+ s_4 (\chi_1, \chi_2) \log (1-\chi_2) \\
&+ s_5 (\chi_1, \chi_2) \log \chi_{21}\,,
\end{split}
\label{eq:FivePoint_Ansatzf3}
\end{equation}
where the functions $p_i$ and $r_j$ are \textit{rational} functions.

The crossing and braiding symmetries result in the following constraints on the rational functions present in \eqref{eq:FivePoint_Ansatzf1} and \eqref{eq:FivePoint_Ansatzf3}:
\begin{equation}
\begin{split}
p_1 (\chi_1, \chi_2) &= - p_1 (1-\chi_1, 1-\chi_2)\,, \\
r_1 (\chi_1, \chi_2) &= - r_3 (1-\chi_1, 1-\chi_2)\,, \\
r_2 (\chi_1, \chi_2) &= - r_4 (1-\chi_1, 1-\chi_2)\,, \\
r_5 (\chi_1, \chi_2) &= - r_5 (1-\chi_1, 1-\chi_2)\,,
\end{split}
\label{eq:FivePoint_Constraints1}
\end{equation}
and
\begin{equation}
\begin{split}
p_3 (\chi_1, \chi_2) &= p_3 (\chi_2, \chi_1)\,, \\
p_3 (\chi_1, \chi_2) &= p_3 (1-\chi_2, 1-\chi_1)\,, \\
s_2 (\chi_1, \chi_2) &= - s_1 (\chi_2, \chi_1)\,, \\
s_3 (\chi_1, \chi_2) &= - s_1 (1-\chi_1, 1-\chi_2)\,, \\
s_4 (\chi_1, \chi_2) &= - s_1 (1-\chi_2, 1-\chi_1)\,, \\
s_5 (\chi_1, \chi_2) &= - s_5 (\chi_2, \chi_1)\,, \\
s_5 (\chi_1, \chi_2) &= - s_5 (1-\chi_2, 1-\chi_1)\,.
\end{split}
\label{eq:FivePoint_Constraints2}
\end{equation}
This already eliminates five functions.
Moreover, we also have relations that relate $f_1$ to $f_3$:
\begin{align}
p_1 (\chi_1, \chi_2) =\ & - \frac{1- \chi_1}{\chi_2^2} p_3 \left( \frac{\chi_1}{\chi_2}, \frac{1}{\chi_2} \right) + \frac{\chi_1}{(1-\chi_2)^2} p_3 \left( \frac{\chi_{12}}{1-\chi_2}, - \frac{\chi_2}{1-\chi_2} \right)\,, \notag \\
r_1 (\chi_1, \chi_2) =\ & - \frac{1- \chi_1}{\chi_2^2} s_1 \left( \frac{\chi_1}{\chi_2}, \frac{1}{\chi_2} \right) + \frac{\chi_1}{(1-\chi_2)^2} s_5 \left( \frac{\chi_{12}}{1-\chi_2}, - \frac{\chi_2}{1-\chi_2} \right)\,, \notag \\
r_2 (\chi_1, \chi_2) =\ & \frac{1-\chi_1}{\chi_2^2} \biggl(
s_1 \left( \frac{\chi_{21}}{\chi_2}, - \frac{1-\chi_2}{\chi_2} \right)
+  s_1 \left( - \frac{1-\chi_2}{\chi_2}, \frac{\chi_{21}}{\chi_2} \right) \notag \\
& +  s_1 \left( \frac{\chi_1}{\chi_2}, \frac{1}{\chi_2} \right) 
+  s_1 \left( \frac{1}{\chi_2}, \frac{\chi_1}{\chi_2} \right)
+  s_5 \left( \frac{\chi_1}{\chi_2}, \frac{1}{\chi_2} \right)
\biggr) \notag  \\
&+ \frac{\chi_1}{(1-\chi_2)^2} s_1 \left( - \frac{\chi_2}{1-\chi_2}, \frac{\chi_{12}}{1-\chi_2} \right)\,, \notag \\
r_5 (\chi_1, \chi_2) =\ & - \frac{1- \chi_1}{\chi_2^2} s_1 \left( \frac{\chi_{21}}{\chi_2}, - \frac{1-\chi_2}{\chi_2} \right)\notag \\
&+ \frac{\chi_1}{(1-\chi_2)^2} s_1 \left( \frac{\chi_{12}}{1-\chi_2}, - \frac{\chi_2}{1-\chi_2} \right)\,,
\label{eq:FivePoint_Constraints3}
\end{align}
from which we can eliminate four more functions.

We are thus left with \textit{three} rational functions to fix, which we choose to be $p_3$, $s_1$, and $s_5$.
Assuming a Regge behavior for the anomalous dimensions, as well as the relations \eqref{eq:FivePoint_Constraints2} and boundary conditions following from the OPE expansion, we can fix these functions up to \textit{two} constants, which correspond to $\Fds$ and $\lambda_{112}$.
These are known from localization, and we use them as input in our bootstrap routine.
$\Fds$ is given in \eqref{eq:FivePointTopologicalSector}, while $\lambda_{112}$ can be computed from \cite{Giombi:2018qox} and reads
\begin{equation}
\lambda_{112} = \sqrt{2} - \frac{3}{2\sqrt{2} \sqrt{\lambda}} - \frac{9}{16 \sqrt{2} \lambda} + \Om(\lambda^{-3/2})\,.
\label{eq:Lambda112_Localization}
\end{equation}
All in all, we find the remaining rational functions to be
\begin{equation}
\begin{split}
p_3 (\chi_1, \chi_2) &= \frac{1}{2\sqrt{2} \sqrt{\lambda}} \left( \frac{4}{\chi_1 \chi_2} + \frac{4}{(1-\chi_1)(1-\chi_2)} + \frac{19}{\chi_{12}^2} \right)\,, \\
s_1 (\chi_1, \chi_2) &= - \frac{\sqrt{2}}{\sqrt{\lambda}} \frac{\chi_1 ( \chi_1 (1 - 3 \chi_2) + 4 \chi_2^2 )}{\chi_2^2 \chi_{12}^3}\,, \\
s_5 (\chi_1, \chi_2) &= \frac{\sqrt{2}}{\sqrt{\lambda}} \left( \frac{1}{\chi_1^2} + \frac{1}{\chi_2^2} + \frac{1}{(1-\chi_1)^2} + \frac{1}{(1-\chi_2)^2}  \right)\,,
\end{split}
\label{eq:FivePoint_FixedFunctions}
\end{equation}
from which the anomalous dimensions can be found to be degenerate and give
\begin{equation}
\gammah^{(1)}_{\Dh} = \frac{\Dh (\Dh+2)}{2 \sqrt{\lambda}}\,.
\label{eq:FivePoint_gamma}
\end{equation}

This concludes our computation of the next-to-leading order using bootstrap techniques.
The full functions $f_j$ can be reconstructed from \eqref{eq:FivePoint_FixedFunctions} and the constraints listed above it.
It is highly likely that the next order should also be bootstrappable, and this computation is the object of current work to appear in \cite{Barrat:2023ta2}.
The first step in this direction consists of extracting the CFT data from the correlator presented in this section, which then can be used as input in the same way as it was done here.

\section{Summary and outlook}
\label{sec:SummaryAndOutlook2}

In this chapter, we considered a family of multipoint correlation functions of scalar fields in the $1d$ supersymmetric Wilson-line defect CFT.
We developed an efficient algorithm that takes the form of the recursion relations given in \eqref{eq:RecursionEvenLO_Diagrams}, \eqref{eq:RecursionOddLO_Diagrams}, \eqref{eq:RecursionNLOBulk_Diagrams} and \eqref{eq:RecursionNLOBdry_Diagrams}, and which are valid up to the next-to-leading order.
This can be used to extract correlation functions of half-BPS operators as well as of certain unprotected operators built of elementary scalars.

Although these new perturbative results are interesting in their own right, they also give access to a more general piece of information: multipoint Ward identities.
We conjectured a set of differential equations that annihilate all our perturbative correlators.
They take the form (see \eqref{eq:MultipointWardIdentities})
\begin{equation*}
\sum_{k=1}^{n-3} \left( \frac{1}{2} \partial_{\chi_k} + \alpha_k \partial_{r_k} - (1-\alpha_k) \partial_{s_k} \right) \, \Fm ( \lbrace \chi; r,s,t \rbrace) \raisebox{-2ex}{$\Biggr 
|$}_{\raisebox{1.5ex}{$\begin{subarray}{l} r_i \to \alpha_i \chi_i \\
 s_i \to (1-\alpha_i)(1-\chi_i)
 \\ t_{ij} \to \alpha_{ij} \chi_{ij}\end{subarray}$}} = 0\,.
\end{equation*}
These equations constitute a natural generalization of the Ward identities for four-point functions, which are known to exist in several superconformal setups.
Ward identities play a crucial role in the superconformal bootstrap program, and in Section \ref{sec:MultipointCorrelatorsAtStrongCoupling} we used them for bootstrapping a five-point function at strong coupling up to the next-to-leading order.

There are many interesting directions in which to further develop the techniques presented here.
It should be possible to push the bootstrap results of Section \ref{sec:MultipointCorrelatorsAtStrongCoupling} up to next-to-next-to-next-to-leading order, similarly to the four-point function story (see \eqref{eq:FourPointFunction_Littlef_Strong} and \cite{Ferrero:2021bsb}).
Moreover, analogous techniques can be used for computing the six-point function $\vev{\Oh_1 \Oh_1 \Oh_1 \Oh_1 \Oh_1 \Oh_1}$, whose structure was introduced in \eqref{eq:SixPoint_ReducedCorrelator}.
In the following, we present in some detail other selected follow-ups, which we feel deserve to be explored.

\subsection{Multipoint correlators of fermions}
\label{subsec:MultipointCorrelators of Fermions}

A natural next step is to include more general operators in our recursion formulas, such as fields that transform non-trivially under transverse rotations.
An example of such operators is fermions.
We discuss here briefly two ways of including fermions in our analysis.

\subsubsection{Sketching a fermionic recursion relation\enspace}
\label{subsubsec:SketchingAFermionicRecursionRelation}

In Section \ref{subsec:CorrelatorsOfFundamentalScalars}, we have discussed the following key aspects of the recursion relations:
\begin{enumerate}
\item at leading order, the basic ingredient of the formulas \eqref{eq:RecursionEvenLO_Diagrams} and \eqref{eq:RecursionOddLO_Diagrams} is the scalar \textit{two-point} function;
\item at next-to-leading order, the basic ingredient of \eqref{eq:RecursionNLOBulk_Diagrams} and \eqref{eq:RecursionNLOBdry_Diagrams} is the scalar \textit{four-point } function.
\end{enumerate}
There is no inherent reason why these principles would not apply to fermionic correlators.
We provide here a sketch of how a recursion relation for fermions would look like.

At the leading order, we expect the building block of the fermionic pendent to \eqref{eq:RecursionEvenLO_StartingValues} to simply be the two-point diagram
\begin{equation}
\DefectTwoFermionsYOne\,.
\label{eq:DefectTwoFermions_LO}
\end{equation}

At the next-to-leading order, the four-point functions of fermions involve both bulk and boundary diagrams, as already seen in \eqref{eq:FourPointFunction_NLODiagramsBulk}-\eqref{eq:FourPointFunction_NLODiagramsBdry} for the scalar case.
Specifically, the bulk diagrams in the channel $(12)(34)$ are
\begin{equation}
\DefectFFFFOneLoopSEOne\ +\
\DefectFFFFOneLoopSETwo\ +\
\DefectFFFFOneLoopHOne\ +\
\DefectFFFFOneLoopHTwo\,.
\label{eq:DefectTwoFermions_NLOBulk}
\end{equation}
Note that similar diagrams should be considered for the channel $(14)(23)$.

Meanwhile, the boundary diagrams contributing to the fermionic four-point functions are given by
\begin{equation}
\DefectFFFFOneLoopYOne\ +\
\DefectFFFFOneLoopYTwo\ +\
\DefectFFFFOneLoopYThree\ +\
\DefectFFFFOneLoopYFour\,.
\label{eq:DefectTwoFermions_NLOBdry}
\end{equation}
These diagrams are anticipated to cancel each other pairwise, similarly to the self-energy diagrams of the Wilson line as observed for instance in \cite{Erickson:2000af}.

We are therefore left with bulk diagrams only.
The integrals associated with these diagrams are expected to all be computable, using techniques such as the fermionic star-triangle identity provided in \eqref{eq:StarTriangle}.
We encounter the first three diagrams in the next chapter.
Furthermore, the fourth one is expected to follow the same structure as the insertion rule (B.7) in \cite{Barrat:2020vch}.

It would be interesting to set up a recursion relation for fermions, building upon the loose ideas presented in this section.
Note that a complete recursion relation should encompass scalar-scalar-fermion-fermion correlators as well.
We do not see additional conceptual challenges in incorporating these correlators, as the associated integrals should be similar to the ones encountered in \eqref{eq:RecursionNLOBulk_Diagrams} and \eqref{eq:RecursionNLOBdry_Diagrams}.

\subsubsection[Correlators \\ of superfields\enspace]{Correlators of superfields}
\label{subsubsec:CorrelatorsOfSuperfields}

There is an alternative perspective for considering the correlation functions of scalars and fermions.
The essence of supersymmetry lies in the interplay between \textit{bosonic} and \textit{fermionic} degrees of freedom.
Fermions are not decoupled from scalar correlators, and it would be desirable to formulate our recursion relations in a way that incorporates this intrinsic property.

This aim can be achieved by upgrading the scalar fields to \textit{superfields}:
\begin{equation}
\phi \longrightarrow \Phi\,.
\label{eq:PromotingphitoPhi}
\end{equation}
Superfields depend on both ordinary spacetime coordinates and additional Grassmann variables associated with the fermionic degrees of freedom.
The matter sector of the theory is described by \textit{chiral} superfields, whereas the gauge fields and their associated superpartners, the gauginos, are encoded in \textit{vector} superfields.

A formulation of correlators at next-to-leading order in terms of superfields involves the following steps.
First, the Feynman rules \eqref{eq:Propagators}-\eqref{eq:QuarticVertices} need to be promoted to a set of Feynman rules for superfields, following the corresponding formulation of the $\Nm=4$ SYM action.
As we are dealing with a defect theory, the second step would be to do the same for the defect Feynman rules \eqref{eq:DefectVertices_OnePoint}.
Perhaps the superfield formulation of the supersymmetric Wilson line presented in \cite{Muller:2013rta,Beisert:2015jxa,Beisert:2015uda} could be relevant for this purpose.
Finally, recursion relations should be developed in the gist of \eqref{eq:RecursionEvenLO_Diagrams}, \eqref{eq:RecursionOddLO_Diagrams}, \eqref{eq:RecursionNLOBulk_Diagrams} and \eqref{eq:RecursionNLOBdry_Diagrams}.
Based on the Feynman diagrammatic approach conducted above, we anticipate that the integrals will not be significantly more challenging, as the correlators of chiral superfields focus on the matter content of the theory.
In other words, the building blocks of a superfield recursion relation should remain closely connected to our recursion relations.

\subsection{More Ward identities}
\label{subsec:MoreWardIdentities}

One highlight of this chapter is the conjecture of superconformal Ward identities for multipoint correlators.
The efficient computation of weak-coupling correlators led to the determination of non-perturbative constraints.
Although these identities can in principle be properly obtained via superspace techniques, we believe that it would have been cumbersome to derive the general formula \eqref{eq:MultipointWardIdentities}.

It is natural to wonder whether this experimental method can be extended to other cases.
We start this discussion by sketching a roadmap for $4d$ $\Nm=4$ SYM, before discussing a possible web of connections between our Ward identities and different setups, which are connected by symmetry.

\subsubsection{$\Nm=4$ Super Yang--Mills\enspace}
\label{subsubsec:Nm4SuperYangMills}

In Section \ref{subsec:MultipointWardIdentities}, we discussed the fact that it was unexpected to find a set of superconformal Ward identities only acting on the highest-weight correlator of the supersymmetric multiplets.
It would be interesting to investigate whether this feature is a consequence of the one-dimensional setup and whether it extends to higher-dimensional systems or not.

One ideal playground for conducting such an experiment is of course the parent theory of the Wilson-line defect CFT: $4d$ $\Nm=4$ SYM.
A similar formula to \eqref{eq:RecursionNLOBulk_Diagrams} and \eqref{eq:RecursionNLOBdry_Diagrams} was developed in \cite{Drukker:2008pi} for multipoint correlators of half-BPS operators, and therefore the pool of results already exists in principle.
However, in this case, the number of cross-ratios is $n(n-3)/2$ both for spacetime and $R$-symmetry.
For instance, four-point functions depend on \textit{four} variables in total, and they satisfy the elegant superconformal Ward identities
\begin{equation}
\begin{split}
\left( \pd_\chi + \pd_w ) \Fm (\chi,\chib, w,\wb) \right|_{w = \chi} = 0\,, \\
\left( \pd_{\chib} + \pd_{\wb} ) \Fm (\chi,\chib, w,\wb) \right|_{\wb = \chib} = 0\,,
\end{split}
\label{eq:N4SYM_WardIdentities}
\end{equation}
reminiscent of \eqref{eq:AlternativeWardIdentities}.

Although the multipoint Ward identities are expected to be more complicated than \eqref{eq:N4SYM_WardIdentities}, perhaps they can be formulated similarly:
\begin{equation}
\sum_{k=1}^{n(n-3)/2} \left( a_k \pd_{\chi_k} + b_k \pd_{\chib_k} + c_k \pd_{r_k} + d_k \pd_{s_k} \right) \Fm ( \lbrace \chi, \chib; r,s,t \rbrace) \raisebox{-2ex}{$\Biggr 
|$}_{\raisebox{1.5ex}{$\begin{subarray}{l} r_i \to \alpha_i \chi_i \\
 s_i \to \beta_i \chib_i
 \\ t_{ij} \to ?\end{subarray}$}} = 0\,.
\label{eq:N4SYM_MultipointWIAnsatz}
\end{equation}
Additional differential operators might be needed to account for the extra cross-ratios, although, in the $1d$ case, there was no need for a derivative with respect to $t$.
It should be straightforward to write an educated ansatz for the Ward identities and use the results of \cite{Drukker:2008pi} to fix the prefactors -- if such identities exist.

\subsubsection[A network of Ward \\ identities?\enspace]{A network of Ward identities?}
\label{subsubsec:ANetworkOfWardIdentities}

\begin{figure}[t!]
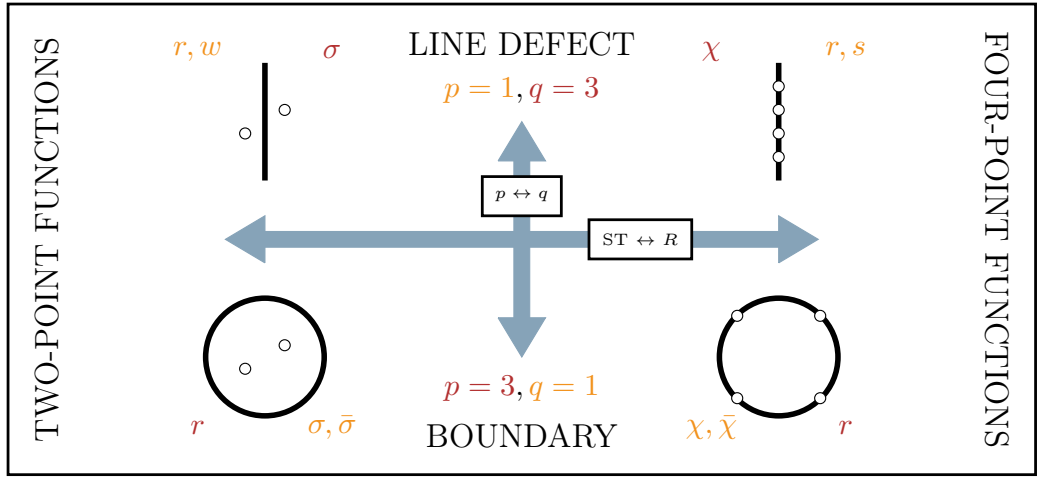

\centering
\AnalyticContinuation
\caption{A network of analytic continuations between the Ward identities of different systems was identified in \cite{Liendo:2016ymz} and is depicted in this figure.
The vertical direction corresponds to exchanging $p \leftrightarrow q$ (i.e., line and boundary defects), while in the horizontal direction, we go from two-point functions in the presence of the defect (left) to four-point functions of defect operators (right).}
\label{fig:AnalyticContinuation}
\end{figure}

We conclude this chapter with an intriguing set of dualities between specific setups.
It was noted below \eqref{eq:AlternativeWardIdentities} that the superconformal Ward identity for four-point functions can be analytically continued between two-point functions of bulk half-BPS operators in the presence of a half-BPS line and four-point functions of defect half-BPS operators along the same line.
The superconformal blocks associated with these setups can also be analytically continued, following a precise but straightforward dictionary.
This connection was established in \cite{Liendo:2016ymz}, where further correspondences were found.
Indeed, the Ward identities \eqref{eq:AlternativeWardIdentities} also capture the superconformal symmetry of two-point functions of half-BPS operators in the presence of a half-BPS \textit{boundary}, as well as four-point functions of half-BPS operators in $3d$ $\Nm=4$ gauge theory, i.e., living on the boundary.
One can navigate between the different setups by exchanging the roles of spacetime and $R$-symmetry variables in the appropriate way (see Figure \ref{fig:AnalyticContinuation}).

This leads to a natural question: can our multipoint Ward identities \eqref{eq:MultipointWardIdentities} also be analytically continued across setups, and if yes where do they take us?
The same question holds for the superblocks of $\vev{\Oh_1 \Oh_1 \Oh_1 \Oh_1 \Oh_2}$, which were derived in Section \ref{subsec:SuperconformalBlocks}.

While we do not provide a definitive answer to these questions here, we can speculate on the outcome and rule out certain scenarios.
Since the analytic continuation between the setup of Chapter \ref{chapter:BootstrappingHolographicDefectCorrelators} and the four-point function of defect operators \eqref{eq:FourPointFunctions_ReducedCorrelator} consists in exchanging spacetime and $R$-symmetry cross-ratios, let us count the cross-ratios for different setups.
For simplicity, we limit ourselves to $R$-symmetry cross-ratios for now, but of course, a full analysis of the setups should eventually be performed.

The first proposal would be to compare $n$-point functions of defect operators and $(n-2)$-point functions of bulk operators in the presence of the Wilson line.
However the counting does not quite work: if we observe an agreement at $n=4$, the number of cross-ratios grows differently for higher $n$:
\begin{align}
\vev{\Oh_{\Dh_1} \ldots \Oh_{\Dh_n}}: & \quad n-3 \text{ spacetime cross-ratios}\,, \\
\vvev{\Op_{\Delta_1} \ldots \Op_{\Delta_{n-2}}}: & \quad \binom{2}{n-2} \text{ $R$-symmetry cross-ratios}\,.
\label{eq:ComparisonOfCrossRatios}
\end{align}

But there are more correlators that one can consider.
Another possibility is to consider a mix of bulk and defect operators.
For example, three-point functions of two bulk and one defect operators depend on \textit{two} $R$-symmetry cross-ratios -- just like our five-point functions depend on two spacetime cross-ratios.
Perhaps this is where our Ward identities and superblocks should be analytically continued.
It would be interesting to investigate whether such a web of analytic continuations exists, and if yes figure out its inner workings.

\chapter{Line defect correlators in fermionic CFT}
\label{chapter:LineDefectCorrelatorsInFermionicCFT}

The study of supersymmetric theories in Chapters \ref{chapter:BootstrappingHolographicDefectCorrelators} and \ref{chapter:MultipointCorrelatorsInTheWilsonLineDefectCFT} has provided valuable insights into line defect CFTs.
We now leave this realm to study models which have closer connections to the real world.
We consider in particular the Yukawa universality classes, which can be expressed as the action \eqref{eq:ActionYukawa}, involving $N$ scalars and $N_f$ fermions.
These models are important quantum field theories, as they possess interesting characteristics reminiscent of the Standard Model, such as asymptotic freedom and dynamical symmetry breaking.
Moreover, they exhibit \textit{emergent supersymmetry}, which has the potential of being observable experimentally \cite{Lee:2006if,Grover:2013rc,Fei:2016sgs,Rong:2019qer,Liendo:2021wpo,Prakash:2023qno}.

There has been significant interest in recent years in studying the phase transitions of these models \textit{across dimensions}.
These universality classes have been suggested to capture a wide range of condensed-matter systems that exhibit emergent Lorentz symmetry.
A notable example is the case of graphene sheets, three-dimensional systems conjectured to be described by \eqref{eq:ActionYukawa} at the critical point.
Graphene undergoes a second-order phase transition, structurally very similar to the weak sector of the Standard Model.
This resemblance has led to the hope that graphene could mimic various symmetry breakings observed in the Standard Model, such as chiral or spontaneous symmetry breaking \cite{Semenoff:2011jf,hatsugai2013chiral,Zhang:2015rme,gutierrez2016imaging,bao2021experimental,pant2021emergence}.
Although an experimental realization of phase transitions in graphene is yet to be achieved, the idea of having a low-energy laboratory for testing phase transitions in the Standard Model is a fascinating prospect.

For these reasons, it is important to determine precisely the critical exponents of the Yukawa universality classes.
Estimations have been made using various techniques, such as the conformal bootstrap \cite{Iliesiu:2015qra,Iliesiu:2017nrv,Bobev:2015vsa,Bashkirov:2013vya}, Monte Carlo simulations \cite{Chandrasekharan:2013aya,Wang:2014cbw,Li:2014aoa,Otsuka:2015iba,Huffman:2017swn,Li:2017sbk,Otsuka:2018kcb,Huffman:2019efk} and the large $N$ expansion \cite{Gracey:1990wi,Gracey:1992cp,Gracey:2017fzu,Manashov:2017rrx,Prakash:2023qno}.
Another powerful method for obtaining analytical results across dimensions is the $\veps$-expansion.
In this context, purely scalar models with $O(N)$ symmetry (obtained by setting $g=0$ in \eqref{eq:ActionYukawa}) have been extensively explored \cite{kleinert2001critical,Kos:2014bka,Kos:2015mba,Kos:2016ysd}, and the interpolation of results between the $2+\veps$ and the $4-\veps$ expansions has yielded consistent results with the other methods aforementioned.

In the context of these scalar models, a defect CFT can be defined by introducing a localized magnetic field, also commonly referred to as a \textit{pinning line} defect.
These defects effectively simulate impurities localized in space, and their presence can be implemented in lattice formulations through the activation of a background field.
These types of magnetic defects are therefore very natural from an experimental standpoint, and indeed they are expected to be observable in nature (see \cite{ParisenToldin:2016szc,Cuomo:2021kfm} and references therein).
It was shown recently that these line defects can also be defined in the universality classes \eqref{eq:ActionYukawa}, and that they can be continued across dimensions \cite{Giombi:2022vnz}.

In this chapter, we focus on what can be considered the two canonical configurations in defect CFT, which have already been explored for $\Nm=4$ SYM: the four-point functions of defect operators, and the two-point functions of bulk operators located outside the defect.
For the $O(N)$ model, perturbative correlators of this type have been computed in \cite{Gimenez-Grau:2022czc,Gimenez-Grau:2022ebb}, and here we extend this analysis to encompass fermions.
The study of fermions across different dimensions is interesting not only from an experimental perspective but also as a captivating theoretical problem.
This is due to the distinct nature of fermions, which are very different objects in three and four dimensions.
Related implications will be explored throughout this chapter.

The content presented here is based on \cite{Barrat:2023ivo} and contains preliminary results to appear in \cite{Barrat:2023ta3}.

\section{An invitation}
\label{sec:AnInvitation3}

As usual, we begin this chapter with an introduction to the subject matter.
It was pointed out in \cite{Giombi:2022vnz} that the universality classes defined in \eqref{eq:ActionYukawa} are physically meaningful if and only if $N \leq 2 N_f + 4$ \cite{Zhou:2022fdn}.
Moreover, it is unclear whether the $O(N)$ symmetry can be preserved in three dimensions beyond $N=3$.
For these reasons, and although the computations will be performed for arbitrary $N$ and $\Nf$, we present in this section three specific models corresponding to the cases $N=1,2,3$.

\subsection{The Gross--Neveu--Yukawa model}
\label{subsec:TheGNYModel}

We begin with the model corresponding to a single scalar field ($N=1$).
The corresponding theory is called the \textit{Gross--Neveu--Yukawa} (GNY) theory, which finds its origin in the study of purely fermionic models.
Here we review how the latter can be UV-completed in the form of a Yukawa quantum field theory, and discuss the RG flow of the model.

\subsubsection{The fermionic model\enspace}
\label{subsubsec:TheFermionicModel}

Scalar quartic vertices play a fundamental role in our understanding of spontaneous symmetry breaking, with significant implications in the Standard Model.
They are central to the Higgs mechanism, responsible for the masses of elementary particles.
One may wonder whether \textit{fermionic} quartic interactions can also lead to interesting physical phenomena.
One of the most studied models in this family is the \textit{Gross--Neveu} (GN) model, a purely fermionic theory featuring a quartic self-interaction \cite{Gross:1974jv}.
In Euclidean space, the GN model is described by the action
\begin{equation}
S_{\text{GN}} = \int d^d x \biggl( i \psib^a \spd \psi^a + \frac{g^2}{2} (\psib^a \psi^a)^2 \biggr)\,,
\label{eq:ActionGN}
\end{equation}
where $a = 1, \ldots, N_f$.
Remarkably, this theory exhibits asymptotic freedom, akin to quantum chromodynamics (QCD).
As the energy scale increases, the interactions in the GN model weaken.
Consequently, it serves as a simpler and more tractable prototype for investigating this feature.

\subsubsection{UV completion\enspace}
\label{subsubsec:UVCompletion}

However, it appears that the GN model is renormalizable only in two dimensions.\footnote{To be more precise, the GN model can be renormalized order by order in the large $N_f$ limit, while for a finite value of $N_f$ it is not renormalizable at $d>2$.}
In four dimensions, it can only be considered an effective field theory, similar to the Fermi model used to describe weak interactions before the discovery of the $W$- and $Z$-bosons \cite{Fermi:1933jpa}.
Fortunately, it is possible to provide an ultraviolet completion of the GN model by introducing an auxiliary Hubbard-Stratonovich scalar field $\phi$ and incorporating a Yukawa interaction \cite{Gross:1974jv}.
This extended model is known as the \textit{Gross--Neveu--Yukawa} (GNY) model, characterized by the following action:
\begin{equation}
S_{\text{GNY}} = \int d^d x \left( \frac{1}{2} \pd_\mu \phi \pd_\mu \phi + \frac{m^2}{2} \phi^2 + i \psib^a \spd \psi^a + g \psib^a \phi \psi^a + \frac{\lambda}{4!} \phi^4 \right)\,.
\label{eq:ActionGNY}
\end{equation}
This action corresponds to the massive case of \eqref{eq:ActionYukawa} with $N=1$, choosing $\phi$ to be a scalar (as opposed to a \textit{pseudo}scalar) by setting $\Sigma^1 = \mathds{1}$ as in \eqref{eq:Sigma1}.

\subsubsection[\\ The critical point\enspace]{The critical point}
\label{subsubsec:TheCriticalPoint3}

There exists a critical value of $m^2$ below which spontaneous symmetry breaking occurs.
The scalar field $\phi$ acquires a non-zero vacuum expectation value (VEV), leading to the fermions becoming massive and consequently to the breaking of parity.
Above the critical value, the VEV of $\phi$ vanishes, the fermions remain massless and parity is preserved.

The phase transition occurs at $m^2 = 0$ and can be studied across dimensions in the context of the $\veps$-expansion.
Setting $d=4-\veps$, the \textit{Wilson--Fisher--Yukawa} (WFY) fixed point (see \eqref{eq:WFY_FixedPoint}) takes \eqref{eq:ActionGNY} from a \textit{trivial} theory at $\veps=0$ to a \textit{strongly-coupled} model at $\veps=1$.
The regime $\veps \sim 0$ has been studied extensively \cite{Zinn-Justin:1991ksq,Herbut:2006cs,Herbut:2009qb,Mihaila:2017ble,Zerf:2017zqi,Herbut:2023xgz}, and the resulting estimations of the critical exponents extrapolated to $\veps = 1$ lie close to results from the conformal bootstrap and Monte Carlo simulations (see for instance Table \ref{table:Universality}).

As mentioned earlier, the GNY model is believed to exhibit emergent supersymmetry.
This phenomenon happens at $N_f = \frac{1}{4}$, where the bosonic and fermionic degrees of freedom are matching.
This feature has been studied through perturbation theory \cite{Fei:2016sgs}, but is believed to hold non-perturbatively.

\begin{figure}[t]
\centering
  \includegraphics[width=.45\linewidth]{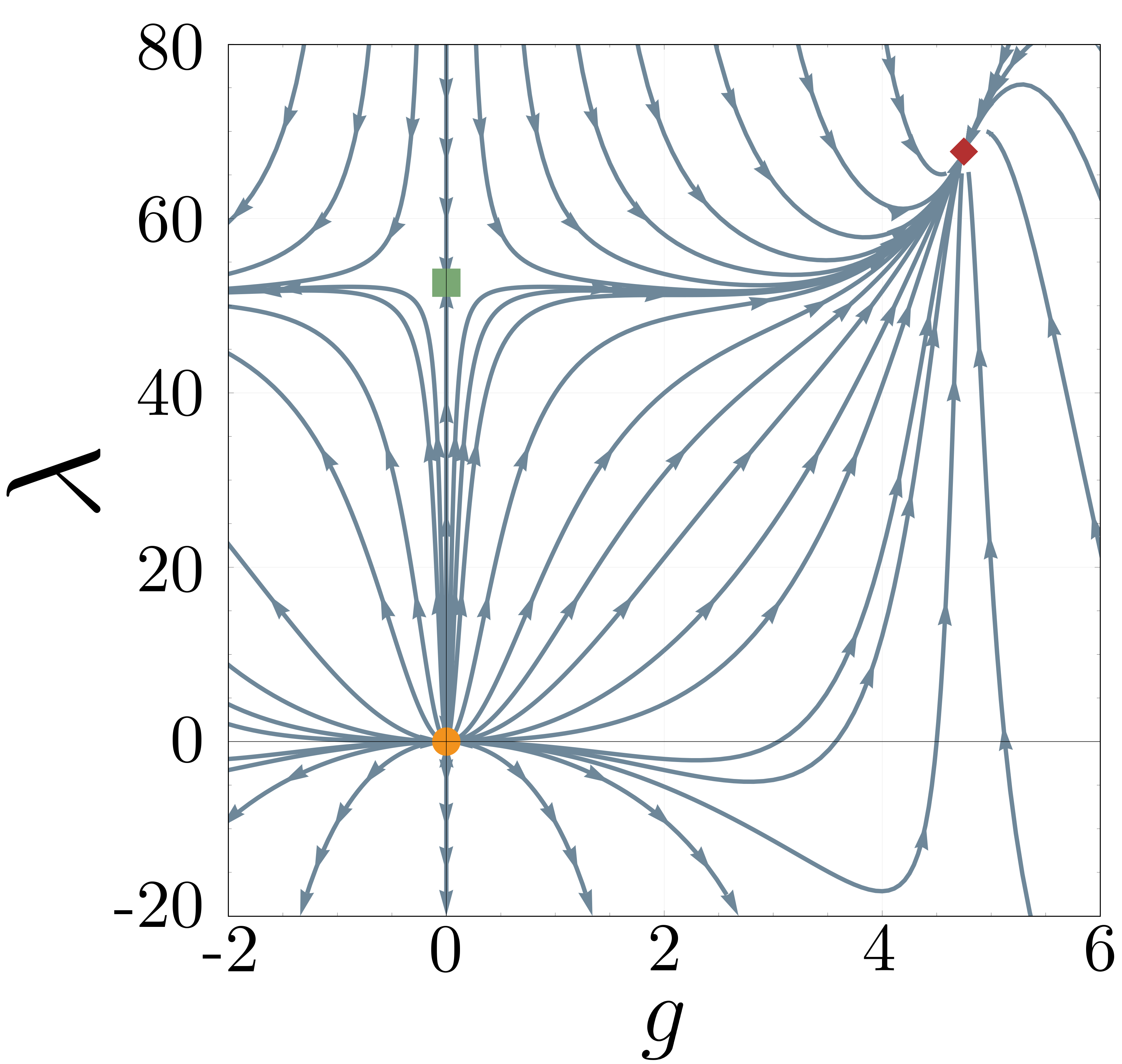}
\caption{This plot shows the RG flow of the GNY model presented in \eqref{eq:ActionGNY} for $N_f = \frac{1}{4}$ at $\Om(\veps)$.
Although perturbative, this plot is expected to describe qualitatively the behavior of the theory in \textit{three} dimensions.
Three fixed points are identified: the trivial UV fixed point at $(g, \lambda) = (0,0)$ (orange circle in the plot), the Ising model at $(0, \frac{16 \pi^2}{3} \veps)$ (green square), while the Wilson--Fisher--Yukawa value is $(\frac{4 \pi}{\sqrt{7}} \veps, \frac{48 \pi^2}{7} \veps)$ (red diamond) and corresponds to the emergent supersymmetry case.
}
\label{fig:PlotRGFlow}
\end{figure}

\subsection{The NJLY and chiral Heisenberg models}
\label{subsec:TheNJLYModelAndChiralHeisenbergModels}

We discuss now two additional models corresponding to the cases $N=2,3$.
These models are the \textit{Nambu--Jona-Lasinio--Yukawa} (NJLY, $N=2$) and the \textit{chiral Heisenberg} (cH, $N=3$) models. 

\subsubsection{The NJLY model\enspace}
\label{subsubsec:TheNJLYModel}

Another notable model featuring four-fermion interactions is the \textit{Nambu--Jona-Lasinio} (NJL) model, which was originally developed as a dynamical framework for understanding the mass generation of nucleons, drawing an analogy with superconductivity \cite{Nambu:1961tp,Nambu:1961fr}.
The NJL model is similar to the GN model \eqref{eq:ActionGN}, but incorporates an additional $\gamma^5$-vertex:
\begin{equation}
S_{\text{NJL}} = \int d^d x \left( i \psib^a \spd \psi^a + \frac{g^2}{2} \left[ (\psib^a \psi^a)^2 - (\psib^a \gamma^5 \psi^a)^2 \right] \right)\,.
\label{eq:ActionNJL}
\end{equation}

Similarly to how the GN model can be UV-completed with an auxiliary scalar field, the universality class of the NJL model near three dimensions can be described by the NJLY model
\begin{equation}
\begin{split}
S_{\text{NJLY}} =\ & \int d^d x \biggl( \frac{1}{2} \pd_\mu \phi^I \pd_\mu \phi^I + \frac{m^2}{2} \phi^I \phi^I + i \psib^a \spd \psi^a \\
&+ g \psib^a \Sigma^I \phi^I \psi^a + \frac{\lambda}{4!} (\phi^I \phi^I)^2 \biggr)\,,
\end{split}
\label{eq:ActionNJLY}
\end{equation}
which corresponds to \eqref{eq:ActionYukawa} for the case $N=2$ with $m \neq 0$.
The matrix $\Sigma^I$ is chosen to be
\begin{equation}
\Sigma^I =
\begin{cases}
\mathds{1} & \text{ for } I=1\,, \\
i \gamma^5 & \text{ for } I=2\,.
\end{cases}
\label{eq:SigmaNJLY}
\end{equation}
In other words, $\phi^1$ is a scalar field, while $\phi^2$ corresponds to a pseudoscalar.

At $N_f=\frac{1}{2}$, this model has also been shown to possess emergent supersymmetry.
In this case, the fermionic sector of \eqref{eq:ActionNJLY} consists of a single Dirac fermion in three dimensions.
The emergence of supersymmetry in the NJLY model has been investigated for instance in \cite{Fei:2016sgs}.

\subsubsection{The chiral Heisenberg model\enspace}
\label{subsubsec:ThecHModel}

We conclude this section with the cH model, an example of a theory described by \eqref{eq:ActionYukawa} with $N=3$.
This model consists of three real scalar fields, resulting in an $O(3)$ symmetry in addition to the $U(N_f)$ symmetry of the fermions.
The action of the cH model can be expressed as \eqref{eq:ActionNJLY}, setting $I=1,2,3$, and with the matrix $\Sigma$ defined as
\begin{equation}
\Sigma^I = \sigma^I \otimes \mathds{1}_2\,.
\label{eq:SigmacH}
\end{equation}
Interestingly, the $O(3)$ symmetry appears to be broken by \eqref{eq:SigmacH} at $d=4$,\footnote{To be more precise, the breaking of symmetry arises from the kinetic term of the fermions, and results from the fact that it is not possible to find a matrix $\Sigma$ that either commutes or anti-commutes with all \textit{four} matrices $\gamma^{I=0,1,2,3}$. This issue is resolved at $d=3$, where $\Sigma$ only needs to (anti)commute with \textit{three} $\gamma$ matrices.} although it is restored at $d=3$.

As for the other models, results for the cH model have been derived from the $\veps$-expansions by interpolating between the estimations for the critical exponents for $d=2+\veps$ and $d=4-\veps$.
Note that, at $d=2+\veps$, it has a description known as the $SU(2)$ GN model, where the fermion bilinear is contracted with a Pauli matrix (see for instance \cite{Gracey:2018qba}).

Supersymmetry is expected to emerge at $N_f = \frac{3}{4}$ \cite{Fei:2016sgs}.
More generally, the supersymmetry can only arise at $N = 4 N_f$ to guarantee that the number of bosonic and fermionic degrees of freedom match.

\subsection{The magnetic-line defect CFT}
\label{subsec:TheMagneticLineDefectCFT}

We consider now the general action \eqref{eq:ActionYukawa} augmented by the inclusion of a magnetic line defect.
This defect consists of a single scalar field, which is chosen to be $\phi^1$ following the conventions spelled out in Section \ref{subsec:TheMagneticLine}.
Throughout this chapter, $\phi^1$ assumes a role analogous to that of $\phi^6$ in the supersymmetric stories of Chapters \ref{chapter:BootstrappingHolographicDefectCorrelators} and \ref{chapter:MultipointCorrelatorsInTheWilsonLineDefectCFT}, with the notable difference that the Yukawa CFTs \eqref{eq:ActionYukawa} are not gauge theories.

\subsubsection{Codimension across dimensions\enspace}
\label{subsubsec:CodimensionAcrossDimensions}

It was recently pointed out that all the models of Sections \ref{subsec:TheGNYModel} and \ref{subsec:TheNJLYModelAndChiralHeisenbergModels} admit a magnetic line defect, which can be studied perturbatively across dimensions.
This defect is characterized by an exponential of a scalar field integrated along a line, akin to the Wilson line and explicitly given in \eqref{eq:Sdefect}.
In four dimensions, a (free) scalar field has dimension $\Delta_\phi \sim 1$, rendering the defect coupling marginal.
Consequently, this defect represents a compelling candidate for the description of a \textit{non-trivial} defect CFT in $d=4-\veps$ dimensions.
It was shown in \cite{Giombi:2022vnz} that this is indeed the case.

On the $2+\veps$ side, the GN model admits a natural scalar line defect, defined as the exponential of a fermion bilinear.
In two dimensions, a free fermion possesses a scaling dimension $\Delta_\psi \sim \frac{1}{2}$, and the defect coupling is again marginal.
However, as one moves away from two dimensions, the defect CFT becomes non-trivial and is anticipated to belong to the same universality class as the defect CFT described above in $4-\veps$ dimensions.
On this basis, the extrapolation of physical quantities between the two perturbative schemes is expected to work similarly in the defect CFT as it does in the bulk theories.

This picture generalizes straightforwardly to the NJLY and cH models presented in Section \ref{subsec:TheNJLYModelAndChiralHeisenbergModels}.
In the case of the NJL model \eqref{eq:ActionNJL} at $d = 2 + \veps$, it is possible to construct two fermion bilinears, $\psib \psi$, and $\psib \gamma^5 \psi$, and the defect will be described by an exponential involving both of these terms, neatly matching the $4-\veps$ analysis \cite{Giombi:2022vnz}.

\subsubsection[Dimension across \\ dimensions\enspace]{Dimensions across dimensions}
\label{subsubsec:DimensionAcrossDimensions}

It is worth noting that, in our setup, the defect remains one-dimensional while the bulk is allowed to undergo dimensional changes.
It is in principle also possible to maintain a fixed codimension and allow the defect to vary in dimension, as is the case for monodromy defects \cite{Billo:2013jda,Gaiotto:2013nva}.\footnote{See \cite{Bianchi:2021snj} for interesting recent progress.}
The interpolation between different dimensions poses several challenges, as it is not clear how to represent the correlators.
While this issue has been recently addressed in the context of boundary CFT \cite{Carmi:2018qzm,Giombi:2021cnr,Herzog:2022jlx}, analyzing higher-codimension defects proves to be more involved.
Although we do not explore defects with varying dimensions here, it would certainly be an interesting direction to pursue in future work.

\section{Correlators of defect operators}
\label{sec:CorrelatorsOfDefectOperators}

This section is dedicated to the computation of correlation functions involving the defect operators presented in Section \ref{subsec:TheMagneticLine}.
Our analysis begins with the two- and three-point functions, where we determine normalization constants, scaling dimensions, renormalization constants, and OPE coefficients.
As opposed to the previous chapters, we include here correlators that involve fermions.
The main part of this section consists of the computation of \textit{four-point} functions of scalar and fermionic operators, separately as well as mixed.
The results are used to extract (defect) CFT data, which can be compared to the perturbative results, similarly to what was done in Chapter \ref{chapter:MultipointCorrelatorsInTheWilsonLineDefectCFT} for the $\Nm = 4$ SYM case.

\subsection{Kinematically-fixed correlators}
\label{subsec:KinematicallyFixedCorrelators2}

We now proceed to compute the two- and three-point functions of the defect operators.
We consider correlators involving the elementary scalars and fermions, as well as the displacement operator presented in \eqref{eq:Displacement}.

\subsubsection{Two scalars\enspace}
\label{subsubsec:TwoScalars}

Let us begin by examining the two-point functions of the fundamental defect scalars $\Oh^{I=1, i}$, with $i=2, \ldots, N$.
They are fixed by conformal symmetry and take the form given in \eqref{eq:ConformalTwoPointFunctions}.
At the fixed point, the Feynman diagrams are
\begin{equation}
\begin{split}
\DefectTwoScalarsYOne\ &= \frac{\delta^{IJ}}{\nh_I} I_{12} \left( 1 - \frac{\veps}{2} \left( 1 - \aleph - \log \tau_{12}^2 \right) \right)\,, \\
\DefectTwoScalarsYTwo\ &= - \frac{g^2 N_f}{4 \pi^2} \delta^{IJ}\, I_{12}
\left( \frac{1}{\veps} + \aleph + \log \tau_{12}^2 \right)\,, \\
\DefectTwoScalarsYThree\ &=  - \frac{\lambda h^2}{64 \pi^2 \nh_I} I_{12} (2 \delta^{I1} \delta^{J1} + \delta^{IJ}) \left( \frac{2}{3 \veps} - \frac{1}{3} + \aleph + \log \tau_{12}^2 \right)\,,
\end{split}
\label{eq:TwoScalars_Diagrams}
\end{equation}
where the first and third diagrams are identical to those in the pure $O(N)$ models \cite{Gimenez-Grau:2022czc}.
$I_{12}$ refers here to the $4d$ propagator.

By imposing finiteness of the correlator, we can determine the renormalization constants for $\Oh^1$ and $\Oh^i$.
For the two scalars $\Oh^1$ and $\Oh^i$, we obtain the expressions
\begin{equation}
\begin{split}
Z_1 &=
1
- \frac{1}{\veps} \frac{\lambda h^2 + 8 g^2 N_f}{64 \pi^2}
+ \Om (\veps^{-2} )\,,  \\
Z_i &=
 1
- \frac{1}{\veps} \frac{\lambda h^2 - 24 g^2 N_f}{192 \pi^2}
+ \Om (\veps^{-2} )\,.
\end{split}
\label{eq:TwoScalars_Z}
\end{equation}

As a sanity check, we can extract the scaling dimensions from the renormalization factors and compare them to \eqref{eq:DefectScalar1_Delta} and \eqref{eq:DefectTilt_Delta}.
We observe a perfect agreement:
\begin{equation}
\begin{split}
\Dh_1 &= \mu \frac{\partial \log Z_1}{\partial \mu} = 1 + \frac{4 - N}{\kappa_1} \veps + \Op(\veps^2)\,,  \\
\Dh_i &= \mu \frac{\partial \log Z_i}{\partial \mu} = 1 + \Op(\veps^2)\,.
\end{split}
\label{eq:TwoScalars_Dh}
\end{equation}

The normalization constants can be obtained from the two-point functions.
We find
\begin{equation}
\begin{split}
\nh_1 &= \frac{1}{4\pi^2} \left\lbrace 1 - \frac{\veps}{2} \left( 2 + \frac{(N - 4)(1 - 2 \aleph)}{\kappa_1} \right)  + \Op(\veps^2) \right\rbrace\,, \\
\nh_i &= \frac{1}{4\pi^2} \left\lbrace 1 - \frac{\veps}{2} \left( 2 + \frac{N - 4}{\kappa_1} \right)  + \Op(\veps^2) \right\rbrace\,.
\end{split}
\label{eq:TwoScalars_nh}
\end{equation}

\subsubsection{Two displacements\enspace}
\label{subsubsec:TwoDisplacements}

The same analysis can be applied to the two-point functions of the displacement operator.
Since the displacement can be expressed as in \eqref{eq:Displacement}, i.e.,  with a transverse derivative acting on the bulk scalars, the diagrams contributing to its two-point function are the same as those in \eqref{eq:TwoScalars_Diagrams}.
This gives
\begin{equation}
\begin{split}
\DefectTwoScalarsYOne\ &= \frac{8 \pi^2}{\nh_{\Disp}} \delta_{\mu\nu} I_{12}^2 \left( 1 - \frac{\veps}{2} \left( 2 - \aleph - \log \tau_{12}^2 \right) \right) \,,  \\
\DefectTwoScalarsYTwo\ &= - \frac{g^2 \Nf}{4 \pi^2 \nh_{\Disp}} \delta_{\mu\nu} I_{12}^2 \delta^{ab} \left( \frac{1}{\veps} + \aleph + \log \tau_{12}^2 \right) + \Om(\veps^2)\,, \\
\DefectTwoScalarsYThree\ &= - \frac{\lambda h^2}{12 \nh_{\Disp}} \delta_{\mu\nu} I_{12}^2 \left( \frac{1}{\veps} - \frac{2}{3} + \frac{3 \aleph}{2} + \frac{3}{2} \log \tau_{12}^2 \right) + \Om(\veps^2)\,,
\end{split}
\label{eq:TwoDisp_Diagrams}
\end{equation}
where here $\mu, \nu = 1,\ldots, d$ are the transverse directions.
The first and third diagrams have been previously computed in \cite{Gimenez-Grau:2022czc}.

Based on these results, we can extract the relevant quantities in the usual way.
We find
\begin{equation}
\begin{split}
Z_\Disp &= 1 
- \frac{1}{\veps} \frac{\lambda h^2 + 24 g^2 N_f }{192 \pi^2}
+ \Om\left(\veps^{-2}\right)\, , \\
\Dh_{\Disp} &= \mu \frac{\partial \log Z_{\Disp}}{\partial \mu} = 2 + \Op(\veps^2)\,, \\
\nh_{\Disp} &= \frac{1}{2 \pi ^2} \left\lbrace
1 -\veps \left(1 - \frac{N-4}{6 \kappa_1 } \right) + \Op(\veps^2) \right\rbrace\,.
\end{split}
\label{eq:TwoDisp_ZDn}
\end{equation}
We observe that the anomalous dimension of this operator vanishes as expected.

\subsubsection[Two \\ fermions\enspace]{Two fermions}
\label{subsubsec:TwoFermions}

Let us now shift our focus to the fermions. 
In $1d$, the two-point function of fermions takes the form
\begin{equation}
\vev{\Oh_{1,0}^a (\bar{s}_1, \tau_1) \Oh_{0,1}^b (s_2, \tau_2)} = \delta^{ab}\, \frac{\bar{s}_1 \gamma^0 s_2}{\tau_{12}^{2 \Dh_a}}\,,
\label{eq:TwoFermions}
\end{equation}
where we have used the polarization spinors $\bar{s}_1$ and $s_2$ defined in \eqref{eq:Spinors_PolarizationSpinors} to streamline the notation.

The diagrams are 
\begin{align}
\vev{\Oh_{1,0}^a (\bar{s}_1, \tau_1) \Oh_{0,1}^b (s_2, \tau_2)} =\ & \DefectTwoFermionsYOne\ +\ \DefectTwoFermionsYTwo\ +\ \DefectTwoFermionsYThree \notag \\
&+\ \DefectTwoFermionsYFour\ + \Om(\veps^2)\,,
\label{eq:TwoFermions_Diagrams}
\end{align}
where the integrals along the line should be understood as going from $-\infty$ to $+\infty$.
As noted in \cite{Giombi:2022vnz}, the second diagram is zero at order $\Om(\veps)$. 
The other integrals can be computed using standard techniques, and we find
\begin{equation}
\begin{split}
\DefectTwoFermionsYThree\ &= \frac{g^2 N}{32\pi^4 \nh_a} \frac{ \bar{s}_1 \gamma^0 s_2}{\tau_{12}^3} \delta^{ab} \left\lbrace \frac{1}{\veps} + \aleph -1 + \log \tau_{12}^2 + \Op(\veps) \right\rbrace\,, \\
\DefectTwoFermionsYFour\ &= - \frac{g^2 h^2}{16 \pi^6 \nh_a} \frac{\bar{s}_1 \gamma^0 s_2}{\tau_{12}^3} \delta^{ab}\,.
\end{split}
\label{eq:TwoFermions_Diagram34}
\end{equation}

In a similar manner to the scalars, we can extract the renormalization constant, scaling dimension, and normalization constant from the two-point function.
We obtain
\begin{equation}
\begin{split}
Z_a &= 1 - \frac{g^2 N}{32 \pi^2 \veps} + \Om\left(\veps^{-2}\right)\,, \\
\Dh_a &= \frac{3}{2} - \frac{\veps}{4} \left(2 - \frac{N}{\kappa_1} \right) + \Op(\veps^2)\,,  \\
\nh_a &= - \frac{1}{2\pi^2} \biggl\lbrace 1 - \frac{\veps}{2 \kappa_1} \biggl( 2 \kappa_1 - \aleph \left( 1 - \frac{N}{2 \kappa_1} \right) \\
&\phantom{=}\ + \frac{4}{\pi^2} \frac{(N-4)(N+8)}{\kappa_1 \kappa_2}  \biggr) + \Op(\veps^2) \biggr\rbrace\,.
\end{split}
\label{eq:TwoFermions_ZDn}
\end{equation}
Note that the renormalization factor, as well as the scaling dimension, agree with the bulk computation, as the diagrams contributing to these results are the same.\footnote{This agreement is expected to be lifted at higher orders of $\veps$.}

\subsubsection[Three scalars\enspace]{Three scalars}
\label{subsubsec:ThreeScalars}

We move now our attention to three-point functions.
The correlators between three scalars $\Oh^{I=1, i}$ are given by a single Feynman diagram at order $\Om(\veps)$:
\begin{align}
 \vev{ \Oh^I (\tau_1) \Oh^J (\tau_2) \Oh^K (\tau_3) } &=\ \DefectThreeScalarsY\  + \Om (\veps^2)\,,
\label{eq:ThreeScalars}
 \end{align}
which should be compared to \eqref{eq:ConformalThreePointFunctions} for extracting the OPE coefficients $\lambda_{IJK}$.
The calculation of this diagram has been carried out in \cite{Gimenez-Grau:2022czc} for the $O(N)$ model, which adapted to our setup yields the OPE coefficients
\begin{equation}
\begin{split}
\lambda_{111} &= \frac{3 \pi}{8}\frac{(4 \kappa_1 - N_f) \sqrt{2 (4-N)(N+8)\kappa_2}}{\kappa_1^2 (N+8)} \veps + \Om (\veps^2) \,, \\
\lambda_{ij1} &= \delta^{ij} \frac{\lambda_{111}}{3} + \Om (\veps^2)\,.
\end{split}
\label{eq:Scalar_lambdas}
\end{equation}
Since they start at $\Om(\veps)$, these OPE coefficients will only appear at order $\Om(\veps^2)$ in the four-point functions of scalars.

The first scalar operators that appear in the OPE $\Oh^I \times \Oh^J$, following $\Oh^1$ itself, are the degenerate operators labeled $s_{\pm}$.
These operators have dimensions close to 2 and can be constructed from $(\phi^1)^2$ and $(\phi^i)^2$ following
\begin{equation}
\begin{pmatrix}
(\phi^1)^2 \\ (\phi^i)^2
\end{pmatrix}
= Z_s
\begin{pmatrix}
s_- \\ s_+
\end{pmatrix}\,.
\label{eq:Definition_spm}
\end{equation}
To determine the correct anomalous dimensions, we impose the condition that the three-point functions involving $\Oh^I$ and $s_{\pm}$ are finite.
The connected correlator consists of a single diagram up to $\Op(\veps)$:
\begin{equation}
\vev{ \Oh^I (\tau_1) \Oh^J (\tau_2) \Oh^K \Oh^K (\tau_3) }_{\text{conn}} =\ \DefectOneScalarTwoDisplacementsYTwo\ +  \Om(\veps^2)\,.
\label{eq:lambdaIJO2}
\end{equation}
This diagram has been computed in \cite{Gimenez-Grau:2022czc}.
The conformal dimension is obtained as $\Dh_{s_{\pm}} = 2 - \veps + \gammah_{s_{\pm}}$, where the anomalous dimensions can be computed by diagonalizing this matrix.
They are given by
\begin{align}
\gammah_{s_{\pm}} =& \frac{4 (N+8) (\kappa_{1}-N+4)-\kappa_{2} (N+4) \mp \kappa_{3}}{4 \kappa_{1} (N+8)} \veps + \Om(\veps^2)\,,
\label{eq:gamma_spm}
\end{align}
where we have defined
\begin{equation}
\kappa_3 := \sqrt{\kappa_{2}^2 N^2+8 \kappa_{2} (N-4) (N-2) (N+8)+16 (N-4)^2 (N+8)^2}\:.
\label{eq:kappa3}
\end{equation}

To complete the computation of the OPE coefficients, we also need to determine the normalization of the two-point functions $\vev{ s_{\pm} (\tau_1) s_{\pm} (\tau_2) }$.
We find the following expression for the normalization factor:
\begin{equation}
\nh_{s_{\pm}} = \pm \frac{(N-1) \left(\kappa_{2} (N-2)+4 (N-4) (N+8) \pm \kappa_3 \right)}{16 \pi ^4 \kappa_3} + \Om(\veps) \,.
\label{eq:Normalization_spm}
\end{equation}
We only display the $\Om(1)$ term here, while the $\Om(\veps)$ term is lengthy and can be found in the ancillary notebook of \cite{Barrat:2023ivo}, along with the renormalization constant.
Note that the operators are properly normalized, i.e., $\vev{ s_{+} (\tau_1) s_{-} (\tau_2) } = 0$.

Combining all the results, we can determine the OPE coefficients:
\begin{equation}
\begin{split}
\lambda_{11 s_{\pm}} &= \pm \frac{2 \kappa_{2} \sqrt{N-1}}{\sqrt{\kappa_{3}^2\pm (\kappa_{2} (N-2)+4 (N-4) (N+8)) \kappa_3}} + \Om(\veps)\,, \\
 \lambda_{ij s_{\pm}} &= \delta^{ij} \frac{\sqrt{\kappa_3 \pm \kappa_{2} (N-2)+4 (N-4) (N+8)}}{\sqrt{\kappa_3}\sqrt{N-1}} + \Om(\veps)\,.
 \end{split}
\label{eq:s_lambdas}
\end{equation}
The $\Om(\veps)$ terms can be found in the \textsc{Mathematica} notebook attached to \cite{Barrat:2023ivo}.

\subsubsection[One scalar and \\ two fermions\enspace]{One scalar and two fermions}
\label{subsubsec:OneScalarAndTwoFermions}

The first example of a mixed correlator involving scalars and fermions is the three-point function $\vev{ \Oh^a_{1,0} \Oh^b_{0,1} \Oh^I }$, which, at the leading order, is given by
\begin{align}
\vev{\Oh^a_{1,0} (\bar{s}_1, \tau_1) \Oh^b_{0,1} (s_2, \tau_2) \Oh^I (\tau_3)} &=\ \DefectTwoFermionsOneScalarY\ + \Om(\veps) \notag \\
&= ( \bar{s}_1 \gamma^0 \Sigma^I s_2 ) \frac{\lambda_{abI}}{\tau_{12}^{2 \Dh_{abI}} \tau_{23}^{2 \Dh_{bIa}} \tau_{13}^{2 \Dh_{Iab}}}\,,
\label{eq:OneScalarTwoFermions}
\end{align}
where the second line indicates the form of such correlators, with $\Dh_{ijk}$ defined below \eqref{eq:ConformalThreePointFunctions}.

This diagram can be evaluated by applying the commutation rules \eqref{eq:Sigma_ScalarsAndPseudoscalars} for $\Sigma^I$ and using the fermionic star-triangle identity \eqref{eq:StarTriangle}.
The results depend on the nature of the field $\phi^I$:
\begin{equation}
\DefectTwoFermionsOneScalarY\ = \pm \frac{\bar{s}_1 \Sigma^I \gamma^0 s_2}{ \nh_a \sqrt{\nh_I} } \frac{g}{64 \pi^4 \tau_{12}^2 \tau_{23} \tau_{31}}\,,
\label{eq:OneScalarTwoFermions_Diagram}
\end{equation}
where the sign is $-$ for $\phi^I$ being a scalar and $+$ for $\phi^I$ being a pseudoscalar.
Upon inserting the normalization constants derived above, we can identify the OPE coefficient to be
\begin{equation}
\lambda_{abI} = \pm \frac{\sqrt{\veps}}{4 \sqrt{2 \kappa_1}} + \Om(\veps)\,.
\label{eq:lambdaabI}
\end{equation}

\subsection{Four-point functions}
\label{subsec:FourPointFunctions}

Four-point functions are the first correlators in our list to have non-trivial kinematics.
To extract defect CFT data, we can expand these correlators in $1d$ conformal blocks, and compare the results with the OPE coefficients derived in the previous section.
We begin by examining correlators consisting solely of scalar operators, then proceed to include fermions.
Finally, we explore an example of a mixed correlator involving both scalars and fermions.

\subsubsection{Four scalars\enspace}
\label{subsubsec:FourScalars}

We begin this section by examining the four-point functions of scalar operators.
As we already know from Chapter \ref{chapter:MultipointCorrelatorsInTheWilsonLineDefectCFT}, such correlators are one-dimensional and kinematically depend on a single cross-ratio $\chi$, defined in \eqref{eq:SpacetimeCrossRatio1d}.

The four-point functions of scalars can be written as
\begin{equation}
\vev{ \Oh^I (\tau_1) \Oh^J (\tau_2) \Oh^K (\tau_3) \Oh^L (\tau_4) } = \Km\,  F^{IJKL} (\chi)\,,
\label{eq:FourScalars}
\end{equation}
where $\Km$ represents the conformal prefactor and is given by
\begin{equation}
\Km := \frac{1}{\raisebox{-.6ex}{$\tau_{12}^{\smash{\Dh_I + \Dh_J}} \tau_{34}^{\smash{\Dh_K+\Dh_L}}$}} \left( \frac{\tau_{24}}{\tau_{14}} \right)^{\smash{\Dh_{IJ}}} \left( \frac{\tau_{14}}{\tau_{13}} \right)^{\smash{\Dh_{KL}}}\,.
\label{eq:FourScalars_Prefactor}
\end{equation}
Expressed in terms of Feynman diagrams, the (connected) four-point function of arbitrary elementary scalars is given by 
\begin{equation}
\vev{ \Oh^I (\tau_1) \Oh^J (\tau_2) \Oh^K (\tau_3) \Oh^L (\tau_4) }
  =\ \DefectSSSSOneLoopX\  + \Om(\veps^2)\,. 
  \label{eq:FourPoint_Diagram}
 \end{equation}
We have already encountered this \textit{$X$-diagram} in Section \ref{subsec:FourPointFunctionsOfHalfBPSOperators}.
Its solution can be found in \eqref{eq:X1234}.
By accounting for the disconnected parts and applying unit-normalization, we find
\begin{equation}
\begin{split}
F^{1111} (\chi) =\ &
1 + \chi^{2 \Dh_1} + \left(\frac{\chi}{1-\chi}\right)^{2 \Dh_1} + \frac{3 \kappa_2\, \veps}{\kappa_1 (N+8)} \chi \ell^{(1)} (\chi) + \Op(\veps^2)\,, \\
F^{1i1j} (\chi) =\ &
\delta^{ij} \chi^{\Dh_1 + 1} 
+ \delta^{ij} \frac{\kappa_2\, \veps}{\kappa_1 (N+8)} \chi \ell^{(1)} (\chi) + \Op(\veps^2)\,, \\
F^{ijkl} (\chi) =\ &
\delta^{ij} \delta^{kl}
+\delta^{ik}\delta^{jl} \chi ^2
+\delta^{il} \delta^{jk} \frac{\chi ^2}{(1-\chi)^2} \\
&+ (\delta^{ij} \delta^{kl} +\delta^{ik} \delta^{jl} +\delta^{il} \delta^{jk}) \frac{\kappa_2\, \veps}{\kappa_1 (N+8)} \chi \ell^{(1)} (\chi) + \Op(\veps^2)\,,
\end{split}
\label{eq:FourScalars_F}
\end{equation}
with $\kappa_1$ and $\kappa_2$ given in \eqref{eq:Kappas}, while $\ell^{(1)} (\chi)$ is the function of transcendentality $1$ that is defined in \eqref{eq:TranscendentalFunctions}.

These four-point functions can be expanded using the $1d$ conformal blocks provided in \eqref{eq:ConformalBlocks_Multipoint_Comb}.
The degenerate operators $s_{\pm}$ are the first ones to appear in the OPE $\Oh^1 \times \Oh^1$.
These operators have been unmixed in Section \ref{subsec:KinematicallyFixedCorrelators2} to obtain the anomalous dimensions $\gammah_{s_{\pm}}$ and the OPE coefficients $\lambda_{IJ s_{\pm}}$.
When considering the conformal-block expansion, we observe only the \textit{average} of the conformal data associated with these operators, starting from
\begin{equation}
\begin{split}
F^{1111} (\chi) =\ & 1 + \left(2-\frac{3 \kappa_{2}\veps }{\kappa_{1}(N+8)} \right) f_{2} (\chi) \\
&+ \veps \left(\frac{3 \kappa_{2} + 4 (N+8) (4-N)}{ \kappa_{1}(N+8)} \right) \partial_{\Delta} f_{2} (\chi) + \ldots \,.
\end{split}
\label{eq:F1111InBlocks}
\end{equation}
This leads to the relations
\begin{align}
& (\Dh_{s_+} -2)\lambda_{1 1 s_+}^2 + (\Dh_{s_{-}} - 2) \lambda_{1 1 s_{-}}^2 = \frac{3 \kappa_{2} + 4 (N+8) (4-N)}{ \kappa_{1}(N+8)} \veps + \Om(\veps^2)\,, \notag \\
& \lambda_{1 1 s_{+}}^2 + \lambda_{1 1 s_{-}}^2 = 2-\frac{3 \kappa_{2}}{\kappa_{1}(N+8)} \veps + \Om(\veps^2)\,,
\label{eq:RelationOPEs}
\end{align}
which are satisfied by \eqref{eq:s_lambdas}.

Similar considerations apply to the correlators $F^{ijkl} (\chi)$ and $F^{1I1J} (\chi)$, which also contain information about the OPE coefficients $\lambda_{ i j s_{\pm}}$.
As expected, the OPE coefficients presented in \eqref{eq:s_lambdas} also satisfy the associated relations.

We can deduce the OPE coefficients and anomalous dimensions associated with additional operators appearing in the OPE $\Oh^1 \times \Oh^i$.
This analysis can be found in \cite{Barrat:2023ivo}.

\subsubsection{Four displacements\enspace}
\label{subsubsec:FourDisplacements}

The diagrams contributing to the four-point function of displacements are identical to those for the four-point function of scalars, with the connected one being depicted in \eqref{eq:FourPoint_Diagram}.
To obtain the result, we differentiate with respect to the transverse coordinates as before.

The diagrams that do not involve fermions have already been calculated for the $O(N)$ model in \cite{Gimenez-Grau:2022czc}, while the fermionic diagram corresponds to the renormalization of the wavefunction.
We find
\begin{equation}
\vev{ \Disp_{\mu} (\tau_1) \Disp_\nu (\tau_2) \Disp_\rho (\tau_3) \Disp_\sigma (\tau_4) } = \frac{F_{\mu\nu\rho\sigma} (\chi)}{\tau_{12}^{2 \Dh_{\Disp}} \tau_{34}^{2 \Dh_{\Disp}}}\,,
\label{eq:FourDisplacements}
\end{equation}
with
\begin{align}
F_{\mu\nu\rho\sigma} (\chi) =\ &
\delta_{\mu\nu} \delta_{\rho\sigma}
+ \delta_{\mu\rho} \delta_{\nu\sigma} \chi^4
+ \delta_{\mu\sigma} \delta_{\nu\rho} \frac{\chi ^4}{(1-\chi)^4}  \notag \\
&+ \veps
(\delta_{il} \delta_{jk}+\delta_{ik} \delta_{jl}+\delta_{ij} \delta_{kl}) \frac{\kappa_2}{10 \kappa_1 (N+8)} \frac{\chi}{(1-\chi)^3} \notag \\
& \phantom{+} \times
\left( 2 \chi (1-\chi) (\chi(1-\chi)-1)
+ \chi^3 (\chi (5 - 2\chi) - 5) H_0 \right.  \notag \\
& \left. \phantom{+ \times (}
+ (1-\chi)^3 (2\chi^2 + \chi + 2) H_1
\right) \notag \\
&
+\Op(\veps^2)\,.
\label{eq:FourDisplacements_Result}
\end{align}
The functions $H_{\vec{a}} := H_{\vec{a}} (\chi)$ are the HPL functions encountered before and defined in \eqref{eq:HPL_Definition}.

\subsubsection{Four fermions\enspace}
\label{subsubsec:FourFermions}

We consider now correlators involving elementary fermions.
The four-point function can be expressed as
\begin{align}
\vev{ \Oh_{1,0}^a \Oh_{0,1}^b \Oh_{1,0}^c \Oh_{0,1}^d } = \Km \left( F^{abcd}_{12,34} (\chi) - \frac{\chi^3}{(1-\chi)^3} F^{adcb}_{14,32} (1-\chi) \right)\,,
\label{eq:FourFermions}
\end{align}
where we have omitted the dependence on $\bar{s}$, $s$ and $\tau$ on the left-hand side for compactness.
Since all the external dimensions are equal, we can set the conformal prefactor to be
\begin{equation}
\Km := \frac{1}{\tau_{12}^{2\Dh_a} \tau_{34}^{2\Dh_a}}\,.
\label{eq:FourFermions_Prefactor}
\end{equation}
In \eqref{eq:FourFermions}, the second term follows from crossing symmetry, and the flavor structure is encoded in $F^{abcd}_{12,34}$.
The subscripts indicate the dependence on the polarization spinors $\bar{s}_1, s_2, \bar{s}_3, s_4$.

The perturbative computation consists of the following diagrams for the connected part:
\begin{align}
\vev{ \Oh_{1,0}^a \Oh_{0,1}^b \Oh_{1,0}^c \Oh_{0,1}^d }_{\text{conn}} =\ \DefectFourFermionsHOne\ +\ \DefectFourFermionsHTwo\ + \Om(\veps^2)\,.
\label{eq:FourFermions_Diagrams}
\end{align}
The disconnected part of the correlator can be easily computed and yields
\begin{align}
\vev{ \Oh_{1,0}^a \Oh_{0,1}^b \Oh_{1,0}^c \Oh_{0,1}^d }_{\text{disc}} =\ & \frac{1}{\tau_{12}^{2\Dh_a} \tau_{34}^{2\Dh_a}}
\biggl\lbrace (\bar{s}_1 \gamma^0 s_2) (\bar{s}_3 \gamma^0 s_4) \delta^{ab} \delta^{cd} \notag \\
&- \frac{\chi^3}{(1-\chi)^3} (\bar{s}_1 \gamma^0 s_4) (\bar{s}_3 \gamma^0 s_2) \delta^{ad} \delta^{bc} \biggr\rbrace\,.
\label{eq:FourFermions_Disconnected}
\end{align}

The connected part of the correlator consists of \textit{$H$-diagrams}.
The first one can be expressed as
\begin{equation}
\DefectFourFermionsHOne\ = g^2 \delta^{ab} \delta^{cd} (\bar{s}_1 \Sigma^I \slashed{\partial}_1 \slashed{\partial}_2 s_2) (\bar{s}_3 \Sigma^I \slashed{\partial}_3 \slashed{\partial}_4 s_4) H_{12,34}\,.
\label{eq:FourFermions_Diagram1}
\end{equation}
We have used the rules given in \eqref{eq:Sigma_Identities} to move the $\Sigma$-matrices in front.
The integral $H_{12,34}$ is defined in \eqref{eq:H1234} and has not been solved analytically yet, even in $1d$. 
In Chapter \ref{chapter:MultipointCorrelatorsInTheWilsonLineDefectCFT}, we encountered the $H$-integrals with derivatives acting on them, and we were able to compute these combinations analytically.
The same magic happens here and, with the help of the fermionic star-triangle relation given in \eqref{eq:StarTriangle}, the sum of the two diagrams yield
\begin{equation}
\begin{split}
F^{abcd}_{12,34} (\chi) =\ & \delta^{ab} \delta^{cd} (\bar{s}_1 \gamma^0 s_2)(\bar{s}_3 \gamma^0 s_4) \\
&+ \frac{\veps}{64 \kappa_1} \delta^{ab} \delta^{cd} (\bar{s}_1 \Sigma^I \gamma^0 s_2)(\bar{s}_3 \Sigma^I \gamma^0 s_4) \frac{\chi}{(1-\chi)^2} \\
& \phantom{-\ } \times \left( (1-\chi)(2-\chi) + \chi^2 (2-\chi) H_0 - \chi (1-\chi)^2 H_1 \right) \\
&+ \Op(\veps^2)\,.
\end{split}
\label{eq:FourFermions_Result}
\end{equation}

We can extract new defect CFT data from this correlator by expanding it in the $1d$ blocks of \eqref{eq:ConformalBlocks_Multipoint_Comb}.
Since we have $N_f$ fermions, there is a $U(N_f)$ flavor symmetry, and we need to decompose the fermions into the \textit{singlet} ($S$) and \textit{adjoint} ($\text{Adj}$) representations:
\begin{align}
F^{abcd}(\chi) =\ & \delta^{ab} \delta^{cd} F^{abcd}_{S} (\chi) + \left(\delta^{ad} \delta^{bc} - \frac{\delta^{ab}\delta^{cd}}{N_f}\right) F^{abcd}_{\text{Adj}} (\chi)\,.
\label{eq:FourFermions_Decomposition}
\end{align}

For simplicity, we focus on the singlet sector and decompose it into conformal blocks.
For the first few operators, we obtain
\begin{align}
F^{abcd}_{S} (\chi) =\ & f_{0} (\chi) + \frac{\veps}{32 \kappa_1} f_{1} (\chi) - \left( \frac{1}{N_f} + \frac{2 (\kappa_1 + N) - 17}{384 N_f \kappa_1} \veps \right) f_{3} (\chi) \notag \\
 &+ \frac{ \kappa_1 + N - 3 }{64 N_f \kappa_1} \veps\, \partial_{\Dh} f_{3} (\chi) + \ldots\,.
\label{FourFermions_S_BlockExpansion}
\end{align}
The coefficient $\lambda^2_{ab1}$ can be read off as the coefficient in front of the block $f_{\Dh = 1} (\chi)$, which matches the expression in \eqref{eq:lambdaabI}.
The absence of a conformal block $f_{\Dh = 2} (\chi) $ indicates that 
\begin{equation}
\lambda_{a b s_{\pm}}^2 = \Om(\veps^2)\,.
\label{eq:lambdaabs}
\end{equation}

\subsubsection[Two scalars and two fermions\enspace]{Two scalars and two fermions}
\label{subsubsec:TwoScalarsAndTwoFermions}

We conclude our analysis of correlators of defect operators with the mixed four-point functions involving two Dirac fermions $\Oh_{1,0}^a$, $\Oh_{0,1}^b$ and two elementary scalars $\Oh^I$, $\Oh^J$, which we assume to be identical for simplicity.
The correlator can be expressed as
\begin{equation}
\vev{ \Oh_{1,0}^a \Oh^I \Oh^J \Oh_{0,1}^b } = \frac{\delta^{ab}}{\tau_{14}^{2\Dh_a} \tau_{23}^{2\Dh_I}} F^{IJ} (\chi)\,,
\label{eq:TwoScalarsTwoFermions}
\end{equation}
where the $O(N)$ tensor structure is encoded in $F^{IJ}$.

As before, the disconnected part of the correlator can be easily obtained and contains only one non-vanishing term:
\begin{align}
\vev{ \Oh_{1,0}^a \Oh^I \Oh^J \Oh_{0,1}^b }_{\text{disc}} = \delta^{ab} \delta^{IJ} \frac{(\bar{s}_1 \gamma^0 s_4)}{\tau_{14}^{2\Dh_a} \tau_{23}^{2\Dh_I}}\,.
\label{eq:TwoScalarsTwoFermions_Disconnected}
\end{align}

The connected part of the correlator consists of two fermion-scalar $H$-diagrams:
\begin{equation}
\vev{ \Oh_{1,0}^a \Oh^I \Oh^J \Oh_{0,1}^b }_{\text{conn}} =\ \DefectFFSSHOne\ + \DefectFFSSHTwo\ + \Om(\veps^2)\,.
\label{eq:TwoScalarsTwoFermions_Diagrams}
\end{equation}
By using the rules \eqref{eq:Sigma_Identities} to commute the $\Sigma$-matrices, we find that the first diagram yields
\begin{equation}
\DefectFFSSHOne\ = \pm g^2 \delta^{ab} (\bar{s}_1\, \Sigma^I \Sigma^J\, \spd_1 (\spd_1 + \spd_4) \spd_3) H_{12,34}\, s_4)\,.
\label{eq:TwoScalarsTwoFermions_HDiagram1}
\end{equation}
The sign is positive if $\phi^I$ is a scalar and negative if it is a pseudoscalar,\footnote{In this formulation, the index $J$ can be kept arbitrary since we have to commute $\Sigma^J$ with $\gamma$ matrices \textit{twice}.} and where the integral can be solved in $1d$ using the identities of Appendix \ref{subsec:OneLoopIntegrals}.

The second diagram gives the same results, and the unit-normalized correlator reads
\begin{equation}
\begin{split}
F^{IJ}(\chi) =\ & \delta^{IJ} (\bar{s}_1 \cdot s_4) \\
&\pm (\bar{s}_1 \Sigma^I \Sigma^J s_4) \frac{\veps}{8 \kappa_1} 
\left(
\frac{\chi^4}{(1-\chi)^2} H_0 + \chi (2+\chi) H_1
\right) \\
&+ \Op(\veps^2)\,.
\end{split}
\label{eq:TwoScalarsTwoFermions_Result}
\end{equation}

Expanding this correlator using the $1d$ blocks \eqref{eq:ConformalBlocks_Multipoint_Comb} for the case of equal external operators, such that 
\begin{equation}
\Sigma^I = \Sigma^J\,, \quad \Sigma^I \Sigma^J = \mathds{1}\,,
\label{eq:TwoScalarsTwoFermions_SigmaCondition}
\end{equation}
we find
\begin{equation}
F^{II} (\chi) =  f_{0}(\chi) +  \frac{\veps}{4 \kappa_1} f_{2} (\chi) + \frac{19\veps}{240 \kappa_1} f_{4} (\chi) - \frac{\veps}{8 \kappa_1} \partial_{\Delta} f_{4} (\chi) + \ldots \:,
\label{eq:TwoScalarsTwoFermions_BlockExpansion}
\end{equation}
where we emphasize that no summation is implied by the repetition of indices on the left-hand side.
Since both the correlator and the block expansion share the same expression for $\Oh^1$ and $\Oh^i$, we obtain identical relations for the OPE coefficients $\lambda_{1 1 \Op}$ and $\lambda_{i j \Op}$, which we denote as $\lambda_{I I \Op}$ for brevity.

For $s_{\pm}$, which has dimension $\Dh_{s_{\pm}} \sim 2$, the following relations hold:
\begin{equation}
\begin{split}
&\lambda_{ab s_{+}} \lambda_{II s_{+}} + \lambda_{ab s_{-}} \lambda_{II s_{-}}  = \frac{\veps}{4 \kappa_1} + \Om(\veps^2)\,, \\
&(\Dh_{s_{+}} - 2)\lambda_{a b s_{+}} \lambda_{II s_{+}} + (\Dh_{s_{-}} - 2)\lambda_{a b s_{-}} \lambda_{II s_{-}} = \Om(\veps^2)\,.
\end{split}
\label{eq:TwoScalarsTwoFermion_Relations}
\end{equation}
By employing the expressions for $\Dh_{s_{\pm}}$ and $\lambda_{11 s_{\pm}}, \lambda_{i j s_{\pm}}$ from \eqref{eq:s_lambdas}, we can deduce the OPE coefficients involving the fermions:
\begin{equation}
\begin{split}
\lambda_{a b s_{\pm}} =\ & \frac{(N\kappa_2 - 4(N-4)(N+8) \pm \kappa_{3})}{16 \kappa_{1} \kappa_{2} \sqrt{\kappa_{3}} \sqrt{N-1}} \notag \\
& \times \sqrt{\kappa_3 \pm 4 (N-4)(N+8) \pm (N-2)\kappa_2} \veps
+ \Om(\veps^2) \,.
\end{split}
\label{eq:lambdaabs2}
\end{equation}
As expected from the result \eqref{eq:lambdaabs} of the conformal block expansion, the OPE coefficients start at $\Om(\veps)$.

\section{Correlators of bulk operators with a defect}
\label{sec:CorrelatorsOfBulkOperatorsWithADefect}

The three- and four-point functions of scalars and fermions on the defect provided us with important defect CFT data.
We can further explore the properties of the defect CFT by considering correlators of \textit{bulk} operators in the presence of the line defect.
These operators were introduced in Section \ref{subsec:TheWilsonFisherYukawaFixedPoint}.
This yields new data, such as the bulk-defect OPE coefficients.
In this section, we consider one- and two-point functions of bulk operators, as well as bulk-defect two-point functions.

\subsection{Kinematically-fixed correlators}
\label{subsec:KinematicallyFixedCorrelators3}

We begin our investigation of bulk operators by examining the kinematically-fixed correlators, namely the one-point functions of bulk operators and the bulk-defect two-point functions.
The OPE coefficients associated with these correlators are the ones appearing in the block expansion of two-point functions of bulk operators, which are studied in detail in Section \ref{subsec:TwoPointFunctionsOfScalars}.

\subsubsection{Squared-scalar one-point function\enspace}
\label{subsubsec:SquaredScalarOnePointFunction}

In Section \ref{subsec:TheMagneticLine}, we discussed how the $\beta$-function of the defect coupling can be extracted through the one-point functions of $\Op^I$.
The corresponding coefficients $a_I$ are the first to appear in the bulk channel expansion of the two-point functions of $\Op^I$ in the presence of the line defect, as we will see in Section \ref{subsec:TwoPointFunctionsOfScalars}.
Other operators are also relevant in this expansion, the lowest-lying ones being the operators $\Op^2$ and $T^{IJ}$ defined in \eqref{eq:BulkSquaredScalar}.

These observables were previously computed for the $O(N)$ model in \cite{Cuomo:2021kfm}.
It is convenient to calculate the general one-point functions $\vvev{\Op^I \Op^J (x)}$ and choose the indices such that they correspond either to $\Op^2$ or $T^{IJ}$. 
Six diagrams contribute to this computation up to $\Om(\veps)$:
\begin{equation}
\begin{split}
\vvev{ \Op^I \Op^J (x) } =\ & \BulkSSOnePointOne\ +\ \BulkSSOnePointTwo\ +\ \BulkSSOnePointThree\ +\ \BulkSSOnePointFour\  \\
& +\ \BulkSSOnePointFive\ +\ \BulkSSOnePointSix\ + \Om(\veps^2)\,.
\end{split}
\label{eq:Bulk_phiIphiJ_Diagrams}
\end{equation}
The diagrams without fermionic contributions were calculated in \cite{Cuomo:2021kfm,Gimenez-Grau:2022ebb,Bianchi:2022sbz}, while the diagrams involving the fermionic loop cancel the wavefunction renormalization of $\Op^I$.
To compute the one-point function coefficient, we require the renormalization factor, anomalous dimension, and normalization factor of $\Op^2$ and $T^{IJ}$.
The anomalous dimensions for the GNY and NJLY models can be found in \cite{Fei:2016sgs}, while the others can be determined by computing the corrections to the propagator $\vev{ \Op^I \Op^J  (x_1) \Op^K \Op^L (x_2) }$.

For the cases of $N = 1,2,3$, the expressions for $\Op^2$ up to $\Om(\veps)$ are given by \cite{Fei:2016sgs,Zerf:2017zqi}
\begin{equation}
\begin{split}
Z_{\Op^2} &= 1 - \frac{\lambda (N+2) + 12 g^2 N_f}{48 \pi^2 \veps} +\Om(\veps^{-2})\,, \\
\gamma_{\Op^2} &= \frac{\lambda(N+2) + 12 g^2 N_f}{48 \pi^2} +\Om(\veps^2)\,,\\
n_{\Op^2} &= \frac{\sqrt{2 N}}{4 \pi^2} \biggl\lbrace
1
- \frac{ 2 (N+8) (2 \kappa_1 + (N-4)(1+\aleph)) + \kappa_2 \aleph ( N + 2 ) }{4 \kappa_1 (N+8)} \veps \\
& \phantom{=\ } + \Om(\veps^2)
\biggr\rbrace\,.
\end{split}
\label{eq:Bulk_O2_ZDn}
\end{equation}
For $T^{IJ}$, we obtain the following results:
\begin{equation}
\begin{split}
Z_{T} &= 1 - \frac{\lambda + 6 g^2 N_f}{24 \pi^2 \veps} + \Om(\veps^{-2})\,, \\
\gamma_{T} &= \frac{\lambda + 6 g^2 N_f}{24 \pi^2} + \Om(\veps^2)\,, \\
n_T &= \frac{1}{2\sqrt{2} \pi^2} \biggl\lbrace
1
- \frac{( N + 8 ) ( 2 \kappa_1 + ( N - 4 ) ( 1 + \aleph ) ) + \kappa_2 \aleph}{2 \kappa_1 (N+8)} \veps \\
& \phantom{=\ } + \Om(\veps^2)
\biggr\rbrace\,.
\end{split}
\label{eq:Bulk_T_ZDn}
\end{equation}
From these expressions, we can extract the one-point function coefficients $a_{\Op^2}$ and $a_{T}$:
\begin{equation}
\begin{split}
a_{\Op^2} &= - \frac{ ( N - 4 )( N + 8 ) }{ 2 \sqrt{2 N} \kappa_2 } + \Om(\veps)\,, \\
a_{T} &= - \frac{ ( N - 4 )( N + 8 )}{ 2 \sqrt{2} \kappa_2 } + \Om(\veps)\,.
\end{split}
\label{eq:Bulk_OnePointFunctions}
\end{equation}
The $\Om(\veps)$ terms are lengthy and can be found in the \textsc{Mathematica} notebook attached to \cite{Barrat:2023ivo}.

\subsubsection[Bilinear one-point \\ function\enspace]{Bilinear one-point function}
\label{subsubsec:BilinearOnePointFunction}

Another one-point function of interest is $\vvev{\Op^{aa}_{1,1} (x)}$, which appears in the block expansion of the two-point function $\vvev{\Op_{1,0}^a (x_1) \Op_{0,1}^b (x_2)}$.
Note that $\Op^{aa}_{1,1}$ is \textit{not} a conformal primary; rather, it is a conformal descendant of $\Op \sim \phi$.
This can be observed by its conformal dimension being $\Delta_I + 2 + \Om(\veps^2)$ \cite{Fei:2016sgs}.

This one-point function receives a contribution at $\Om(\sqrt{\veps})$:
\begin{equation}
\vvev{\Op^{aa}_{1,1} (x)} =\ \BulkFFOnePoint\ + \Om(\veps)\,.
\label{eq:OnePointBilinear_Diagrams}
\end{equation}
This diagram is new and can be expressed as
\begin{equation}
\BulkFFOnePoint\ = \frac{1}{n_{\Op^{aa}_{1,1}}} \frac{g h N_f}{16 \pi^3 |x|^3}\,,
\label{eq:OnePointBilinear_Diagram1}
\end{equation}
which translates to
\begin{equation}
a_{\Op^{aa}_{1,1}} =-  \frac{N_f \sqrt{(4-N) (N+8)}}{2 \sqrt{\kappa_1 \kappa_2} } + \Om(\veps)\,.
\label{eq:aaa}
\end{equation}

\subsubsection{Bulk-defect two-point functions\enspace}
\label{subsubsec:BulkDefectTwoPointFunctions}

As discussed above, the OPE coefficients arising in the defect channel of the block expansion for the two-point functions of bulk operators correspond to bulk-defect two-point functions.
In this section, we compute the correlators between one bulk scalar $\Op^I$ and one defect scalar, which can either be $\Oh^1$ or the tilt $\Oh^i$.

The form of the bulk-defect two-point functions can be found in \eqref{eq:BulkDefectTwoPoint}.
At $\Om(\veps)$, they are given by the following Feynman diagrams:
\begin{align}
\vvev{ \Op^I (x_1) \Oh^J (\tau_2) } =\ \BulkDefectTwoPointOne\ +\ \BulkDefectTwoPointTwo\ +\ \BulkDefectTwoPointThree\ + \Om(\veps^2)\,.
\label{eq:BulkDefectTwoPoint_Diagrams}
\end{align}

The two first diagrams have been computed in \cite{Gimenez-Grau:2022ebb}, while the last one corresponds to a simple self-energy integral.
By summing all the diagrams and accounting for the renormalization terms, we obtain the following expressions for the bulk-defect OPE coefficients:
\begin{equation}
b_{IJ} = 1 + \frac{3 (N - 4)(1-\log 2)}{2 \kappa_1} (3 \delta_{J1} + \delta_{Jj}) \veps + \Om(\veps^2)\,,
\label{eq:BulkDefect_bs}
\end{equation}
with $J=\lbrace 1,j \rbrace$ as usual. 

\subsection{Two-point functions of scalars}
\label{subsec:TwoPointFunctionsOfScalars}

In the presence of a defect, the two-point functions of bulk operators are no longer kinematically fixed.
Rather, they depend on two cross-ratios that encode the distances between the two operators and between the defect and each operator.
A detailed introduction to this setup can be found in Section \ref{subsec:ConformalDefects}.
In this section,we compute the correlator $\vvev{ \Op^I (x_1) \Op^J (x_2) }$ up to $\Om(\veps)$.
The result is then expanded in conformal blocks for both the defect and bulk channels.

\subsubsection{The perturbative result\enspace}
\label{subsubsec:ThePerturbativeResult}

There is only one diagram contributing to the connected two-point functions up to $\Om (\veps)$:
\begin{equation}
\vvev{ \Op^I (x_1) \Op^J (x_2) }_{\text{conn}} =\ \TwoPointNLO\ + \Om(\veps^2)\,.
\label{eq:BulkTwoPoint_Diagrams}
\end{equation}
After Wick contractions, this diagram translates to the following integral:
\begin{equation}
\TwoPointNLO\ = - \frac{\lambda h^2}{2 n_I} \int d\tau_3 \int d\tau_4\, X_{1234} + \Om(\veps^2)\,,
\label{eq:BulkTwoPoint_Diagram}
\end{equation}
where $X_{1234}$ corresponds to the conformal integral defined in \eqref{eq:X1234}.

While an expression for $X_{1234}$ in terms of cross-ratios has been known for a long time \cite{Usyukina:1994iw,Usyukina:1994eg}, no solution in terms of analytical functions exists in the literature for \eqref{eq:BulkTwoPoint_Diagram}.
Recent works \cite{Gimenez-Grau:2022ebb,Bianchi:2022sbz} have shown however that it is given by a derivative acting on a single bulk block.
Consequently, the integral in \eqref{eq:BulkTwoPoint_Diagram} can be expressed as a series expansion of the following form:
\begin{align}
\frac{ H(r,w) }{ |x_1^\perp| |x_2^\perp| } :=\ & - 128 \pi^4 \int d\tau_3 \int d\tau_4\, X_{1234} \notag \\
=\ & \frac{1}{ |x_1^\perp| |x_2^\perp| } \sum_{n,s=0}^{\infty} \sum_{m=0}^s r^{2n+s+1} w^{2m-s} \frac{ \left( \frac{1}{2} \right)_m \left( \frac{1}{2} \right)_n (-s)_m (s+1)_n}{ m! n! \left( \frac{1}{2}-2 \right)_m \left( \frac{1}{2}-2 \right)_{n+1} } \notag \\
& \times \left( \log 4r + H_{n+s} - H_{n+s+1/2} + H_{-1/2} \right)\,,
\label{eq:Hrw}
\end{align}
where $H_n$ refers here to harmonic numbers \cite{conway1998book}.

Taking into account the disconnected terms as well as the renormalization constant, we obtain
\begin{align}
F^{IJ} (r,w) =\ & \delta^{IJ} \chi^{-\Delta_I}+ \delta^{I1} \delta^{J1} a_I^2 \notag \\
& - (\delta^{IJ} + 2 \delta^{I1} \delta^{J1}) \frac{3 \veps (N - 4)}{4 \kappa_1} H(r,w ) + \Om(\veps^2)\,.
\label{eq:BulkTwoPoint_Result}
\end{align}
Here, $a_I$ corresponds to the one-point function coefficients given in \eqref{eq:OnePoint_PhiBulk}.
The general form of \eqref{eq:BulkTwoPoint_Result} remains the same as in the $O(N)$ model \cite{Gimenez-Grau:2022ebb,Bianchi:2022sbz},\footnote{This agreement is anticipated to be lifted at the next order in $\veps$ due to the inclusion of connected diagrams involving fermions, such as the spider diagram of Section \ref{subsec:TheFourPointFunctionAtNNLO}.} with the exception of additional fermionic contributions to the coefficient multiplying $H(r,w)$, which arise from the couplings evaluated at the fixed point.

In the following sections, we expand this expression in both the defect and bulk channels to extract valuable CFT data.
The OPE coefficients associated with the low-lying operators can then be compared to the explicit calculations conducted in the previous section.

\subsubsection{Defect channel\enspace}
\label{subsubsec:DefectChannel}

In the defect channel, the correlator $F^{IJ} (r,w)$ contains two types of operators: $O(N)$ singlets $\Oh^{S}_{s,n}$ and $O(N)$ vectors $\Oh^{V}_{s,n}$, with $s$ the number of single derivatives and $n$ the number of (Euclidean) d'Alembert operators.
Their conformal dimensions are defined through
\begin{equation}
\Dh_{\Oh^{S,V}_{s,n}} = \Dh_I + s + n + \gammah_{\Oh^{S,V}_{s,n}} \,.
\label{eq:DefectChannel_ScalingDims}
\end{equation}
In general, these operators are degenerate, except for the case $n=0$.

For higher values of $n$, it becomes necessary to solve a mixing problem.
This has been accomplished in \cite{Cuomo:2021kfm,Gimenez-Grau:2022czc} for the $O(N)$ model and it was reiterated in Section \ref{subsec:KinematicallyFixedCorrelators2} for the specific instance $n=1$ and $s=0$ to determine the anomalous dimension $\gammah_{s_{\pm}}$.
For more general cases $n,s>0$, we provide the \textit{averaged} CFT data.
We reserve solving the mixing problem for future work.

The decomposition of the correlator in the two symmetry channels is
\begin{align}
F^{IJ} (r,w) = \delta^{I1} \delta^{J1} \Fh_{S} (r,w) + (\delta^{IJ} - \delta^{I1} \delta^{J1}) \, \Fh_{V} (r,w)\,,
\label{eq:DefectChannel_DecompositionChannels}
\end{align}
with
\begin{equation}
\begin{split}
\Fh_S (r,w) &= a_I^2 + \xi^{-\Delta_I} - \frac{3(N-4)}{4 \kappa_1} \veps H(r,w) + \Om(\veps^2)\,, \\
\Fh_V (r,w) &= \xi^{-\Delta_I} - \frac{N-4}{4 \kappa_1} \veps H(r,w) + \Om(\veps^2)\,,
\end{split}
\label{eq:DefectChannel_FhSAndFhV}
\end{equation}
where we have defined
\begin{equation}
\xi := \frac{(1 - r w)(w - r)}{ r w } = \frac{x_{12}^2}{ |x_1^\perp| |x_2^\perp| }\,.
\label{eq:xi}
\end{equation}
This can be compared to the expansion in defect blocks.
An elegant expression for the decomposition of $H(r,w)$ was derived in \cite{Gimenez-Grau:2022ebb,Bianchi:2022sbz}:
\begin{equation}
H(r,w) = - \sum_{s=0}^{\infty} \frac{2}{1+2s} \left( \frac{2}{1+2s} - H_s  + H_{s - \frac{1}{2}} - \partial_{\hat{\Delta}} \right) \hat{f}_{s+1,s} (r,w)\:.
\label{eq:DefectChannel_Hrw}
\end{equation}
The derivative of the block corresponds to the anomalous dimension of the corresponding operator, and hence \eqref{eq:DefectChannel_Hrw} provides a straightforward method for extracting the defect CFT data.

Note that the constant terms in \eqref{eq:DefectChannel_FhSAndFhV} correspond to the defect identity given by $\hat{f}_{0,0} (r,w)$.
This leaves us with the factors $\xi^{- \Delta_I}$, whose expansion in defect blocks is well-known \cite{Lemos:2017vnx}.

Combining all the pieces, we are ready to extract the CFT data and we obtain the following OPE coefficients:
\begin{align}
b_{\Op \Oh^S_{s,0}} =\ &2^{\frac{s}{2}} \biggl\lbrace
1 + \frac{\veps}{ 4 \kappa_1  }
\biggl(
\frac{ 6 (N-4)}{(1 + 2s )^2} - \frac{\kappa_1 (1 + 2s) + 3 (N-4)}{(1 + 2s )^2} H_s \notag \\
&+ 3 (N-4) H_{s-\frac{1}{2}}
\biggr)
+ \Om(\veps^2)
\biggr\rbrace\,,
\label{eq:DefectChannel_BOOhS}
\end{align}
for $n=0$.
Setting $s = 0$, we observe that this result matches exactly the bulk-defect OPE coefficient $b_{I 1}$ given in \eqref{eq:BulkDefect_bs}.
As mentioned earlier, we only need to focus on the derivative term in \eqref{eq:DefectChannel_Hrw} to extract the anomalous dimension.
This leads to
\begin{align}
\Dh_{\Oh^S_{s,0}} = 1 + s + \frac{(N-4) (s-1)}{\kappa_1 (1+2 s)} \veps + \Om(\veps^2)\,.
\label{eq:DhS}
\end{align}
For $s = 0$, this corresponds exactly to $\Dh_1$, as shown in \eqref{eq:DefectScalar1_Delta}.
On the other hand, for $s = 1$, we expect it to yield the scaling dimension of the displacement $\Dh_\Disp = 2$.
Indeed, we observe that the $\Om(\veps)$ correction vanishes and thus matches \eqref{eq:Displacement_Delta}.

A similar analysis can be carried out for the vector channel, and we find the bulk-defect OPE coefficients
\begin{align}
b_{\Op \Oh^{V}_{s,0}} = 2^{\frac{s}{2}} \biggl\lbrace
1 + \frac{N-4}{4 \kappa_1 (1+2s) } \Bigg(
\frac{3+2s}{1+2s} +  2 s H_s+H_{s-\frac{1}{2}}
\Bigg) \veps
+ \Om(\veps^2)
\biggr\rbrace\,,
\label{eq:DefectChannel_bOOhV}
\end{align}
for $n=0$.
Again, this can be compared for $s = 0$ with $b_{I j}$ in \eqref{eq:BulkDefect_bs}, and we find a perfect match.
The anomalous dimensions are once again obtained from the derivative term in the expansion of $H(r,w)$, yielding the following conformal dimensions:
\begin{align}
\Dh_{\Oh^{V}_{s,0}} = 1 + s + \frac{s (N-4)}{\kappa_1 (1+2 s)} \veps +\Om(\veps^2)\,.
\label{eq:DefectChannel_DhOV}
\end{align}
As a check, we observe that, for $s = 0$, the $\Om(\veps)$ term vanishes and reproduces the protected dimension $\Dh_i = 1$ of the tilt operator.

\subsubsection[\\ Bulk channel\enspace]{Bulk channel}
\label{subsubsec:BulkChannel}

In the bulk channel, the $\Op^I \times \Op^J$ OPE involves $O(N)$ singlet operators $\Op^{S}_{\ell,n}$, with the first one being the operators $\Op^2$ and $T^{IJ}$ defined in \eqref{eq:BulkSquaredScalar}.
Beyond that, the operators belonging to the lowest-twist family can be expressed as 
\begin{equation}
\Op^{S}_{\ell,0} \sim \partial_{\mu_1} \cdots \partial_{\mu_\ell} \phi^2\,, \qquad \Op^{T}_{\ell,0} \sim \partial_{\mu_1} \cdots \partial_{\mu_\ell} \left( \phi^I \phi^J  - \frac{\delta^{IJ}}{N} \phi^2 \right)\,,
\label{eq:BulkChannel_Operators}
\end{equation}
where $\ell \geq 2$.
These operators correspond to higher-spin currents.
As a consequence, their conformal dimensions and OPE coefficients are protected up to $\Om(\veps)$ and given by the sum of the conformal dimension of $\Op^I$ and their spin.
More generally, the CFT data is given by 
\begin{equation}
\begin{split}
\Delta_{\Op^{S,T}_{\ell,0}} &= 2 \Delta_I + \ell + \Om(\veps^2)\,, \\
\lambda^{2}_{I J \Op^{S}_{\ell,0}}  &= \frac{2^{\ell + 1} (\Delta_I)_{\ell}^2}{N \ell! (2 \Delta_I + \ell - 1)_{\ell}} + \Om(\veps^2)\,, \\
\lambda^{2}_{IJ \Op^{T}_{\ell,0}} &= N \lambda^{2}_{I J \Op^{S}_{\ell,0}} + \Om(\veps^2)\,.
\end{split}
\label{eq:BulkChannel_OneDeltaTwoLambda}
\end{equation}
Operators with $n>0$ are not protected up and exhibit degeneracy.

The correlators can be decomposed into two symmetry channels, following
\begin{align}
F^{IJ} (r,w) = \delta^{IJ} F_{S} (r,w) + \left(\delta^{I1} \delta^{J1} - \frac{\delta^{IJ}}{N}\right) F_{T} (r,w)\,, 
\label{eq:BulkChannel_Decomposition}
\end{align}
with
\begin{equation}
\begin{split}
F_{S} (r,w) &= \xi^{-\Delta_I} + \frac{a_I^2}{N} - \frac{(N-4)(N-2)}{4 N \kappa_1} \veps H(r,w)\,,  \\
F_{T} (r,w) &= a_I^2 - \frac{N-4}{2\kappa_1} \veps H(r,w)\,.
\end{split}
\label{eq:BulkChannel_FSAndFT}
\end{equation}
The decomposition in bulk channel blocks presents additional challenges, as their closed form is currently unknown.
However, \eqref{eq:BulkTwoPoint_Result} shares a similar form with the correlator $\vvev{ \Op^I \Op^J }$ computed for the $O(N)$ model in \cite{Gimenez-Grau:2022ebb,Bianchi:2022sbz}, allowing us to reuse some of the known results in our analysis.
In particular, as mentioned earlier, $H(r,w)$ can be expressed in terms of bulk blocks as
\begin{equation}
\xi H(r,w) = (\partial_{\Delta} - 1 - \log 2) f_{2,0}(r,w) + \Om(\veps)\,,
\label{eq:BulkChannel_HrwAsDerivative}
\end{equation}
where $f_{\Delta,\ell}(r,w)$ represents the bulk conformal blocks given in \eqref{eq:ConformalBlocks_TwoPoint_Bulk}.

By examining \eqref{eq:BulkChannel_HrwAsDerivative}, we observe that $H(r,w)$ only corrects the operator $\Op^2$.
We can therefore directly use the results from \cite{Gimenez-Grau:2022ebb,Bianchi:2022sbz} for the CFT data of the other operators.
The remaining terms in \eqref{eq:BulkChannel_HrwAsDerivative} consist of a constant term corresponding to the bulk identity $f_{0,0} (r,w)$ and a term proportional to $\xi^{-\Delta_I}$, whose expansion in bulk blocks is provided in \cite{Gimenez-Grau:2022ebb}.

For the $O(N)$ singlets, 
we obtain the following one-point functions of twist-two operators ($n=0$):
\begin{align}
a_{\Op^{S}_{\ell,0}} =\ & - \frac{ 2^{\frac{1-\ell}{2}} (N-4) (N+8)}{\pi^{3/4} \kappa_2 \ell^2} \frac{\Gamma \left(\frac{1+\ell}{2} \right) \Gamma(1+\ell)^{1/2} }{\Gamma(\ell/2)^2 \Gamma (\ell + 1/2)^{1/2} } \biggl\lbrace
1 - \biggl(
\frac{2 \kappa_2\, a_I^{(1)}}{(N-4)(N+8)} \notag \\
&- \frac{(N-4)}{2 \kappa_{1}} \left( 2 H_{\frac{\ell-1}{2}} + H_{2 \ell} - 2 H_\ell -  H_{\ell-\frac{1}{2}} + 2 \log 2 \right)
\biggr) \veps \notag \\
&+ \Om(\veps^2)
\biggr\rbrace\,.
\label{eq:BulkChannel_aOS}
\end{align}
In the equation above, we denote as $a_I^{(1)}$ the $\Om(\veps)$ correction to $a_I$, which can be found in the \textsc{Mathematica} notebook attached to \cite{Barrat:2023ivo}.

The averaged CFT data for the higher-twist $n=1$ operators is given by
\begin{align}
\lambda_{I J \Op^{S}_{\ell,1}} a_{\Op^{S}_{\ell,1}} = - \frac{(1+\ell)^2 (N-4) (N+8)  \Gamma \left(\frac{1+\ell}{2}\right)^3}{64 \pi  \kappa_2 N\, \Gamma \left(\frac{\ell}{2}+2\right) \Gamma \left(\ell+\frac{3}{2}\right)} \veps + \Om(\veps^2)\,.
\label{eq:BulkChannel_lambdaIJOSaOS}
\end{align}
For the traceless symmetric operators, we obtain the following conformal dimensions and OPE coefficients:
\begin{align}
a_{\Op^{T}_{\ell,0}} =\ & - \frac{ 2^{\frac{1-\ell}{2}} (N-4) (N+8)}{\pi^{3/4} \kappa_2 \ell^2} \frac{\Gamma \left(\frac{1+\ell}{2} \right) \Gamma(1+\ell)^{1/2} }{\Gamma(\ell/2)^2 \Gamma (\ell + 1/2)^{1/2} } \notag \\
& \times
\biggl\lbrace
1
- \biggl(
\frac{4 \kappa_1 \kappa_2\, a_I^{(1)}}{(N-4)(N+8)} \notag \\
&- (N-4) ( 2 H_{\frac{l-1}{2}}-2 H_l+H_{2 l}-H_{l-\frac{1}{2}}+ 2 \log 2 )
\biggr) \frac{\veps}{2 \kappa_1} \notag \\
&+ \Om(\veps^2)
\biggr\rbrace\,, \label{eq:BulkChannel_aOT} \\
\lambda_{IJ \Om^{T}_{\ell,1}} a_{\Om^{T}_{\ell,1}} =\ &\frac{(1 + \ell)^2 (N+8)  \Gamma \left(\frac{1 + \ell}{2}\right)^3}{128 \pi\,  \Gamma \left(\frac{\ell}{2}+2\right) \Gamma \left(\ell+\frac{3}{2}\right)} \veps + \Om(\veps^2)\,.
  \label{eq:BulkChannel_lambdaOOOTaOT}
\end{align}
These expressions can be compared with the one-point functions of the operators $\Op^2$ and $T^{IJ}$ computed in Section \ref{subsec:KinematicallyFixedCorrelators3}, but first we require the bulk OPE coefficients $\lambda_{I J \Op^2}$ and $\lambda_{I J T}$.
These coefficients are easy to calculate and give
\begin{align}
\lambda_{I J \Op^2} 
&= \delta^{IJ} \sqrt{\frac{2}{N}} 
\biggl(
1
-\frac{\kappa_2 (N+2)  }{4 \kappa_1 (N+8)} \veps + \Om(\veps^2)
\biggr)\,, \label{eq:BulkChannel_lambdaIJO2} \\
\lambda_{I J T} &= \sqrt{2}
\biggl (1
-\frac{\kappa_2}{2 \kappa_1 (N+8)} \veps + \Om(\veps^2)
\biggr)\,.
\label{eq:BulkChannel_lambdaIJT}
\end{align}
With these OPE coefficients and the one-point functions, we can verify the block expansion and confirm that it correctly reproduces the conformal data of $\Op^2$ and $T^{IJ}$. 

\section{Summary and outlook}
\label{sec:SummaryAndOutlook3}

This chapter has been devoted to the investigation of the magnetic line defect CFT in Yukawa models characterized by $O(N)$ and $U(N_f)$ symmetries.
These theories are particularly relevant for understanding chiral and spontaneous symmetry breaking, and they find practical applications in condensed-matter systems such as graphene sheets.
Moreover, they manifest emergent supersymmetry, providing a fascinating avenue for the observation of supersymmetry in laboratories.

This setup represents a natural extension of the line defect CFT previously explored in $O(N)$ models.
The main difference is the presence of fermions in the bulk, which give rise to new fermionic excitations on the one-dimensional defect, akin to those discussed in Section \ref{subsec:MultipointCorrelators of Fermions} for $\Nm = 4$ SYM.
In Section \ref{sec:CorrelatorsOfDefectOperators}, we examined correlators of defect operators using the $\veps$-expansion, putting special emphasis on the new fermionic excitations.
We provided in particular explicit expressions for the $H$-diagrams involving four fermions and represented as
\begin{equation*}
\DefectFourFermionsHOne\,.
\end{equation*}
The case of two scalars and two fermions was also considered, and diagrams such as
\begin{equation*}
\DefectFFSSHOne
\end{equation*}
have been calculated.
These correlators serve as a valuable resource for extracting CFT data through conformal block expansions, and can potentially be used as input for the numerical bootstrap.\footnote{It should however be noted that, for values of $\veps$ between $0$ and $1$, unitarity is violated. This is discussed for instance in \cite{Hogervorst:2015akt,Ji:2018yaf}.}
Such an analysis was initiated in \cite{Gimenez-Grau:2022czc}, and it is likely that the models considered here are also contained in the results of Section 4.
It might be feasible to use the data provided here for steering the numerics, and hopefully solve specific models of interest.

In addition to correlators confined to the line defect, we studied how the excitations within the bulk are impacted by the defect.
Specifically, we directed our attention toward two-point functions of scalars, which exhibit non-trivial kinematics and depend on two conformal invariants.
For $O(N)$ models, this analysis has been done recently in \cite{Gimenez-Grau:2022ebb,Bianchi:2022sbz}.
Due to the similarity of the diagrammatic calculation, we were able to recycle a substantial portion of the results and extract novel CFT data specific to the fermionic CFTs.

In the following, we explore potential avenues for future research.
Having understood the two-point functions of bulk scalars, the next natural step is to consider two-point functions of bulk \textit{fermions} in the presence of the magnetic line.
This analysis introduces several conceptual and technical challenges, which we outline in Section \ref{subsec:TwoPointFunctionsOfFermions}.
Notably, the question of how to analytically continue fermions across dimensions has not yet been studied systematically, and a naive counting of tensor structures already reveals discrepancies between three and four dimensions.

This aspect is not specific to defects and can be analyzed within the bulk theory itself.
In Section \ref{subsec:TakeAWalkOnTheBulkSide}, we provide a preliminary analysis of a four-point function involving two scalars and two fermions.
It should be possible to continue this correlator from $d=4$ to $d=3$.
To the best of our knowledge, this particular type of correlator has never been investigated using the $\veps$-expansion.

\subsection{Two-point functions of fermions}
\label{subsec:TwoPointFunctionsOfFermions}

One motivation for studying the Yukawa action \eqref{eq:ActionYukawa} across different dimensions lies in the fact that fermions exhibit distinct characteristics in three and four dimensions.
To gain insights into how fermionic correlators in $d=3$ are captured by the $\veps$-expansion, we evaluate them using perturbation theory.
We provide here an outline of the computation involving two fundamental fermions in the presence of the line defect.

\subsubsection{The leading order\enspace}
\label{subsubsec:TheLeadingOrder}

We begin by calculating the leading order of the correlator.
The connected part is given by a single $Y$-diagram of order $\Om(\sqrt{\veps})$, which reads
\begin{equation}
\TwoPointFFLO\ = \frac{g h}{n_a} \delta^{ab} \left( \bar{s}_1 \int d\tau_3 \int dx_4\, \spd_1 I_{14} \spd_2 I_{24} I_{34}\, s_2 \right)\,.
\label{eq:BulkTwoFermions_LODiagram}
\end{equation}
This integral can be evaluated using the fermionic star-triangle identity \eqref{eq:StarTriangle}, yielding
\begin{equation}
\TwoPointFFLO\ = - \frac{g h}{8 \pi} \frac{\delta^{ab}}{x_{12}^2 (|x_1^\perp| + |x_2^\perp|)} \bar{s}_1 \left( \frac{ \sx_1 \sx_2 }{|x_1^\perp| |x_2^\perp|} + \mathds{1} \right) s_2\,.
\label{eq:BulkTwoFermions_LOResults}
\end{equation}
One may wonder why pinching $x_2 \to x_1$ does not seem to reproduce the result given in \eqref{eq:OnePointBilinear_Diagram1}.
In particular, it generates a quadratic divergence due to the presence of $x_{12}^2$ in the denominator.
The reason is that the star-triangle identity is strictly valid at $d=4$ and that in the course of the computation \eqref{eq:OnePointBilinear_Diagram1} we dropped the quadratic divergence to keep the next term.
Therefore, the right way to pinch \eqref{eq:BulkTwoFermions_LOResults} is to start with \eqref{eq:BulkTwoFermions_LODiagram} and use integration by parts to recover \eqref{eq:OnePointBilinear_Diagram1}.

\subsubsection{A challenging integral\enspace}
\label{subsubsec:AChallengingIntegral}

The order $\Om(\veps)$ contribution consists of one $H$-diagram connecting two scalar insertions on the line with the external fermions through two Yukawa vertices.
This results in a challenging finite \textit{ten}-dimensional integral that we can only partially solve at the moment.
After performing Wick contractions, the diagram yields
\begin{equation}
\begin{split}
\TwoPointFFNLO\ =\ & - \frac{g^2 h^2}{n_a} \delta^{ab} \biggl(
\bar{s}_1 \int d\tau_3 \int d\tau_4 \\
&\times \int d^4 x_5 \int d^4 x_6\,
\spd_1 I_{15} \spd_6 I_{56} \spd_6 I_{26} I_{35} I_{46}\, s_2
\biggr)\,.
\end{split}
\label{eq:BulkTwoFermions_NLODiagram1}
\end{equation}
This computation is reminiscent of the one carried out in $\Nm=4$ SYM, and outlined in Section \ref{subsec:AFirstGlimpseAtWeakCoupling} (see also \cite{Barrat:2020vch}).
In particular, closed-form expressions could not be determined for the diagrams
\begin{equation}
\TwoPointNoClosedFormOne \qquad \text{and} \qquad \TwoPointNoClosedFormTwo\,.
\label{eq:BulkTwoFermions_ExampleSYM}
\end{equation}

Let us see how far we can get.
One four-dimensional integral can be lifted by applying the fermionic star-triangle identity \eqref{eq:StarTriangle}, yielding
\begin{equation}
\TwoPointFFNLO\ \sim
\bar{s}_1\, \spd_1 \int d\tau_3 \int d\tau_4\, I_{24}\, J_{123;4}\, \sx_{24}\, s_2\,,
\label{eq:BulkTwoFermions_NLOIntegral}
\end{equation}
where we have introduced the tensor integral
\begin{equation}
J_{123;4} = \int d^4 x_5\, \sx_{54}\, I_{15} I_{25} I_{35} I_{45}\,.
\label{eq:BulkTwoFermions_J1234Definition}
\end{equation}
The integral $J_{123;4}$ can be solved by applying tensor decomposition. There are several automated tools available for this purpose, and here we use the \textsc{X} package \cite{Patel:2016fam}.
We find
\begin{equation}
J_{123;4} = \frac{2}{\phi_K} j_{123;4}\,,
\label{eq:BulkTwoFermions_J1234Result}
\end{equation}
where $\phi_K$ corresponds to the Kibble function defined as
\begin{equation}
\begin{split}
\phi_K &:= \Phi_{1234} + \Phi_{1324} + \Phi_{1423} + \Psi_{123} + \Psi_{124} + \Psi_{134} + \Psi_{234}\,,  \\
\Phi_{1234} &:= - \frac{1}{64 \pi^6 I_{12} I_{34}} \left( \frac{1}{I_{12}} + \frac{1}{I_{34}} - \frac{1}{I_{13}} - \frac{1}{I_{14}} - \frac{1}{I_{23}} - \frac{1}{I_{24}} \right)\,, \\
\Psi_{123} &:= - \frac{1}{64 \pi^6 I_{12} I_{13} I_{23}}\,.
\end{split}
\label{eq:BulkTwoFermions_Kibble}
\end{equation}
In addition, we introduced
\begin{equation}
j_{123;4} := \slashed{f}_{1234}\, X_{1234} + \slashed{g}_{123;4} Y_{123} + \slashed{g}_{124;3} Y_{124} + \slashed{g}_{134;2} Y_{134} + \slashed{g}_{234;1} Y_{234}\,.
\label{eq:BulkTwoFermions_j1234}
\end{equation}
The solutions for the $X$- and $Y$-integrals can be found in Appendix \ref{subsec:FiniteIntegrals}.  The prefactor function $\slashed{f}_{1234}$ can be expressed in terms of propagators as
\begin{equation}
\slashed{f}_{1234} = a_{1234}\, \slashed{x}_1 + a_{2341}\, \slashed{x}_2 + a_{3412}\, \slashed{x}_3 + (a_{4123}\, - 1) \slashed{x}_4\,,
\label{eq:BulkTwoFermions_Slashedf}
\end{equation}
where the coefficient functions are defined as
\begin{align}
a_{1234} :=\ & - \frac{1}{I_{23} I_{24} I_{34}} \biggl\lbrace 2
+ \frac{I_{24} I_{34}}{I_{14} I_{23}}
+ \frac{I_{23} I_{34}}{I_{13} I_{24}}
+ \frac{I_{23} I_{24}}{I_{12} I_{34}} - I_{34} \left( \frac{1}{I_{13}} + \frac{1}{I_{14}} \right) \notag \\
& - I_{24} \left( \frac{1}{I_{12}} + \frac{1}{I_{14}} \right)
- I_{23} \left( \frac{1}{I_{12}} + \frac{1}{I_{13}} \right)
 \biggr\rbrace\,.
 \label{eq:BulkTwoFermions_a1234}
\end{align}
The function $\slashed{g}_{123;4}$ can also be expressed in an elegant way:
\begin{equation}
\slashed{g}_{123;4} := b_{123;4} \slashed{x}_1
+ b_{231;4} \slashed{x}_2
+ b_{312;4} \slashed{x}_3
+ c_{123} \slashed{x}_4\,,
\label{eq:BulkTwoFermions_Slashedg}
\end{equation}
with
\begin{equation}
\begin{split}
b_{123;4} &:= \frac{1}{I_{23}} \biggl(
\frac{1}{I_{12}}
+ \frac{1}{I_{13}}
- \frac{1}{I_{23}}
+ \frac{1}{I_{24}}
+ \frac{1}{I_{34}}
- \frac{2}{I_{14}}
\biggr) \\
&\phantom{:=}
- \left( \frac{1}{I_{12}} - \frac{1}{I_{13}} \right) \left( \frac{1}{I_{24}} - \frac{1}{I_{23}} \right)\,,  \\
c_{123} &:= \frac{1}{I_{12}^2} + \frac{1}{I_{13}^2} + \frac{1}{I_{23}^2}
-2 \left( \frac{1}{I_{12} I_{13}} + \frac{1}{I_{12} I_{23}} + \frac{1}{I_{13} I_{23}} \right)\,.
\end{split}
\label{eq:BulkTwoFermions_bAndc}
\end{equation}
This is as far as we can go for now, and we are left with a difficult two-dimensional integral as well as a slashed derivative with respect to $x_1$.
We note however that numerical computation of this integral can now be efficiently performed.

For scalars, the breakthrough for solving \eqref{eq:BulkTwoPoint_Diagram} was achieved through the block expansion.
Similarly, in \cite{Barrat:2020vch} the diagrams \eqref{eq:BulkTwoFermions_ExampleSYM} were circumvented by the use of symmetry to extract the CFT data and obtain a series expansion of the correlator.
Perhaps there exists a similar formula to \eqref{eq:BulkChannel_HrwAsDerivative} for the fermionic case, where the integral can be expressed as a finite sum of bulk blocks.
The block expansions of two-point functions of fermions are not known yet, and even the tensor structure of the correlator has not been analyzed.
It would be interesting to better understand this correlator and explore the possibility of solving the integral \eqref{eq:BulkTwoFermions_NLOIntegral} by using the underlying symmetry.

\subsection{Take a walk on the bulk side}
\label{subsec:TakeAWalkOnTheBulkSide}

The computations in the previous section led us to consider a tangential setup: the four-point functions of two scalars and two fermions in the \textit{bulk} theory.
To conclude this chapter, we calculate this correlator up to the next-to-leading order, gaining a better understanding of the behavior of fermions as we deviate from the four-dimensional case.
We discuss in particular the open problem of continuing the tensor structures from $d=4$ to $d=3$.

\subsubsection{A perturbative result\enspace}
\label{subsubsec:APerturbativeResult2}

In the course of the computation above, we solved the diagram
\begin{equation*}
\FourPointSSFFNLOOne\,.
\end{equation*}
Even in the absence of a defect, this configuration already exhibits non-trivial characteristics across dimensions.
Surprisingly, correlators of this nature, involving a combination of two scalars and two fermions, seem to have never been studied in the context of the $\veps$-expansion.
We initiate here the analysis of these mixed correlators, while the outcome will be presented in \cite{Barrat:2023ta3}.

To gain insight into the behavior of fermionic degrees of freedom across dimensions, we use the Weyl fermions defined in \eqref{eq:Spinors_DefinitionChiXi}.
We limit ourselves for now to the correlator
\begin{equation}
\vev{ \chib^a \chi^b \Op^I \Op^J } = \frac{\delta^{ab} \delta^{IJ}}{x_{12}^{2 \Delta_a + 1} x_{34}^{ 2 \Delta_I }} \left( 
\Ids^{12}\, f_1 (u,v) + \Ids^{12}_{34}\, f_2 (u,v)
\right)\,,
\label{eq:BulkStory_TensorStructures}
\end{equation}
where the tensor structures $\Ids$ have been worked out with the help of the \textsc{Mathematica} notebook attached to \cite{Cuomo:2017wme}.
As usual, we suppress the dependence on the kinematic variables on the left-hand side when they become too cumbersome.
These functions are defined as
\begin{equation}
\begin{split}
\Ids^{12} &:= \bar{s}_1\, \sx_{12}\, s_2\,,  \\
\Ids^{12}_{34} &:= ( \bar{s}_1 \sigmab^\mu s_2 ) \frac{I_{34}}{2}
\biggl(
\frac{x_{23}^\mu}{I_{14}}
- \frac{x_{24}^\mu}{I_{13}}
+ \frac{x_{13}^\mu}{I_{24}}
- \frac{x_{14}^\mu}{I_{23}}
- \frac{x_{43}^\mu}{I_{12}}
- \frac{x_{12}^\mu}{I_{34}} \\
&\phantom{:=}\
+ 8 \pi^2 \eps^{\mu\nu\rho\sigma} x_{14}^{\nu} x_{32}^{\rho} x_{34}^{\sigma}
\biggr)\,.
\end{split}
\label{eq:BulkStory_I12AndI1234}
\end{equation}
Note that these tensor structures are in principle strictly applicable to the case $d=4$, even though $\veps$ appears in the scaling dimensions $\Delta_I$, $\Delta_a$.
We will come back to this problem shortly.

For now, let us compute the correlator \eqref{eq:BulkStory_TensorStructures} at next-to-leading order, where two diagrams contribute:
\begin{equation}
\vev{ \chib^a \chi^b \Op^I \Op^J } = \FourPointSSFFNLOOne\ +\ \FourPointSSFFNLOTwo\,.
\label{eq:BulkStory_Diagrams}
\end{equation}
Following \eqref{eq:BulkTwoFermions_J1234Result}, the correlator can be expressed as
\begin{equation}
\begin{split}
\frac{f_i (u,v)}{\kappa(\chi,\chib)} &=
a_i (\chi,\chib)
\frac{\log \chi \chib}{(1-\chi)(1-\chib)}
+
b_i (\chi,\chib)
\frac{\log (1-\chi)(1-\chib)}{\chi\chib} \\
&\phantom{=\ }
+ i\, c_i (\chi,\chib) \ell(\chi,\chib)\,,
\end{split}
\label{eq:BulkStory_Results}
\end{equation}
where we have defined the auxiliary functions
\begin{equation}
\begin{split}
\kappa(\chi,\chib) &:=
\frac{(\chi \chib)^2}{512 \pi^6 |\chi-\chib|^2}\,, \\
i |\chi-\chib| \ell (\chi,\chib) &:= \frac{\pi^2}{3}
+ \log (-x_1) \log (-x_2)
+ 2 (\text{Li}_2 (x_1) + \text{Li}_2 (x_2)) \\
&\phantom{:=\ }
+ \log ((1-\chi)(1-\chib)) \log \biggl(
- \frac{\chi +\chib - i |\chi-\chib| - 2 \chi\chib}{\chi +\chib + i |\chi-\chib|}
\biggr)\,, \\
x_1 &:= \frac{2}{2-\chi-\chib-i|\chi-\chib|}\,, \\
x_2 &:= (1-\chi)(1-\chib) x_1\,.
\end{split}
\label{eq:BulkStory_AuxiliaryFunctions}
\end{equation}
The cross-ratios $\chi$ and $\chib$ are related to $u$ and $v$ via \eqref{eq:SpacetimeCrossRatiosChi}.

\subsubsection{Tensor structures\enspace}
\label{subsubsec:TensorStructures}

Numerous interesting computations can be performed using the result above.
In particular, we can in principle expand it in blocks and extract the CFT data.
To that purpose, we need to answer the question asked earlier: what happens to the tensor structures across different dimensions?
Indeed, the spacetime indices of \eqref{eq:BulkStory_I12AndI1234} correspond to $\mu = 0,1,2,3$.
Moreover, the counting of all possible tensor structures does not align between $4d$ and $3d$.
In $4d$, each of the following configurations yields two tensor structures \cite{Cuomo:2017wme}:
\begin{align}
&\vev{ (1,0)\, (1,0)\, (0,0)\, (0,0) }\,, \vev{ (1,0)\, (0,1)\, (0,0)\, (0,0) }\,, \notag \\
&\vev{ (0,1)\, (1,0)\, (0,0)\, (0,0) }\,, \vev{ (0,1)\, (0,1)\, (0,0)\, (0,0) }\,,
\label{eq:BulkStory_TensorStructures4d}
\end{align}
summing up to a total of \textit{eight} structures.
In $3d$, in the absence of chirality, only the following structure is allowed \cite{Iliesiu:2015qra}:
\begin{equation}
\vev{ (1)\, (1)\, (0)\, (0) }\,,
\label{eq:BulkStory_TensorStructures3d}
\end{equation}
which can be decomposed into \textit{four} distinct structures.

At present, it is unclear how to properly continue these tensor structures across dimensions.
One observation is that an analytic continuation can be performed between the structures
\begin{equation}
\begin{matrix} \vev{ (1,0)\, (0,1)\, (0,0)\, (0,0) } \\ \vev{ (0,1)\, (1,0)\, (0,0)\, (0,0) } \end{matrix} \longleftrightarrow \vev{ (1)\, (1)\, (0)\, (0) }\,,
\label{eq:BulkStory_Analytic Continuation}
\end{equation}
while the missing structures \textit{do not arise} in the perturbative calculation.
At the Lagrangian level, this is due to the form of the vertices for Weyl fermions.
Perhaps this feature extends to all Lagrangian models where parity is preserved.
It would be interesting to put this statement on firmer grounds and see whether there exist certain models which remain confined to $4d$, while half of the tensor structures naturally vanish for those that can be continued to $3d$.

\chapter{Conclusions}
\label{chapter:Conclusions}

We conclude with a summary of the results, before providing suggestions for future research directions related to the themes encountered throughout this thesis.

\section{Summary of results}
\label{sec:SummaryOfResults}

This thesis has been dedicated to the study of defects in conformal field theories, exploring both the weak- and strong-coupling regimes.
The two canonical configurations of correlation functions have been considered: on one hand those involving bulk operators in the presence of a defect, and on the other one those involving excitations of the defect.
Chapters \ref{chapter:BootstrappingHolographicDefectCorrelators} and \ref{chapter:MultipointCorrelatorsInTheWilsonLineDefectCFT} focus on the case of the $\Nm=4$ SYM model, while Chapter \ref{chapter:LineDefectCorrelatorsInFermionicCFT} departs from supersymmetry to study magnetic line defects in fermionic models that might be observable experimentally.
We now provide a concise summary of the key findings from these chapters.

For the scenario of two bulk operators in the presence of a defect, we have introduced a novel dispersion relation, applicable to general defects.
A method has been presented to derive strong-coupling results in the Wilson-line defect CFT.
Subsequent works based on this method have expanded its applicability to non-Lagrangian models such as $6d$ $\Nm=(2,0)$ with surface defects \cite{Meneghelli:2022gps} and to the non-supersymmetric $O(N)$ model featuring a magnetic line \cite{Gimenez-Grau:2022ebb,Bianchi:2022sbz}.
We have also examined multipoint correlators of defect operators within the weak-coupling regime, using this pool of results to propose non-perturbative constraints identified as superconformal Ward identities.
Exploiting these constraints, we have initiated a strong-coupling computation of a five-point function using bootstrap techniques.
We expect that such progression from weak to strong coupling should apply to many other models, hopefully leading to new interesting results in the near future.
Furthermore, we have conducted a perturbative analysis of non-supersymmetric fermionic CFTs with a magnetic line, employing the $\veps$-expansion.
This has led to novel results for correlation functions of scalars and fermions, providing approximate descriptions of strongly-coupled $3d$ models and initiating a conformal-block analysis.
These theories are relevant in condensed-matter and high-energy physics, making the defect bootstrap program a promising avenue for future investigations in this context.

\section{Future directions}
\label{sec:FutureDirections}

Each chapter of this thesis was concluded with a detailed sketch of potential future prospects for the specific topics.
We focus now on more general goals relevant to the study of conformal field theory and conformal defects.

\bigskip

One interesting line of research involves combining the conformal bootstrap ideas with \textit{multipoint} correlators.
This should result in additional constraints on scalar OPE coefficients while giving access to \textit{spinning} operators, as illustrated in Section \ref{subsec:Applications} with the snowflake channel of the six-point functions.
Such a study is challenging due to the proliferation of cross-ratios.
In Section \ref{sec:MultipointCorrelatorsAtStrongCoupling}, we initiated an analytic bootstrap computation of a five-point function in the one-dimensional Wilson-line defect CFT, which may offer better control over the constraints due to slower cross-ratio growth.
If a systematic study of the superconformal blocks is undertaken, this model could serve as a valuable laboratory for the development of analytic and numerical multipoint bootstrap techniques, which have the potential to apply to higher-dimensional models.

Chapter \ref{chapter:BootstrappingHolographicDefectCorrelators} presents a dispersion relation for two-point functions of bulk operators in a defect CFT, analogous to the dispersion relation for four-point functions in bulk theories of \cite{Carmi:2019cub}.
A generalization of this relation to multipoint correlators and spinning operators would play an important role in the bootstrap program.
Additionally, correlators involving both bulk and defect operators could have interesting applications, especially in conjunction with the Ward identities anticipated in Section \ref{subsec:BulkDefectDefectCorrelators}.

A systematic framework for organizing the space of conformal defects is currently lacking.
The associated fundamental question is the following: given a particular bulk CFT, what is the space of admissible conformal defects?
While progress has been made in classifying defect CFTs in the context of free theories \cite{Lauria:2020emq,Behan:2020nsf,Behan:2021tcn}, there is still much ground to cover for interacting models.
The bootstrap program has been useful in classifying bulk theories, and perhaps a similar development can be achieved for defect theories.
A comprehensive classification would likely necessitate deriving the conformal blocks for additional setups that were not explored in this thesis, particularly those involving multiple bulk and defect operators.

Finite-temperature quantum field theories are closely connected to the study of defects \cite{Dowker:1978md,kennedy1982finite}.
When studying a QFT at finite temperature, the system is typically considered in thermal equilibrium, where the concept of temperature emerges from the presence of a heat bath interacting with the field theory.
We can interpret this heat bath as a \textit{boundary} defect that affects the behavior of the fields near the boundary.
Similarly to the line defects studied here, the boundary defect introduces additional constraints and boundary conditions on the fields, leading to modified dynamics compared to the zero-temperature case.
This viewpoint allows to draw concrete connections between the two fields, and many of the methods presented here should apply for instance to thermal $\Nm=4$ SYM.

While this thesis focuses on single defects in bulk theories, considering setups with two or more defects is also important.
However, little is known about \textit{multidefect} setups, and understanding the OPE of two defects is crucial for making progress.
The fusion of conformal defects in interacting theories was considered in \cite{SoderbergRousu:2023zyj}.
For $\Nm=4$ SYM, special geometries have been considered.
Exact results were obtained for coincident Wilson loops \cite{Beccaria:2020ykg,Galvagno:2021bbj,Muck:2021vyx}, while the case of two parallel circular Wilson loops was considered at weak coupling in \cite{Plefka:2001bu,Arutyunov:2001hs}.
These correlators exhibit complex kinematics and require new tools due to the complete breaking of conformal symmetry in general geometries.

\bigskip

To conclude, we propose several more exotic directions aimed at relating the field of conformal defects to neighboring areas of theoretical physics.
For instance, one-point functions of spin-$1$ fields in gauge theories have applications in the double-copy formalism, where they can be related to a metric \cite{Shi:2021qsb}.\footnote{We are thankful to J. Plefka for discussions about unpublished results.}
It would be interesting to explore whether the conformal bootstrap ideas can be combined with frameworks like the worldline formalism of \cite{Jakobsen:2022psy,Jakobsen:2023ndj}.
In the context of the $\Nm=4$ SYM theory, light-like non-supersymmetric Wilson loops are dual to scattering amplitudes \cite{Drummond:2007aua,Brandhuber:2007yx,Drummond:2007au,Drummond:2008vq,Arkani-Hamed:2010zjl,Alday:2011ga,Alday:2012hy,Chicherin:2022nqq,Chicherin:2022bov}, and this duality has been used for instance to study five-particle amplitudes \cite{Chicherin:2022zxo}.\footnote{We thank D. Chicherin for interesting discussions about this topic.}
A defect CFT can also be formulated for this case, with a $SO(3) \times SO(6)_R$ internal symmetry, and it would be interesting to see if a connection can be established between the techniques developed here and the computation of scattering amplitudes in $\Nm˜=4$ SYM.
Furthermore, the conformal window in quantum chromodynamics (QCD) has been a subject of long-standing interest.
The conformal window is spanned by the interval $N_f^\star \leq N_f \leq \frac{11}{2} N$, and while the upper edge $N_f = \frac{11}{2} N$ can be accessed with perturbative techniques, the lower bound $N_f^\star$  has been so far elusive and remains an open problem.
Lattice simulations have been the favored approach thus far (see \cite{DeGrand:2015zxa} and references therein) because non-Abelian gauge CFTs coupled to fermion matter are expected to sit well within the region allowed by the numerical bootstrap.
It is conceivable that the novel analytic bootstrap techniques could shed light on certain aspects of the conformal window, perhaps in combination with the modern incarnation of the $S$-matrix bootstrap \cite{Paulos:2016fap,Paulos:2016but,Paulos:2017fhb,Homrich:2019cbt,Kruczenski:2022lot,Albert:2022oes,Albert:2023jtd}.

\begin{appendices}

\chapter{Spinors}
\label{app:Spinors}

\section{Dirac and Weyl spinors}
\label{sec:DiracAndWeylSpinors}

We provide here our conventions for the spinor fields with respect to the action \eqref{eq:ActionYukawa} and the operators studied in Chapter \ref{chapter:LineDefectCorrelatorsInFermionicCFT}.
The Dirac fields with spinor indices $A=1, \ldots, 4$ can be decomposed into two basic \textit{Weyl} spinors as follows:
\begin{equation}
\psi^A = \binom{\chi_\alpha}{\xi^{\dagger \dot{\alpha}}}\,, \quad \bar{\psi}^A = \biggl( \xi^\alpha\ \chi^{\dagger}_{\dot{\alpha}} \biggr)\,,
\label{eq:Spinors_DefinitionChiXi}
\end{equation}
with $\alpha, \dot{\alpha} = 1,2$.
The Weyl spinors are two-component vectors defined as
\begin{equation}
\chi = \binom{\chi_1}{\chi_2}\,, \quad \xi^\dagger = \left( \xi_1\ \xi_2 \right)\,.
\label{eq:Spinors_WeylAsTwoComponentVectors}
\end{equation}
Spinors with an undotted index $\alpha$ transform as left-handed spinors $(1,0)$, while right-handed spinors $(0,1)$ are complex conjugates of the $(1,0)$ representation and carry a dotted index $\dot{\alpha}$.
The dot is here to indicate the transformation property, i.e.
\begin{equation}
\chi^\dagger_{\smash{\dot{\alpha}}} = (\chi_\alpha)^\dagger\,.
\label{eq:Spinors_TransformationProperty}
\end{equation}
Indices can be raised and lowered in the following way:
\begin{equation}
\chi^\alpha = \eps^{\alpha\beta} \chi_\beta = - \eps^{\beta\alpha} \chi_\beta\,,
\label{eq:Spinors_RaisingAndLowering}
\end{equation}
which implies
\begin{equation}
\chi^\alpha \xi_\alpha = - \chi_\alpha \xi^\alpha\,.
\label{eq:Spinors_RaisingAndLowering2}
\end{equation}
Here the tensor $\eps^{\alpha\beta}$ is defined as
\begin{equation}
\eps^{12} = - \eps^{21} = \eps_{21} = - \eps_{12} = +1\,,
\label{eq:Spinors_EpsMatrix}
\end{equation}
and a similar definition can be formulated for dotted indices:
\begin{equation}
\eps_{\smash{\dot{\alpha} \dot{\beta}}} = \eps_{\alpha\beta}\,, \qquad \eps^{\smash{\dot{\alpha} \dot{\beta}}} = \eps^{\alpha\beta}\,.
\label{eq:Spinors_RaisingAndLowering3}
\end{equation}

For external operators, it is convenient to use \textit{polarization spinors} $s^A$, $\bar{s}^{A}$ to avoid cluttering the indices.
We define
\begin{equation}
\psi (s,x) := s^A \psi^{A} (x)\,, \qquad \bar{\psi} (s,x) := \bar{s}^A \bar{\psi}^{A} (x)\,,
\label{eq:Spinors_PolarizationSpinors}
\end{equation}
and a similar definition holds for the Weyl fermions as well.

\section{Dirac and Pauli matrices}
\label{sec:DiracAndPauliMatrices}

The four-dimensional (Euclidean) Dirac matrices are defined in the chiral representation as
\begin{equation}
(\gamma^{\mu})^{AB} := \begin{pmatrix}
0 & (\sigma^\mu)_{\alpha \dot{\beta}} \\
(\bar{\sigma}^\mu)^{\dot{\alpha} \beta} & 0
\end{pmatrix}\,,
\label{eq:Spinors_GammaMatrices}
\end{equation}
where we have introduced
\begin{equation}
(\sigma^\mu)_{\alpha\dot{\beta}} := \left( \sigma^0\,, i \sigma^i \right)\,, \qquad (\bar{\sigma}^\mu)^{\dot{\alpha}\beta} := \left( \sigma^0\,, - i \sigma^i \right)\,.
\label{eq:Spinors_GammaInPauli}
\end{equation}
The Pauli matrices $\sigma^0\,, \sigma^i$ are defined as
\begin{equation}
\sigma^0 = \mathds{1}_2\,, \quad
\sigma^1 =  \renewcommand{\arraystretch}{.85}\begin{pmatrix} 0 & 1 \\ 1 & 0 \end{pmatrix}\,, \quad
\sigma^2 = \begin{pmatrix} 0 & -i \\ i & 0 \end{pmatrix}\,, \quad
\sigma^3 = \begin{pmatrix} 1 & 0 \\ 0 & -1 \end{pmatrix}\,.
\label{eq:Spinors_PauliMatrices}
\end{equation}
The $\gamma$ matrices satisfy the Euclidean Clifford algebra
\begin{equation}
\lbrace \gamma^\mu, \gamma^\nu \rbrace^{AB} = 2 \delta^{\mu\nu}\, \mathds{1}^{AB}\,,
\label{eq:Spinors_Clifford}
\end{equation}
and we can define an additional $\gamma$ matrix as
\begin{equation}
(\gamma^5)^{AB} := \begin{pmatrix}
\mathds{1}_\alpha^{\phantom{\alpha}\beta} & 0 \\
0 & - \mathds{1}^{\smash{\dot{\alpha}}}_{\phantom{\alpha}\smash{\dot{\beta}}}
\end{pmatrix}\,.
\label{eq:Spinors_Gamma5}
\end{equation}
This definition ensures that $\gamma^5$ satisfies the following properties:
\begin{equation}
\lbrace \gamma^5\,, \gamma^\mu \rbrace = 0\,, \qquad (\gamma^5)^\dagger = \gamma^5\,, \qquad (\gamma^5)^2 = \mathds{1}\,.
\label{eq:Spinors_Gamma5Properties}
\end{equation}

\chapter{Conformal blocks}
\label{app:ConformalBlocks}

In this appendix, we gather the conformal blocks necessary for the computations performed throughout this thesis.

\section{Two-point functions of half-BPS operators}
\label{sec:TwoPointFunctionsOfHalfBPSOperators}

We begin with the two-point functions of half-BPS operators in the presence of a half-BPS line defect studied in Chapter \ref{chapter:BootstrappingHolographicDefectCorrelators}.
We first provide the bosonic blocks, before giving the superblocks that are relevant for the bootstrap computations performed in Section \ref{sec:TwoPointFunctionsAtStrongCoupling}.

\subsection{Bosonic blocks}
\label{subsec:BosonicBlocks}

\subsubsection{Bulk channel\enspace}
\label{subsubsec:BulkChannel2}

\begingroup
\allowdisplaybreaks

For an exchanged operator of scaling dimension $\Delta$ and spin $\ell$, the conformal blocks in the \textit{bulk channel} expressed in \eqref{eq:DefectTwoPoint_BlockExpansionBulk2} take the form \cite{Billo:2016cpy,Isachenkov:2018pef,Liendo:2019jpu}
\begin{align}
f_{\Delta,\ell}(z,\bar{z}) 
=\ & \sum_{m=0}^\infty \sum_{n=0}^\infty \frac{4^{m-n}}{m! n!} \frac{\left( - \frac{\ell}{2} \right)_m^2 \left( \frac{2 - \ell - \Delta}{2} \right)_m}{\left( - \ell \right)_m \left( \frac{3 - \ell - \Delta}{2} \right)_m} \frac{\left( \frac{\Delta - 1}{2} \right)_n^2 \left( \frac{\Delta + \ell }{2} \right)_n}{\left( \Delta - 1 \right)_n \left( \frac{\Delta + \ell + 1}{2} \right)_n} \frac{\left( \frac{\Delta + \ell}{2} \right)_{n-m}}{\left( \frac{\Delta + \ell - 1}{2} \right)_{n-m}} \notag \\
& \times (1-z\bar{z})^{\ell - 2m} 
\ _4F_3 \left( \begin{array}{c}
    -n, -m , \frac{1}{2} , \frac{\Delta - \ell - 2}{2} \\
    \frac{2 - \Delta - \ell - 2n}{2}, 
    \frac{\Delta + \ell - 2m}{2} , 
    \frac{\Delta - \ell - 1}{2}
    \end{array} ; 1 \right) \notag  \\
& \times \left[ (1-z)(1-\bar{z}) \right]^{\frac{\Delta - \ell}{2} + m + n} \notag \\
& \times\ _2F_1 \left(  \begin{array}{c}
    \frac{\Delta + \ell}{2} -m+n , \frac{\Delta + \ell}{2} -m+n  \\
    \Delta + \ell - 2(m-n) 
    \end{array} ; 1 - z\bar{z} \right)\, .
\label{eq:ConformalBlocks_TwoPoint_Bulk}
\end{align}
As discussed in Section \ref{subsec:TwoPointFunctionsOfScalars}, no closed form is known for these infinite sums.

\endgroup

\subsubsection{Defect channel\enspace}
\label{subsubsec:DefectChannel2}

For an exchanged \textit{defect} operator of $1d$ scaling dimension $\Dh$ and spin $s$, the conformal blocks take a simple factorized form \cite{Billo:2016cpy,Isachenkov:2018pef,Liendo:2019jpu}:
\begin{equation}
 \fh_{\Dh,s} (z, \zb)
 = z^{\frac{\Dh-s}{2}} \zb^{\frac{\Dh+s}{2}} 
 \, _2F_1\left(\frac{1}{2},-s;\frac{1}{2}-s;\frac{z}{\zb}\right)\ _2F_1\left(\frac{1}{2},\Dh;\Dh+\frac{1}{2};z \zb\right)\, .
\label{eq:ConformalBlocks_TwoPoint_Defect}
\end{equation}

\subsection{Half-BPS superblocks}
\label{subsec:HalfBPSSuperblocks}

We give here the superconformal blocks relevant to the bootstrap computation of the correlators $\vvev{\Op_{\Delta_1} \Op_{\Delta_2}}$ presented in Chapter \ref{chapter:BootstrappingHolographicDefectCorrelators}.
The analysis in Section \ref{subsec:AnExample} shows that only the exchange of half-BPS operators $\Bm_{[0,\Delta,0]}$ needs to be considered.
These blocks have been derived in \cite{Liendo:2016ymz} for identical external operators, and the generalization to arbitrary operators gives
\begin{equation}
\begin{split}
\Gm^{\Delta_{12}}_{[0,\Delta,0]} (z, \zb; \sigma) =\ &
c_0\, h^{\Delta_{12}}_\Delta(\sigma) f^{\Delta_{12}}_{\Delta,0}(z,\zb)
+c_1\, h^{\Delta_{12}}_{\Delta-2}(\sigma) f^{\Delta_{12}}_{\Delta+2,2}(z,\zb) \\
&+ c_2\, h^{\Delta_{12}}_{\Delta-4}(\sigma) f^{\Delta_{12}}_{\Delta+4,0}(z,\zb)\,,
\end{split}
\label{eq:ConformalBlocks_TwoPoint_BulkHalfBPS}
\end{equation}
with the $R$-symmetry blocks given by
\begin{equation}
h_\Delta (\OR) = \OR^{-\Delta/2} \, _2F_1\left(-\frac{\Delta}{2},-\frac{\Delta}{2};-\Delta-1; \frac{\OR}{2} \right)\,.
\label{eq:ConformalBlocks_BulkRSymmetryBlocks}
\end{equation}
The coefficients are fixed using the Ward identities \eqref{eq:WardIdentities}.
They read
\begin{equation}
\begin{split}
c_0 &= 1\,, \\
c_1 &= \frac{(\Delta_{12}^2 - \Delta^2)(\Delta_{12}^2 - (\Delta+2)^2)}{128 \Delta (\Delta+1)^2(\Delta+3)}\,,  \\
c_2 &= \frac{(\Delta_{12}^2 - \Delta^2)^2 [(\Delta_{12}^2 - 4)^2-2\Delta^2 (\Delta_{12}^2+4)+\Delta^4]}{16384 (\Delta-2)(\Delta-1)^2\Delta^2(\Delta+1)(\Delta+2)(\Delta+3)}\,.
\end{split}
\label{eq:ConformalBlocks_TwoPoint_BulkHalfBPS_Coefficients}
\end{equation}

\section{Multipoint conformal blocks}
\label{sec:MultipointConformalBlocks}

We consider now the conformal blocks for multipoint correlators.
For the comb channel, an arbitrary number of external operators is first treated for the bosonic case, before moving our attention to the superconformal blocks of the five-point function considered in Section \ref{sec:MultipointCorrelatorsAtStrongCoupling}.
We then discuss the case of the snowflake channel for six-point functions.

\subsection{Comb channel}
\label{subsec:CombChannel}

The comb channel is introduced in Section \ref{subsec:CorrelatorsOfHalfBPSOperators}, following the expansion in blocks given in \eqref{eq:BlockExpansionComb}.

\subsubsection{Bosonic blocks\enspace}
\label{subsubsec:BosonicBlocks}

For an arbitrary number of external scalar operators, the conformal blocks in the comb channel have been derived in \cite{Rosenhaus:2018zqn} and read
\begin{equation}
\begin{split}
&f_{\Dh_1 \ldots \Dh_{n-3}} (\eta_1, \ldots, \eta_{n-3}) := \prod_{k=1}^{n-3}\, \eta_k^{\Dh_k} \\
& \times F_{K}\left[\begin{array}{c}
\left.  \Dh_1, \Dh_1+\Dh_2-\Dh_{\phi}, \ldots, \Dh_{n-4}+\Dh_{n-3}-\Dh_{\phi},\Dh_{n-3}\right.\\
 2 \Dh_1, \ldots, 2 \Dh_{n-3}\end{array} ; \eta_{1}, \ldots, \eta_{n-3} \right]\,,
 \end{split}
\label{eq:ConformalBlocks_Multipoint_Comb}
\end{equation} 
where the cross-ratios are defined in \eqref{eq:CrossRatiosComb}, and where the function $F_K$ is a multivariable hypergeometric function defined by the following expansion:
\small
\begin{align}
& F_{K} \Bigg[\begin{array}{c}a_{1}, b_{1}, \ldots, b_{k-1}, a_{2} \\ c_{1}, \ldots, c_{k}\end{array} ; \, x_1, \ldots, x_{k}\Bigg]  \notag \\
& \qquad = \sum_{n_{1}, \ldots, n_{k}=0}^{\infty} \frac{\left(a_{1}\right)_{n_{1}}\left(b_{1}\right)_{n_{1}+n_{2}}\left(b_{2}\right)_{n_{2}+n_{3}} \cdots\left(b_{k-1}\right)_{n_{k-1}+n_{k}}\left(a_{2}\right)_{n_{k}}}{\left(c_{1}\right)_{n_{1}} \cdots\left(c_{k}\right)_{n_{k}}} \frac{x_{1}^{n_{1}}}{n_{1} !} \cdots \frac{x_{k}^{n_{k}}}{n_{k} !}\,.
\label{eq:ConformalBlocks_GeneralizedHyper}
\end{align}
\normalsize
Here $(a)_{n}=\Gamma(a+n) / \Gamma(a)$ refers to the Pochhammer symbol.
These blocks are used in Section \ref{subsec:Applications} to check our results and initiate a bootstrap analysis.

\subsubsection{Five-point superblocks\enspace}
\label{subsubsec:FivePointSuperblocks}

We discuss here the superblocks of the five-point function $\vev{\Oh_1 \Oh_1 \Oh_1 \Oh_1 \Oh_2}$, for which a bootstrap analysis is initiated in Section \ref{sec:MultipointCorrelatorsAtStrongCoupling}.
Two comb channels (\textit{symmetric} and \textit{asymmetric}) are relevant and presented in Figure \ref{fig:OPESymmetricAsymmetric}.
The expressions are lengthy, and thus we refer to \cite{Barrat:2023ta3} for the full expressions.
Here we limit ourselves to sketching the method followed for deriving them.

Similarly to \eqref{eq:ConformalBlocks_TwoPoint_BulkHalfBPS}, a superblock for two exchanged operators $\Oh_1$ and $\Oh_2$ consists of a linear combination of bosonic conformal blocks.
The spacetime blocks $f_{\Dh_1, \Dh_2} (\chi_1, \chi_2)$, defined in \eqref{eq:ConformalBlocks_Multipoint_Comb}, depend on two scaling dimensions corresponding to the content of the multiplets.
Similarly, the $R$-symmetry blocks $h_{[a_1, b_1],[a_2,b_2]} ( \lbrace r,s,t \rbrace )$ carry Dynkin labels fixed by the two exchanged multiplets.
Although they are not known in closed form, it is easy to generate them case by case using the appropriate Casimir equations.

Concretely, the superblocks take the form
\begin{equation}
\Gm_{\Oh_1, \Oh_2} ( \lbrace \chi; r,s,t \rbrace ) = \sum\, c_{[a_1,b_1],[a_2,b_2]}^{\Dh_1, \Dh_2}\, h_{[a_1,b_1],[a_2,b_2]} ( \lbrace r,s,t \rbrace ) f_{\Dh_1, \Dh_2} (\chi_1, \chi_2)\,,
\label{eq:ConformalBlocks_FivePointForm}
\end{equation}
where the range of the sum is fixed by the scalar content of the multiplets, and by selection rules that can be analyzed using, for instance, the package \textsc{LieART} \cite{Feger:2019tvk}.
Superconformal blocks are fixed by superconformal symmetry, and remarkably the conjectured multipoint Ward identities \eqref{eq:MultipointWardIdentities} fix \textit{all} the coefficients for \textit{all} the blocks, enabling the CFT data analysis of Section \ref{subsec:BootstrappingTheCorrelator}.

\subsection{Snowflake channel}
\label{subsec:SnowflakeChannel}

We now discuss the conformal blocks for the snowflake channel of six-point functions defined in \eqref{eq:BlockExpansionSnowflake}.
For our purposes, we write a series expansion of the form\footnote{We thank L. Quintavalle for sharing notes with us on this topic.}
\begin{equation}
f_{\Delta_1 \Delta_2 \Delta_3}\left(z_1\,, z_2\,,z_3 \right) = z_1 ^{\Delta_1} z_2^{\Delta_2} z_3^{\Delta_3} \sum_{n_1, n_2, n_3} c_{n_1, n_2, n_3} z_1^{n_1} z_2^{n_2} z_3^{n_3}\,,
\label{eq:ConformalBlocks_Multipoint_Snowflake}
\end{equation}
where we only need coefficients $c_{n_1, n_2, n_3}$ for low values of $n_1$, $n_2$, $n_3$.\footnote{In \cite{Fortin:2020zxw}, a different Taylor expansion is used with a closed-form expression for the corresponding coefficients $c_{n_1, n_2, n_3}$.}
The cross-ratios are defined in \eqref{eq:CrossRatiosSnowflake}.
It is easy to determine the coefficients up to an overall normalization by applying the Casimir equations on the blocks order by order (see Appendix D of \cite{Barrat:2022eim}).
We find the form given in \eqref{eq:Snowflake_Expansion}.

\chapter{Integrals}
\label{app:Integrals}

This appendix is dedicated to the computation of the integrals needed to derive the results presented in this thesis. We provide one- and two-loop Feynman integrals, as well as boundary integrals specific to line defects.

\section{Preliminaries}
\label{sec:Preliminaries}

The first section of this appendix contains the basic background useful for expressing the integrals throughout this thesis.
We begin by discussing regularization schemes, before introducing the harmonic polylogarithms, which form a natural basis of functions in perturbative results.

\subsection{Regularization schemes}
\label{subsec:RegularizationSchemes}

We present the two regularization methods that are used throughout this thesis, which are the \textit{point-splitting} and \textit{dimensional} regularization schemes.

\subsubsection{Point splitting\enspace}
\label{subsubsec:PointSplitting}

The gist of the point-splitting regularization scheme is to insert a small distance $\veps$ between two coincident points.
In this way, the coincident points are \textit{split} and one may start with a finite integral to keep control of the divergences.

In practice, the regularization scheme is introduced by setting
\begin{equation}
I_{11} := \frac{1}{4 \pi^2 \veps^2}
\label{eq:Definition_PointSplitting}
\end{equation}
in finite expressions.
Note that in this scheme, both logarithmic and unphysical quadratic divergences become apparent.
An example of a $\log$-divergent integral in point-splitting regularization is given in \eqref{eq:YDivergent_PointSplitting}.

\subsubsection{Dimensional regularization\enspace}
\label{subsubsec:DimensionalRegularization}

Dimensional regularization on the other hand makes use of the fact that divergences arise in integrals at a certain number of spacetime dimensions.
The idea is to set $d = 4 - \veps$ and to extract the divergences by considering the limit $\veps \to 0$.
Note that the divergences are scheme-invariant, i.e.,  point-splitting and dimensional regularization yield the \textit{same} divergences when identifying $\log \veps^2 \sim \frac{1}{\veps}$.

The finite parts may however differ, but the differences should cancel when considering the full \textit{observable} quantity.
One artifact often encountered is the combination
\begin{equation}
\aleph := 1 + \log \pi + \gamma_{\text{E}}\,,
\label{eq:Aleph}
\end{equation}
with $\gamma_{\text{E}} = 0.57722 \ldots$ the \textit{Euler--Mascheroni} constant.
An example of an integral in dimensional regularization is provided in \eqref{eq:YDivergent_DimReg}.

\subsection{Harmonic polylogarithms}
\label{subsec:HarmonicPolylogarithms}

We find it convenient to express perturbative results using \textit{harmonic polylogarithms} (HPLs).
These single-variable functions are denoted as $H_{\vec{a}} (\chi)$, with $\vec{a} := (a_1, \ldots, a_n)$ and $a_k = \lbrace -1, 0, 1 \rbrace$.
They are defined recursively via \cite{Remiddi:1999ew,Maitre:2005uu,Maitre:2007kp}
\begin{equation}
H_{a_1, \ldots, a_n} (\chi) = \int_0^\chi dt\, f_{a_1} (t)\, H_{a_2, \ldots, a_n} (t)\,,
\label{eq:HPL_Definition}
\end{equation}
with
\begin{equation}
f_{-1} (\chi) := \frac{1}{1+\chi}\,, \quad f_{0} (\chi) := \frac{1}{\chi}\,, \quad f_{1} (\chi) := \frac{1}{1-\chi}\,.
\label{eq:HPL_BasisFunctions}
\end{equation}
The recursion can be started by setting $H_{\vec{a}} (\chi)$ with $\vec{a}$ collapsed to zero argument to $1$.
The \textit{transcendental weight} of the HPLs corresponds then to the number of elements in $\vec{a}$.
For instance, the transcendental functions of weight $1$ are
\begin{equation}
\begin{split}
H_{-1} (\chi) &= \log (1 + \chi)\,, \\
H_0 (\chi) &= \log \chi\,, \\
H_1 (\chi) &= - \log (1-\chi)\,.
\end{split}
\label{eq:HPL_Weight1}
\end{equation}
These functions form a useful basis when working in perturbation theory.
Note that we often use the shorthand notation $H_{\vec{a}} := H_{\vec{a}} (\chi)$ throughout this appendix and the thesis.

\section{Feynman integrals}
\label{sec:FeynmanIntegrals}

We provide results for four-dimensional massless Feynman integrals.
These integrals have been studied extensively both in Euclidean and Minkowski spacetimes.
We focus first on finite Feynman integrals, before showing integrals that exhibit logarithmic divergences.

\subsection{Finite integrals}
\label{subsec:FiniteIntegrals}

We give here a list of finite four-dimensional massless integrals at one- and two-loop and show how to extend these results to fermionic integrals.

\subsubsection{$4d$ integrals\enspace}
\label{subsubsec:4dIntegrals}

The first integral that we look at is a well-known four-point massless integral that arises in computations involving a $\phi^4$ vertex \cite{Usyukina:1994iw,Usyukina:1994eg}:\footnote{For the modern notation, see for instance \cite{Beisert:2002bb,Drukker:2008pi}.}
\begin{align}
X_{1234} = \XIntegral\ &:= \int d^4 x_5\, I_{15} I_{25} I_{35} I_{45} \notag \\
&\phantom{:}= \frac{I_{12} I_{34}}{16\pi^2}\ \chi \chib\, D( \chi, \chib)\,,
\label{eq:X1234}
\end{align}
where we have defined the \textit{Bloch-Wigner} function \cite{bloch1978applications}
\begin{equation}
D(\chi, \chib) := \frac{1}{\chi - \chib} \left( 2 \Li_2 (\chi) - 2 \Li_2 (\chib) + \log \chi \chib\, \log \frac{1-\chi}{1-\chib} \right)\,,
\label{eq:BlochWigner}
\end{equation}
with the cross-ratios $\chi$ and $\chib$ defined in \eqref{eq:SpacetimeCrossRatiosChi}.
Note that the Bloch-Wigner function is crossing symmetric and, as such, satisfies the identity
\begin{equation}
D(\chi, \chib) = D(1-\chi\,, 1-\chib)\,.
\end{equation}

From the $X$-integral, it is possible to obtain the three-point integral defined as
\begin{align}
Y_{123} = \YIntegral\ & := \int d^4 x_4\, I_{14} I_{24} I_{34} \notag \\
&\phantom{:}= \lim\limits_{\tau_4 \to \infty} I_{34}^{-1} X_{1234} = \frac{I_{12}}{16\pi^2} \chi \chib\, D( \chi, \chib)\,,
\label{eq:Y123}
\end{align}
where now the cross-ratios are defined as the following limits of the four-point variables defined in \eqref{eq:SpacetimeCrossRatiosuv}:
\begin{equation}
\chi \chib := \lim\limits_{\tau_4 \to \infty} u = \frac{I_{13}}{I_{12}}\,, \qquad (1-\chi)(1-\chib) := \lim\limits_{\tau_4 \to \infty} v = \frac{I_{13}}{I_{23}}\,.
\label{eq:CrossRatiosY}
\end{equation}

It is often useful to consider derivatives of the $Y$-integral, for which the following identities hold:
\begin{equation}
\begin{split}
\partial_{1 \mu} Y_{123} &= - (\partial_{2 \mu} + \partial_{3 \mu}) Y_{123}\,, \\[.6em]
\partial_1^2\, Y_{123} &= - I_{12} I_{13}\,, \\
\left( \partial_1 \cdot \partial_2 \right) Y_{123} &= \frac{1}{2} \left( I_{12} I_{13} + I_{12} I_{23} - I_{13} I_{23} \right)\,.
\end{split}
\label{eq:YIdentities}
\end{equation}
All these identities are elementary to prove, using integration by parts and the scalar Green's equation given in \eqref{eq:GreensEq}.

Another four-point integral of interest is the $H$-integral, defined as
\begin{equation}
H_{12,34} = \HIntegral\ := \int d^4 x_5\, I_{15} I_{25}\, Y_{345}\,.
\label{eq:H1234}
\end{equation}
This integral is \textit{eight}-dimensional and is in principle a two-loop integral.
It satisfies
\begin{equation}
H_{12,34} = H_{21,34} = H_{12,43} = H_{34,12}\,.
\label{eq:H_Permutations}
\end{equation}
The $H$-integral is \textit{not} conformal, and as a matter of fact, no exact solution has been obtained yet to the best of our knowledge.
However, it appears in one-loop computations only associated with \textit{derivatives}, which effectively turn it into a four-dimensional integral.
Using the identities given in \eqref{eq:YIdentities}, it is easy to obtain the following relations:
\begin{equation}
\begin{split}
\partial_1^2 H_{12,34} &= - I_{12}\, Y_{134}\,, \\
\left( \partial_1 \cdot \partial_2 \right) H_{12,34} &= \frac{1}{2} \left[ I_{12} ( Y_{134} + Y_{234} ) - X_{1234} \right]\,.
\end{split}
\label{eq:H_Identities}
\end{equation}
Unfortunately, as far as we know there is no identity known for $\left( \partial_1 \cdot \partial_3 \right) H_{12,34}$.

These identities can be used to determine integrals that arise in the computation of Feynman diagrams. For example, diagrams with two scalar-scalar-gluon vertices give rise to the expression
\begin{equation}
F_{12,34} := \frac{\pd_{12} \cdot \pd_{34}}{I_{12} I_{34}} H_{12,34}\,,
\label{eq:F1234_Definition}
\end{equation}
with $\pd_{ij}^\mu := \pd_i^\mu - \pd_j^\mu$.
The result can be elegantly expressed in terms of $X$- and $Y$-integrals \cite{Beisert:2002bb}:
\begin{equation}
\begin{split}
F_{12,34} =\ & \frac{X_{1234}}{I_{13}I_{24}} - \frac{X_{1234}}{I_{14}I_{23}} + \left( \frac{1}{I_{14}} - \frac{1}{I_{13}} \right) Y_{134} + \left( \frac{1}{I_{23}} - \frac{1}{I_{24}} \right) Y_{234} \\
& + \left( \frac{1}{I_{23}} - \frac{1}{I_{13}} \right) Y_{123} + \left( \frac{1}{I_{14}} - \frac{1}{I_{24}} \right) Y_{124}\,.
\end{split}
\label{eq:F1234_Result}
\end{equation}

For fermionic integrals, there exists a special star-triangle identity \cite{d1971theoretical,Vasiliev:1981yc,Baxter:1997tn}:
\begin{equation}
\VertexFermionFermionScalar = \spd_1 \spd_3\, Y_{123} = - 4 \pi^2 \sx_{12} \sx_{23} I_{12} I_{13} I_{23}\,,
\label{eq:StarTriangle}
\end{equation}
which is used throughout Chapter \ref{chapter:LineDefectCorrelatorsInFermionicCFT} to compute integrals in Yukawa models.

\subsubsection{$1d$ limits\enspace}
\label{subsubsec:1dLimits}

We now consider the $1d$ limit of the Feynman integrals considered in the previous section, i.e.,  the limit where all the external points are aligned.
Such expressions are needed for the computations of $1d$ correlators of operators inserted on line defects.

For the $X$-integral \eqref{eq:X1234}, we find
\begin{align}
X_{1234}^{1d} &= \lim\limits_{\chib \to \chi} X_{1234}^{4d} = \frac{I_{12} I_{34}}{16\pi^2} \chi^2 D(\chi)\,,
\label{eq:X1234_1d}
\end{align}
where we define the $1d$ Bloch-Wigner function $D(\chi)$ as
\begin{equation}
\chi^2 D(\chi) := \lim\limits_{\chib \to \chi} \chi \chib\, D(\chi, \chib) = - \frac{2\chi}{1-\chi} \left(\chi H_0 (\chi) - (1-\chi) H_1 (\chi) \right)\,.
\label{eq:BlochWigner_1d}
\end{equation}
Note that in $1d$, $X_{1234}$ is one degree of transcendentality lower than in the higher-$d$ result \eqref{eq:BlochWigner}. 

Similarly, the $Y$-integral reads in the $1d$ limit
\begin{equation}
Y_{123} = - \frac{I_{12}}{8\pi^2} \left( \frac{\tau_{12}}{\tau_{13}} \log \frac{\tau_{12}}{\tau_{13}} + \frac{\tau_{23}}{\tau_{13}} \log \frac{\tau_{23}}{\tau_{13}} \right)\,,
\label{eq:Y123_1d}
\end{equation}
where we note that the polylogarithms dropped as for $X_{1234}$.

We now consider the $1d$ limit of the $H$-integral and the related integrals.
If we choose the conformal frame $(\tau_1\,, \tau_2\,, \tau_3\,, \tau_4) \to (0\,, \chi\,, 1\,, \infty)$, \eqref{eq:F1234_Result} becomes
\begin{equation}
F_{12,34} = \frac{1}{8\pi^2} \left( \log (1-\chi) + i \pi (3-\chi) \right)\,.
\label{eq:F1234_1d}
\end{equation}
Note that this integral is \textit{not} conformal, but since we encounter it in conformal correlators, we are only interested in its expression in the conformal frame.

In combination with the star-triangle identity \eqref{eq:StarTriangle}, all these results can be used to compute the fermionic integrals of Section \ref{subsec:FourPointFunctions}.
Note also that certain finite limits of the $H$-integral can be computed in $1d$.
For instance, we have
\begin{equation}
H_{12,13} = \frac{1}{512 \pi^6 \chi} \bigl(
2 H_{1,0,0} - H_{0,1,0} + 5 H_{0,0,1}
\bigr)\,,
\label{eq:H1213_1d}
\end{equation}
which has to be understood in the context of four-point functions.
The transcendentality weight of this expression is $3$, suggesting as anticipated above that $H$-integrals are more complicated objects than one-loop integrals.
There exists a beautiful identity relating the $H$-integrals together in the conformal frame:
\begin{equation}
\chi H_{12,13} + (1-\chi) H_{13,23} - \chi(1-\chi) H_{12,23} = \frac{3 \zeta_3}{512 \pi^6}\,.
\label{eq:H_BeautifulIdentity}
\end{equation}

We conclude by giving the result for a two-loop ladder integral called the \textit{kite} integral, which in $1d$ is given by the expression \cite{Usyukina:1994eg,Drummond:2006rz}
\begin{align}
K_{13,24} &= \KiteIntegral := \frac{1}{I_{13}} \int d^4 x_5\, I_{15} I_{25} I_{35}\, X_{1345} \notag \\
&= \frac{I_{13} I_{24}}{128 \pi^4 \chi (1-\chi)}
\bigl(
5 ( H_{0,0,1} + H_{0,1,0} + 2 H_{0,1,1} ) \notag \\
&\phantom{=\ }+ 2 ( H_{1,0,0} + 2 H_{1,1,0} )
+ 7 H_{1,0,1} + 6 H_{1,1,1}
\bigr)\,.
\label{eq:KiteIntegral}
\end{align}
This integral is encountered in the fermion loop \eqref{eq:TwoLoopsSpiderDiagram}.
Note that the result \eqref{eq:H1213_1d} can also be seen as a special limit of \eqref{eq:KiteIntegral}.

\subsection{Divergent integrals}
\label{subsec:DivergentIntegrals}

In this section, we focus on certain \textit{pinching} limits of the bulk integrals, i.e.,  cases where two (or more) external points coincide.
All the cases of interest can be written in terms of the $Y$-integral, which we present using the two regularization schemes of Section \ref{subsec:RegularizationSchemes}: point-splitting and dimensional regularization.

The $Y$-integral becomes $\log$-divergent when two points are pinched together:
\begin{equation}
Y_{112} = Y_{122} = \IntegralYOneTwoTwo\ = - \frac{I_{12}}{16\pi^2} \left( \log \frac{I_{12}}{I_{11}} - 2 \right)\,.
\label{eq:YDivergent_PointSplitting}
\end{equation} 
In dimensional regularization, we have
\begin{equation}
Y_{112} = - \frac{I_{12}}{8 \pi^2} \left( \frac{1}{\varepsilon} + \aleph + \log x_{12}^2 + \Op(\varepsilon) \right)\,.
\label{eq:YDivergent_DimReg}
\end{equation}

The pinching of two points in the $X$-integral can be related to $Y$-integrals:
\begin{equation}
X_{1233} = \IntegralXOneTwoThreeThree\ = \frac{1}{2} ( I_{13} Y_{223} + I_{23} Y_{113} ) - \frac{I_{13} I_{23}}{32 \pi^2} \log \frac{I_{13} I_{23}}{I_{12}^2}\,.
\label{eq:X1233}
\end{equation}
One can also consider the limit where the external points coincide pairwise.
This is equivalent to a two-point integral with doubled propagators. In this case, we have
\begin{equation}
X_{1122} = \IntegralXOneOneTwoTwo\ = 2 I_{12} Y_{112} - \frac{I_{12}^2}{8 \pi^2}\,.
\label{eq:X1122}
\end{equation}

Note that the $F$-integral also reduces to a nice expression when two external points coincide:
\begin{equation}
\begin{split}
F_{13,23} =\ \IntegralFOneThreeTwoThree\ =\ &  \frac{1}{2}
\left(\frac{Y_{113}}{I_{13}} + \frac{Y_{223}}{I_{23}} \right) + Y_{123} \left( \frac{1}{I_{13}} + \frac{1}{I_{23}} - \frac{2}{I_{12}} \right) \\
&+ \frac{1}{32 \pi^2} \log \frac{I_{13} I_{23}}{I_{12}^2}\,.
\end{split}
\label{eq:F1323}
\end{equation}

\section{Defect integrals}
\label{sec:DefectIntegrals}

This section is dedicated to \textit{defect} integrals, i.e.,  integrals that arise in perturbation theory when a field couples to the defect.
We first describe elementary integrals that do not involve a bulk vertex, before moving our attention to the less trivial case of boundary integrals involving a four-dimensional vertex.
We also consider the case of integrals involving one and two couplings on the line.

\subsection{Elementary integrals}
\label{subsec:ElementaryIntegrals}

The most general elementary integral that can be considered is
\begin{equation}
\int_{\tau_2}^{\tau_3} d\tau_4\, I_{14} = \DefectVertexOnePointScalarCenterBis\ = \frac{1}{|x_1^\perp|} \left\lbrace
\tan^{-1} \left( \frac{\tau_{12}}{ |x_1^\perp| } \right)
- \tan^{-1} \left( \frac{\tau_{13}}{ |x_1^\perp| } \right)
\right\rbrace\,.
\label{eq:ElementaryIntegral}
\end{equation}
This result can be used for instance to derive the value of all the vertices listed in \eqref{eq:DefectVertices_OnePoint}.
These integrals are particularly useful to derive the expression given in \eqref{eq:TwoLoopsXYDiagramOneb}.

\subsection{One-loop integrals}
\label{subsec:OneLoopIntegrals}

We now consider the defect integrals encountered in the one-loop diagrams of Section \ref{sec:CorrelationFunctionsOfScalarsAtWeakCoupling}.
We first show how to solve the $T$-integrals, before listing the results for $U$-integrals, which are relevant for correlators of unprotected operators.

\subsubsection{$T$-integrals\enspace}
\label{subsubsec:TIntegrals}

The $T$-integrals are defined via\footnote{This class of integrals also appears in \cite{Kiryu:2018phb}, where they are defined slightly differently and labelled as $B_{ij;kl}$.}
\begin{equation}
T_{ij,kl} := \pd_{ij} \int_{\tau_k}^{\tau_l} d\tau_m\, \veps(\tau_i\, \tau_j\, \tau_m)\, Y_{ijm}\,.
\label{eq:T_Definition}
\end{equation}
It is the integral along the line of a $Y$-vertex of the type scalar-scalar-gluon.
Here $\veps(\tau_i\, \tau_j\, \tau_k)$ encodes the change of sign due to the path ordering and the color trace.
It is formally defined as
\begin{equation}
\veps(\tau_i\, \tau_j\, \tau_k) := \sgn \tau_{ij}\, \sgn \tau_{ik}\, \sgn \tau_{jk}\,.
\label{eq:eps_Definition}
\end{equation}
An important identity satisfied by $\veps(\tau_i\, \tau_j\, \tau_k)$ is
\begin{equation}
\pd_i\, \veps(\tau_i\, \tau_j\, \tau_k) = 2 ( \delta( \tau_{ij} ) - \delta( \tau_{ik} ) )\,,
\label{eq:eps_Identity}
\end{equation}
which allows to extract the divergences from the integrals encountered for instance in the diagrams
\begin{align}
\TDivergent &= \int_{\tau_k}^{\tau_l} d\tau_m\, \veps(\tau_i\, \tau_j\, \tau_m)\, \pd_{ij} Y_{ijm} = T_{ij,kl} - 4 Y_{iij}\,.
\label{eq:TDivergent}
\end{align}
Note that this identity is valid only for $\tau_k < \tau_i < \tau_j < \tau_l$ and must be adapted for other cases.

The $T$-integrals are finite and can be evaluated.
When the range of integration is the entire line, the integral is easy to perform and results in
\begin{equation}
T_{ij,(-\infty)\infty} = - \frac{I_{ij}}{12}\,.
\label{eq:T_Basic}
\end{equation}
This follows from
\begin{equation}
T_{ij,(-\infty)i} = - T_{ij,ij} = T_{ij,(-\infty)i} = - \frac{I_{ij}}{12}\,.
\label{eq:T_Identity1}
\end{equation}
Another direct consequence of \eqref{eq:T_Identity1} is
\begin{equation}
\left. T_{ij,kl} \right|_{\tau_k < \tau_i < \tau_j < \tau_l} = - \frac{I_{ij}}{12} - T_{ij,(-\infty)k} - T_{ij,l\infty}\,.
\label{eq:T_Identity2}
\end{equation}
There also exists a beautiful relation between $T$- and $Y$-integrals:
\begin{equation}
I_{ik} T_{jk,ki} + I_{jk} T_{ik,jk} = - \frac{I_{ik} I_{jk}}{12} + I_{ik} I_{jk} \left( \frac{1}{I_{ik}} + \frac{1}{I_{jk}} - \frac{2}{I_{ij}} \right) Y_{ijk}\,.
\label{eq:T_Identity3}
\end{equation}

The $T$-integrals can be performed explicitly for different orderings of the $\tau$'s, and here we give the results assuming $\tau_1 < \tau_2 < \tau_3 < \tau_4$ \cite{Kiryu:2018phb}:
\begin{equation}
\begin{split}
T_{12,34} &= \frac{I_{12}}{8\pi^2} \left( 4 L_R \left( \frac{\tau_{12}}{\tau_{14}} \right) - 4 L_R \left( \frac{\tau_{12}}{\tau_{13}} \right) - C_{123} + C_{124} \right)\,, \\
T_{34,12} &= \frac{I_{34}}{8\pi^2} \left( 4 L_R \left( \frac{\tau_{34}}{\tau_{14}} \right) - 4 L_R \left( \frac{\tau_{34}}{\tau_{24}} \right) - C_{341} + C_{342} \right)\,, \\
T_{14,23} &= \frac{I_{14}}{8\pi^2} \left( 4 L_R \left( \frac{\tau_{24}}{\tau_{14}} \right) - 4 L_R \left( \frac{\tau_{34}}{\tau_{14}} \right) - C_{412} - C_{143} \right)\,, \\
T_{23,41} &= \frac{I_{23}}{8\pi^2} \left( - 4 L_R \left( \frac{\tau_{23}}{\tau_{13}} \right) - 4 L_R \left( \frac{\tau_{23}}{\tau_{24}} \right) - C_{234} - C_{123} \right)\,,
\end{split}
\label{eq:T_Results}
\end{equation}
where we defined the help function
\begin{equation}
C_{ijk} := - 32 \pi^4\, \tau_{ij} (\tau_{ik} + \tau_{jk})\, Y_{ijk}\,.
\label{eq:T_HelpFunction}
\end{equation}
The \textit{Rogers dilogarithm} $L_R(\chi)$ is defined as \cite{rogers1907function}
\begin{equation}
L_R (\chi) := - \frac{1}{2} ( H_{1,0} - H_{0,1} )\,,
\label{eq:RogersDilog}
\end{equation}
with $H_{\vec{a}} := H_{\vec{a}} (\chi)$.

It is easy to take pinching limits of the integrals given above.
For example, we have
\begin{equation}
T_{12,23} = \frac{I_{12}}{8\pi^2} \left( 4 L_R \left( \frac{\tau_{12}}{\tau_{13}} \right) - \frac{2\pi^2}{3} + C_{123} \right) + Y_{112}\,,
\label{eq:T_PinchingLimit}
\end{equation}
using the fact that $L_R(1) = \frac{\pi^2}{6}$.
All the other pinching limits can be obtained in the same way.

\subsubsection{$U$-integrals\enspace}
\label{subsubsec:UIntegrals}

Correlation functions involving the scalar coupling to the line defect generate another type of boundary integral.
We denote them by $U_{i,jk}$, and they are defined as
\begin{equation}
U_{i,jk} =\ \UIntegral\ := \int_{\tau_j}^{\tau_k} d\tau_n\, I_{in}\,,
\label{eq:U_Definition}
\end{equation}
where the interrupted scalar line should be connected to the orange dots on the line.
These integrals are easy to perform explicitly and give
\begin{equation}
U_{i,jk} = - \frac{1}{4\pi^2} \left( \frac{1}{\tau_{ij}} - \frac{1}{\tau_{ik}} \right)\,,
\label{eq:U_Result}
\end{equation}
which is valid for both the orderings $\tau_a < \tau_i < \tau_j$ and $\tau_i < \tau_j < \tau_a$.
Special cases, such as $\tau_a = \tau_i$, can be obtained by regularizing the divergences in the integral using point-splitting.

At the next-to-leading order, we also encounter the integrals
\begin{equation}
U^{(2)}_{ij,kl} := \int_{\tau_k}^{\tau_l} d\tau_m\, I_{im} U_{j,nl}\,.
\label{eq:U2_Definition}
\end{equation}
We compute them explicitly for three different configurations, from which the other relevant ones can be obtained by taking the appropriate limits.
Assuming the ordering $\tau_1 < \tau_2 < \tau_3 < \tau_4$, we get
\begin{equation}
\begin{split}
U^{(2)}_{12,34} =\ & \frac{I_{12}}{4 \pi^2} \left\lbrace - \frac{\tau_{12}^2 \tau_{34}} {\tau_{23}\tau_{24}\tau_{14}} + \frac{1}{\tau_{23}\tau_{24}} \left((\tau_2^2+ \tau_3 \tau_4)  \log \frac{\tau_{23}\tau_{14}}{\tau_{13}\tau_{24}} \right. \right.  \\
& \left. \left. \phantom{\frac{I_{12}}{4 \pi^2}\ }+\tau_2(\tau_3 +\tau_4)\log \frac{\tau_{13}\tau_{24}}{\tau_{23}\tau_{14}} + \tau_{12} \tau_{34} \right) \right\rbrace \,,  \\
U^{(2)}_{34,12} =\ & \frac{I_{34}}{4 \pi^2 } \left\lbrace - \frac{\tau_{34}^2 \tau_{12}} {\tau_{13}\tau_{14}\tau_{23}} + \frac{1}{\tau_{13}\tau_{23}} \left( (\tau_3^2+ \tau_1 \tau_2) \log \frac{\tau_{14}\tau_{23}}{\tau_{13}\tau_{24}}  \right. \right.  \\
& \left. \left. \phantom{\frac{I_{12}}{4 \pi^2}\ }+\tau_3(\tau_1 +\tau_2) \log \frac{\tau_{13}\tau_{24}}{\tau_{14}\tau_{23}} + \tau_{34} \tau_{12} \right) \right\rbrace \,,\\
U^{(2)}_{14,23} =\ & \frac{I_{14}}{4 \pi^2} \left\lbrace \frac{\tau_{14}^2 \tau_{23}} {\tau_{12}\tau_{13}\tau_{34}} + \frac{1}{\tau_{12}\tau_{13}} \left( (\tau_1^2+ \tau_2 \tau_3)  \log \frac{\tau_{12}\tau_{43}}{\tau_{42}\tau_{13}} \right. \right.  \\
&  \left. \left. \phantom{\frac{I_{12}}{4 \pi^2}\ }+\tau_1 (\tau_2 +\tau_3) \log \frac{\tau_{42}\tau_{13}}{\tau_{12}\tau_{43}} + \tau_{41} \tau_{23} \right) \right\rbrace\,.
\end{split}
\label{eq:U2_Results}
\end{equation}

\end{appendices}

\cleardoublepage
\phantomsection
\addcontentsline{toc}{chapter}{\uppercase{Bibliography}}
\bibliography{auxi/biblio}
\bibliographystyle{./auxi/JHEP}

\makeatletter
\@openrightfalse
\makeatother

\includepdf[pages=-]{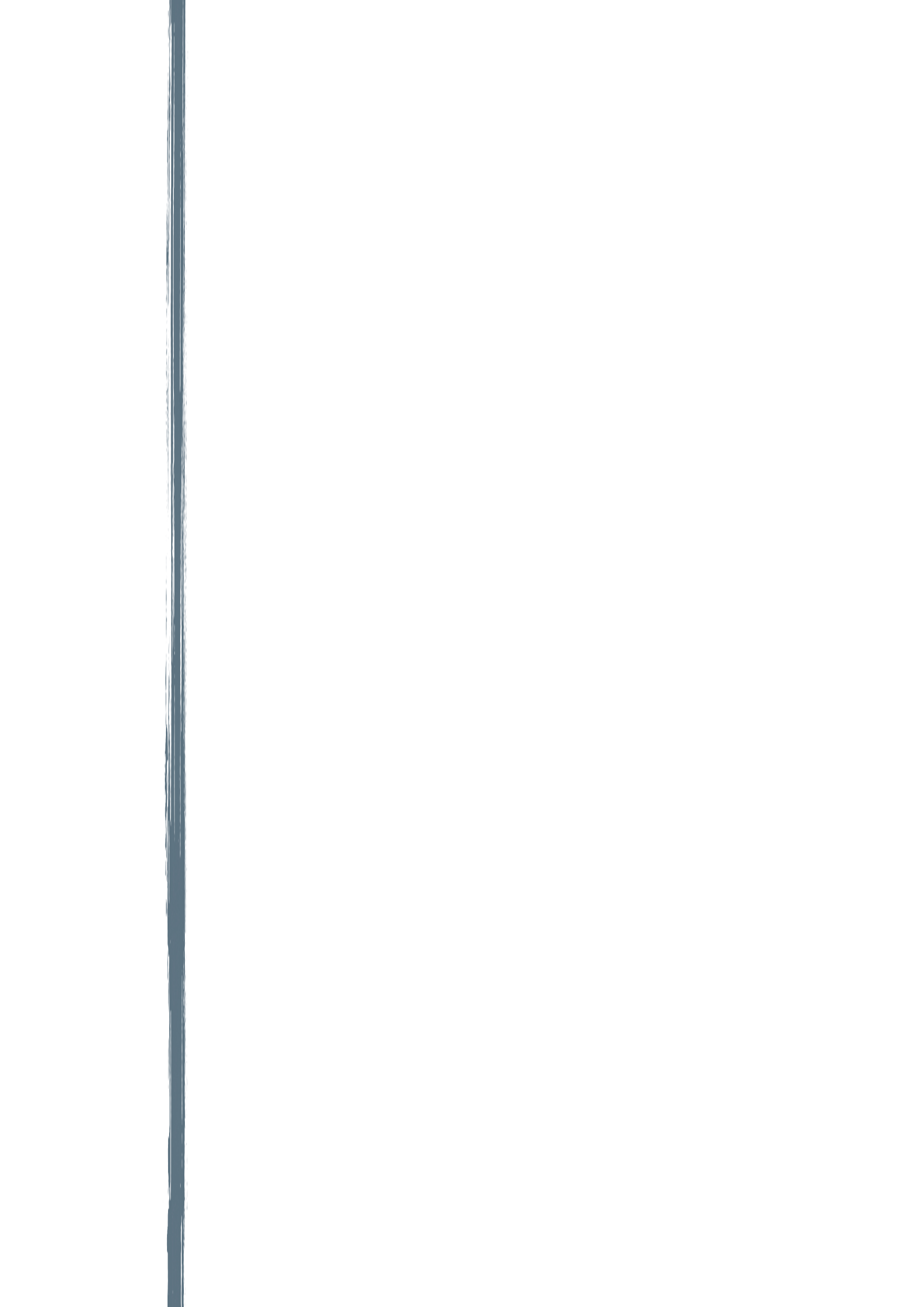}

\end{document}